\documentclass[12pt,dvips,twoside,a4paper]{report}

\usepackage{graphics,epsfig}
\usepackage{float}
\usepackage{subfigure}
\usepackage{geometry}
\usepackage{fancyhdr}
\usepackage[switch]{lineno}
\restylefloat{figure}
\restylefloat{table}

\newcommand{\ptjet}{p_{ T}}

\newcommand{\phijet}{\phi}
\newcommand{\rapjet}{y}

\newcommand{\akt}{\hbox{anti-${k_t}$} }
\newcommand{\njet}{N^{\rm jet}}

\include{my_definitions}
\begin{document}
\thispagestyle{empty}
\setlength{\topmargin}{0mm}
\setlength{\voffset}{0mm}

\begin{figure}[!htb]
\begin{center} 
\includegraphics[width=1\textwidth]{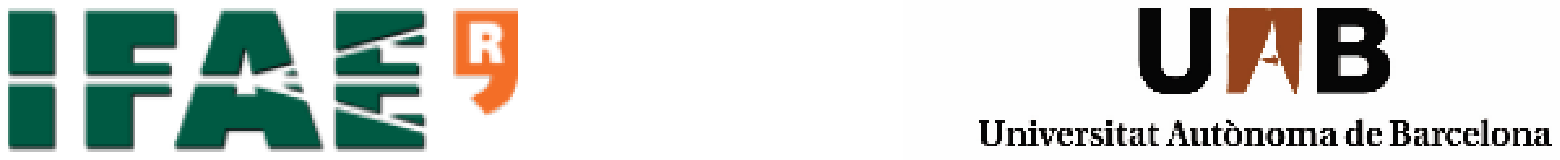}
\end{center}
\end{figure}

\begin{center}

\LARGE
\textbf{Study of Inclusive Jet Production and Jet Shapes in proton-proton collisions at $\mathrm{\sqrt{s} = 7}$~TeV using the ATLAS Detector\footnote{Ph.D. Dissertation}} \\
\vspace{1.5cm}

\large
Francesc Vives Vaqu\'e \\
Institut de F\'isica d'Altes Energies \\
Universitat Aut\`onoma de Barcelona  \\
Departament de F\'isica  \\
Edifici Cn, Campus UAB     \\
E-08193 Bellaterra (Barcelona) \\

\vspace{1.5cm}
\large
Barcelona, July 2011\\

\vspace{1.5cm}
\large
\textit{supervised by}  \\
Prof. Mario Mart\'inez P\'erez \\
ICREA / Institut de F\'isica d'Altes Energies /\\
Universitat Aut\`onoma de Barcelona
\end{center}

\newpage
$\,$
\thispagestyle{empty}
\clearpage
\newpage
\thispagestyle{empty}
\normalsize
\vspace{10cm}
\noindent The studies presented in this thesis are part of the following publications:
\begin{itemize}
\item The ATLAS Collaboration, Study of jet shapes in inclusive jet production in $pp$ collisions at $\sqrt{s} = 7$~TeV 
using the ATLAS detector, Physical Review D 83, 052003 (2011)

\item The ATLAS Collaboration, Measurement of inclusive jet and dijet cross sections in proton-proton collisions 
at 7 TeV centre-of-mass energy with the ATLAS detector, European Physical Journal C71 1512 (2011)

\end{itemize}
\vspace{2cm}
and of the following public ATLAS notes:
\begin{itemize}
\item The ATLAS Collaboration, Jet Shapes in ATLAS and Monte Carlo modeling, ATL-PUB-2011-010 (2011)  
\item The ATLAS Collaboration, Measurement of inclusive jet and dijet cross sections in proton-proton collision data at 7~TeV 
centre-of-mass energy using the ATLAS detector, ATLAS-CONF-2011-047 (2011)
\item The ATLAS Collaboration, Properties and internal structure of jets produced in soft proton-proton collisions 
at $\sqrt{s} = 900$~GeV, ATLAS-CONF-2010-018 (2010)
\end{itemize}

\newpage
$\,$
\thispagestyle{empty}
\clearpage
\newpage

\newpage
\chapter*{Acknowledgments}
\thispagestyle{empty}

First of all, I would like to thank my supervisor Mario Mart\'inez. His broad knowledge of physics, 
involvement in the analysis I was performing and love for rigorous work have played a crucial role in this thesis. 
I am grateful to Enrique F\'ernandez, Matteo Cavalli and Martine Bosman, who welcomed me at IFAE, and 
from whom I admire their passion for physics.

I also want to thank the ATLAS collaboration. In particular, I want to mention Fabiola Gianotti, the 
spokesperson, Kevin Einsweiler and Jon Butterworth, the Standard Model group conveners, and Tancredi Carli and Richard
Teuscher, conveners of the Jet/EtMiss group.

I started the jet shapes analysis with Monica D'Onofrio. I would like to thank her for this collaboration,  
but mainly for her multiple advises and friendship. Other postdocs have helped me during my Ph.D., and 
I am grateful to all of them: Bilge Demirk\"oz, with whom I analyzed the first ATLAS data, Christophe Ochando, 
Ilya Korolkov, Luca Fiorini and Sebastian Grinstein.

Estel, Evelin and Valerio have been excellent office-mates, and actually more than that.  
Just to put an example, they were the first friends to visit my daughter Gemma when she was born! 
I also want to thank my other colleagues at IFAE, in particular Machi, Jordi and Volker.  

I am grateful to Juan and Sonia, for their encouragement during the Ph.D., and to all my friends working at 
CERN: Ila, Peppe, Alessio, Ciccio, Simone, Lidia, Delo, Paola...

Finally, I want to thank the most important people in my life: my family. 
I want to particularly mention my wife, Maria Laura, not only for moving to life with me close to CERN, 
but mainly for her daily support. 

\pagenumbering{roman}
\newpage
$\,$
\thispagestyle{empty}
\clearpage
\newpage

\tableofcontents
\newpage
$\,$
\clearpage
\newpage

\addcontentsline{toc}{chapter}{List of Figures}
\listoffigures
\newpage
$\,$
\clearpage
\newpage

\addcontentsline{toc}{chapter}{List of Tables}
\listoftables
\newpage
$\,$
\clearpage
\newpage

\chapter*{Introduction}
  \addcontentsline{toc}{chapter}{Introduction}

\pagenumbering{arabic}

The Standard Model (SM) is the theory that provides the best description of the properties and interactions 
of elementary particles. The strong interaction between quarks and gluons is described by the 
Quantum Chromodynamics (QCD) field theory. Jet production is the high-$\ptjet$ process with the largest cross section at 
hadron colliders. The jet cross section measurement is a fundamental test of QCD and it is
sensitive to the presence of new physics. It also provides information on the parton distribution
functions and the strong coupling. 
One of the fundamental elements of jet measurements is the proper understanding of the energy flow
around the jet core and the validation of the QCD description contained in the event generators, such as
parton shower cascades, and the fragmentation and underlying event models. Jet shapes observables are
sensitive to these phenomena and thus very adequate to this purpose. The first measurement of the inclusive 
jet cross section in $pp$ collisions at $\sqrt{s} = 7$~TeV delivered by the LHC was done using 
an integrated luminosity of 17~nb${}^{-1}$ recorded by the ATLAS experiment. 
The measurement was performed for jets with $\ptjet > 60$~GeV and $|y| < 2.8$, reconstructed with the $\akt$ 
algorithm with radius parameters $R = 0.4$ and $R = 0.6$.

This Ph.D. Thesis presents the updated measurement of the inclusive jet cross section using the full 2010 data set, 
corresponding to 37~pb${}^{-1}$ collected by ATLAS. Jets with $\ptjet > 20$~GeV and $|y| < 4.4$ are considered in this 
analysis. The measurement of the jet shapes using the first 
3~pb${}^{-1}$ is also presented, for jets with $\ptjet > 30$~GeV and $|y| < 2.8$. 
Both measurements are unfolded back to the particle level. The inclusive jet cross section measurement is compared to 
NLO predictions corrected for non-perturbative effects, and to predictions from an event generator that 
includes NLO matrix elements. Jet shapes measurements are compared to the predictions from several LO 
matrix elements event generators. 

The contents of this Thesis are organized as follows: Chapter~1 contains a description 
of the strong interaction theory and jet phenomenology.
The LHC collider and the ATLAS experiment are described in Chapter~2. The inclusive jet cross section 
measurement is described in detail in Chapter~3, and the jet shapes measurements in Chapter~4. Additional 
comparison of the jet shapes measurement to Monte Carlo event generator predictions are shown in Chapter~5.
There are two appendixes at the end of the document. The first one contains additional jet shapes studies, 
and the second one is devoted to energy flow studies at calorimeter level. 

\newpage
\chapter{QCD at Hadron Colliders}
\label{chap1}

\section{The Standard Model}
\label{standard_model}

The Standard Model (SM)~\cite{halzen} is the most successful theory describing the properties and
interactions (electromagnetic, weak and strong) of the elementary particles. 
The SM is a gauge quantum field theory based in the symmetry group $SU(3)_{C} \times SU(2)_{L} \times U(1)_{Y}$, 
where the electroweak sector is based in the $SU(2)_{L} \times U(1)_{Y}$ group, 
and the strong sector is based in the $SU(3)_{C}$ group.

Interactions in the SM occur via the exchange of integer spin bosons. The mediators of the electromagnetic and strong interactions, the photon
and eight gluons respectively, are massless. The weak force acts via the exchange of three massive
bosons, the $W^{\pm}$ and the $Z$.

The other elementary particles in the SM are half-integer spin fermions: six quarks and six leptons. Both
interact electroweakly, but only quarks feel the strong interaction.
Electrons ($e$), muons($\mu$) and taus($\tau$) are
massive leptons and have electrical charge Q = -1. Their associated neutrinos ($\nu_{e}$, $\nu_{\mu}$, $\nu_{\tau}$) 
do not have electrical charge. Quarks can be classified in up-type ($u$, $s$, $t$) and down-type ($d$,$s$,$b$)
depending on their electrical charge (Q = 2/3 and Q = -1/3 respectively).
For each particle in the SM, there is an anti-particle with opposite quantum numbers.

The SM formalism is written for massless particles and the Higgs mechanism 
of spontaneous symmetry breaking is proposed for generating non-zero boson and fermion 
masses. The symmetry breaking requires the introduction of a new field that leads to 
the existence of a new massive boson, the Higgs boson, that has still not been observed.

\section{Quantum Chromodynamics Theory}

Quantum Chromodynamics (QCD)~\cite{ellis} is the renormalizable gauge field theory that describes 
the strong interaction between colored particles in the SM. 
It is based in the $SU(3)$ symmetric group, and its lagrangian reads:

\begin{equation}
\mathcal{L}_{QCD} = -\frac{1}{4}F_{\alpha\beta}^{A}F_{A}^{\alpha\beta} + \sum_{flavors}\bar{q}(i\gamma^{\mu}D_{\mu}-m)q
\end{equation}

\noindent where the sum runs over the six different types of quarks, $q$, that have mass $m$. 
The field strength tensor, $F_{\alpha\beta}^{A}$ is derived from the gluon field $\mathcal{A}_{\alpha}^{A}$:

\begin{equation}
F_{\alpha\beta}^{A} = [\partial_{\alpha}\mathcal{A}_{\beta}^{A} - \partial_{\beta}\mathcal{A}_{\alpha}^{A} -gf^{ABC}\mathcal{A}_{\alpha}^{B}\mathcal{A}_{\beta}^{C}]
\end{equation}

\begin{figure}[tbh]
\begin{center}
\includegraphics[width=8cm]{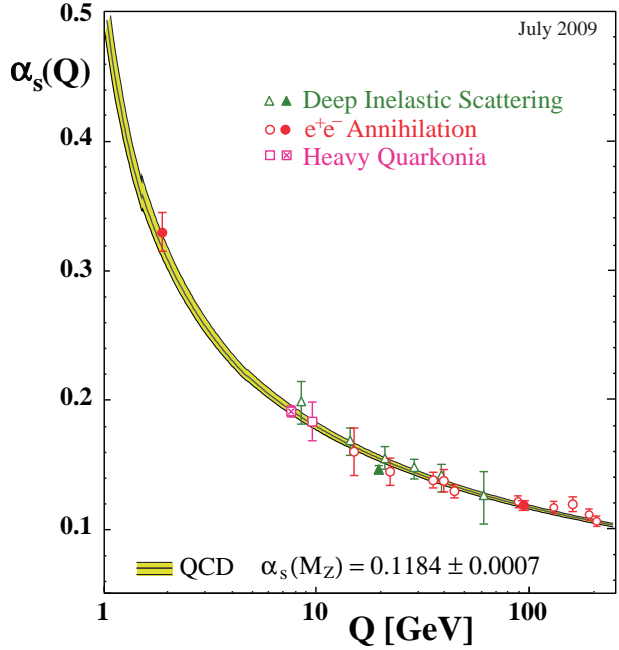}
\caption[Summary of measurements of $\alpha_{s}$ as a function of the energy scale Q]{Summary of measurements of $\alpha_{s}$ as a function of the energy scale Q, from ~\cite{PDG}.}
\label{fig_alpha}
\end{center}
\end{figure}

\noindent $f^{ABC}$ are the structure constants of $SU(3)$, and the indices A, B, C run over the eight color 
degrees of freedom of the gluon field. The third term originates from the non-abelian character of the $SU(3)$ group, 
and is the responsible of the gluon self-interaction, giving rise to triple and quadruple gluon vertexes. This leads to 
a strong coupling, $\alpha_{s} = g^{2}/4\pi$ that is large at low energies and small at high energies (see Figure~\ref{fig_alpha}). 
Two consequences follow from this: 
\begin{itemize}
\item Confinement: The color field potential increases linearly with the distance, and therefore quarks and gluons can never be 
observed as free particles. They are always inside hadrons, either mesons (quark-antiquark) or baryons 
(three quarks each with a different color). 
If two quarks separate far enough, the field energy increases 
and new quarks are created forming colorless hadrons.

\item Asymptotic freedom: At small distances the strength of the strong coupling is that low that quark and gluons behave as essentially 
free. This allows to use the perturbative approach in this regime, where $\alpha_{s} \ll 1$.

\end{itemize}

\section{Deep inelastic scattering}

\begin{figure}[tbh]
\begin{center}
\includegraphics[width=5cm]{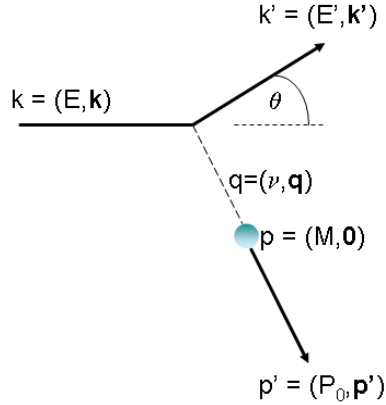}
\caption{Electron scattering from a proton.}
\label{fig_epscattering}
\end{center}
\end{figure}

The scattering of electrons from protons, as illustrated in Figure~\ref{fig_epscattering}, 
has played a crucial role in the understanding of the proton structure. 
If the energy of the incoming electron ($E$) is low enough, the proton can be considered as a point charge (without structure). 
The differential cross section with respect to the solid angle of the scattered electron is\footnote{The mass of the electron 
is neglected in all formulas in this Section by assuming $E >> m$.}:

\begin{equation}
\frac{d\sigma}{d\Omega} = \left(\frac{\alpha}{2Esin^{2}(\theta/2)}\right)^{2}\frac{E'}{E}\left(cos^{2}(\theta/2) + \frac{2EE'sin^{4}(\theta/2)}{M^{2}}\right)
\end{equation}

\noindent where $\alpha$ ($\sim1/137$) is the fine structure constant, $\theta$ is the angle at which the electron is scattered, 
$E'$ is the outgoing electron energy and $M$ the mass of the proton. $E'$ is kinematically determined by $\theta$.

For higher energies of the incoming electrons, the interaction is sensitive to the proton structure, 
and the cross section becomes:

\begin{equation}
\frac{d\sigma}{d\Omega} = \left(\frac{\alpha}{4MEsin^{2}(\theta/2)}\right)^{2}\frac{E'}{E}[2K_{1}sin^{2}(\theta/2) + K_{2}cos^{2}(\theta/2)]
\end{equation}

\noindent $K_{1}$ and $K_{2}$ are functions that contain information on the proton structure and should be determined experimentally. 
Given that $E'$ is kinematically determined by $\theta$, $K_{1}$ and $K_{2}$ only dependent on one variable.

\begin{figure}[tbh]
\begin{center}
\includegraphics[width=5cm]{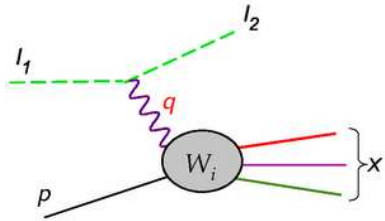}
\caption{Electron-proton deep inelastic scattering.}
\label{fig_DIS}
\end{center}
\end{figure}

Finally, for even higher electron energies, the proton breaks in a multi-hadron final state as illustrated in Figure~\ref{fig_DIS}. 
The cross section is then:

\begin{equation}
\frac{d\sigma}{dE'd\Omega} = \left(\frac{\alpha}{2Esin^{2}(\theta/2)}\right)^{2}[2W_{1}sin^{2}(\theta/2) + W_{2}cos^{2}(\theta/2)]
\end{equation}

\noindent Now $p'$ is the sum of the momenta of the hadrons originating from the proton, and it is not constrained by $p'^{2} = M^{2}$. 
Therefore, $W_{1}$ and $W_{2}$ are functions of two independent variables, $E'$ and $\theta$. Theoretically it is more convenient to use 
the Lorentz-invariant variables $q^{2} = -(k - k')^{2}$ and $x = q^{2}/2qp$, where $p$ is the momenta of the incoming proton. 

The Parton Model describes the proton as built out of three point-like quarks (`valence quarks') with spin 1/2, 
and interprets $x$ as the fraction of the proton momentum carried by the quark.
From the idea that at high $q^{2}$ the virtual photon interacts with a quark essentially free, Bjorken predicted that 
$W_{1}$ and $W_{2}$ depend only on $x$ at large $q^{2}$ ($q^{2} \geq 1$~GeV):

\begin{equation}
MW_{1}(q^{2},x) \to F_{1}(x)
\end{equation}
\begin{equation}
\frac{q^{2}}{2Mx}W_{2}(q^{2},x) \to F_{2}(x)
\end{equation}

\begin{figure}[tbh]
\begin{center}
\includegraphics[width=8cm]{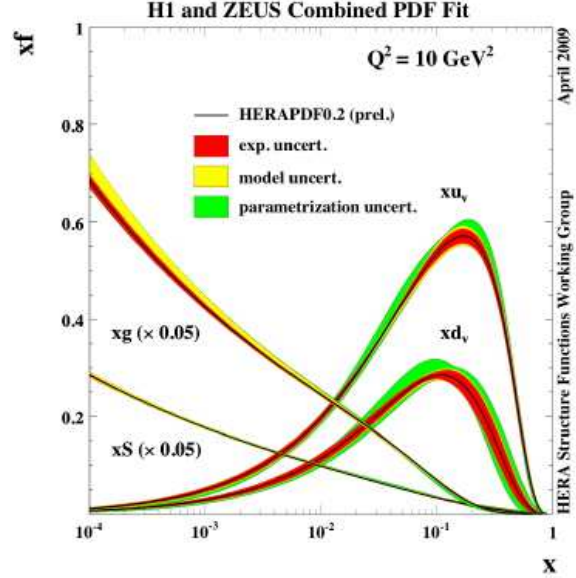}
\caption{Example of PDFs of the valence quarks of the proton, the gluon, and the sea quarks as a function of $x$.}
\label{fig_proton_pdfs}
\end{center}
\end{figure}

\noindent According to the Parton Model:

\begin{equation}
F_{1}(x) = \frac{1}{2}\sum_{i}Q_{i}^{2}f_{i}(x)
\end{equation}

\noindent where $f_{i}(x)$, called Parton Distribution Function (PDF), is the probability that the $i$th 
quark carries a fraction of the proton momentum $x$, and $Q_{i}$ is the electrical charge of the quark. 
Therefore, it is expected that 

\begin{equation}
\int_{0}^{1}x\sum_{i}f_{i}(x)dx = 1
\end{equation}

\noindent but it was found experimentally that the result of this integral is 0.5. 
The rest of the proton momentum is carried by gluons. The introduction of gluons leads to a more complex 
description of the protons structure: quarks radiate gluons, and gluons produce $q\bar{q}$ pairs (`sea quarks') 
or radiate other gluons. 
Figure~\ref{fig_proton_pdfs} shows the PDFs of the valence quarks of the proton, the gluon, and the sea quarks. 
The valence quarks dominate at large $x$, whereas the gluon dominates at low $x$.

The radiation of gluons results in a violation of the scaling behavior of $F_{1}$ and $F_{2}$, introducing a logarithmic dependence on 
$q^{2}$, which is experimentally observed (see Figure~\ref{fig_F2}). 
The functional form of the PDFs can not be predicted from pQCD, but it is possible to predict their evolution 
with $q^{2}$. 

The parton interactions at first order in $\alpha_{s}$ are gluon radiation ($q \rightarrow qg$), gluon splitting 
($g \rightarrow gg$) and quark pair production ($g \rightarrow q\bar{q}$). 
The probability that a parton of type $p$ radiates a quark or gluon and becomes a parton of type $p'$, carrying fraction 
$y = x/z$ of the momentum of parton $p$ (see Figure~\ref{fig_split}) is given by the splitting functions:

\begin{figure}[tbh]
\begin{center}
\includegraphics[width=10cm]{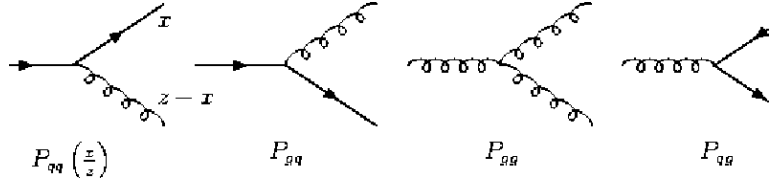}
\caption{Diagrams at LO of the different parton interactions.}
\label{fig_split}
\end{center}
\end{figure}

\begin{equation}
P_{gg}(y) = 6\left[\frac{1-y}{y} + \frac{y}{1-y} + y(1-y)\right]
\end{equation}

\begin{equation}
P_{gq}(y) = \frac{4}{3}\frac{1+(1-y)^{2}}{y}
\end{equation}

\begin{equation}
P_{qg}(y) = \frac{1}{2}[y^{2}+(1-y)^{2}]
\end{equation}

\begin{equation}
P_{qq}(y) = \frac{4}{3}\frac{1+y^{2}}{1-y}
\end{equation}

The evolution of the PDFs as a function of $q^{2}$ follow the DGLAP 
(Dokshitzer, Gribov, Lipatov, Altarelli and Parisi) equations~\cite{dglap}: 

\begin{equation}
\frac{dq_{i}(x,q^{2})}{d\log(q^{2})} = \frac{\alpha_{s}}{2\pi}\int_{1}^{x}\left(q_{i}(z,q^{2})P_{qq}(\frac{x}{z}) +g(z,q^{2})P_{qg}(\frac{x}{z})\right)\frac{dz}{z}
\label{dglap1}
\end{equation}
\begin{equation}
\frac{dg(x,q^{2})}{d\log(q^{2})} = \frac{\alpha_{s}}{2\pi}\int_{1}^{x}\left(\sum_{i}q_{i}(z,q^{2})P_{gq}(\frac{x}{z}) +g(z,q^{2})P_{gg}(\frac{x}{z})\right)\frac{dz}{z}
\label{dglap2}
\end{equation}

The first equation describes the evolution of the quark PDF with $q^{2}$ due to gluon radiation and quark pair production, whereas 
the second equation describes the change of the gluon PDF with $q^{2}$ due to gluon radiation and gluon splitting. 
The equations assume massless partons and therefore are only valid for gluons and the light quarks (u, d and s).

\begin{figure}[tbh]
\begin{center}
\includegraphics[width=8cm]{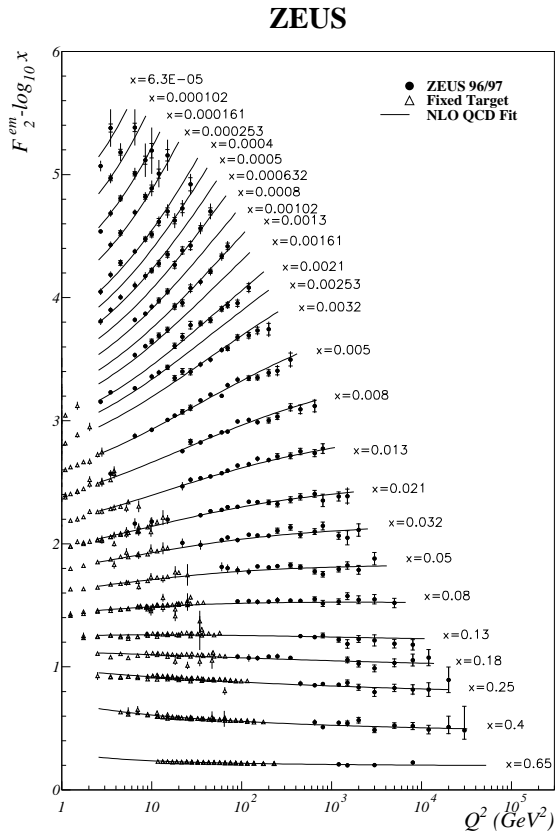}
\caption{Structure function $F_{2}$ of the proton as measured by ZEUS, BCDMS, E665 and NMC experiments.}
\label{fig_F2}
\end{center}
\end{figure}

\section{Perturbative QCD}

\subsection{The factorization theorem}

The QCD factorization theorem is a crucial concept of QCD, that states that cross sections in hadron-hadron interactions can be separated into a  
a hard partonic cross section (short-distance) component and a long-distance component, described by universal PDFs:

\begin{equation}
\sigma(P_{1},P{2}) = \sum_{i,j}\int dx_{1}dx_{2}f_{i}(x_{1},\mu_{F}^{2})f_{j}(x_{2},\mu_{F}^{2}) \times \sigma_{ij}(x_{1},x_{2},\alpha_{s}(\mu_{F}^{2},\mu_{R}^{2}),q^{2}/\mu_{F}^{2})
\end{equation}

\noindent where $P_{1}$, $P_{2}$ are the momenta of the interacting hadrons, the sum runs over all parton types, 
and $\sigma_{ij}$ is the partonic cross 
section of the incoming partons with hadron momenta fraction $x_{1}$, $x_{2}$. $\mu_{R}$ is the scale at which the renormalization 
is performed, and $\mu_{F}$ is an arbitrary parameter that separates the hard from the soft component. 
Both scales are typically chosen to be of the order of $q^{2}$.

\begin{figure}[tbh]
\begin{center}
\includegraphics[width=12cm]{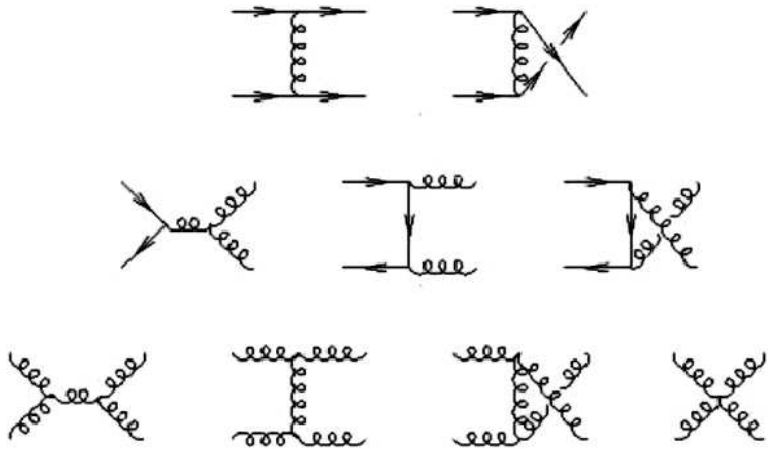}
\caption{Leading order diagrams for $2 \rightarrow 2$ parton interactions.}
\label{fig_diagrams}
\end{center}
\end{figure}

Partonic cross sections in leading order (LO) calculations for jet production are O($\alpha_{s}^{2}$), 
since they are based on $2 \rightarrow 2$ parton interactions ($gg \rightarrow gg, qg \rightarrow qg, qq \rightarrow qq$), as
shown in Figure~\ref{fig_diagrams}. The dominant process is the $gg$ scattering because of the larger color charge of the gluons. 
 
Next-to-leading-order (NLO) diagrams 
include contributions from gluon initial- or final-state radiation and loops on the diagrams already shown. 
The partonic cross sections at NLO reduce the dependence on the normalization and factorization scales, 
and are calculable using programs such as JETRAD~\cite{jetrad} and NLOJET++~\cite{nlojet}.
Predictions at higher orders are not yet available due to the large number of diagrams involved.

\subsection{Parton Distribution Functions}

As already explained, perturbative QCD (pQCD) can predict the evolution of the PDFs~\cite{forte} with respect to $q^{2}$ 
using the DGLAP equations, but not their functional form. 
Therefore, PDFs should be extracted from experimental data at a given $q^{2} = Q_{0}^{2}$. In particular, 
seven functions should be determined, one for the gluon and the others for each one of the light quarks and 
anti-quarks. Experimental data from a large variety of processes is used to constrain several aspects of the 
PDFs: measurements of Drell-Yan production, inclusive jet cross sections and W-asymmetry in $p\bar{p}$ 
collisions, and deep-inelastic $e$, $\mu$ or $\nu$ scattering.

Typically, specific functional forms are postulated for the PDFs with a set of free parameters. These parameters 
are obtained optimizing the comparison between experimental data and predictions using the PDFs, for example by 
minimizing  a $\chi^{2}$. The functional form assumed for several sets of PDFs is:

\begin{equation}
f_{i}(x,Q_{0}^{2}) = x^{\alpha_{i}}(1-x)^{\beta_{i}}g_{i}(x)
\end{equation}

\noindent where $\alpha_{i}$ and $\beta_{i}$ are the free fit parameters and $g_{i}(x)$ is a function that 
tends to a constant in the limits $x \rightarrow 0$ and $x \rightarrow 1$. 
This functional form is motivated by counting rules~\cite{crules} and Regge theory~\cite{regge}, that suggest that 
$f_{i}(x) \sim (1-x)^{\beta_{i}}~\mathrm{when}~x \rightarrow 1$  
and $f_{i}(x) \sim x^{\alpha_{i}}~\mathrm{when}~x \rightarrow 0$ respectively. Both limits are approximate, 
and even if these theories predict the values of $\beta_{i}$ and $\alpha_{i}$, they are taken as free fit parameters 
when computing the PDFs. This approach is used by three of the PDFs used in the analyses presented in this Thesis: 
CTEQ~\cite{CTEQ}, MSTW~\cite{MSTW} and HERA~\cite{HERA} PDFs. For example in the case of HERAPDFs, $g_{i}(x)$ is:

\begin{equation}
g_{i}(x) = 1 + \epsilon_{i}x^{1/2} + D_{i}x + E_{i}x^{2}
\end{equation}

NNPDFs~\cite{NNPDF} follow a different approach, using neural networks as a parton parametrization. 
Neural networks are functional forms that can fit a large variety of functions. 

\subsection{Uncertainties}

\begin{figure}[tbh]
\begin{center}
\includegraphics[width=10cm]{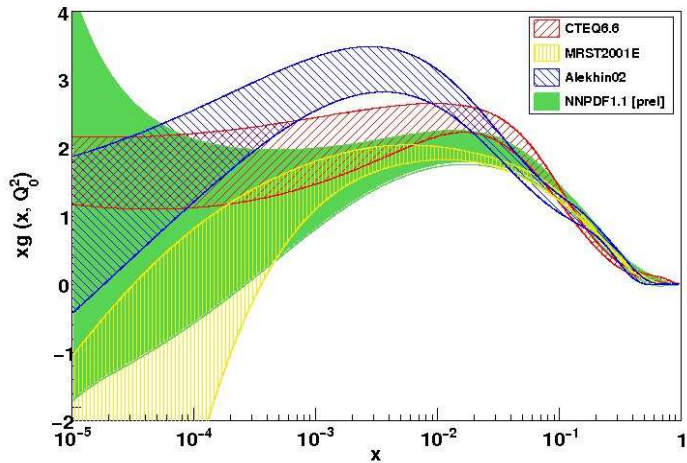}
\caption{PDF of the gluon as a function of $x$ according to different PDF groups at $q^{2} = 2$~GeV${}^{2}$.}
\label{fig_gluon_pdf}
\end{center}
\end{figure}

There are three main sources of uncertainties in the calculation of pQCD observables:
\begin{itemize}
\item The lack of knowledge of higher order terms neglected in the calculation. 
It is estimated by varying the renormalization scale, $\mu_{R}$, usually by a 
factor of two with respect to the default choice. The factorization scale, $\mu_{F}$, is independently varied to evaluate 
the sensitivity to the choice of scale where the PDF evolution is separated from the partonic cross section. The envelope 
of the variation that these changes introduce in the observable is taken as a systematic uncertainty.

\item Uncertainties on parameters of the theory, like the $\alpha_{s}$ and the heavy quark masses, that are 
propagated into the observable. 

\item PDFs have another uncertainty coming from the way the experimental data is used to determine the PDFs. 
This uncertainty is typically estimated using the Hessian method. 
If $a_{0}$ is the vector of the PDF parameters where 
$\chi^{2}(a_{0})$ is minimized, all parameters such that $\chi^{2} - \chi^{2}_{0} < T$ are considered 
acceptable fits, where $T$ is a parameter called tolerance. 
PDF parameters are expressed in terms of an orthogonal basis, and variations along the positive and negative 
directions of each eigenvector ($a_{i}^{+}$, $a_{i}^{-}$) such that $\chi^{2} - \chi^{2}_{0} = T$ are performed. 
The uncertainty in the observable $\Gamma$ is:

\begin{equation}
\delta\Gamma^{+} = \sqrt{\sum_{i}max({\Gamma(a_{i}^{+})-\Gamma(a_{0}),\Gamma(a_{i}^{-})-\Gamma(a_{0}),0})^{2}}
\end{equation}
\begin{equation}
\delta\Gamma^{-} = \sqrt{\sum_{i}min({\Gamma(a_{i}^{+})-\Gamma(a_{0}),\Gamma(a_{i}^{-})-\Gamma(a_{0}),0})^{2}}
\end{equation}

\noindent where $\Gamma(a)$ is the observable computed using the PDFs with the parameters in vector $a$.
NNPDF use a Monte Carlo approach to evaluate the uncertainties, in which the probability distribution in parameter 
space derives from a sample of MC replicas of the experimental data. Figure~\ref{fig_gluon_pdf} shows the PDF of the 
gluon with its uncertainties obtained following different approaches.
\end{itemize}

\section{Monte Carlo simulation}

Complete pQCD calculations are always performed only up to a fixed order in $\alpha_{s}$, but the 
enhanced soft-gluon radiation and collinear configurations at higher orders can not be neglected. 
They are taken into account in the parton shower (PS) approximation, that sum the leading contributions of 
these topologies to all orders. Monte Carlo (MC) generator programs include the PS approximation, 
as well as models to reproduce non-perturbative effects, such as the hadronization of 
the partons to colorless hadrons and the underlying event (UE). 

\subsection{Parton Shower}
\label{sec_ps}
\begin{figure}[tbh]
\begin{center}
\includegraphics[width=8cm]{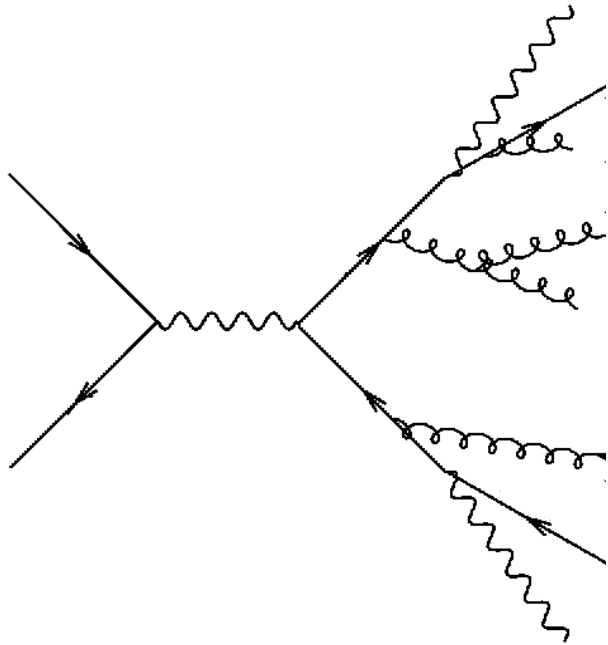}
\caption{Illustration of the parton shower from the outgoing partons of the hard interaction.}
\label{fig_ps}
\end{center}
\end{figure}

The PS approximation describes successive parton emission from the partons in the hard interaction, as illustrated 
in Figure~\ref{fig_ps}. 
The evolution of the showering is governed by DGLAP equations~\ref{dglap1} and~\ref{dglap2}, from which 
the Sudakov form factors~\cite{sudakov} are derived for the numerical implementation of the parton shower. 
These factors represent the probability that a 
parton does not branch between an initial scale ($t_{i}$) and a lower scale ($t$). Once a branching occurs 
at a scale $t_{a}$, $a \rightarrow bc$, subsequent branchings are derived from the scales $t_{b}$ and $t_{c}$. 
They can be angle-, $Q^{2}$- or $\ptjet$-ordered. In the first case subsequent branchings have smaller 
opening angles than this between $b$ and $c$, whereas in the second, parton emissions are produced in 
decreasing order of intrinsic $\ptjet$.

Successive branching stops at a cutoff scale, $t_{0}$, of the order of $\Lambda_{QCD}$, 
after producing a high-multiplicity partonic state.  
Since quark and gluons can not exist isolated, MC programs contain models for the hadronization of the 
partons into colorless hadrons.

\subsection{Hadronization}

The hypothesis of local parton-hadron duality states that the momentum and quantum numbers of the hadrons follow those 
of the partons. This hypothesis is the general guide of all hadronization models, but do not give details on the formation 
of hadrons, that is described in models with parameters that are tuned to experimental data. 
There are two main models of hadron production.

\textbf{The string model}~\cite{string} describes the behavior of $q\bar{q}$ pairs using string dynamics. 
The field between each $q\bar{q}$ pair is represented by a string with uniform energy per unit length.
As the $q$ and the $\bar{q}$ move apart from each other and thus the energy of the color field increases, 
the string connecting the two is tightened, until it breaks into a new $q'\bar{q}'$ pair. 
If the invariant mass of either of these string pieces is large enough, further breaks may occur. 
In the string model, the string break-up process is assumed to proceed until only on-mass-shell hadrons remain.
In the simplest approach of baryon production, a diquark is treated 
just like an ordinary antiquark. A string can break either by quark-antiquark or antidiquark-diquark pair production, 
leading to three-quark states. There are more sophisticated models, but the formation of baryons is still 
poorly understood.

\textbf{The cluster model}~\cite{hcluster} is based on the confinement property of perturbative QCD, exploited to form 
color-neutral clusters. 
After the perturbative parton showering, all gluons are split into light quark-antiquark or diquark-antidiquark pairs. 
Color-singlet clusters are formed from the quarks and anti-quarks. 
The clusters thus formed are fragmented into two hadrons. If a cluster is too light 
to decay into two hadrons, it is taken to represent the lightest single hadron of its
flavor. Its mass is shifted to the appropriate value by an exchange of momenta with a neighboring cluster. 
If the cluster is too heavy, it decays into two clusters, that are further fragmented into hadrons.

\subsection{Underlying Event}

The UE comes from the partons that do not participate in the hard interaction. They contribute to the final 
state via their color-connection to the hard interaction, and via extra parton-parton interactions. 
Its simulation is based on the eikonal model, that describes the underlying event activity as additional 
uncorrelated partonic scatters. 
The number of interactions per event $<n>$ depends on the impact parameter $b$. A small $b$ value corresponds 
to a large overlap between the two colliding hadrons, and therefore a higher probability for multiple interactions.  
For a given $b$, the parton-parton cross section $\sigma_{hard}$ is computed as a function of the transverse momentum in 
the center-of-mass frame of the scattering process $\hat{p}_{T}$. Since this cross section 
diverges as $\hat{p}_{T} \rightarrow 0$ a cut-off parameter $\hat{p}_{T}^{min}$ is introduced, where 
experimentally $\hat{p}_{T}^{min} \sim 2$~GeV. $<n>$ is extracted from the ratio between 
the total hadron cross section $\sigma_{nd}$ and the parton-parton cross section,  
$<n> = \sigma_{hard}/\sigma_{nd}$, and assumed to be Poisson-distributed. 

The UE models are tuned using experimental data, such as the jet shapes described in Chapters~4 and~5.

\subsection{Monte Carlo Generator Programs}

\subsubsection{PYTHIA Monte Carlo}

The PYTHIA~\cite{PYTHIA} MC event generator includes hard processes at LO, and uses the PS model  
for initial- and final-state radiation. The hadronization is performed using the string model. 
It includes an underlying event model to describe the interactions between the proton remnants. 

The PYTHIA tunes \textbf{DW}~\cite{DW} and \textbf{Perugia2010}~\cite{Perugia2010} use CTEQ5L PDFs, and both have been produced using Tevatron data. 
In the former the PS is $Q^{2}$-ordered, whereas in the latter it is $\ptjet$-ordered.

In autumn 2009, the MRST LO* PDFs~\cite{mrst2007lo} were used in PYTHIA for the first time in ATLAS. This required to tune
the PYTHIA model parameters, resulting in the \textbf{MC09}~\cite{MC09} tune. 
It was done using Tevatron data, mainly from underlying event and minimum bias analyzes. The PS is $\ptjet$-ordered.

The PYTHIA-\textbf{AMBT1}~\cite{AMBT1} tune followed the MC09 one, and also uses MRST LO* PDFs and $\ptjet$-ordered PS. 
It was derived using ATLAS data, in particular charged particle multiplicities in $pp$ interactions at 0.9 and 7~TeV 
center-of-mass energy.

\subsubsection{HERWIG}

HERWIG~\cite{herwig} is a general-purpose MC event generator for hard processes in particle colliders.
It uses an angular-ordered parton-shower for initial- and final-state QCD radiation, and a cluster model to reproduce 
the hadronization of the partons. The Fortran version of HERWIG is interfaced with JIMMY~\cite{Jimmy} to simulate 
multiple parton-parton interactions.

HERWIG++~\cite{HERWIG++} is the C++ version of HERWIG, that is expected to replace the Fortran one at a given point. The 
underlying event is modeled inside the program, that therefore do not use JIMMY.

\subsubsection{ME + Parton Shower: Alpgen and Powheg}

Alpgen~\cite{ALPGEN} is an event generator of multi-parton hard processes in hadronic collisions,  
that performs the calculation of the exact LO matrix elements for a large set of parton-level processes.
It uses the ALPHA algorithm~\cite{alpha} to compute the matrix elements for large parton 
multiplicities in the final state. The advantage of this algorithm is that its complexity increases 
slower than the Feynman diagrams approach when increasing the particles in the final state.
Powheg~\cite{POWHEG} is a MC event generator that includes NLO matrix elements. 
Alpgen and Powheg contain an interface to both PYTHIA and HERWIG for the parton showering, 
the hadronization and the underlying event simulation. 

\section{Jet Algorithms}

Quarks and gluons from the hard scattering result on a collimated flow of particles due to 
parton shower and hadronization. This collimated flow of particles is called jet.
There are several jet definitions~\cite{salam} with the main purpose of reconstructing jets 
with kinematics that reflect that of the initial parton. 
These definitions can be classified in two main types of jet algorithms: cone algorithms and sequential recombination algorithms.
 
\subsection{Cone algorithms}

Typically, cone jet algorithms start by forming cones of radius $R$ in 
the $y-\phi$ space around a list of seeds, that can be all particles in the final state or those above 
a given energy threshold. The center of the cone is recomputed from the particles inside by following 
either the snowmass or the four-momentum recombination. 
In the four-momenta recombination, the jet momenta is the sum of the momenta of its constituents:

\begin{equation} 
(E,p_{x},p_{y},p_{z})^{jet} = \sum_{const.}(E,p_{x},p_{y},p_{z})^{i}
\end{equation}

\noindent whereas in the snowmass scheme, the jet is considered massless, its transverse energy is the sum 
of the transverse energy of its constituents and the jet ($\eta,\phi$) are the average of 
the ($\eta,\phi$) of the constituents weighted by its transverse energy:

\begin{equation} 
E_{T}^{jet} = \sum_{const.}E_{T}^{i}
\end{equation}
\begin{equation}
(\eta,\phi)^{jet} = \frac{1}{E_{T}^{jet}}\sum_{const.}(\eta,\phi)^{i}E_{T}^{i}
\end{equation}
\begin{equation}
m^{jet} = 0
\end{equation}

A cone is formed from the new center and the process 
repeated until the particles inside the cone are no longer changed by further iterations. 
Usually the algorithm is allowed to form overlapping cones and then decides whether to merge or split them
depending on the fraction of energy they share. 

This last step makes the cone algorithms collinear or infrared unsafe, 
and affects the definition of the parton-level jet cross section to all orders in pQCD.
A jet algorithm is infrared safe if the addition of an extra 
particle with infinitesimal energy do not change the jet configuration in the final state. 
If the replacement of a particle by two collinear particles (which momenta sum is equal to that of the original particle) 
do not change the jet configuration in the final state, the jet algorithm is collinear safe. 

In order to solve this, cone-based jet algorithms have been formulated such that they find all stable cones through 
some exact procedure, avoiding the use of seeds. 
These algorithms are very time-consuming from the computational point of view, which constitutes a disadvantage 
in high-multiplicity events such as those at the LHC.

\subsection{Sequential recombination jet algorithms}

Sequential recombination jet algorithms cluster particles according to their relative transverse momentum, 
instead of spacial separation. This is motivated by the parton shower evolution as described in Section~\ref{sec_ps}.  
For all particles in the final state, the algorithm computes the following distances:

\begin{equation}
d_{ij} = \mathrm{min}(k_{ti}^{2p},k_{tj}^{2p})\frac{\Delta R_{ij}^{2}}{R^{2}}
\label{eq_akt1}
\end{equation}
\begin{equation}
d_{iB} = k_{ti}^{2p}
\label{eq_akt2}
\end{equation}

\noindent where $k_{ti}$ is the transverse momentum of particle $i$, $R_{ij} = \sqrt{\Delta y^{2} + \Delta \phi^{2}}$ 
between particles $i$ and $j$, $R$ a parameter of the algorithm that approximately controls the size of the jet, 
and $p$ depends on the jet algorithm: 
$p = 1$ for the $k_{t}$ algorithm, $p = 0$ for the Cambridge/Aachen algorithm, and $p = -1$ for the $\akt$ algorithm.
The distance $d_{iB}$ is introduced in order to separate particles coming from the hard interaction than those 
coming from the interaction between remnants. 
The smallest distance is found, and if it is $d_{ij}$, particles $i$ and $j$ are combined 
into one single object. If instead it is $d_{iB}$, particle $i$ is considered a jet an removed from the list. 
The distances are recalculated with the remaining objects, and the process repeated until no particle is left in the list. 
Jets are defined as those objects with $\ptjet$ above a given threshold. 

\begin{figure}[tbh]
\begin{center}
\includegraphics[width=12cm]{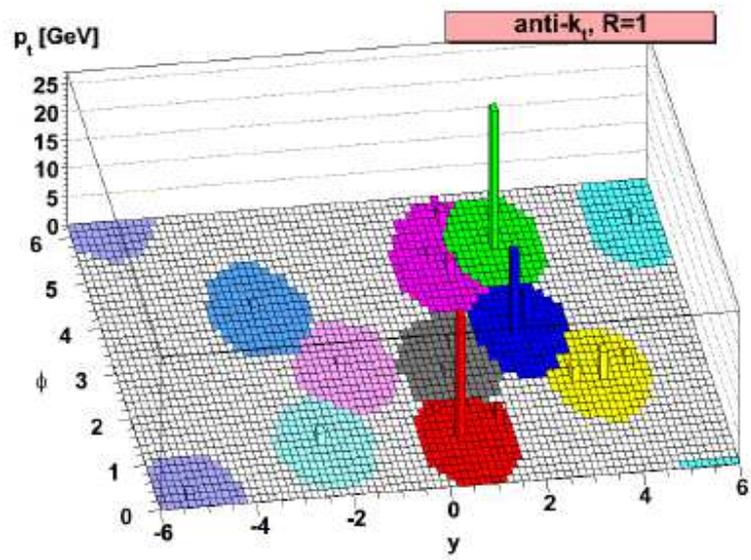}
\caption{A sample parton-level event clustered with the $\akt$ algorithm.}
\label{fig_akt}
\end{center}
\end{figure}

These algorithms are very convenient, mainly because they are infrared and collinear safe and computationally fast. 
In particular, the $\akt$ algorithm~\cite{aktal} produces jets with a conical structure in ($y,\phi$), as illustrated in 
Figure~\ref{fig_akt}, that facilitates dealing with pile-up. It is the default jet finding algorithm in the LHC experiments.

\newpage
$\,$
\clearpage
\newpage

\chapter{The ATLAS Detector at the Large Hadron Collider}
\label{chap2}

The analyses described in this Thesis are performed using proton-proton collision data produced by the Large Hadron Collider (LHC) 
and collected by the ATLAS detector. In this Chapter, the LHC and the ATLAS detector are described, giving more emphasis to the 
elements that are relevant for the analyses.

\section{The Large Hadron Collider}

\begin{figure}[tbh]
\begin{center}
\mbox{
\includegraphics[width=0.495\textwidth]{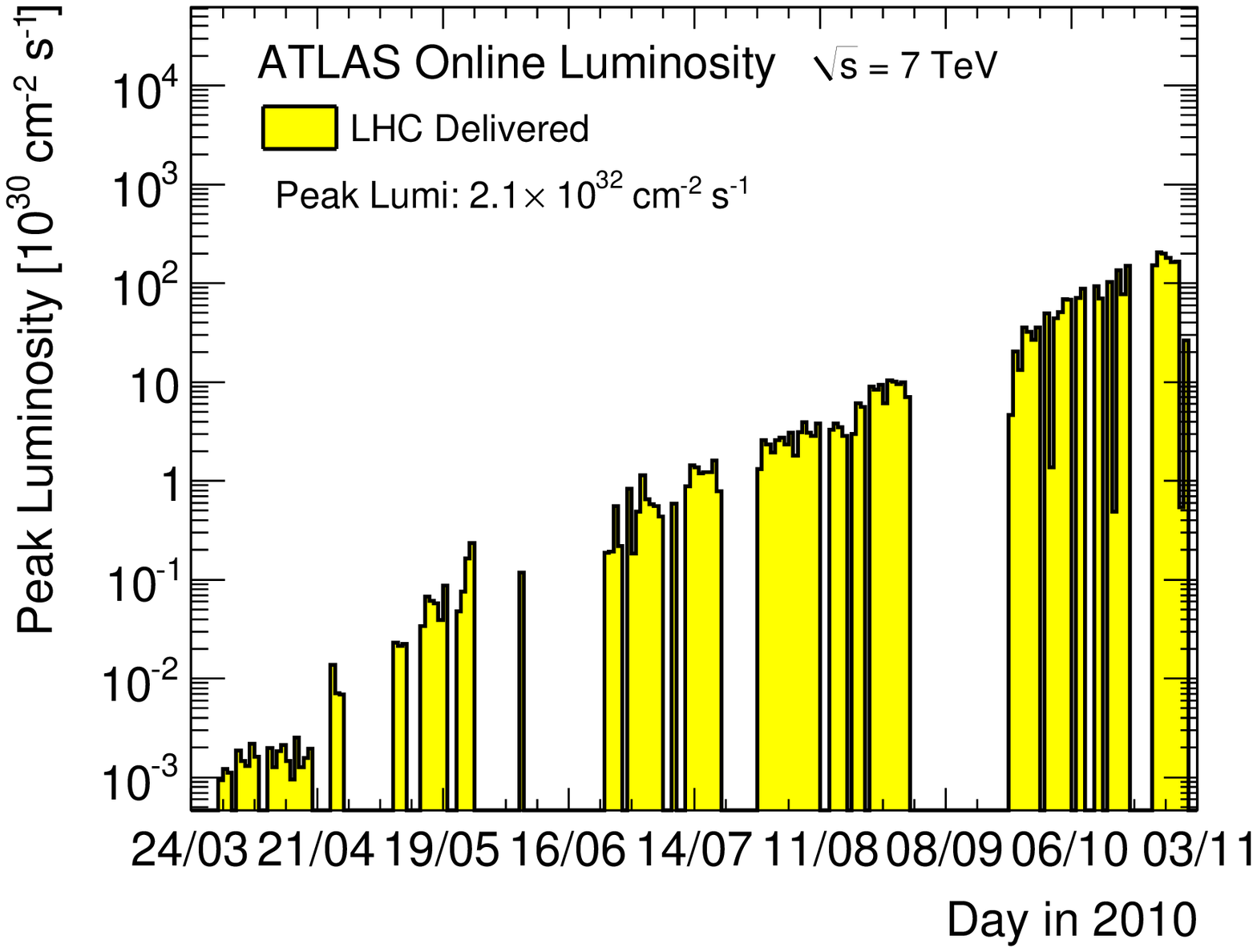}
\includegraphics[width=0.495\textwidth]{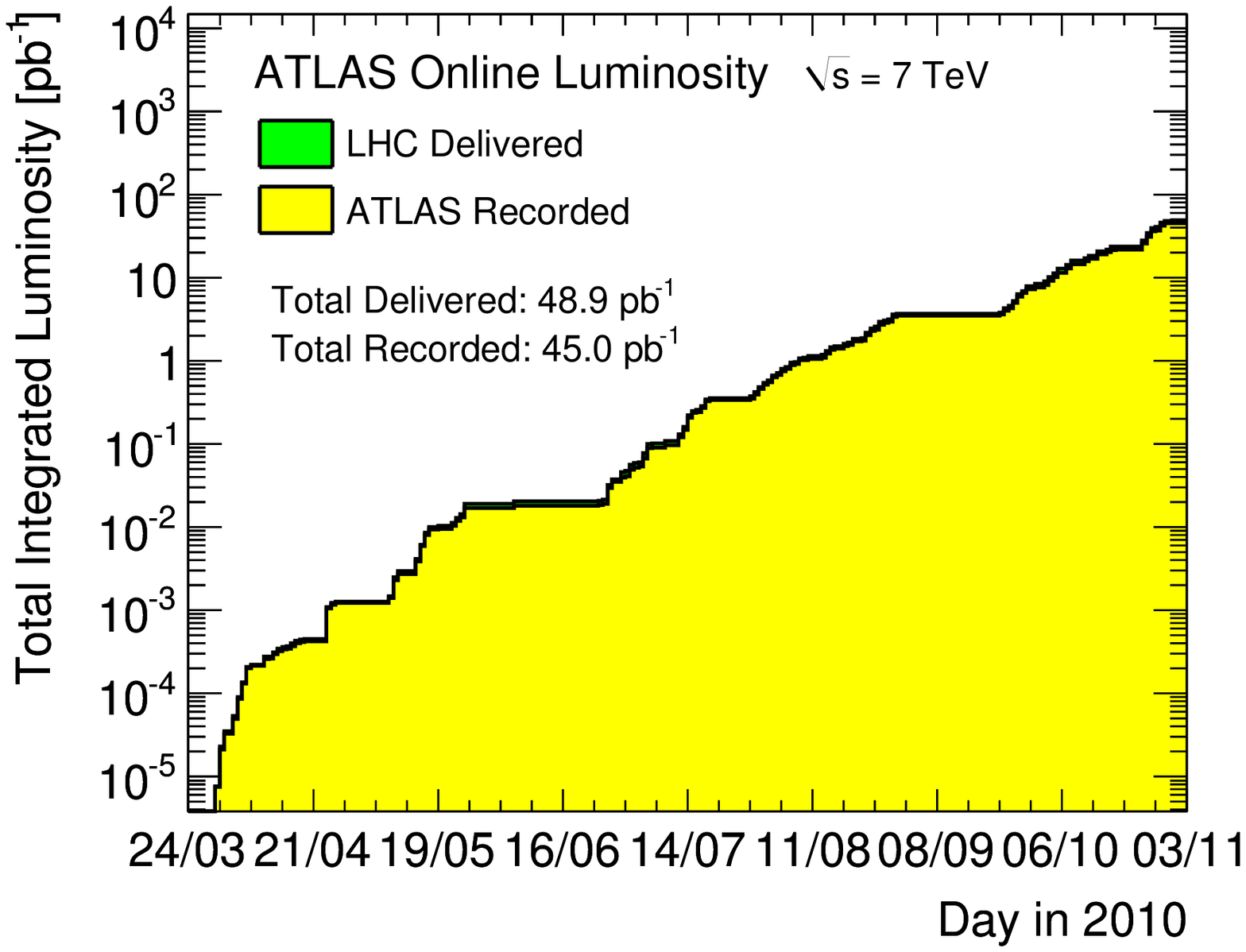}
}
\end{center}
\vspace{-0.7 cm}
\caption[Maximum instantaneous and integrated luminosity versus day delivered by the LHC]{\small
Maximum instantaneous luminosity (left) and cumulative integrated luminosity (right) versus day delivered by the LHC 
and recorded by ATLAS for pp collisions at 7 TeV center-of-mass energy during stable beams in 2010.
}
\label{fig_lumi}
\end{figure}

The LHC~\cite{LHC} is a superconducting accelerator built in a circular tunnel of 27~km in circumference that is located at CERN. The tunnel is 
situated between 45~to 170~m underground, and straddles the Swiss and French borders on the outskirts of Geneva. 
Two counter rotating proton beams injected into the LHC from the SPS accelerator at 450~GeV are 
further accelerated to 3.5~TeV while moving around the LHC ring guided by magnets inside a continuous vacuum.
During 2010, the instantaneous luminosity was increased over time, with a maximum peak at $\mathrm{2\cdot10^{32}~cm^{-2}s^{-1}}$, 
and the total integrated luminosity delivered by the LHC was of 48~pb${}^{-1}$ from which ATLAS recorded 45~pb${}^{-1}$ 
(see Figure~\ref{fig_lumi}).

There are four main detectors placed along the accelerator line: ATLAS and CMS, that are general-purpose detectors, 
ALICE, dedicated to heavy-ions physics, and LHCb, dedicated to B-physics.

\section{The ATLAS experiment}

\begin{figure}[tbh]
\begin{center}
\includegraphics[width=14cm]{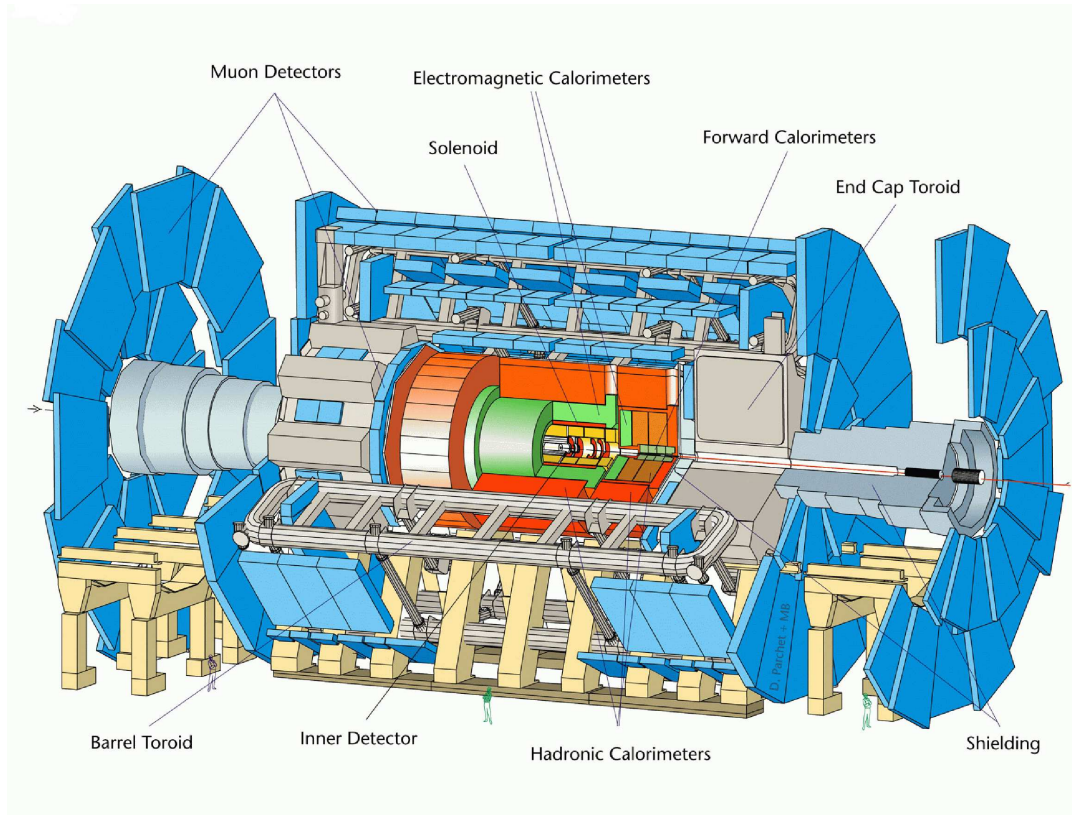}
\caption{View of the full ATLAS detector.} 
\label{fig_ATLAS}
\end{center}
\end{figure}

The ATLAS detector~\cite{ATLAS} is an assembly of several sub-detectors arranged in consecutive layers around the beam axis, as shown 
in Figure~\ref{fig_ATLAS}. The main sub-detectors are the Inner Detector, the Calorimeters and the Muon System, that are described 
in the next Sections. ATLAS is 46~m long, 25~m wide and 25~m high, and weights 7000~t.

The ATLAS coordinate system and its nomenclature will be used repeatedly throughout this Thesis, and is thus described here. 
The ATLAS reference system is a Cartesian
right-handed coordinate system, with the nominal collision point at the origin. The anti-clockwise beam
direction defines the positive $z$-axis, while the positive $x$-axis is defined as pointing from the collision
point to the center of the LHC ring and the positive $y$-axis pointing upwards.  The azimuthal angle $\phi$ is
measured around the beam axis, and the polar angle $\theta$ is measured with respect to the $z$-axis.
The pseudorapidity is defined as $\eta = - {\rm ln}({\rm tan}(\theta/2))$.
The rapidity is defined as $y = 0.5 \times {\rm ln}[(E+p_z)/(E-p_z)]$, where $E$ denotes the energy 
and $p_z$ is the component of the momentum along the beam direction.

The ATLAS detector was designed to optimize the search for the Higgs boson and a large variety of physics phenomena 
at the TeV~scale proposed by models beyond the Standard Model. The main requirements that follow from these goals are:

\begin{itemize}
\item Given the high LHC luminosity, detectors require fast, radiation-hard electronics and sensor elements. 
In addition, high detector granularity is needed to handle the particle fluxes and to reduce the influence of overlapping events.

\item Large acceptance in pseudorapidity with almost full azimuthal angle coverage.

\item Good charged-particle momentum resolution and reconstruction efficiency in the inner tracker. 

\item Very good electromagnetic calorimetry for electron and photon identification and measurements, complemented by full-coverage 
hadronic calorimetry for accurate jet and missing transverse energy measurements. 

\item Good muon identification and momentum resolution over a wide range of momenta and the ability to determine unambiguously the charge 
of high $\ptjet$ muons.

\item Highly efficient triggering on low transverse-momentum objects with sufficient background rejection is a prerequisite to achieve 
an acceptable trigger rate for most physics processes of interest.
\end{itemize}

\section{Inner Detector}

The Inner Detector (ID) was designed in order to perform high precision measurements 
with fine detector granularity in the very large track density events produced by the LHC. 
The ID, that is $\pm 3512$~m long and 1150~mm in radius, is built 
out of three components, in increasing order of distance with respect to beam axis: the Pixel detector, 
the Semiconductor Tracker (SCT) and the Transition Radiation Tracker (TRT).
The precision tracking detectors (pixels and SCT) cover the region
$|\eta| < 2.5$, and are segmented in $r - \phi$ and z, whereas the TRT cover the region $|\eta| < 2$ and is only 
segmented in $r - \phi$. 
The ID has around 87 million readout channels, 80.4~millions 
in the pixel detector, 6.3~millions in the SCT and 351~thousand in the TRT.
All three are immersed in a 2~T magnetic field generated by the central solenoid, which extends over a length of
5.3~m with a diameter of 2.5~m.

The ID is used to reconstruct tracks and production and decay vertices, 
and provides a position resolution of 10, 17 and 130~$\mu$m (Pixel, SCT, TRT) in the 
$r - \phi$ plane as well as 115 and 580~$\mu$m (Pixel, SCT) in the $r - z$ plane. 
The momentum resolution as a function of $\ptjet$ of the track is parametrized as:

\begin{equation}
\frac{\sigma_{p_{T}}}{p_{T}} = P_{1} \oplus P_{2} \times p_{T}
\end{equation}

\noindent and the values $P_{1} = 1.6 \pm 0.1\%$ and $P_{2} = (53 \pm 2) \times 10^{-5}$~GeV${}^{-1}$ were 
determined using cosmic rays~\cite{cosmic}. 
Extrapolation of the fit result yields to a momentum resolution of about 1.6\% at low momenta and of about 50\% at 1~TeV.


\section{Calorimeters}

\begin{figure}[tbh]
\begin{center}
\includegraphics[width=10cm]{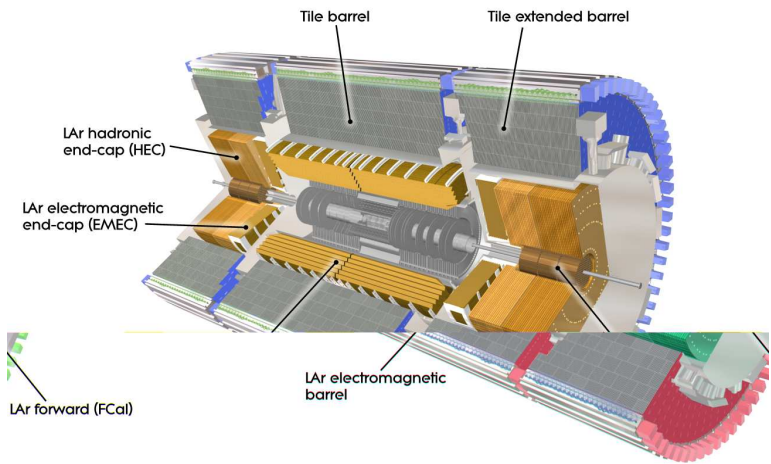}
\caption{View of the calorimeter system.}
\label{fig_calorimeter}
\end{center}
\end{figure}

The calorimeter systems of ATLAS, illustrated in Figure~\ref{fig_calorimeter} surround the Inner Detector system and cover
the full $\phi$-space and $|\eta| <$~4.9, extending radially 4.25~m. The calorimeter systems 
can be classified in electromagnetic calorimeters, designed for precision measurements of electrons and photons, 
and hadronic calorimeters, that collect the energy from hadrons.  
Calorimeter cells are pseudo-projective towards the interaction region in $\eta$.
The granularity of the electromagnetic calorimeter is typically $0.025\times0.025$ in $|\Delta\eta|\times|\Delta\phi|$, whereas 
the hadronic calorimeters have granularity of $0.1\times0.1$ in most of the regions. 
The energy response of the calorimeter to single particles is discussed in the next Chapter.

\subsection{Liquid Argon Calorimeter}

\begin{figure}[tbh]
\begin{center}
\includegraphics[width=10cm]{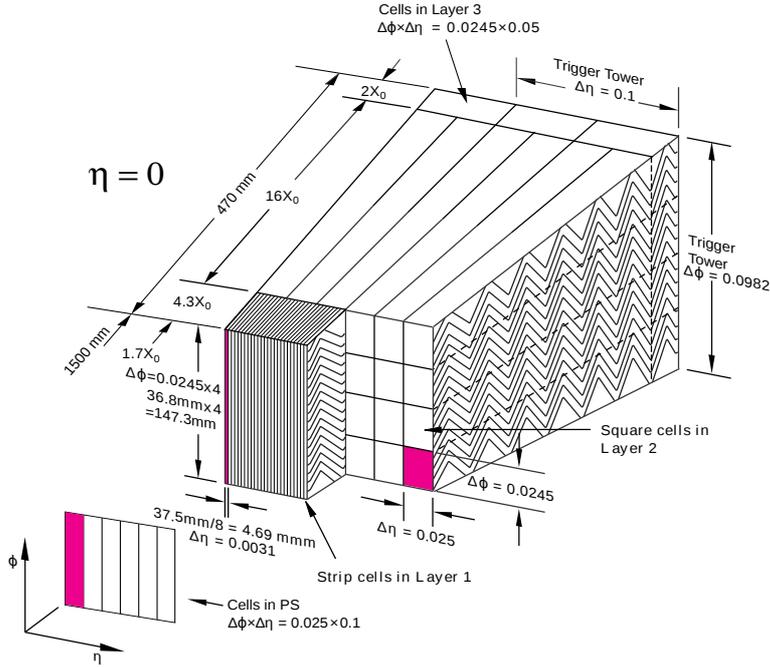}
\caption{Sketch of a module of the LAr calorimeter.}
\label{fig_lar_module}
\end{center}
\end{figure}

The electromagnetic calorimeter is a lead-LAr detector with accordion-shaped kapton electrodes and lead absorber plates over its full
coverage. The accordion geometry provides complete $\phi$ symmetry without azimuthal cracks. 
The calorimeter is divided into a barrel part ($|\eta| < 1.475$) and two end-cap components
($1.375 < |\eta| < 3.2$), each housed in their own cryostat. 
Over the central region ($|\eta| < 2.5$), the EM calorimeter is segmented in three layers in depth, whereas 
in the end-cap it is segmented in two sections in depth. Figure~\ref{fig_lar_module} shows an sketch of 
a module of the LAr calorimeter.

\subsection{Hadronic calorimeters}

\begin{figure}[tbh]
\begin{center}
\includegraphics[width=12cm]{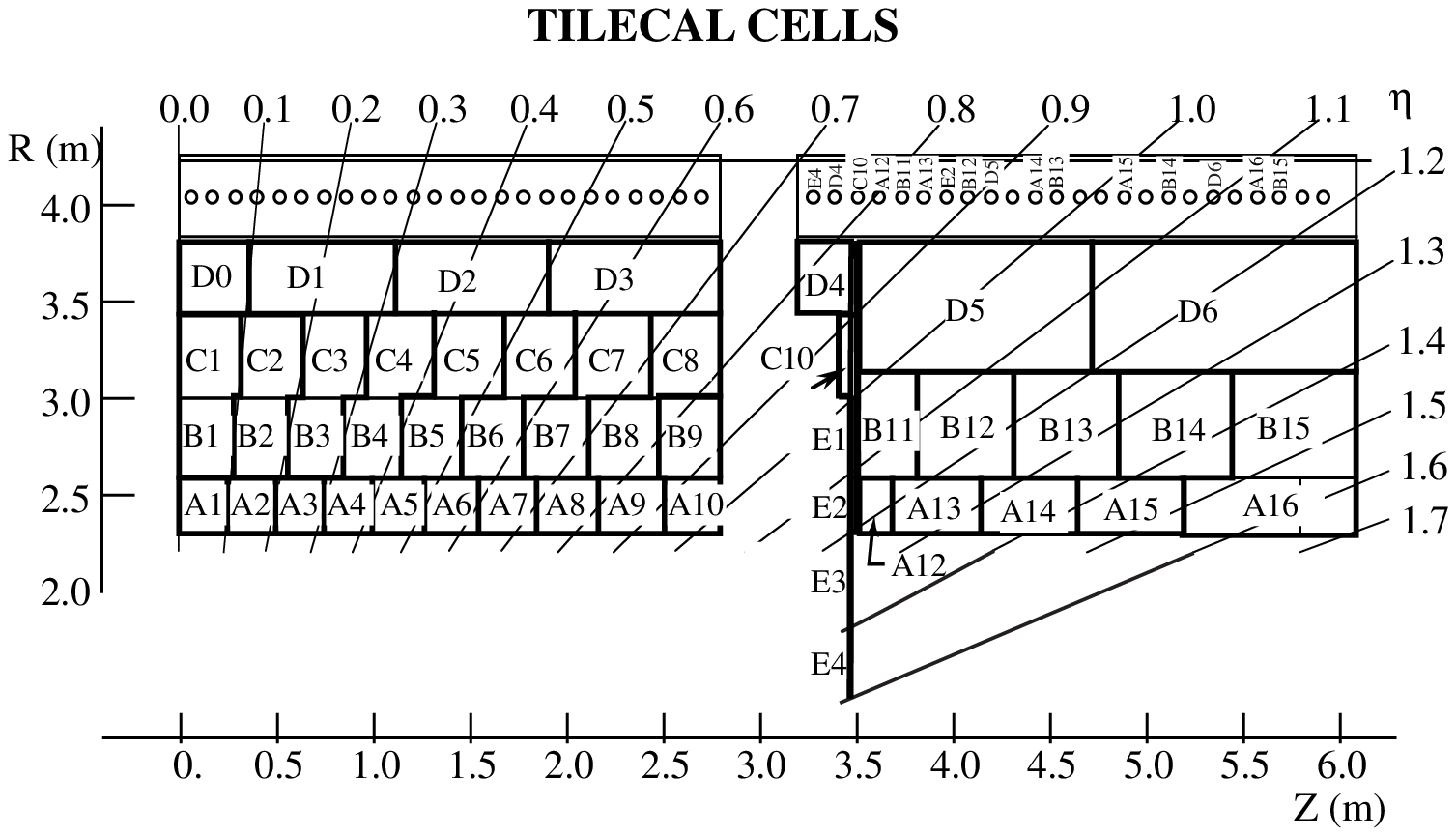}
\caption{Barrel and extended barrel sections of the Tile Calorimeter.}
\label{fig_tilecal_module}
\end{center}
\end{figure}

The \textbf{Tile Calorimeter} is placed directly outside the electromagnetic calorimeter envelope. Its
barrel covers the region $|\eta| < 1.0$, and its two extended barrels the range $0.8 < |\eta| < 1.7$. It is a
sampling calorimeter using steel as the absorber and scintillating tiles as the active material, 
and it is segmented in depth in three layers, approximately 1.5, 4.1 and 1.8 interaction lengths ($\lambda$) 
thick for the barrel and 1.5, 2.6, and 3.3 $\lambda$ for the extended barrel, as illustrated in Figure~\ref{fig_tilecal_module}.
Two sides of the scintillating tiles are read out by wavelength shifting
fibers into two separate photomultiplier tubes.

The \textbf{Hadronic End-cap Calorimeter (HEC)} consists of two
independent wheels per end-cap, located directly behind the end-cap electromagnetic calorimeter
and sharing the same LAr cryostats. They cover the region $1.5 < |\eta| < 3.2$, 
and each wheel is divided into two segments in depth, for a total of four layers per end-cap.
The HEC is a sampling calorimeter built out of copper plates intervealed with LAr gaps, that are the active medium.

The \textbf{Forward Calorimeter (FCal)} is integrated into the end-cap cryostats, and it 
extends in $|\eta|$ from~3.1 to~4.9.
It consists of three modules in each end-cap: the first, made of copper, is optimized for
electromagnetic measurements, while the other two, made of tungsten, measure predominantly the
energy of hadronic interactions. In all three modules LAr is the sensitive medium. 

\subsection{Calorimeter Topological Clusters}
\label{sec_cluster}

Calorimeter clusters~\cite{topocluster} are used as input to the jet finding algorithm in the studies presented in this Thesis. 
They are built out of neighboring calorimeter cells with significant signal over noise. Therefore, 
the 3-D shape and the number of cells of clusters are not fixed. The noise is computed for each cell independently, and it 
is defined as the expected RMS of the electronic noise for the current gain and conditions plus the contribution from 
pileup added in quadrature.

In order to make clusters, all cells with a signal to noise ratio above~4 are taken as seed cells. These cells  
are considered in descending order of signal to noise ratio, adding all neighboring cells to them forming the so called proto-clusters. 
Neighboring cells with signal to noise ratio between 2~and~4 are taken as seed cells in the next iteration. 
If a cell is adjacent to more than one proto-cluster and its signal to noise ratio is above~2 the proto-clusters are merged, whereas 
if it is smaller than~2 the cell is only added to the first proto-cluster. Once there are no more cells in the seed list,
an splitting algorithm based on local maxima is run over the proto-clusters in order to separate clusters that are not isolated. 

Final clusters are treated as massless and their energy, at the electromagnetic (EM) scale, 
is the sum of the energies of the cells belonging to the cluster. 
The EM scale is the appropriate scale for the 
reconstruction of the energy deposited by electrons or photons in the calorimeter.
The EM energy scale was first determined using electron test-beam
measurements in the barrel and endcap calorimeters~\cite{testbeam}~\cite{electrons}. Muons from test-beams and produced in cosmic rays
were used to validate the EM scale in the hadronic calorimeter~\cite{testbeam}~\cite{muons}.
Recently, a recalibration of the electromagnetic calorimeters has been derived from $Z \to ee$ events in pp collisions. 
The EM scale uncertainty is 1.5\% in the EM LAr barrel calorimeter, and 3\% in the Tile calorimeter. 

\section{Muon System}

The calorimeter is surrounded by  the Muon Spectrometer. The
air-core toroid system, with a long barrel and two inserted end-cap
magnets, generates a large magnetic field volume with strong bending
power within a light and open structure. Multiple-scattering effects
are thereby minimized, and excellent muon
 momentum resolution is achieved with three stations of high-precision
tracking chambers. The muon  instrumentation includes trigger chambers 
with very fast time response.

\section{Luminosity measurement}

The luminosity, $\mathcal{L}$, of a $pp$ collider that operates at a revolution frequency $f_{r}$ and $n_{b}$ bunches cross 
at the interaction point can be expressed as

\begin{equation}           
\mathcal{L} = \frac{n_{b}f_{r}n_{1}n_{2}}{2\pi\Sigma_{x}\Sigma_{y}}
\label{eq_vdm}
\end{equation}

\noindent where $n_{1}$ and $n_{2}$ are the numbers of particles in the two colliding bunches and $\Sigma_{x}$ and $\Sigma_{y}$ 
characterize the widths of the horizontal and vertical beam profiles, that are measured 
using van der Meer scans. The observed event rate is recorded while scanning the two beams across each other
first in the horizontal ($x$), then in the vertical ($y$) direction. This measurement yields two bell-shaped curves, with the 
maximum rate at zero separation, from which one extracts the values of $\Sigma_{x}$ and $\Sigma_{y}$. The luminosity at zero separation can
then be computed using Equation~\ref{eq_vdm}.

ATLAS measures the luminosity~\cite{lum_vdm} in inelastic interactions  
using different detectors and algorithms, all based on event-counting techniques, that is, on determining the 
fraction of bunch crossings (BCs) during which a specified detector registers an event satisfying a given selection requirement:

\begin{equation}
\mathcal{L} = \frac{\mu n_{b}f_{r}}{\sigma_{inel}} = \frac{\mu_{vis} n_{b}f_{r}}{\epsilon\sigma_{inel}} = \frac{\mu_{vis} n_{b}f_{r}}{\sigma_{vis}}
\label{eq_lum2}
\end{equation}

\noindent where $\mu$ is the average number of inelastic interactions per BC, 
$\sigma_{inel}$ is the $pp$ inelastic cross section, $\epsilon$ is the efficiency for one 
inelastic $pp$ collision to satisfy the event-selection criteria, 
and $\mu_{vis} \equiv \epsilon\mu$ is the averaged number of visible (passing that event selection criteria) 
inelastic interactions per BC. The visible cross section $\sigma_{vis} \equiv \epsilon\sigma_{inel}$ 
is the calibration constant that relates the measurable quantity $\mu_{vis}$ to the luminosity $\mathcal{L}$ . Both $\epsilon$ and 
$\sigma_{vis}$ depend on the pseudorapidity distribution and particle composition of the collision products, and are 
therefore different for each luminosity detector and algorithm.

In the limit $\mu_{vis} \ll 1$, the average number of visible inelastic interactions per BC is given by the intuitive expression

\begin{equation}           
\mu_{vis} \approx \frac{N}{N_{BC}}
\end{equation}

\noindent where $N$ is the number of events passing the selection criteria that are observed during a given time interval, 
and $N_{BC}$ is the number of bunch crossings in that same interval. When $\mu$ increases, the probability that two or 
more $pp$ interactions occur in the same BC is not negligible, and $\mu_{vis}$ is no longer linearly related to 
the raw event count $N$. Instead $\mu_{vis}$ must be calculated taking into account Poisson statistics, 
and in some cases, instrumental or pile-up related effects.

$\sigma_{vis}$ can be extracted from Equation~\ref{eq_lum2} using the measured values of $\mu_{vis}$ 
and $\mathcal{L}$, computed with the van der Meer technique, that allows the determination of 
$\sigma_{vis}$ without a priori knowledge of the inelastic $pp$ cross section or of detector efficiencies.

\section{Trigger}
\label{sec_trig}

The trigger system has three different levels: Level-1 trigger (LVL1), Level-2 trigger (LVL2) and the Event Filter (EF). 
Each trigger level refines the decisions made at the previous step and, when necessary, applies additional selection 
criteria. The rate of selected events~\footnote{The LHC provides a collision rate of 40~MHz within 
full operational state.} is reduced to around 200~Hz for permanent storage trough the trigger chain, with an 
event size is approximately 1.3~Mbyte.

The L1 trigger searches for high transverse-momentum muons, electrons, photons, jets, and $\tau-$leptons 
decaying into hadrons, as well as large missing and total transverse energy. Its selection is
based on information from a subset of detectors, and reduces the event rate to about 75~kHz in less than 2.5~$\mu$s.
In each event, the L1 trigger defines one or more regions in $\eta$ and $\phi$ from those regions within 
the detector where its selection process has identified interesting features. 
Data in these regions, called Regions-of-Interest (RoI’s), include information on the type of feature identified
and the criteria passed. This information is subsequently used by the high-level trigger.

The L2 selection is seeded by the RoI information provided by the L1 trigger over a dedicated
data path. L2 selections use, at full granularity and precision, all the available detector data within
the RoI’s (approximately 2\% of the total event data). The L2 menus are designed to reduce the
trigger rate to approximately 3.5~kHz, with an event processing time of about 40~ms, averaged over
all events. The final stage of the event selection is carried out by the event filter, which reduces
the event rate to roughly 200~Hz. Its selections are implemented using offline analysis procedures
within an average event processing time of the order of 4~s.

In the measurements presented in this Thesis, three different triggers have been used: the Minimum Bias Trigger Scintillators
(MBTS), the central jet trigger, covering $|\eta| < 3.2$, and the forward jet trigger, spanning $3.1 < |\eta| < 4.9$.
The MBTS trigger requires at least one hit in the minimum bias scintillators located in front of the
endcap cryostats, covering $2.09 < |\eta| < 3.84$, and is the primary trigger used to select minimum-bias
events in ATLAS. The central and forward jet triggers independently select data using several jet $E_{T}$ 
thresholds at the EM scale. More details on the trigger paths used to select events are given in next Chapter.

\newpage

\chapter{Inclusive Jet Cross Section}
\label{chap3}

The jet cross section measurement is a fundamental test of QCD and it is sensitive to the presence 
of new physics. It also provides information on the parton distribution functions and the strong coupling.
Since jet production has the largest cross section of all high $\ptjet$ processes at LHC, the related observables 
have the highest reach in energy at any given integrated luminosity. 

\begin{figure}[tbh]
\begin{center}
\includegraphics[width=0.65\textwidth]{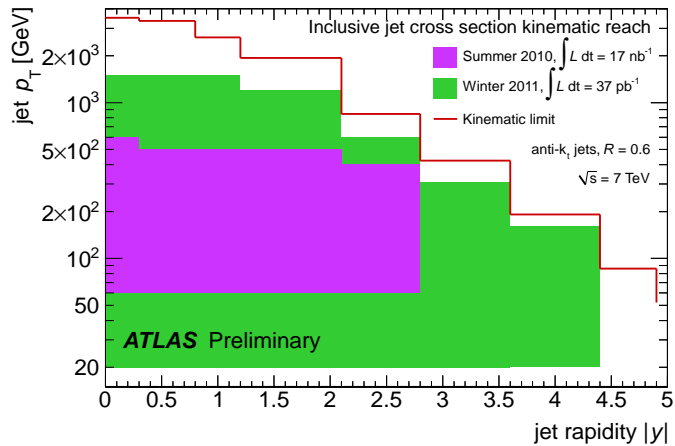}
\end{center}
\vspace{-0.7 cm}
\caption[Kinematic range of the inclusive jet cross section measurements]{\small
Kinematic range of the inclusive jet cross section measured in this analysis~\cite{jet_conf} compared to that
of the previous study reported in~\cite{jet_pro_atlas} for jets identified using the $\akt$ algorithm with R = 0.6.
}
\label{fig_extension}
\end{figure}

The first measurement of the inclusive jet cross section in ATLAS was done using an integrated luminosity of $17$~nb${}^{-1}$ 
for jets with $\ptjet > 60$~GeV and $|\rapjet| < 2.8$, and was published in the EPJC~\cite{jet_pro_atlas}.
In this Chapter, a measurement of the inclusive jet cross section for jets with $\ptjet > 20$~GeV and $|\rapjet| < 4.4$~\cite{jet_conf} 
is presented, based on $37.3 \pm 1.2$~pb${}^{-1}$ of integrated luminosity collected by ATLAS. The kinematic range of 
these two measurements is shown in Figure~\ref{fig_extension} as a function of the jet $\ptjet$ and $|y|$. 
In this Chapter, the MC samples used in the jet measurements are presented, followed by a detailed discussion of the 
jet reconstruction and calibration, the jet and event selection criteria, and the unfolding procedure to correct 
the measurement for detector effects. Finally, the measurement is compared to theoretical predictions.

\section{Monte Carlo simulation}
\label{sec_mc}

Samples of inclusive jet events in proton-proton collisions at $\sqrt{s} = 7$~TeV are used in 
the inclusive jet cross section analysis. PYTHIA 6.423 with the AMBT1 tune MC samples are used to correct for 
detector effects, to compute the non-perturbative corrections applied to the NLO predictions, and to estimate part 
of the systematic uncertainties of the measurements. 

The MC generated samples are passed through a full simulation~\cite{atlas_sim} of the ATLAS detector and trigger, based on GEANT4~\cite{geant}, 
that simulates the interactions of the generated particles with the detector material.
The Quark Gluon String Precompound (QGSP) model~\cite{qgs} is used for the fragmentation of the nucleus, and
the Bertini cascade (BERT) model~\cite{bertini} for the description of the interactions of the hadrons in the medium of the nucleus.
Test-beam measurements for single pions have shown that these simulation settings best describe
the response and resolution in the barrel~\cite{sim_barrel} and end-cap~\cite{sim_endcap} calorimeters.
 

\section{Jet reconstruction and calibration}

In the analysis presented in this Chapter, jets are reconstructed using the $\akt$ algorithm, with jet radius parameter $R = 0.6$ 
(results with $R = 0.4$ can be found in~\cite{jet_conf}). 
The jet finding algorithm is run over calorimeter clusters at the EM scale,
and the four-momenta recombination is used. 

The EM scale needs to be further calibrated to account for calorimeter non-compensation (the energy 
response to hadrons is lower than the response to electrons of the same energy), dead material (inactive regions of the detector where 
energy is lost) and leakage (energy deposits from particles which shower is not fully contained in the calorimeter). Moreover, corrections 
are needed for particles inside the truth jet but not the calorimeter jet, and energy losses in calorimeter clustering and jet reconstruction.
In order to correct for all these effects, a jet energy scale (JES) correction as a function of the 
jet energy and pseudorapidity is applied to jets at the EM scale~\cite{JES}.

The corrections are derived from Monte Carlo (MC) samples produced using PYTHIA 6.423 with the AMBT1 tune. 
Truth jets are reconstructed from stable particles in the final state, excluding 
muons and neutrinos, that are taken into account when unfolding the measurement to the hadron level (see Section~\ref{sec_unfolding}). 
Corrections are derived from pairs of calorimeter and truth jets matched within $\Delta$R $< 0.3$, where 
$\Delta$R $= \sqrt{\Delta \eta^2 + \Delta \phi^2}$. 
The energy response $\mathcal{R}$ is defined as the ratio of the calorimeter 
and the truth jet energy, $\mathcal{R} = E^{calo}/E^{truth}$.
Both calorimeter and truth jets are required to be isolated, that is, only jets with no other
jet with $\ptjet > 7$~GeV at $\Delta$R $< 2.5$R are accepted. 
The motivation for this requirement is that non-isolated jets have lower response than isolated jets, 
and the fraction of non-isolated jets in a given sample is physics dependent. The cut $\Delta$R $< 2.5$R 
is chosen because $\akt$ jets may extend a bit further than R.
The average jet energy response, $<\mathcal{R}>$, and the average jet energy at the EM scale, 
$<E{}^{calo}>$, are measured in bins of truth jet energy, $E^{truth}$. 
A function is derived by fitting $<\mathcal{R}>$ as a function of $<E^{calo}>$ with the following parametrization: 

\begin{equation}
\mathcal{R}{}_{calib}(E^{calo}) = \sum_{i}a_{i}(ln(E^{calo}))^{i}
\end{equation}

\noindent where $a_{i}$ are free parameters of the fit. The JES correction is the inverse of this function, and it is 
applied to the jet four-momenta at the EM scale.

\begin{figure}[tbh]
\begin{center}
\includegraphics[width=0.65\textwidth]{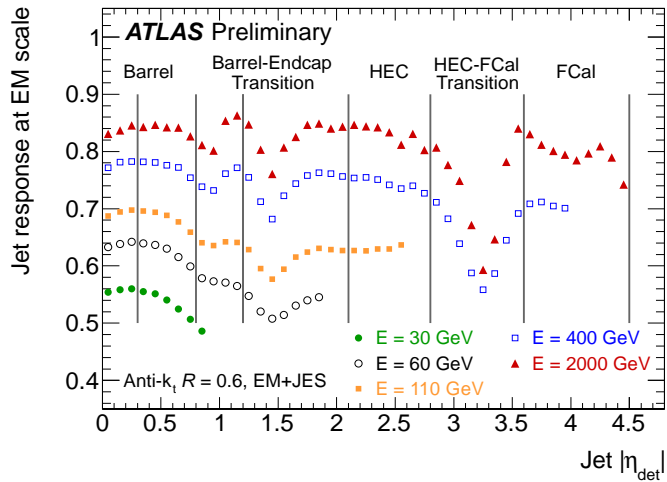}
\end{center}
\vspace{-0.7 cm}
\caption[Jet energy response at the EM scale as a function of the $\eta_{det}$]{\small
Jet energy response at the EM scale as a function of the $\eta_{det}$, for several calibrated jet energy values. 
The intervals in $\eta_{det}$ used to evaluate the JES uncertainty are also shown.
}
\label{fig_jet_response}
\end{figure}

Figure~\ref{fig_jet_response} shows the calorimeter EM response as a function of detector pseudo-rapidity, 
$\eta_{det}$\footnote{$\eta_{det}$ is used for the jet pseudo-rapidity with respect to the center of the ATLAS detector.}, 
for several jet energies. The response has large
fluctuations depending on the $\eta_{det}$ region. Therefore, 
the procedure to derive the JES corrections is performed independently in different $\eta_{det}$ bins.

\begin{figure}[tbh]
\begin{center}
\includegraphics[width=0.65\textwidth]{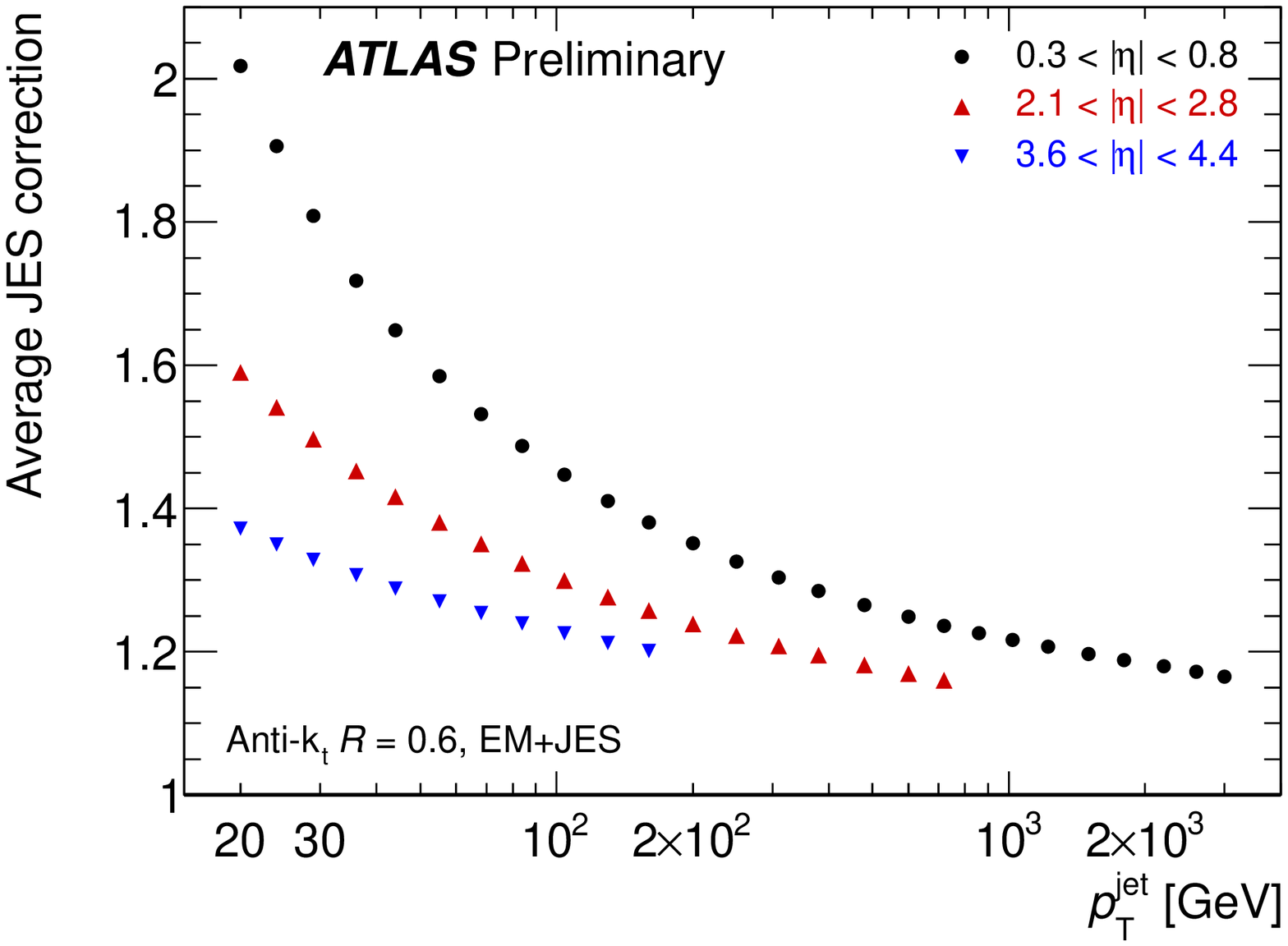}
\end{center}
\vspace{-0.7 cm}
\caption[Average JES correction as a function of calibrated $\ptjet$]{\small
Average JES correction as a function of calibrated $\ptjet$ in three representative $\eta$ regions. The correction is only shown over the valid 
kinematic range, E = 3.5 TeV at maximum.
}
\label{fig_jet_correction}
\end{figure}

Figure~\ref{fig_jet_correction} shows the average JES 
corrections as a function of $\ptjet$ in three $\eta$ regions. Corrections are larger in the central region (up to 2.1 at 
low $\ptjet$), and decrease as the $\ptjet$ increases in all $\eta$ regions.

\begin{figure}[tbh]
\begin{center}
\includegraphics[width=0.65\textwidth]{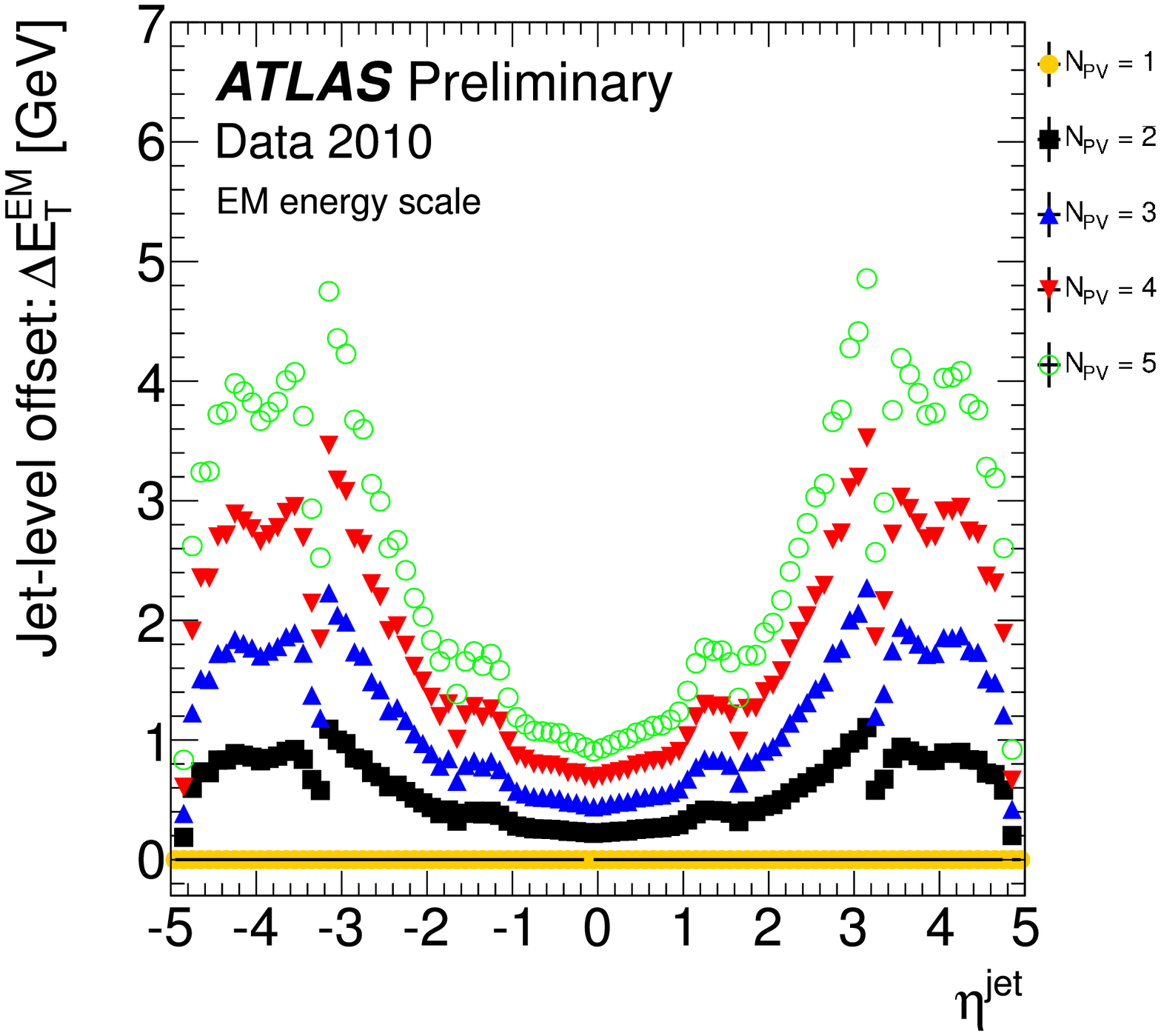}
\end{center}
\vspace{-0.7 cm}
\caption{\small
Jet offset at the EM scale shown as a function of pseudorapidity and the number of reconstructed primary vertices. 
}
\label{fig_pileup}
\end{figure}

An additional correction is required in cases where there are more than one proton-proton interaction 
in the same event (pile-up), because the jet energy includes other contributions apart from the one coming from 
the collision of interest. 
These corrections, called offset corrections~\cite{pileup}, have been derived by estimating the average extra-amount 
of energy per jet in data as a function of the number of primary vertices and $\eta$.
The offset corrections, shown in Figure~\ref{fig_pileup}, are subtracted from the jet energy at the EM scale before 
applying the JES correction.

There are two corrections of the jet position that leave the jet energy unaffected. The first is 
needed because calorimeter clusters kinematics refer to the center of the ATLAS detector. In order to correct jets back to point to 
the primary vertex, each jet constituent kinematics is recalculated using the vector from the primary vertex to the cluster centroid 
as its direction. The jet four-momenta is then re-set to the sum of the recalculated four-momenta of its constituents. This correction 
improves slightly ($< 1\%$) the jet $\ptjet$ response.

Finally, the jet $\eta$ is further corrected for the bias introduced by the variation of the jet energy response in the different 
parts of the calorimeter. This last correction, added to the jet $\eta$, is smaller than 0.01 in most parts of the calorimeter, 
and goes up to 0.05 in the transition regions between the different calorimeters.

\section{Jet and event selection}
\label{sec_criteria}

The analysis is based in the full 2010 data set of pp collisions at $\sqrt s = 7$~TeV collected by the ATLAS detector, 
corresponding to an integrated luminosity of $37$~pb${}^{-1}$. Events are 
considered when the trigger system, tracking detectors, calorimeters 
and magnets were operating at the nominal conditions. 

Events are then required to have at least one primary vertex with five 
or more tracks pointing to it in order to remove beam-related backgrounds and cosmic rays. 
For events containing a jet with $\ptjet > 20$~GeV, this requirement has an efficiency well above $99\%$.

\begin{figure}[tbh]
\begin{center}
\mbox{
\includegraphics[width=0.495\textwidth]{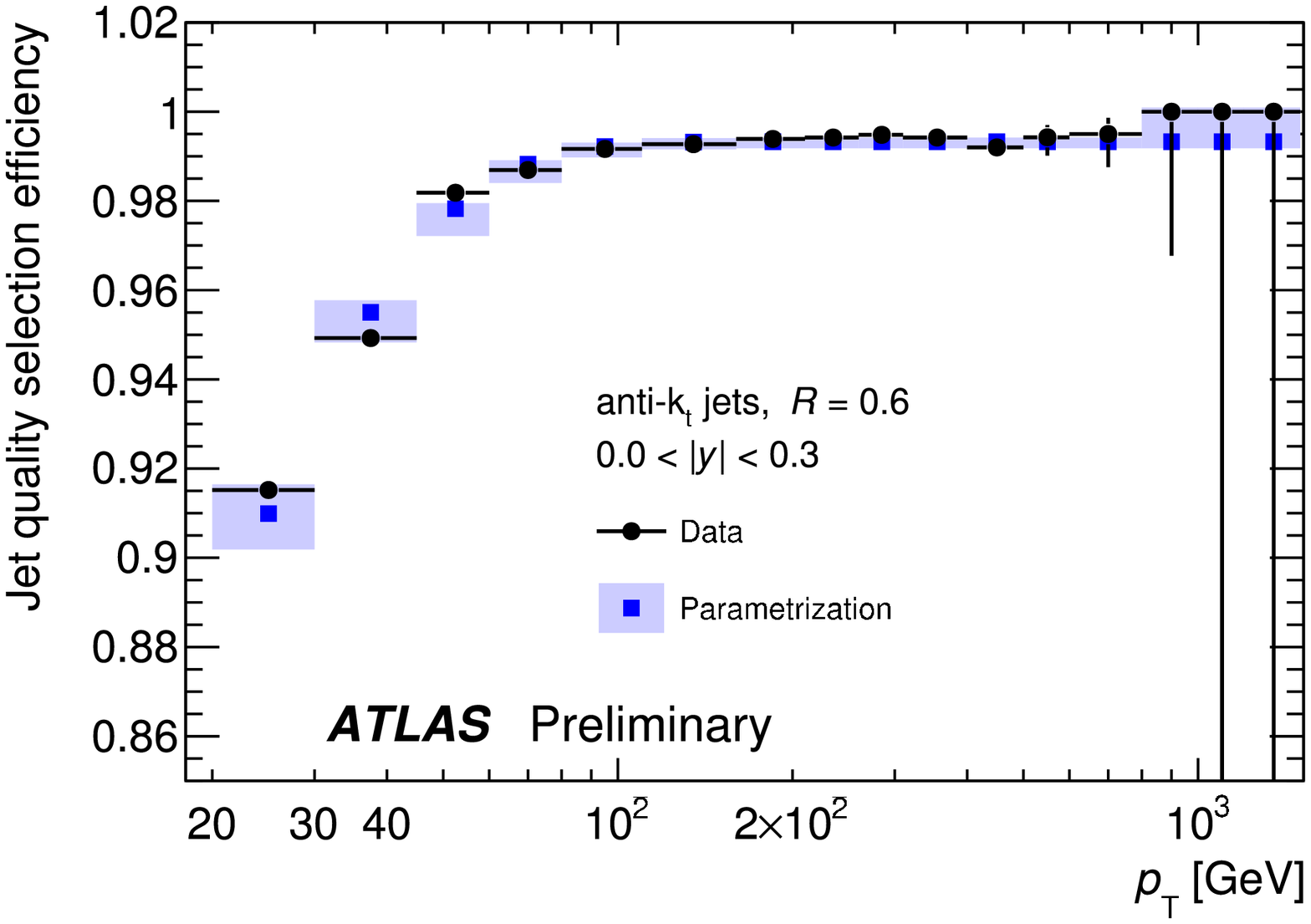}
\includegraphics[width=0.495\textwidth]{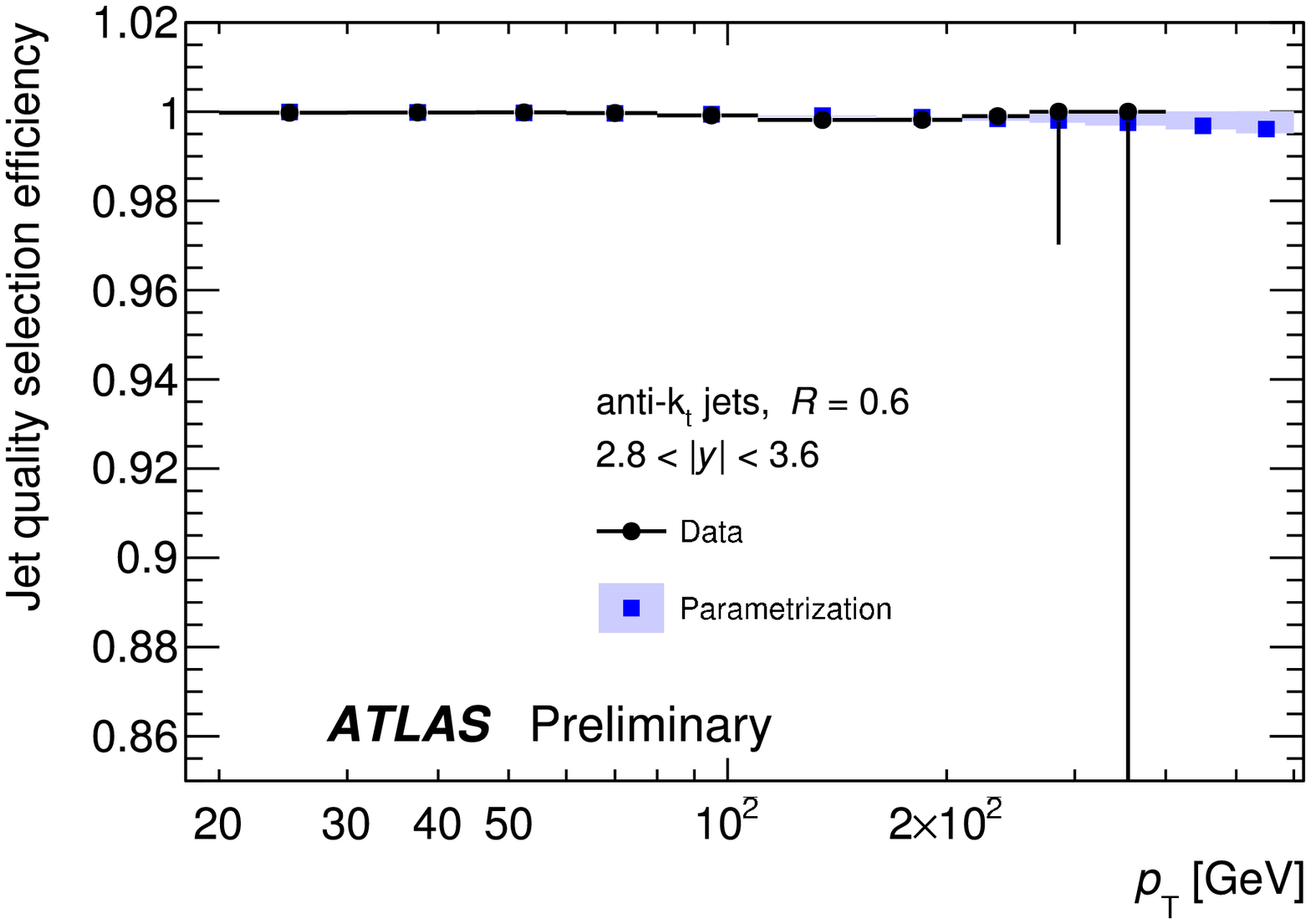}
}
\end{center}
\vspace{-0.7 cm}
\caption[Efficiency for jet identification as a function of $\ptjet$]{\small
Efficiency for jet identification as a function of $\ptjet$ for $\akt$ jets with R = 0.6 for the rapidity
regions $|y| < 0.3$ (left plot) and $2.8 < |y| < 3.6$ (right plot). The black circles indicate the efficiency
measured in-situ using a tag-probe method, while the blue squares and the shaded band indicates the
parametrized central value and the systematic uncertainty on the efficiency obtained by varying the tag
jet selection. The turn-on is due to the fact that jet cleaning cuts are harder for low-$\ptjet$ jets.
}
\label{fig_qual}
\end{figure}

Jets with $\ptjet > 20$~GeV and $|\rapjet| < 4.4$ are selected, and required to pass quality criteria established to reject jets 
not coming from a proton-proton collision. Fake jets may originate from calorimeter noise, mainly noise bursts in the hadronic 
endcap calorimeter electronics and coherent noise from the electromagnetic calorimeter, or from cosmic rays or beam background. 
For each source of fake jets a quality criteria has been designed by combining some of the following variables (for more 
details see~\cite{fake}):
\begin{itemize}
\item The fraction of the jet energy deposited in the electromagnetic calorimeter.
\item The maximum energy fraction in one calorimeter layer. 
\item The energy fraction in the hadronic end-cap calorimeter.
\item Jet quality, defined as the fraction of LAr cells with a cell Q-factor greater than 4000. 
The cell Q-factor measures the difference between the measured pulse shape and the predicted pulse shape 
that is used to reconstruct the cell energy. 
\item HEC jet quality, defined as the LArQuality except that is calculated only with the HEC calorimeter.
\item Negative energy in the jet, due to an offset correction higher than the jet energy.
\item Jet time computed as the energy squared cells mean time.
\item Minimum number of cells containing at least 90\% of the jet energy.
\item Jet charged fraction, defined as the ratio of the sum of the $\ptjet$ of the tracks associated to 
a jet divided by the calibrated jet $\ptjet$.
\end{itemize}
A tag-and-probe method was use to determine the efficiency of the jet quality criteria, as shown in Figure~\ref{fig_qual} for two 
representative rapidity regions. For jets with $\ptjet > 100$~GeV, the efficiency is above $99\%$ in all rapidity regions. 
The maximum inefficiency, around 8\%, is for central jets at low $\ptjet$.
Jet cross sections are corrected for the inefficiency when it is larger than 1\%.

\begin{figure}[tbh]
\begin{center}
\mbox{
\includegraphics[width=0.495\textwidth]{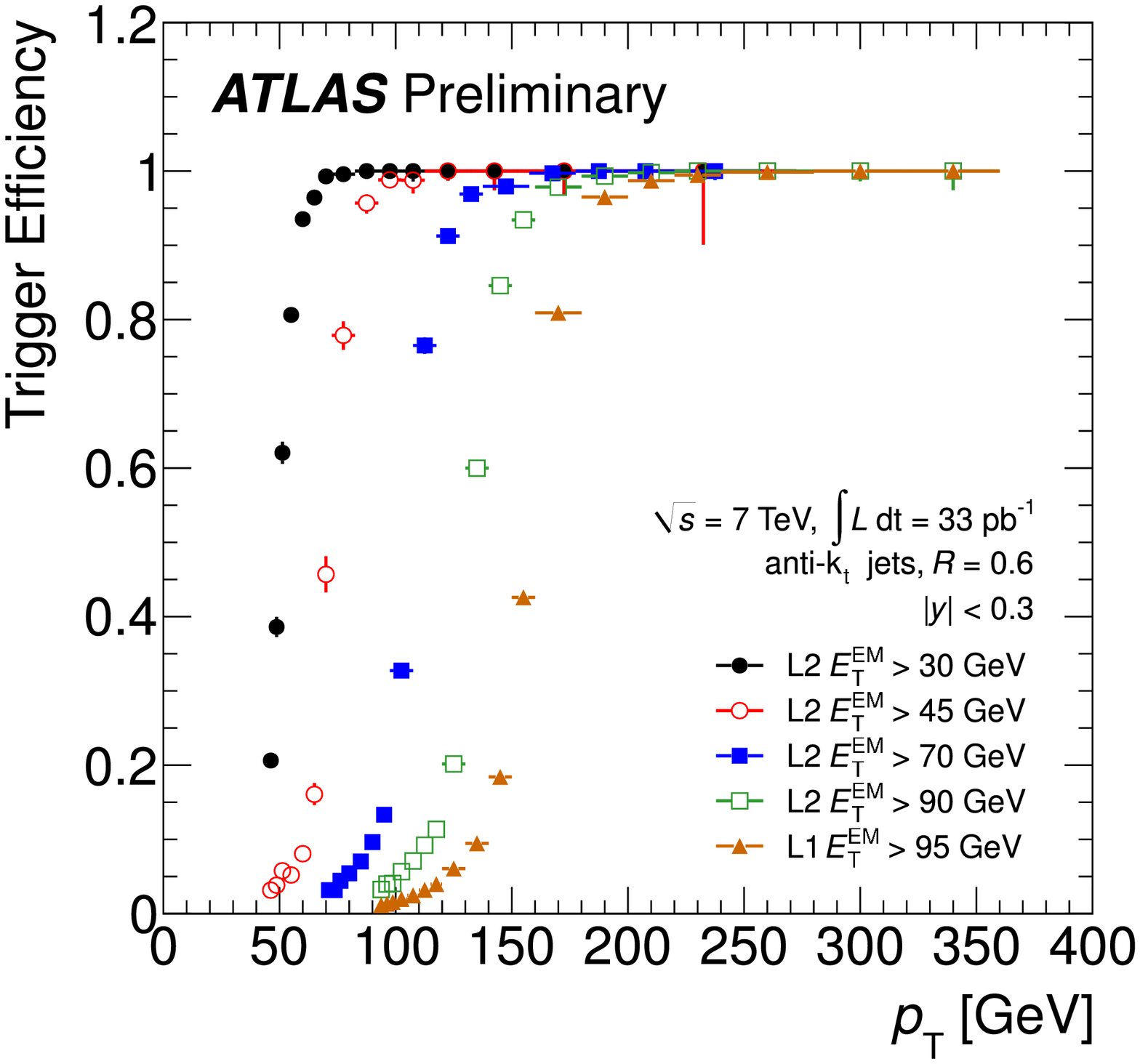}
\includegraphics[width=0.495\textwidth]{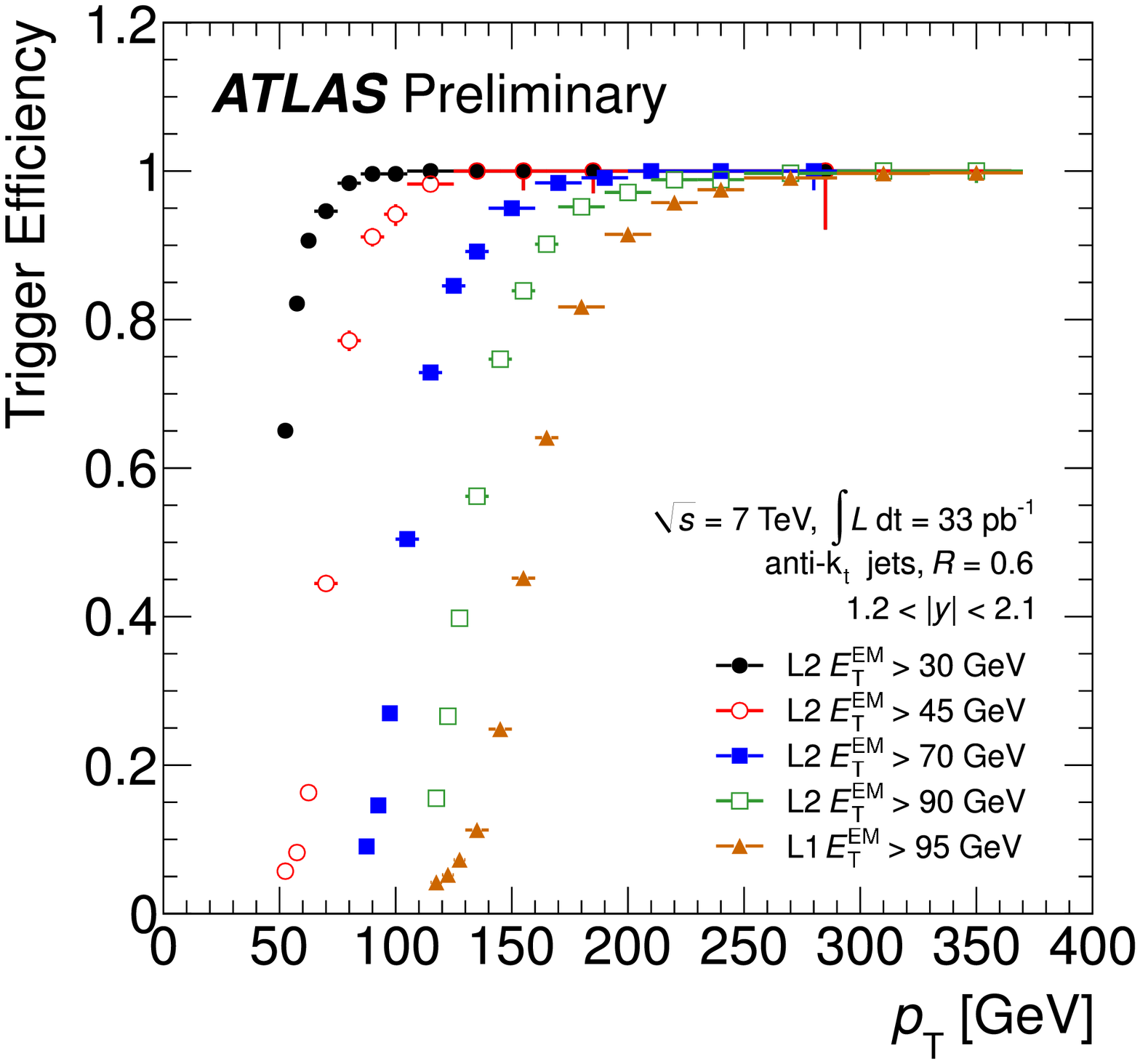}
}
\mbox{
\includegraphics[width=0.495\textwidth]{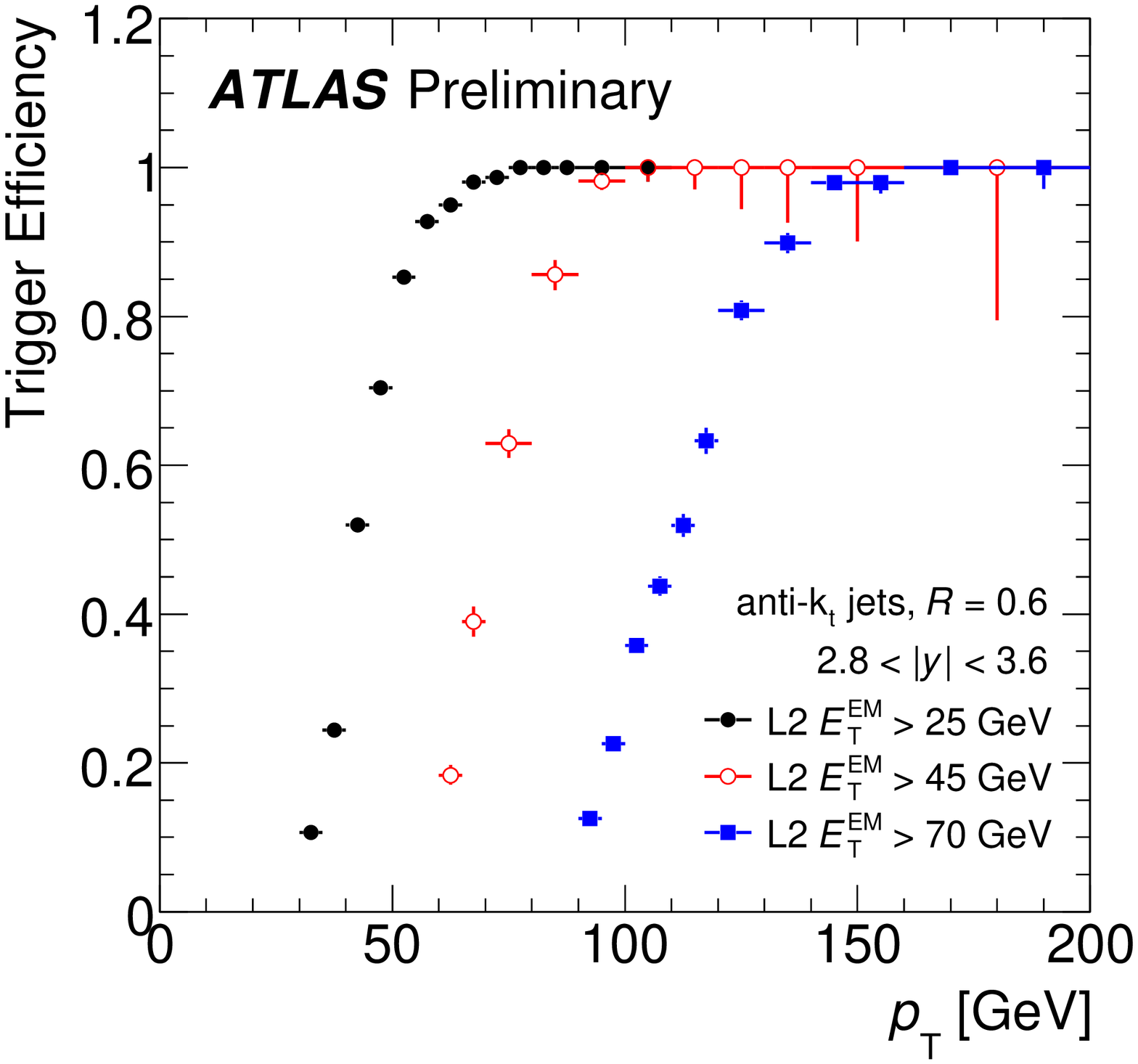}
\includegraphics[width=0.495\textwidth]{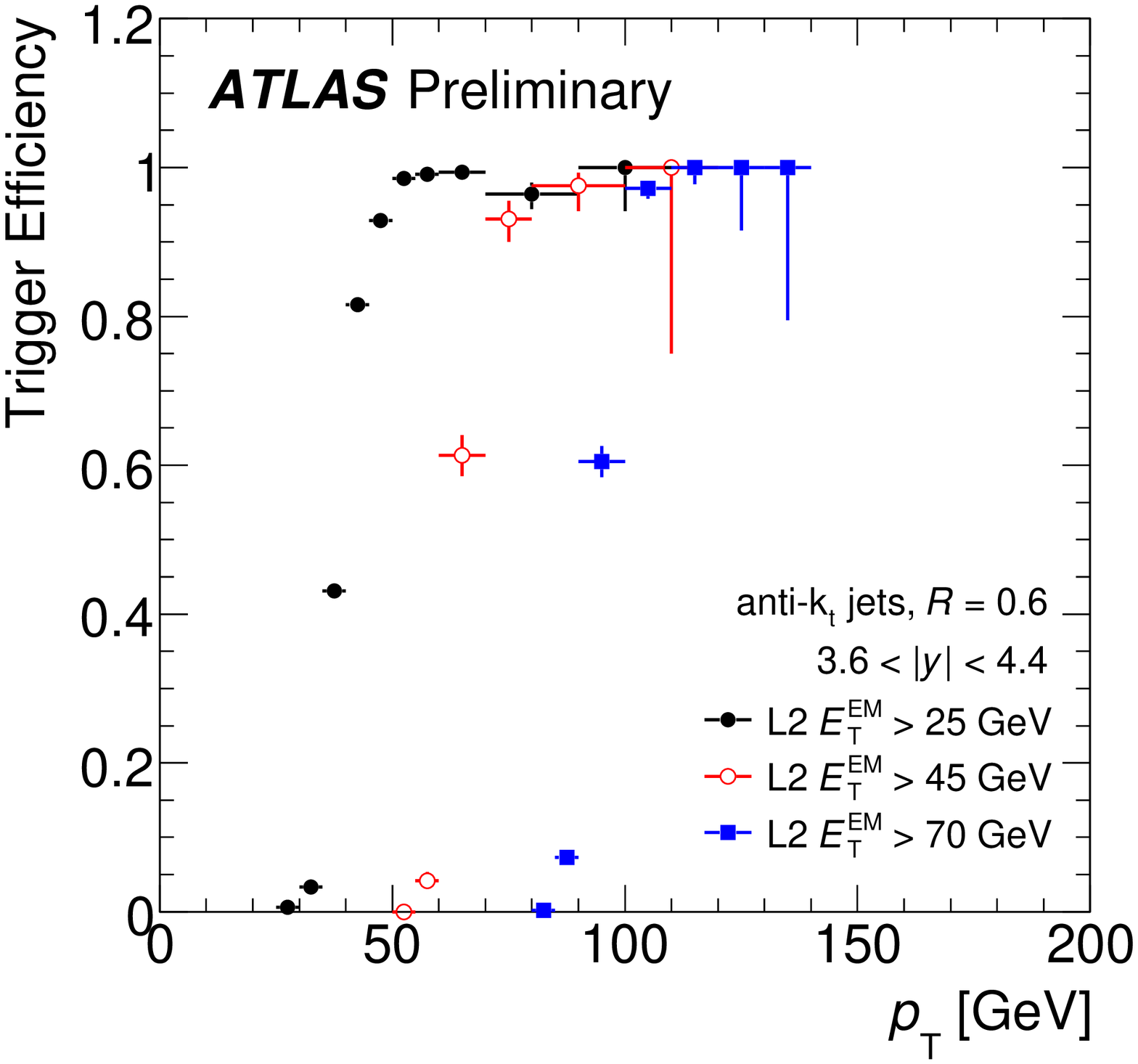}
}
\end{center}
\vspace{-0.7 cm}
\caption[Jet trigger efficiency as a function of reconstructed jet $\ptjet$]{\small
Combined L1+L2 jet trigger efficiency as a function of reconstructed jet $\ptjet$ for anti-kt jets with R = 0.6 
in four  representative $|y|$ regions are shown for different L2 trigger thresholds. The trigger thresholds are at the electromagnetic scale, 
while the jet $\ptjet$ is at the calibrated scale. The highest trigger chain does not apply a threshold at L2, so its L1 threshold is listed. 
}
\label{fig_trig}
\end{figure}

The MBTS trigger is used to select jets between 20~GeV and 60~GeV, whereas central and forward jet triggers 
are used to select jets with $\ptjet > 60$~GeV. In the first runs (Period 1), the jet trigger was using only the L1 decision 
to reject events, whereas starting from September (Period 2) the L2 decision was also used for rejection. The threshold in the L2 
is typically 15~GeV larger than the L1 threshold in the same trigger chain, except for the L1 10~GeV threshold 
that corresponds to the L2 15~GeV threshold. As explained in Section~\ref{sec_trig}, the trigger thresholds are 
at the EM scale. 
Figure~\ref{fig_trig} shows the trigger efficiency as a function of the trigger threshold and the 
calibrated jet $\ptjet$ in four representative $|y|$ regions.

Given the increase of the instantaneous luminosity during 2010, prescales have been applied 
to jet triggers with low $E_{T}$ thresholds in order to control the trigger rate. 
For each jet $\ptjet$ and rapidity bin in this analysis, a trigger is used to select jets such   
that is fully efficient ($> 99\%$) and that the prescale is as low as possible. 
The trigger efficiency for a given energy threshold, $E_{T}^{1}$, is computed with a sample 
of the events passing a trigger with a lower threshold, $E_{T}^{2} < E_{T}^{1}$. 
The efficiency is defined as the fraction of jets passing the trigger with $E_{T}^{1}$ threshold 
with respect to the number of jets passing the trigger with $E_{T}^{2}$ threshold. 

Tables~3.1 and~3.2 show the trigger used in each jet $\ptjet$ and $|y|$ bin. 
In the region $|y| < 2.8$, the central jet trigger (L1\_J, L2\_J) is used. 
The trigger has lower efficiency in $1.2 < |y| < 2.1$ than in the rest of the central region. 
This is due to the fact that between $\eta$ of 1.3 and 1.5, there is a transition between the
barrel and the end-cap of the liquid argon calorimeter. The electronic signals from
these two parts of the calorimeter are not added up at the trigger level to integrate over the full jet
energy deposition.
The logical OR of central and forward (L1\_JF, L2\_JF) jet triggers is used in the rapidity range $2.8 < |y| < 3.6$, since 
none of them are fully efficient alone. The forward jet trigger is used in the bin $3.6 < |y| < 4.4$.

\clearpage

\clearpage

\begin{table}[b!]
\begin{center}
\caption{Triggers used in the central region, $|y| < 2.8$.}
\begin{tabular}{|c|c|c|c|c|}
\hline
 & \textbf{Period 1} & \textbf{Period 1} & \textbf{Period 2} & \textbf{Period 2}\\
\hline
$\ptjet$ (GeV) & $|y| < 1.2~\|~|y| > 2.8$ & $1.2 < |y| < 2.1$ & $|y| < 1.2~\|~|y| > 2.8$ & $1.2 < |y| < 2.1$\\
\hline
60-80   & L1\_J5   & L1\_J5   & L2\_J15  &  L2\_J15\\
\hline
80-110  & L1\_J15  & L1\_J5   & L2\_J30  &  L2\_J15\\
\hline
110-160 & L1\_J30  & L1\_J15  & L2\_J45  &  L2\_J30\\
\hline
160-210 & L1\_J55  & L1\_J30  & L2\_J70  &  L2\_J45\\
\hline
210-260 & L1\_J75  & L1\_J55  & L2\_J90  &  L2\_J70\\
\hline
260-310 & L1\_J95  & L1\_J75  & L1\_J95  &  L2\_J90\\
\hline
310-400 & L1\_J115 & L1\_J95  & L1\_J115 &  L1\_J95\\
\hline
$> 400$ & L1\_J115 & L1\_J115 & L1\_J115 &  L1\_J115\\
\hline
\end{tabular}
\end{center}
\label{tab_trig1}
\end{table}

\begin{table}[b!]
\begin{center}
\caption{Triggers used in the forward region, $2.8 < |y| < 4.4$.}
\begin{tabular}{|c|c|c|c|c|}
\hline
 & \textbf{Period 1} & \textbf{Period 1} & \textbf{Period 2} & \textbf{Period 2}\\
\hline
$\ptjet$ (GeV) & $2.8 < |y| < 3.6$ & $3.6 < |y| < 4.4$ & $2.8 < |y| < 3.6$ & $3.6 < |y| < 4.4$\\
\hline
60-80   & L1\_J10 $\|$ L1\_FJ10  & L1\_J10  & not used              &  L2\_FJ25\\
\hline
80-110  & L1\_J10 $\|$ L1\_FJ10  & L1\_J30  & L2\_J25 $\|$ L2\_FJ25 &  L2\_FJ25\\
\hline
110-160 & L1\_J30 $\|$ L1\_FJ30  & L1\_J55  & L2\_J45 $\|$ L2\_FJ45 &  L2\_FJ45\\
\hline
160-210 & L1\_J55 $\|$ L1\_FJ55  & L1\_J55  & L2\_J45 $\|$ L2\_FJ45 &  L2\_FJ70\\
\hline
$> 210$ & L1\_J55 $\|$ L1\_FJ55  & L1\_J55  & L2\_J70 $\|$ L2\_FJ70 &  L2\_FJ70\\
\hline
\end{tabular}
\end{center}
\label{tab_trig2}
\end{table}

\clearpage

\section{Unfolding to the particle level}
\label{sec_unfolding}

The measured differential cross section with respect to the jet $\ptjet$ and $y$ is defined as follows:

\begin{equation}
\frac{d\sigma}{d\ptjet dy} = \frac{N^{jets}}{\mathcal{L} \Delta \ptjet \Delta y}
\label{eq_xs}
\end{equation}

\noindent where $N^{jets}$ is the number of jets in the $\Delta \ptjet$ and $\Delta y$ bins, and 
$\mathcal{L}$ is the integrated luminosity.

This cross section is corrected for detector effects back to the particle level with a bin-by-bin unfolding procedure, using  
PYTHIA~6.423 with the AMBT1 tune simulated samples. In order to improve the agreement between the MC and the data, 
the MC is reweighted in order to scale its prediction, that uses a modified LO PDF, to that of a NLO PDF.
Corrections are derived in each $\ptjet$ and $\rapjet$ bin by computing the ratio between 
the inclusive jet $\ptjet$ distributions using truth jets (including muons and neutrinos) and using calorimeter simulated jets.
These corrections are multiplied to equation~\ref{eq_xs}, and have values between 0.8~and~1.0 in most jet $\ptjet$ regions, and around~0.6 
at very low and very high jet $\ptjet$.

\section{Systematic Uncertainties}

In this section several sources of systematic uncertainty on the measurement are considered. More emphasis is given to 
the explanation of the JES uncertainty, since it is the main contribution to the total uncertainty on the inclusive jet 
cross section measurement.

\subsection{JES uncertainty}

The JES uncertainty comes mainly from the uncertainty on the single particle response, 
limitations in the detector knowledge (such as the amount of dead material), 
and the physics models and parameters (mainly fragmentation and underlying event) in the MC event 
generator used to derive the JES corrections. 
Seven $\eta_{det}$ bins have been selected to estimate the JES uncertainty, taking into account the position 
of the different calorimeters (see Figure~\ref{fig_jet_response}).
The JES uncertainty is first calculated in the region $|\eta_{det}| < 0.8$ and then propagated forward in rapidity using 
a jet $\ptjet$ balance technique, explained later in this section. The reason is 
that the detector is better known in the central region, and that test-beam measurements 
to estimate the uncertainty due to the calorimeter response to single particles 
were only performed in the range $|\eta_{det}| < 0.8$. 

\subsubsection{Uncertainty on the calorimeter response to single particles}

The calorimeter response to single hadrons has been studied comparing the calorimeter energy response to the 
momentum of isolated tracks~\cite{E_over_p}. Results have been obtained from pp collisions at 
$\sqrt{s} =$~0.9 and 7~TeV, finding an agreement between data and MC within 2-5\% for tracks with momentum 
between 0.5 to 20~GeV. 
The calorimeter response to particles with momentum in the range between 20~GeV and 350~GeV has been derived from test-beam measurements, 
in which pion beams with momentum up to 350~GeV were injected into a slice of the ATLAS detector. 
For particles with larger momentum, an additional uncertainty of 10\% on top of the 350~GeV measured uncertainty is 
applied in order to take into account longitudinal leakage and calorimeter non-linearities. 
The uncertainty of the single particle response in the calorimeter has been used to derive the related jet energy scale uncertainty. 
This is done using MC, and varying randomly the energy of the particles of the jets separately, and according to the 
values of the single particle response uncertainty. The JES calibration constants are then computed 
with these modified jets and the results are compared to the nominal constants in order to derive the uncertainty. 
The uncertainty on the jet $\ptjet$ increases from $1.5\%$ to $4\%$ with increasing jet $\ptjet$.

\subsubsection{Uncertainties due to the detector simulation}

Since the JES calibration is derived using MC samples, deviations in the description of the 
detector in the MC introduce an uncertainty on the JES. This has been estimated by varying the material budget in the 
simulation. In the particle range of the E/p studies using isolated tracks only the dead material in the inner 
detector has been varied, since the measurement is already performed with the ATLAS detector.
For other particles, also changes in the calorimeter material have been simulated.
The jet response in these modified samples is compared to the nominal one in order to estimate 
the systematic uncertainty, that is always below $1\%$.

As explained in Section~\ref{sec_cluster}, clusters are build by adding neighboring cells with signal over noise above 
certain thresholds. Therefore, miss-modeling of the calorimeter cell noise in the MC would lead to different cluster 
formation that eventually would affect the jet reconstruction. The impact of possible miss-modeling of the noise has 
been evaluated by varying the signal over noise thresholds in the cluster reconstruction. 
The difference of the jet energy response for jets reconstructed with 
the modified noise thresholds and the nominal jets has been taken as a systematic uncertainty. 
This goes up to $2\%$ for jets with $\ptjet < 45$~GeV, and is negligible at higher $\ptjet$.

\subsubsection{Uncertainties due to the event modelling in the Monte Carlo generators}

The jet energy response obtained using PYTHIA with the AMBT1 tune, used to derive the corrections, 
has been compared with those obtained using ALPGEN+HERWIG+JIMMY and PYTHIA with the Perugia2010 tune. 
This estimates the uncertainty due to the different QCD description contained in the MCs, such as 
fragmentation or underlying event. It goes from $2\%$ at low $\ptjet$ to $1\%$ for $\ptjet > 100$~GeV.

\subsubsection{Uncertainties due to the relative calibration in the forward regions}

As already mentioned, the JES uncertainties are evaluated for jets with $|\eta| < 0.8$. These uncertainties are transferred to the forward 
region by using a $\ptjet$ balance technique in which the $\ptjet$ of forward jets is measured relative to that of 
central jets in dijet events. The difference of the $\ptjet$ balance of a central and a forward jet in data and in MC 
is taken as a systematic uncertainty, that is added in quadrature to the JES uncertainty in the central region.  
This is the largest source of uncertainty at low $\ptjet$ in the forward region, 
going up to around 12\%. For jets with $\ptjet > 100$~GeV it is always smaller than 2\%.

\subsubsection{Uncertainty of the method to derive the JES calibration constants}

There is an uncertainty associated to the method used to derive the JES calibration constants. It comes mainly 
from the assumption that every jet needs the same average compensation and from the fact that the same correction factor 
is used for the energy and the $\ptjet$, whereas the jet mass is not always zero. 
Two statistical independent MC samples were used to derive the uncertainty associated to the method itself. One was used to 
derive the corrections, that were applied to the calorimeter jets of the other. Deviations of the jet response from unity in this 
second sample are taken as systematic uncertainties, that are generally below $1\%$, apart from the central region, where 
the uncertainty amounts up to $2\%$ at low $\ptjet$.

\subsubsection{Total uncertainty on the jet $\ptjet$}

\begin{figure}[tbh]
\begin{center}
\mbox{
\includegraphics[width=0.65\textwidth]{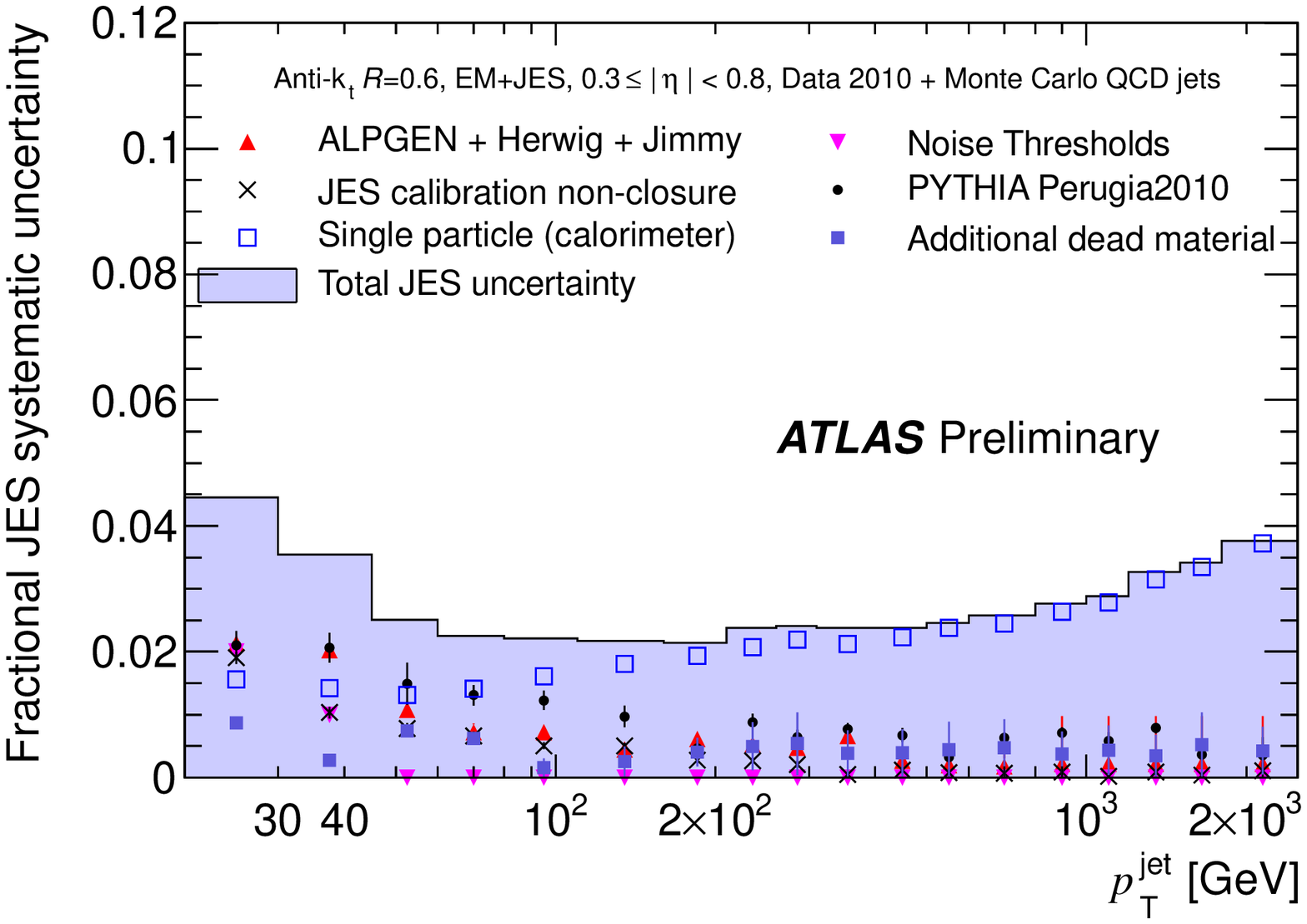}
}
\mbox{
\includegraphics[width=0.65\textwidth]{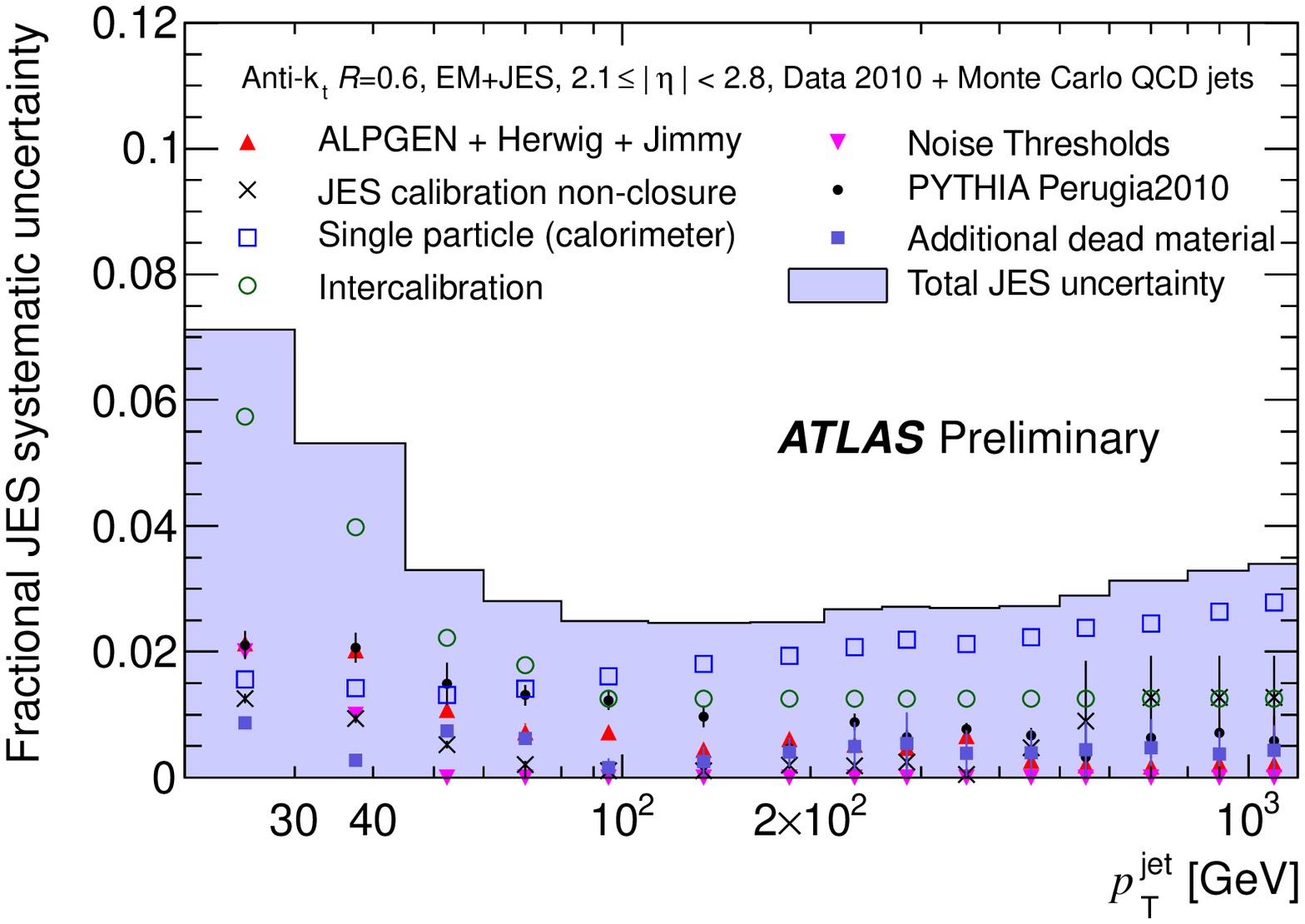}
}
\mbox{
\includegraphics[width=0.65\textwidth]{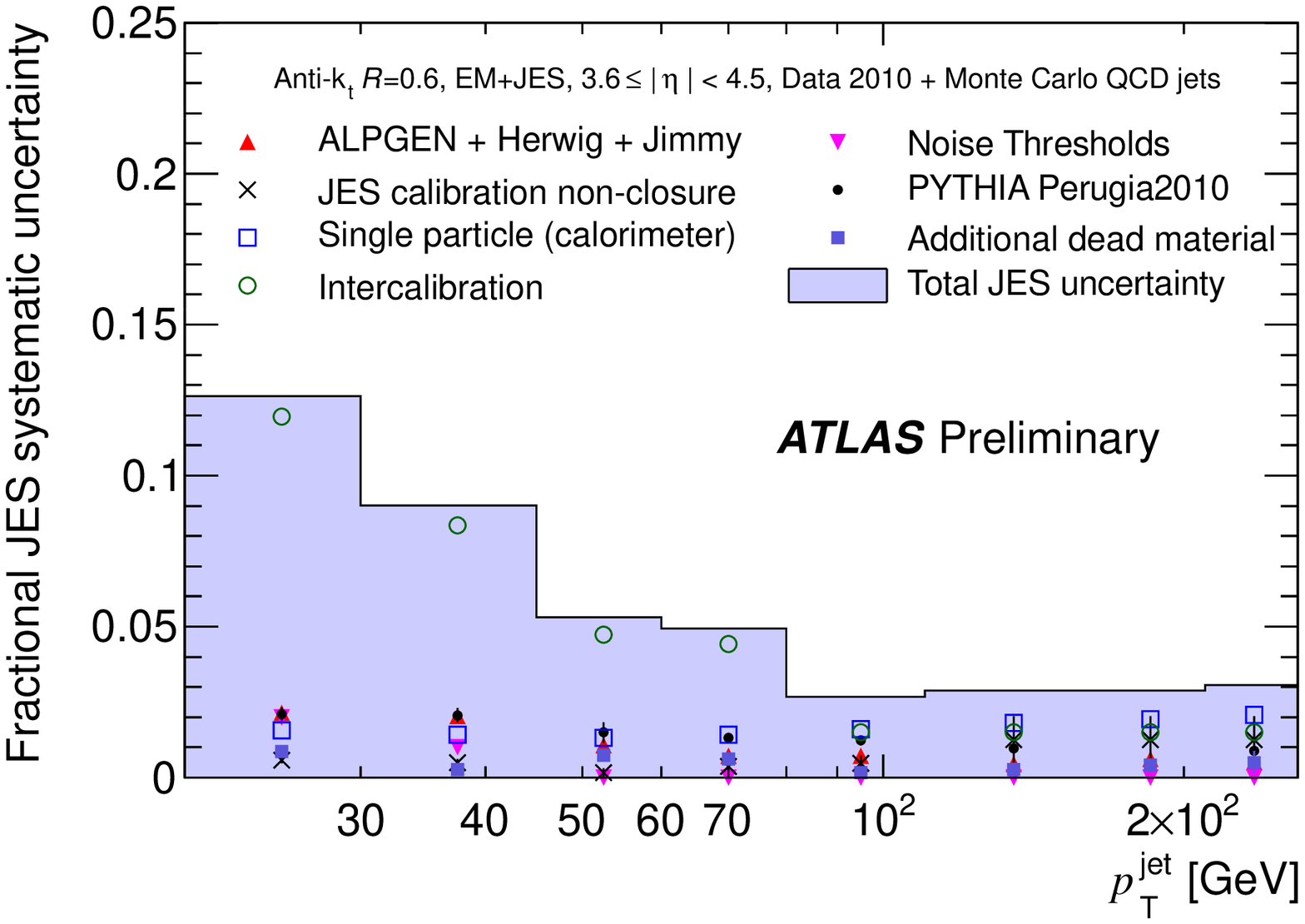}
}
\end{center}
\vspace{-0.7 cm}
\caption[Fractional JES uncertainty as a function of $\ptjet$]{\small
Fractional JES uncertainty as a function of $\ptjet$ for jets in different $\eta$ regions. 
The JES uncertainty for regions $|\eta| > 0.8$ 
contains the contribution from the $\eta$ inter-calibration between central and forward jets in
data and Monte Carlo added in quadrature. The total uncertainty is shown as the solid light blue area, and
the individual sources are also shown.
}
\label{fig_JESUncertainty}
\end{figure}

In Figure~\ref{fig_JESUncertainty}, the fractional JES uncertainty and its components are shown as a function of $\ptjet$ in three 
representative $\eta$ regions. The fractional JES uncertainty is below $4\%$ for $\ptjet > 80$~GeV in all $\eta$ regions, it is 
smaller than $5\%$ for all $\ptjet$ in the central region, and amounts up to $13\%$ at low $\ptjet$ in the most forward region. 

\subsubsection{Uncertainty due to multiple interactions}

\begin{figure}[tbh]
\begin{center}
\includegraphics[width=0.65\textwidth]{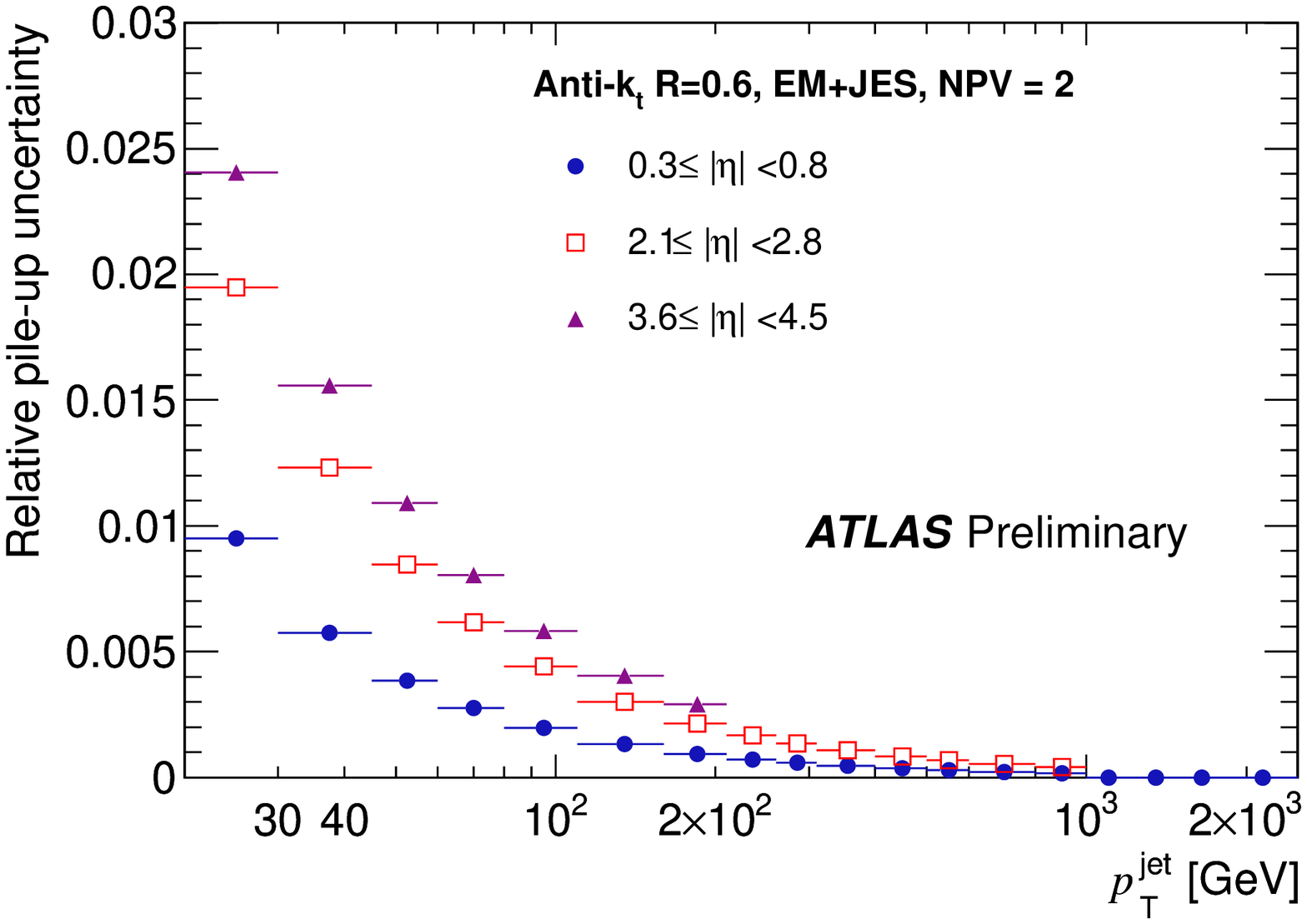}
\end{center}
\vspace{-0.7 cm}
\caption[Relative pile-up uncertainty in the case of two measured primary vertices]{\small
Relative pile-up uncertainty in the case of two measured primary vertices, for central ($0.3 < | \eta | < 0.8$, full circles), 
endcap ($2.1 < | \eta | < 2.8$, open squares) and forward ($3.6 < | \eta | < 4.5$, full triangles) jets as a function of $\ptjet$.
}
\label{fig_jes_pileup}
\end{figure}

A final contribution to account for the uncertainty due to pile-up is added in quadrature in events with more
than one primary vertex. This uncertainty is estimated mainly from the variation of the offset corrections derived 
using tracks instead of calorimeter towers. 
It is shown in Figure~\ref{fig_jes_pileup} for events with two primary vertices.
In this case, the uncertainty due to pile-up for low $\ptjet$ jets is around 1\% in the central region and almost 
2.5\% in the most forward region. The uncertainty is always below 1\% for jets with $\ptjet > 60$~GeV. 
In the case of three primary vertices, the pile-up uncertainty is approximately twice that of two primary vertices, 
and with four primary vertices the uncertainty for central, endcap
and forward jets is less than 3\%, 6\% and 8\%, respectively. The relative uncertainty due to pile-up for
events with 5 additional interactions becomes less than 1\% for all jets with $\ptjet > 200$~GeV.

\subsubsection{Uncertainty on the measurement due to the JES uncertainty}

The JES uncertainty is the largest contribution to the uncertainty of the inclusive jet cross section measurement,  
that goes from around +30\%/-20\% in the central region to approximately +80\%/-50\% in the most forward region.

\subsection{Other sources of systematic uncertainties}

\begin{itemize}
\item The \textbf{unfolding factors} have been recomputed varying the jet $\ptjet$ in the MC in order to take into account 
the jet energy and angular resolution, and the differences on the cross section shape in data and in MC. These 
unfolding factors have been compared to estimate the systematic uncertainty, that is typically between $2\%$ and $5\%$, 
except for the lowest $\ptjet$ bin where it goes up to 20\%.

\item Matching of track jets to calorimeter jets has been performed in both data and Monte Carlo to 
estimate the modeling of the calorimeter \textbf{jet reconstruction efficiency} in the MC. The difference is smaller than 
2\% (3\%) for jets with $\ptjet > 20$~GeV (30 GeV), and is taken as a systematic uncertainty.

\item In this analysis, jets are selected with \textbf{triggers} that are 99\% efficient. A conservative 1\% 
uncertainty overall has been considered to account for the trigger inefficiency.

\item The systematic uncertainty on the efficiency of the \textbf{jet quality criteria} is taken as a systematic uncertainty on 
the cross section, and it is always below 1\%.

\item Finally, the uncertainty on the \textbf{luminosity} measurement is 3\%.

\end{itemize}

\clearpage

\section{Results}

\begin{figure}[tbh]
\begin{center}
\includegraphics[width=0.85\textwidth]{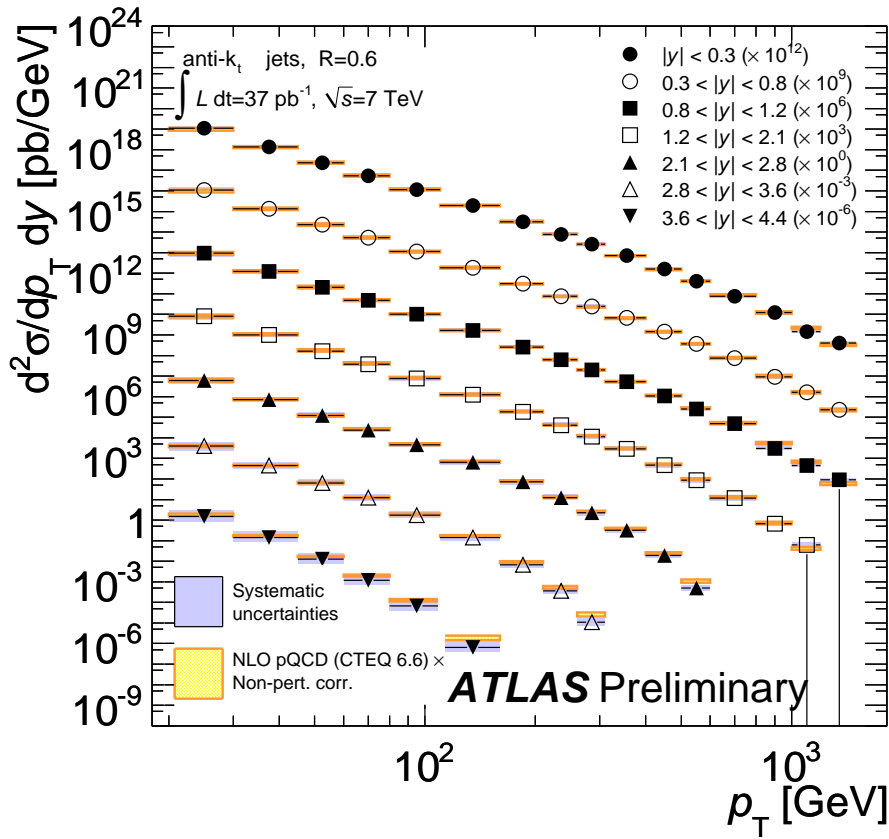}
\end{center}
\vspace{-0.7 cm}
\caption[Inclusive jet cross section as a function of $\ptjet$ in different regions of $|y|$]{\small
Inclusive jet cross section as a function of $\ptjet$ in different regions of
$|y|$ for jets identiﬁed using the $\akt$ algorithm with R = 0.6. For convenience, the cross sections are
multiplied by the factors indicated in the legend. The data are compared to NLO QCD calculations to
which non-perturbative corrections have been applied. The error bars indicate the statistical uncertainty
on the measurement, and the shaded band indicates the quadratic sum of the systematic uncertainties. 
There is an additional overall uncertainty of $3\%$ due to the luminosity measurement that is not shown.
}
\label{fig_results}
\end{figure}

The inclusive jet cross section unfolded to the hadron level is shown in Figure~\ref{fig_results} for different $|y|$ regions up to 
$|y| < 4.4$. It is measured from 20~GeV to 1.5~TeV and spans up to ten orders of magnitude. 
The data are compared to NLO predictions corrected for non-perturbative effects.
The NLO predictions are obtained from NLOJET++4.1.2 program with CTEQ 6.6 PDFs, and the  
corrections for non-perturbative effects are derived using samples produced with PYTHIA~6.423 with the AMBT1 tune and applied to NLO predictions. 
These corrections are obtained bin-by-bin, comparing the cross section with and without hadronization and underlying event. 
The correction is dominated by the underlying event at low 
$\ptjet$ (1.5 at 20 GeV), and tends to 1 as the $\ptjet$ increases. 
The uncertainty on the NLO predictions is obtained adding in quadrature 
uncertainties from the PDFs, the choice of factorization and renormalization scales, and the value of the strong coupling constant. 
It is combined with the uncertainty on the non-perturbative corrections, estimated from the maximum difference 
of the corrections with AMBT1 with respect to other PYTHIA tunes and to HERWIG++. The total uncertainty in the predictions 
is typically around 20\% at low and high $\ptjet$ and 10\% at intermediate $\ptjet$ values.

\begin{figure}[tbh]
\begin{center}
\mbox{
\includegraphics[width=0.65\textwidth]{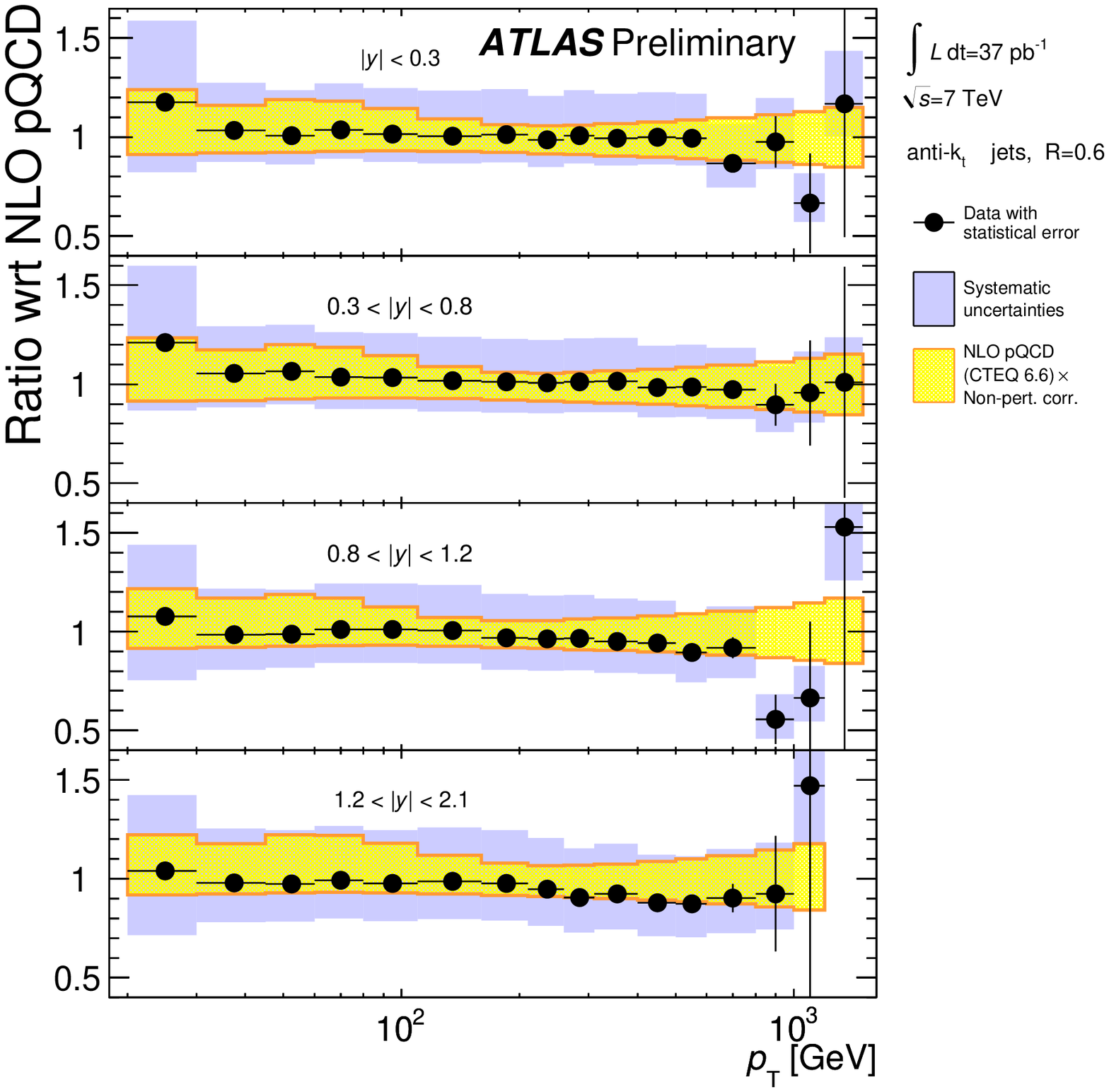}
}
\mbox{
\includegraphics[width=0.65\textwidth]{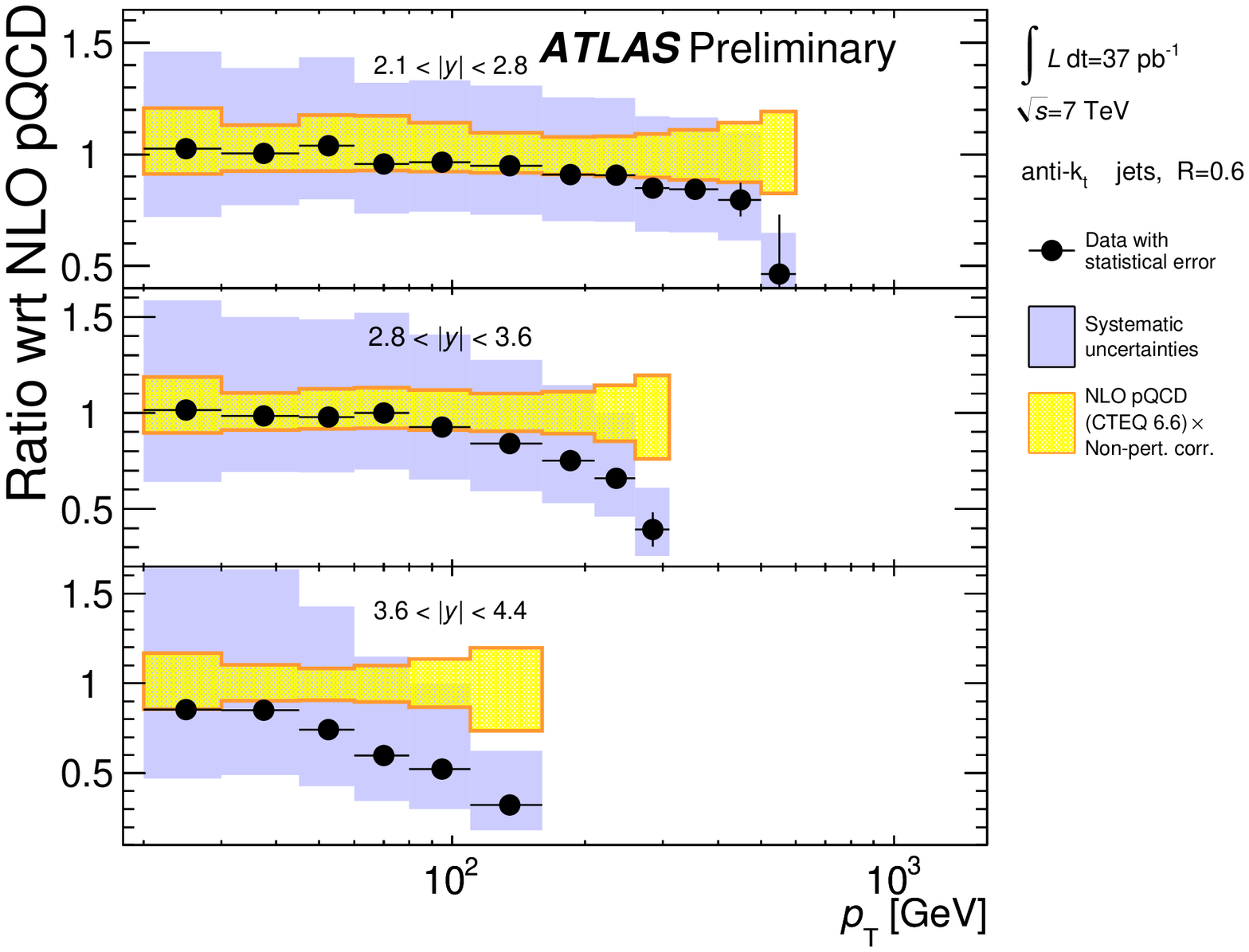}
}
\end{center}
\vspace{-0.7 cm}
\caption[Ratio of the inclusive jet cross section as a function of $\ptjet$ in different regions of $|y|$ of the 
data to the theoretical predictions]{\small
Inclusive jet cross section as a function of $\ptjet$ in different regions of
$|y|$ for jets identiﬁed using the $\akt$ algorithm with R = 0.6. The ratio of the data to the theoretical
prediction is shown. The total systematic uncertainties on the theory and measurement, calculated as 
described in Figure~\ref{fig_results}, are indicated. 
}
\label{fig_ratio}
\end{figure}

Figure~\ref{fig_ratio} shows the ratio of the measured cross section in 
data and the theory prediction, that are in agreement within uncertainties. The cross section 
in data is lower than the cross section predicted by NLO predictions in the forward region and at high $\ptjet$. 

\begin{figure}[tbh]
\begin{center}
\mbox{
\includegraphics[width=0.65\textwidth]{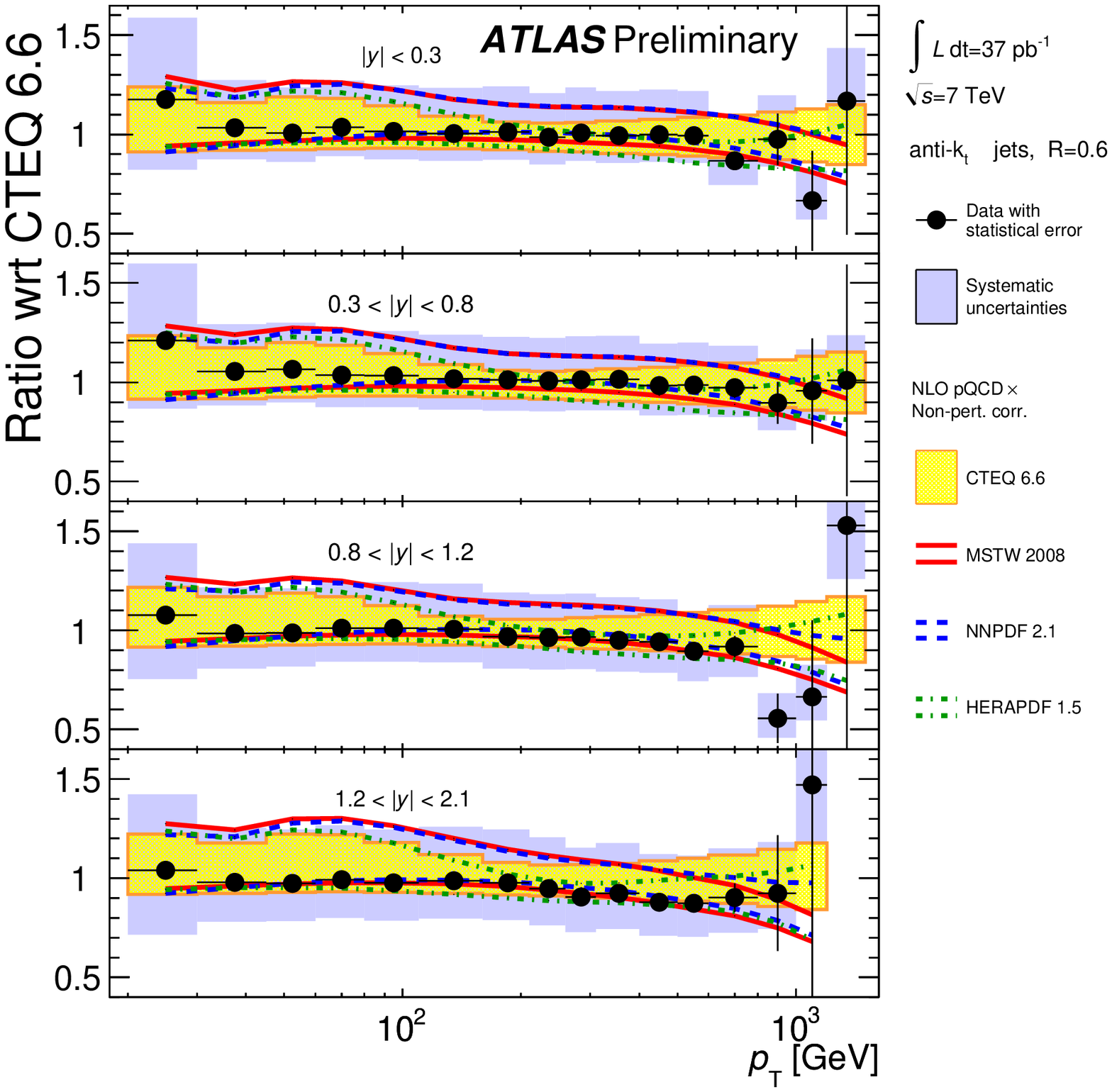}
}
\mbox{
\includegraphics[width=0.65\textwidth]{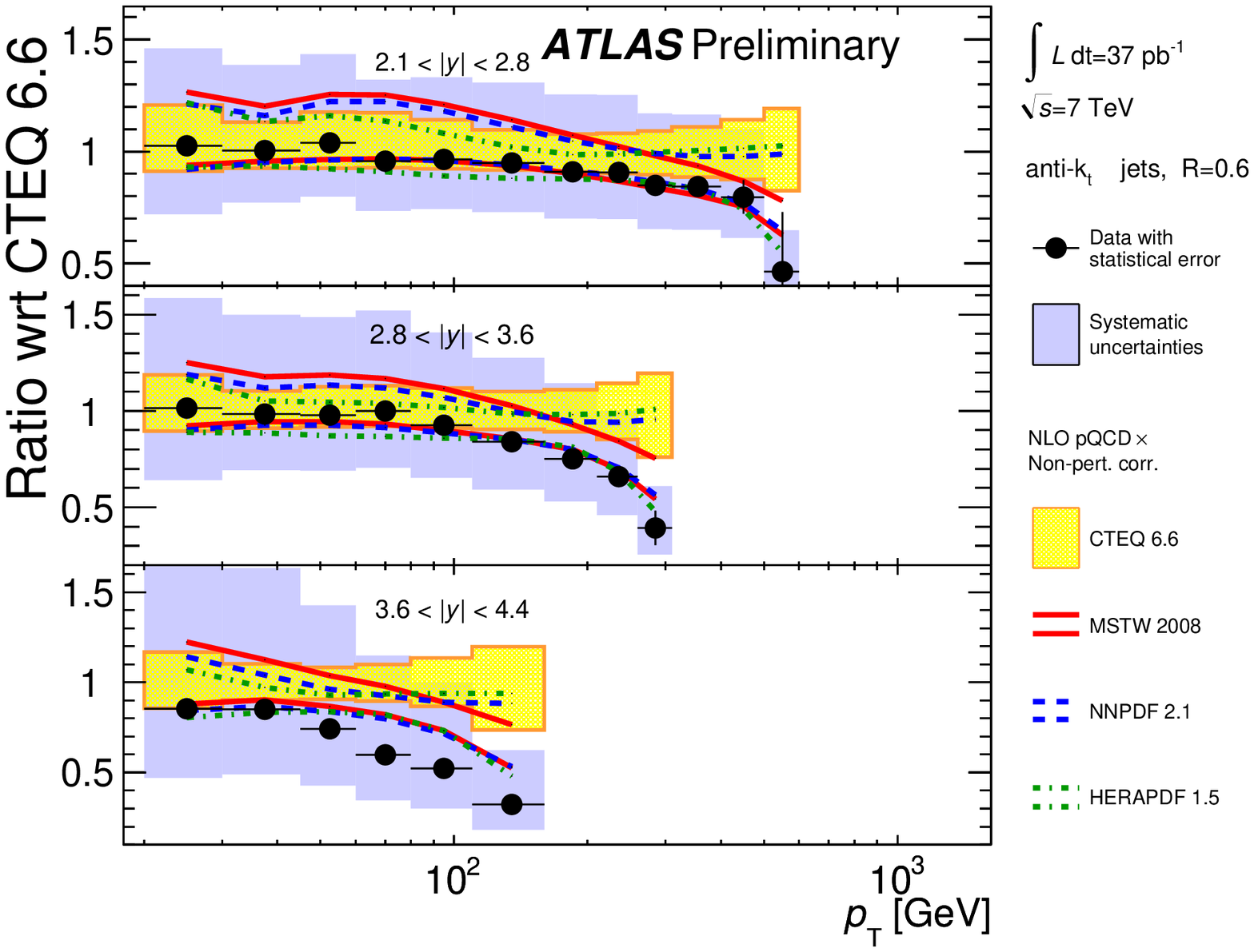}
}
\end{center}
\vspace{-0.7 cm}
\caption[Ratio of the inclusive jet cross section as a function of $\ptjet$ in different regions of $|y|$ of the 
data to the theoretical predictions (inclusion of results with other PDFs)]
{\small Inclusive jet cross section as a function of $\ptjet$ in different regions
of $|y|$ for jets identiﬁed using the $\akt$ algorithm with R = 0.6. The theoretical error bands obtained by
using different PDF sets (CTEQ 6.6, MSTW 2008, NNPDF 2.1, HERA 1.5) are shown. The data points
and the error bands are normalized to the theoretical estimates obtained by using the CTEQ 6.6 PDF set.
}
\label{fig_pdfs}
\end{figure}

In Figure~\ref{fig_pdfs}, data are compared to results using CTEQ 6.6, MTSW 2008, NNPDF 2.1, and HERAPDF 1.5. All cross sections are normalized 
to that obtained with CTEQ 6.6. Predictions using MTSW 2008, NNPDF 2.1, and HERAPDF 1.5 are closer to data than those using CTEQ 6.6, but 
all have the tendency to produce higher cross sections than data at high $\ptjet$ in the forward region. This may be due to the fact that 
PDFs are currently poorly constrained in this kinematic region.

\begin{figure}[tbh]
\begin{center}
\mbox{
\includegraphics[width=0.65\textwidth]{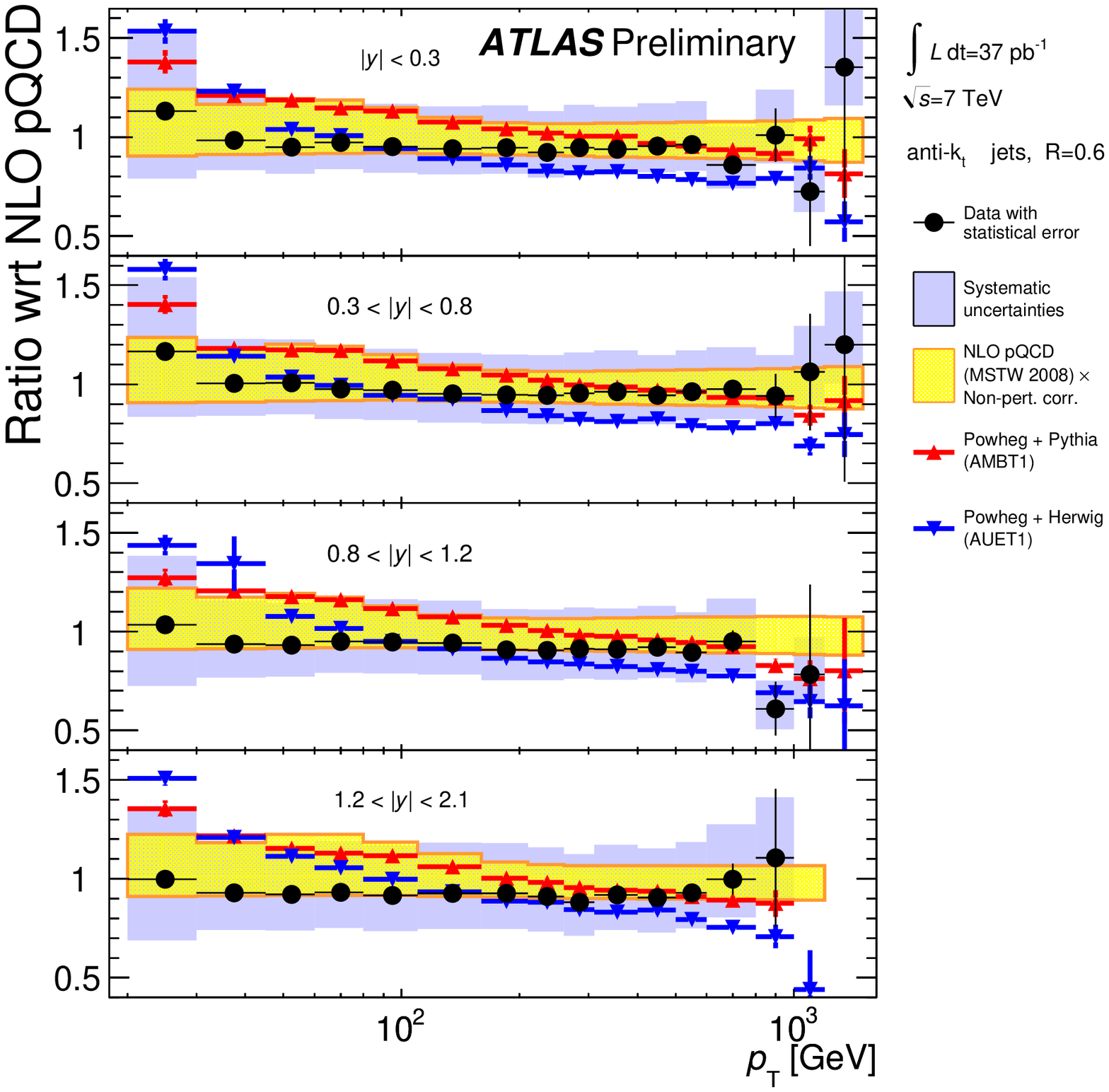}
}
\mbox{
\includegraphics[width=0.65\textwidth]{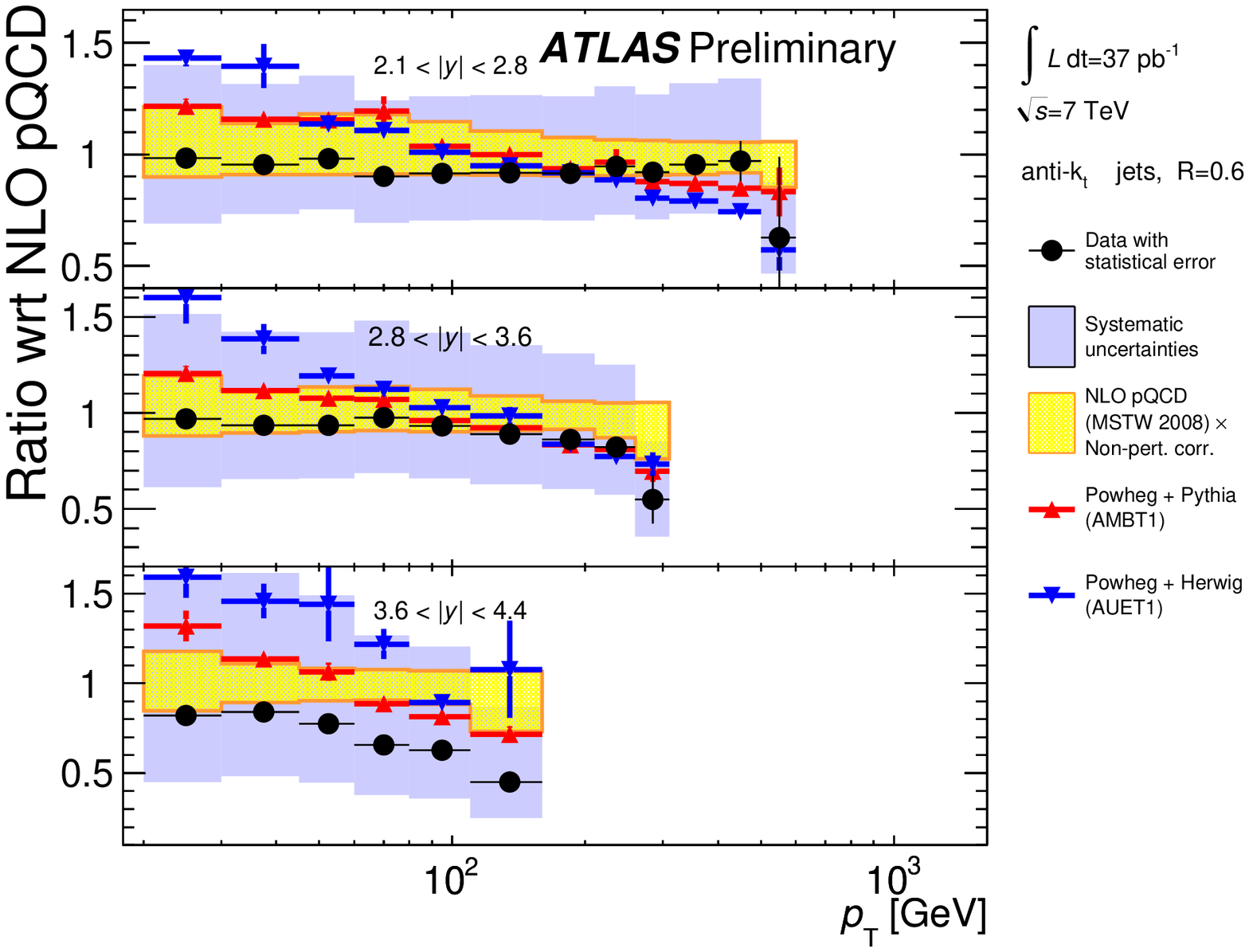}
}
\end{center}
\vspace{-0.7 cm}
\caption[Ratio of the inclusive jet cross section as a function of $\ptjet$ in different regions of $|y|$ of the
data to the theoretical predictions (inclusion of Powheg results)]
{\small
Inclusive jet cross section as a function of $\ptjet$ in different regions of $|y|$
for jets identiﬁed using the $\akt$ algorithm with R = 0.6. 
The ratios of the POWHEG predictions showered
by PYTHIA and HERWIG and the data to the NLO predictions corrected for the non-perturbative effects is shown. The
ratio shows only the statistical uncertainty on the POWHEG prediction. The total systematic
uncertainties on the theory and measurement are indicated. 
}
\label{fig_powheg}
\end{figure}

Finally, Figure~\ref{fig_powheg} shows the comparison of data and POWHEG predictions using MTSW 2008, 
both normalized to the NLO with MTSW 2008 cross section shown in Figure~\ref{fig_pdfs}. 
POWHEG is interfaced with either PYTHIA or HERWIG for the parton showering, the hadronization 
and the underlying event modeling. 
POWHEG predictions agree with data within uncertainties, but tend to produce larger cross section 
at low $\ptjet$ and, mainly POWHEG+HERWIG, smaller cross sections at high $\ptjet$. 
POWHEG gives different results depending on whether PYTHIA or HERWIG is used for the showering, 
which may be taken as an indication of the level of uncertainty related to the parton shower implementation. 

\clearpage

\clearpage

\clearpage

\newpage
\chapter{Jet Shapes}
\label{chap4}

One of the fundamental elements of jet measurements is the proper understanding of the energy flow
around the jet core and the validation of the QCD description contained in the event generators, such as
parton shower cascades, and the fragmentation and underlying event models. Jet shapes~\cite{jetshape} observables are
sensitive to these phenomena and thus very adequate to this purpose. They have been measured in 
$p\bar{p}$~\cite{ppbar}, $e^{\pm}p$~\cite{ep}, and $e^+ e^-$~\cite{ee} colliders. 

Before the first pp collisions delivered by the LHC, the sensitivity of jet shapes to the 
UE and pile-up in ATLAS was shown using MC simulation 
at 10~TeV~\cite{mc_shapes} (see Appendix~\ref{appendix_MC}). 
Jet shapes studies at detector level were performed with the very
first ATLAS data at a center of mass energy of 900~GeV~\cite{900GeV_shapes}, \cite{int_shapes} 
(see Appendix~\ref{appendix_900}).

Results at calorimeter level using 7~TeV data were used in the first measurement of the inclusive 
jet cross section in ATLAS~\cite{jet_pro_atlas} based on 17~nb${}^{-1}$ to validate the QCD description in the event generators 
and to consolidate the confidence in the corrections and calibrations applied. Jet shapes at calorimeter level in the forward 
region ($|y| > 2.8$) have been computed with the same purpose for the inclusive jet cross section measurement using the full 2010 
dataset~\cite{jet_conf}, corresponding to 37~pb${}^{-1}$.

In this Chapter, measurements of differential and integrated jet shapes 
in proton-proton  collisions at $\sqrt{s}=7$~TeV are presented~\cite{pp}, using data 
collected by the  ATLAS experiment corresponding to $3$~pb${}^{-1}$ of total integrated luminosity.
The definition of jet shapes observables and the event selection criteria are explained, followed 
by a description of Monte Carlo samples used and a detailed discussion on the corrections for detector effects
and systematic uncertainties. 
Finally, results are presented and compared to predictions from several MC event generators.


\section{Jet shape definition}

The differential jet shape $\rho(r)$ as a function of the distance  $r=\sqrt{\Delta y^2 + \Delta \phi^2}$ to the jet axis is
defined as the average fraction of the jet $\ptjet$ that lies inside an annulus of inner radius
${r - \Delta r/2}$ and outer radius ${r + \Delta r/2}$ around the jet axis:

\begin{equation}
\rho (r) = \frac{1}{\Delta r}\frac{1}{\njet}\sum_{\rm jets}\frac{p_{T}(r-\Delta r/2,r+\Delta r/2)}{p_{T}
(0,R)},\ \Delta r/2 \le r \le R-\Delta r/2, 
\end{equation}

\noindent
where $  p_{T}(r_1,r_2)$ denotes the scalar sum of the $p_{T}$ of the jet constituents in the annulus between 
radius $r_1$ and $r_2$, $\njet$ is the number of jets, and $R=0.6$ and $\Delta r = 0.1$ are used. 
 The points from the differential jet shape 
at different $r$ values are correlated since, by definition, $\sum_0^R \rho (r) \ \Delta r = 1$. 
Alternatively, the integrated jet shape $\Psi(r)$ is defined as the 
average fraction of the jet $\ptjet$ that lies inside a cone of radius $r$ 
concentric with the jet cone:

\begin{equation}
\Psi (r) = \frac{1}{\njet}\sum_{\rm jets}\frac{p_{T}(0,r)}{p_{T}(0,R)},\ 0 \le r \le R, 
\end{equation}

\noindent
where, by definition, $\Psi (r = R) = 1$, and the points at different $r$ values are strongly correlated. 
This geometrical definition of the shape of the jet, based on concentric cones around the jet axis, is 
particularly adequate for jets reconstructed using the \akt algorithm.

In this analysis, the jet finding algorithm is run over uncorrected energy clusters or towers (calorimeter level), 
Inner Detector tracks, or final-state particles in  the MC generated events (particle level).
The jet shape measurements are performed in different regions of jet $\ptjet$ and $|\rapjet|$, and a minimum 
of 100 jets in data are required in each region to limit the statistical 
fluctuations on the measured values.

\section{Event selection}

The event and jet selection criteria, close to that of the inclusive jet cross section measurement 
detailed in Section~\ref{sec_criteria}, are:

\begin{itemize}
\item Detector components relevant for this analysis, such as the calorimeter and the Inner Detector, operating at the nominal conditions.
\item Only data from runs until the end of August are used, in order to avoid bunch train 
pile-up~\footnote{Bunch train pile-up refers to the energy deposit coming from pp collisions in a 
different bunch crossing than the one of the event under consideration.}.
This corresponds to $3$~pb${}^{-1}$ of total integrated luminosity collected by the  ATLAS experiment.
\item The trigger is required to be fully efficient in each $\ptjet$ and $y$ bin.
\item The events are required to  
have one and only one reconstructed primary vertex with 
a $z$-position within 10~cm of the origin of the coordinate system, which
suppresses pile-up contributions from multiple proton-proton interactions in the same bunch crossing, beam-related backgrounds and cosmic rays. The tight cut on the $z$-position is performed in order to keep the projective geometry of the calorimeter. 
\item Only jets with corrected transverse momentum $\ptjet > 30$~GeV and rapidity $|\rapjet| < 2.8$ are considered.
\end{itemize}

  
\section{Monte Carlo simulation}

Monte Carlo simulated  samples are used to correct the jet shapes for detector effects 
back to the hadron level, and to estimate part of the systematic uncertainties. 
PYTHIA~6.4.21, HERWIG++~2.4.2 and ALPGEN 2.13 event generators are used to produce 
inclusive jet events in proton-proton collisions at $\sqrt{s} = 7$~TeV.
In the case of PYTHIA, samples are generated using three  tuned sets of parameters denoted as 
ATLAS-MC09, DW, and Perugia2010. 
In addition, a special PYTHIA-Perugia2010 sample without UE contributions is generated using the RIVET package~\cite{rivet}. 
HERWIG++ and PYTHIA-MC09 samples are generated with MRST2007LO${}^{*}$~\cite{MSTW}~\cite{mrst2007lo} PDFs, 
PYTHIA-Perugia2010 and PYTHIA-DW with CTEQ5L~\cite{cteq} PDFs, and ALPGEN with CTEQ61L~\cite{cteq6} PDFs.

\section{Jet shapes at calorimeter level}

\begin{figure}[tbh]
\begin{center}
\mbox{
\includegraphics[width=0.495\textwidth]{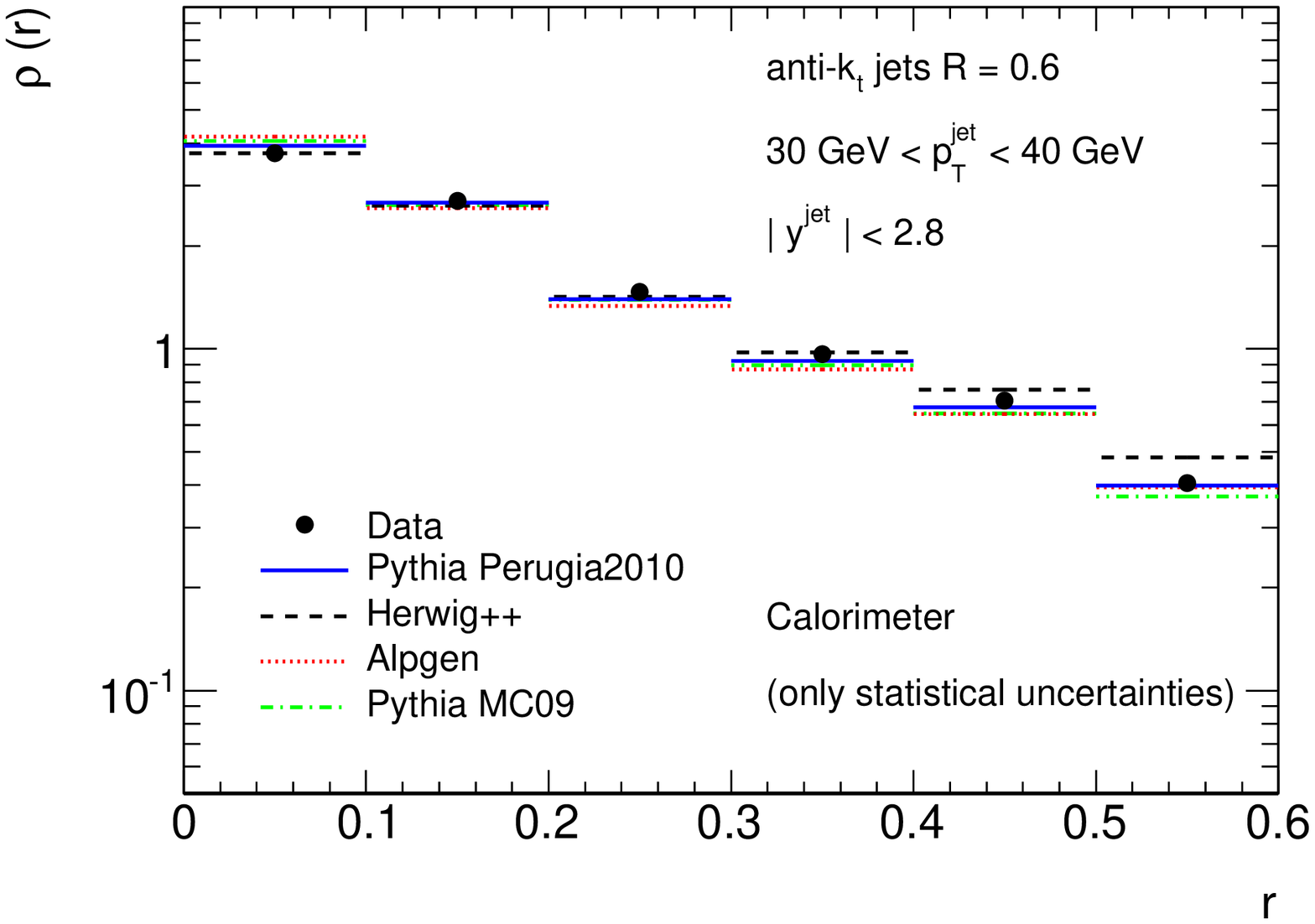}
\includegraphics[width=0.495\textwidth]{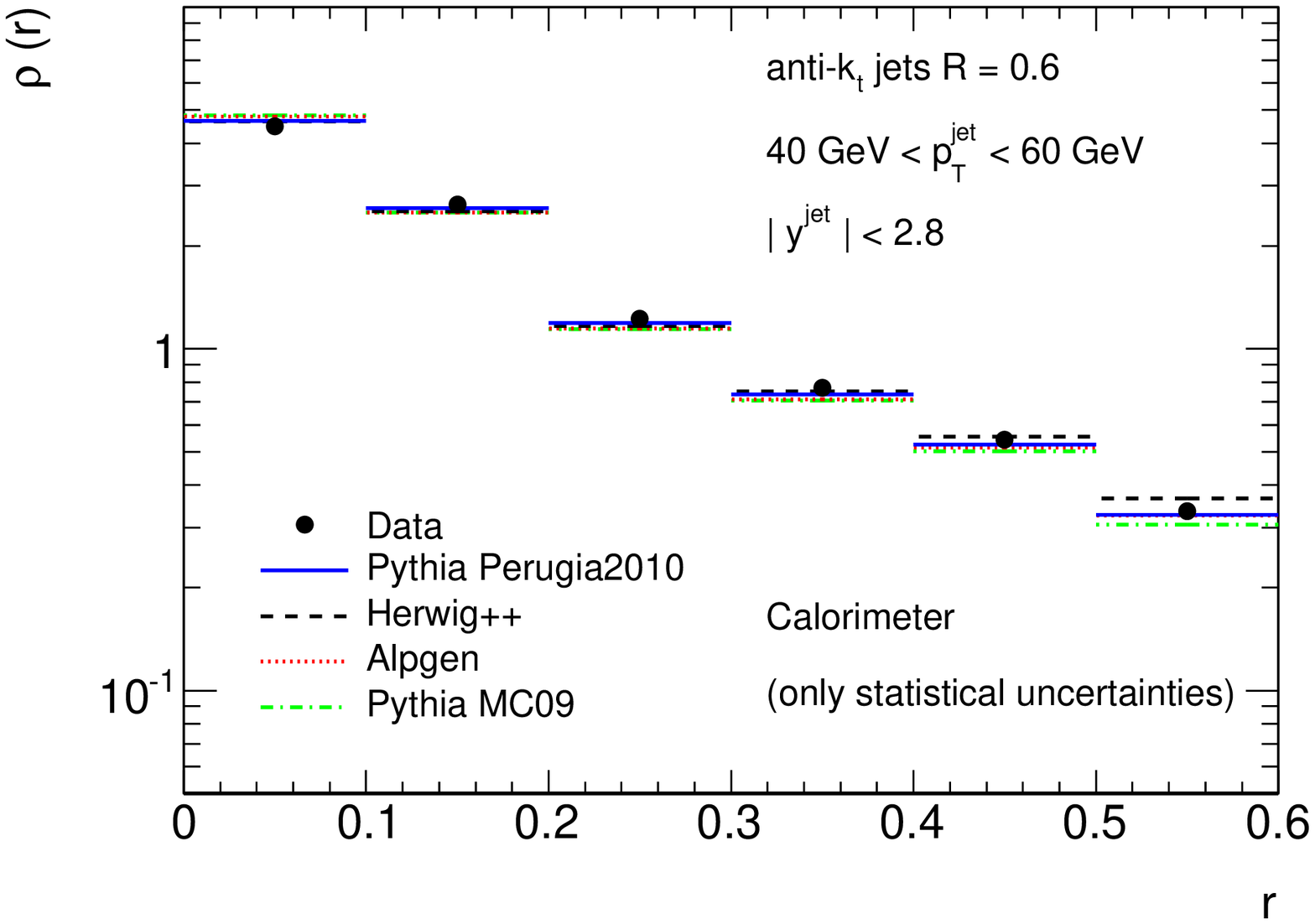}
}
\mbox{
\includegraphics[width=0.495\textwidth]{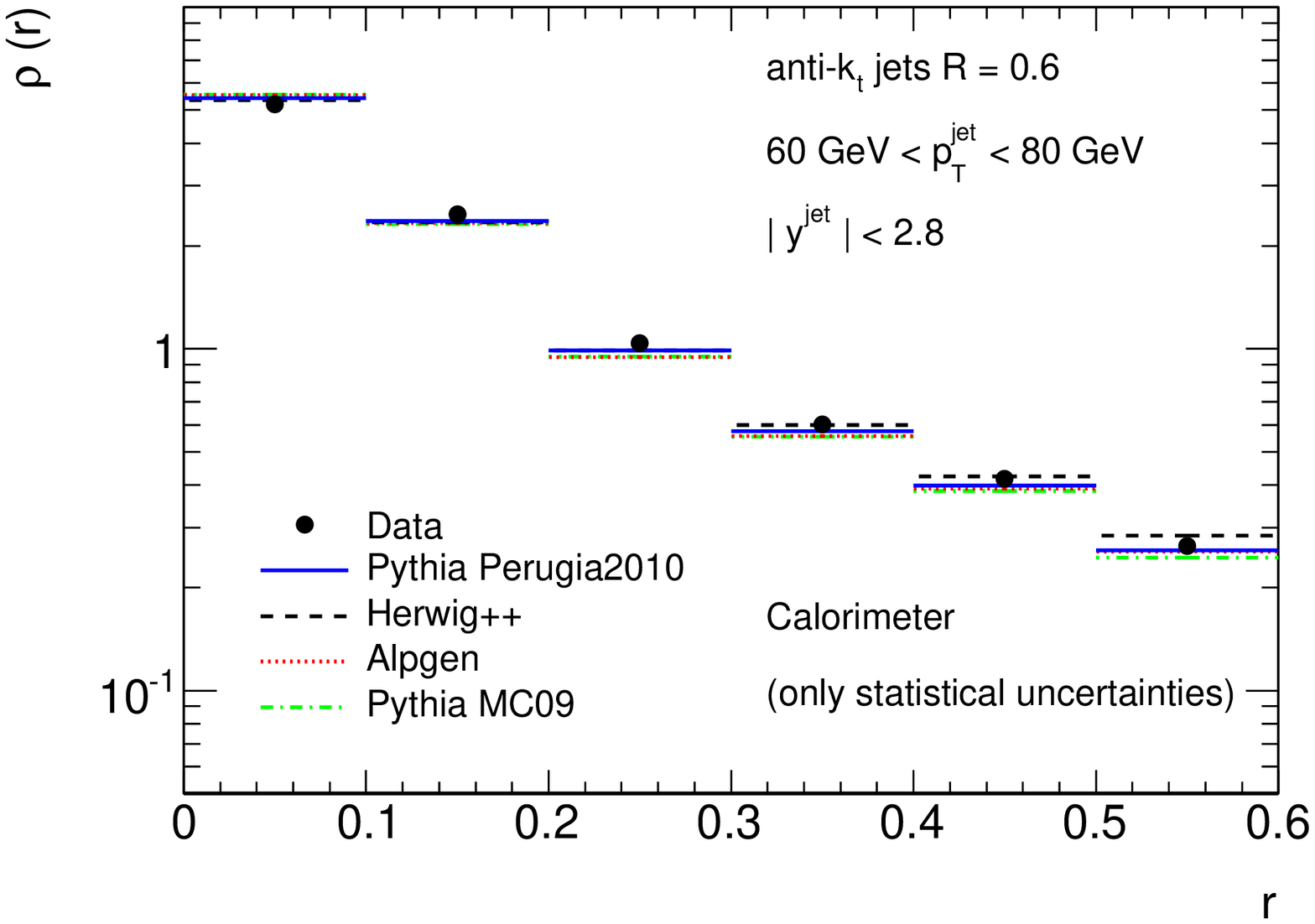}
\includegraphics[width=0.495\textwidth]{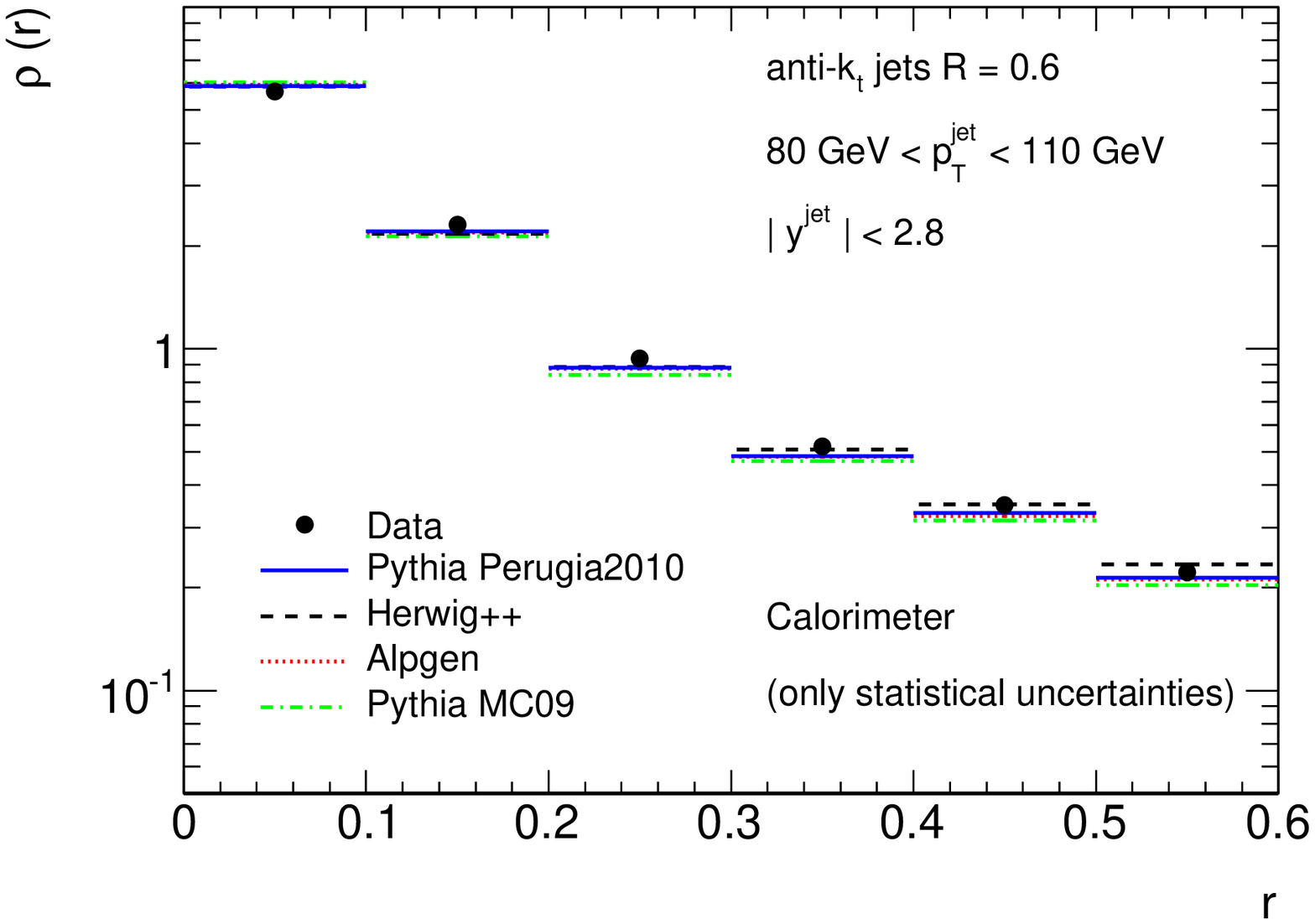}
}
\mbox{
\includegraphics[width=0.495\textwidth]{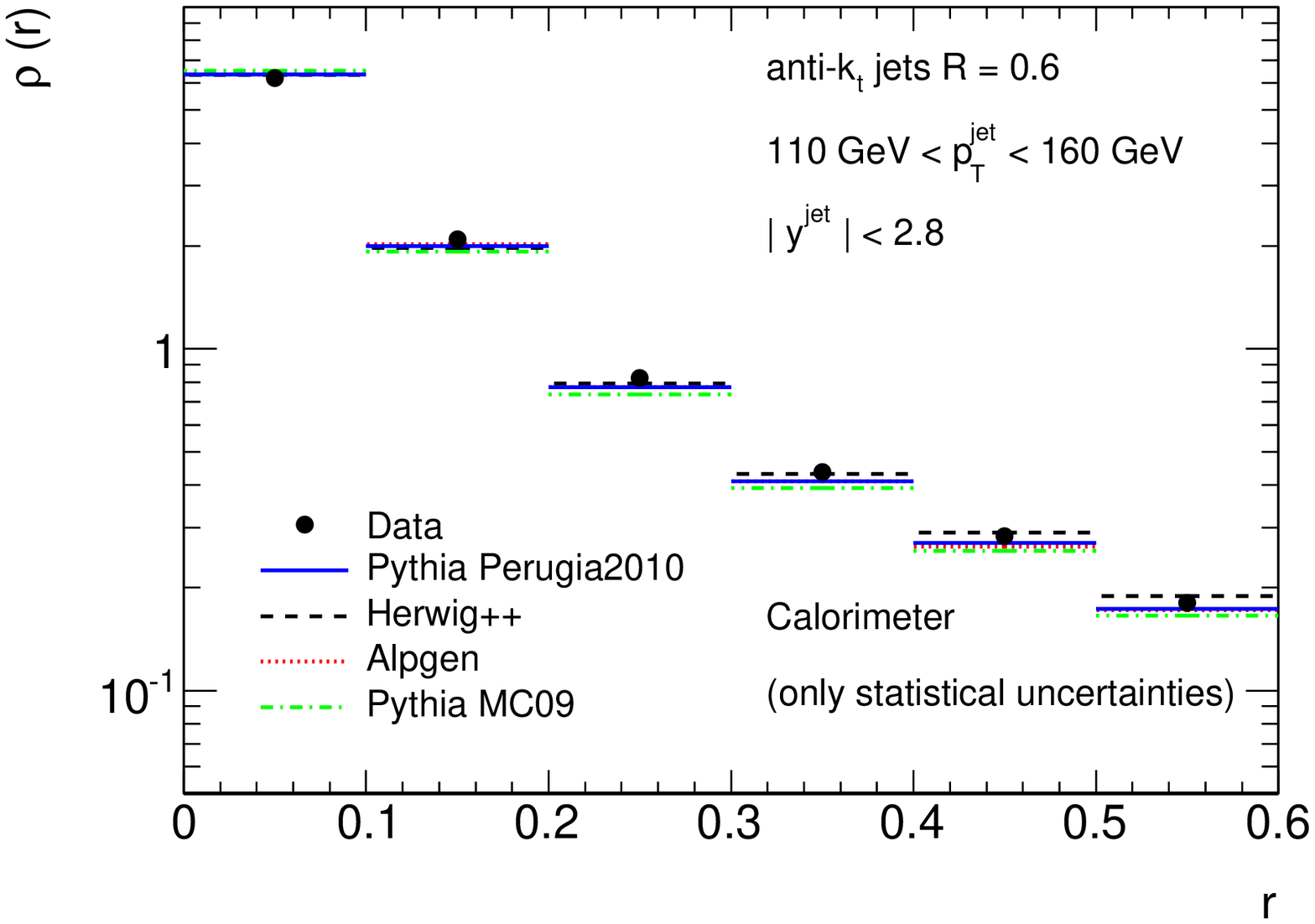}
\includegraphics[width=0.495\textwidth]{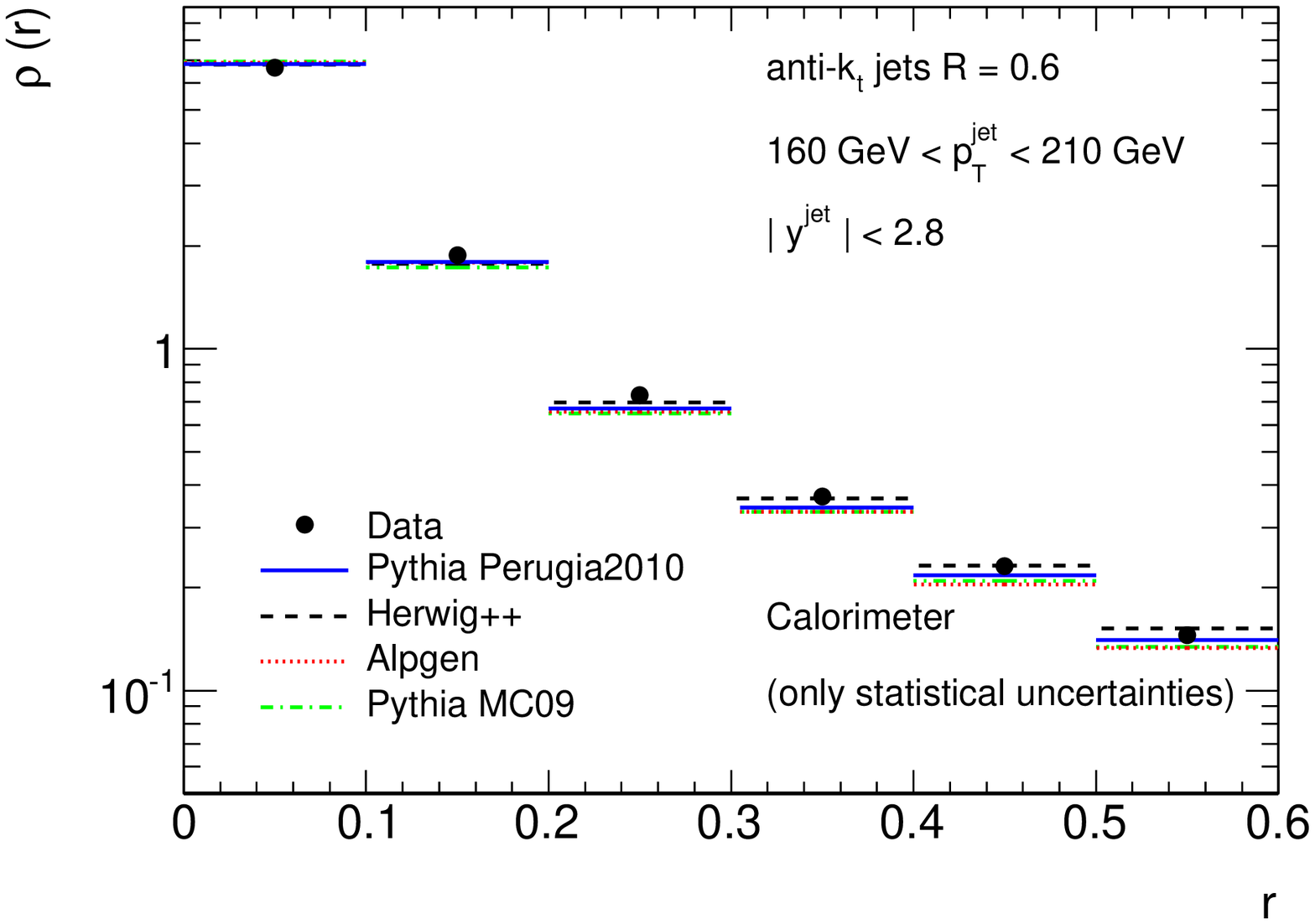}
}
\end{center}
\vspace{-0.7 cm}
\caption[The measured differential jet shape using calorimeter clusters for jets
with $|\rapjet| < 2.8$ and $30 \ {\rm GeV} < \ptjet < 210  \ {\rm GeV}$]
{\small
The measured uncorrected differential jet shape using calorimeter clusters for jets
with $|\rapjet| < 2.8$ and $30 \ {\rm GeV} < \ptjet < 210  \ {\rm GeV}$
is shown in different $\ptjet$ regions.
The predictions of   PYTHIA-Perugia2010 (solid lines),   HERWIG++ (dashed lines),   ALPGEN interfaced with HERWIG and JIMMY (dotted lines),
and PYTHIA-MC09 (dashed-dotted lines) are shown for comparison. Only statistical uncertainties are considered.}
\label{fig_calo1}
\end{figure}

\begin{figure}[tbh]
\begin{center}
\mbox{
\includegraphics[width=0.495\textwidth]{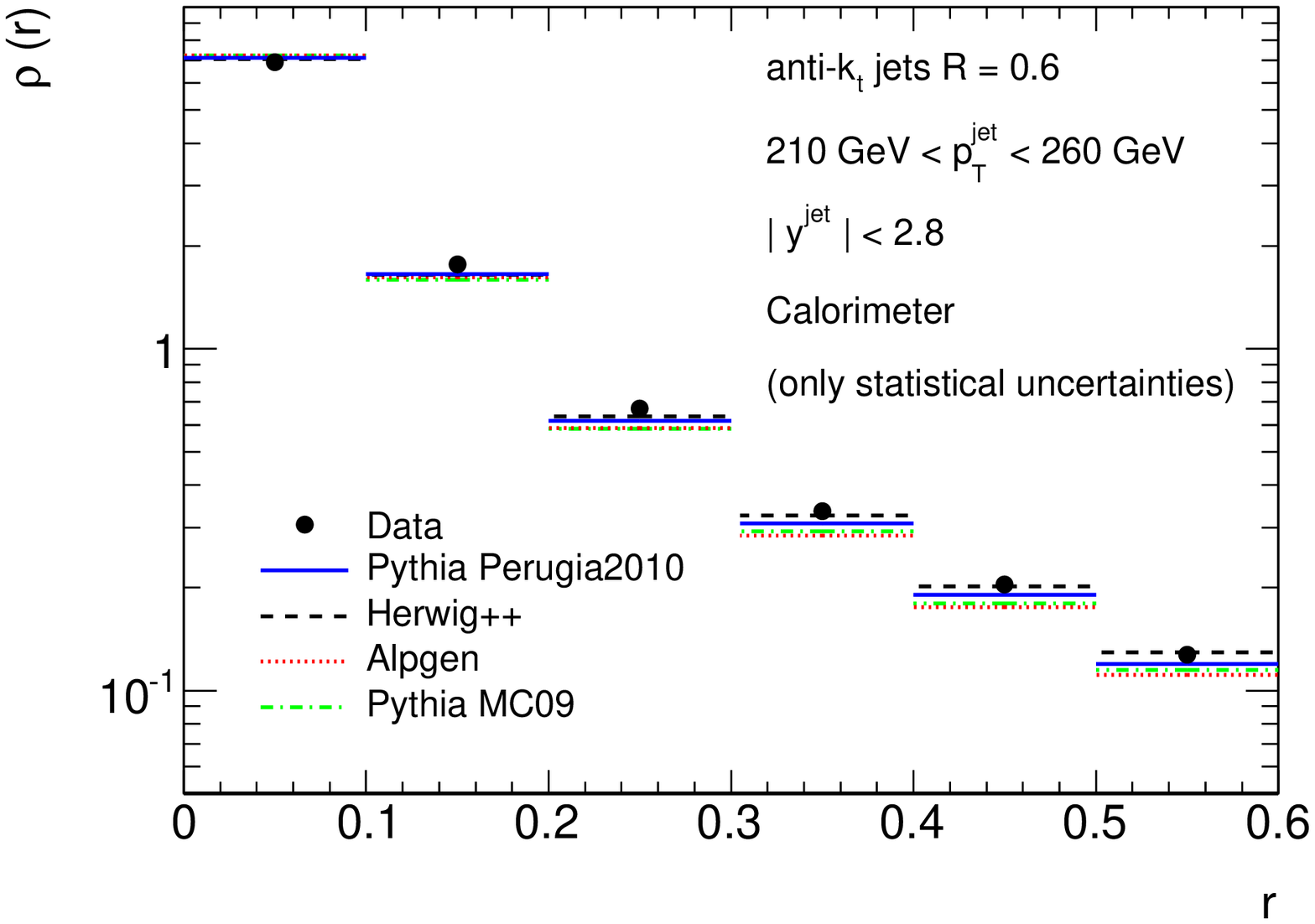}
\includegraphics[width=0.495\textwidth]{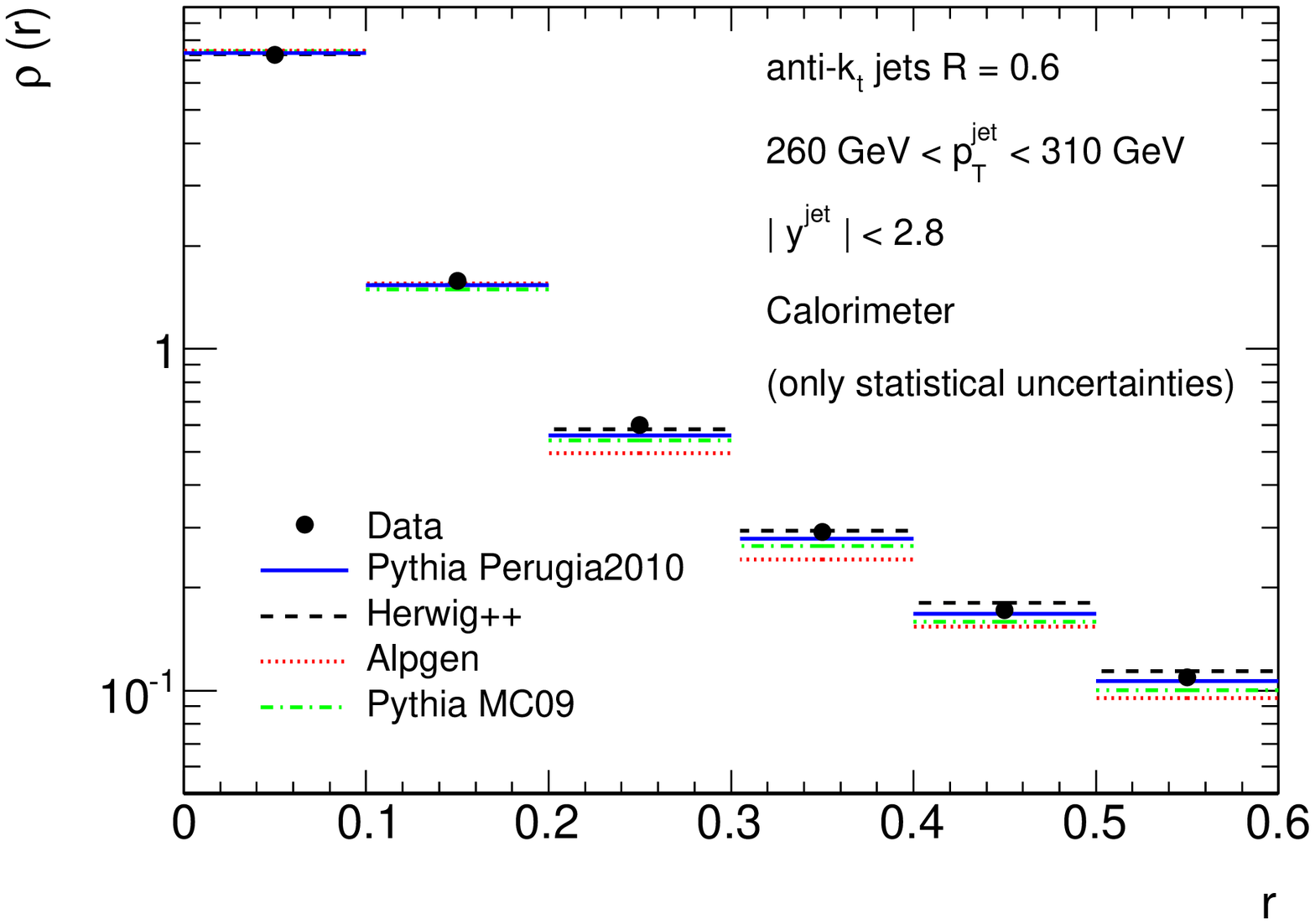}
}
\mbox{
\includegraphics[width=0.495\textwidth]{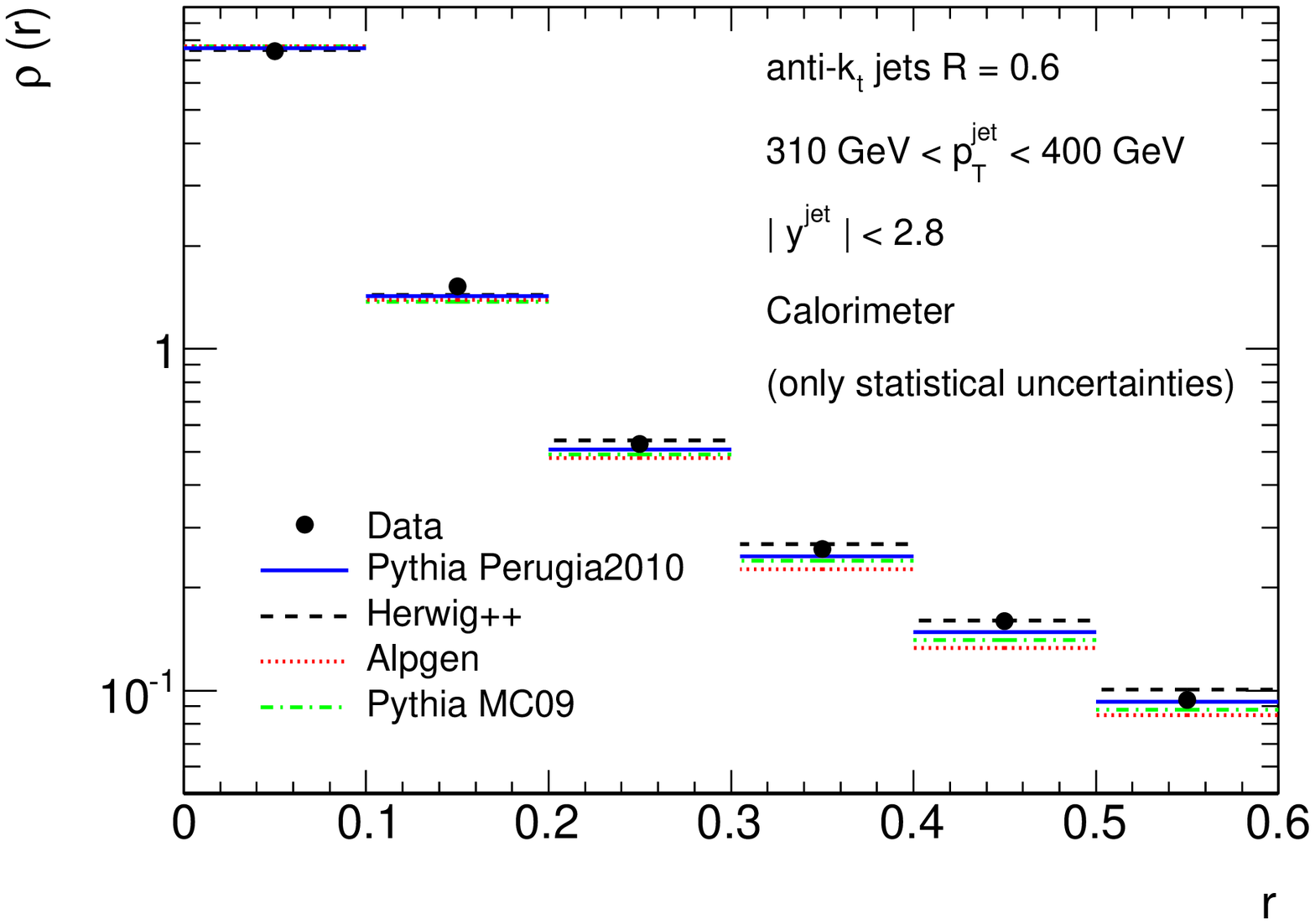}
\includegraphics[width=0.495\textwidth]{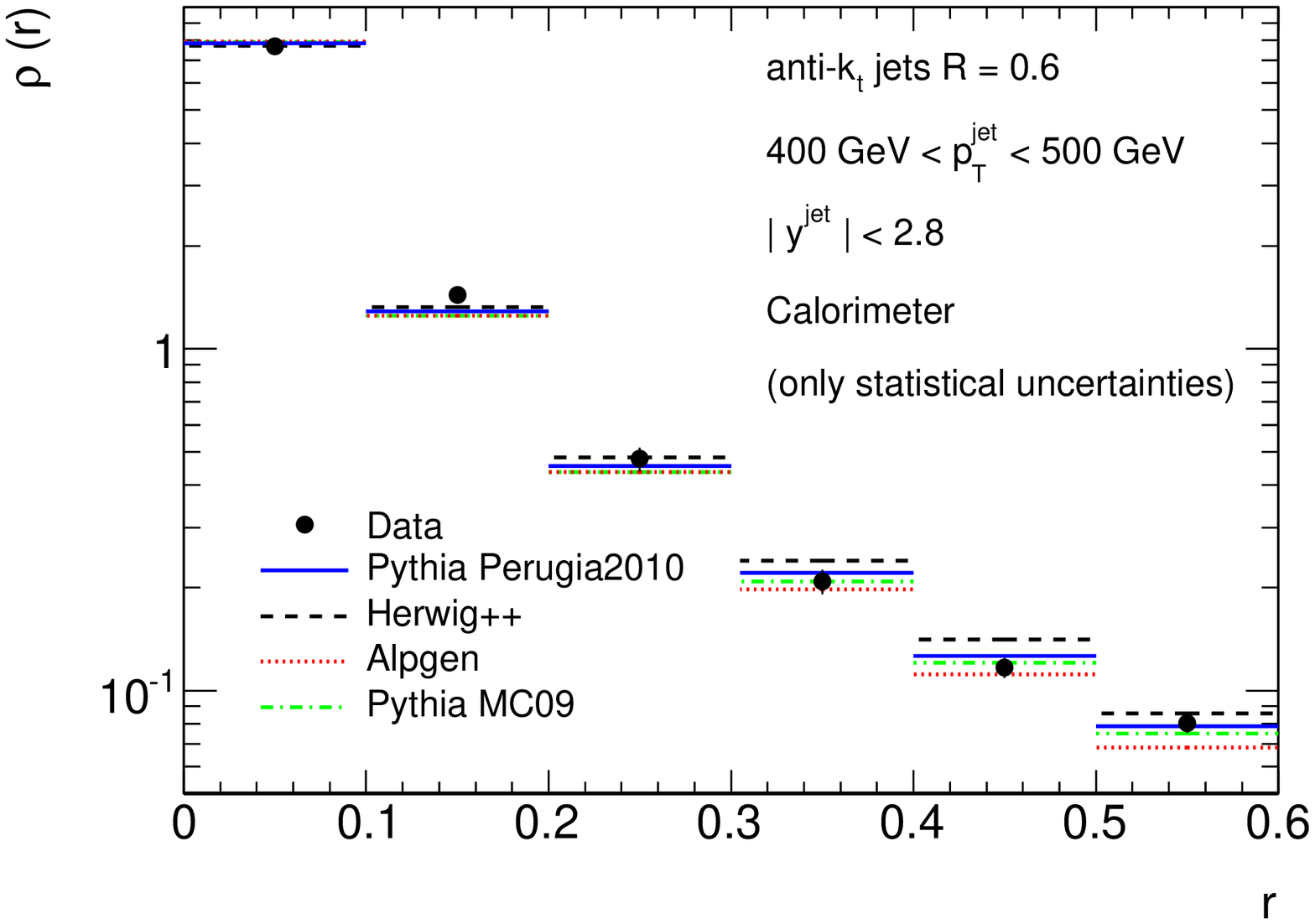}
}
\mbox{
\includegraphics[width=0.495\textwidth]{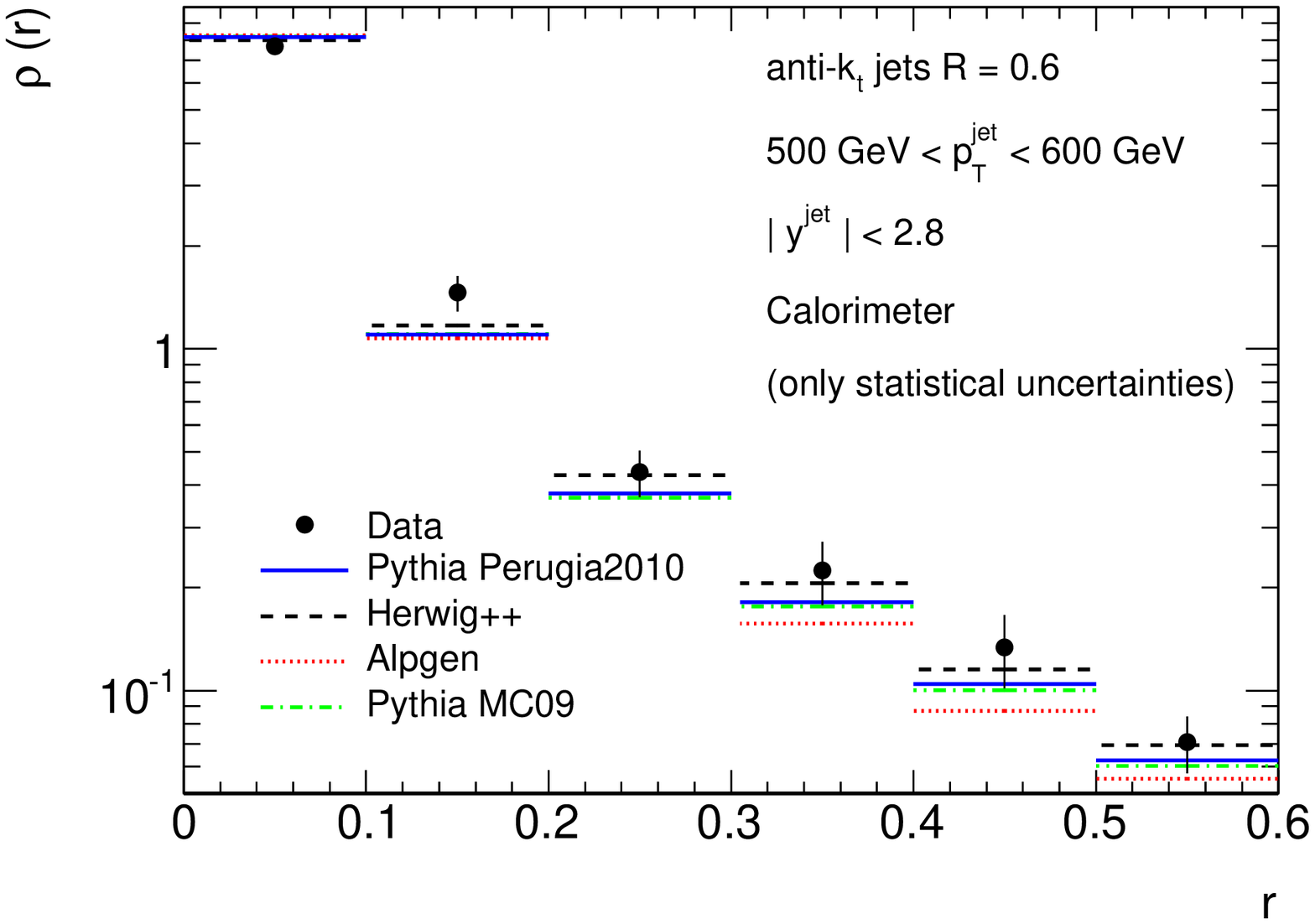}
}
\end{center}
\vspace{-0.7 cm}
\caption[The measured differential jet shape using calorimeter clusters for jets
with $|\rapjet| < 2.8$ and $210 \ {\rm GeV} < \ptjet < 600  \ {\rm GeV}$]
{\small
The measured uncorrected differential jet shape using calorimeter clusters for jets
with $|\rapjet| < 2.8$ and $210 \ {\rm GeV} < \ptjet < 600  \ {\rm GeV}$
is shown in different $\ptjet$ regions. 
The predictions of   PYTHIA-Perugia2010 (solid lines),   HERWIG++ (dashed lines),   ALPGEN interfaced with HERWIG and JIMMY (dotted lines),
and PYTHIA-MC09 (dashed-dotted lines) are shown for comparison. Only statistical uncertainties are considered.
}
\label{fig_calo2}
\end{figure}

\begin{figure}[tbh]
\begin{center}
\mbox{
\includegraphics[width=0.495\textwidth]{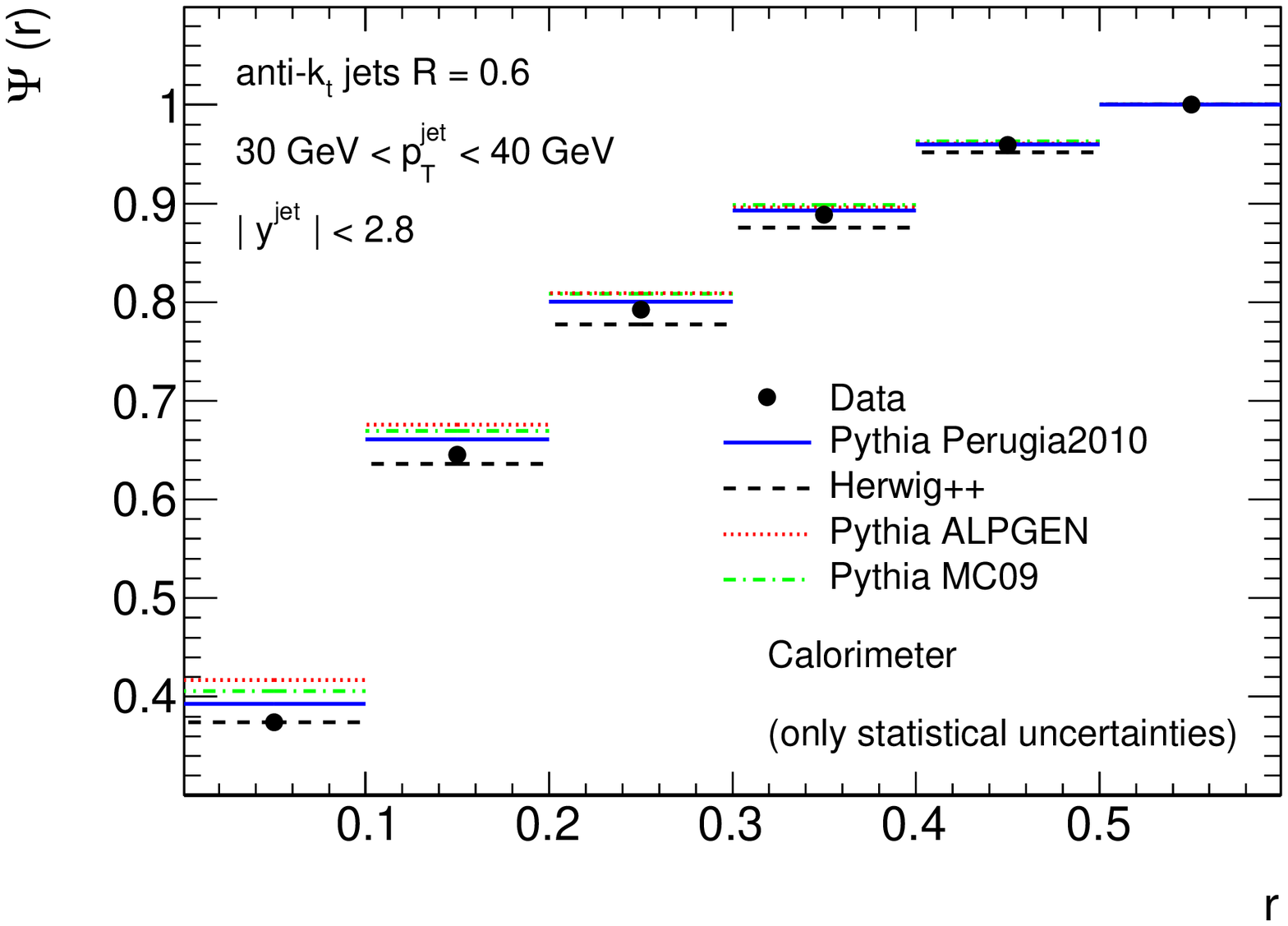}
\includegraphics[width=0.495\textwidth]{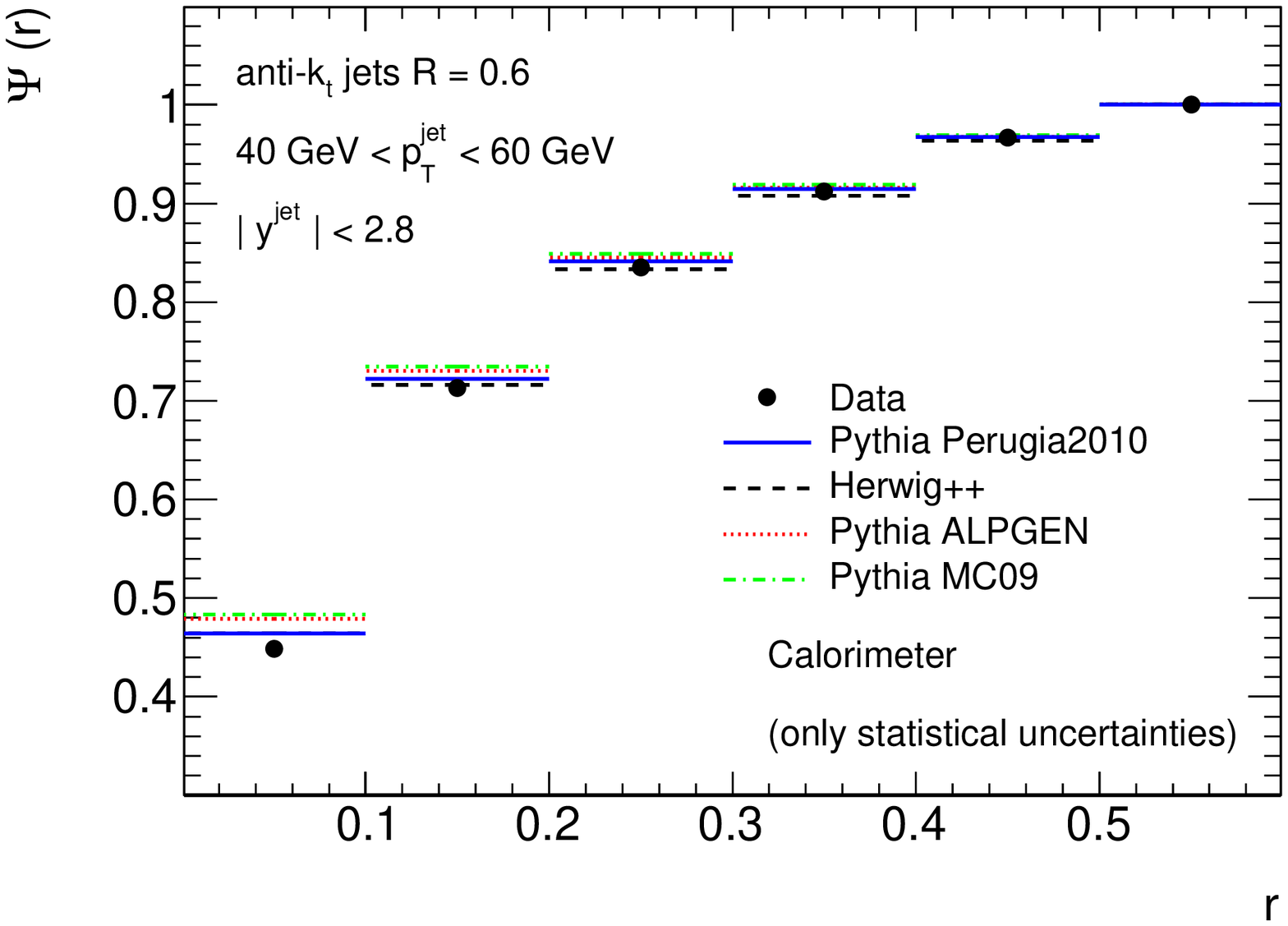}
}
\mbox{
\includegraphics[width=0.495\textwidth]{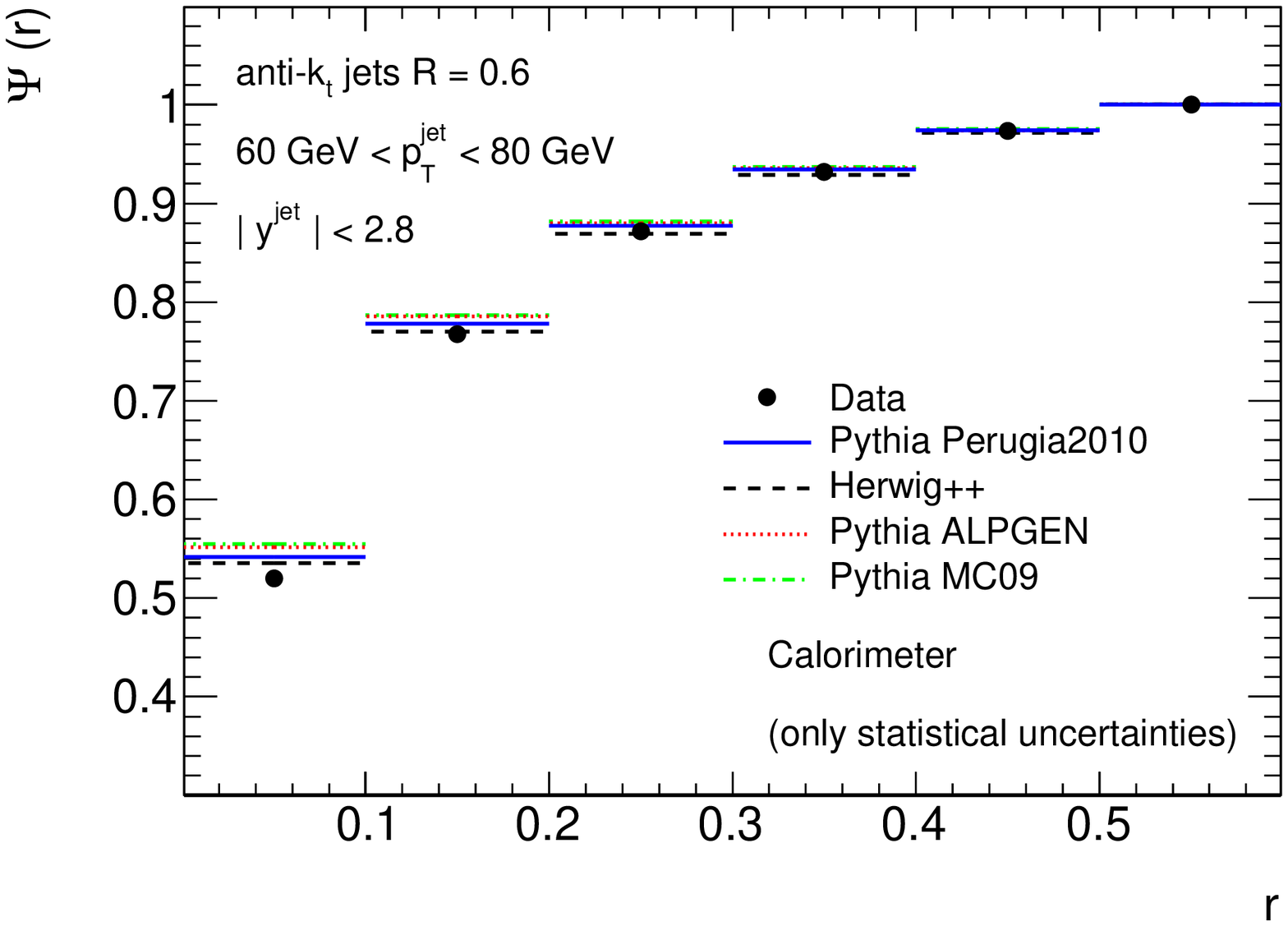}
\includegraphics[width=0.495\textwidth]{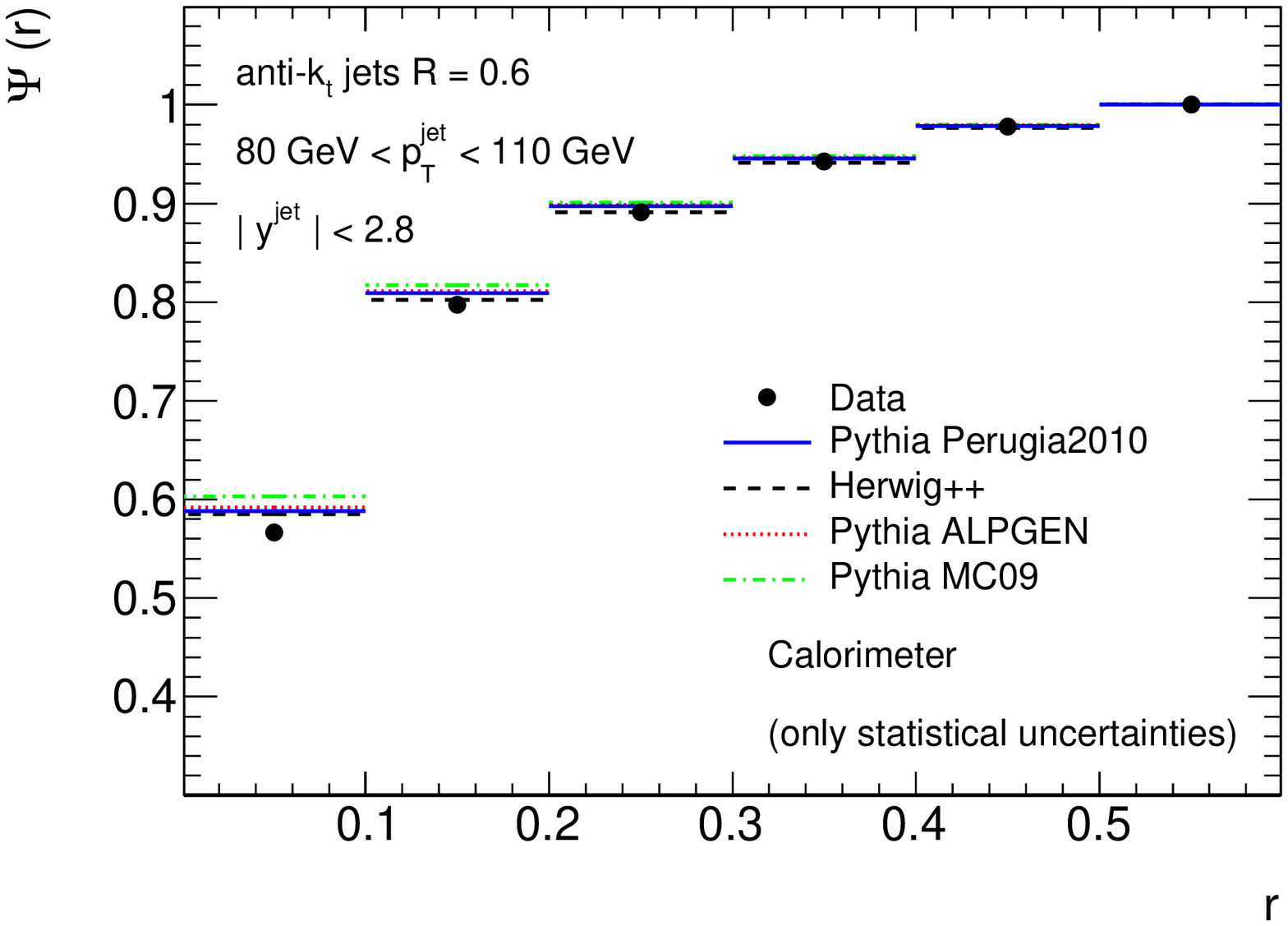}
}
\mbox{
\includegraphics[width=0.495\textwidth]{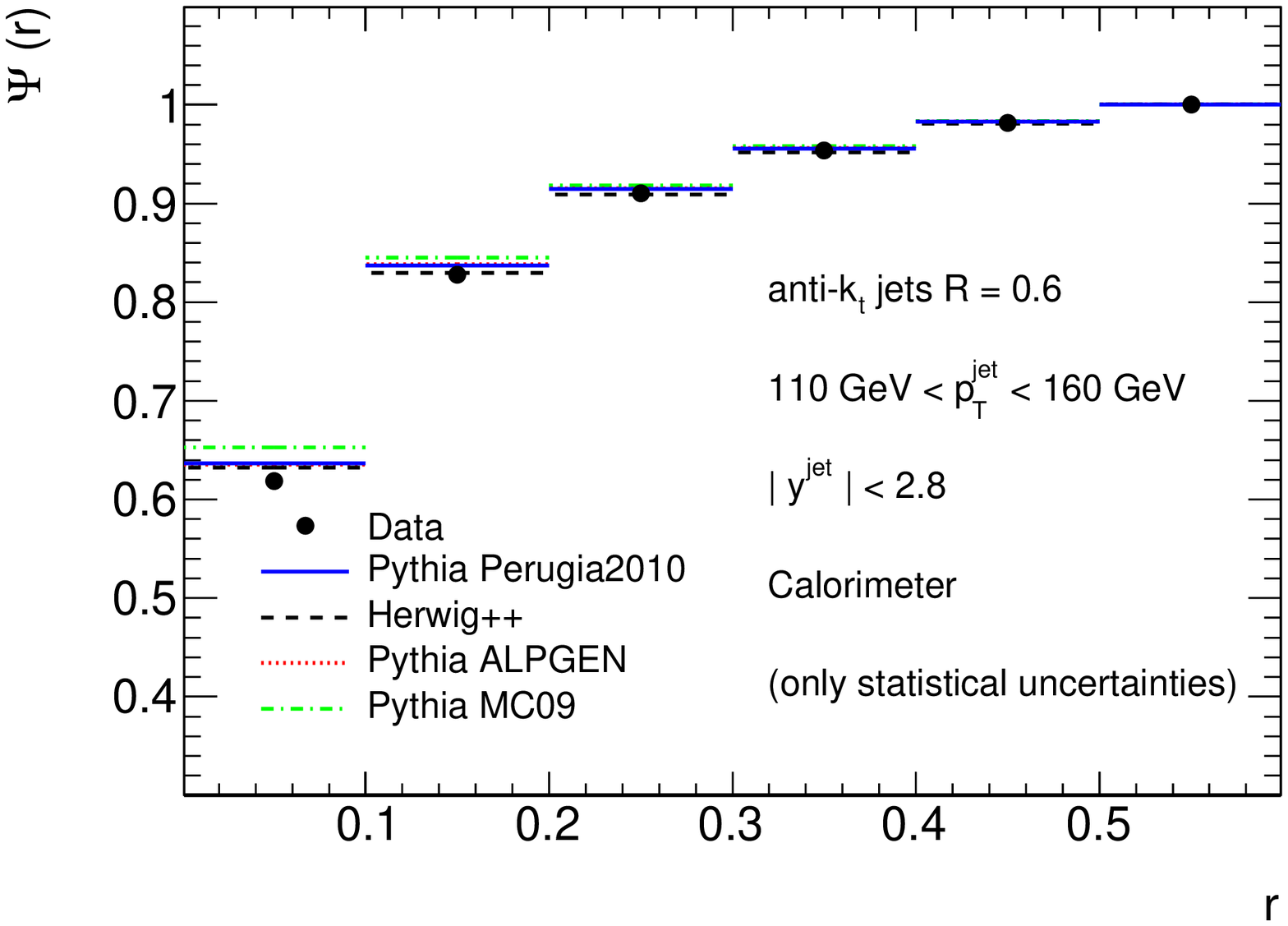}
\includegraphics[width=0.495\textwidth]{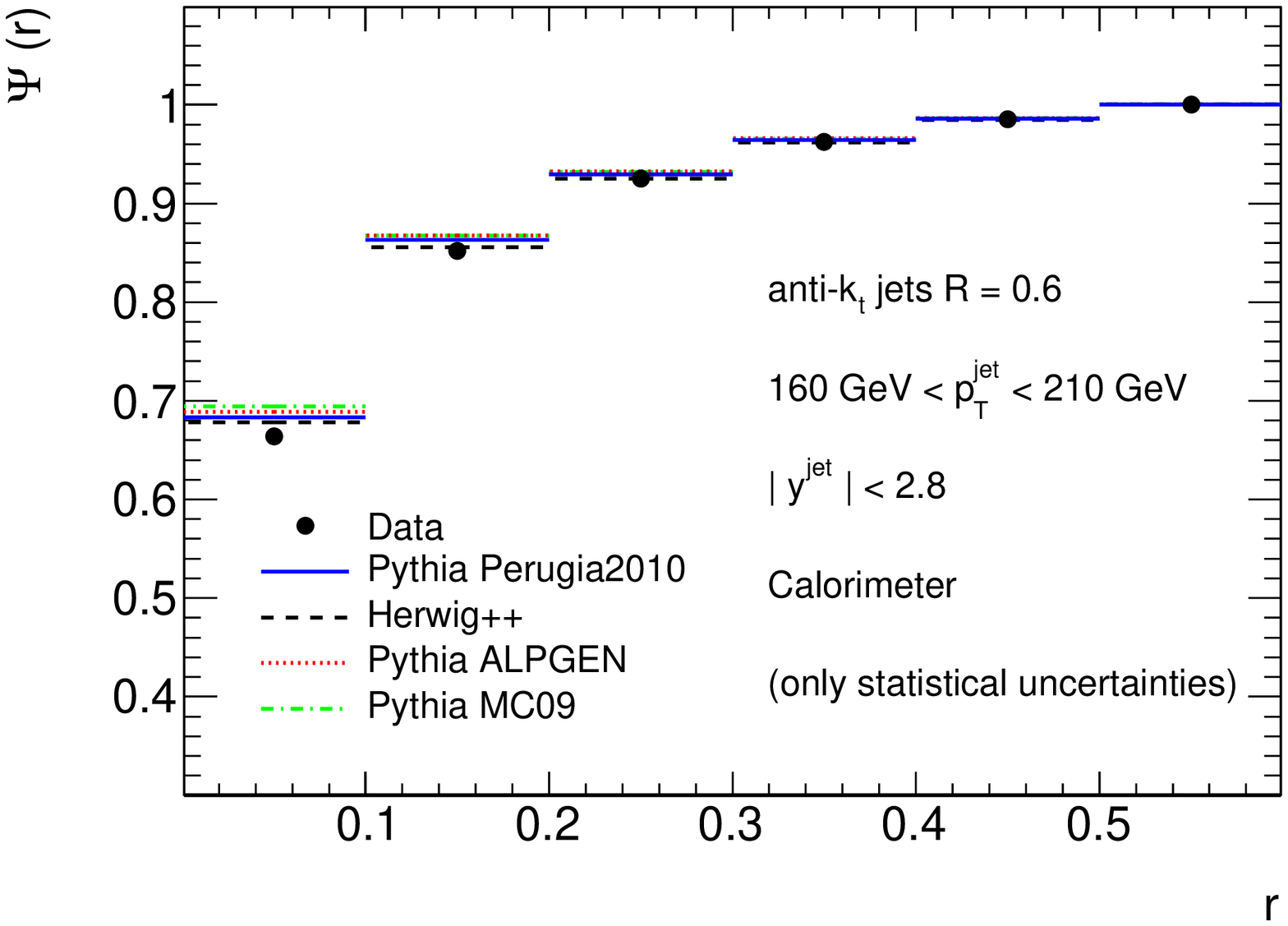}
}
\end{center}
\vspace{-0.7 cm}
\caption[The measured integrated jet shape using calorimeter clusters for jets
with $|\rapjet| < 2.8$ and $30 \ {\rm GeV} < \ptjet < 210  \ {\rm GeV}$]
{\small
The measured uncorrected integrated jet shape using calorimeter clusters for jets
with $|\rapjet| < 2.8$ and $30 \ {\rm GeV} < \ptjet < 210  \ {\rm GeV}$
is shown in different $\ptjet$ regions.
The predictions of   PYTHIA-Perugia2010 (solid lines),   HERWIG++ (dashed lines),   ALPGEN interfaced with HERWIG and JIMMY (dotted lines),
and PYTHIA-MC09 (dashed-dotted lines) are shown for comparison. Only statistical uncertainties are considered.
}
\label{fig_calo3}
\end{figure}

\begin{figure}[tbh]
\begin{center}
\mbox{
\includegraphics[width=0.495\textwidth]{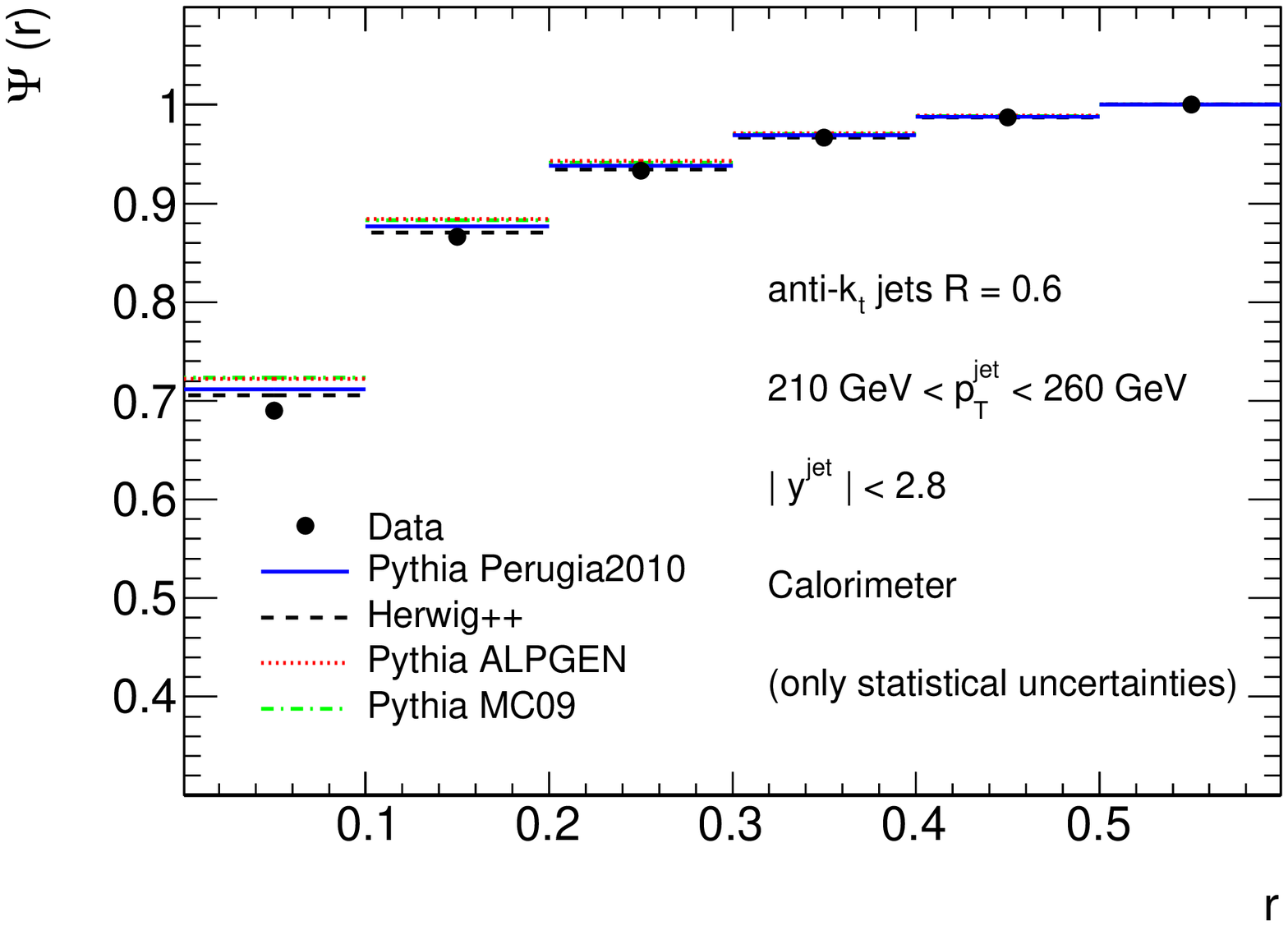}
\includegraphics[width=0.495\textwidth]{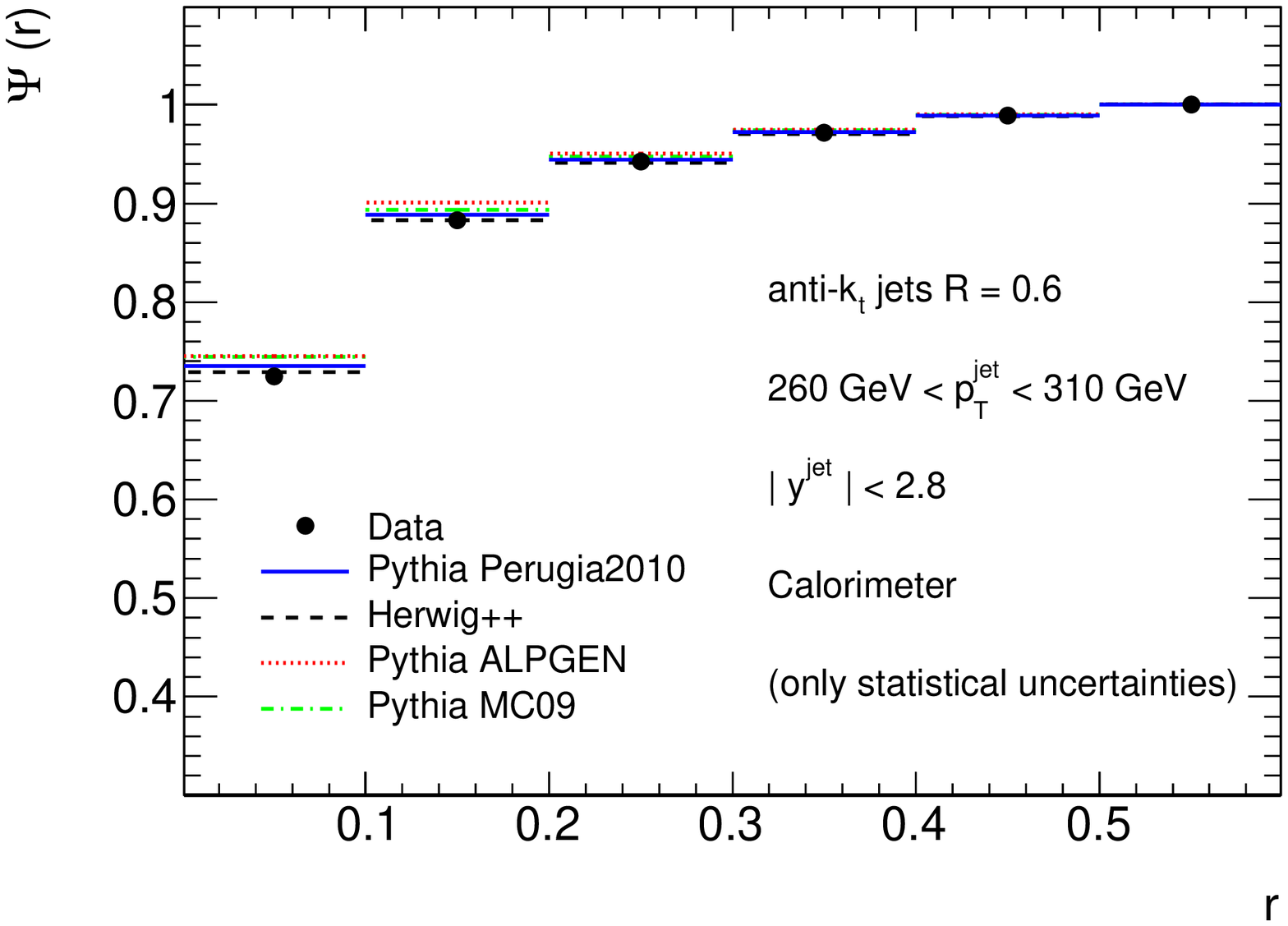}
}
\mbox{
\includegraphics[width=0.495\textwidth]{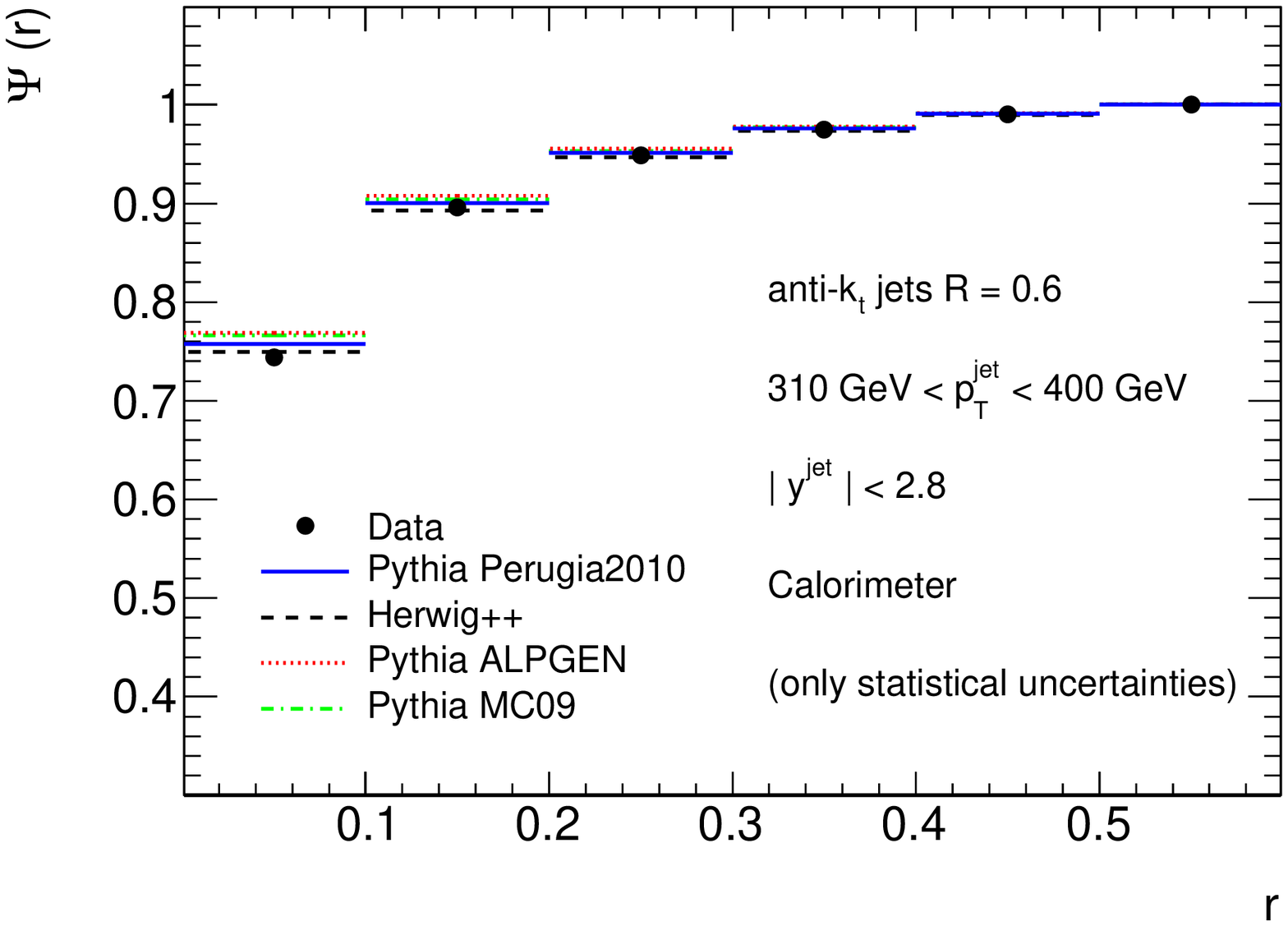}
\includegraphics[width=0.495\textwidth]{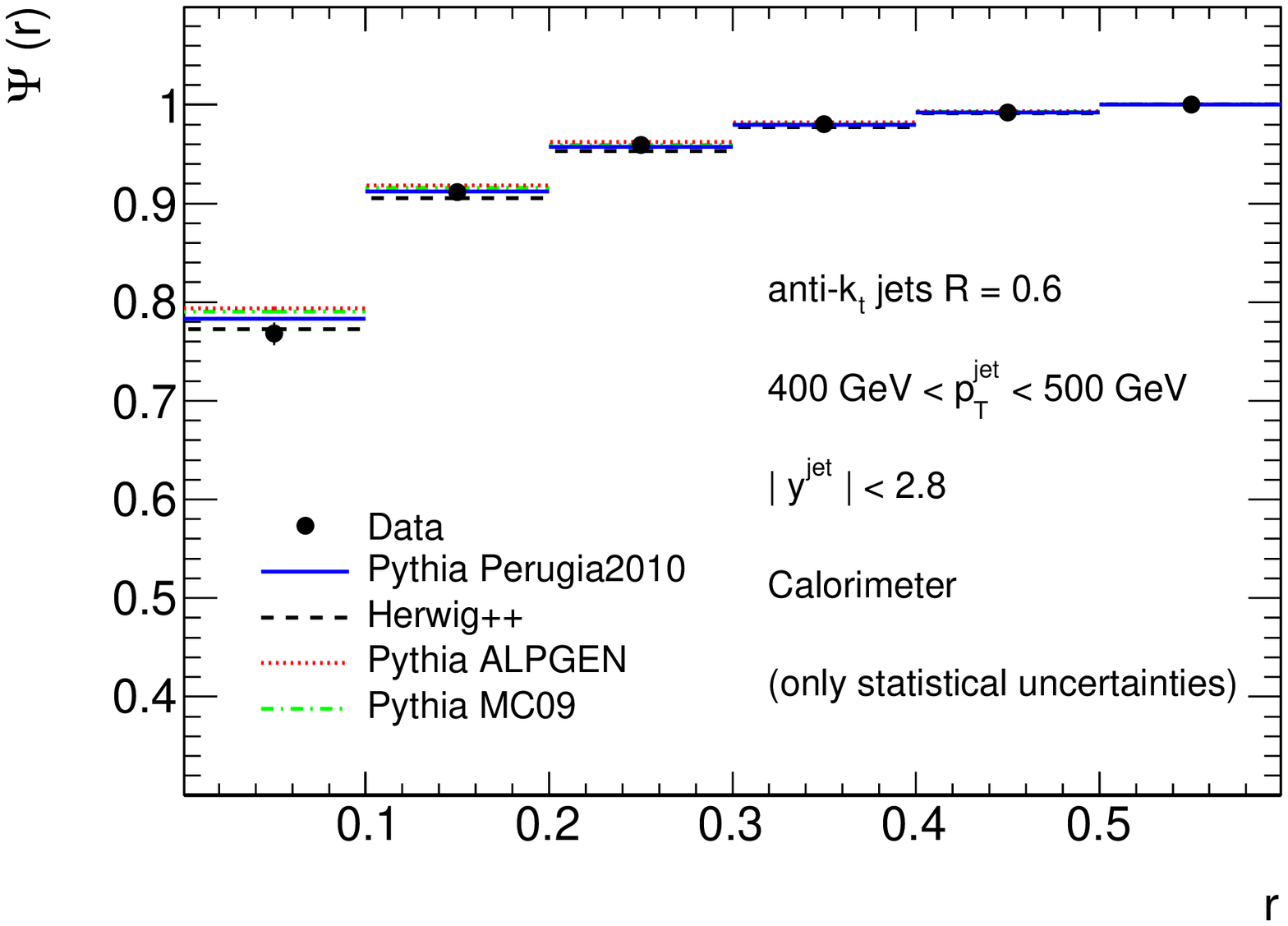}
}
\mbox{
\includegraphics[width=0.495\textwidth]{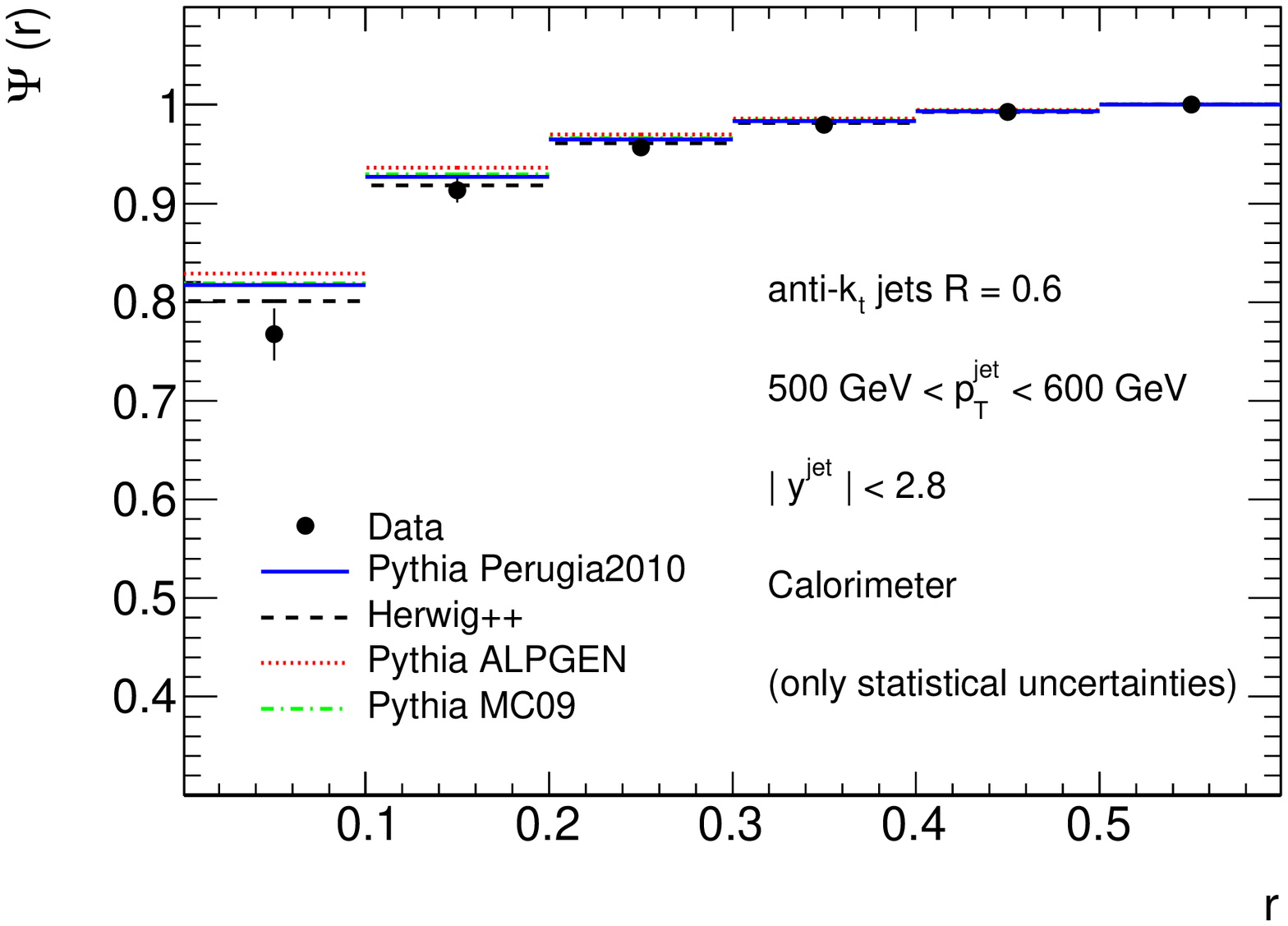}
}
\end{center}
\vspace{-0.7 cm}
\caption[The measured integrated jet shape using calorimeter clusters for jets
with $|\rapjet| < 2.8$ and $210 \ {\rm GeV} < \ptjet < 600  \ {\rm GeV}$]
{\small
The measured uncorrected integrated jet shape using calorimeter clusters for jets
with $|\rapjet| < 2.8$ and $210 \ {\rm GeV} < \ptjet < 600  \ {\rm GeV}$
is shown in different $\ptjet$ regions.
The predictions of   PYTHIA-Perugia2010 (solid lines),   HERWIG++ (dashed lines),   ALPGEN interfaced with HERWIG and JIMMY (dotted lines),
and PYTHIA-MC09 (dashed-dotted lines) are shown for comparison. Only statistical uncertainties are considered.
}
\label{fig_calo4}
\end{figure}

The measured differential (integrated) jet shapes, as determined by using calorimeter clusters, 
are shown in Figures~\ref{fig_calo1}~and~\ref{fig_calo2} (Figures~\ref{fig_calo3}~and~\ref{fig_calo4}). 
The data present a prominent peak at 
low $r$, which indicates that the majority of the jet momentum is concentrated around the jet axis. 
As expected, jets get narrower as the $\ptjet$ increases. The data are compared to several MC predictions, 
being PYTHIA-Perugia2010 the one that describes the data best. A reasonable agreement between jet shapes data and 
PYTHIA-Perugia2010 has been also observed in the different $|\rapjet|$ regions considered in this analysis. Therefore, 
PYTHIA-Perugia2010 is used to correct both differential and integrated jet shapes for detector effects back to the particle level. 

\section{Correction for detector effects}

The correction for detector effects is done using a bin-by-bin 
procedure that also accounts for the efficiency of the selection 
criteria and of the jet reconstruction in the calorimeter. 
Here, the method is described in detail for the differential case.
A similar procedure is employed to correct independently the integrated measurements.
The correction factors $U(r,\ptjet,|\rapjet|)$ are computed 
separately in each jet $\ptjet$ and $|\rapjet|$ region. They are defined as the ratio between the 
jet shapes at the particle level  $\rho(r)^{par}_{mc}$, 
obtained using particle-level jets in the kinematic range under consideration, 
and the reconstructed jet shapes at the 
calorimeter level  $\rho(r)^{cal}_{mc}$, after the selection criteria are applied and using calorimeter-level jets
in the given $\ptjet$ and $|\rapjet|$ range.

\begin{figure}[tbh]
\begin{center}
\mbox{
\includegraphics[width=0.495\textwidth]{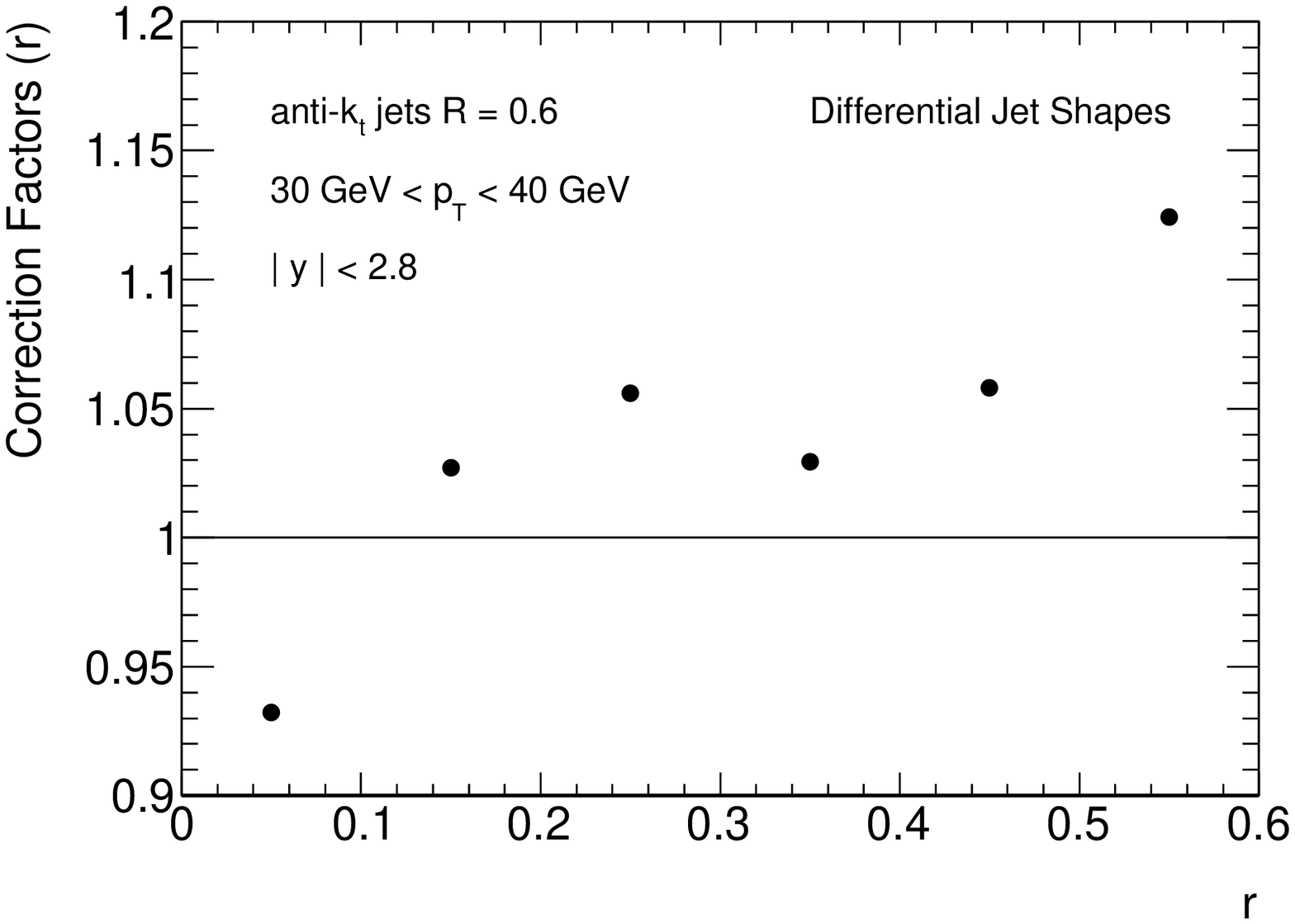}
\includegraphics[width=0.495\textwidth]{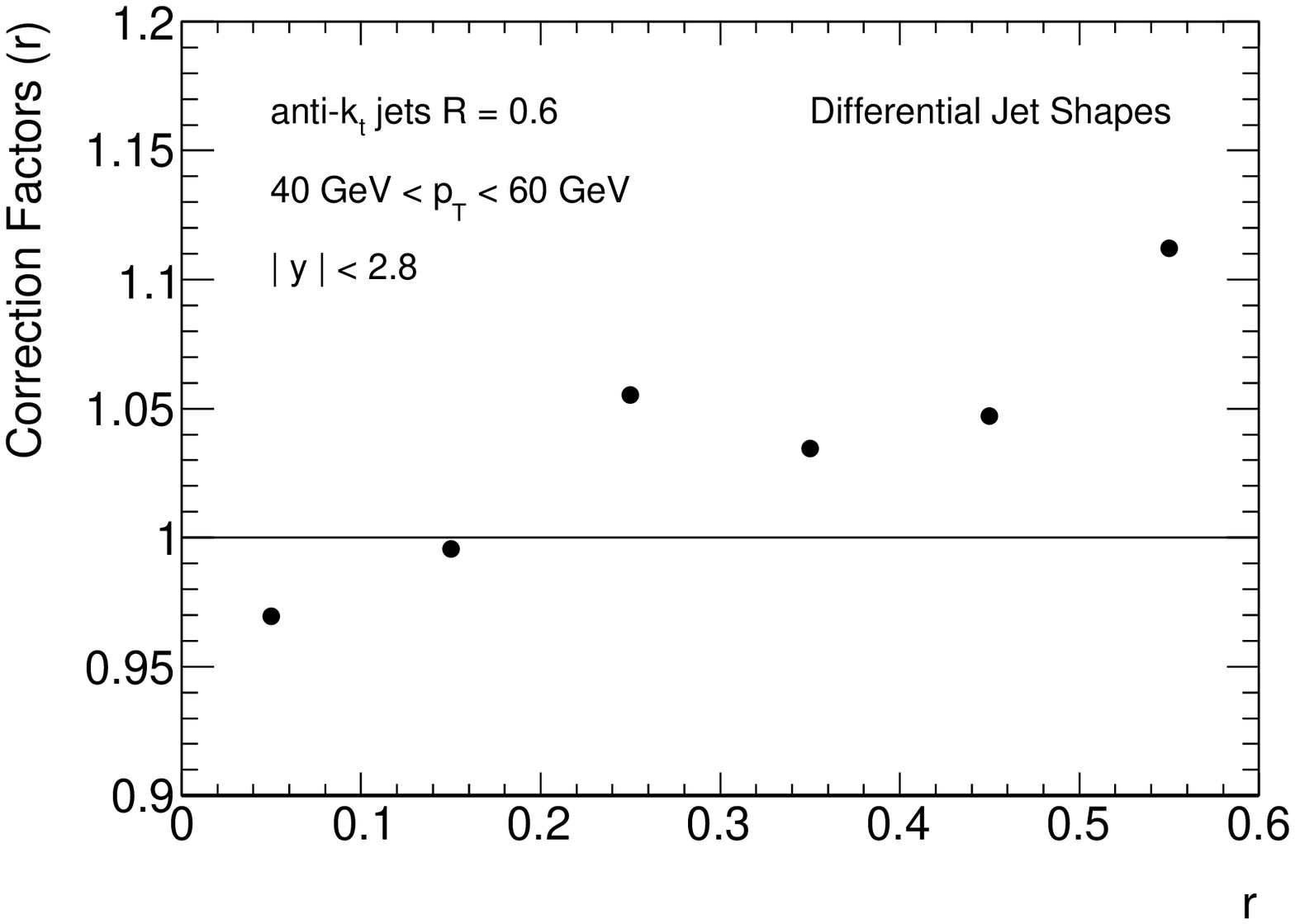}
}
\mbox{
\includegraphics[width=0.495\textwidth]{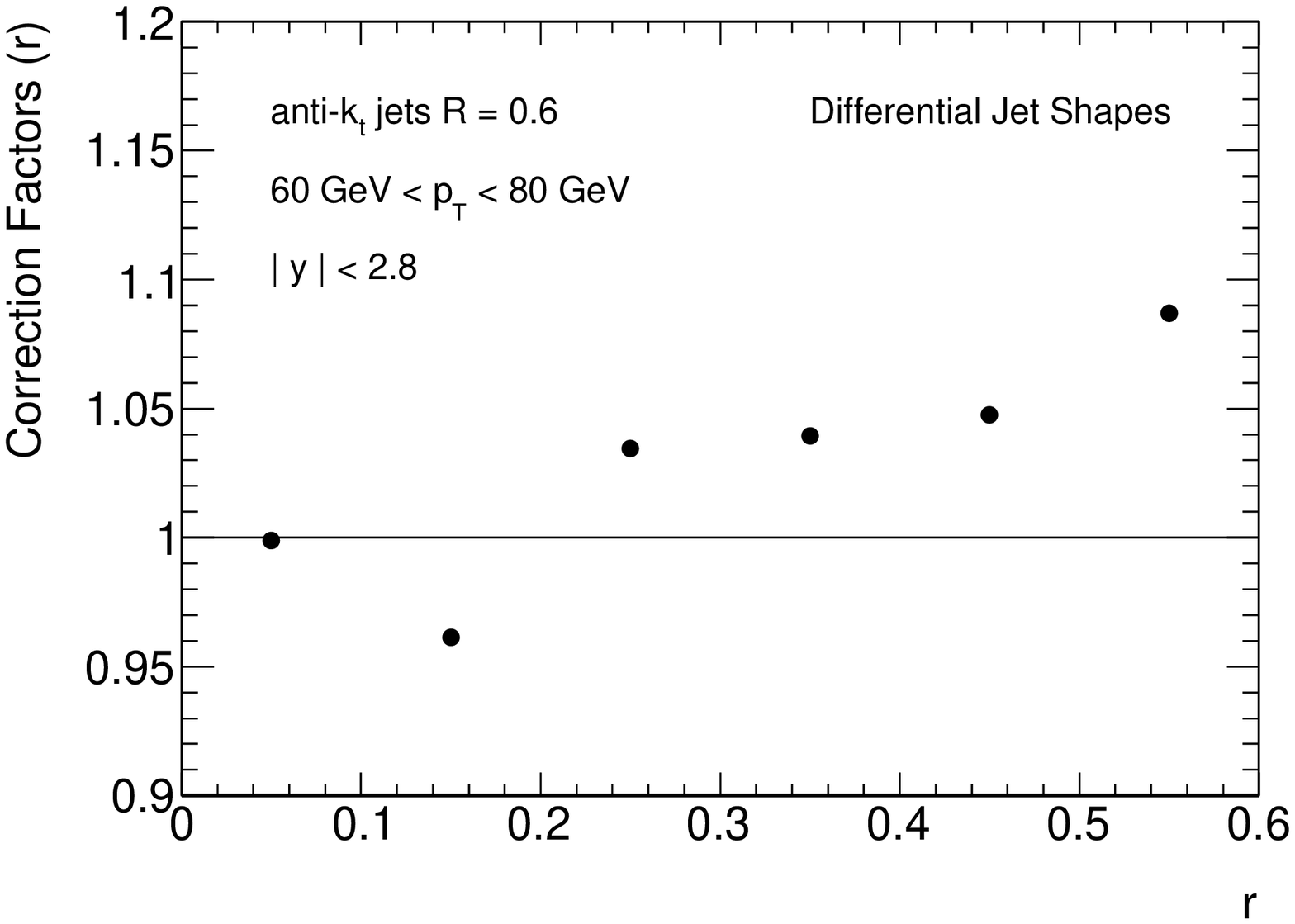}
\includegraphics[width=0.495\textwidth]{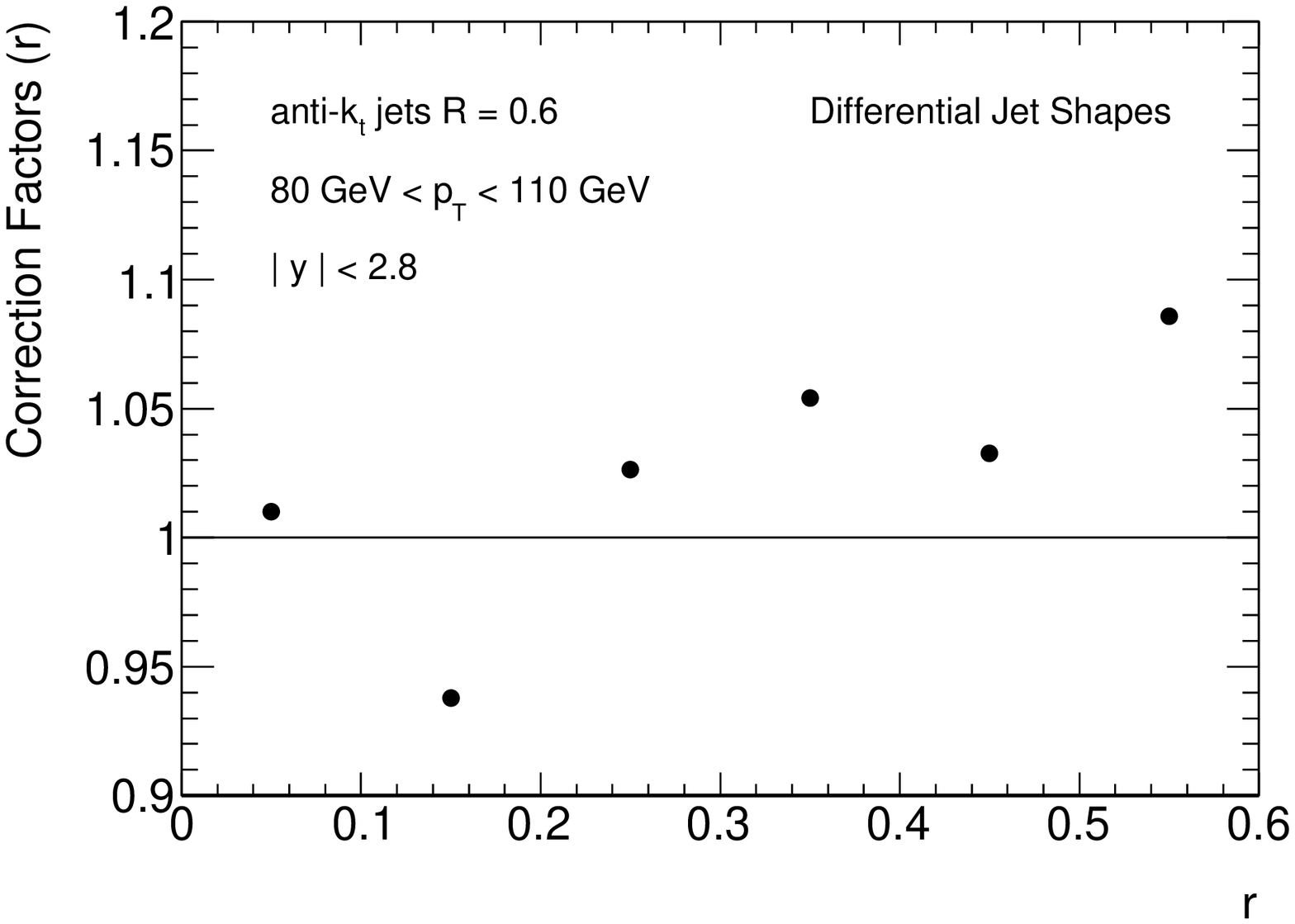}
}
\mbox{
\includegraphics[width=0.495\textwidth]{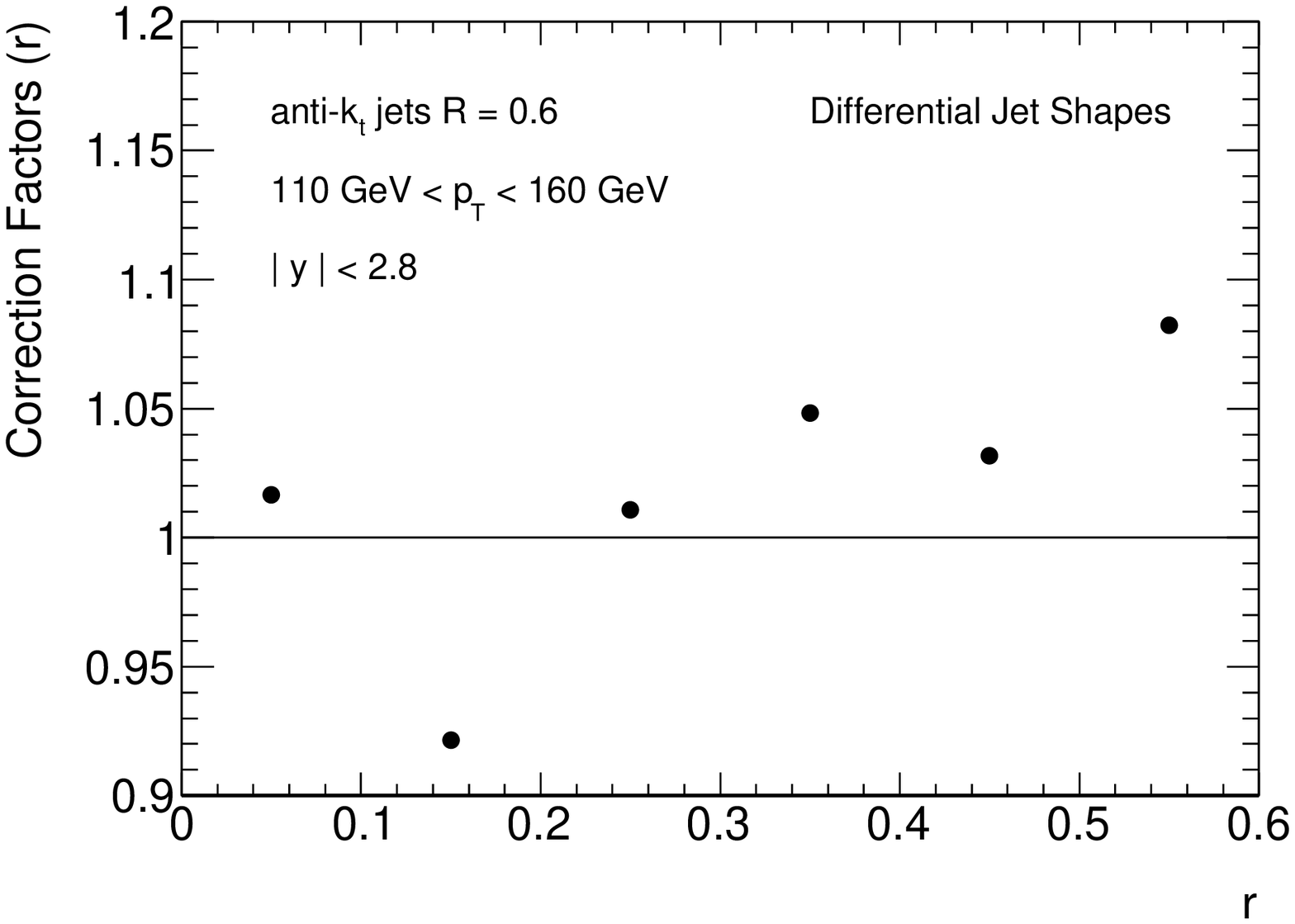}
\includegraphics[width=0.495\textwidth]{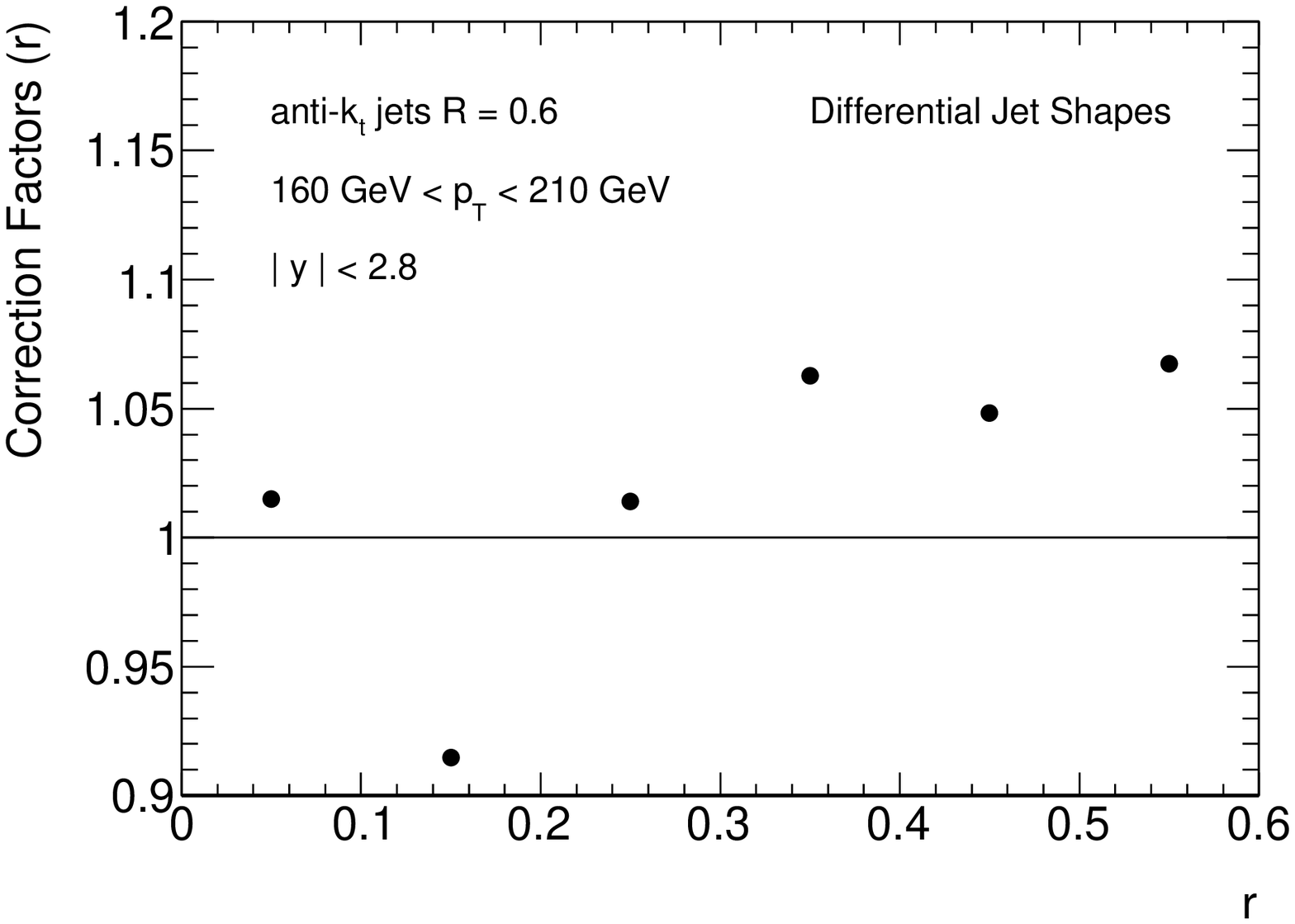}
}
\end{center}
\vspace{-0.7 cm}
\caption{\small
Correction factors applied to the measured differential jet shapes to correct the measurements for detector effects
for jets with $|\rapjet| < 2.8$ and $30 \ {\rm GeV} < \ptjet < 210  \ {\rm GeV}$.
}
\label{fig_unf1}
\end{figure}

\begin{figure}[tbh]
\begin{center}
\mbox{
\includegraphics[width=0.495\textwidth]{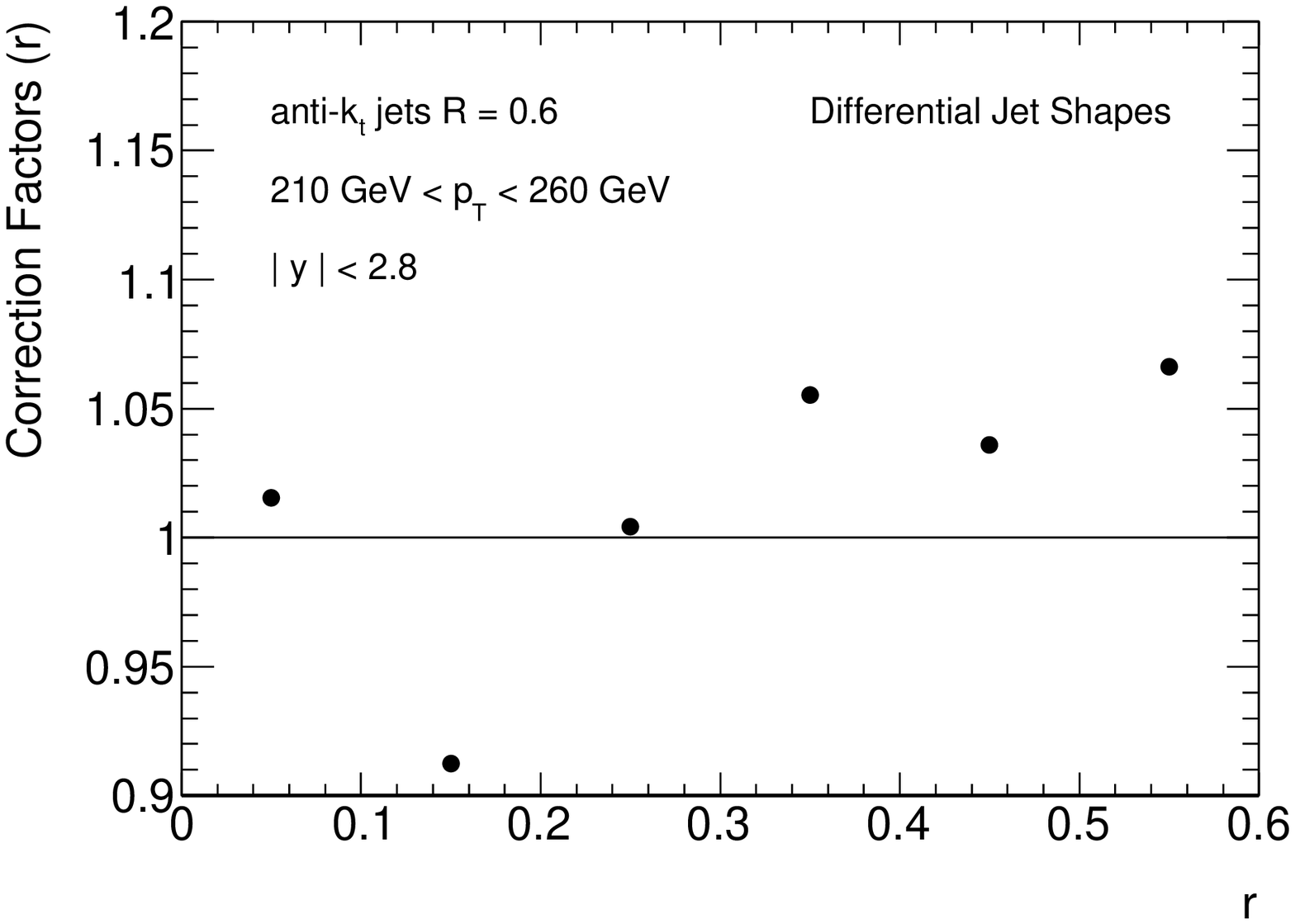}
\includegraphics[width=0.495\textwidth]{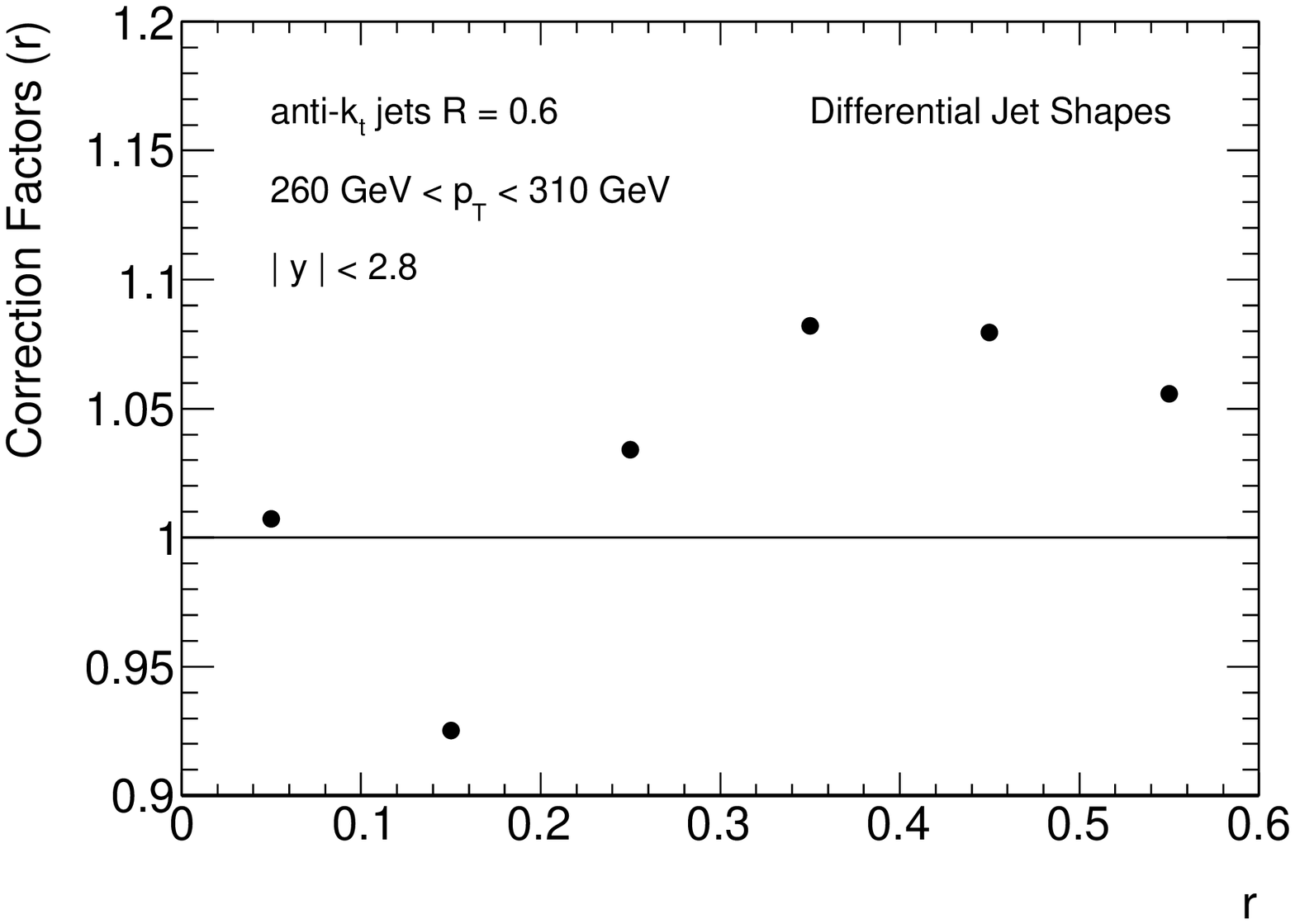}
}
\mbox{
\includegraphics[width=0.495\textwidth]{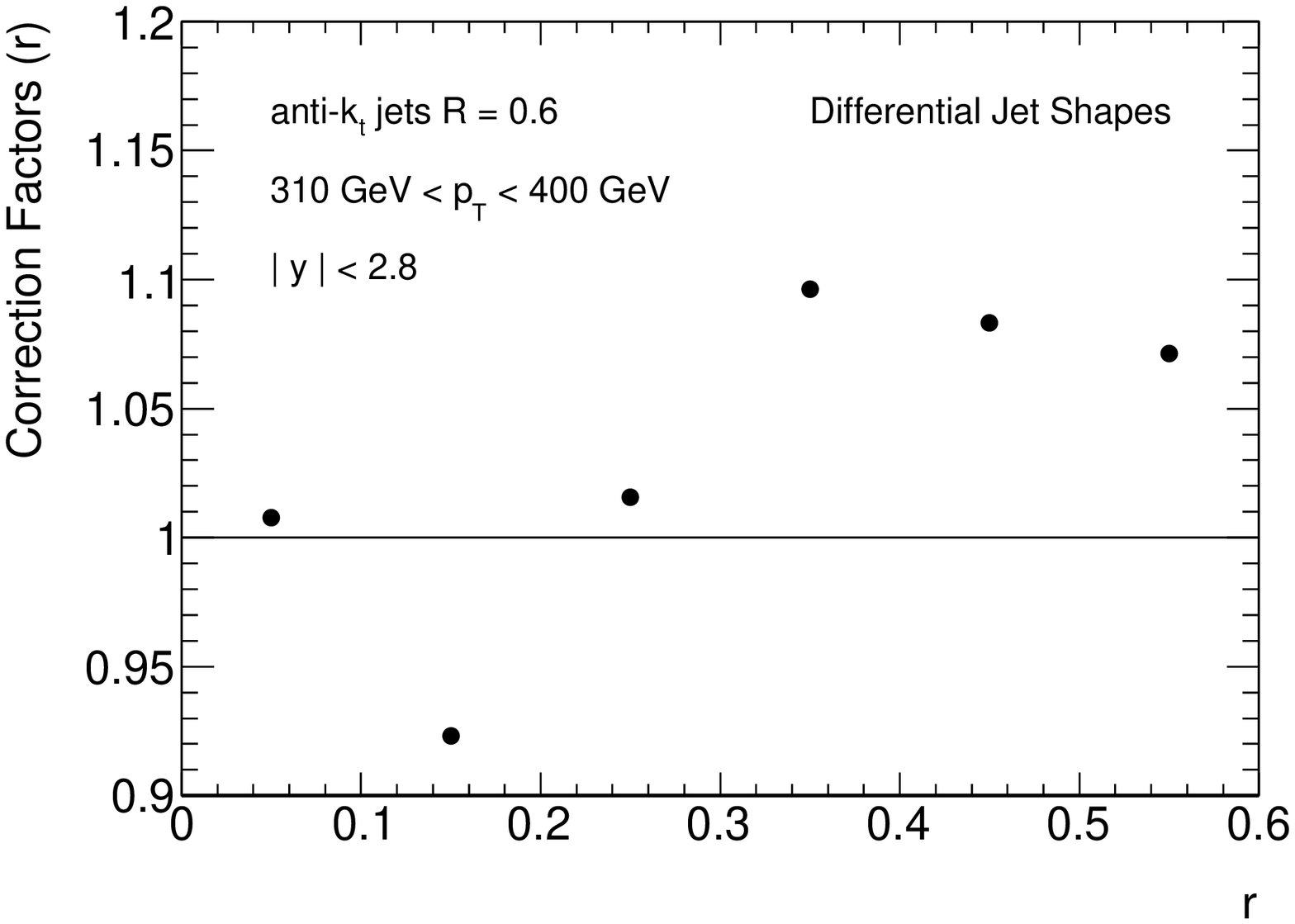}
\includegraphics[width=0.495\textwidth]{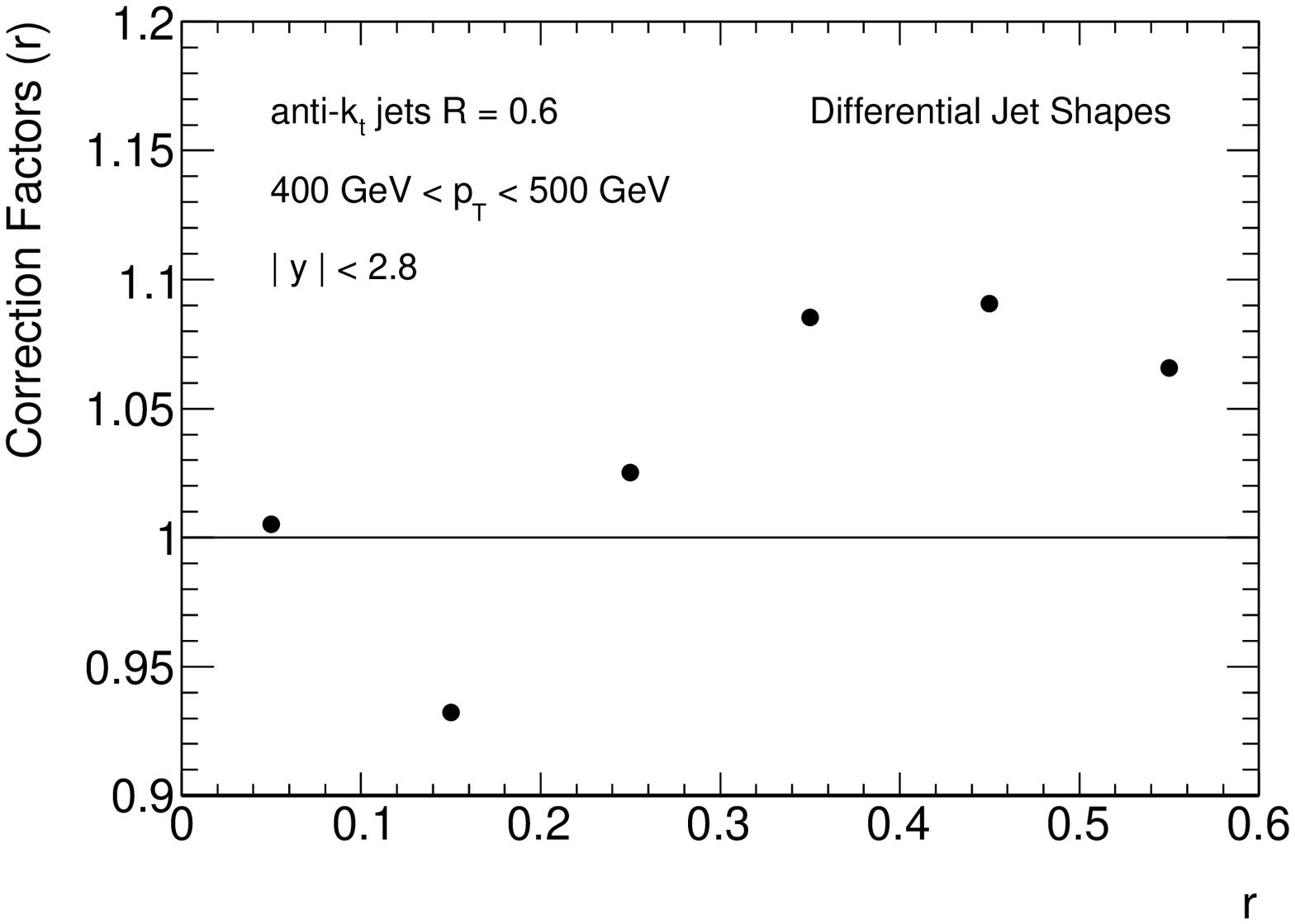}
}
\mbox{
\includegraphics[width=0.495\textwidth]{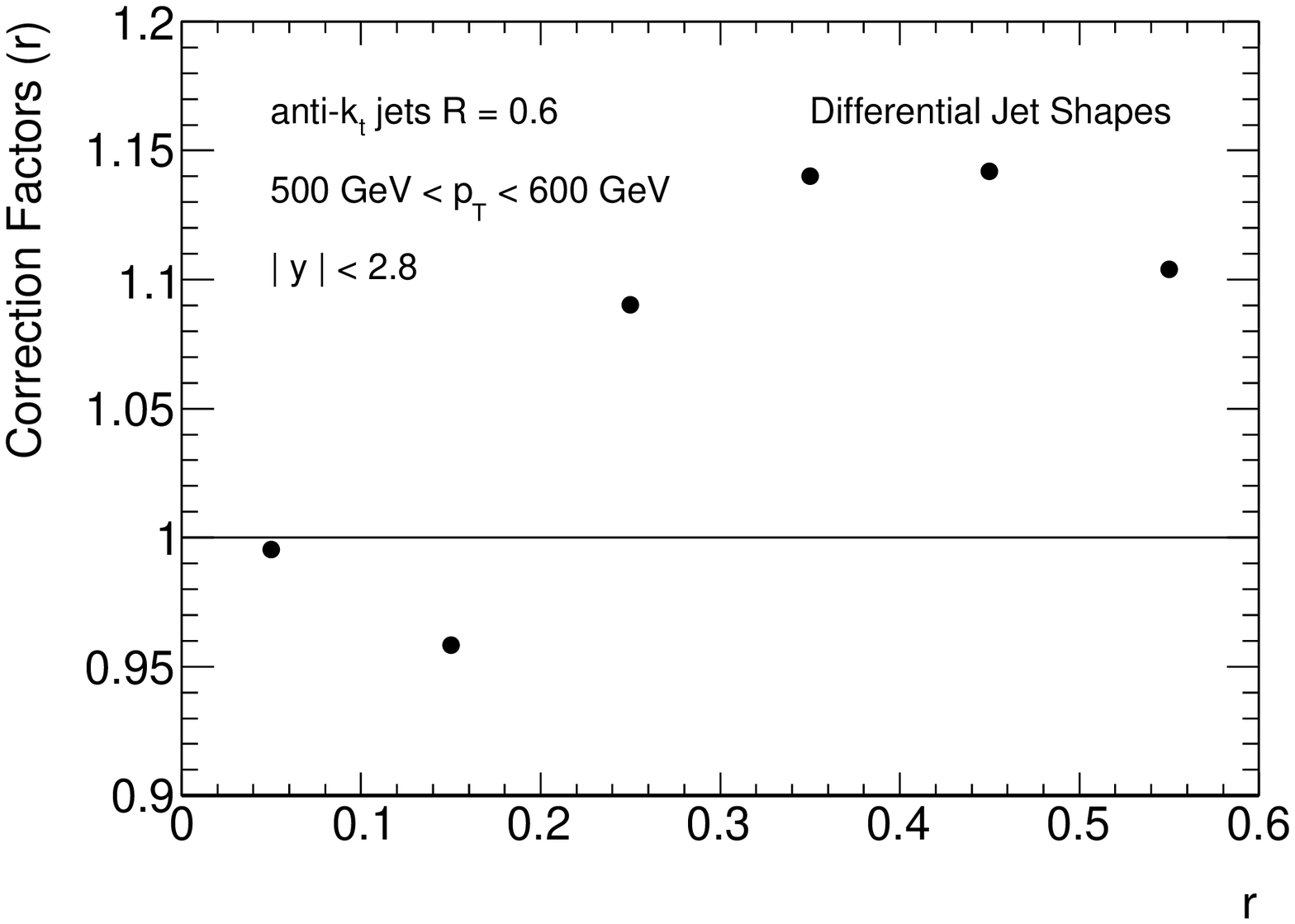}
}
\end{center}
\vspace{-0.7 cm}
\caption{\small
Correction factors applied to the measured differential jet shapes to correct the measurements for detector effects
for jets with $|\rapjet| < 2.8$ and $210 \ {\rm GeV} < \ptjet < 600  \ {\rm GeV}$
}
\label{fig_unf2}
\end{figure}

\begin{figure}[tbh]
\begin{center}
\mbox{
\includegraphics[width=0.495\textwidth]{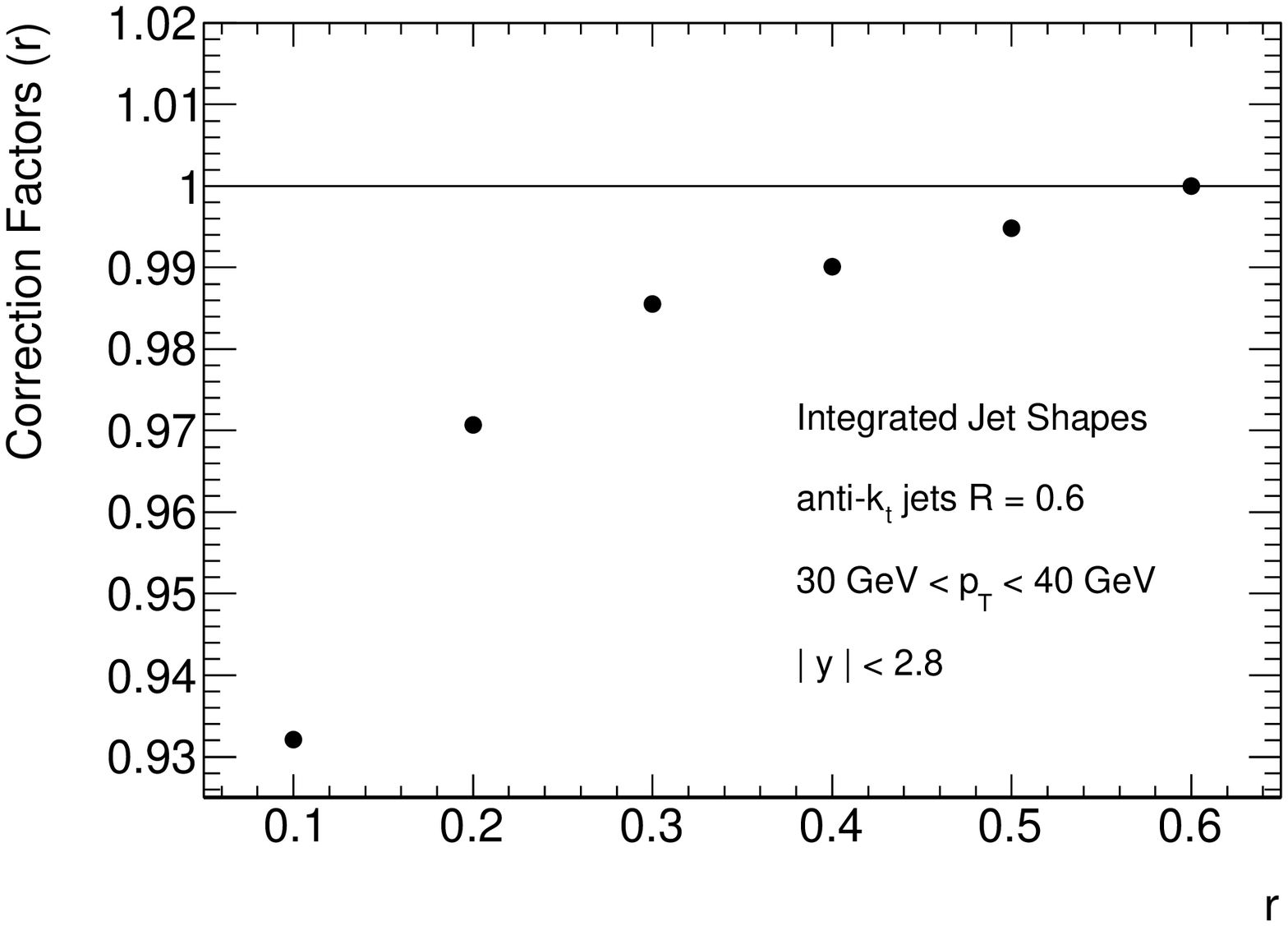}
\includegraphics[width=0.495\textwidth]{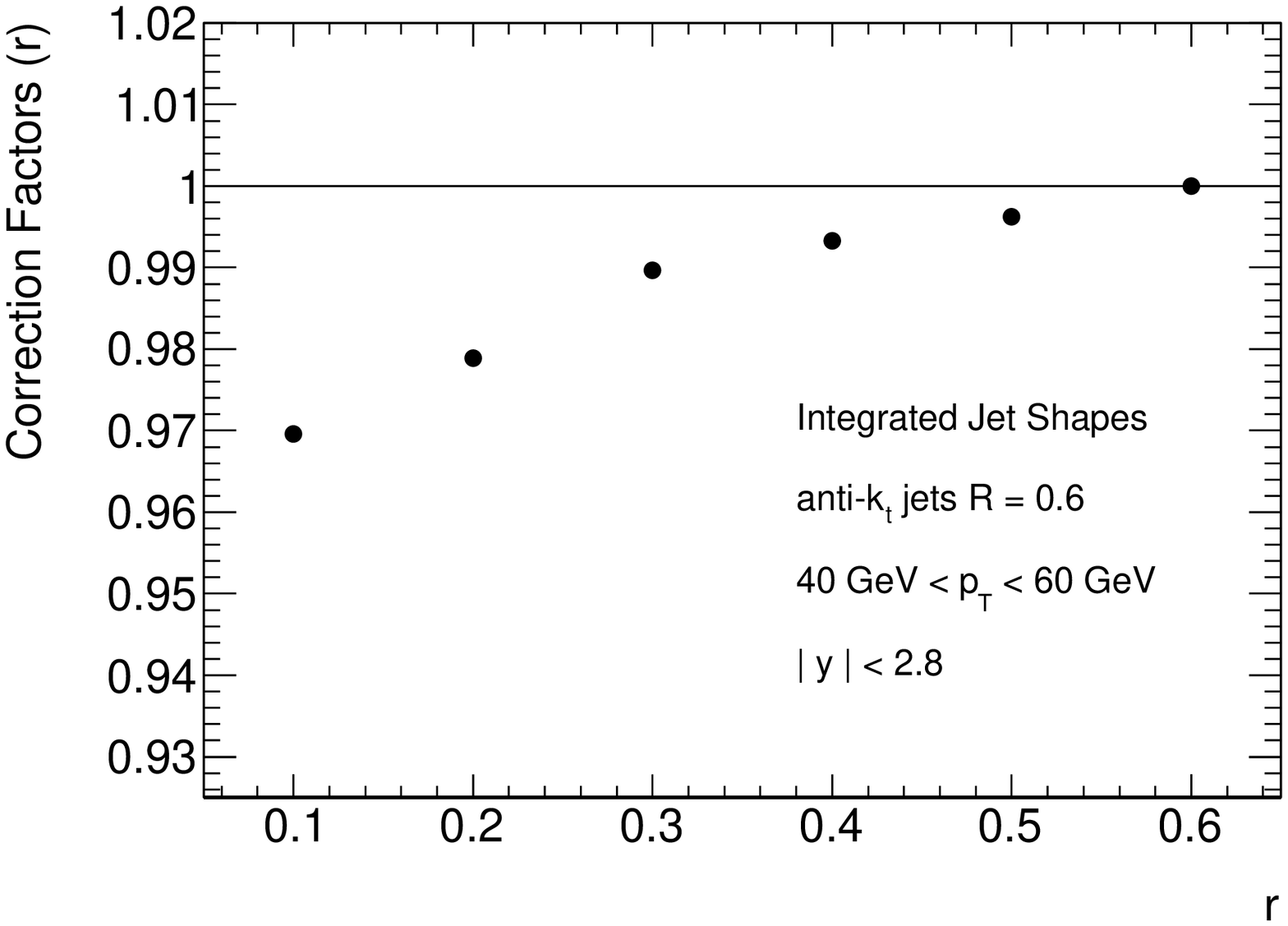}
}
\mbox{
\includegraphics[width=0.495\textwidth]{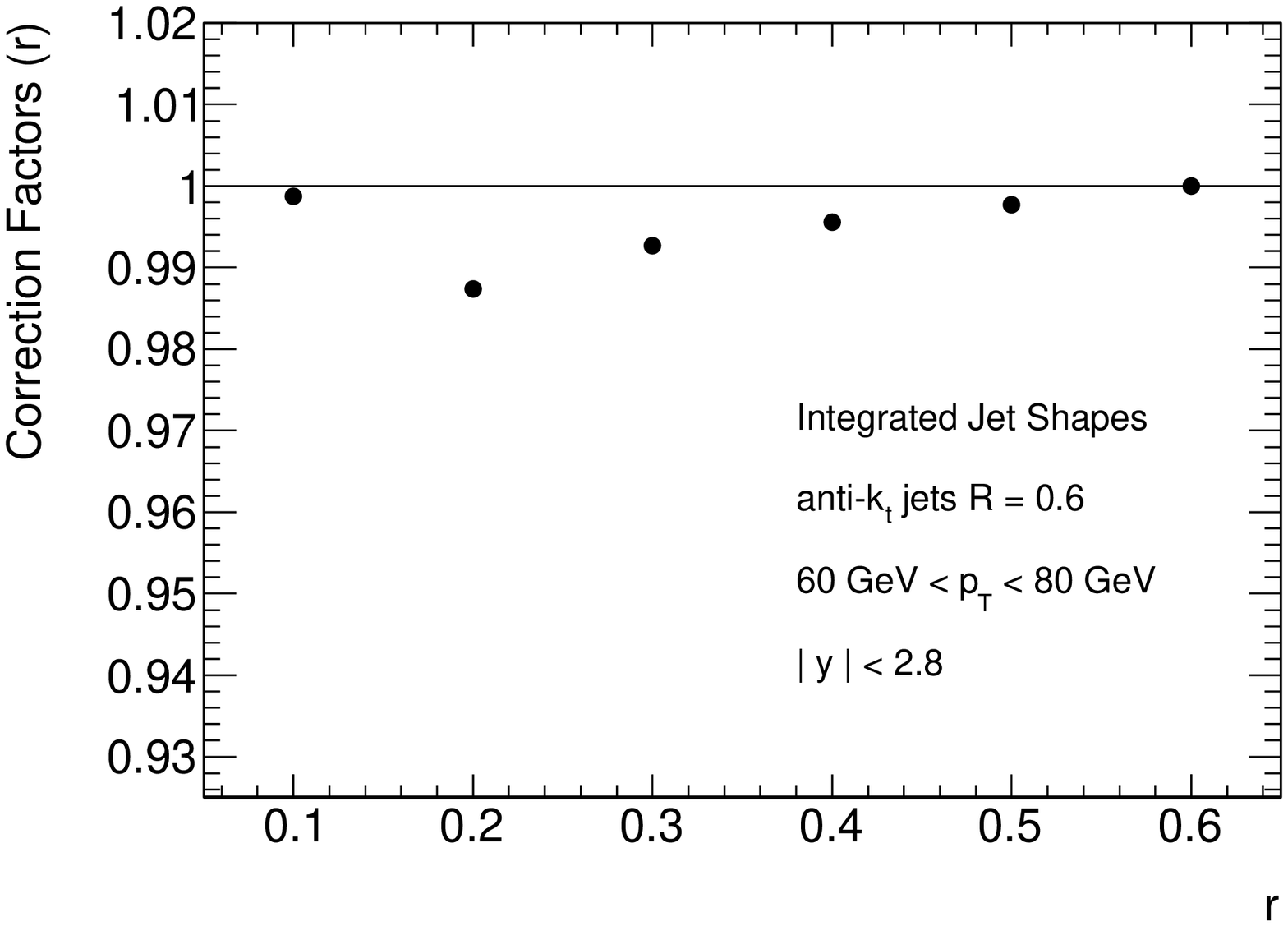}
\includegraphics[width=0.495\textwidth]{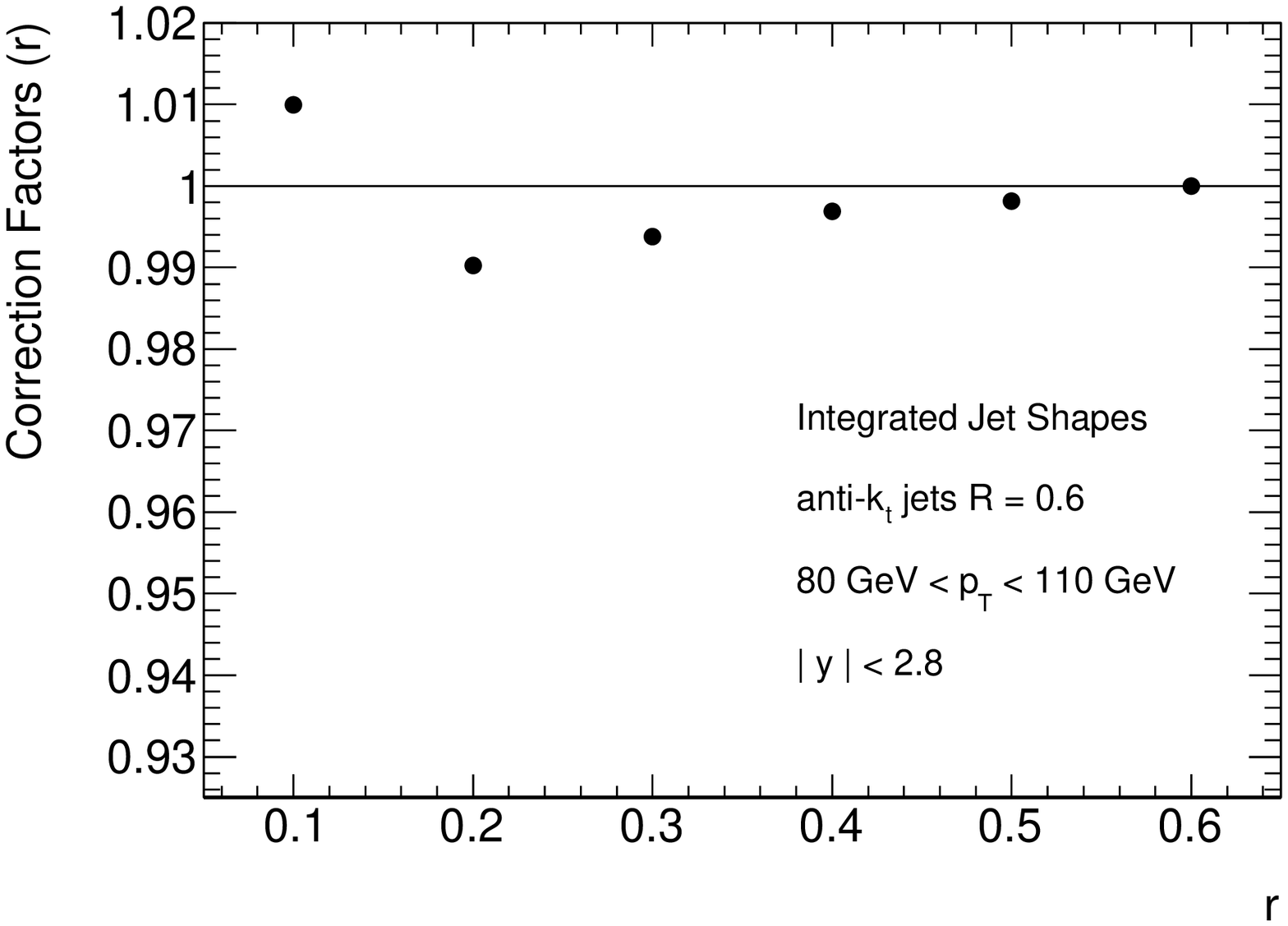}
}
\mbox{
\includegraphics[width=0.495\textwidth]{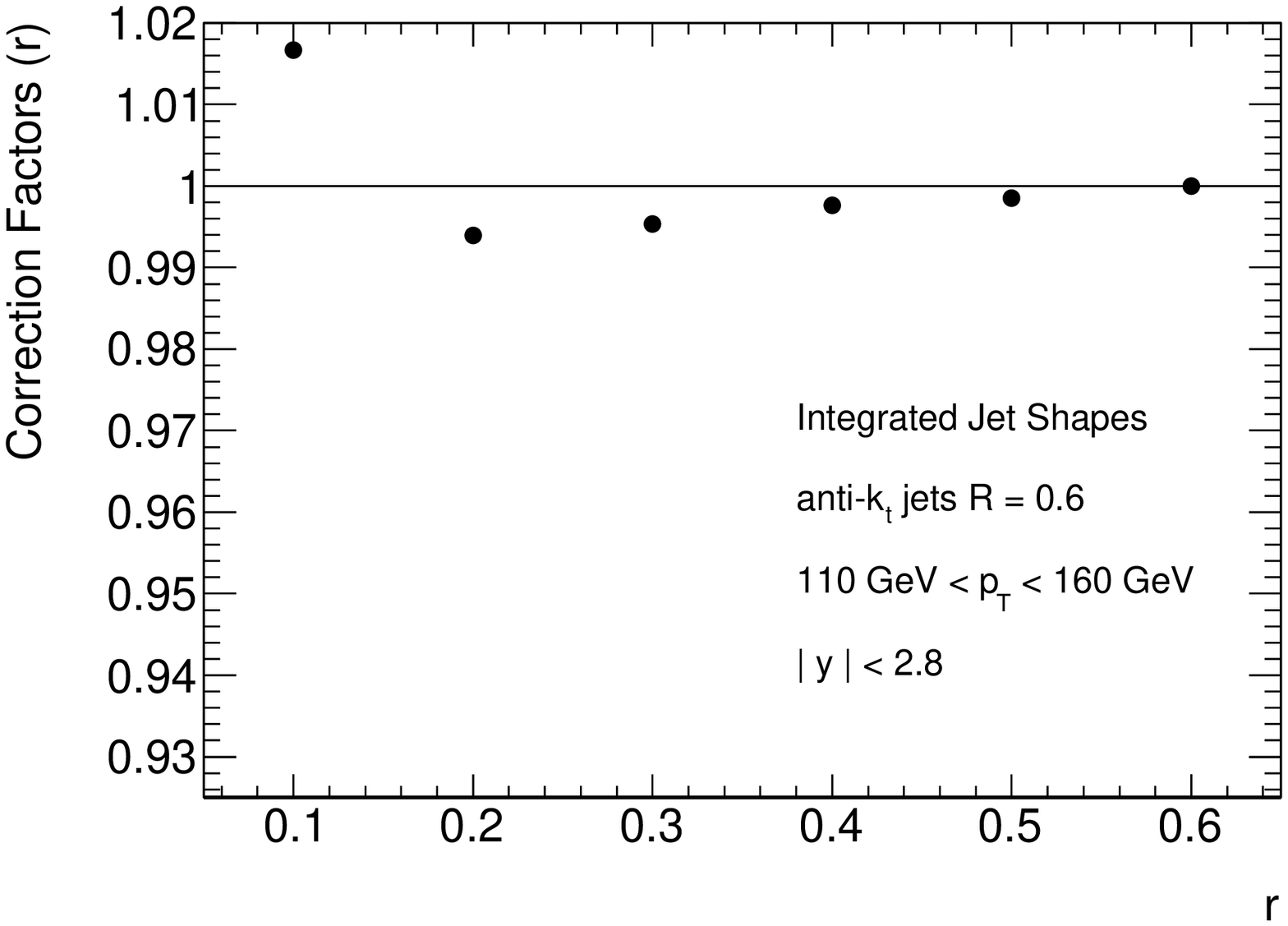}
\includegraphics[width=0.495\textwidth]{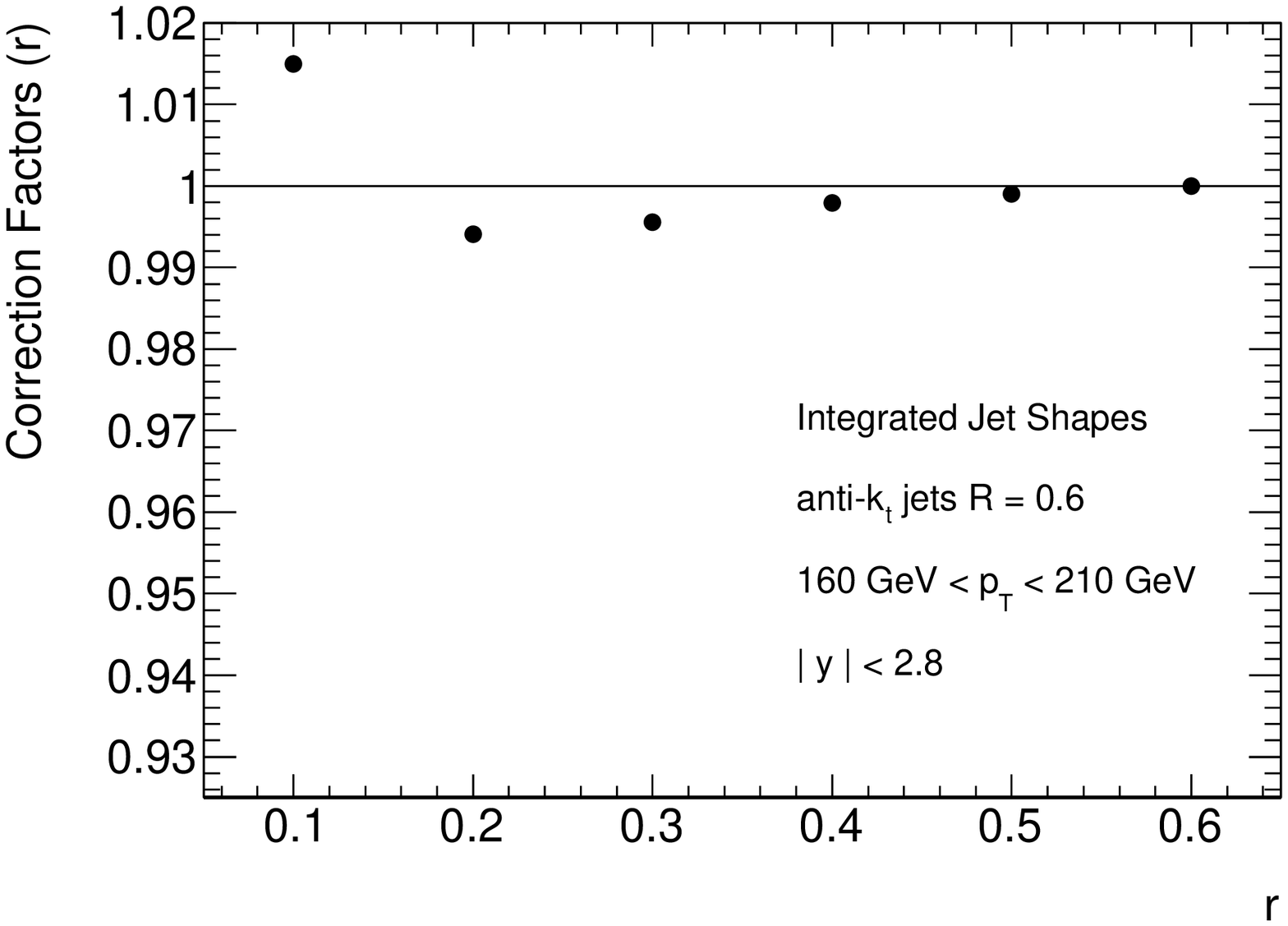}
}
\end{center}
\vspace{-0.7 cm}
\caption{\small
Correction factors applied to the measured integrated jet shapes to correct the measurements for detector effects 
for jets with $|\rapjet| < 2.8$ and $30 \ {\rm GeV} < \ptjet < 210  \ {\rm GeV}$.
}
\label{fig_unf3}
\end{figure}

\begin{figure}[tbh]
\begin{center}
\mbox{
\includegraphics[width=0.495\textwidth]{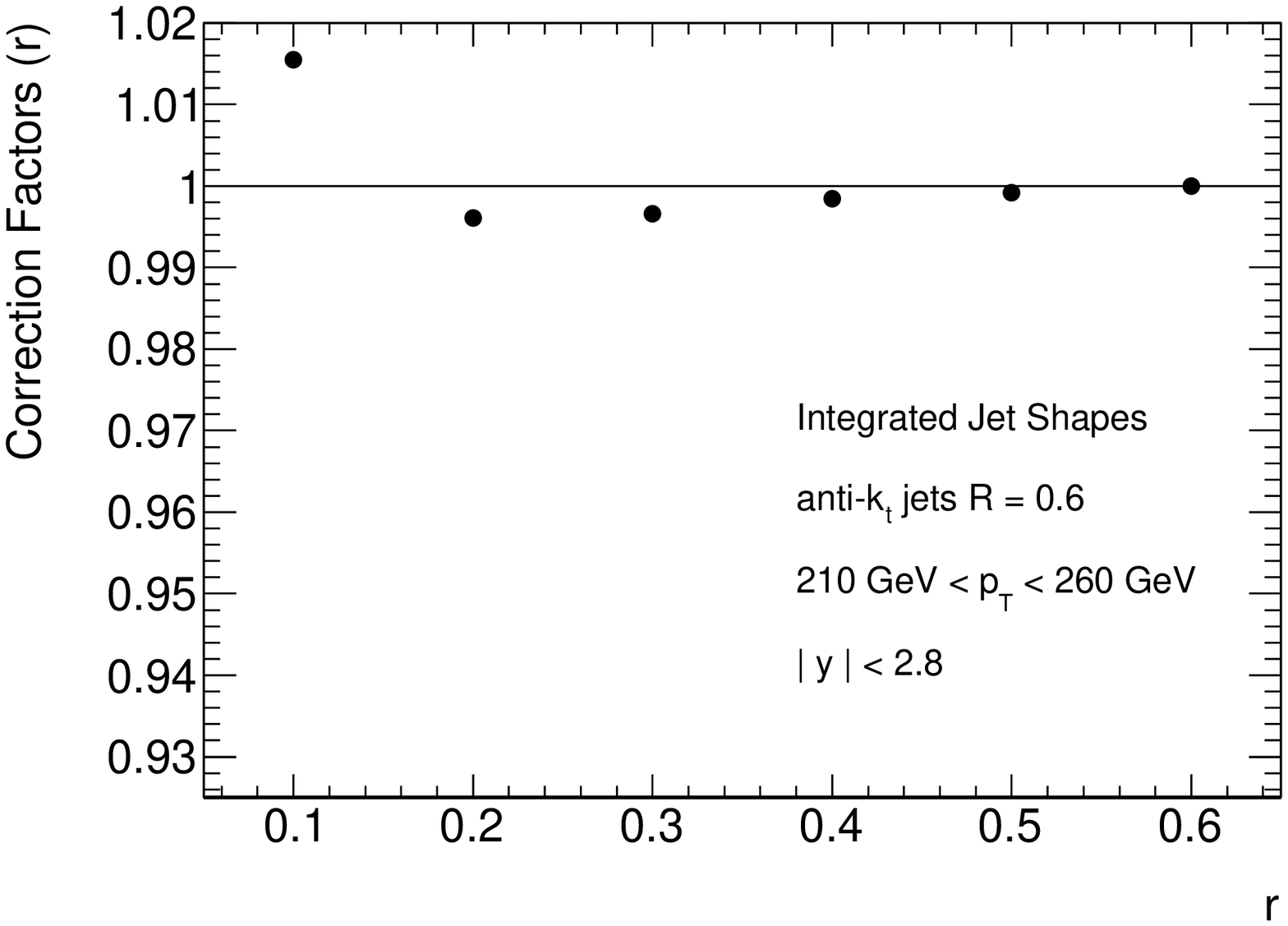}
\includegraphics[width=0.495\textwidth]{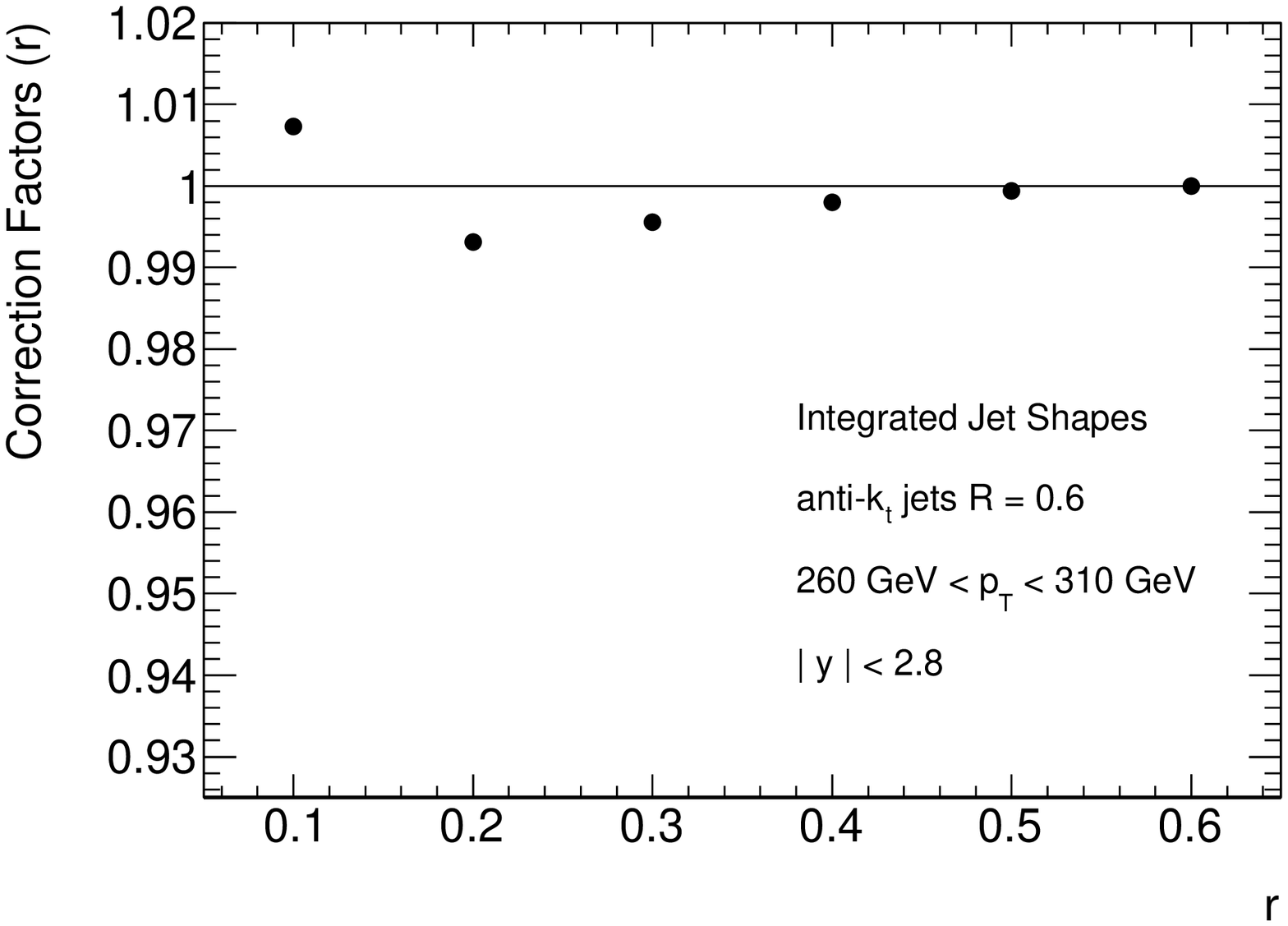}
}
\mbox{
\includegraphics[width=0.495\textwidth]{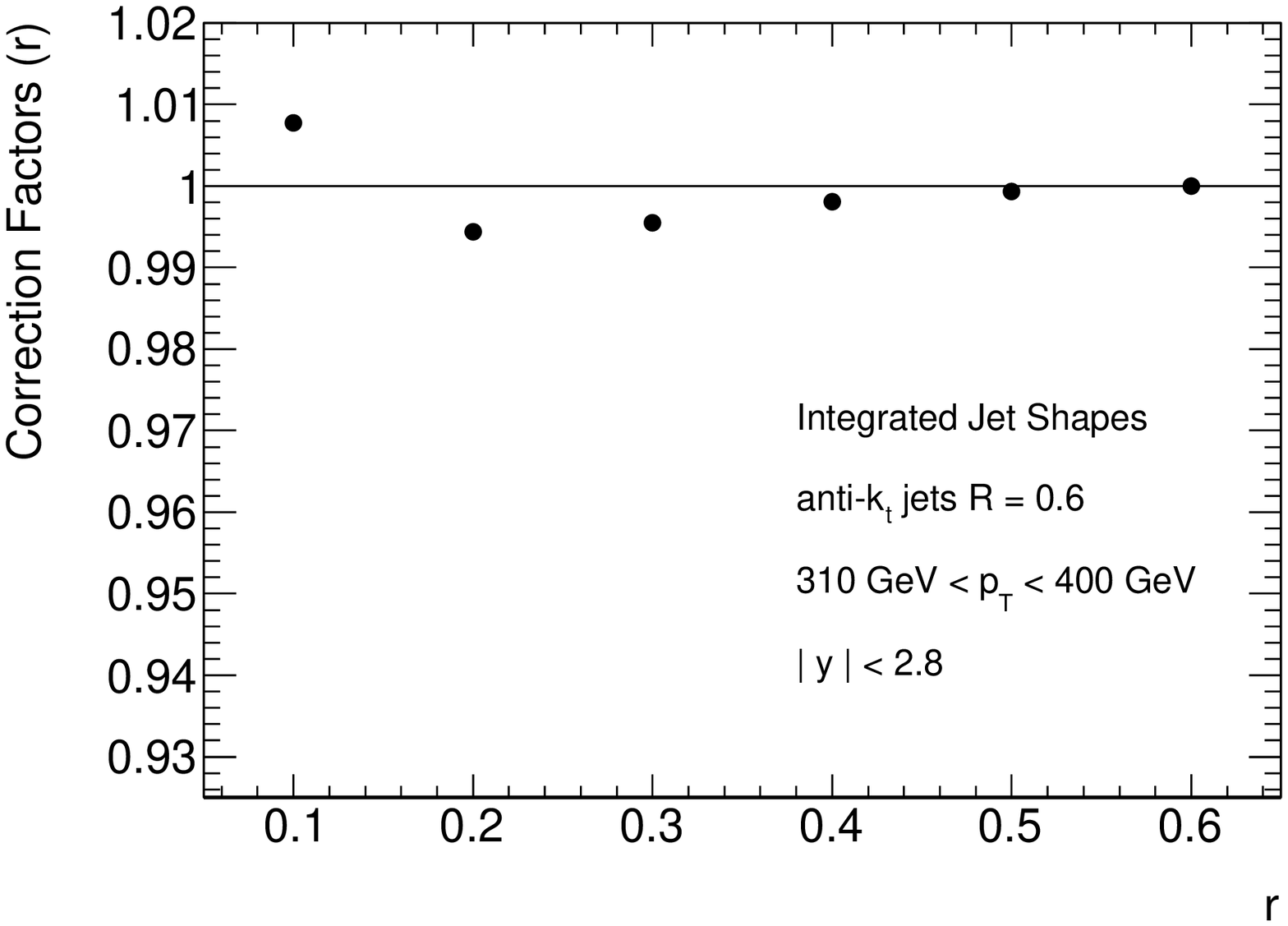}
\includegraphics[width=0.495\textwidth]{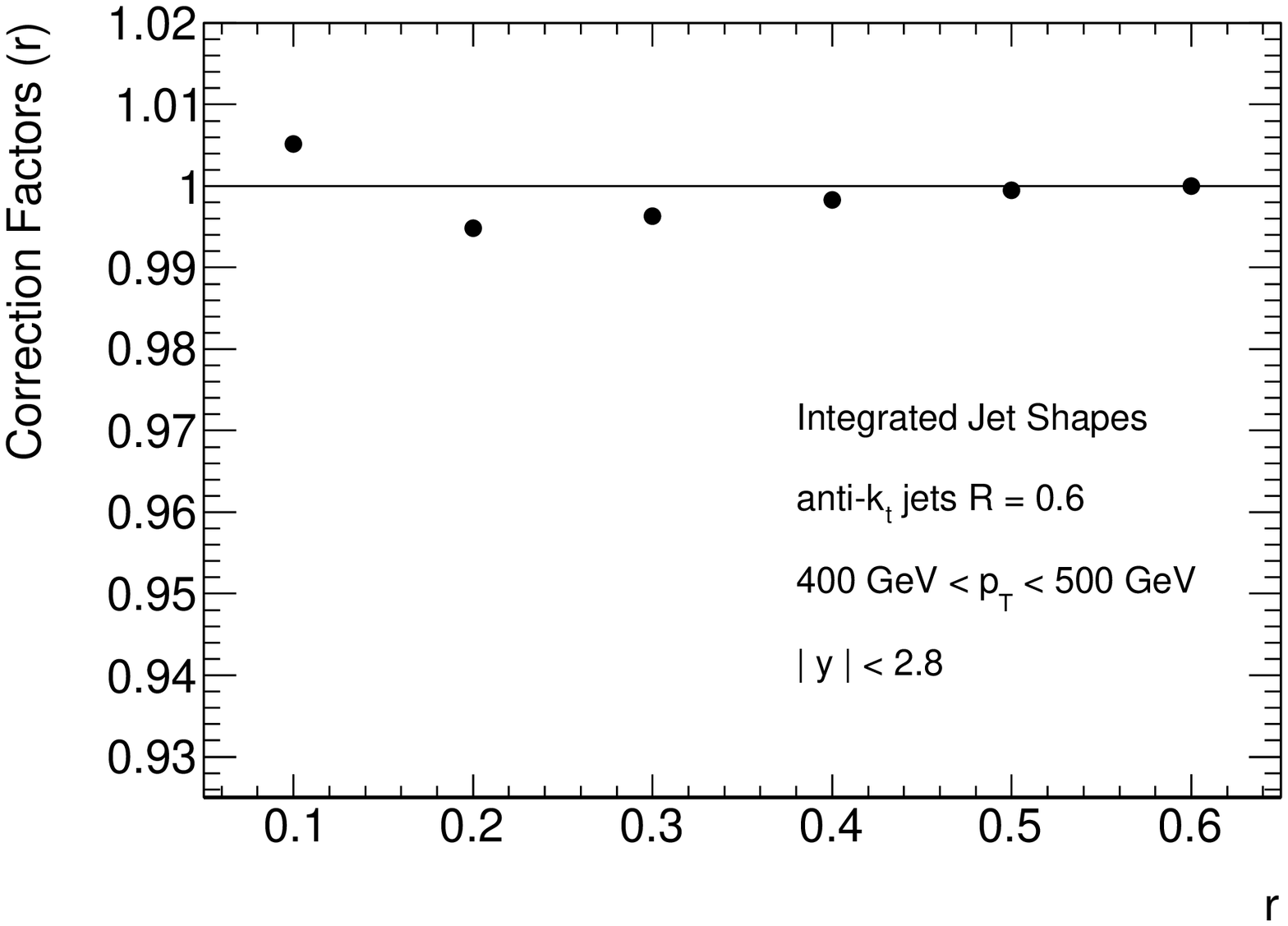}
}
\mbox{
\includegraphics[width=0.495\textwidth]{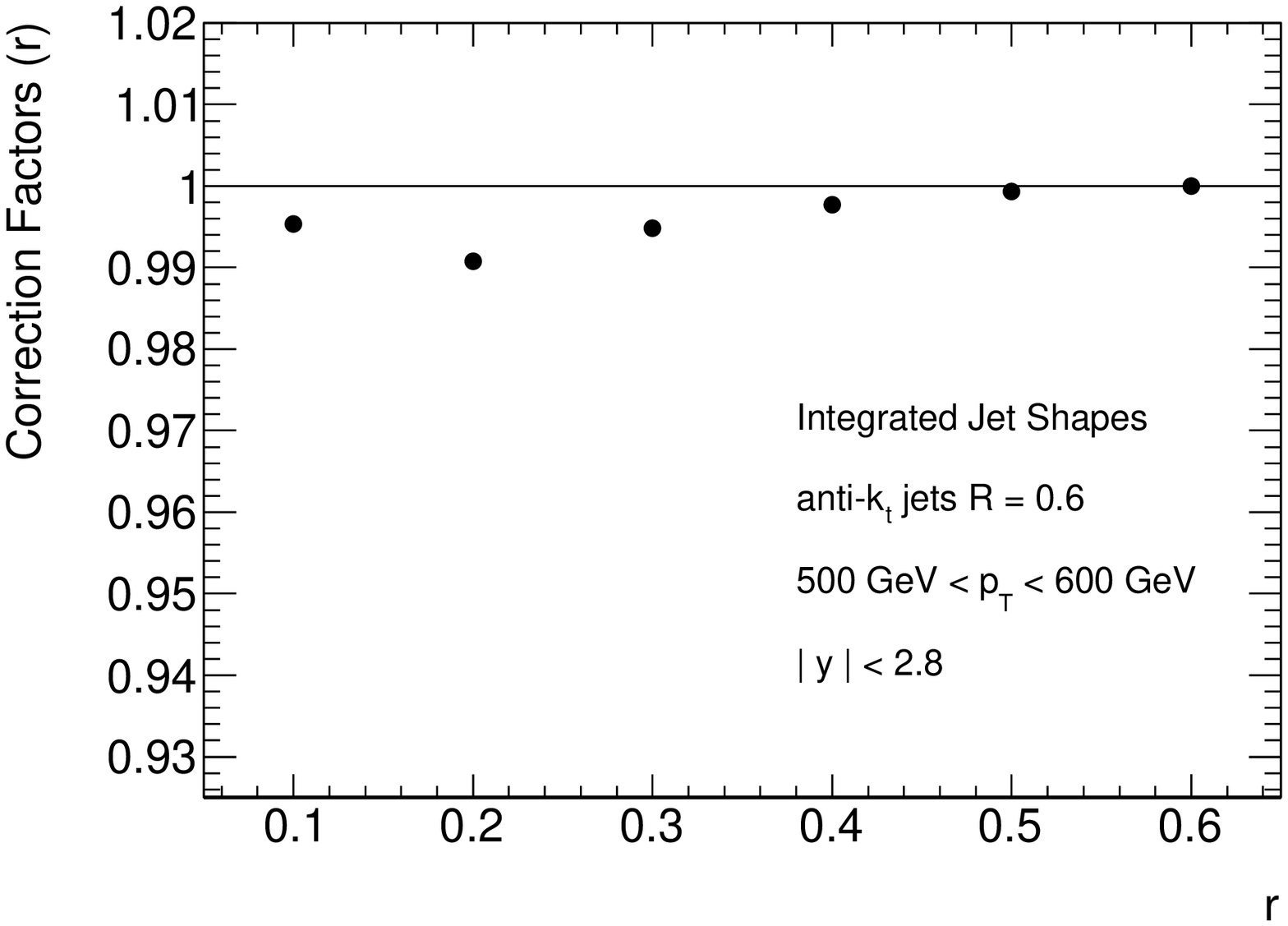}
}
\end{center}
\vspace{-0.7 cm}
\caption{\small
Correction factors applied to the measured integrated jet shapes to correct the measurements for detector effects 
for jets with $|\rapjet| < 2.8$ and $210 \ {\rm GeV} < \ptjet < 600  \ {\rm GeV}$
}
\label{fig_unf4}
\end{figure}

The correction factors $U(r,\ptjet,|\rapjet|) = \rho(r)^{par}_{mc}/\rho(r)^{cal}_{mc}$ present a moderate $\ptjet$ and $|\rapjet|$
dependence and vary between 0.95 and 1.1 as $r$ increases. For the integrated jet shapes, the correction factors differ from unity by less than $5 \%$. 
This is shown in Figures~\ref{fig_unf1}~and~\ref{fig_unf2} (\ref{fig_unf3}~and~\ref{fig_unf4}) for the differential (integrated) jet shapes.
The corrected jet shape measurements in each $\ptjet$ and $|\rapjet|$ region 
are computed by multiplying bin-by-bin the measured uncorrected jet shapes in data 
by the corresponding correction factors.

\clearpage

\clearpage

\clearpage

\clearpage


\section{Systematic uncertainties}

A detailed study of systematic uncertainties on the measured differential and integrated 
jet shapes has been performed. The impact on the differential measurements is described here.

\begin{figure}[tbh]
\begin{center}
\mbox{
\includegraphics[width=0.495\textwidth]{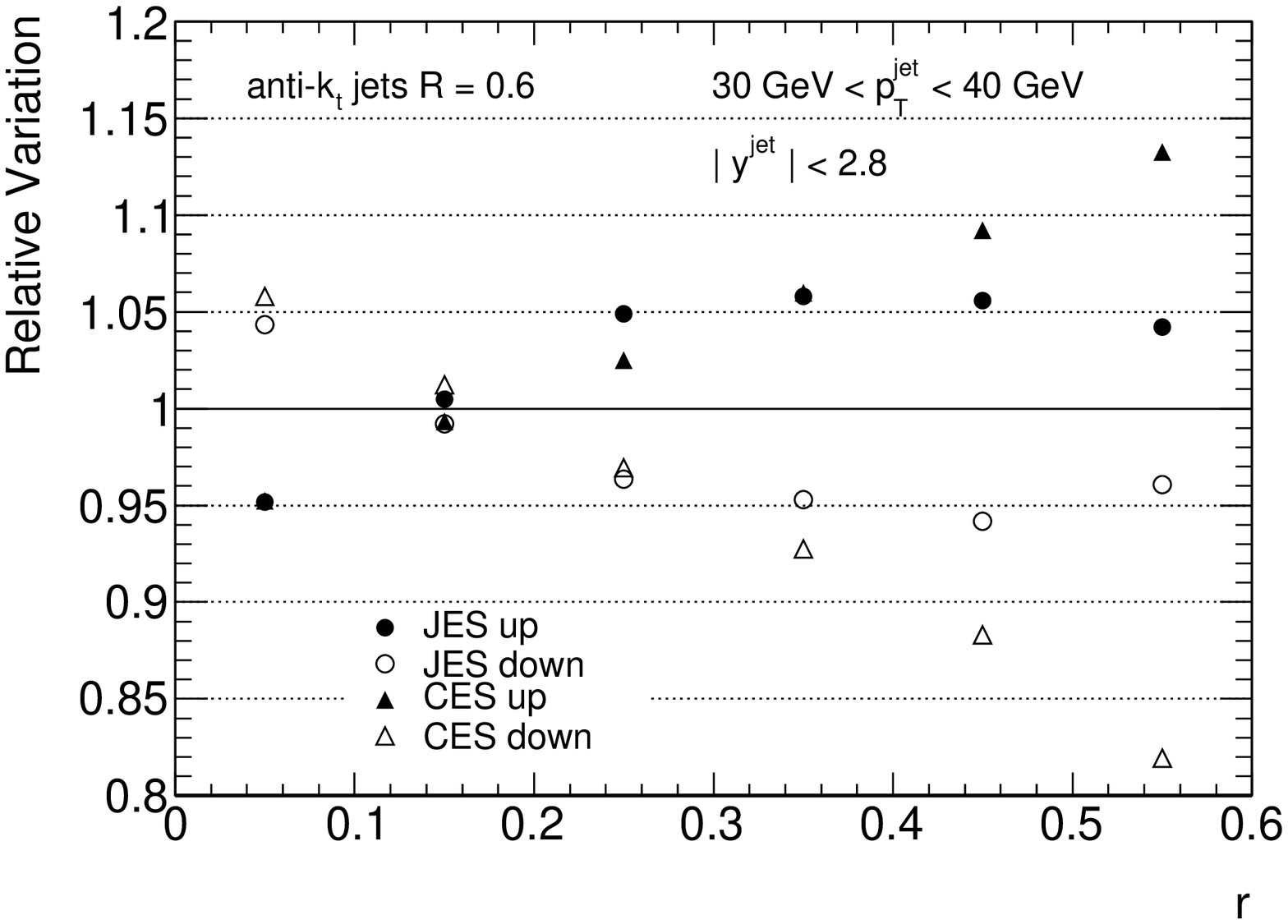}
\includegraphics[width=0.495\textwidth]{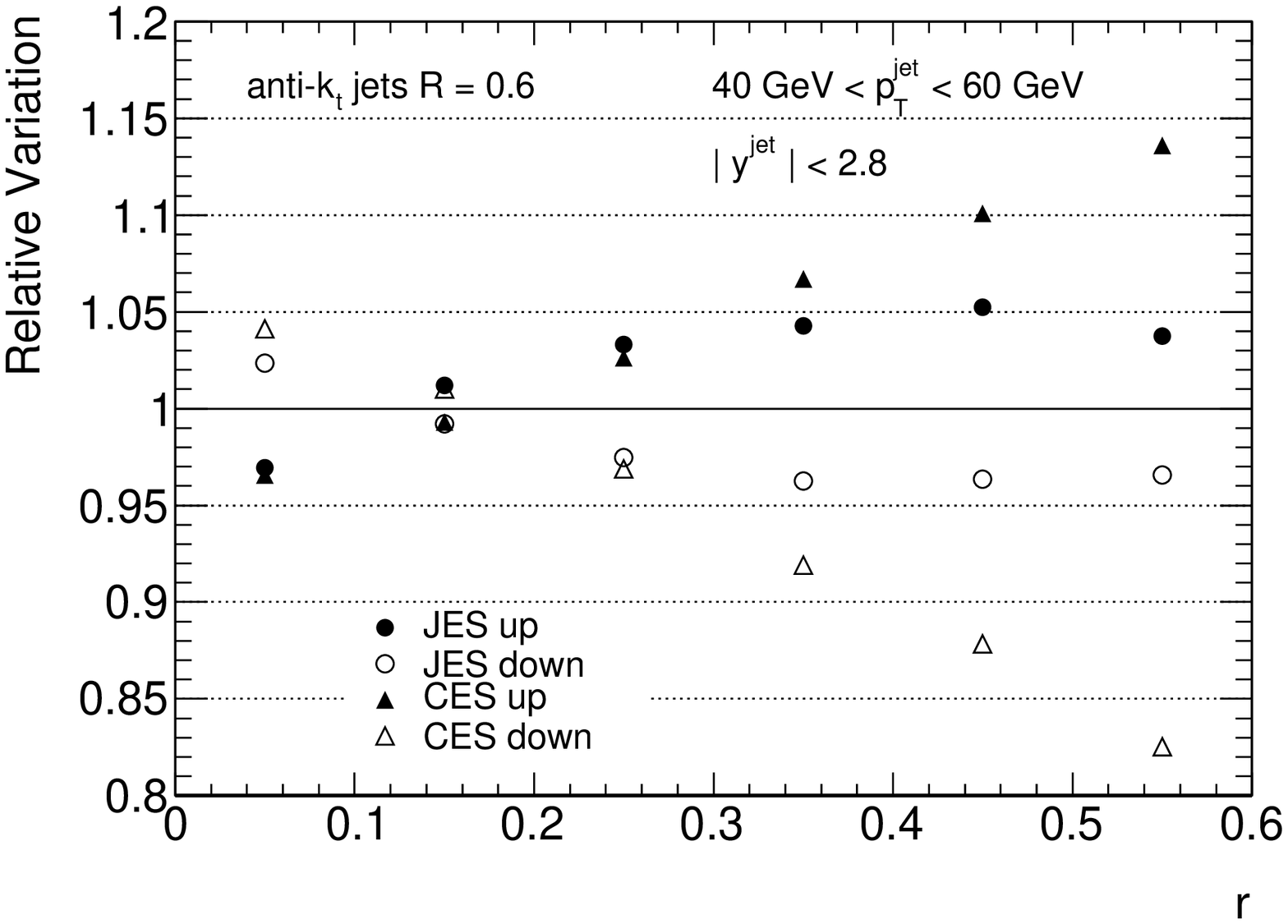}
}
\mbox{
\includegraphics[width=0.495\textwidth]{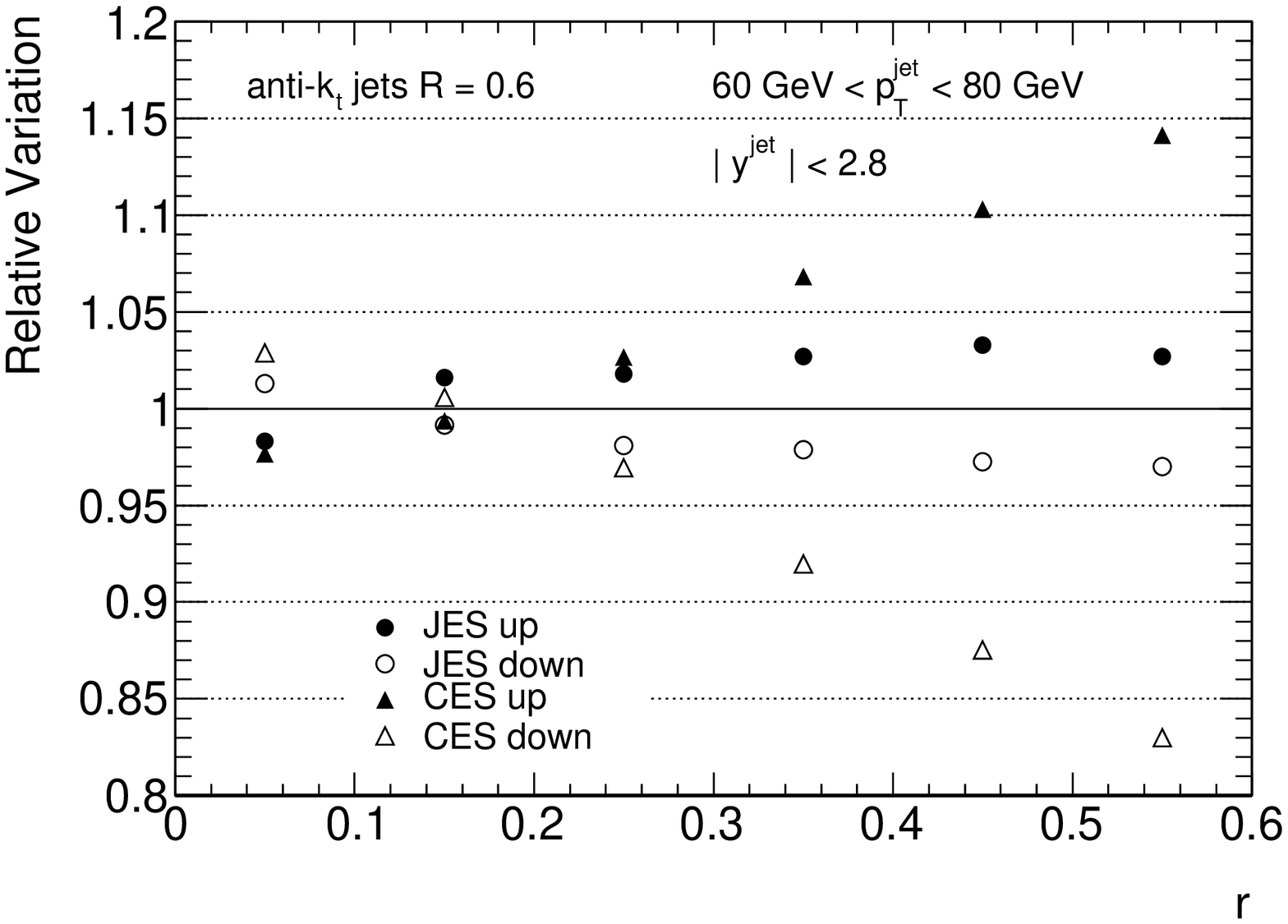}
\includegraphics[width=0.495\textwidth]{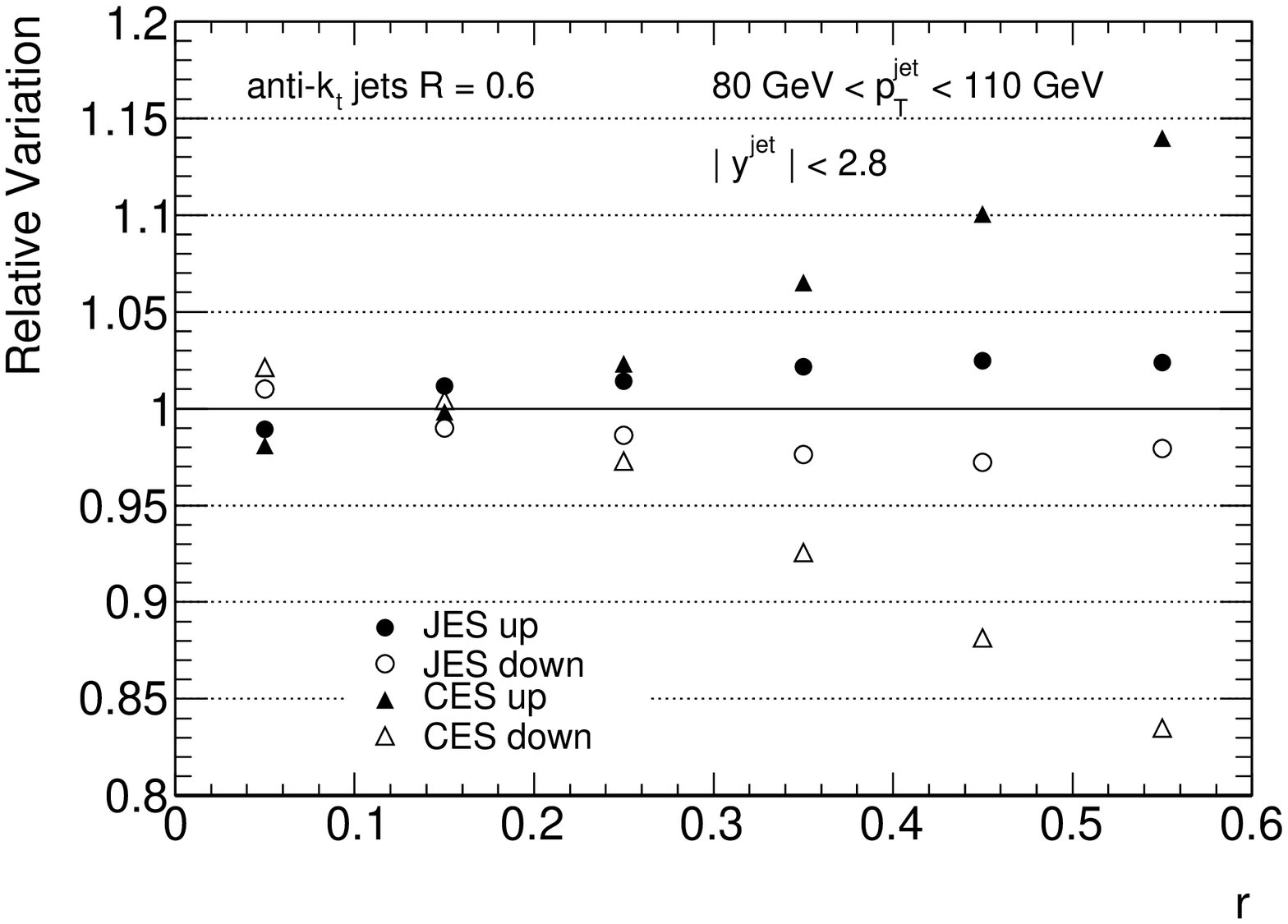}
}
\mbox{
\includegraphics[width=0.495\textwidth]{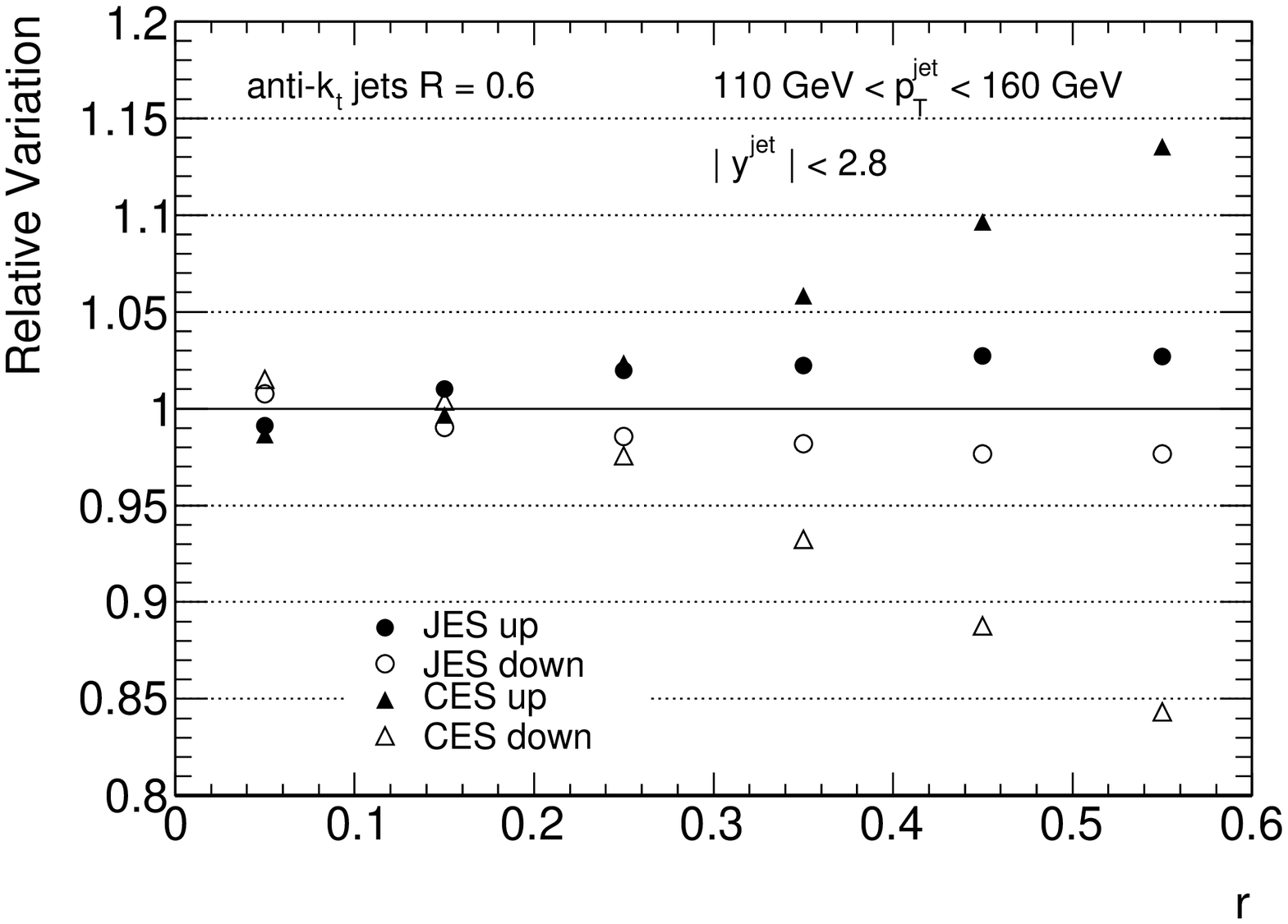}
\includegraphics[width=0.495\textwidth]{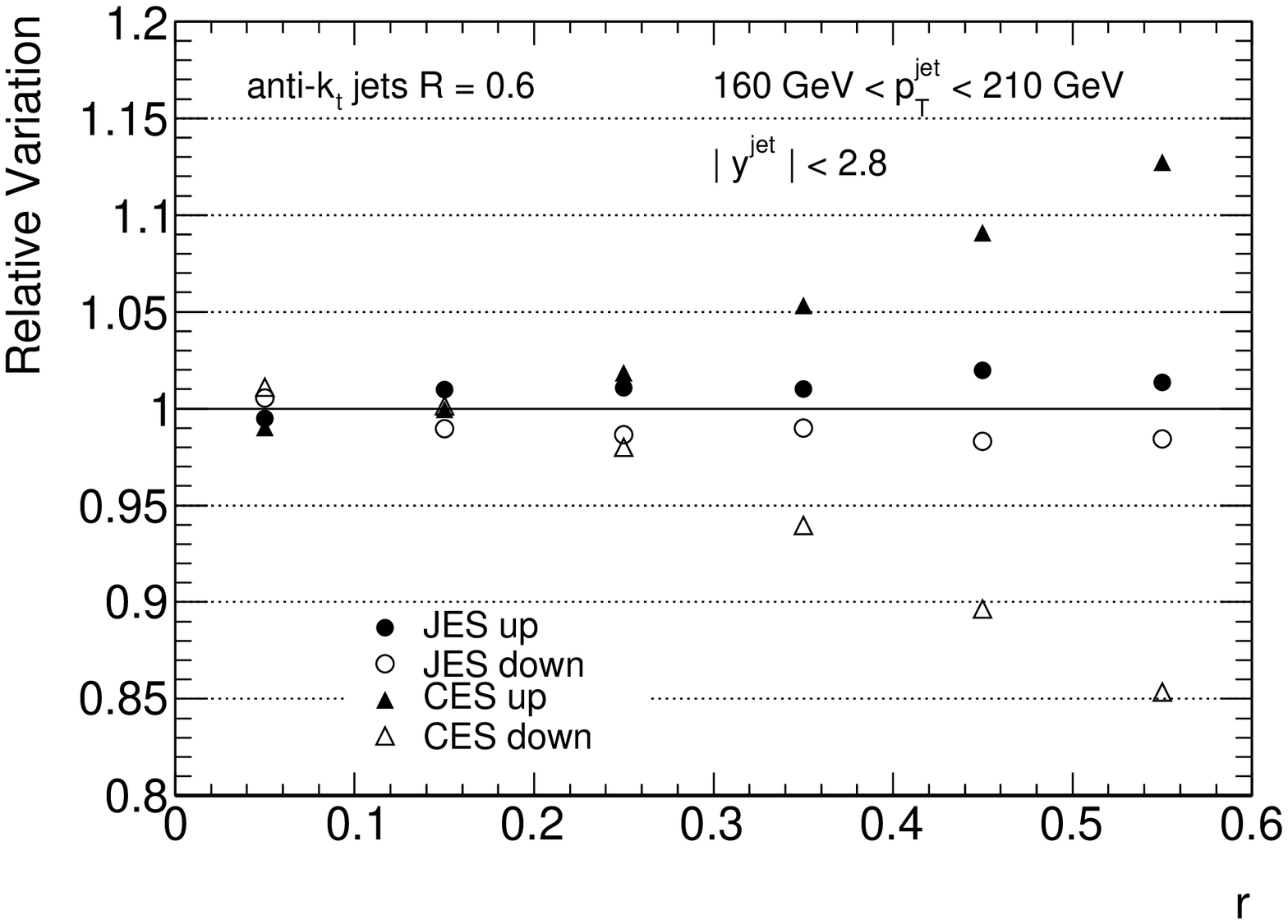}
}
\end{center}
\vspace{-0.7 cm}
\caption{\small
Systematic uncertainty on the differential jet shape related to the absolute energy scale uncertainties on clusters and jets, 
for jets with $|\rapjet| < 2.8$ and $30 \ {\rm GeV} < \ptjet < 210  \ {\rm GeV}$.
}
\label{fig_CJES1}
\end{figure}

\begin{figure}[tbh]
\begin{center}
\mbox{
\includegraphics[width=0.495\textwidth]{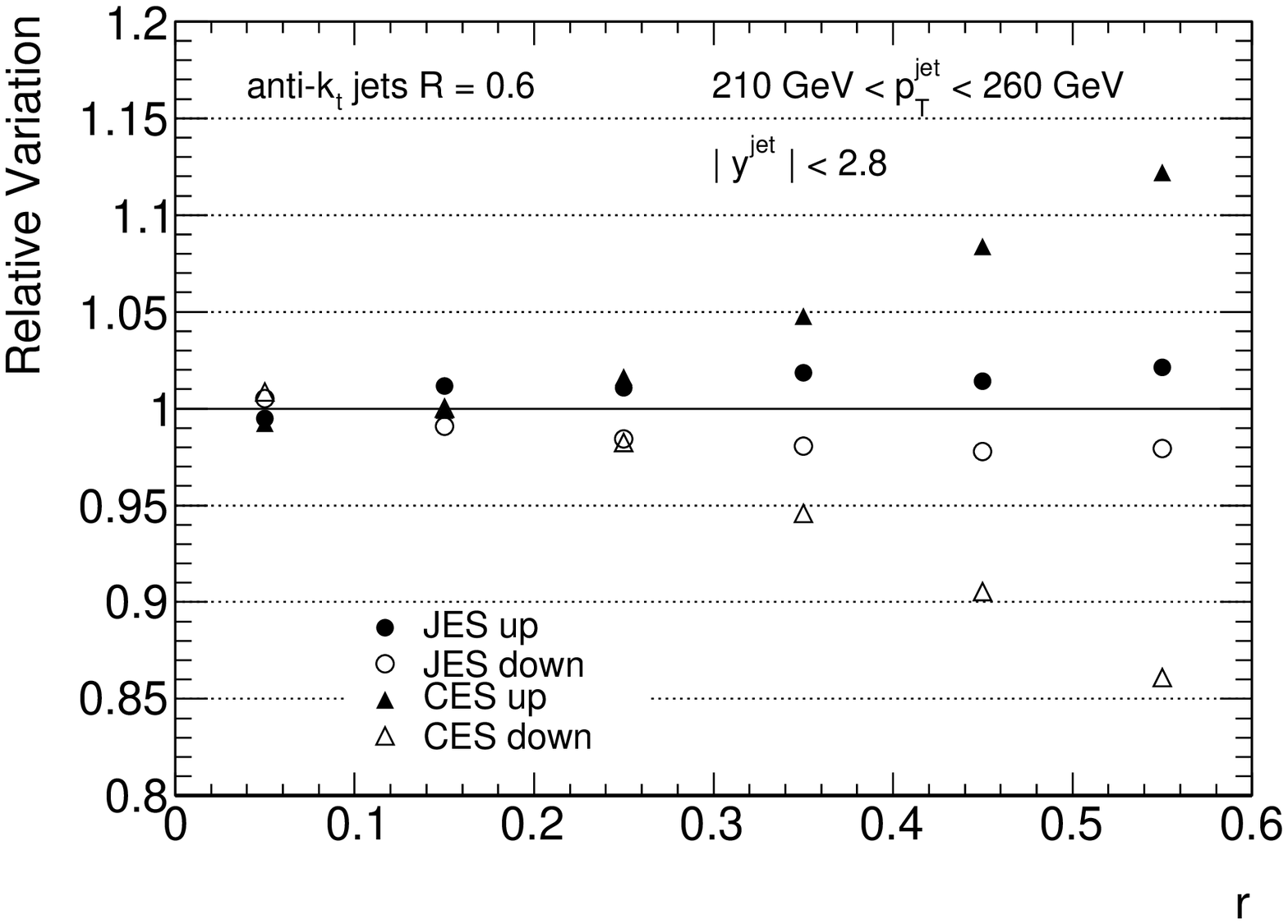}
\includegraphics[width=0.495\textwidth]{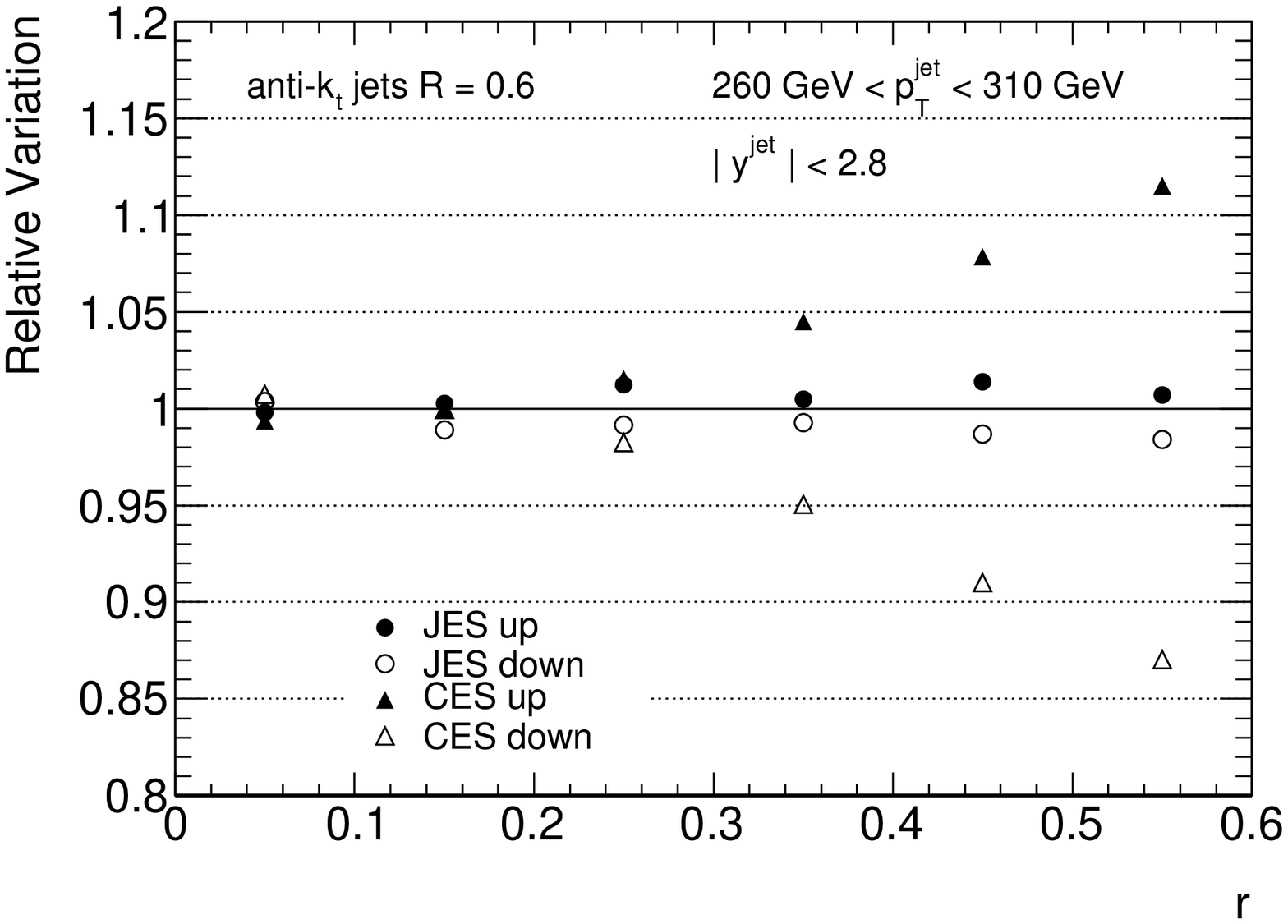}
}
\mbox{
\includegraphics[width=0.495\textwidth]{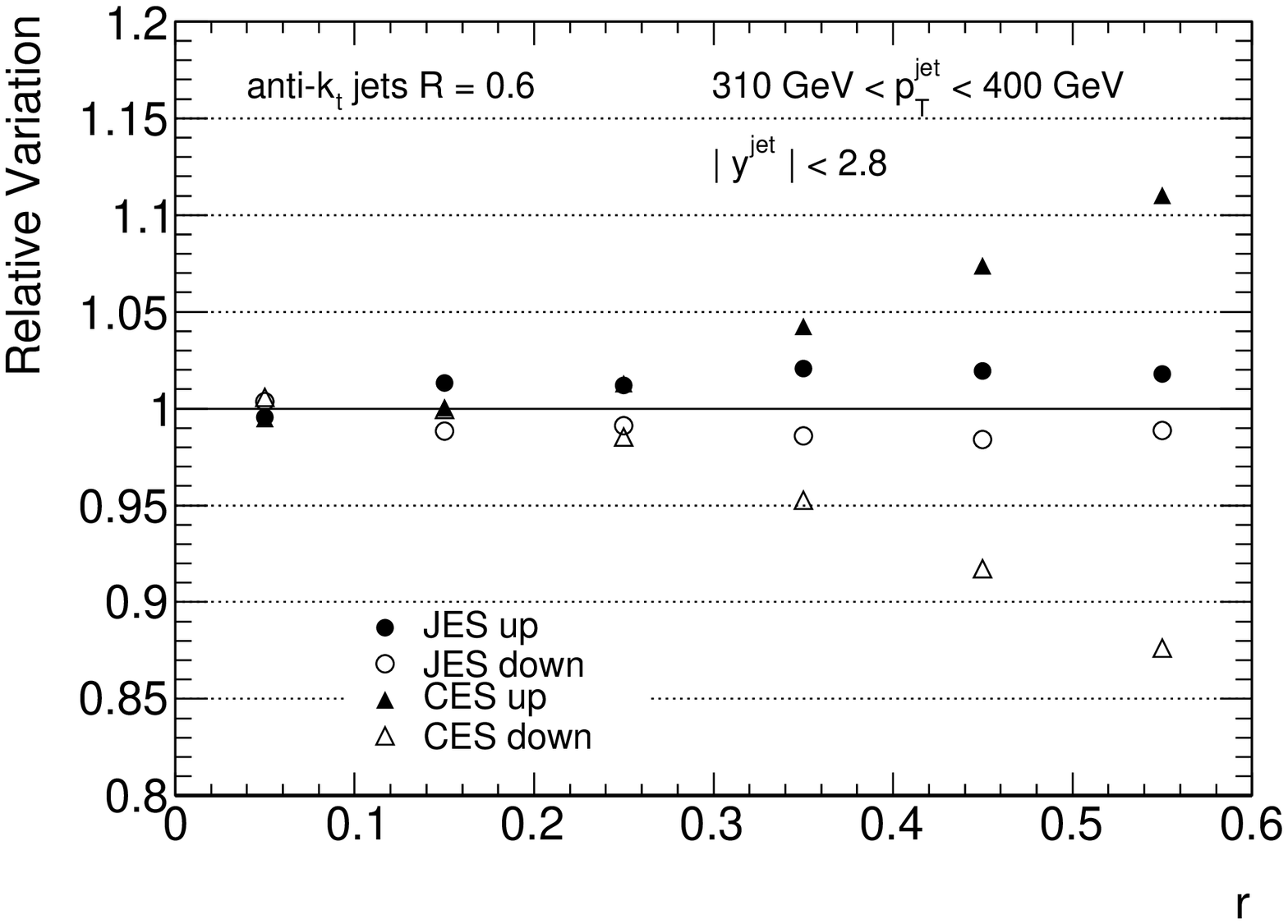}
\includegraphics[width=0.495\textwidth]{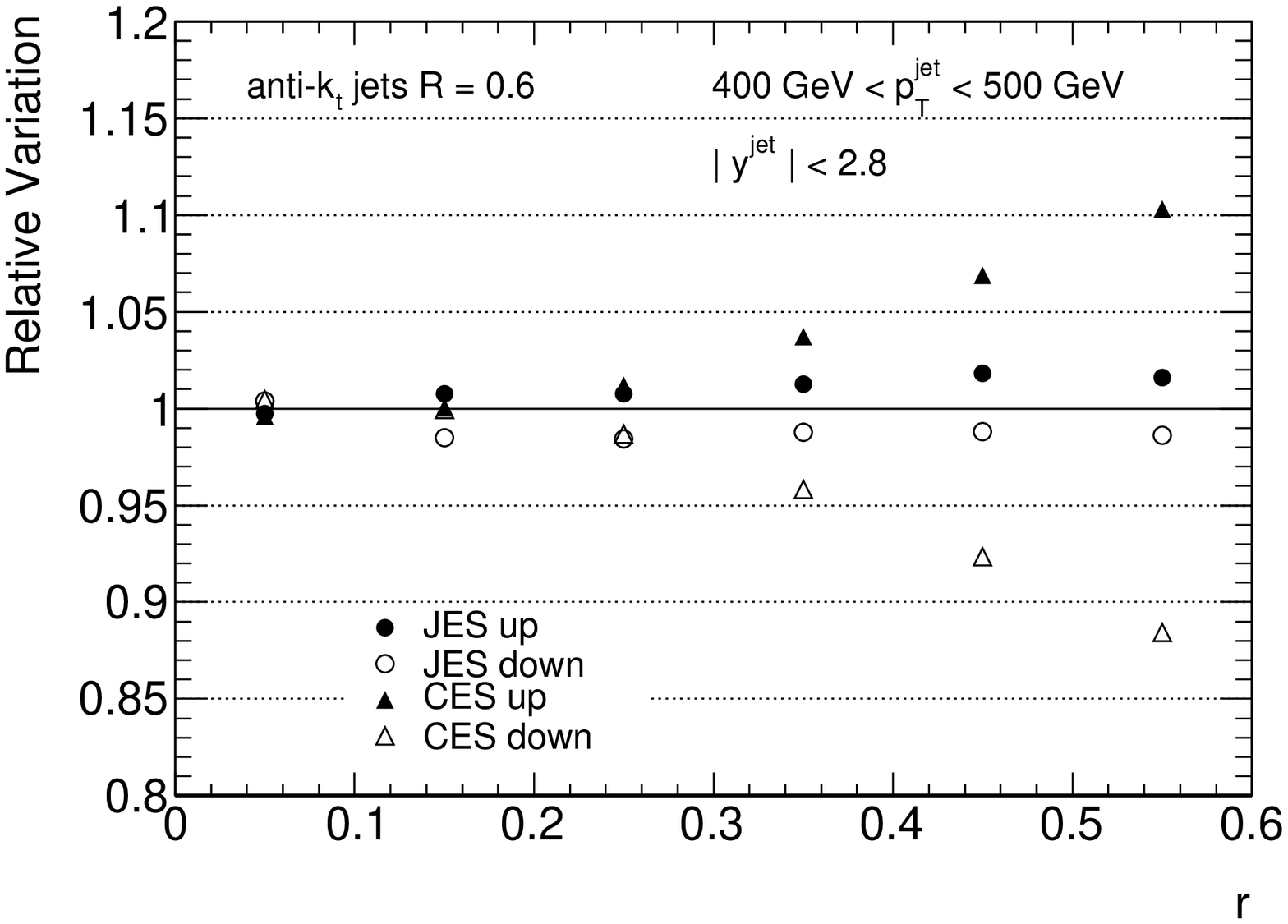}
}
\mbox{
\includegraphics[width=0.495\textwidth]{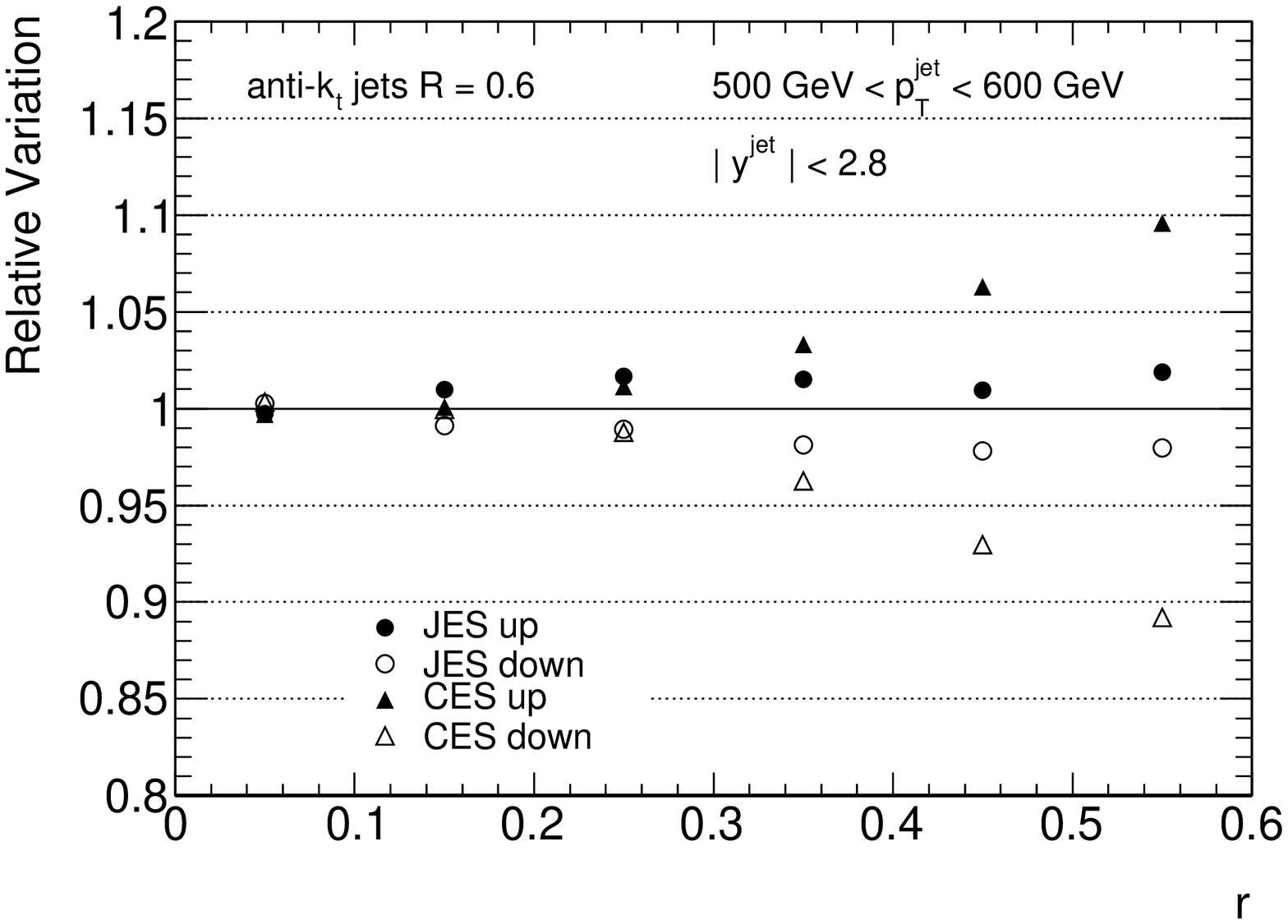}
}
\end{center}
\vspace{-0.7 cm}
\caption{\small
Systematic uncertainty on the differential jet shape related to the absolute energy scale uncertainties on clusters and jets, 
for jets with $|\rapjet| < 2.8$ and $210 \ {\rm GeV} < \ptjet < 600  \ {\rm GeV}$.
}
\label{fig_CJES2}
\end{figure}

\begin{figure}[tbh]
\begin{center}
\mbox{
\includegraphics[width=0.495\textwidth]{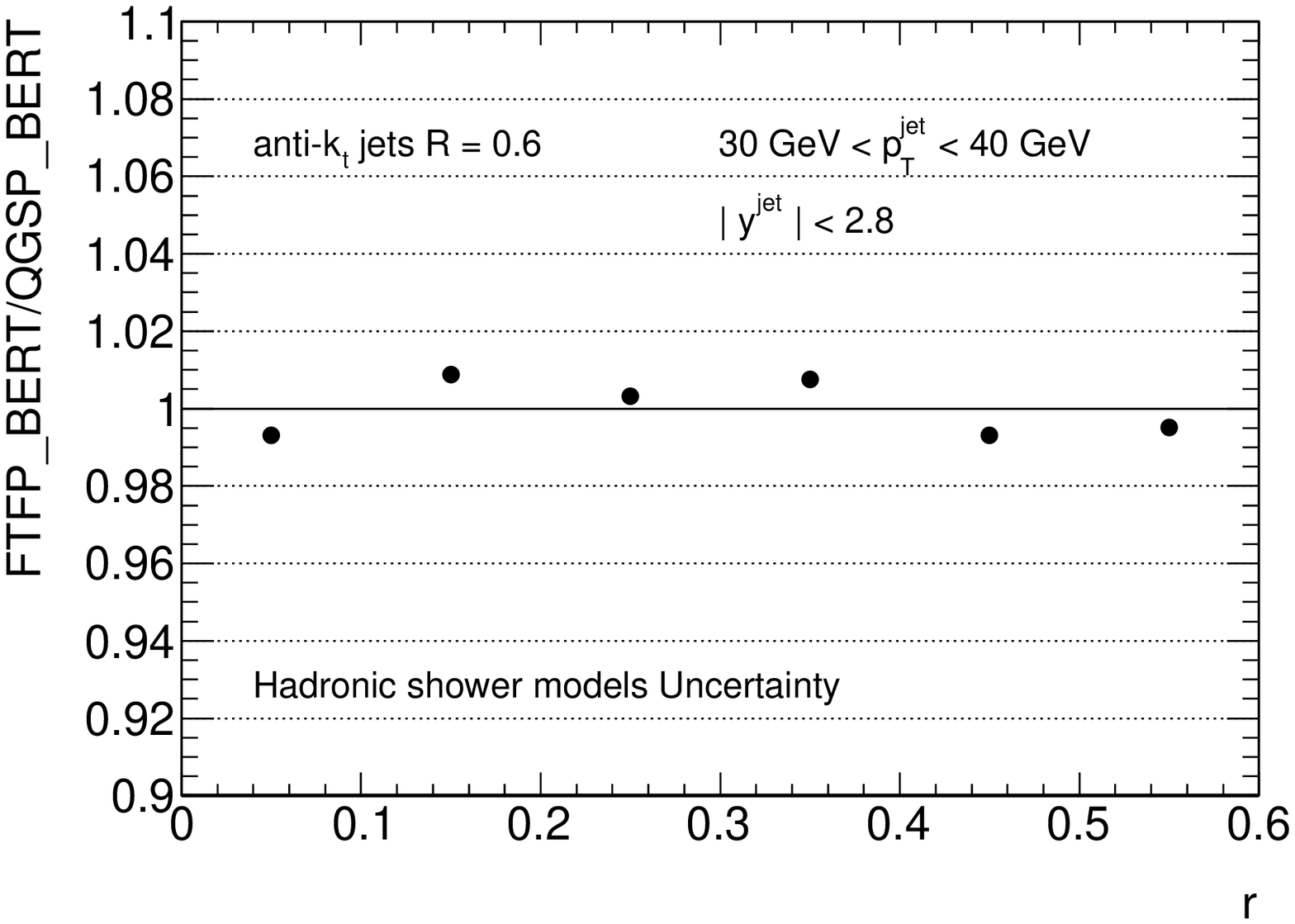}
\includegraphics[width=0.495\textwidth]{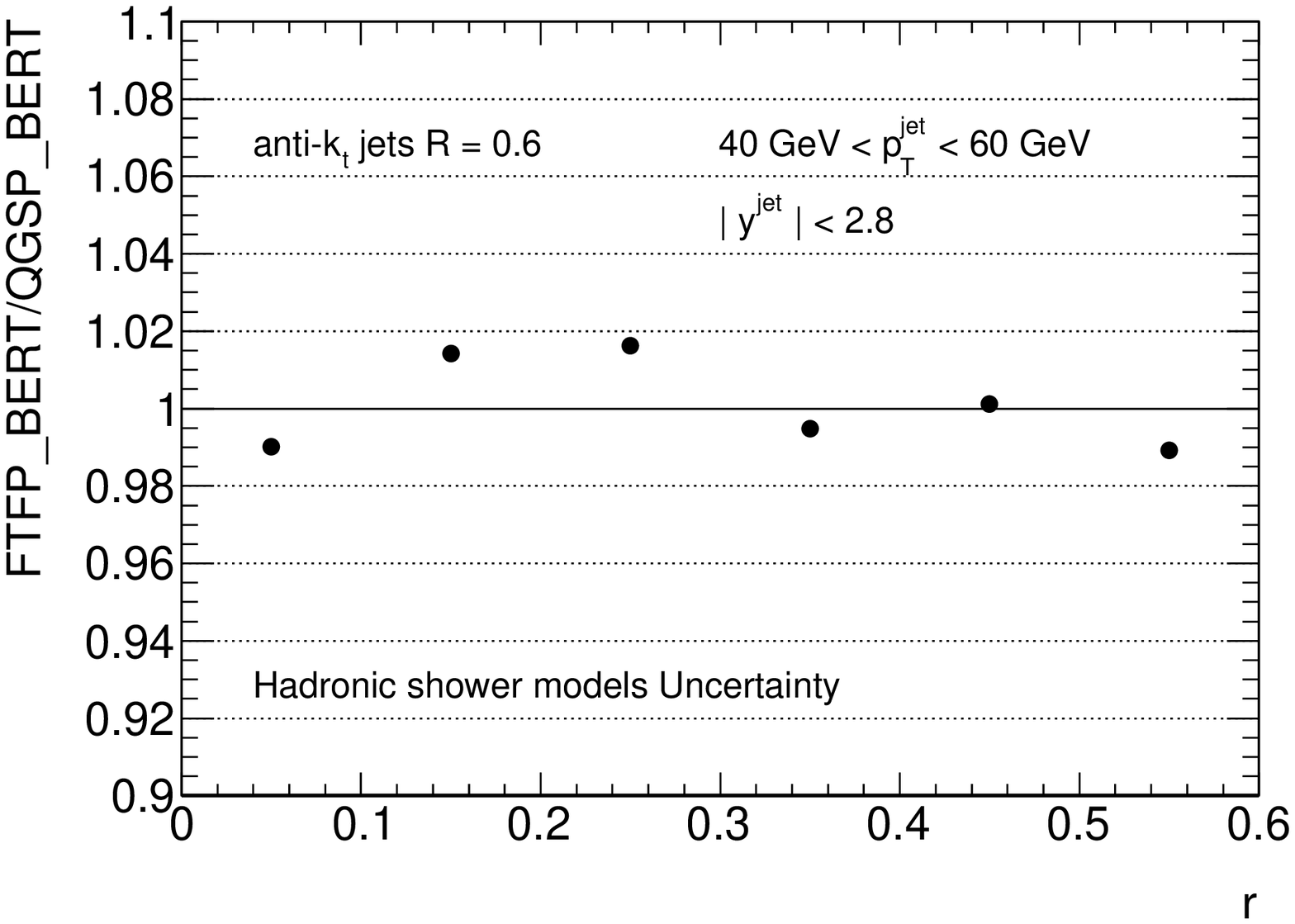}
}
\mbox{
\includegraphics[width=0.495\textwidth]{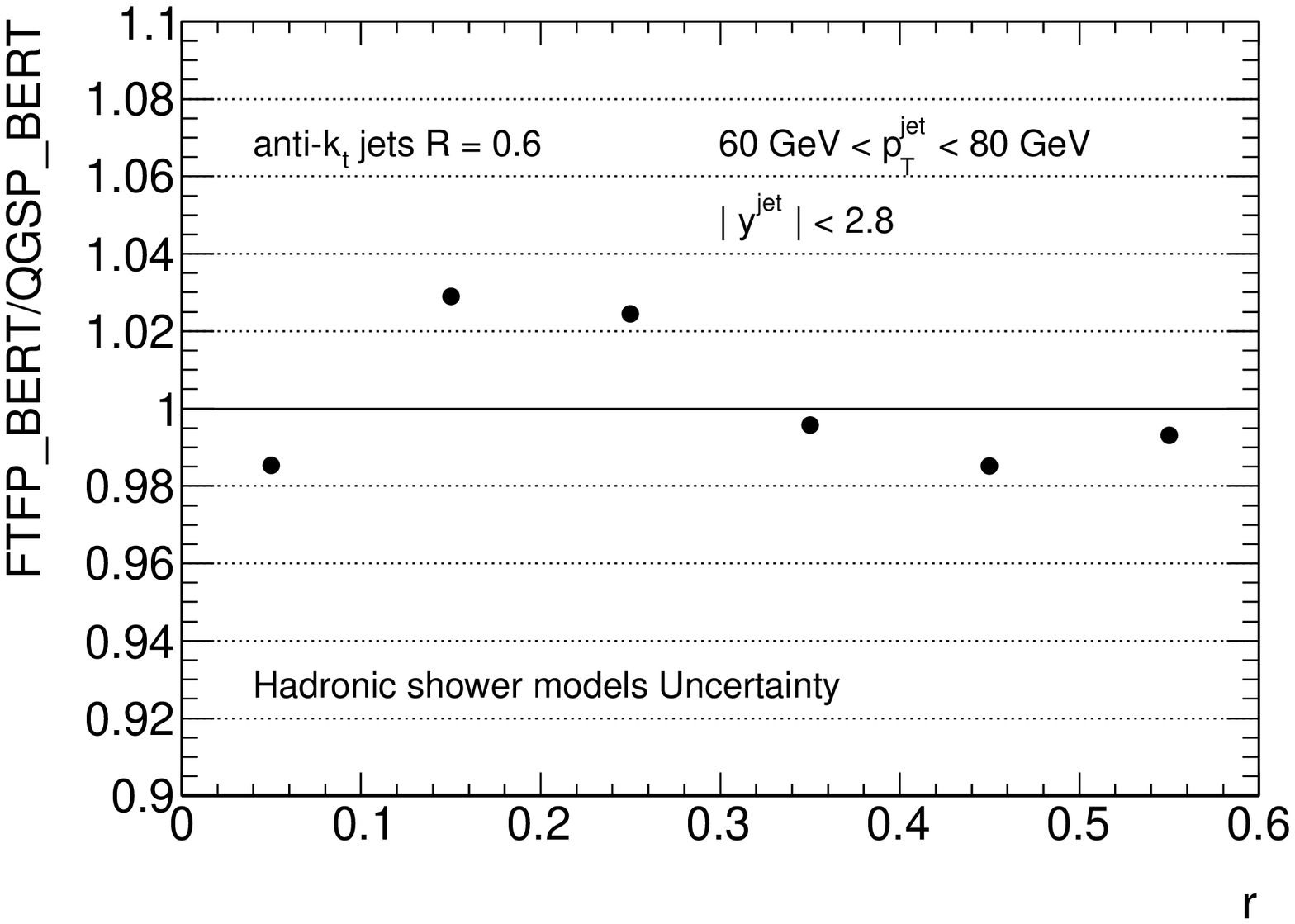}
\includegraphics[width=0.495\textwidth]{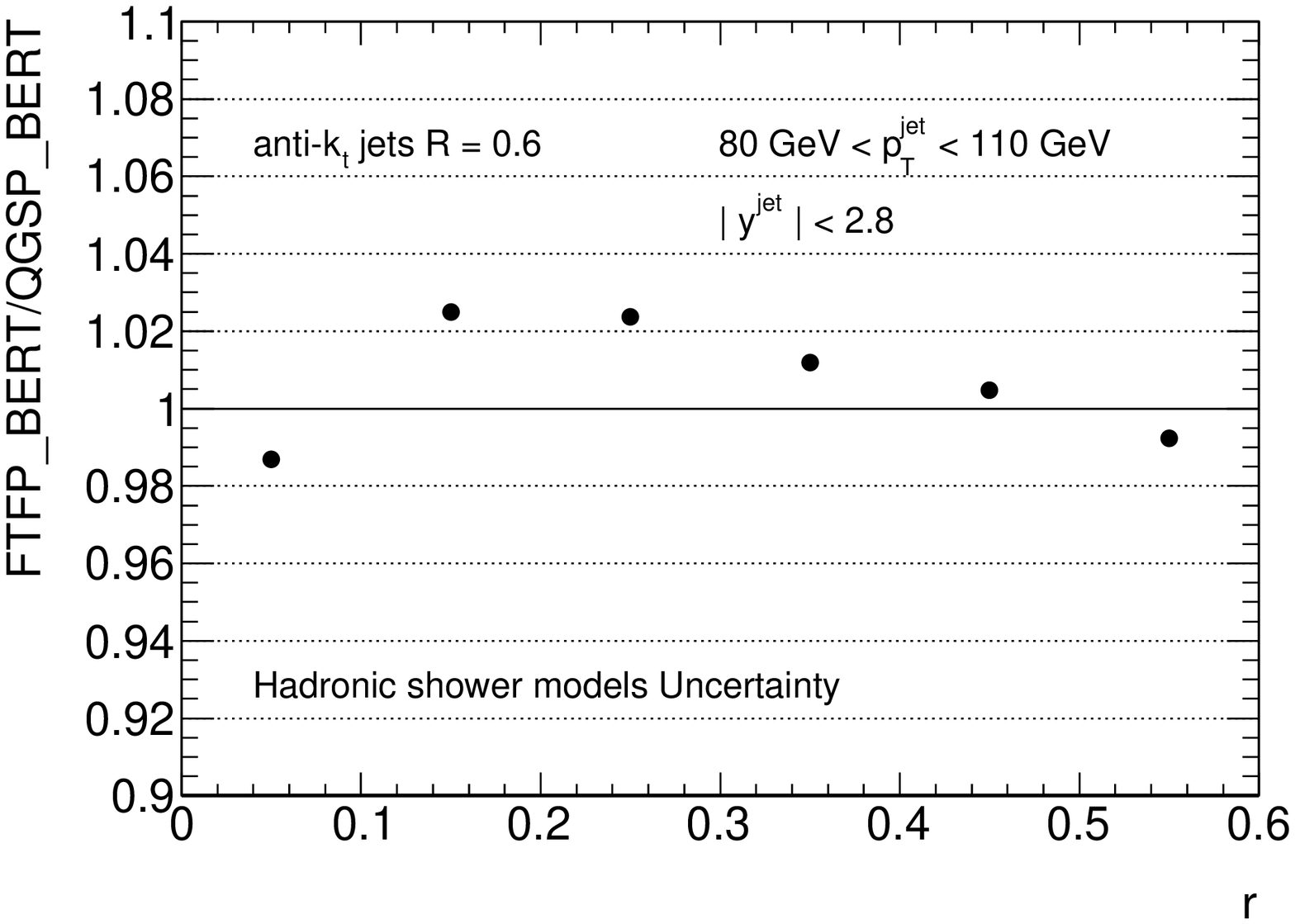}
}
\mbox{
\includegraphics[width=0.495\textwidth]{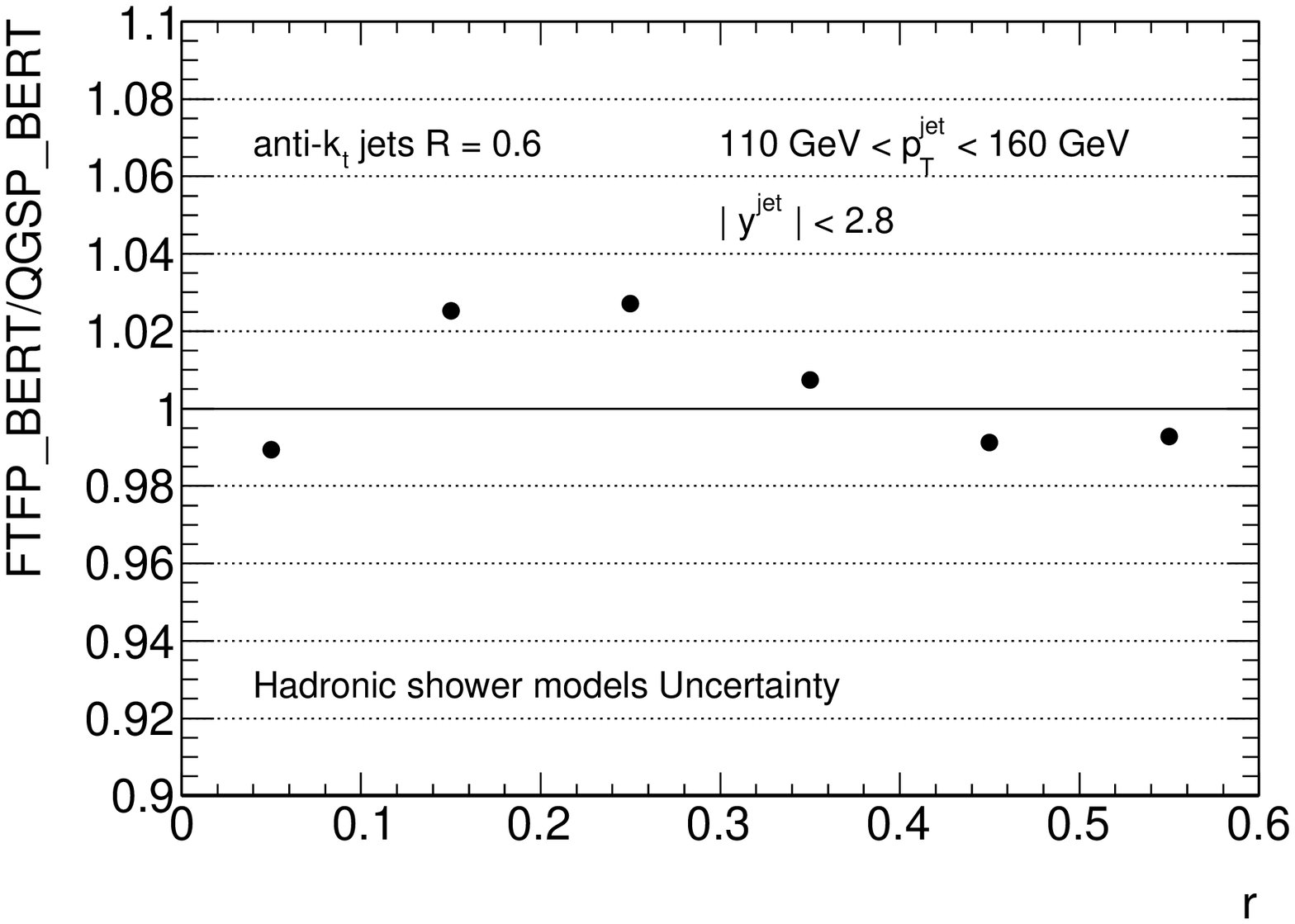}
\includegraphics[width=0.495\textwidth]{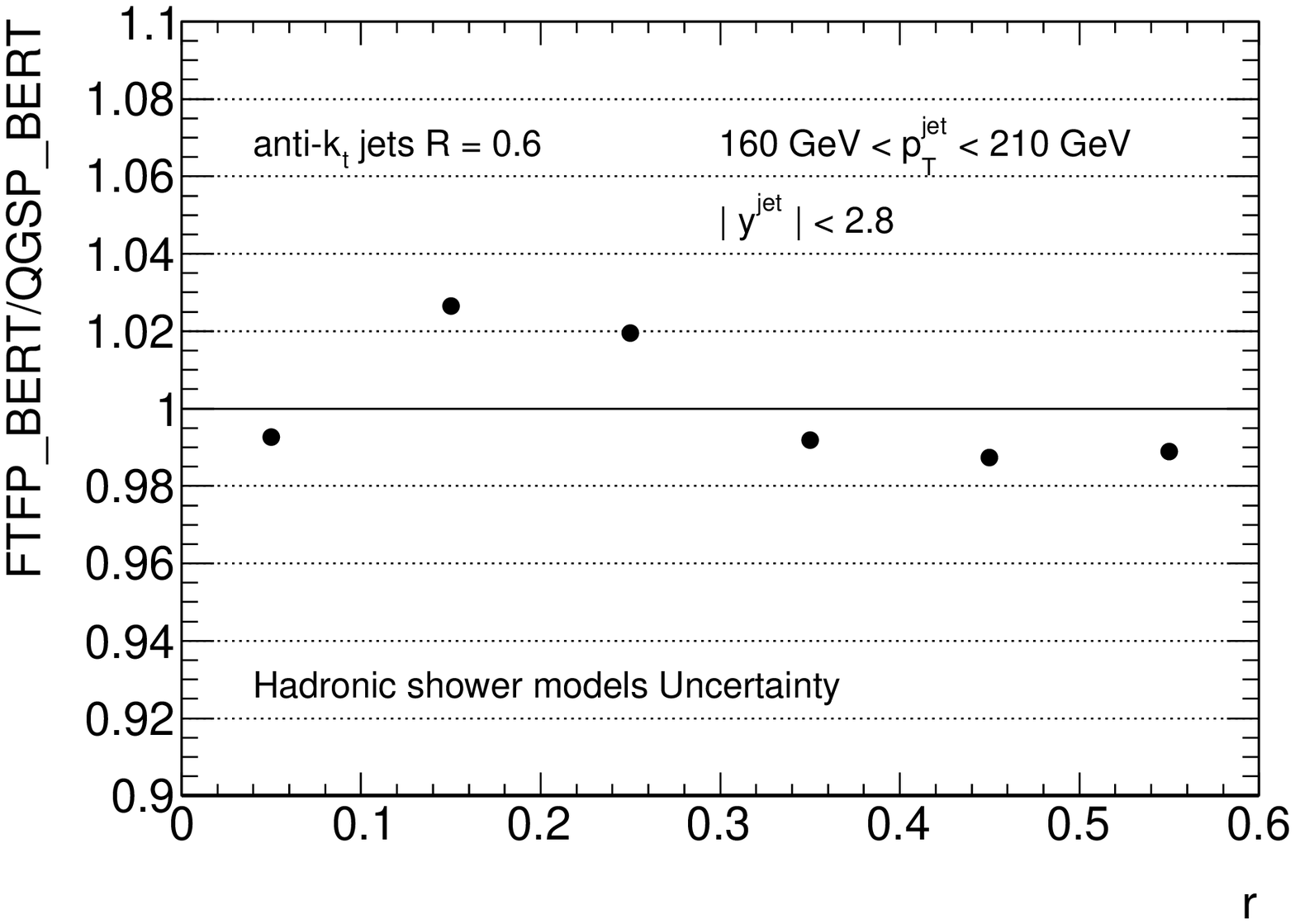}
}
\end{center}
\vspace{-0.7 cm}
\caption{\small
Systematic uncertainty on the differential jet shape related to the calorimeter showering model, 
for jets with $|\rapjet| < 2.8$ and $30 \ {\rm GeV} < \ptjet < 210  \ {\rm GeV}$.
}
\label{fig_FTFP1}
\end{figure}

\begin{figure}[tbh]
\begin{center}
\mbox{
\includegraphics[width=0.495\textwidth]{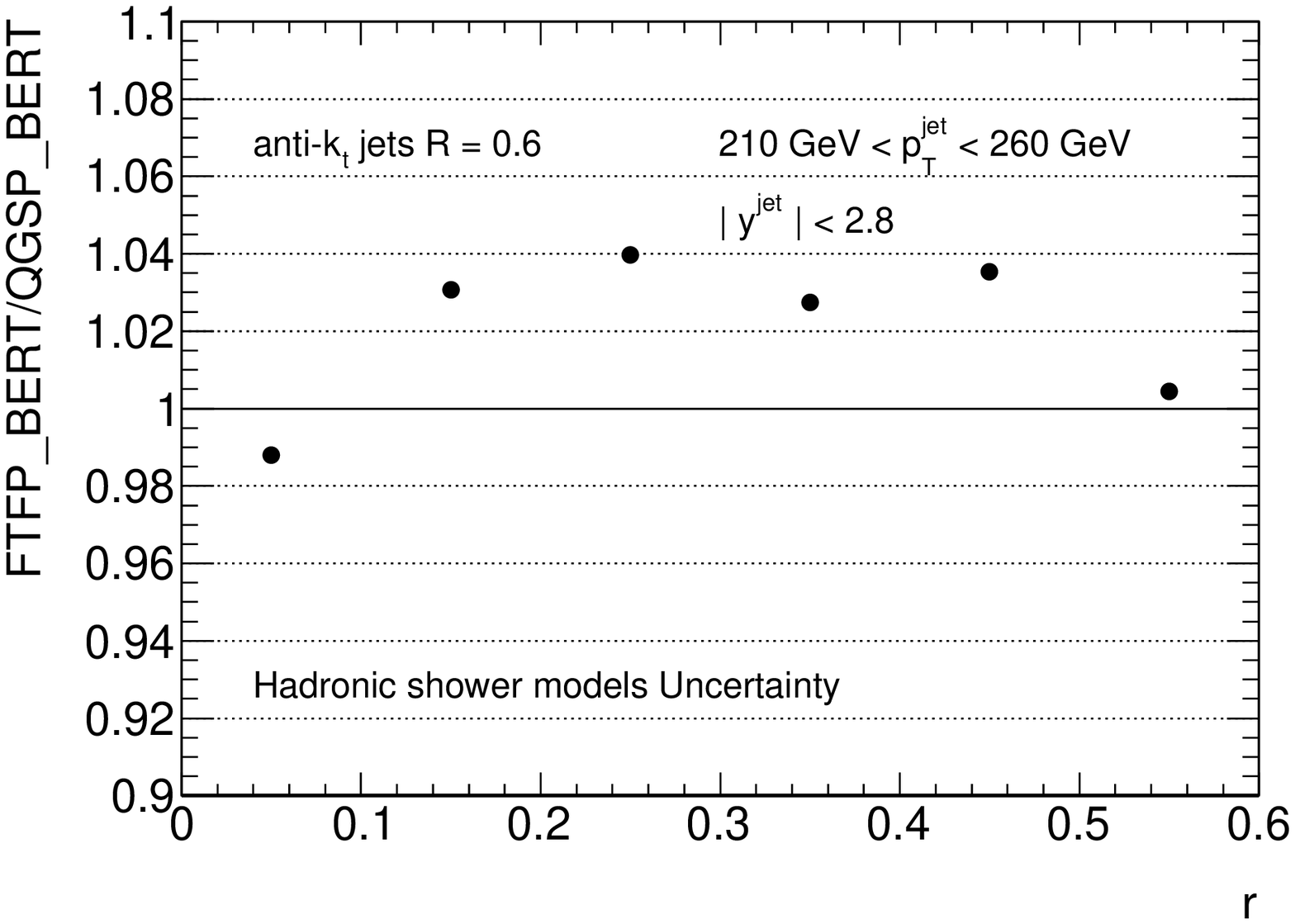}
\includegraphics[width=0.495\textwidth]{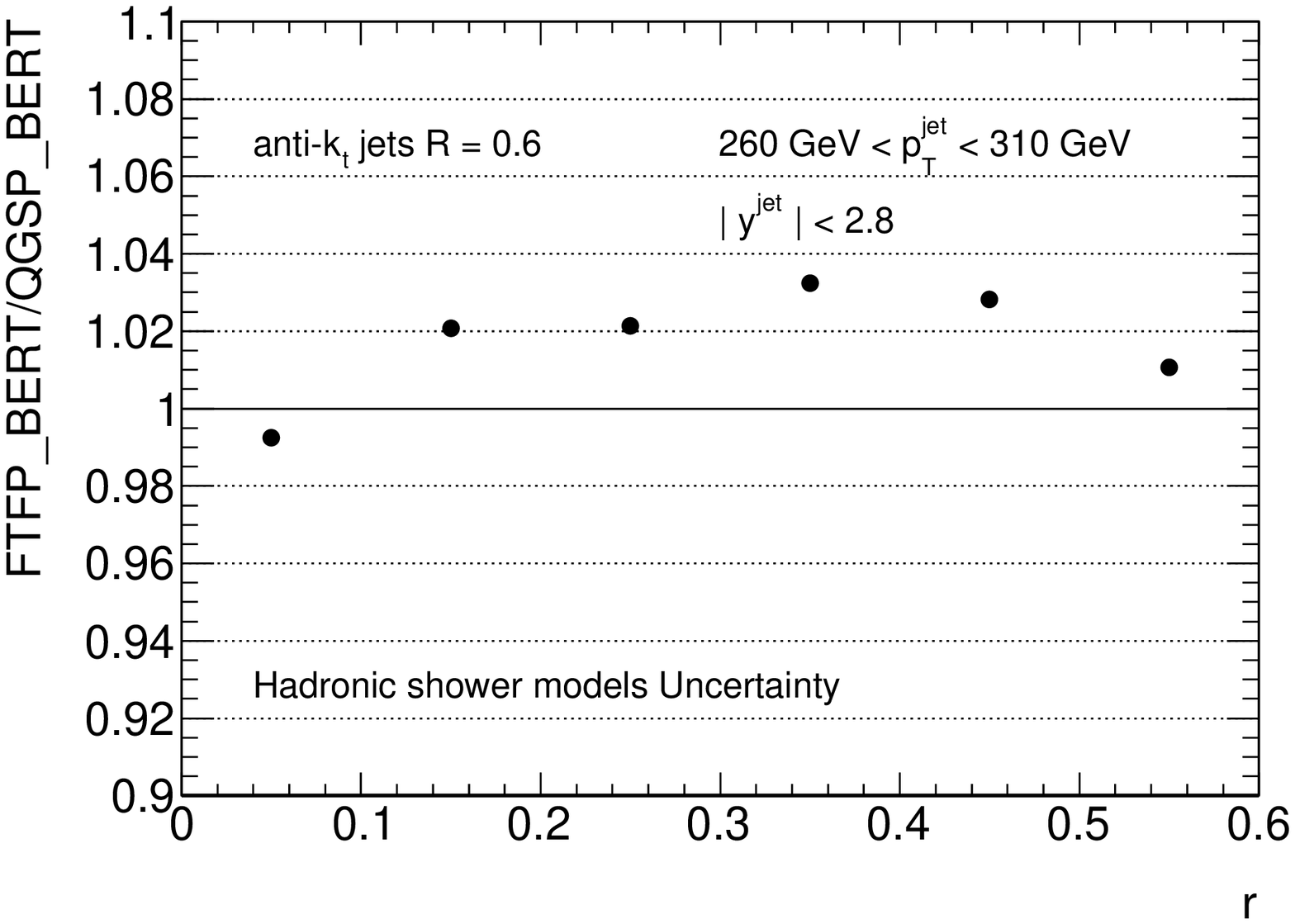}
}
\mbox{
\includegraphics[width=0.495\textwidth]{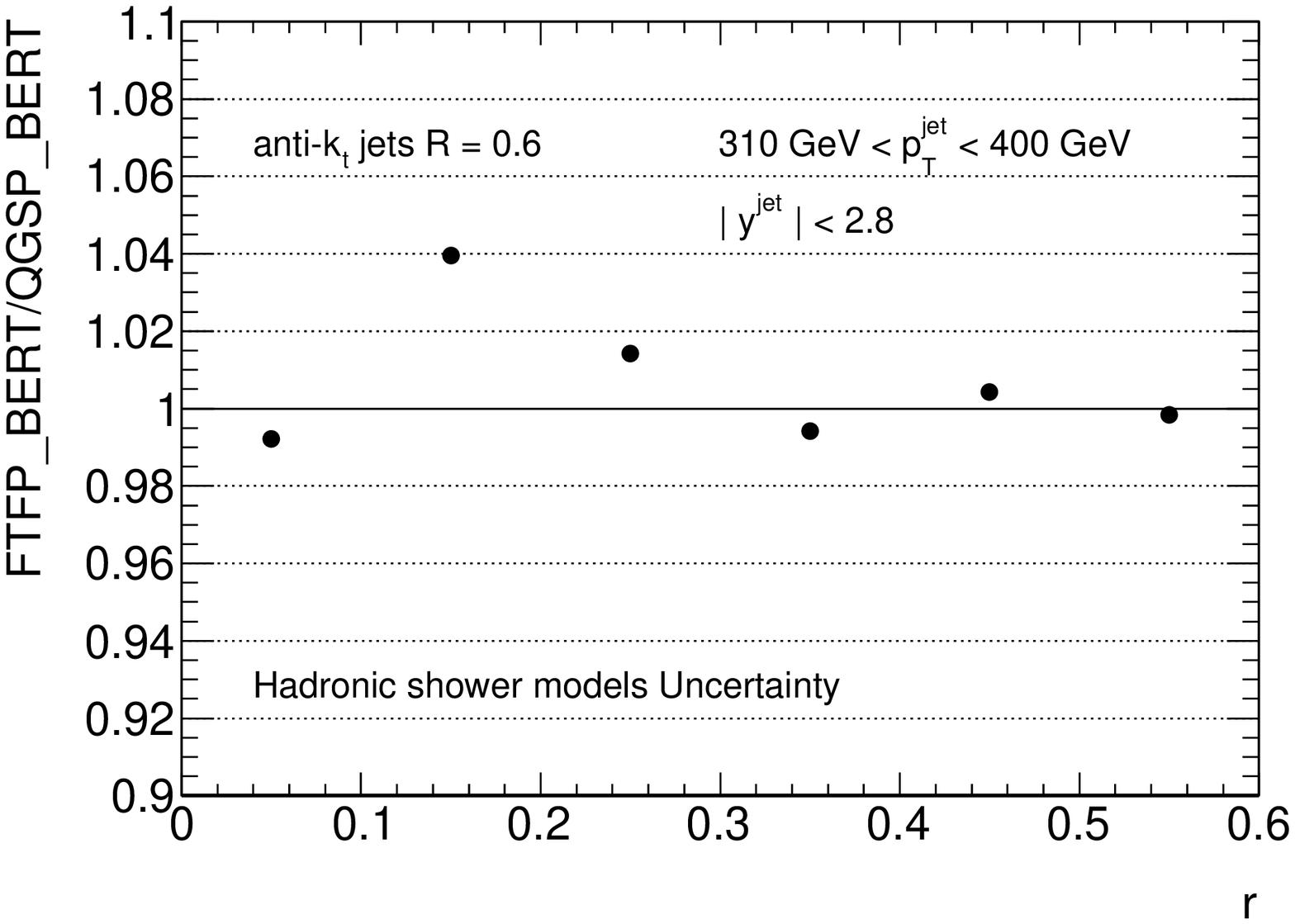}
\includegraphics[width=0.495\textwidth]{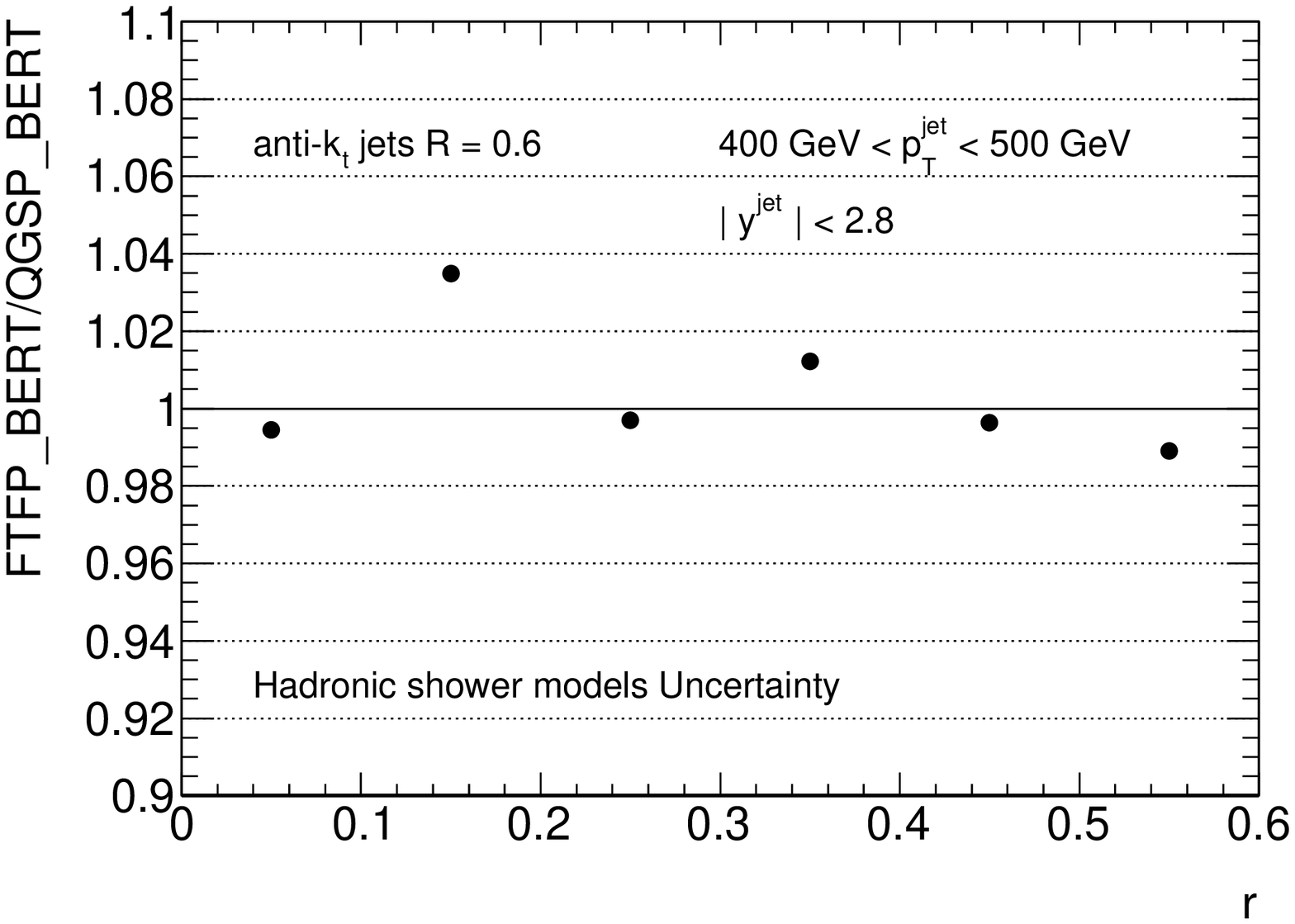}
}
\mbox{
\includegraphics[width=0.495\textwidth]{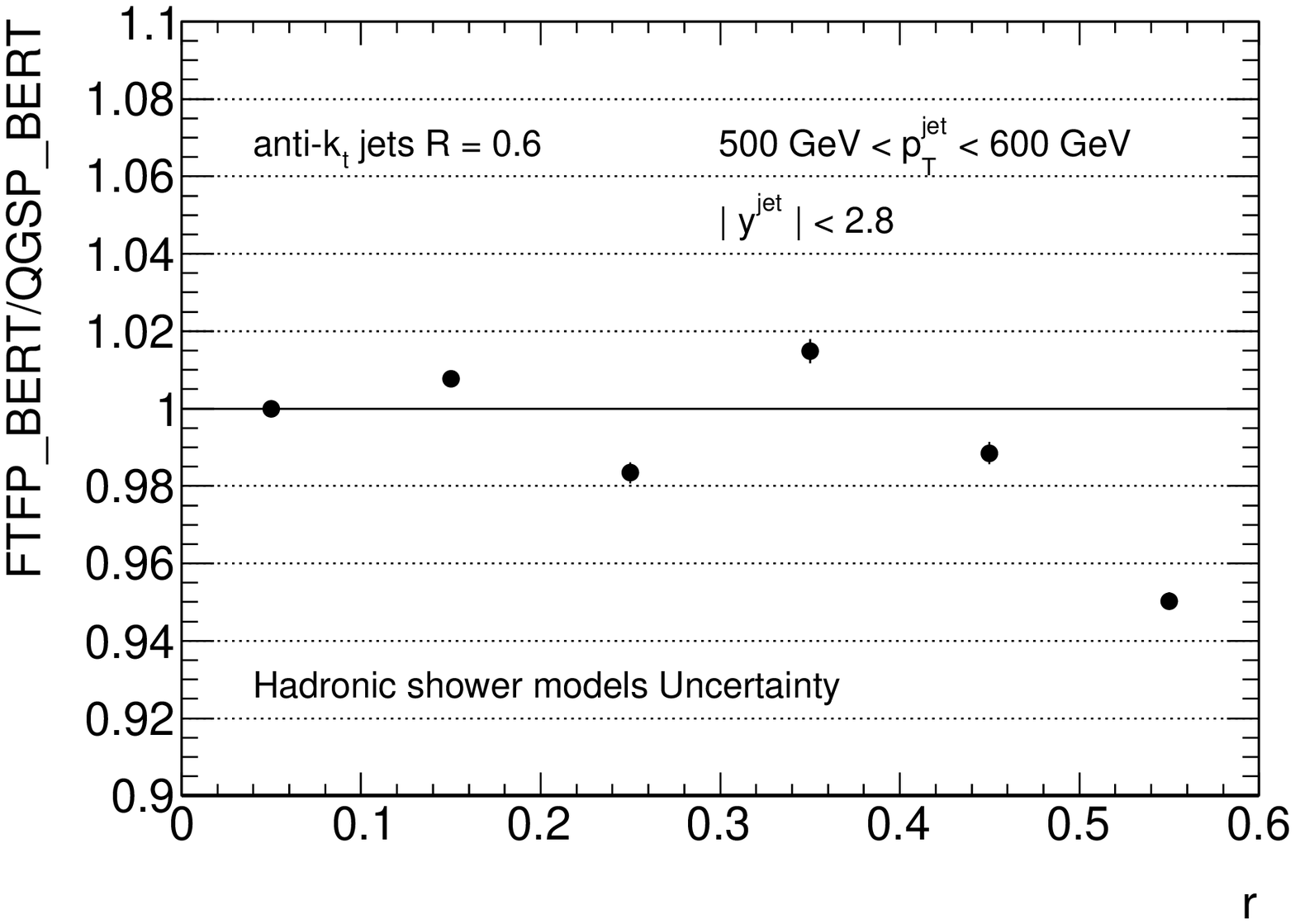}
}
\end{center}
\vspace{-0.7 cm}
\caption{\small
Systematic uncertainty on the differential jet shape related to the calorimeter showering model,
for jets with $|\rapjet| < 2.8$ and $210 \ {\rm GeV} < \ptjet < 600  \ {\rm GeV}$.
}
\label{fig_FTFP2}
\end{figure}

\begin{figure}[tbh]
\begin{center}
\mbox{
\includegraphics[width=0.495\textwidth]{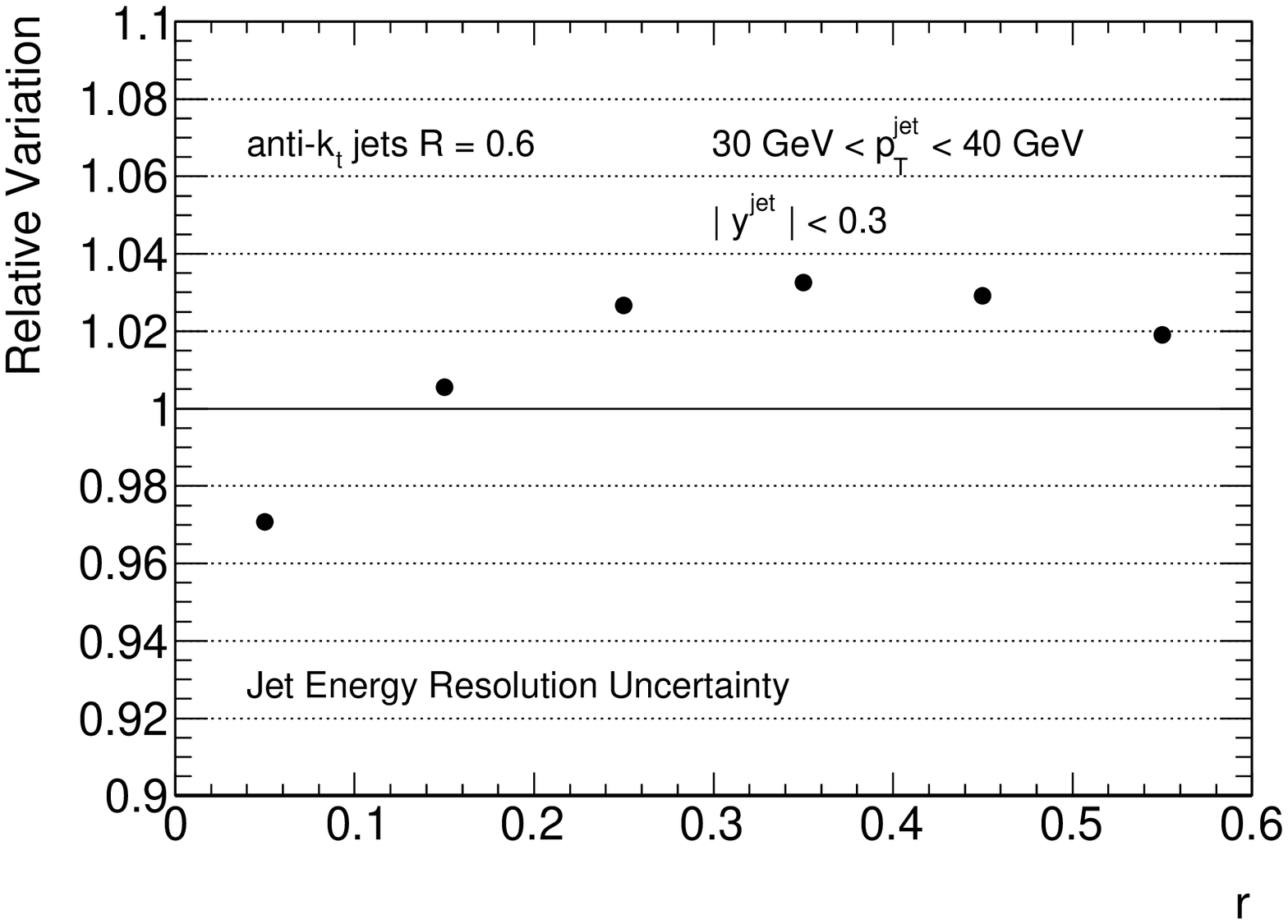}
\includegraphics[width=0.495\textwidth]{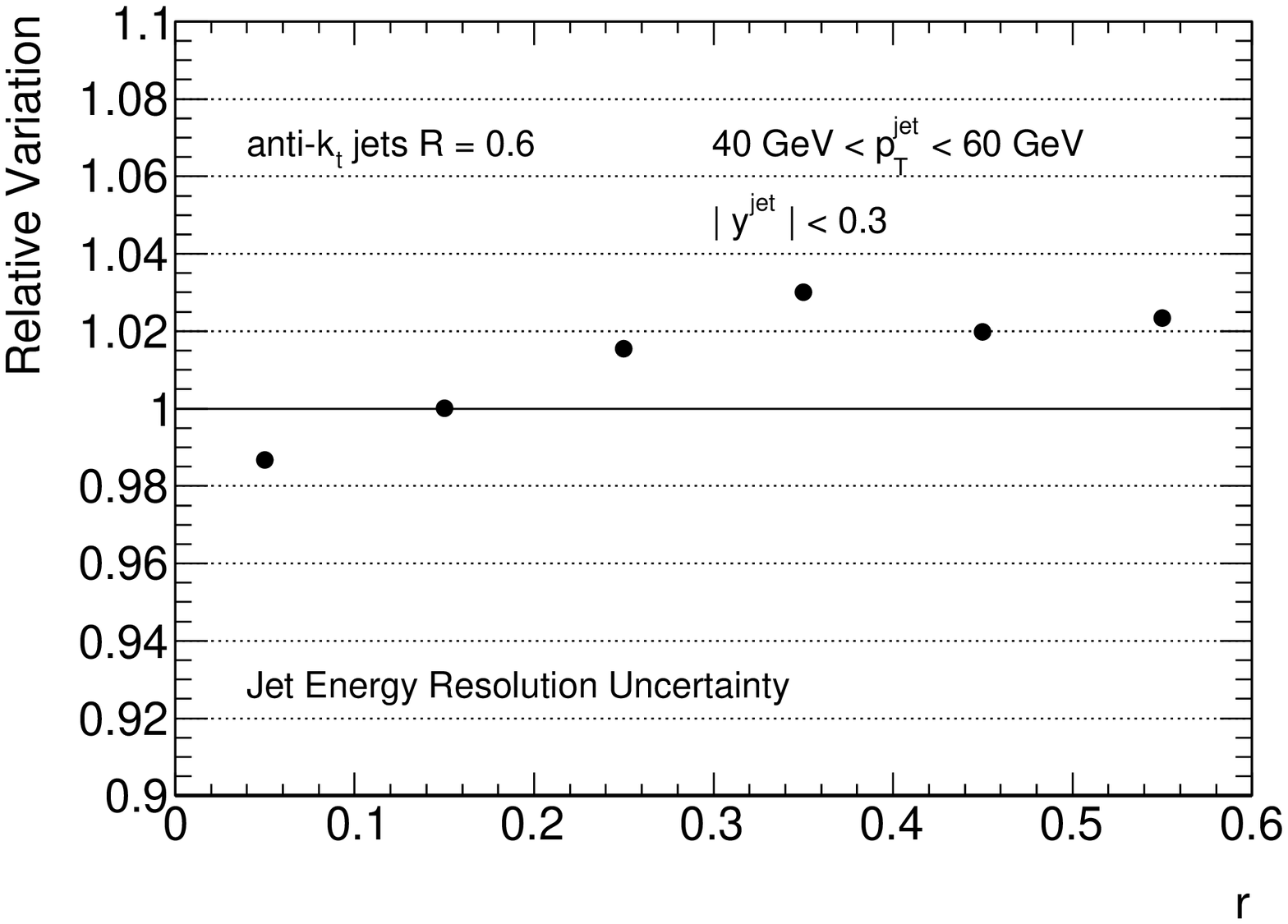}
}
\mbox{
\includegraphics[width=0.495\textwidth]{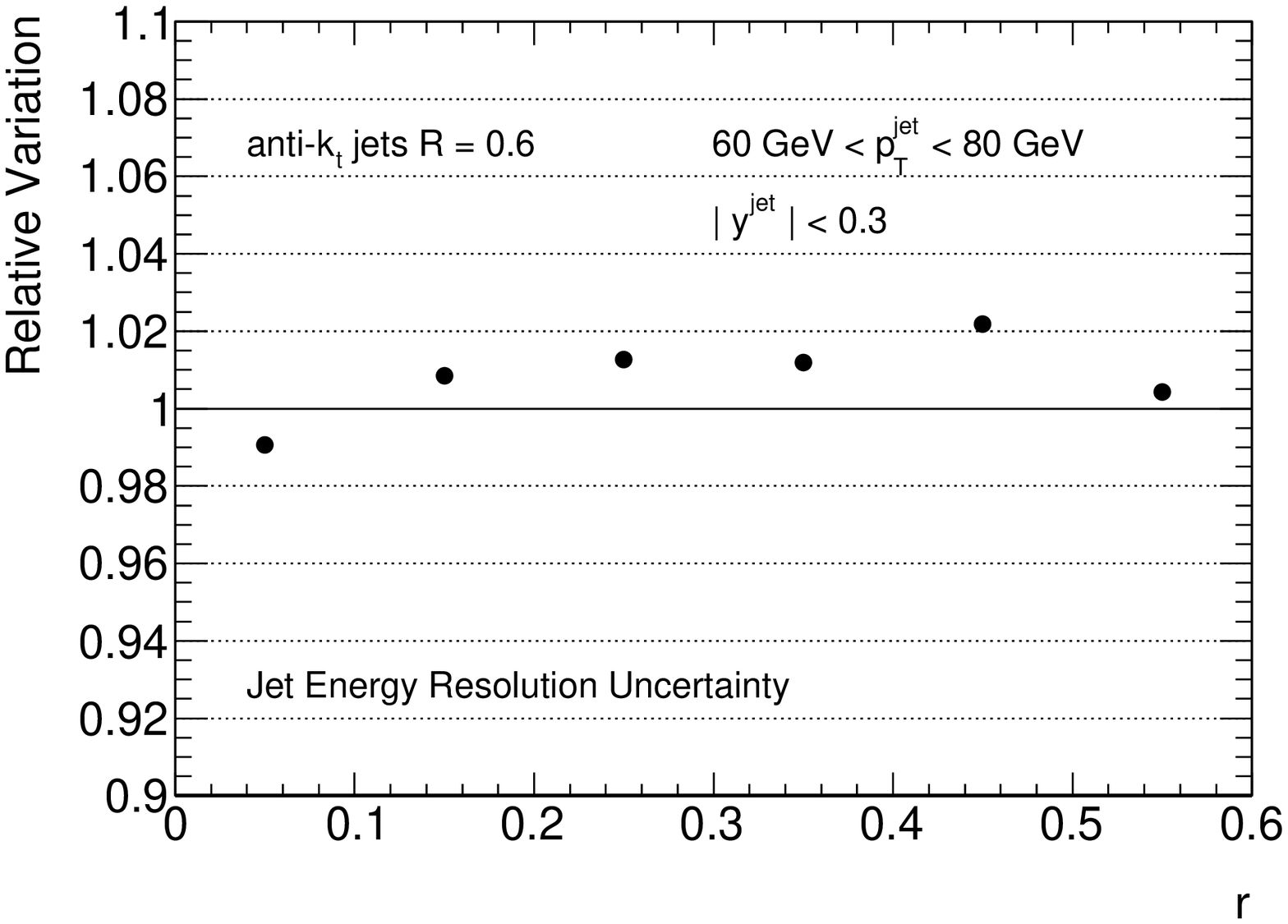}
\includegraphics[width=0.495\textwidth]{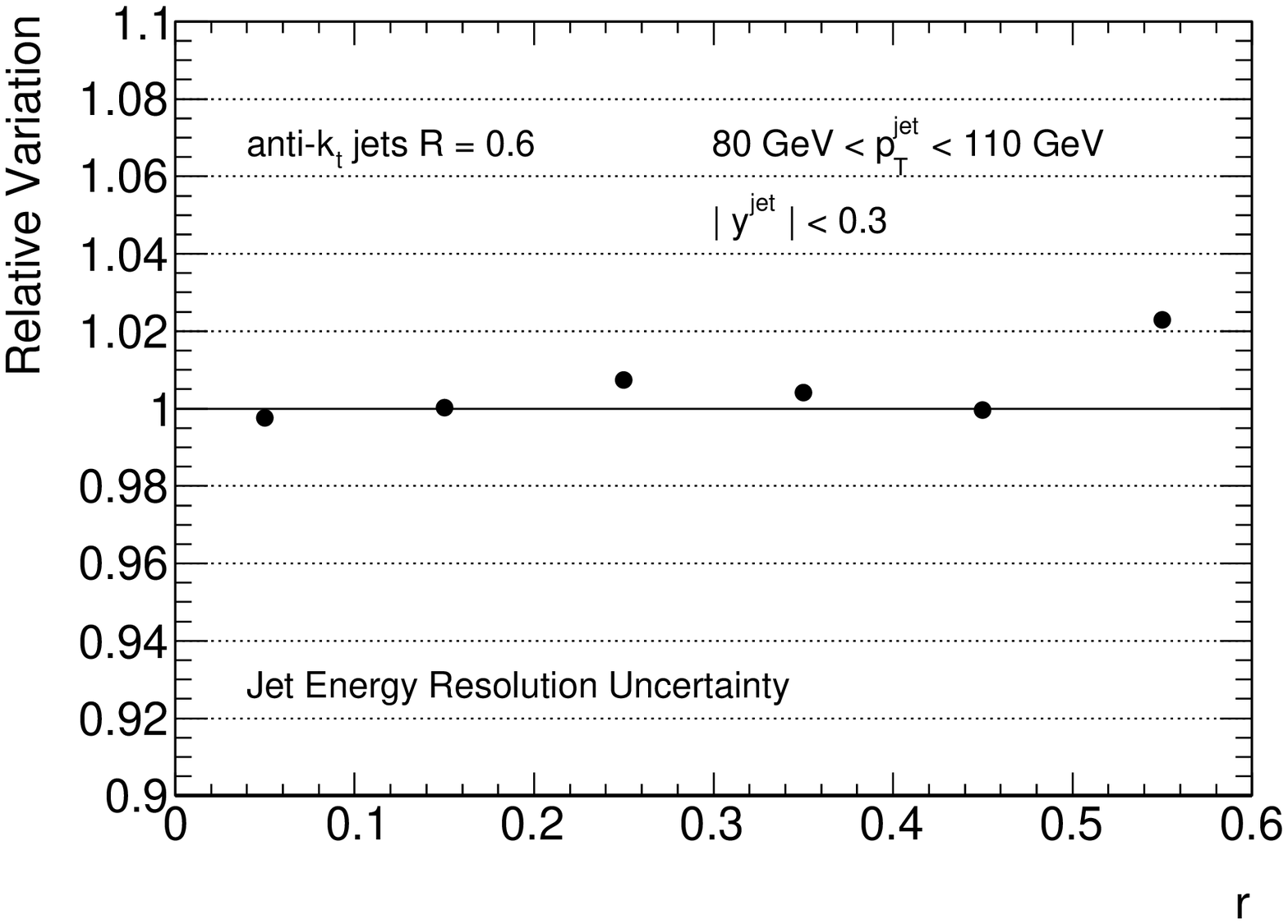}
}
\mbox{
\includegraphics[width=0.495\textwidth]{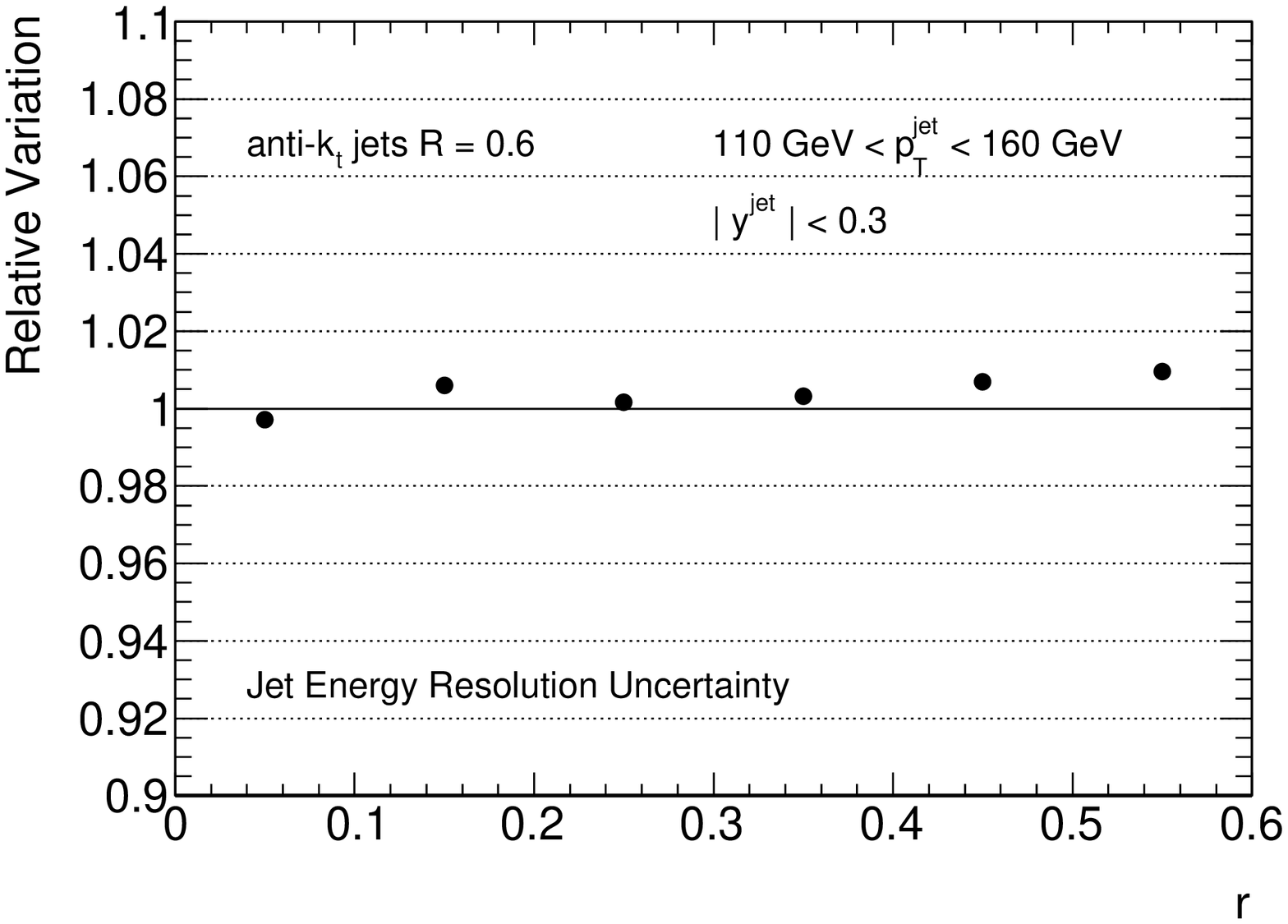}
\includegraphics[width=0.495\textwidth]{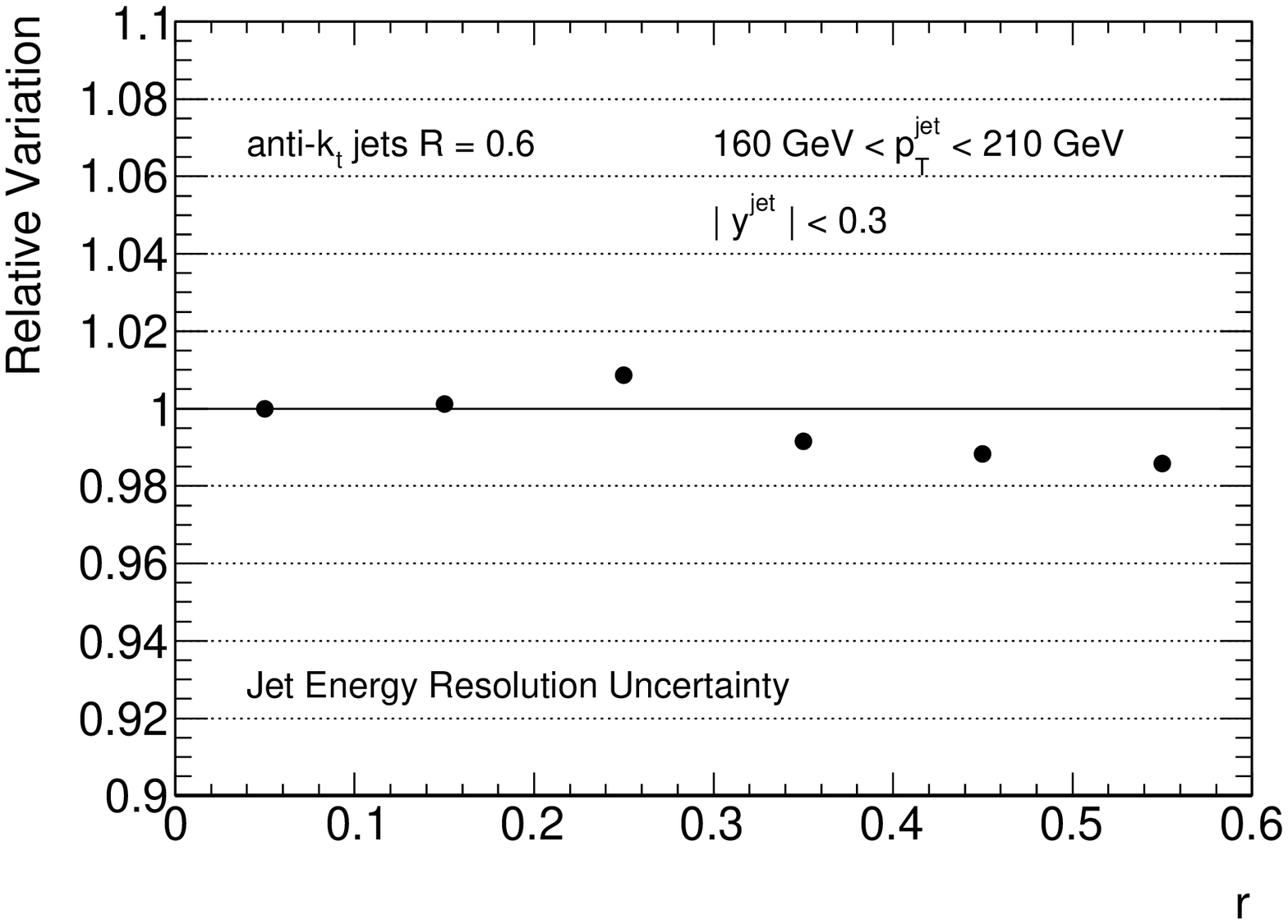}
}
\end{center}
\vspace{-0.7 cm}
\caption{\small
Systematic uncertainty on the differential jet shape related to the jet $\ptjet$ resolution, 
for jets with $|\rapjet| < 2.8$ and $30 \ {\rm GeV} < \ptjet < 210  \ {\rm GeV}$.
}
\label{fig_RES1}
\end{figure}

\begin{figure}[tbh]
\begin{center}
\mbox{
\includegraphics[width=0.495\textwidth]{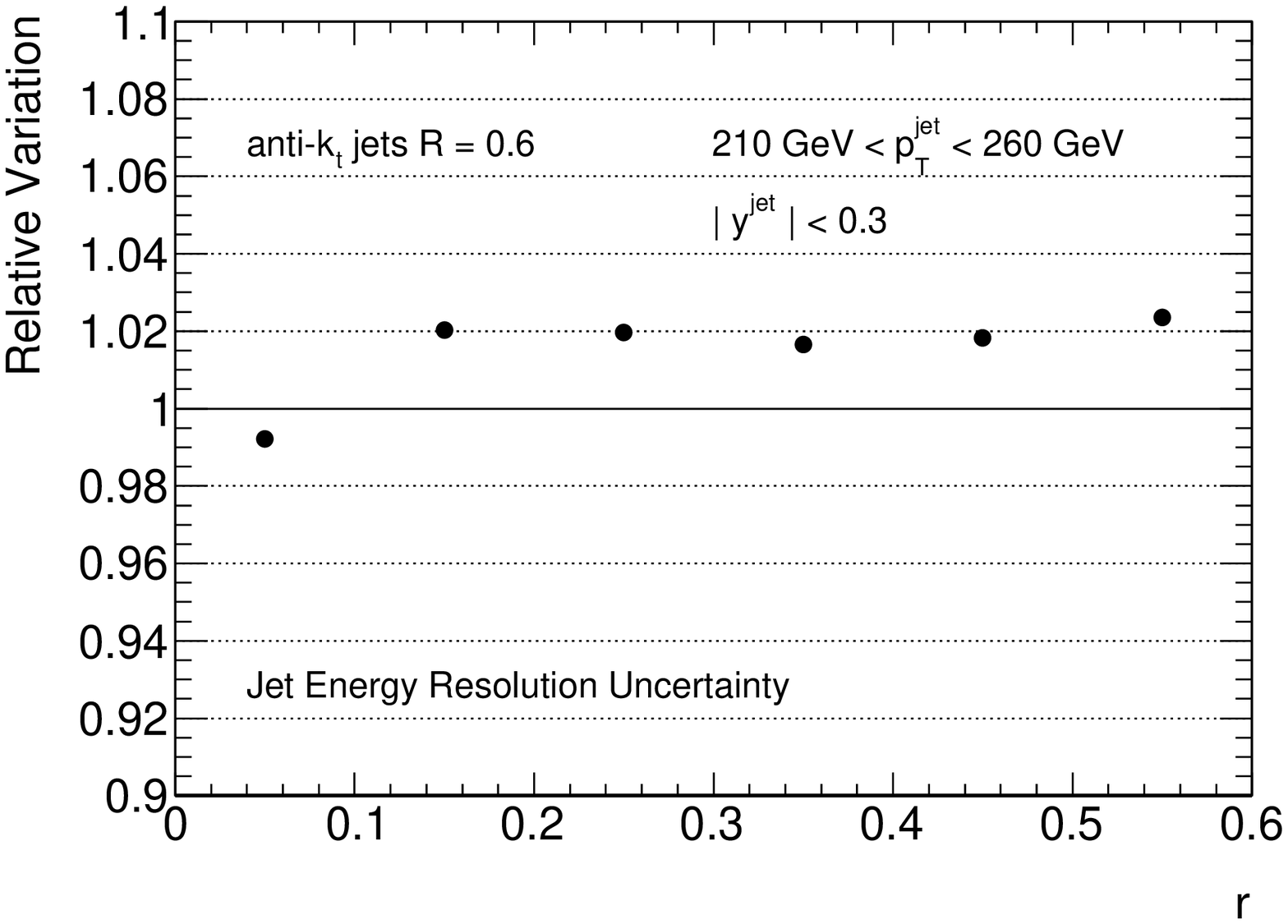}
\includegraphics[width=0.495\textwidth]{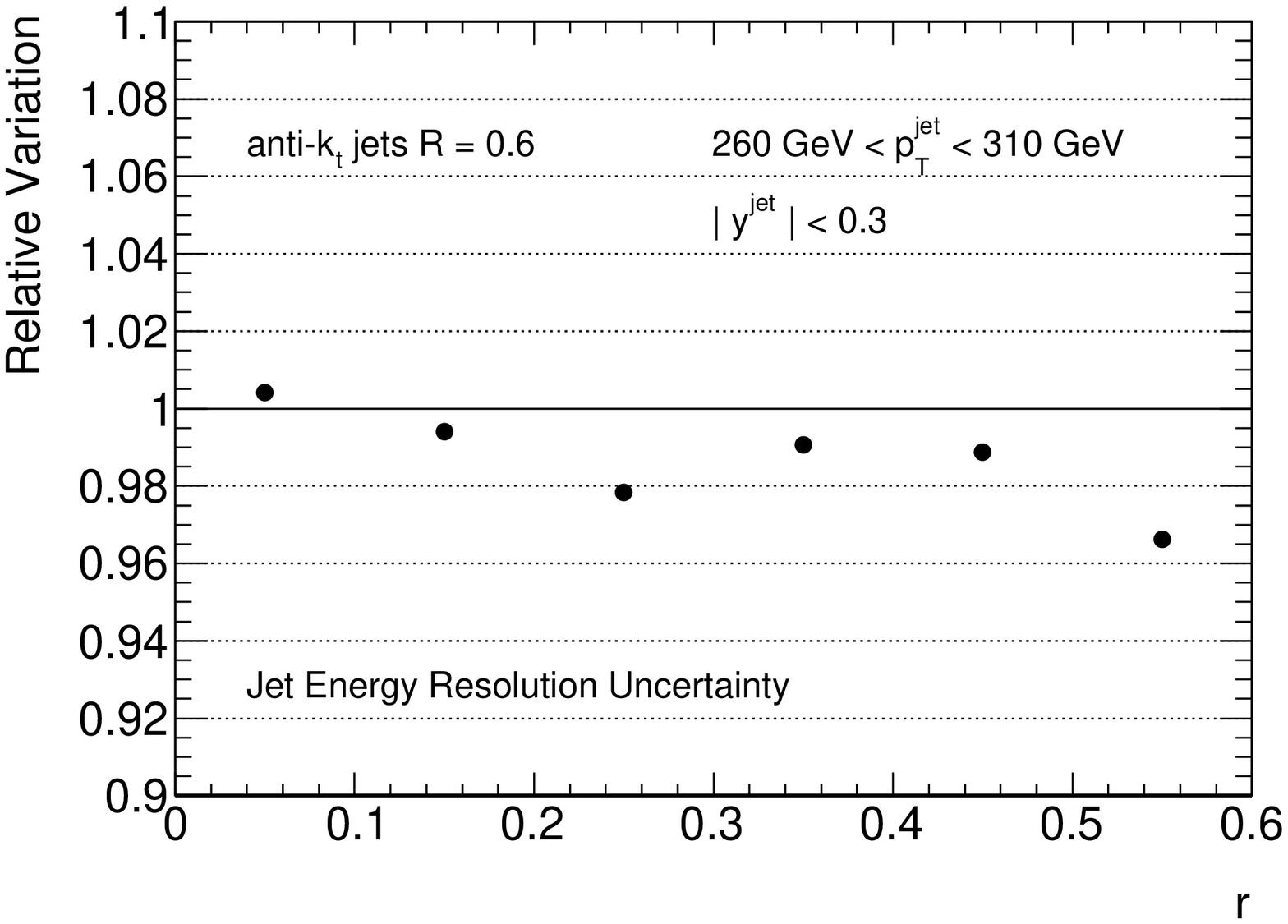}
}
\mbox{
\includegraphics[width=0.495\textwidth]{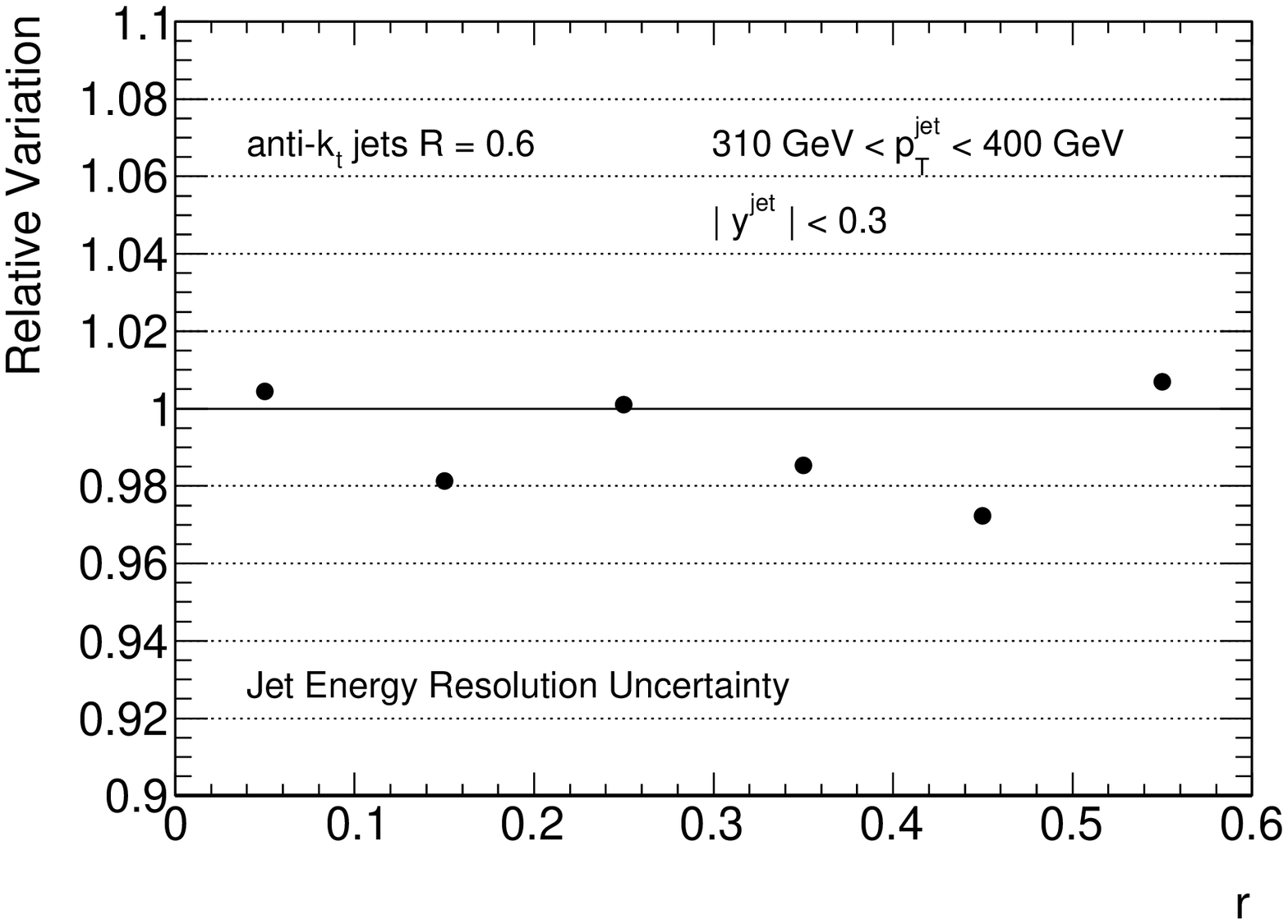}
\includegraphics[width=0.495\textwidth]{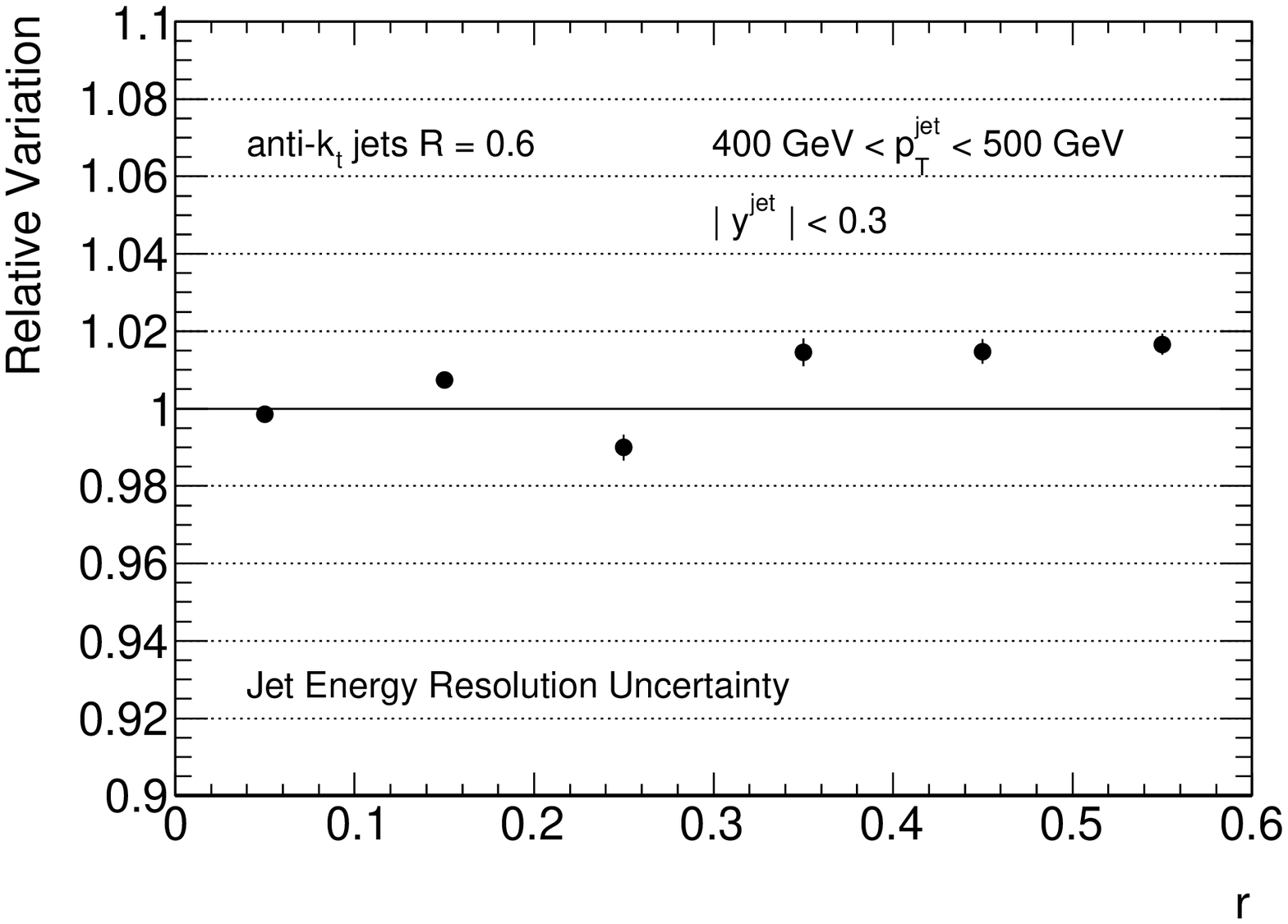}
}
\mbox{
\includegraphics[width=0.495\textwidth]{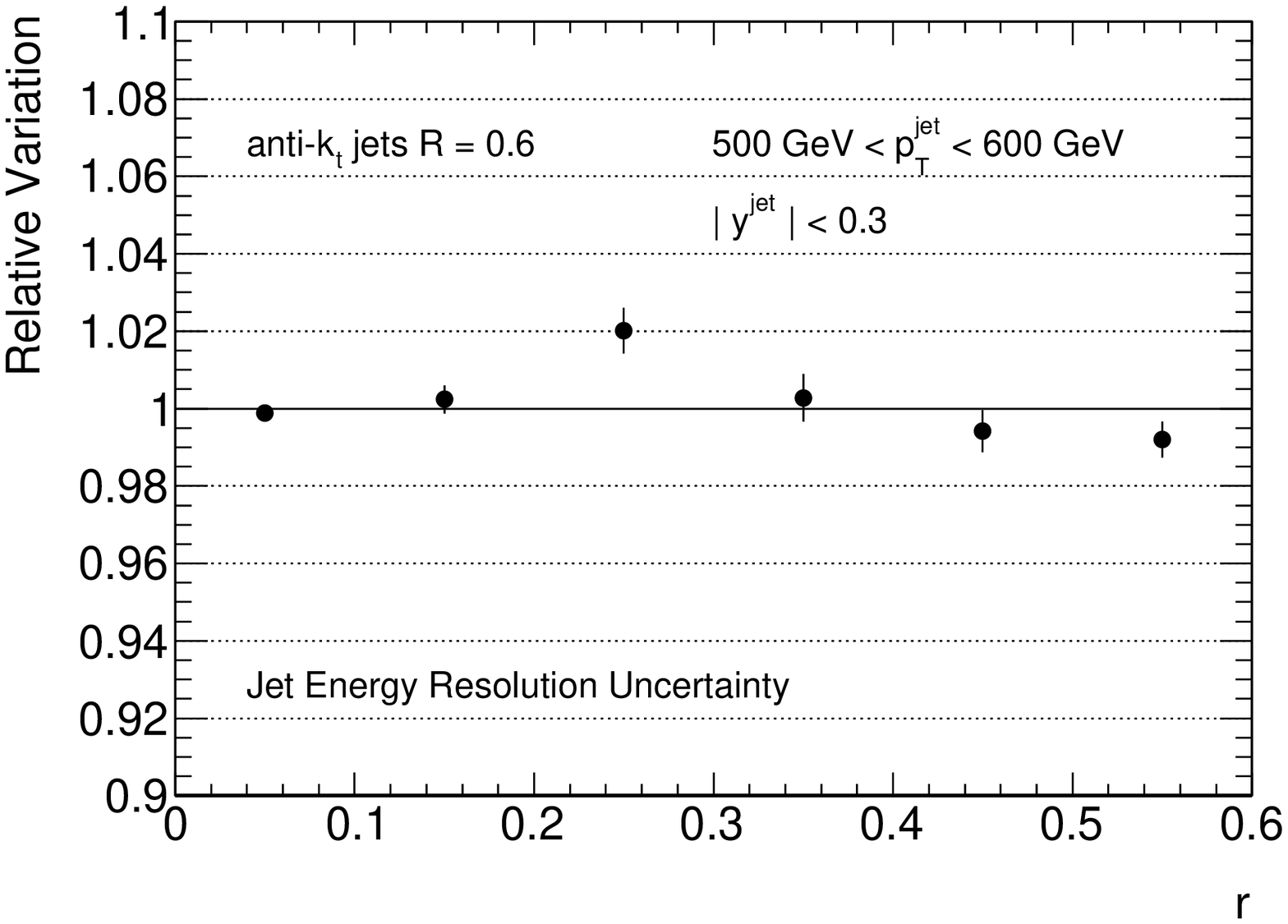}
}
\end{center}
\vspace{-0.7 cm}
\caption{\small
Systematic uncertainty on the differential jet shape related to the jet $\ptjet$ resolution, 
for jets with $|\rapjet| < 2.8$ and $210 \ {\rm GeV} < \ptjet < 600  \ {\rm GeV}$.
}
\label{fig_RES2}
\end{figure}

\begin{figure}[tbh]
\begin{center}
\mbox{
\includegraphics[width=0.495\textwidth]{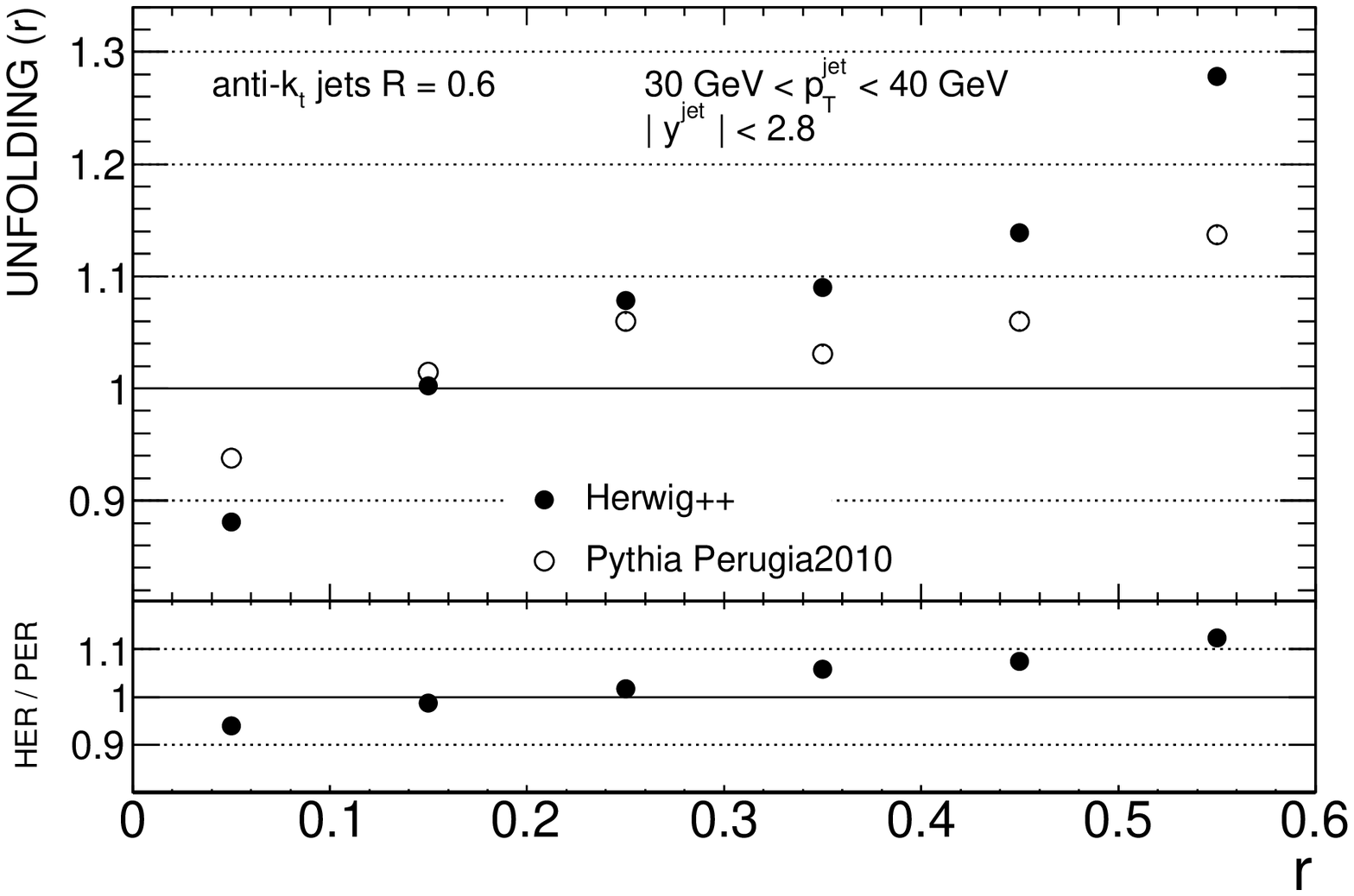}
\includegraphics[width=0.495\textwidth]{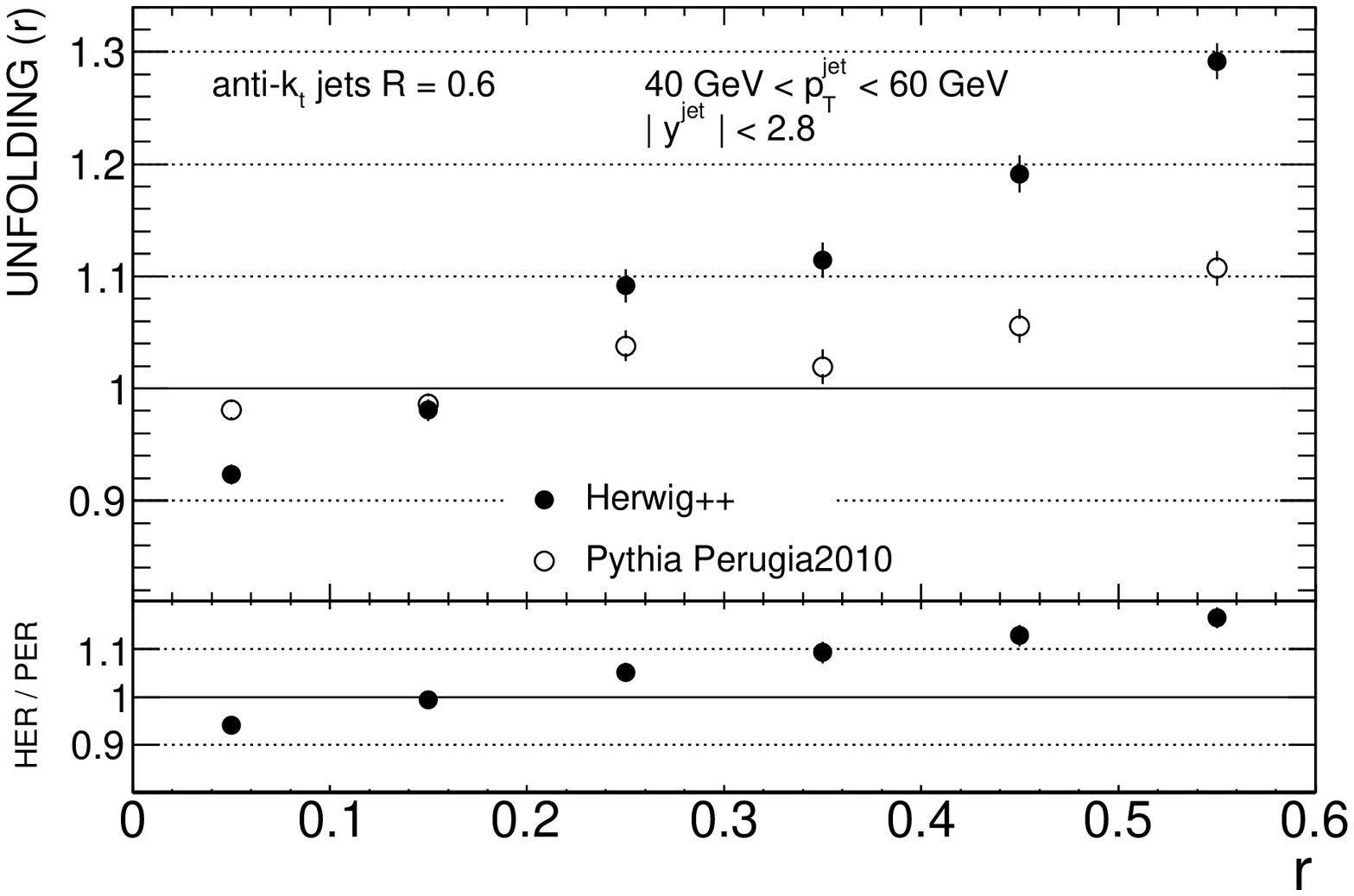}
}
\mbox{
\includegraphics[width=0.495\textwidth]{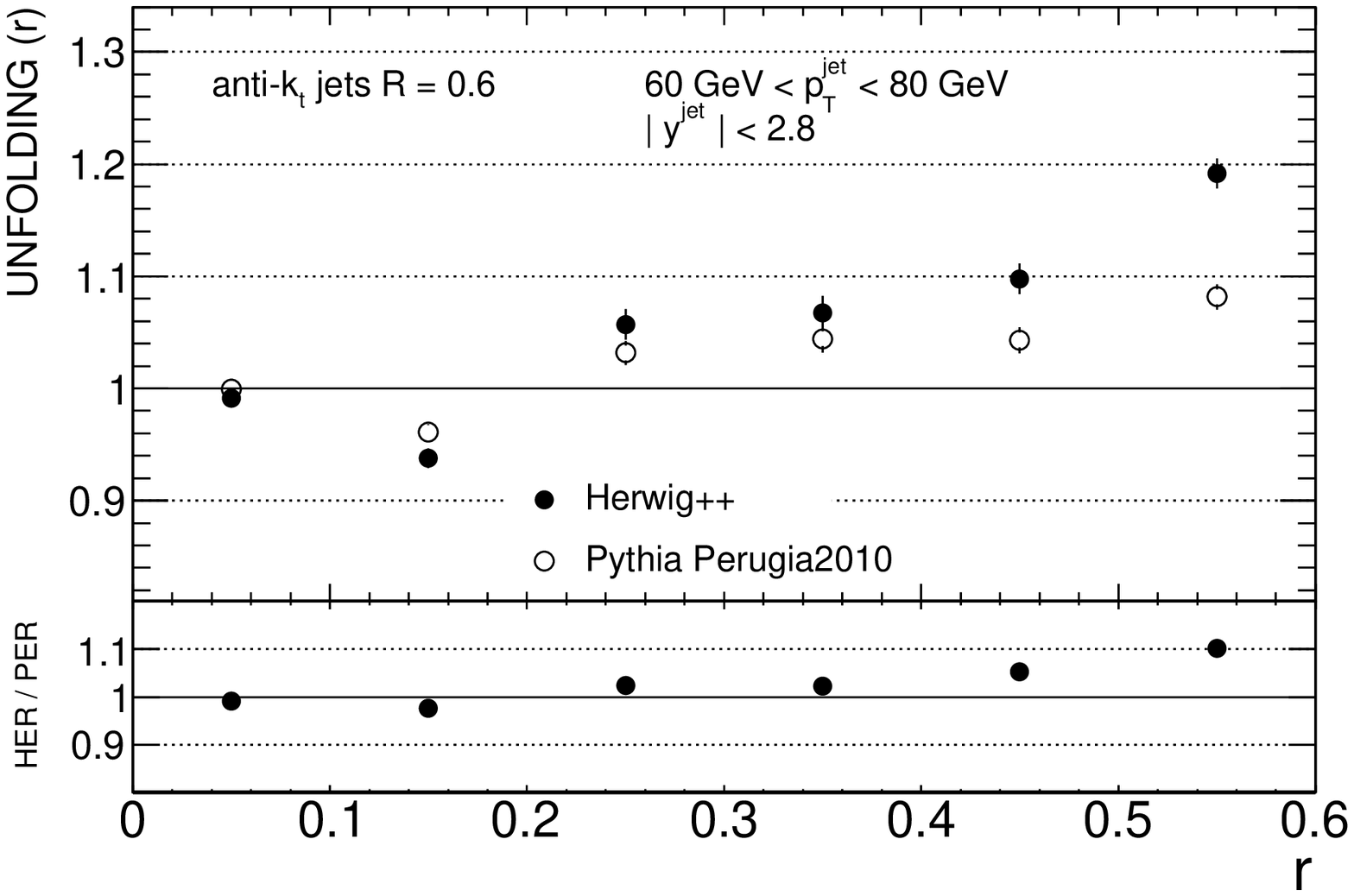}
\includegraphics[width=0.495\textwidth]{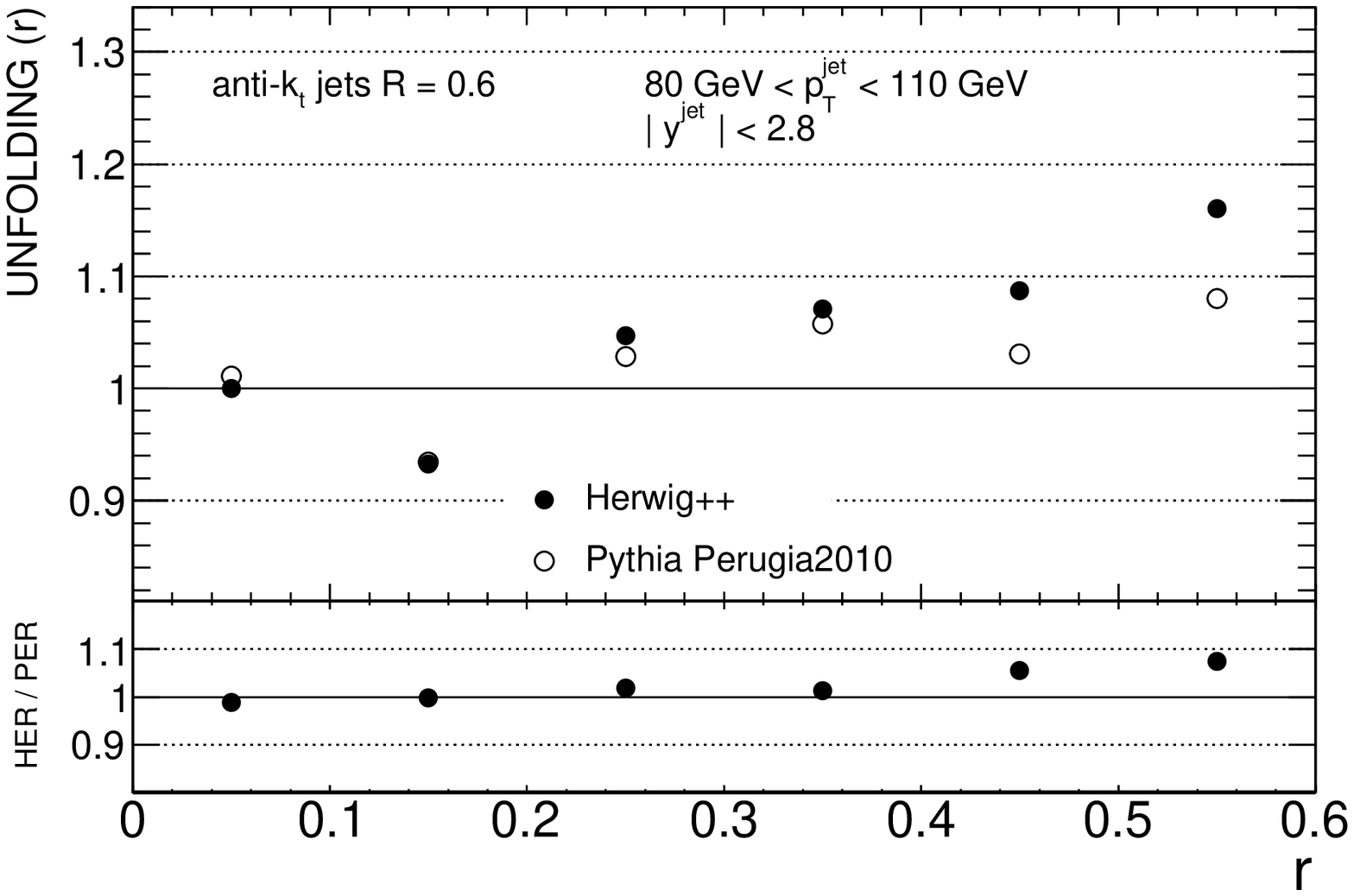}
}
\mbox{
\includegraphics[width=0.495\textwidth]{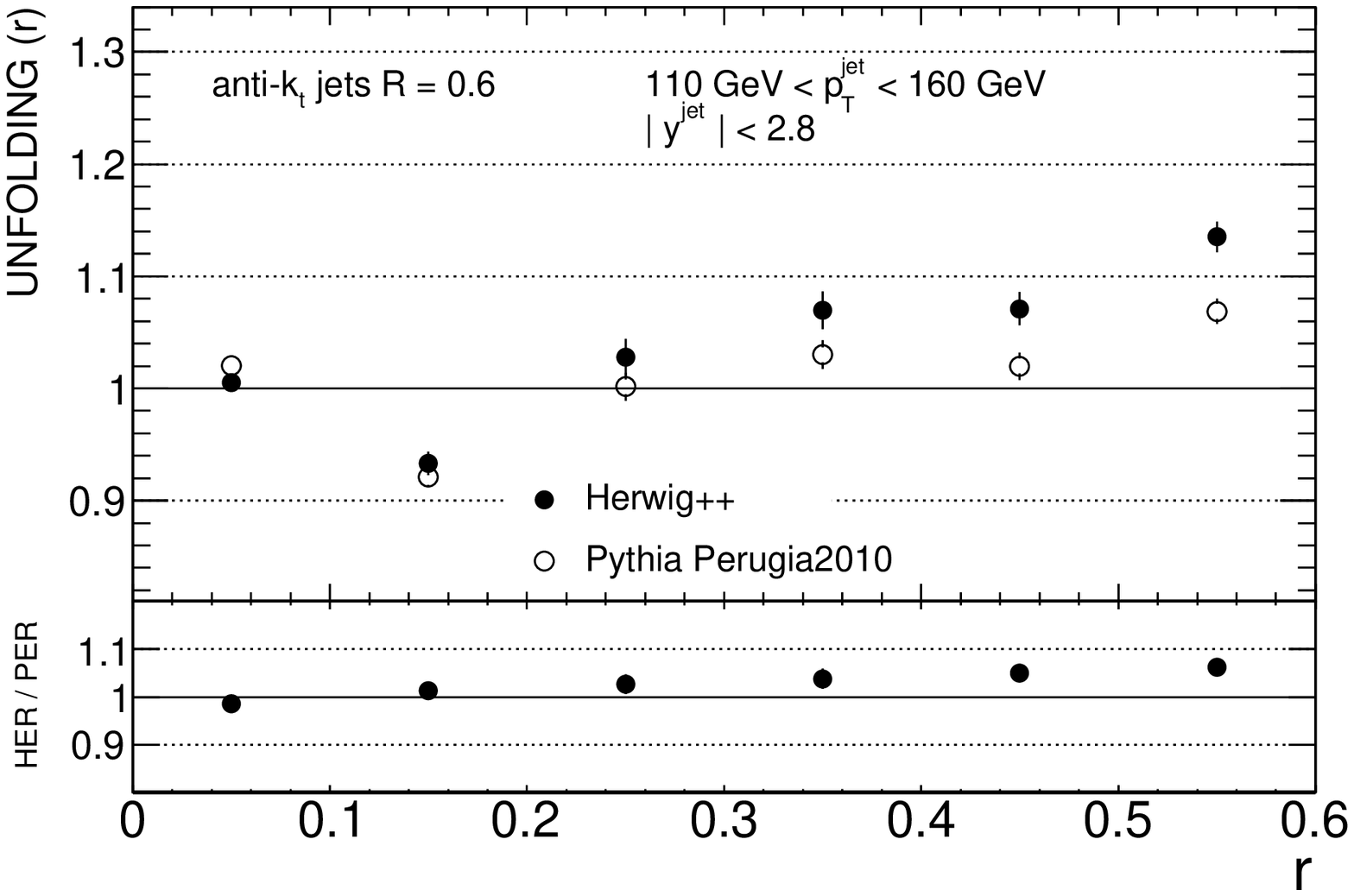}
\includegraphics[width=0.495\textwidth]{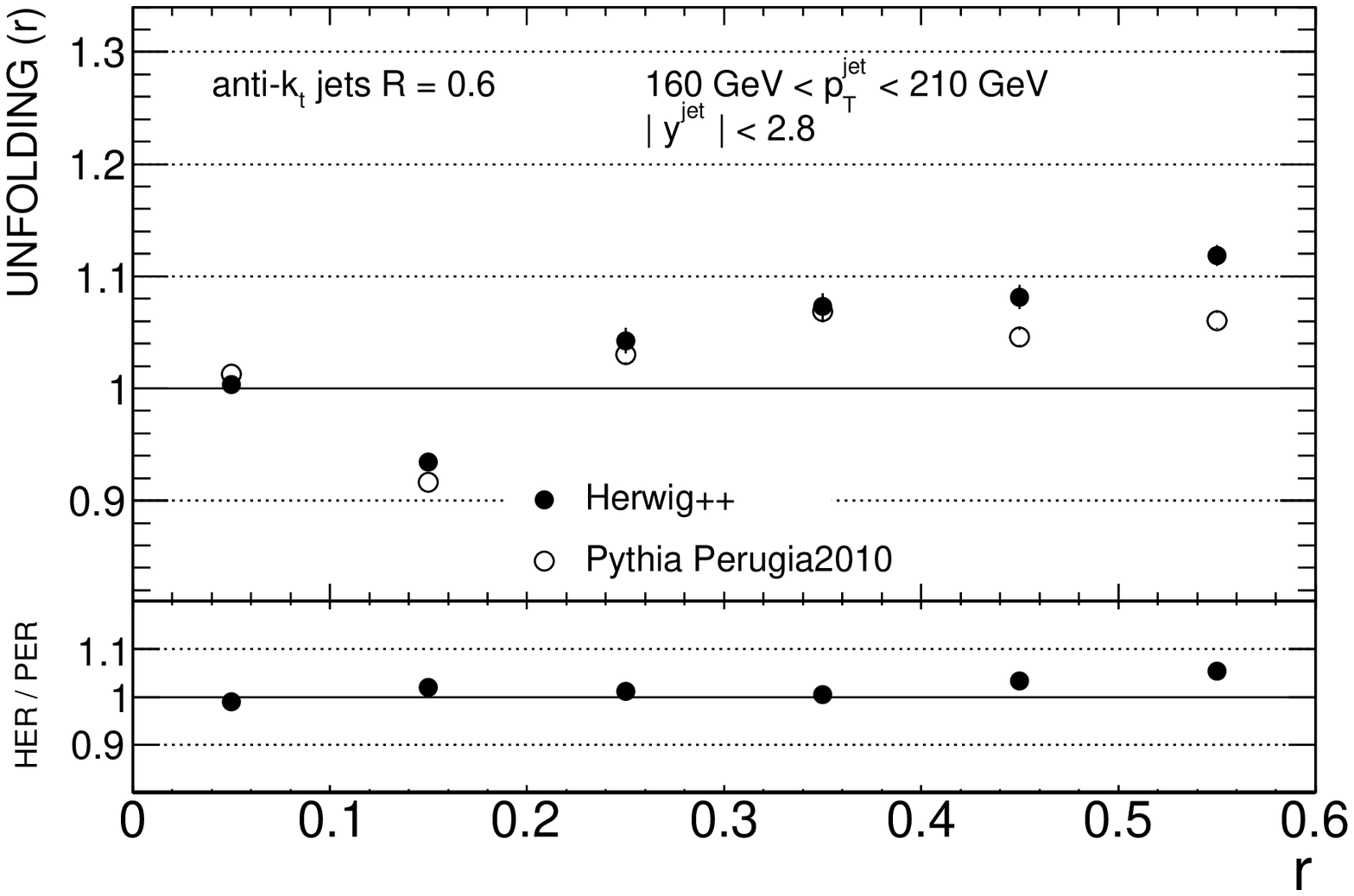}
}
\end{center}
\vspace{-0.7 cm}
\caption{\small
Systematic uncertainty on the differential jet shape related to the correction for detector effects with different physics models assumptions, 
for jets with $|\rapjet| < 2.8$ and $30 \ {\rm GeV} < \ptjet < 210  \ {\rm GeV}$.
}
\label{fig_Unfolding1}
\end{figure}

\begin{figure}[tbh]
\begin{center}
\mbox{
\includegraphics[width=0.495\textwidth]{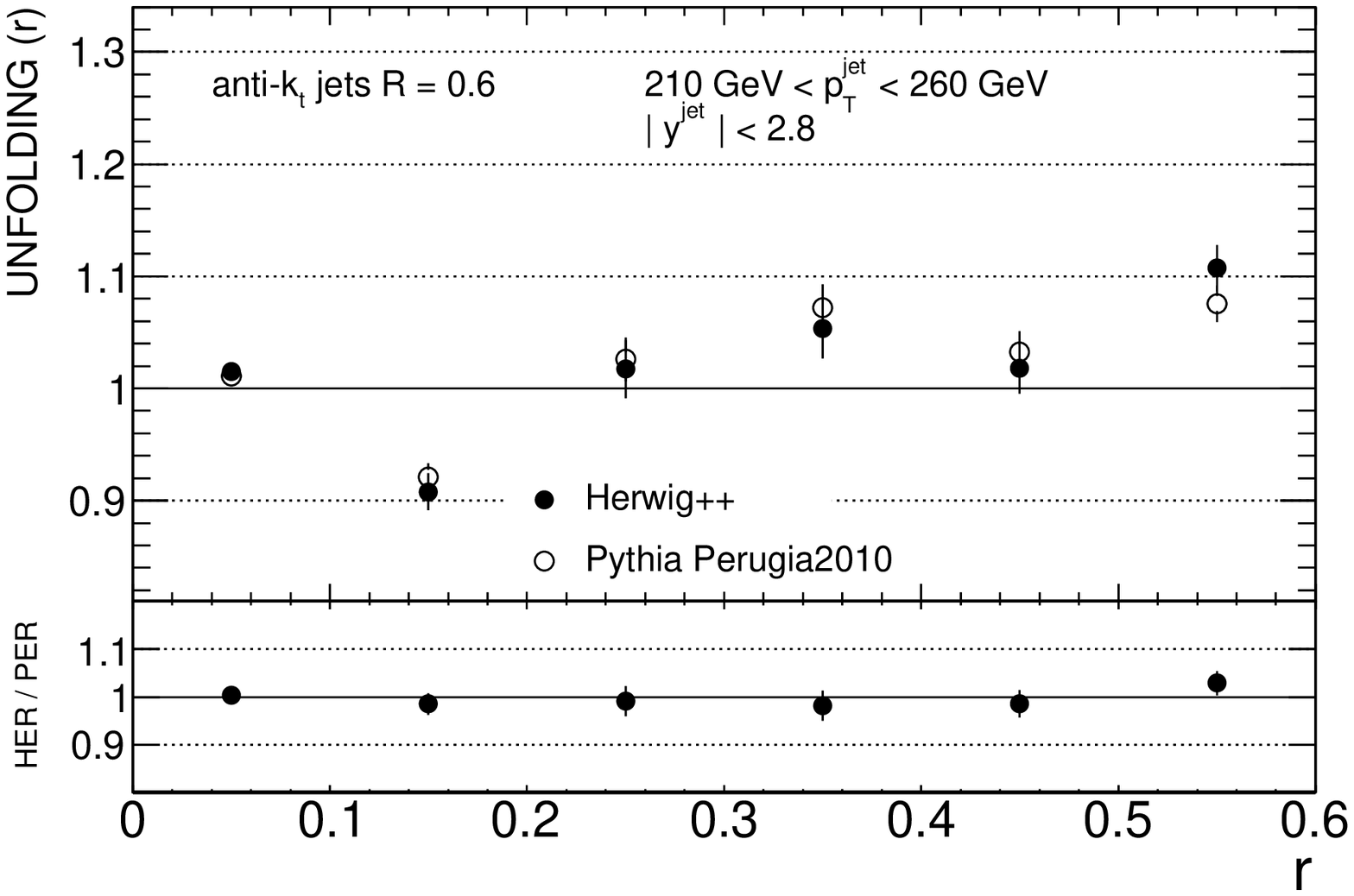}
\includegraphics[width=0.495\textwidth]{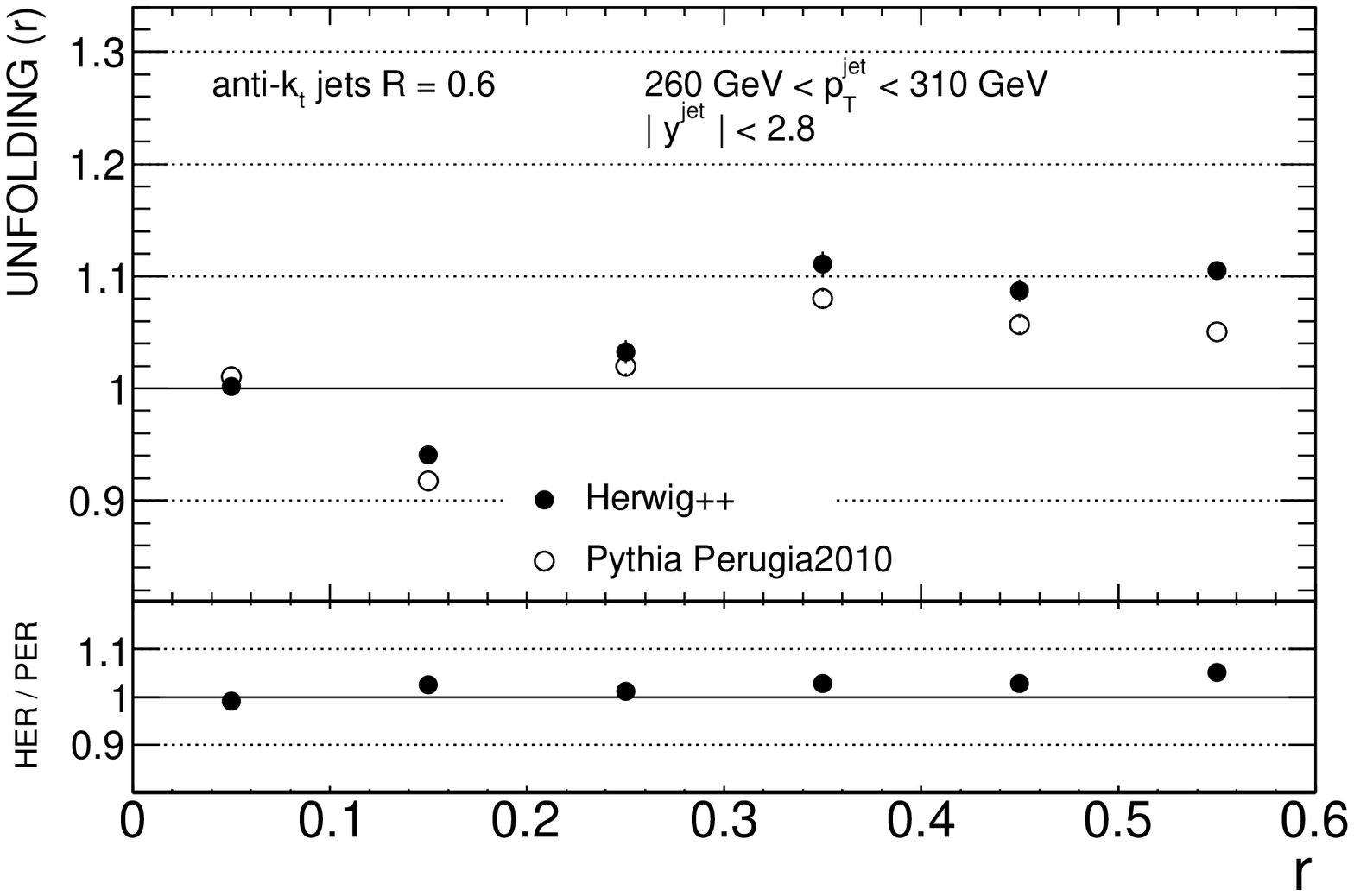}
}
\mbox{
\includegraphics[width=0.495\textwidth]{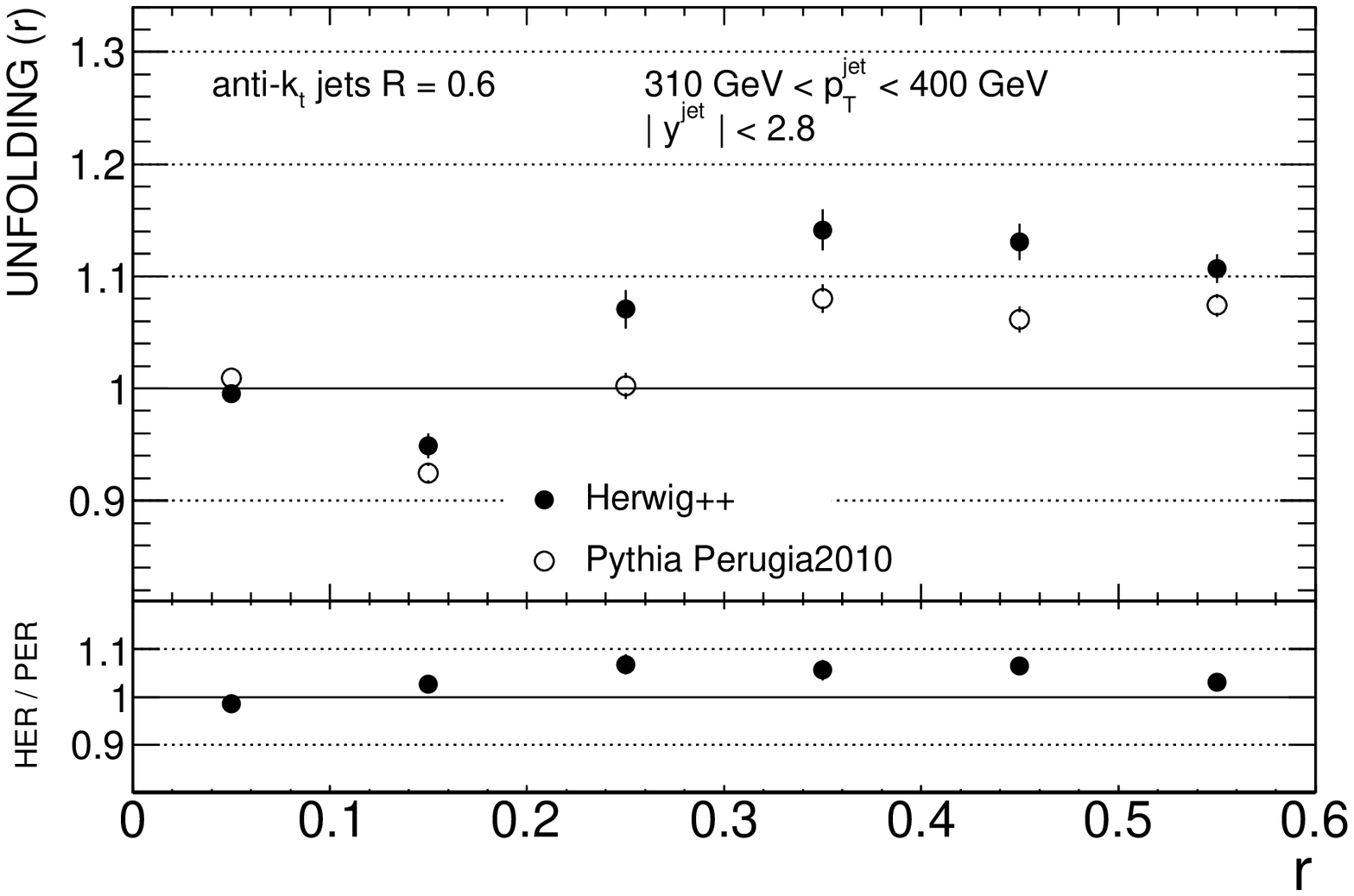}
\includegraphics[width=0.495\textwidth]{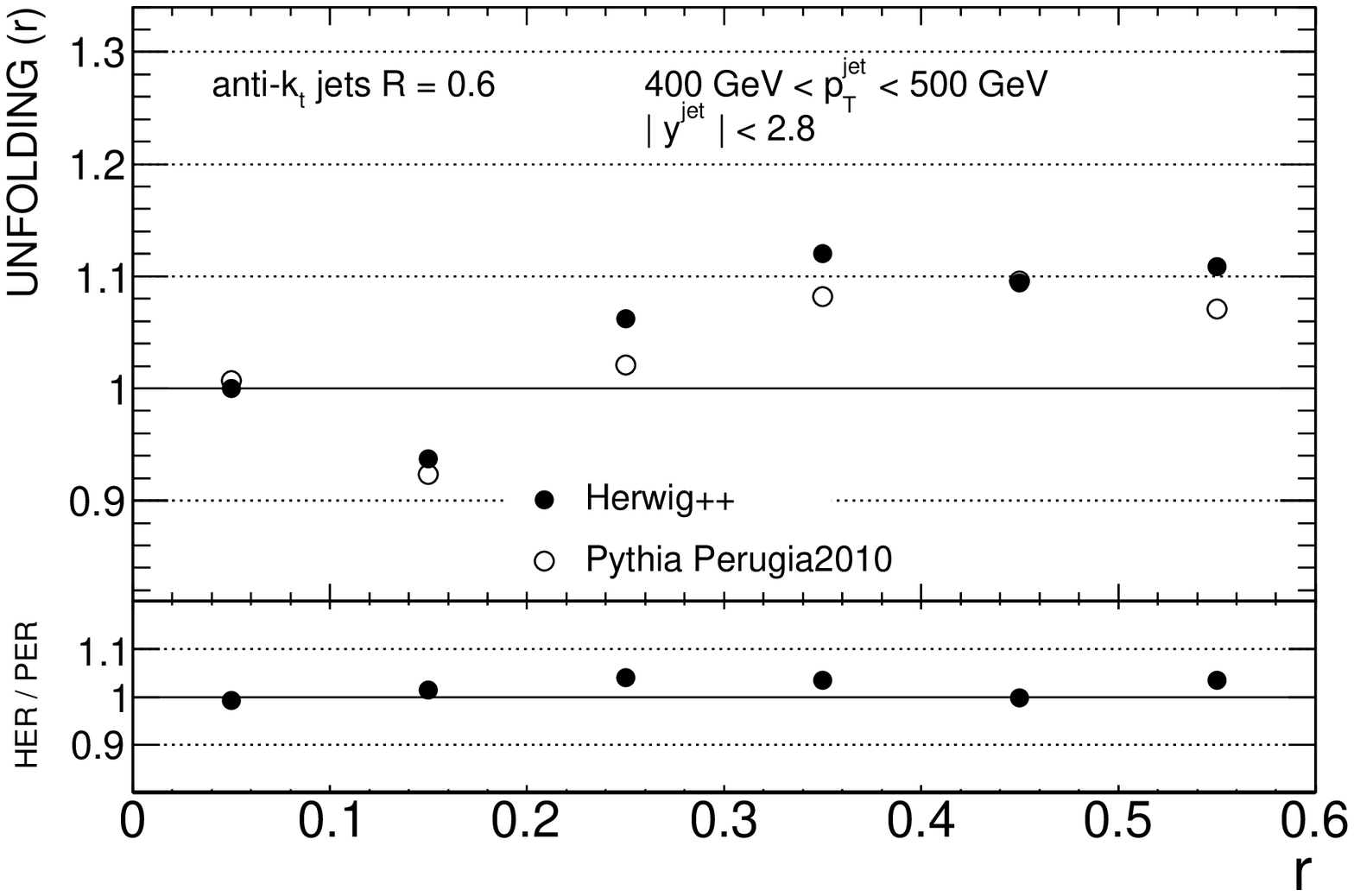}
}
\mbox{
\includegraphics[width=0.495\textwidth]{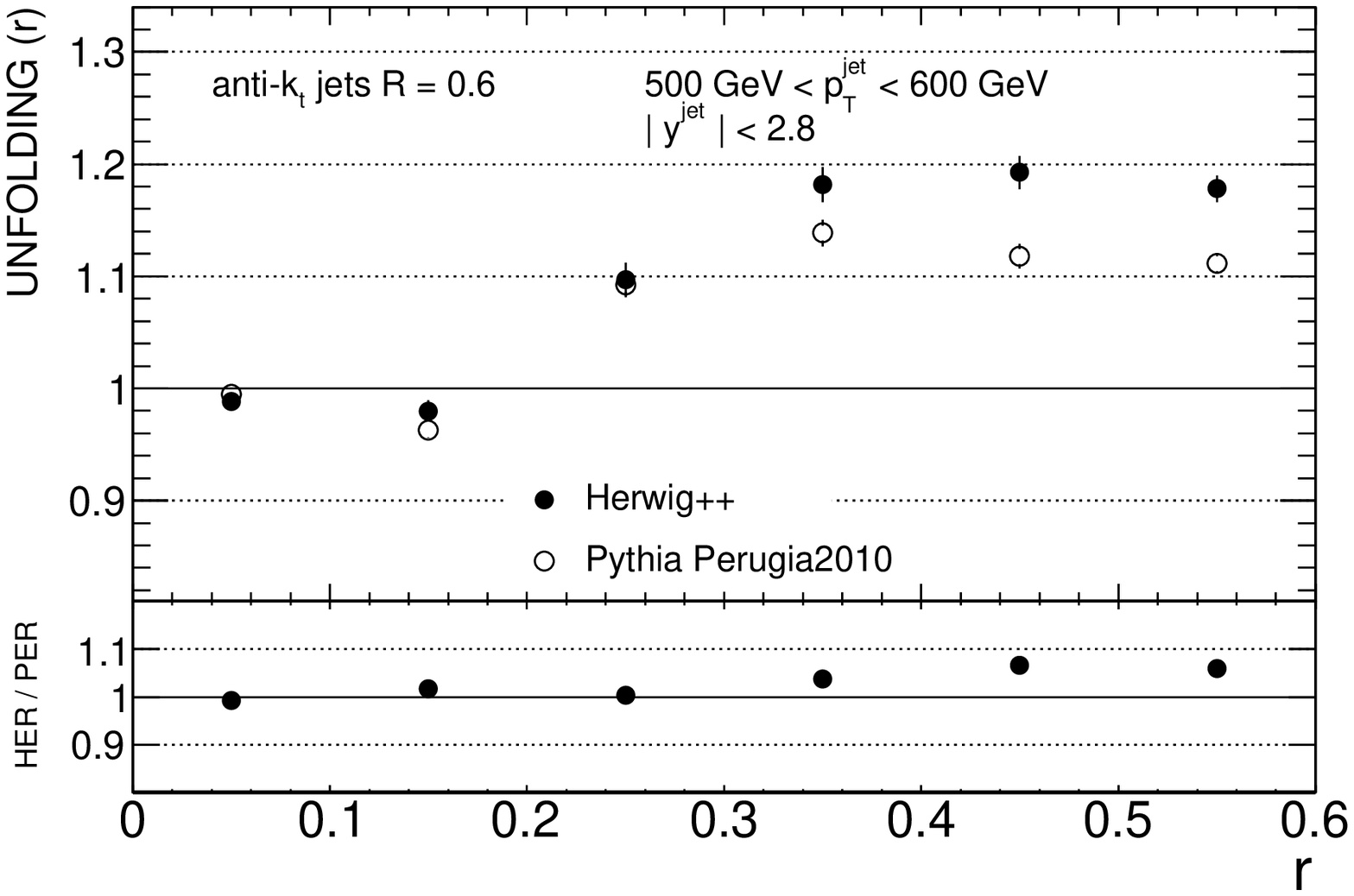}
}
\end{center}
\vspace{-0.7 cm}
\caption{\small
Systematic uncertainty on the differential jet shape related to the correction for detector effects with different physics models assumptions,
for jets with $|\rapjet| < 2.8$ and $210 \ {\rm GeV} < \ptjet < 600  \ {\rm GeV}$.
}
\label{fig_Unfolding2}
\end{figure}

\begin{figure}[tbh]
\begin{center}
\mbox{
\includegraphics[width=0.495\textwidth]{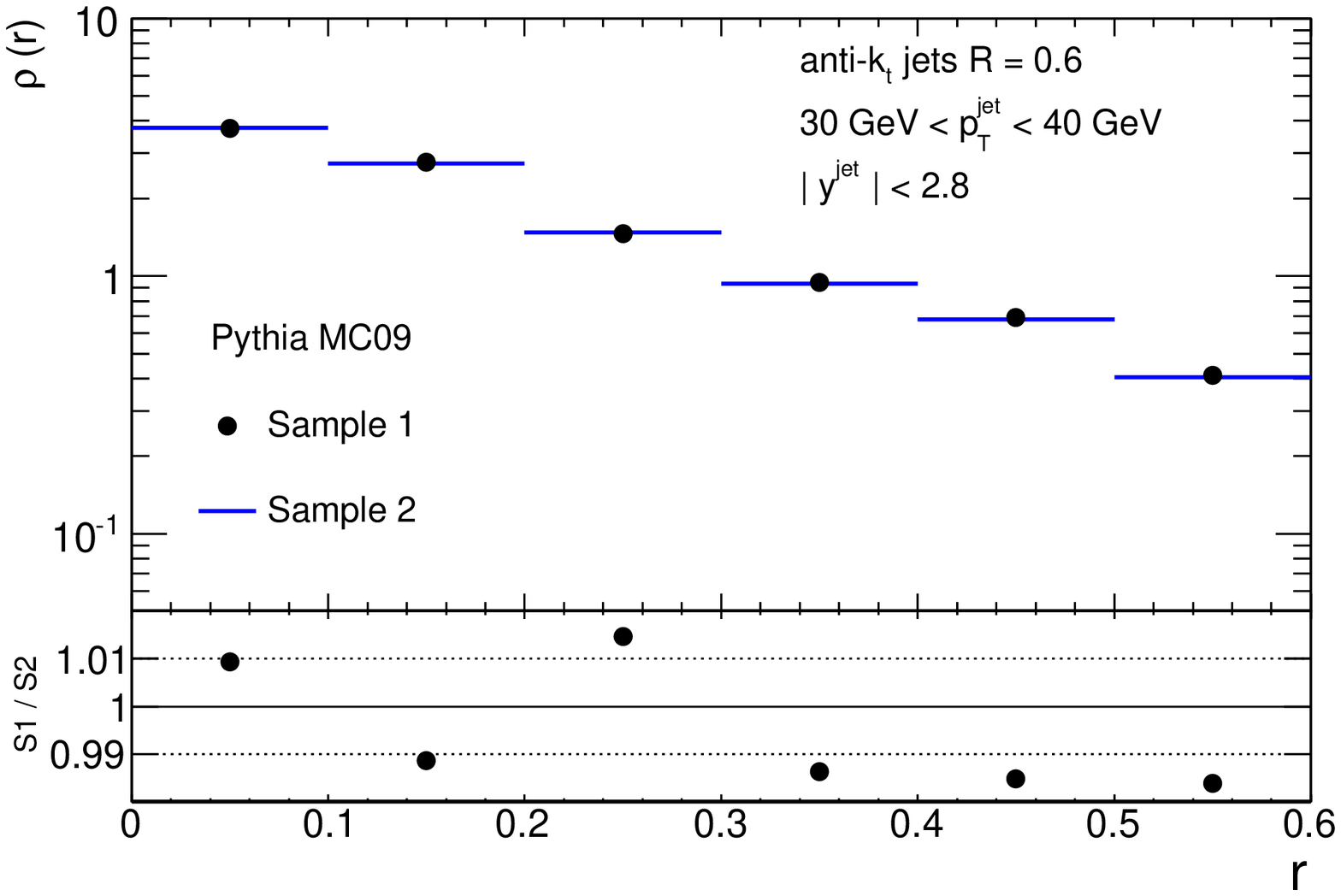}
\includegraphics[width=0.495\textwidth]{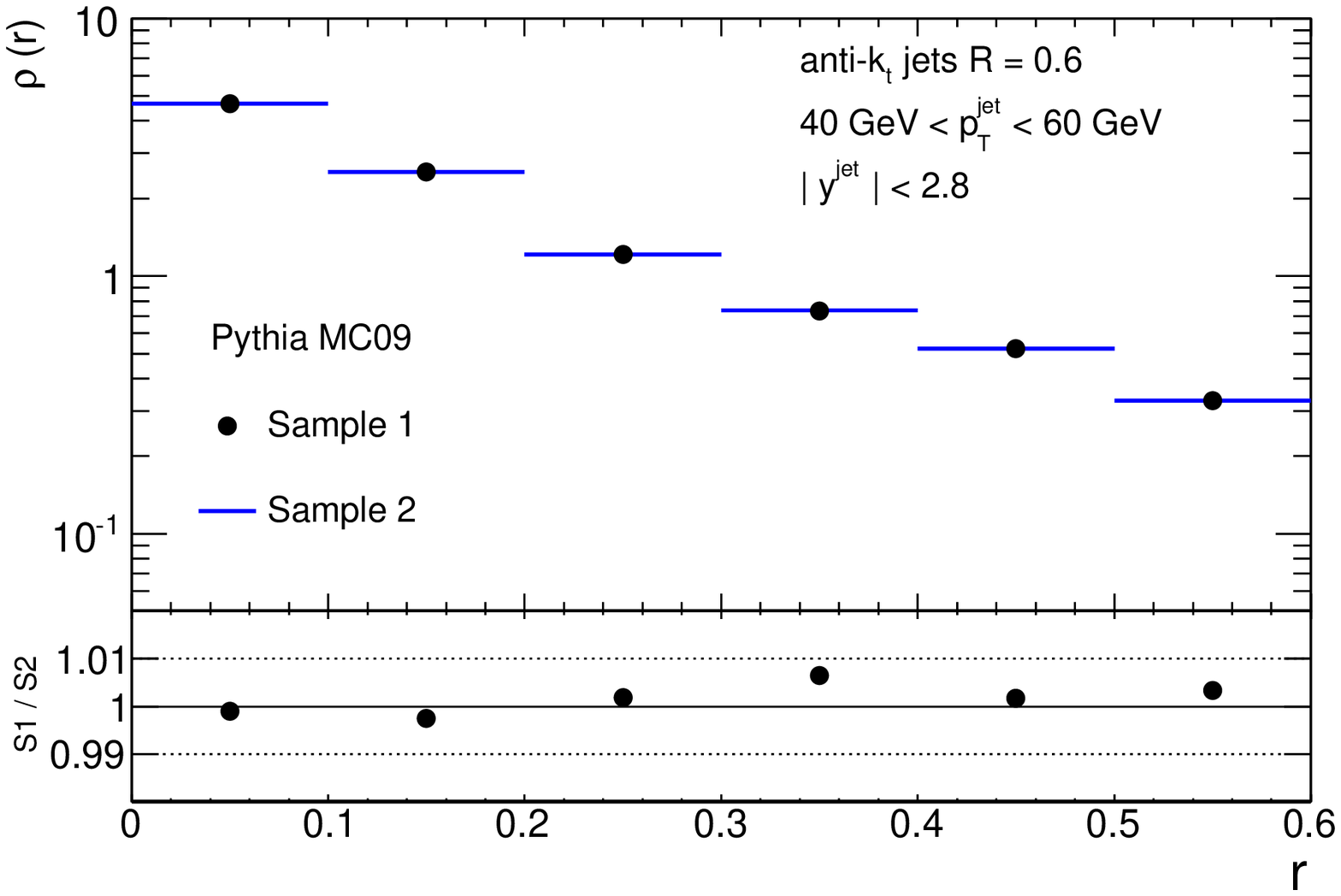}
}
\mbox{
\includegraphics[width=0.495\textwidth]{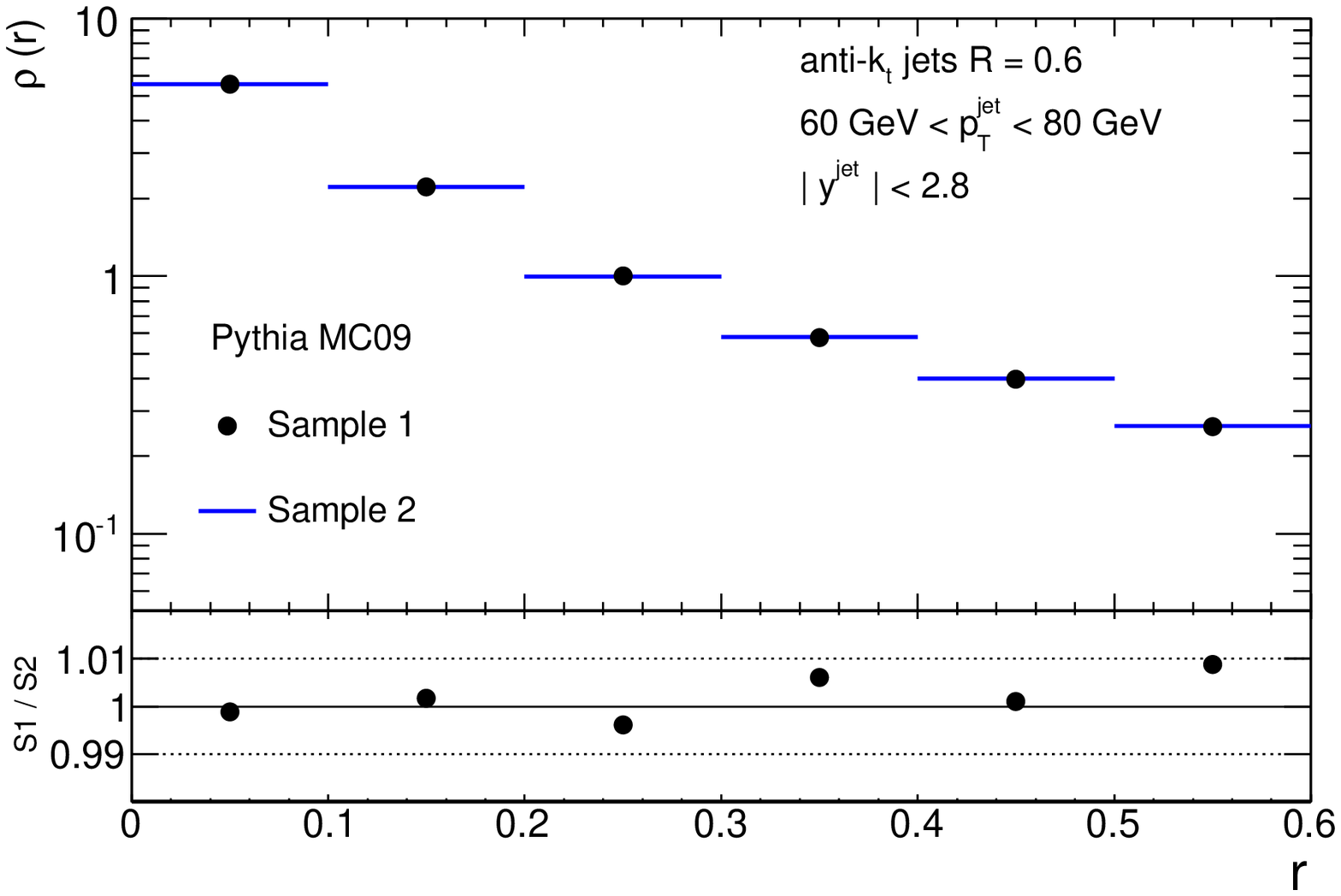}
\includegraphics[width=0.495\textwidth]{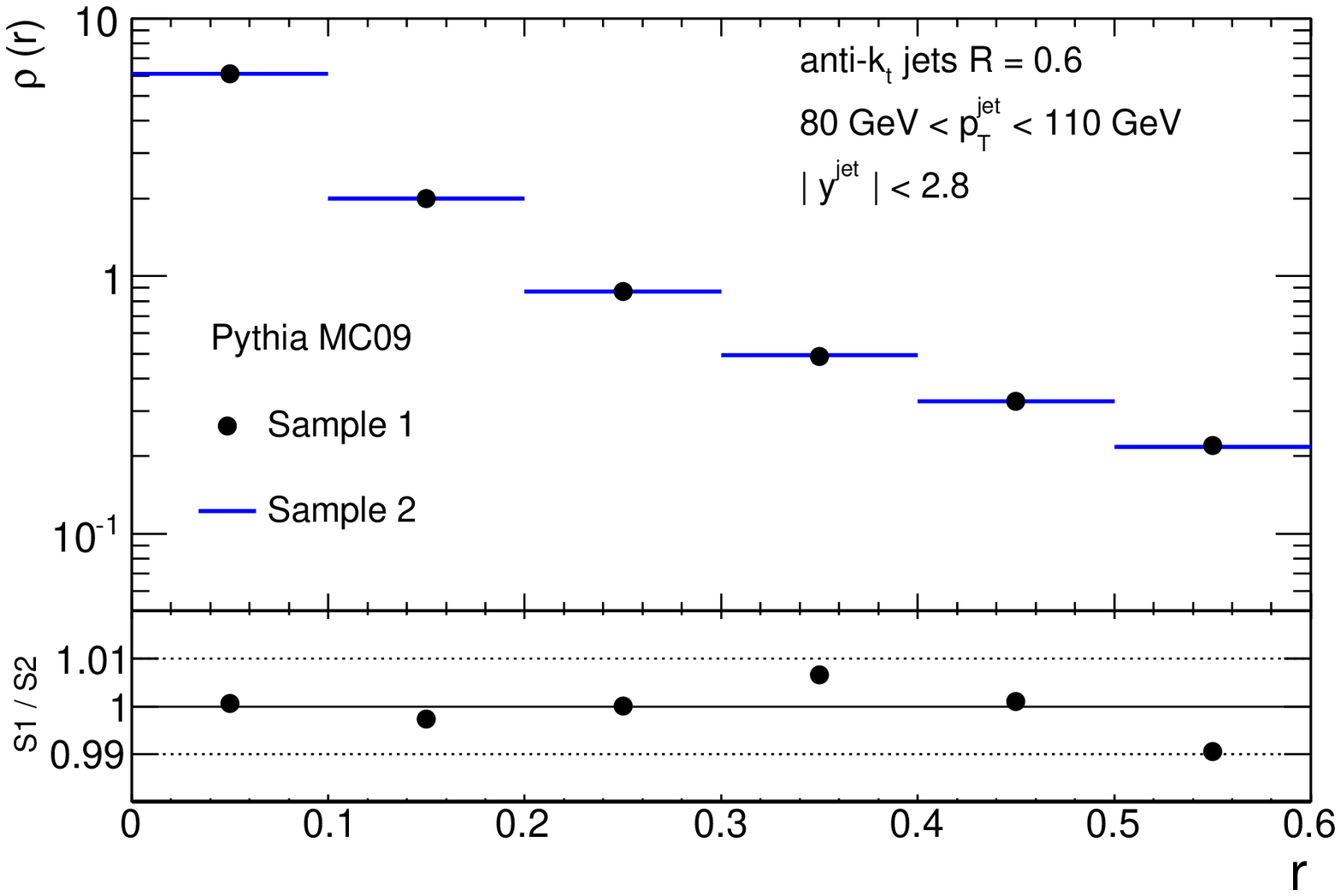}
}
\mbox{
\includegraphics[width=0.495\textwidth]{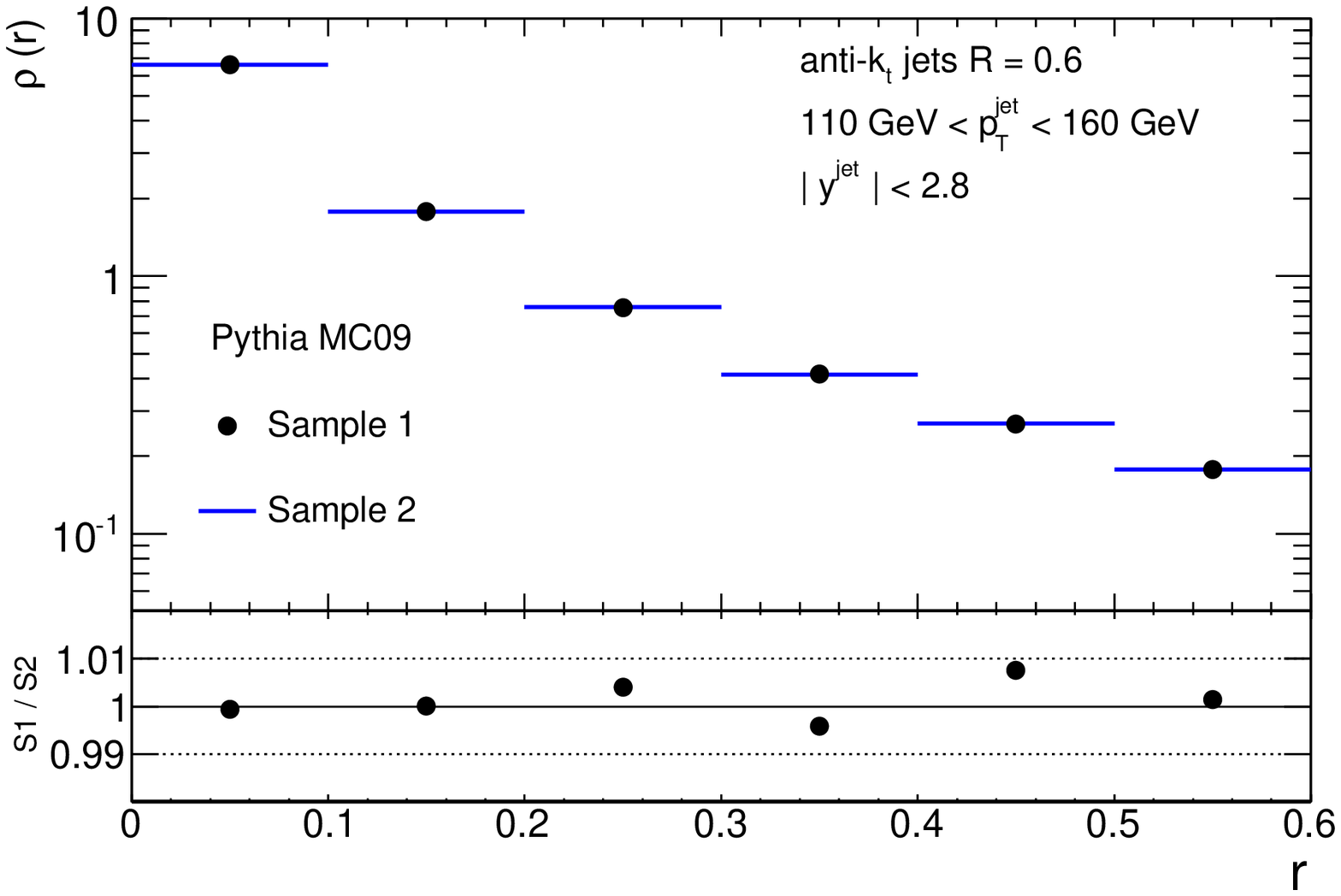}
\includegraphics[width=0.495\textwidth]{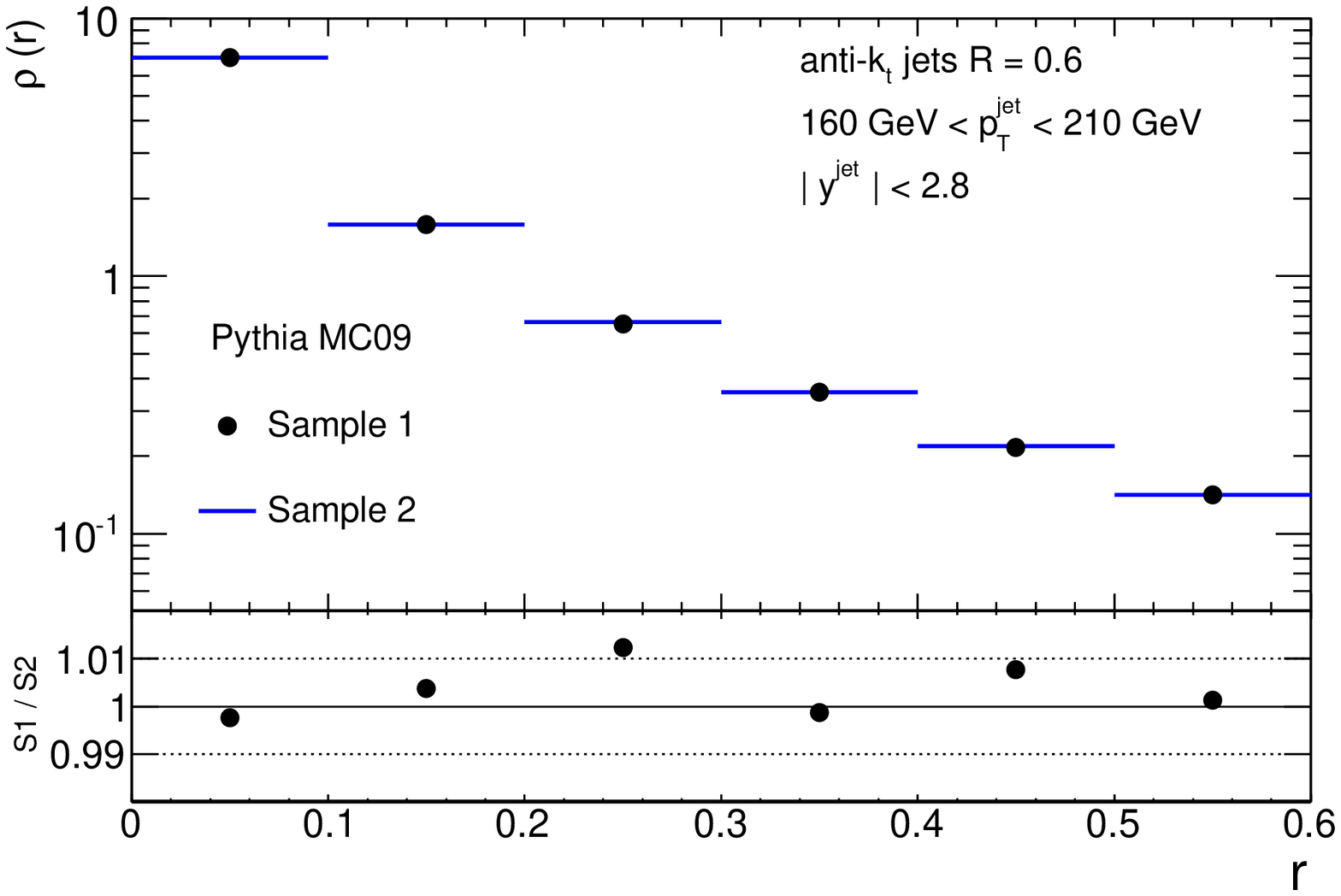}
}
\end{center}
\vspace{-0.7 cm}
\caption{\small
Systematic uncertainty on the differential jet shape related to the non-closure of the correction for detector effects procedure,
for jets with $|\rapjet| < 2.8$ and $30 \ {\rm GeV} < \ptjet < 210  \ {\rm GeV}$.
}
\label{fig_NonClosure1}
\end{figure}

\begin{figure}[tbh]
\begin{center}
\mbox{
\includegraphics[width=0.495\textwidth]{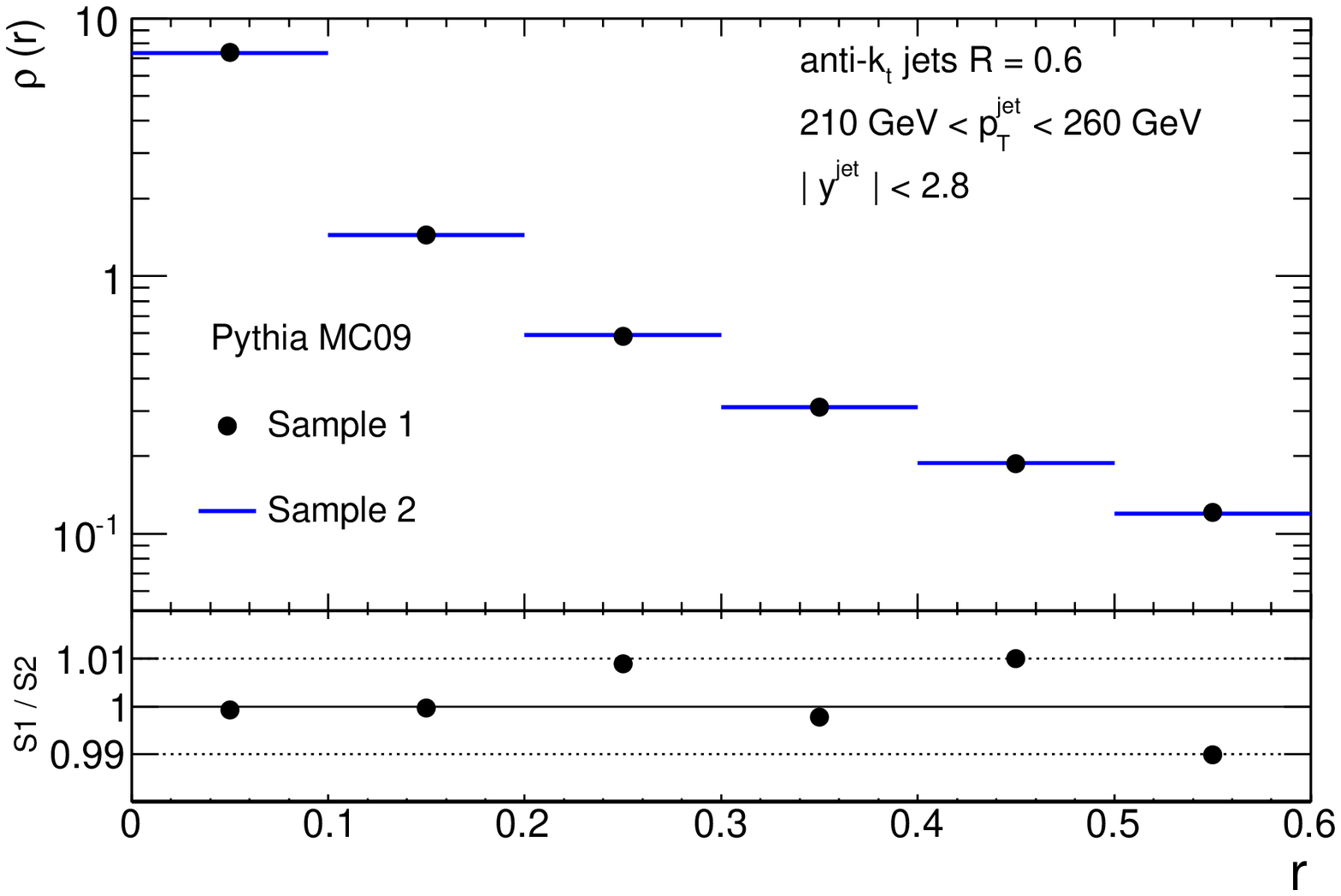}
\includegraphics[width=0.495\textwidth]{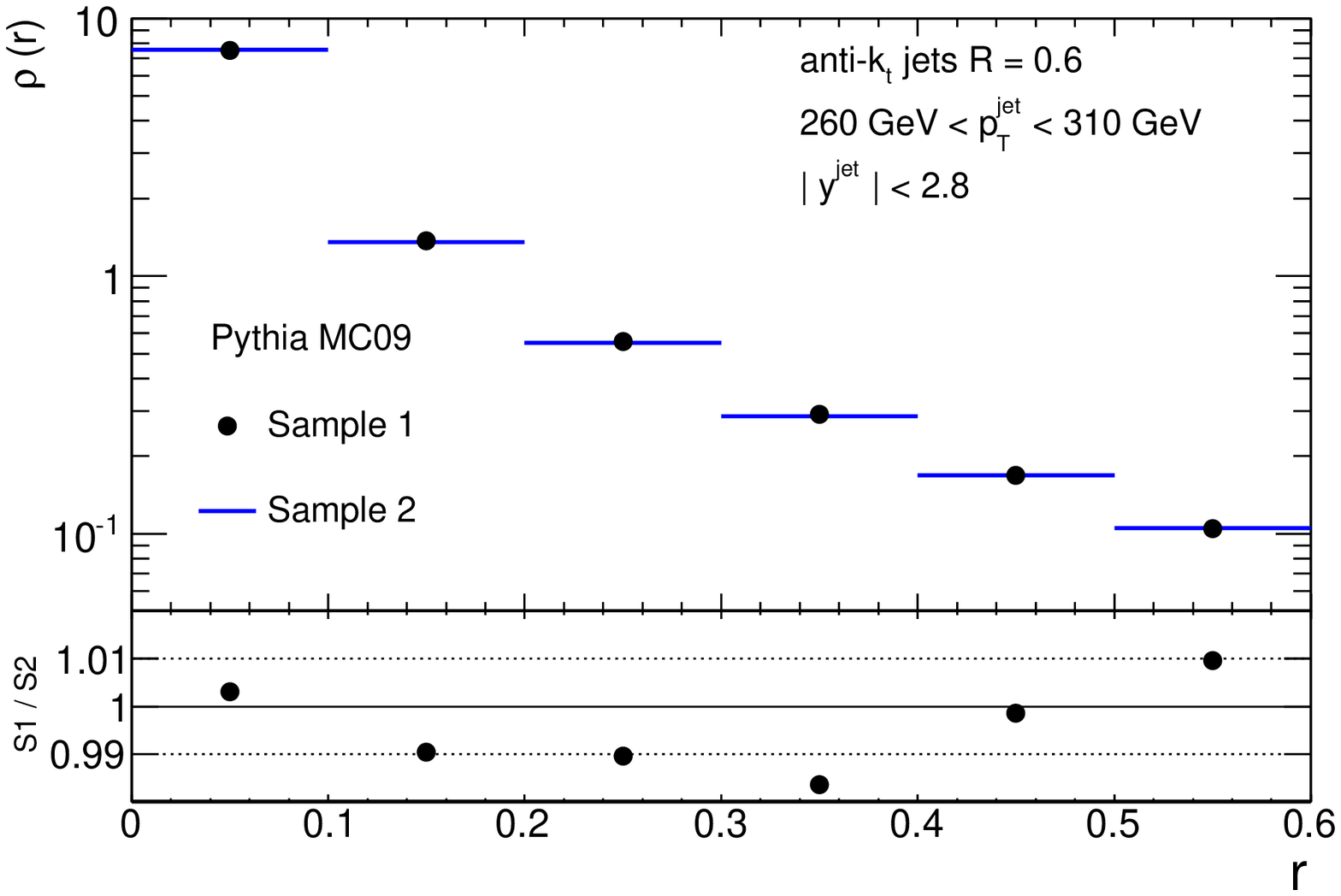}
}
\mbox{
\includegraphics[width=0.495\textwidth]{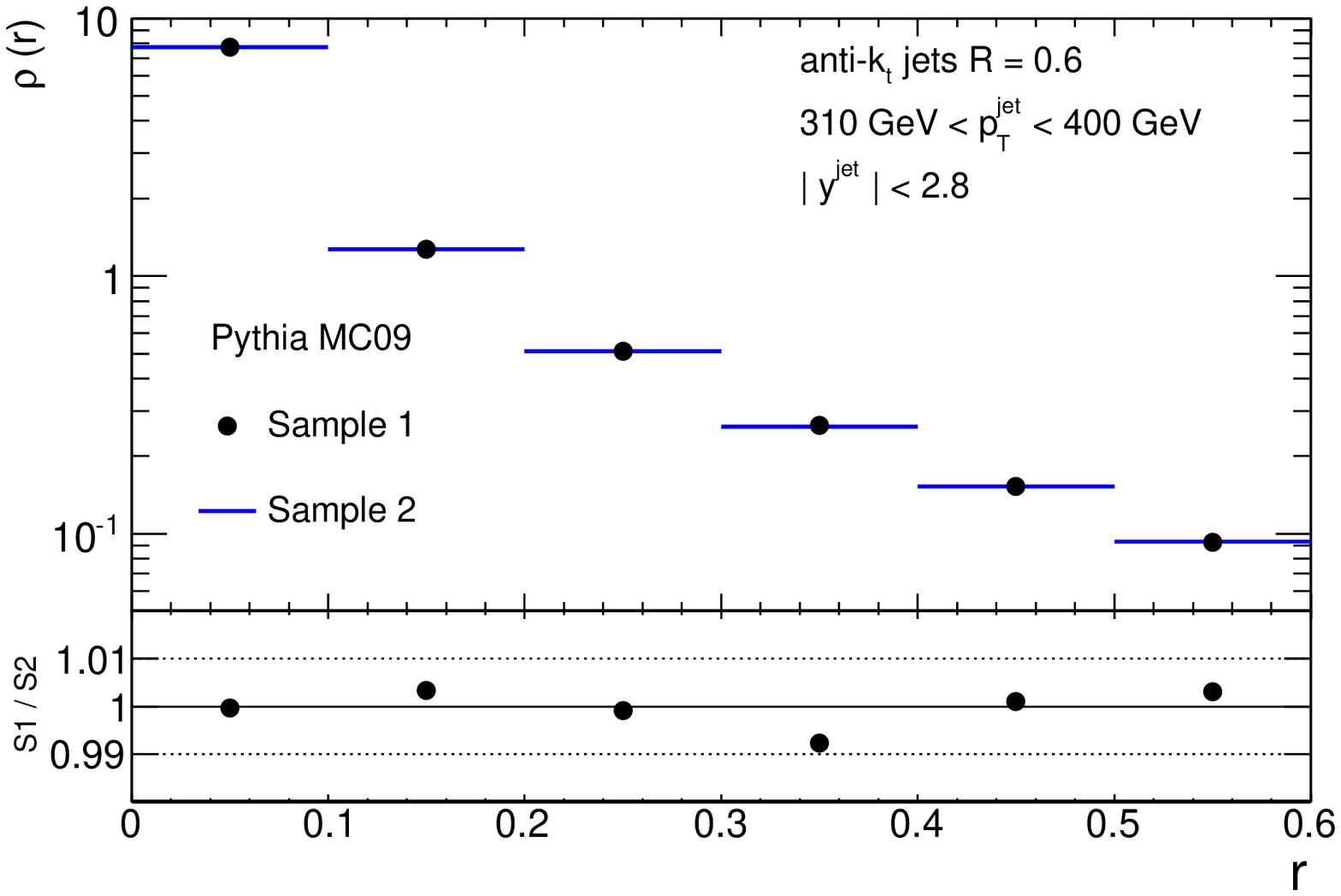}
\includegraphics[width=0.495\textwidth]{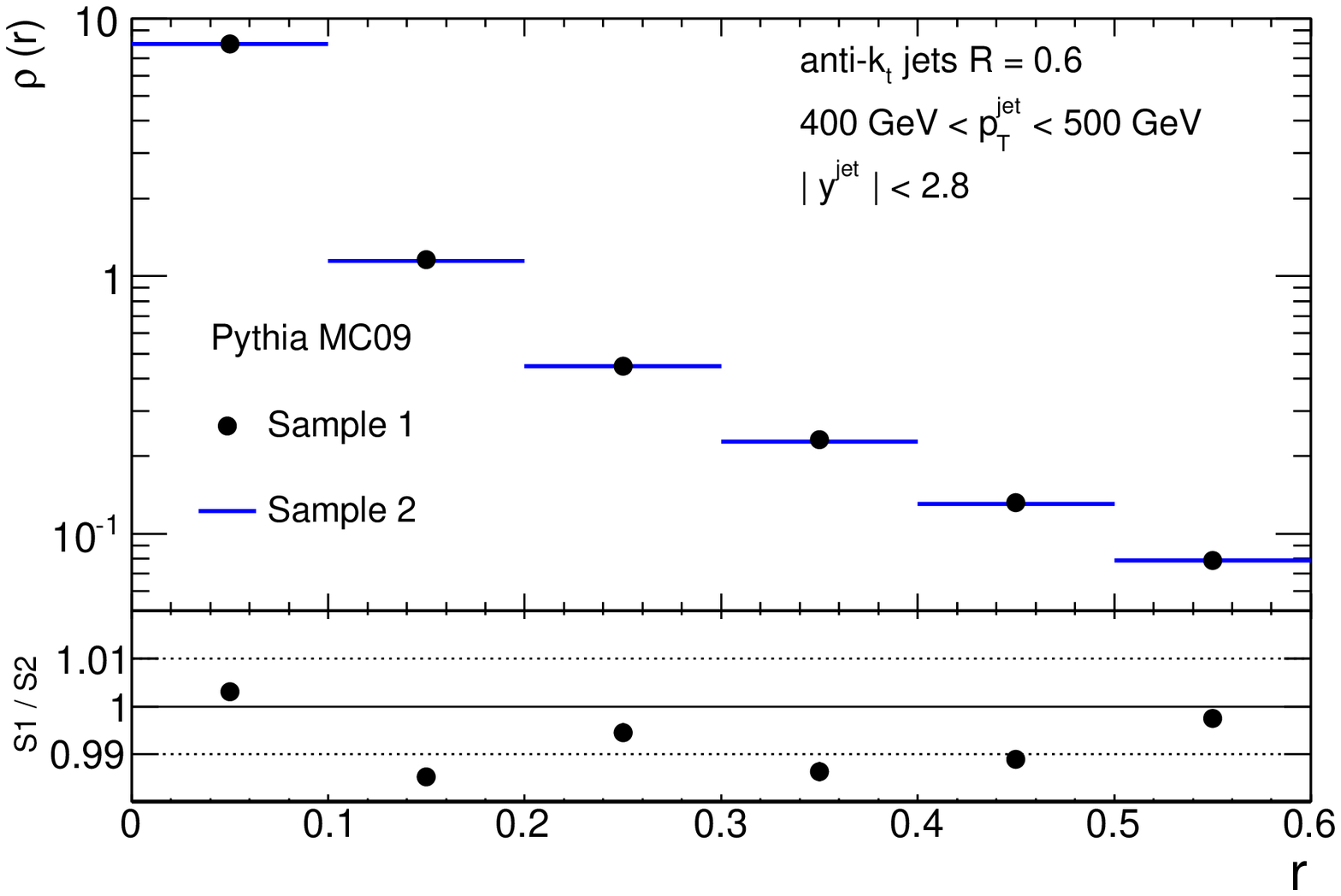}
}
\mbox{
\includegraphics[width=0.495\textwidth]{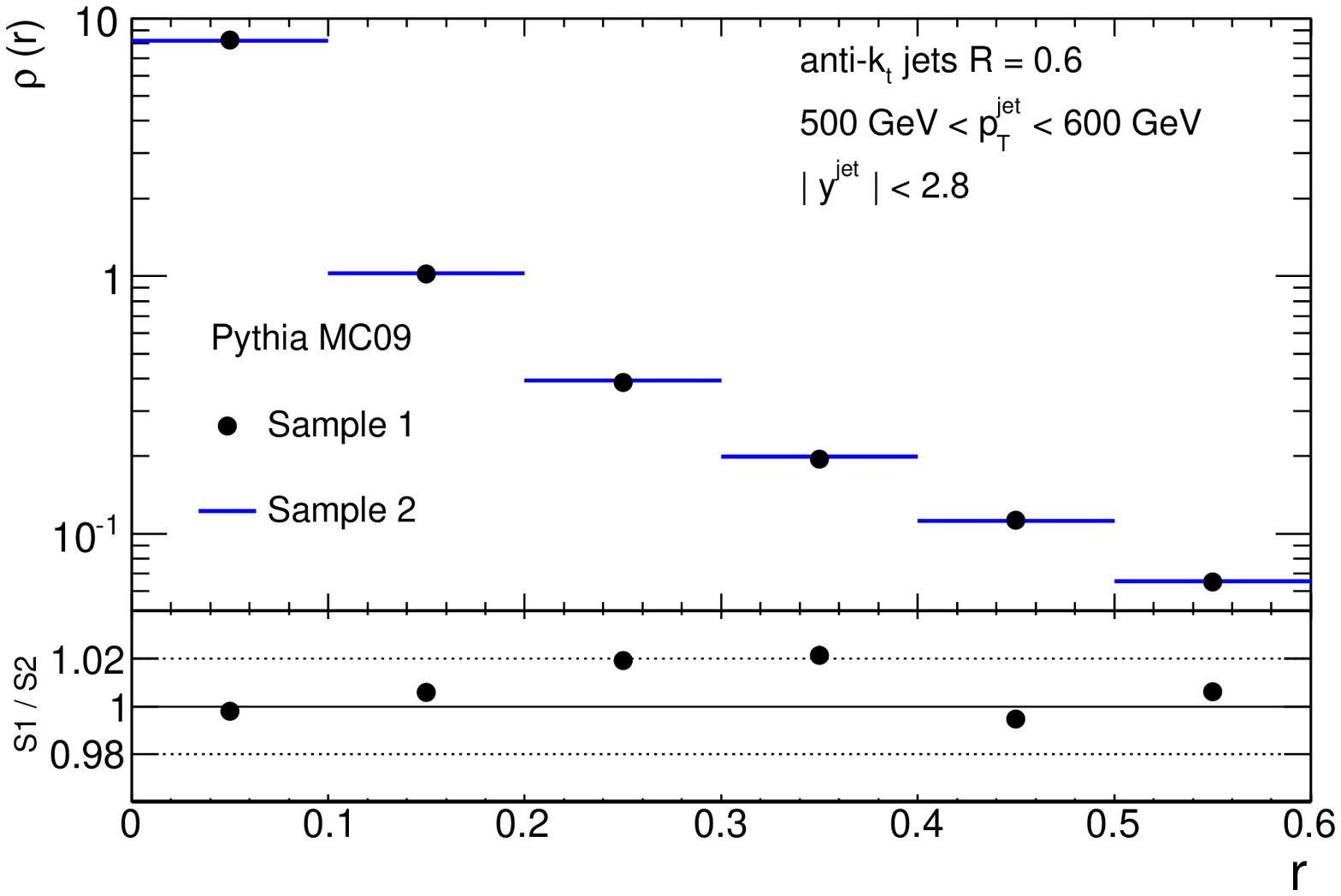}
}
\end{center}
\vspace{-0.7 cm}
\caption{\small
Systematic uncertainty on the differential jet shape related to the non-closure of the correction for detector effects procedure,
for jets with $|\rapjet| < 2.8$ and $210 \ {\rm GeV} < \ptjet < 600  \ {\rm GeV}$.
}
\label{fig_NonClosure2}
\end{figure}

\begin{itemize}

\item The absolute energy scale of the individual clusters belonging to the jet
is varied in the data according to studies using
isolated tracks~\cite{jet_pro_atlas}, which parametrize the uncertainty on
the calorimeter cluster energy as a function of $\ptjet$ and $\eta$ 
of the cluster. 
This variation is performed by multiplying the cluster energies by 
$1 \pm a \times (1 + b/\ptjet^{cluster})$ for up (down) variation, with a = 5\% if 
$|\eta| < 3.2$ (a = 10\% if $|\eta| > 3.2$), b = 1.5, and $\ptjet^{cluster}$ in~GeV. 
The jet position is then recalculated 
according to the new cluster four-momenta. 
The maximum variation on the differential jet shapes is taken as a systematic
uncertainty, that varies typically 
between 3$\%$ to 15$\%$ as $r$ increases, and constitutes the dominant 
systematic uncertainty in this analysis (see Figures~\ref{fig_CJES1}~and~\ref{fig_CJES2}).

\item The systematic uncertainty on the measured jet shapes  
arising from the details of the model used to simulate calorimeter showers in the MC  events is
studied.  A different simulated sample is considered, where the
FRITIOF~\cite{ftfp} plus BERT showering model is employed instead of the QGSP plus BERT model.
FRITOF+BERT provides the second best description of the test-beam results~\cite{sim_barrel} after QGSP+BERT.
This introduces an uncertainty on the measured differential jet shapes that varies
between $1\%$ to 4$\%$, and is approximately independent of  $\ptjet$
and $|\rapjet|$ (see Figures~\ref{fig_FTFP1}~and~\ref{fig_FTFP2}).

\item The measured jet $\ptjet$ is varied by 2$\%$ to 8$\%$,  
depending on $\ptjet$ and $|\rapjet|$, to account for the remaining 
uncertainty on the absolute jet energy scale~\cite{jet_pro_atlas},  
after removing contributions already accounted for and related to the 
energy of the single clusters and  the calorimeter shower modeling, as discussed above.
This introduces an uncertainty of about 3$\%$ to 5$\%$ 
in the measured differential jet shapes (see Figures~\ref{fig_CJES1}~and~\ref{fig_CJES2}). 

\item The 14$\%$ uncertainty on the jet energy resolution~\cite{jet_pro_atlas} translates into a smaller than 2$\%$
effect on the measured differential jet shapes (see Figures~\ref{fig_RES1} and~\ref{fig_RES2}).

\item The correction factors are recomputed using HERWIG++, which implements different parton shower, fragmentation and UE models 
than PYTHIA, and compared to   PYTHIA-Perugia2010 (see Figures~\ref{fig_Unfolding1} and~\ref{fig_Unfolding2}).
In addition, the correction factors are also computed using ALPGEN and PYTHIA-DW 
for $\ptjet < 110$~GeV, where these MC samples provide a reasonable description of the uncorrected shapes in the data. 
The results from HERWIG++ encompass the variations obtained using all the above generators and are conservatively
adopted in all $\ptjet$ and $|\rapjet|$ ranges to compute systematic uncertainties on the differential jet shapes. These
uncertainties increase between 2$\%$ and 10$\%$ with increasing r.

\item Two statistically independent samples are used to test the closure of the 
correction for detector effects procedure. One was used to compute the shapes using calorimeter clusters, 
while the other was used to correct these shapes for detector effects. In Figures~\ref{fig_NonClosure1}~and~\ref{fig_NonClosure2} 
corrected results from the first sample are compared to results at particle level from the second one. The ratio between the 
two deviates from unity about 1$\%$, value that is included as uncertainty on the differential measurements.

\item No significant dependence on instantaneous luminosity
is observed in the measured jet shapes, indicating that residual
pile-up contributions are negligible after selecting events with 
only one reconstructed primary vertex. 

\item It was verified that the presence of dead calorimeter regions in the data does not affect 
the measured jet shapes, since only affected a small fraction of jets.  

\end{itemize}

\begin{figure}[tbh]
\begin{center}
\mbox{
\includegraphics[width=0.475\textwidth]{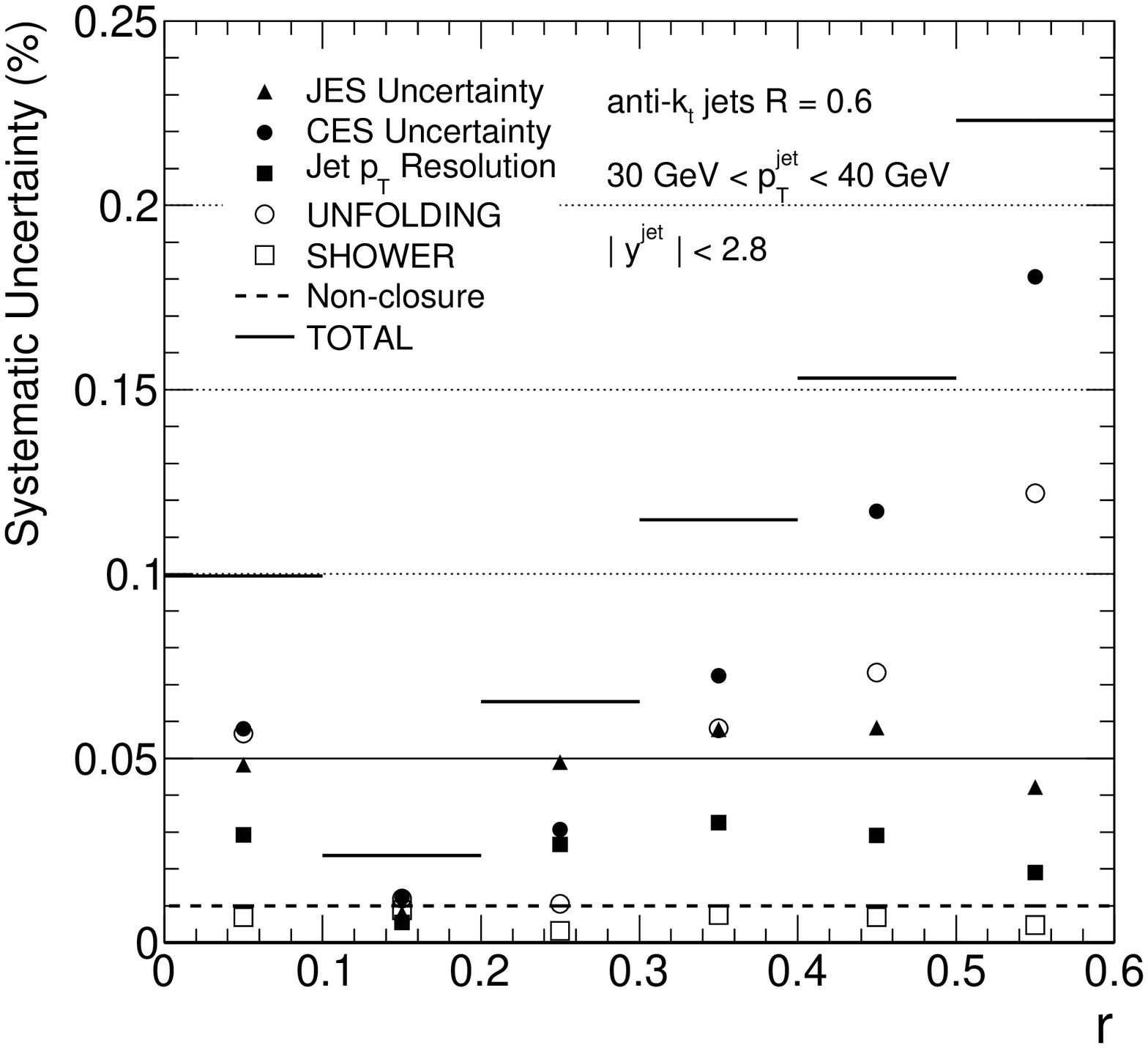}
\includegraphics[width=0.475\textwidth]{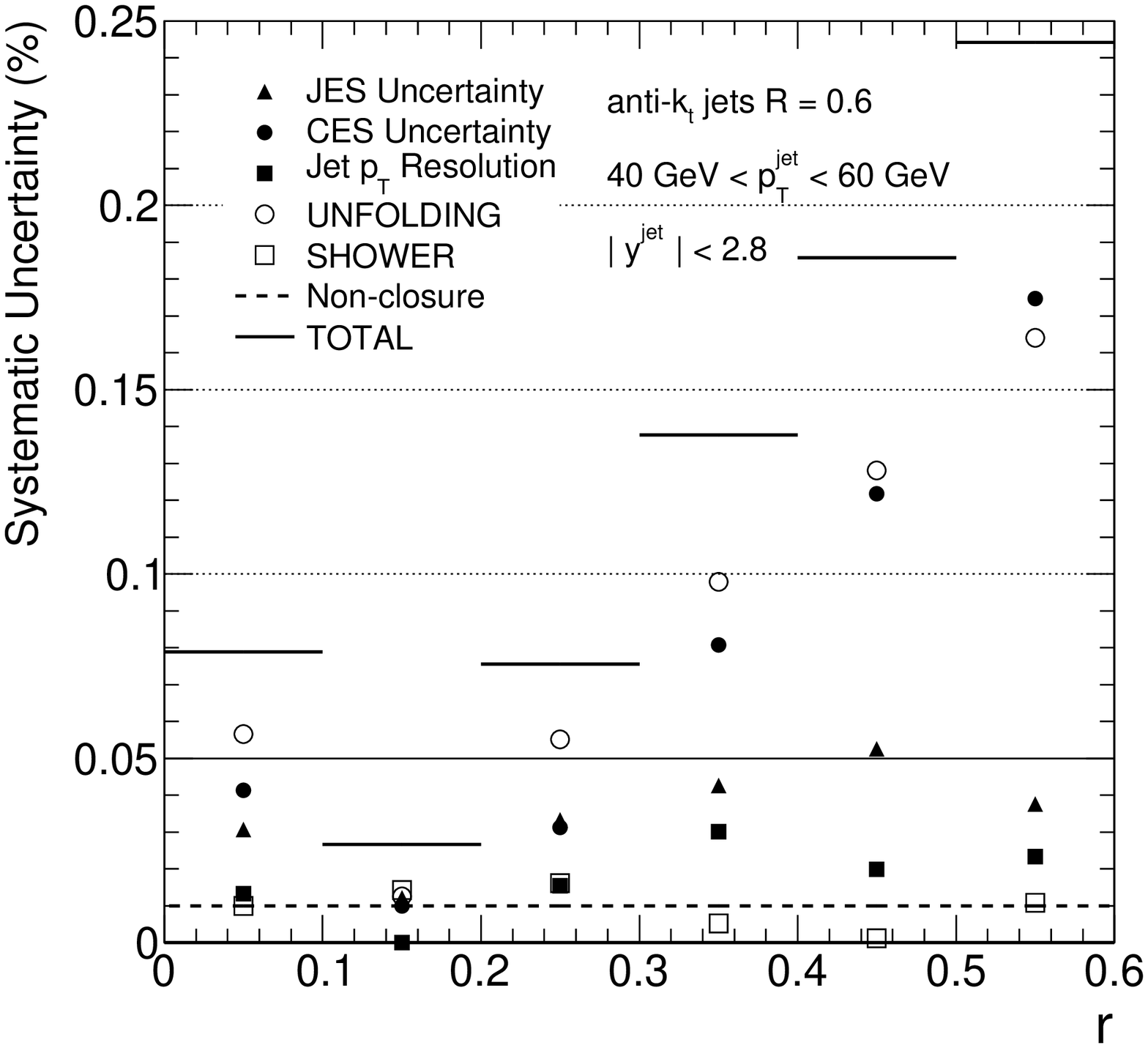}
}
\mbox{
\includegraphics[width=0.475\textwidth]{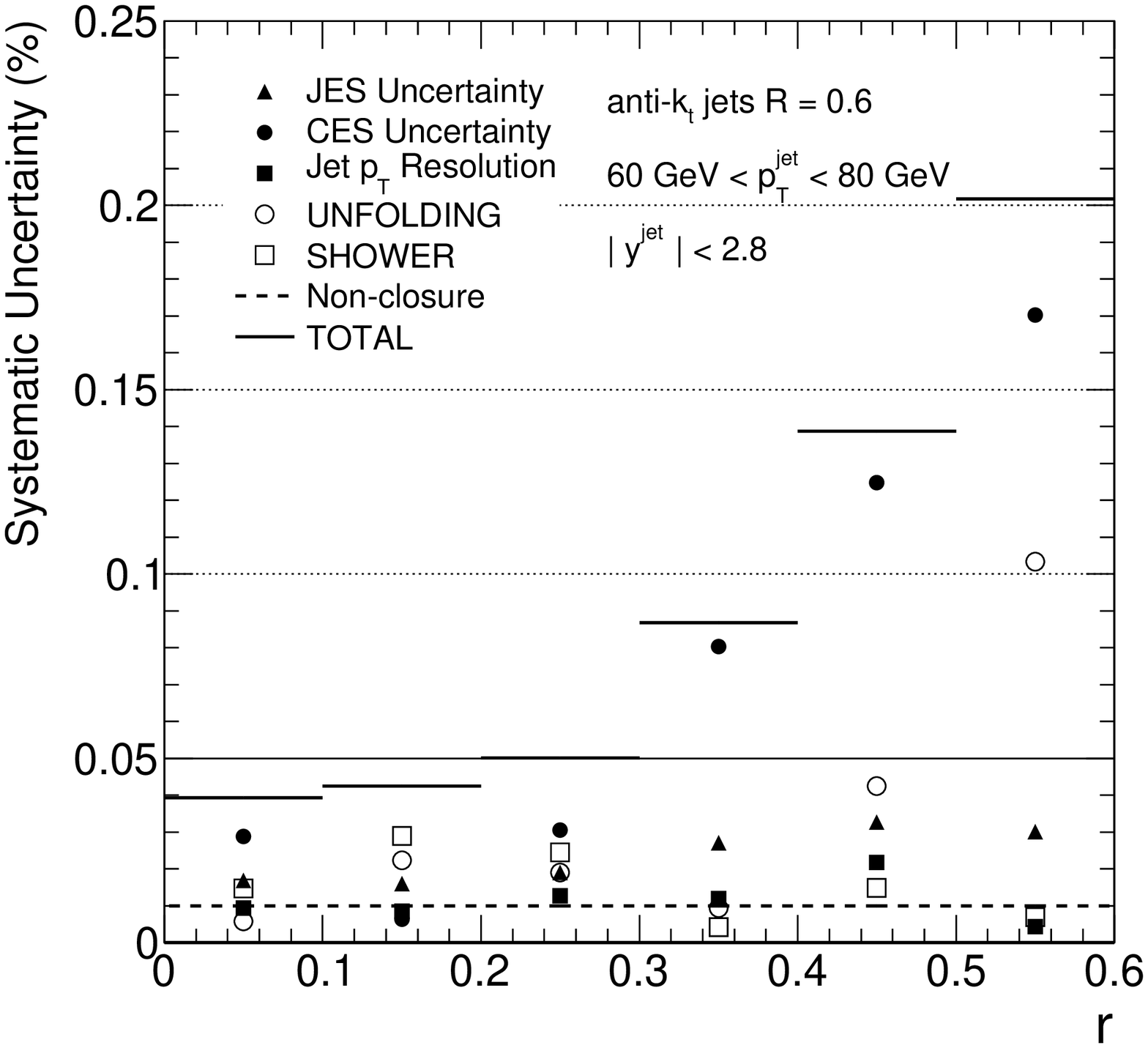}
\includegraphics[width=0.475\textwidth]{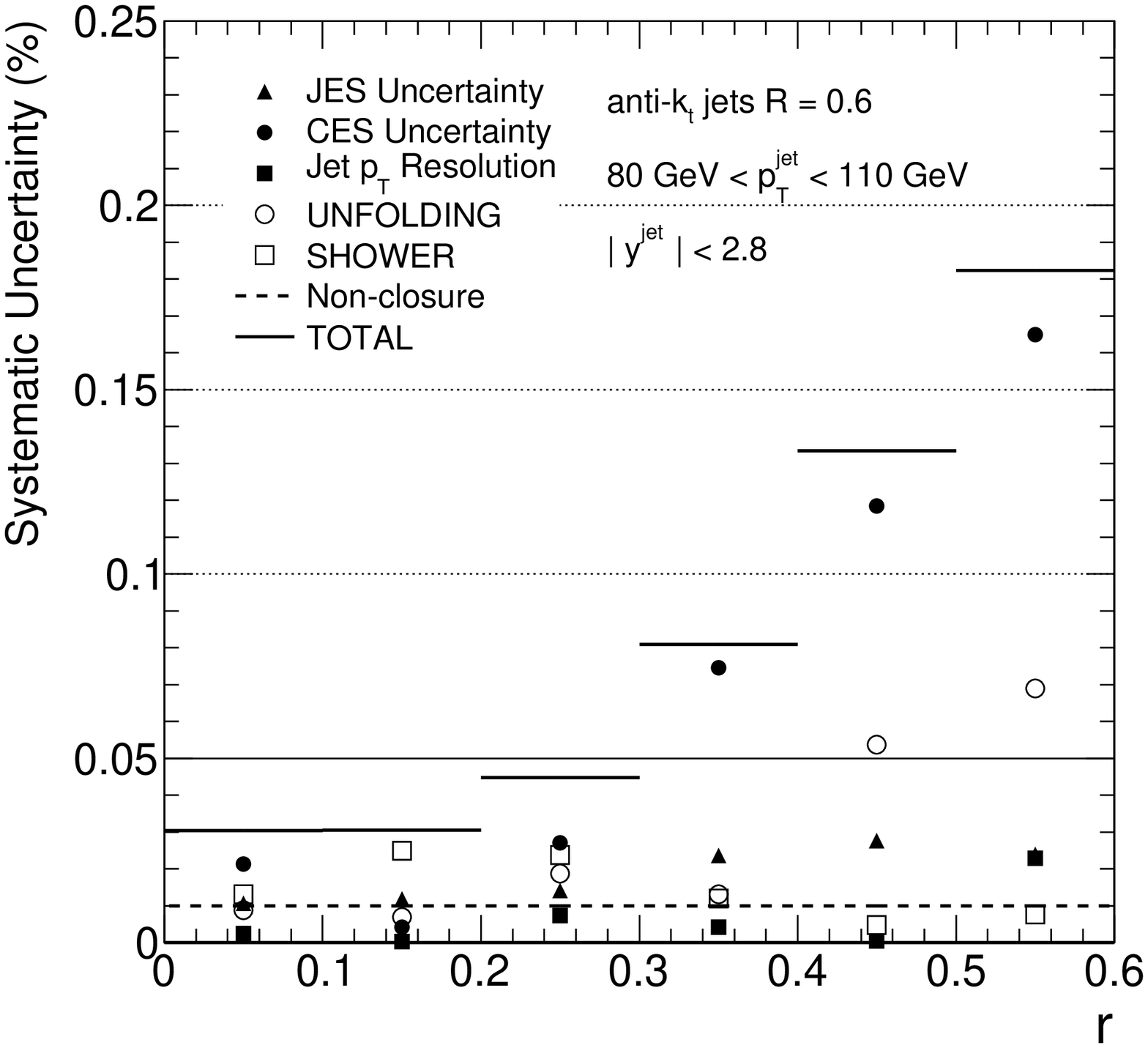}
}
\mbox{
\includegraphics[width=0.475\textwidth]{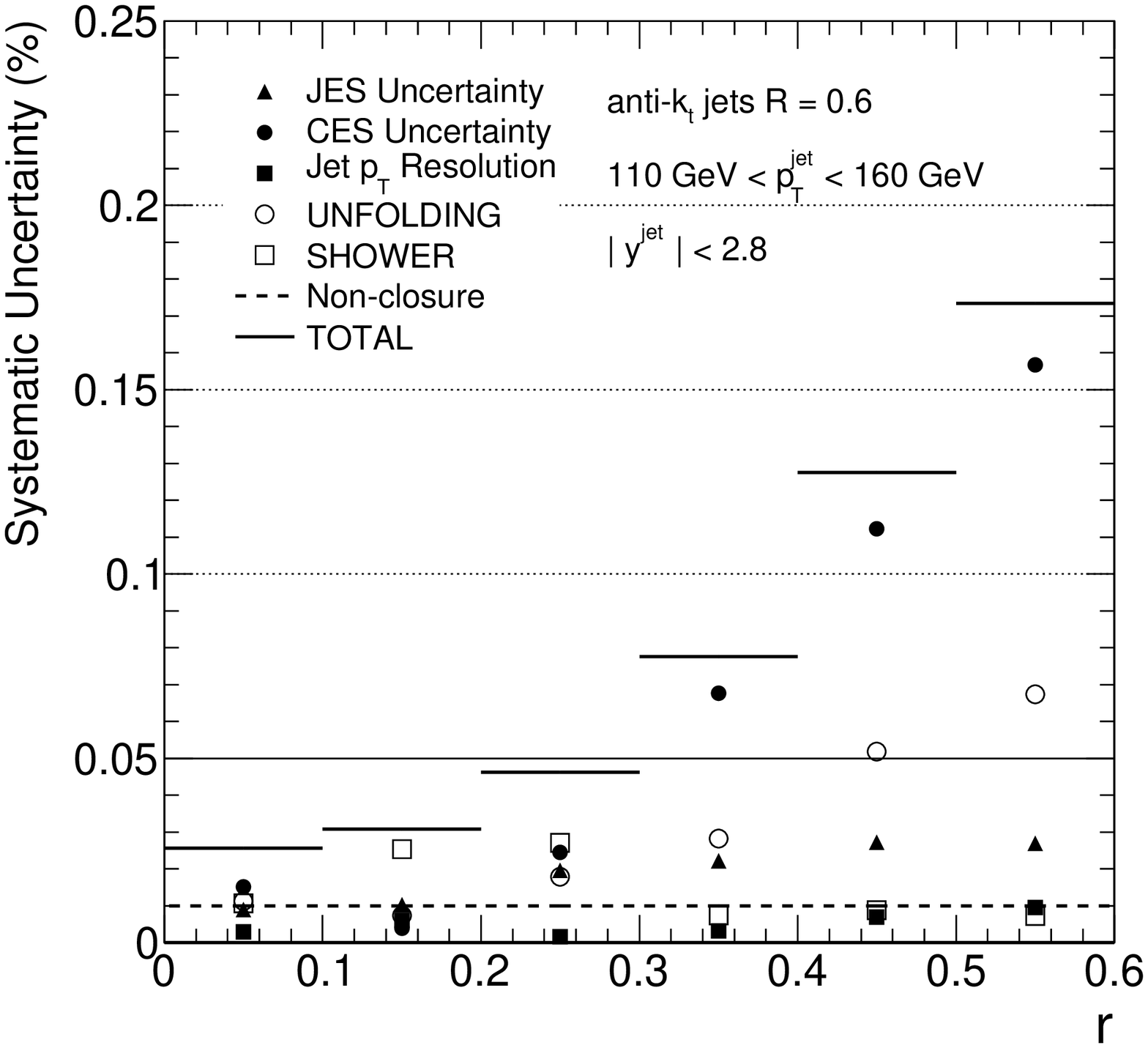}
\includegraphics[width=0.475\textwidth]{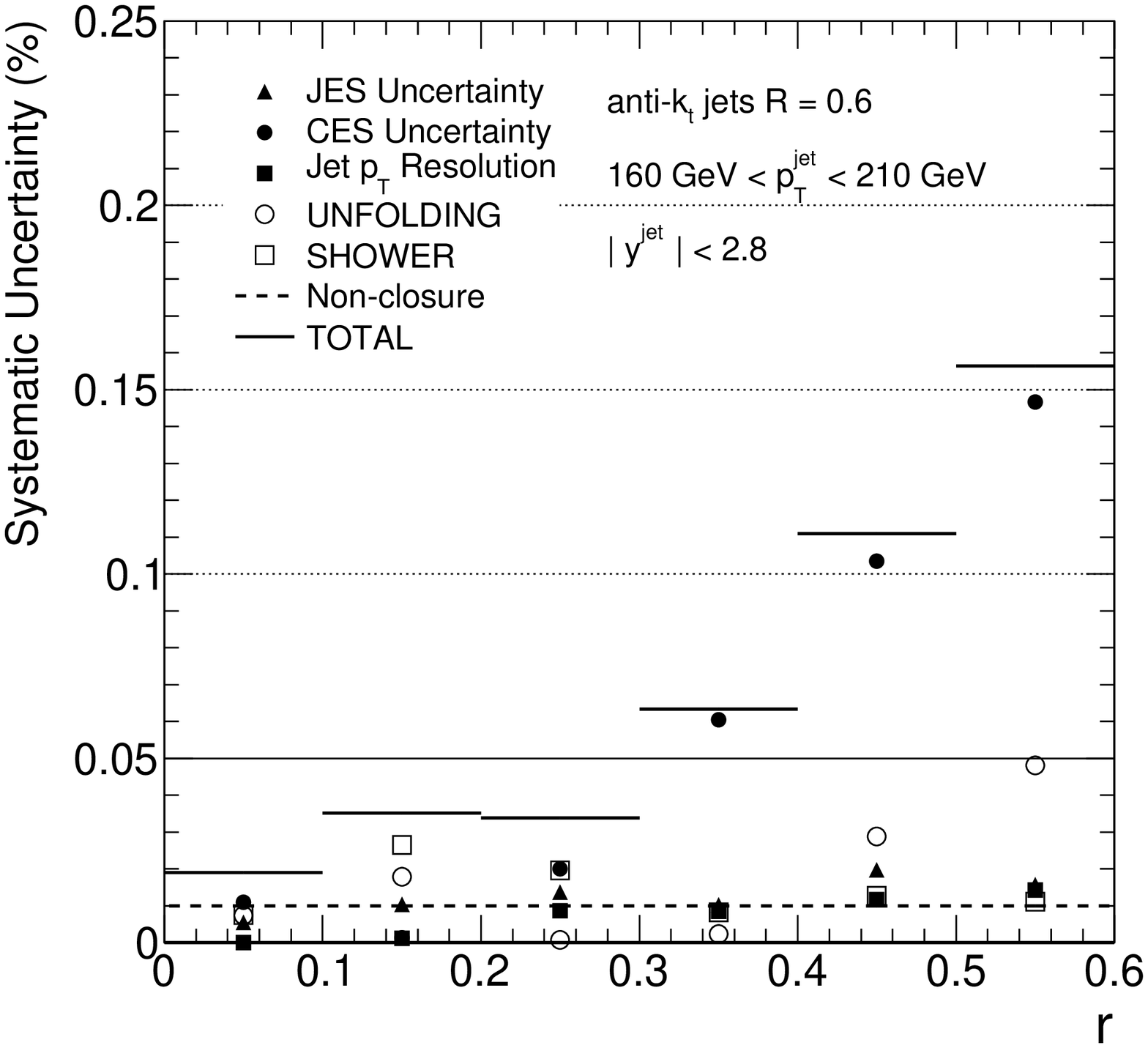}
}
\end{center}
\vspace{-0.7 cm}
\caption{\small
Summary of systematic uncertainties for the differential jet shape measurements for jets
with $|\rapjet| < 2.8$ and $30 \ {\rm GeV} < \ptjet < 210  \ {\rm GeV}$.
}
\label{fig_dif_sys1}
\end{figure}

\begin{figure}[tbh]
\begin{center}
\mbox{
\includegraphics[width=0.475\textwidth]{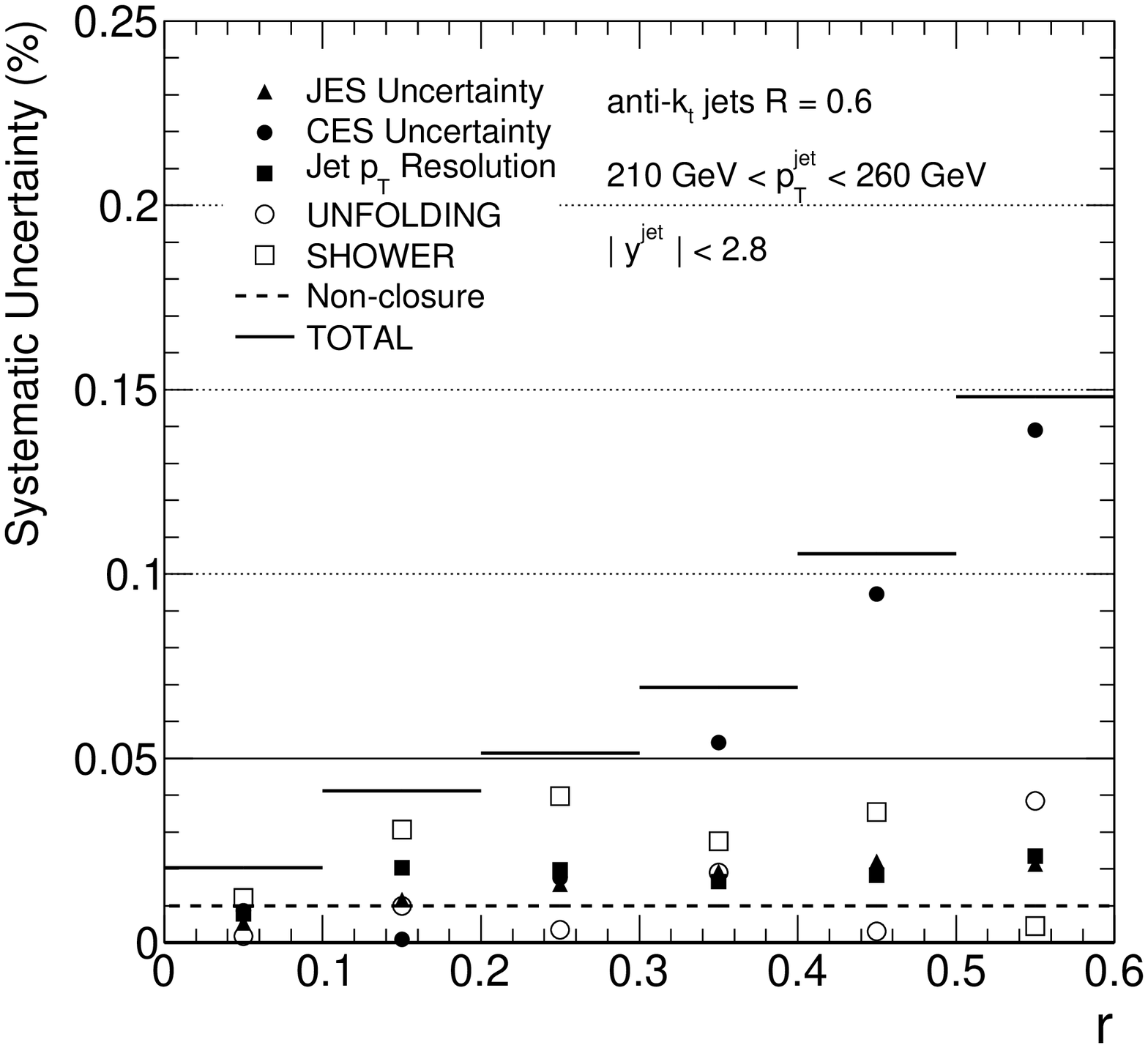}
\includegraphics[width=0.475\textwidth]{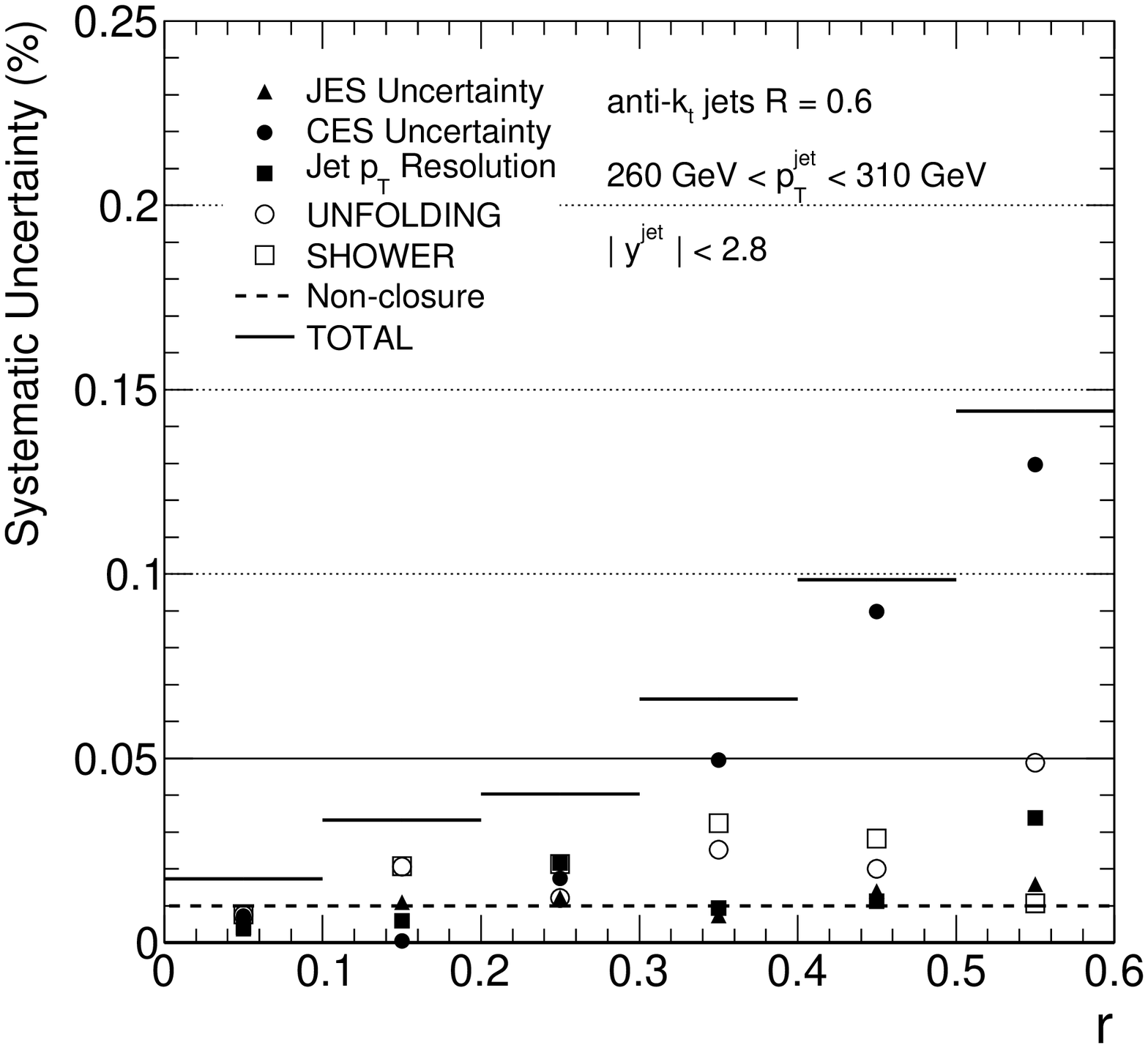}
}
\mbox{
\includegraphics[width=0.475\textwidth]{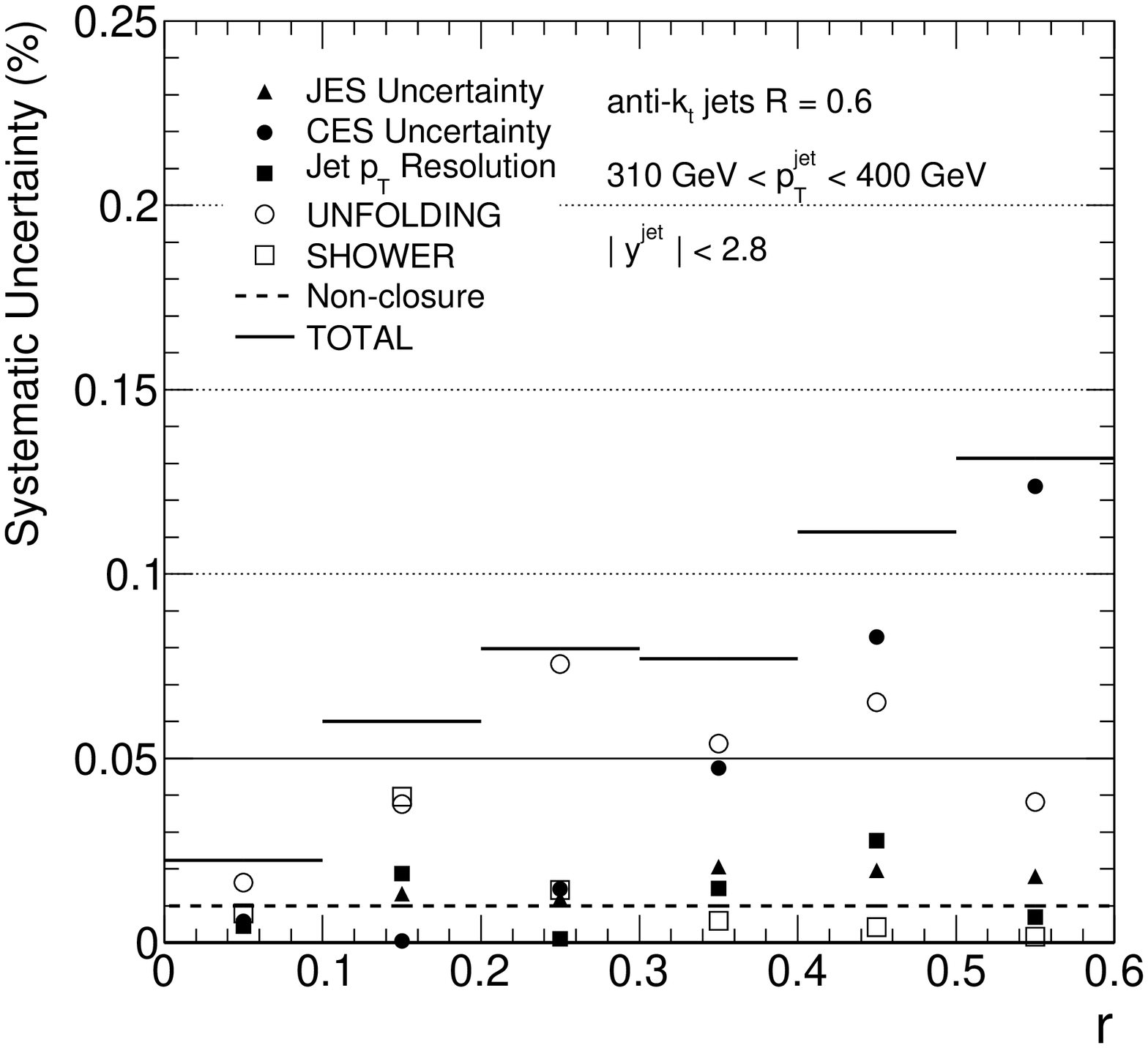}
\includegraphics[width=0.475\textwidth]{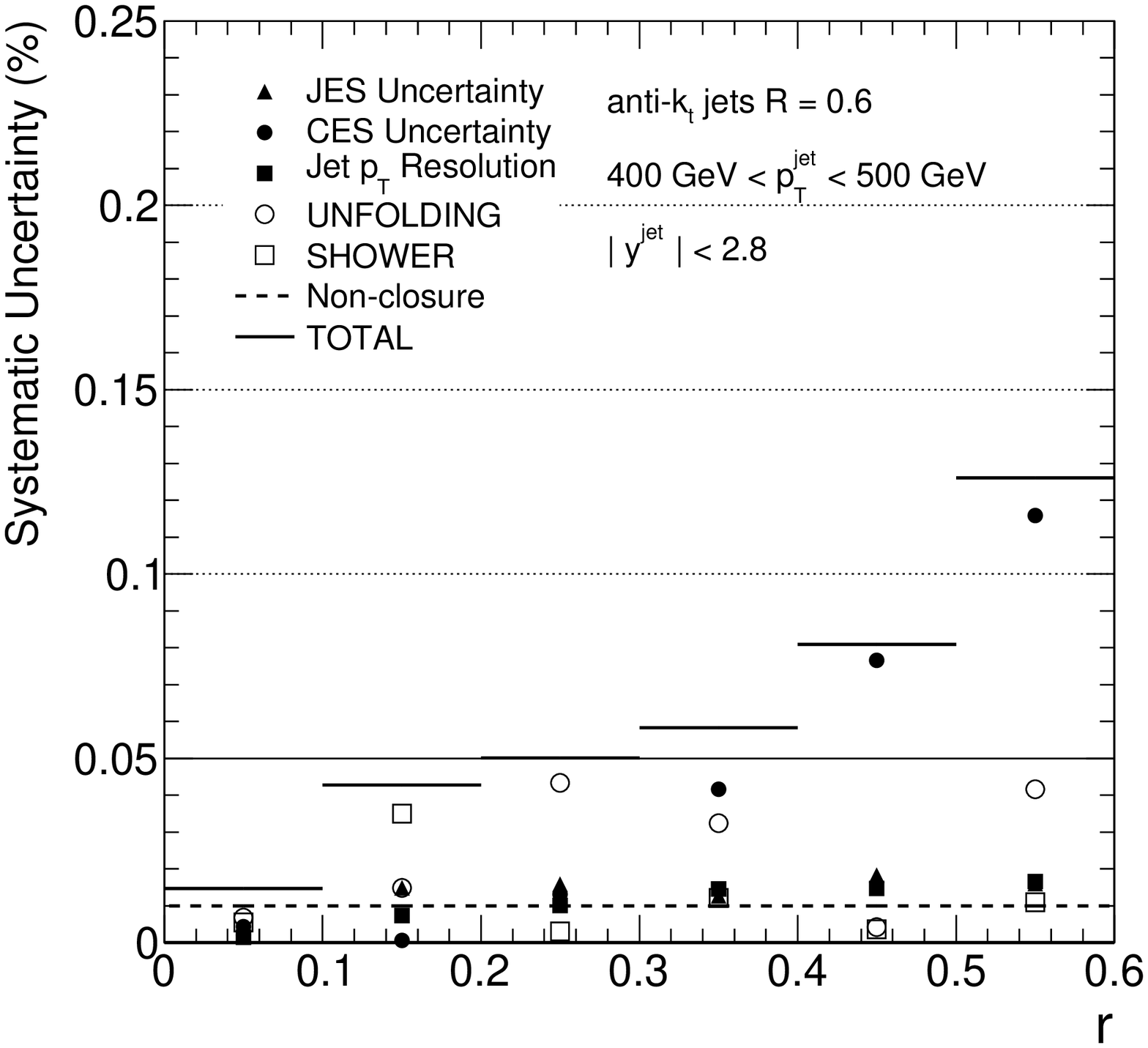}
}
\mbox{
\includegraphics[width=0.475\textwidth]{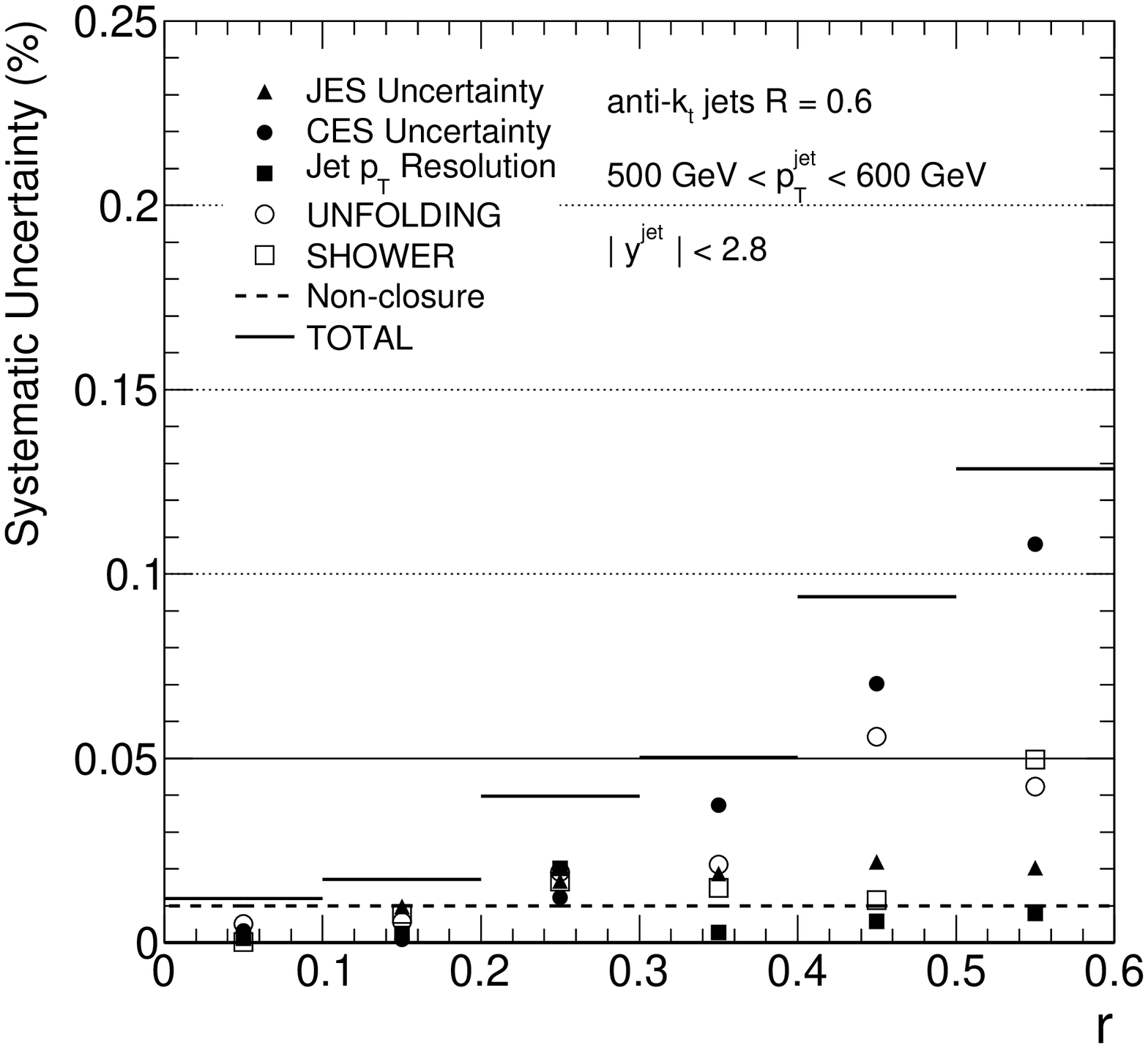}
}
\end{center}
\vspace{-0.7 cm}
\caption{\small
Summary of systematic uncertainties for the differential jet shape measurements for jets
with $|\rapjet| < 2.8$ and $210 \ {\rm GeV} < \ptjet < 600  \ {\rm GeV}$.
}
\label{fig_dif_sys2}
\end{figure}

\begin{figure}[tbh]
\begin{center}
\mbox{
\includegraphics[width=0.475\textwidth]{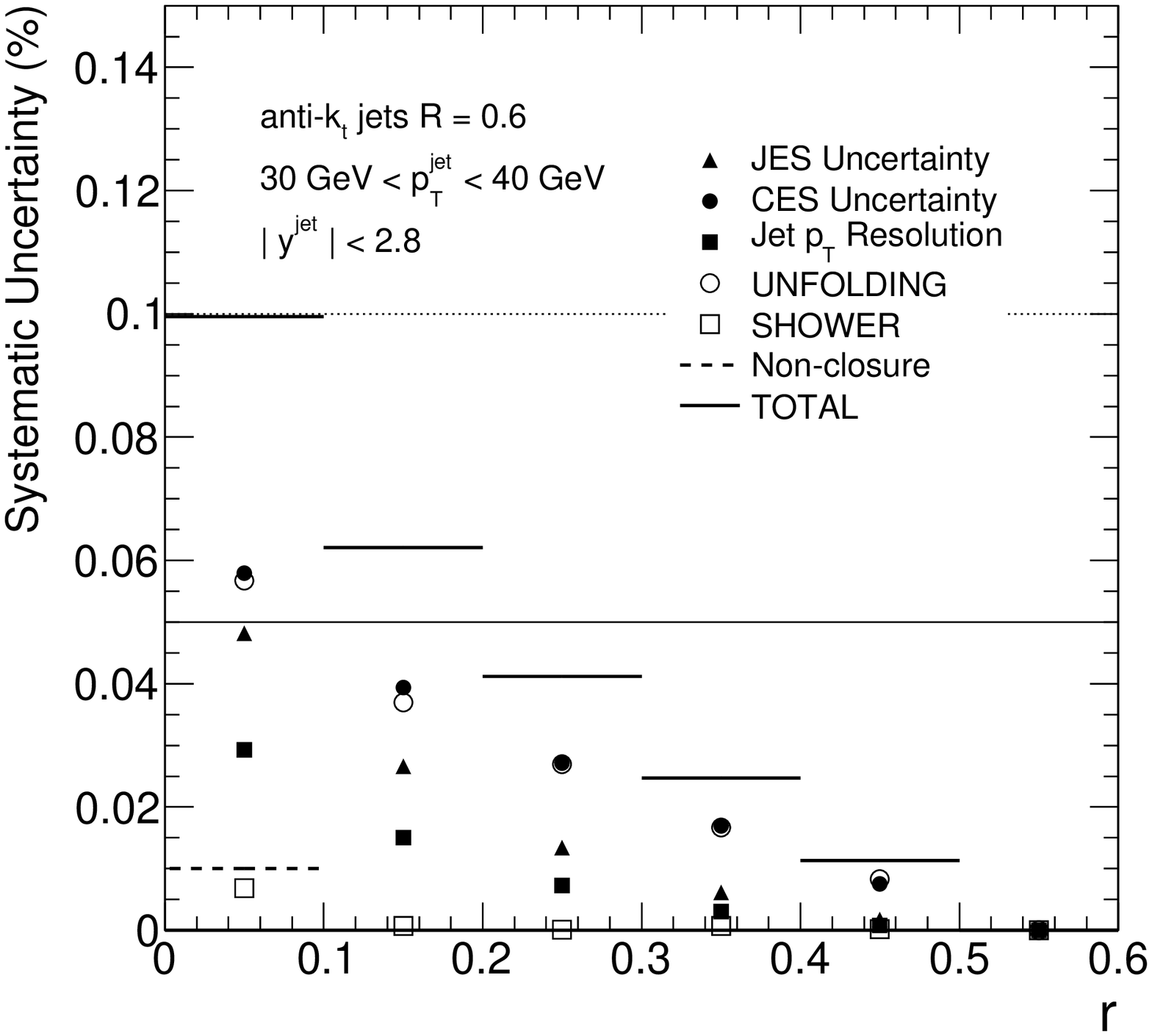}
\includegraphics[width=0.475\textwidth]{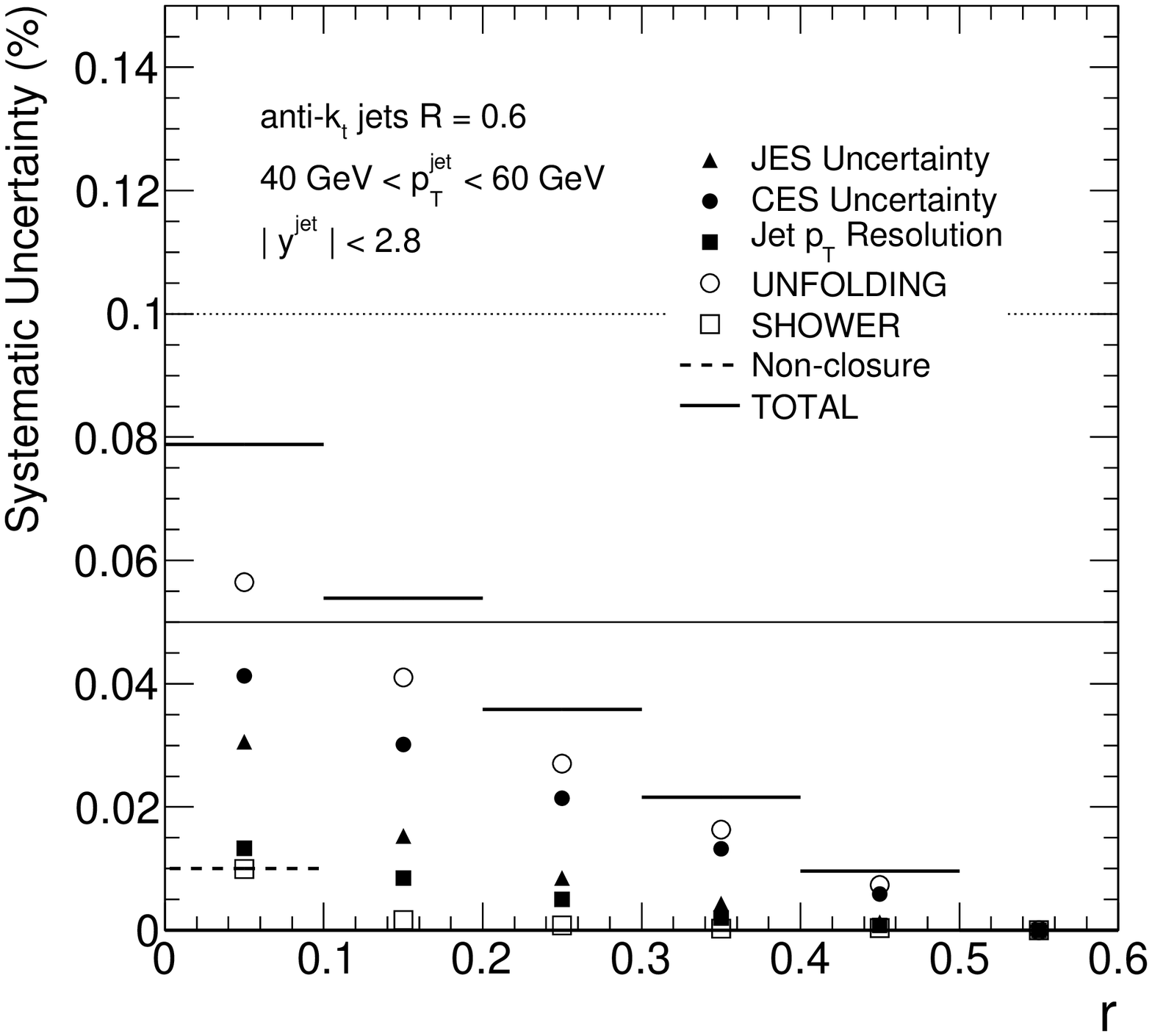}
}
\mbox{
\includegraphics[width=0.475\textwidth]{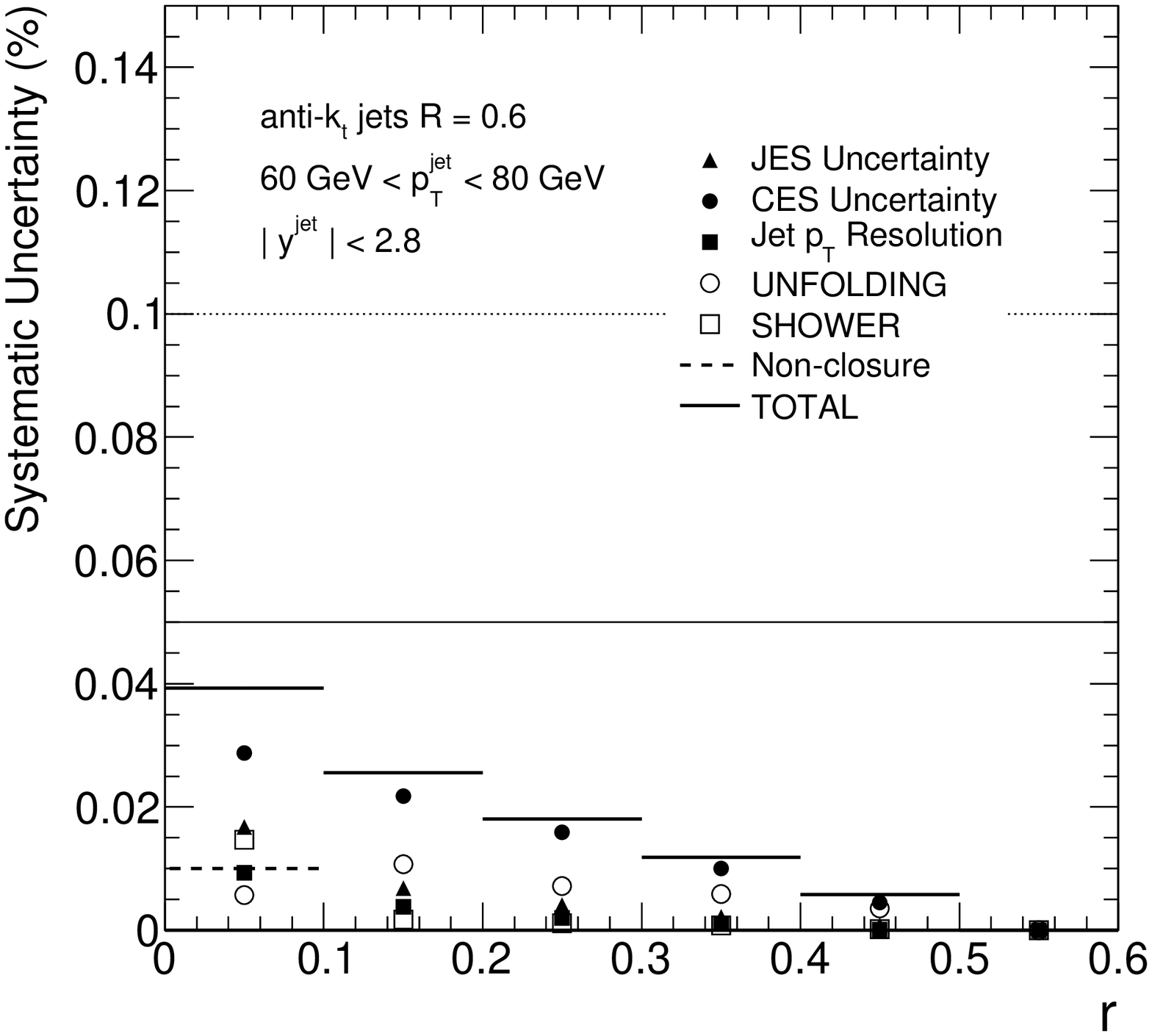}
\includegraphics[width=0.475\textwidth]{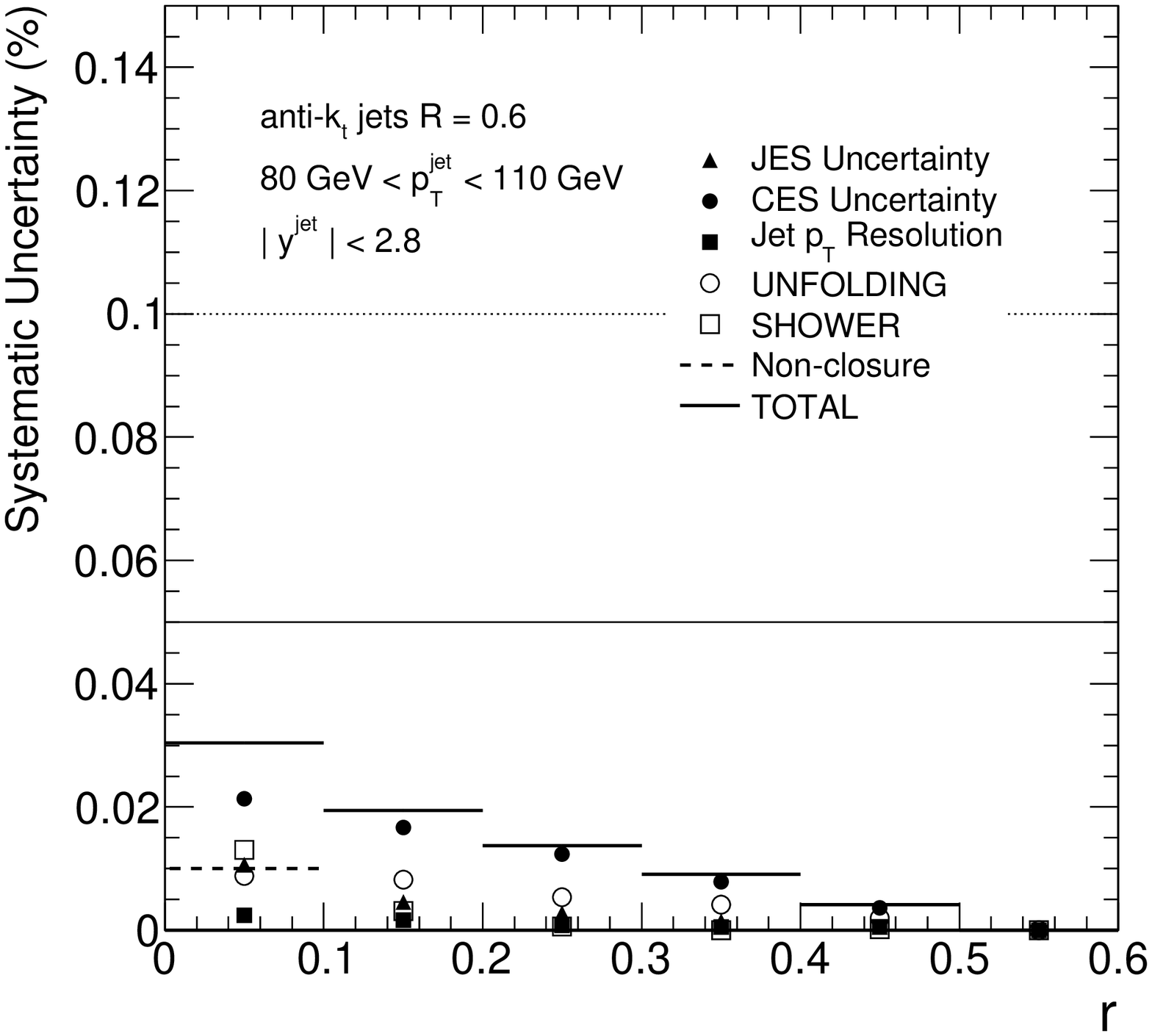}
}
\mbox{
\includegraphics[width=0.475\textwidth]{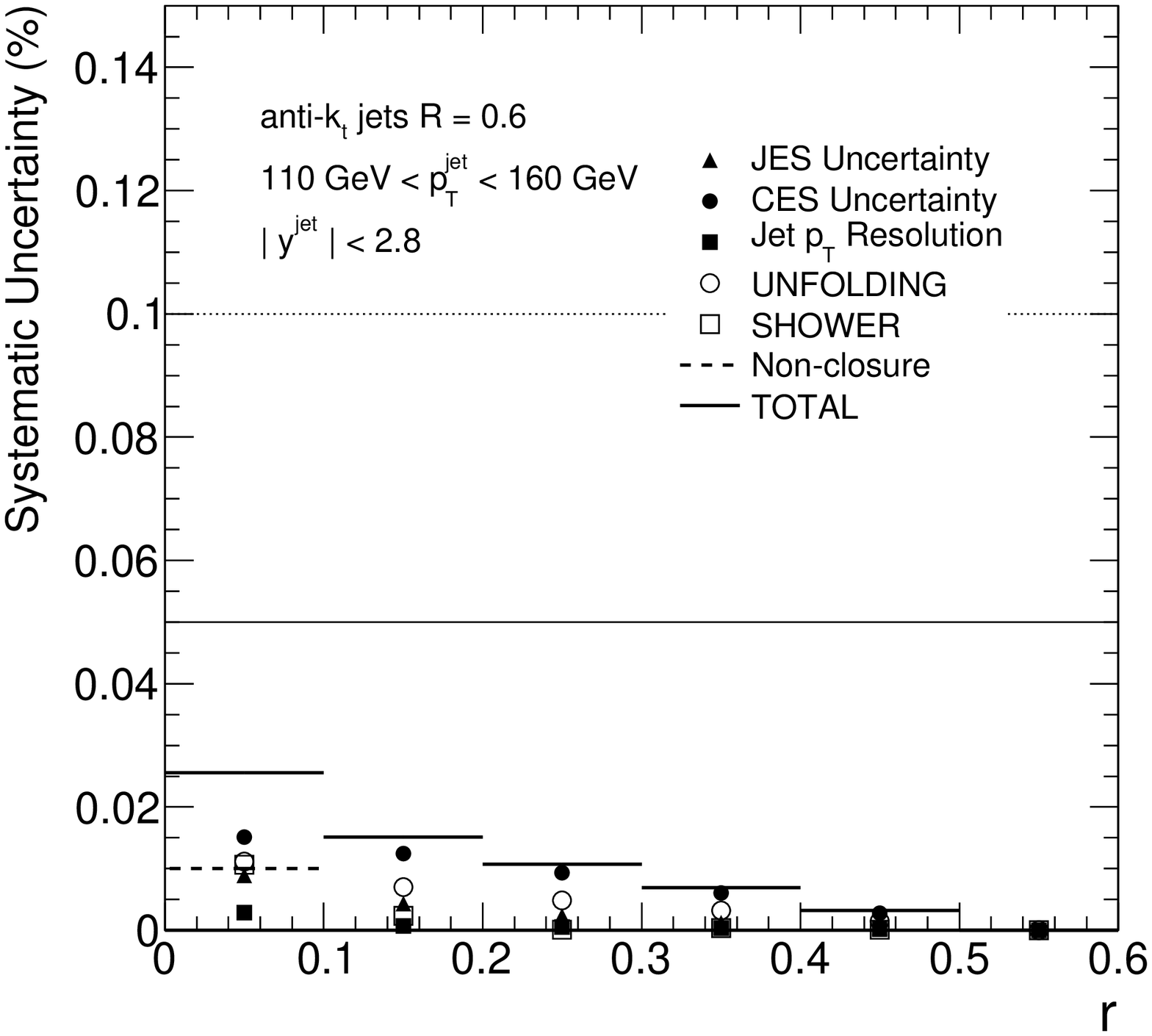}
\includegraphics[width=0.475\textwidth]{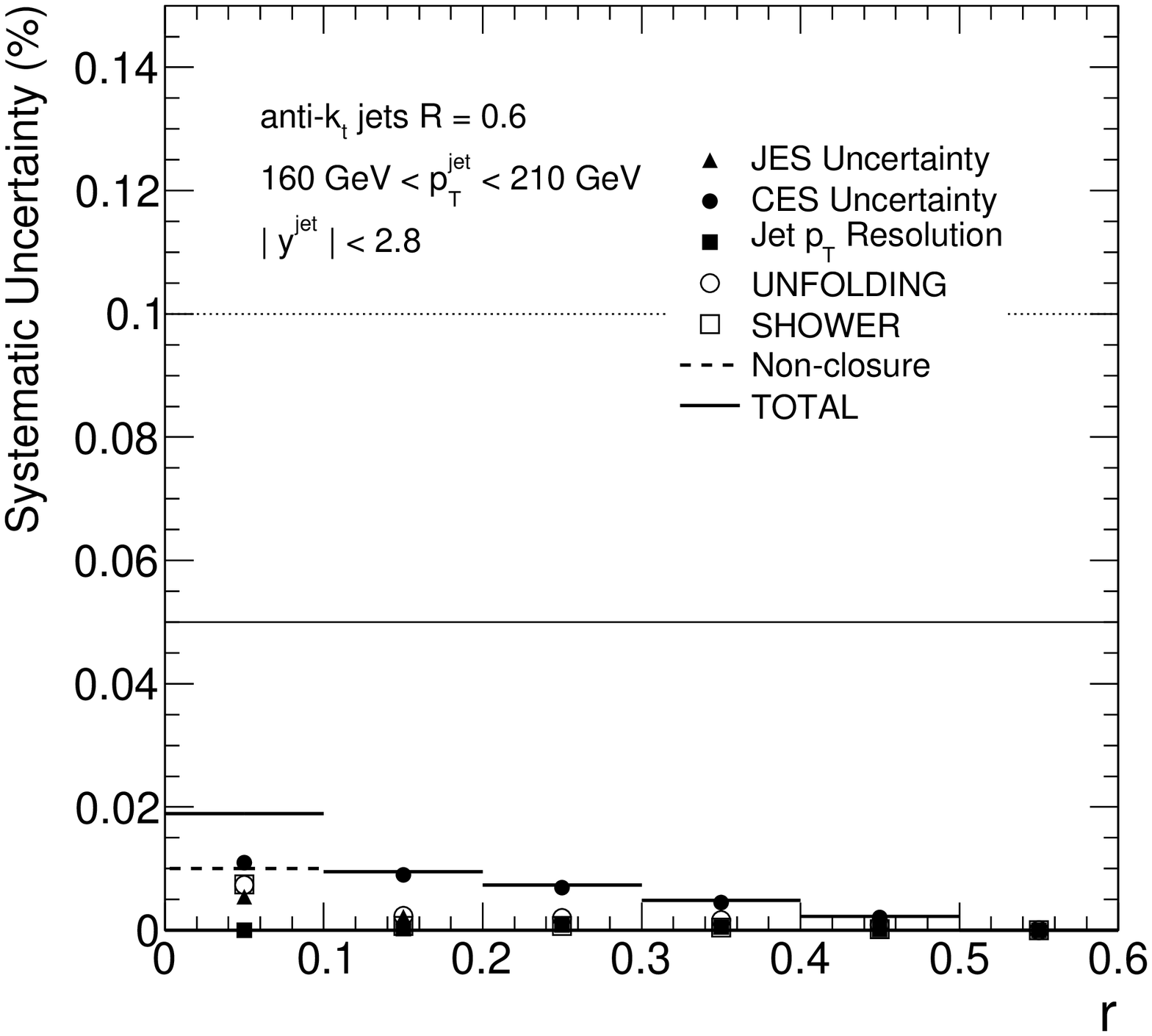}
}
\end{center}
\vspace{-0.7 cm}
\caption{\small
Summary of systematic uncertainties for the integrated jet shape measurements for jets
with $|\rapjet| < 2.8$ and $30 \ {\rm GeV} < \ptjet < 210  \ {\rm GeV}$.
}
\label{fig_int_sys1}
\end{figure}

\begin{figure}[tbh]
\begin{center}
\mbox{
\includegraphics[width=0.475\textwidth]{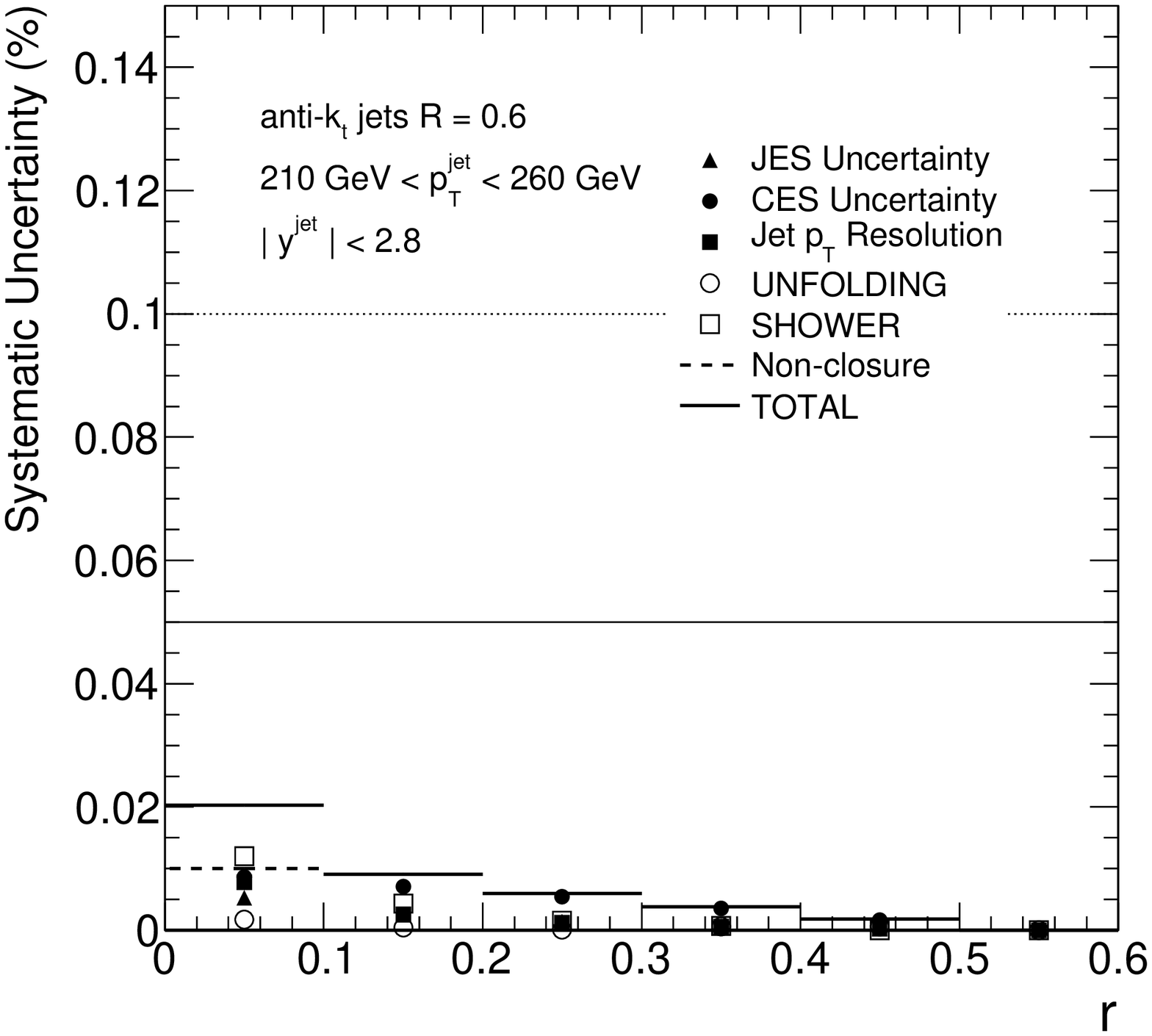}
\includegraphics[width=0.475\textwidth]{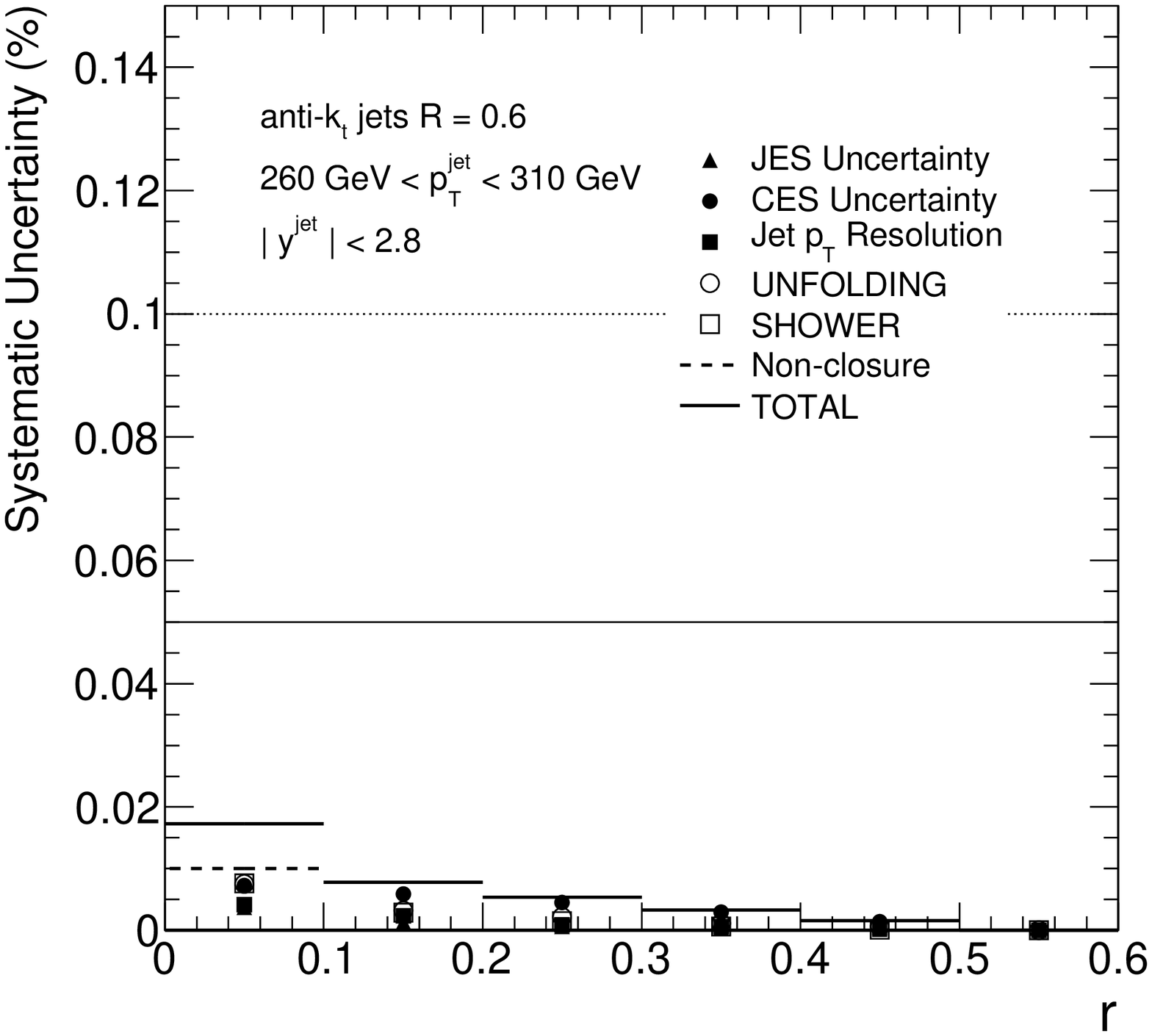}
}
\mbox{
\includegraphics[width=0.475\textwidth]{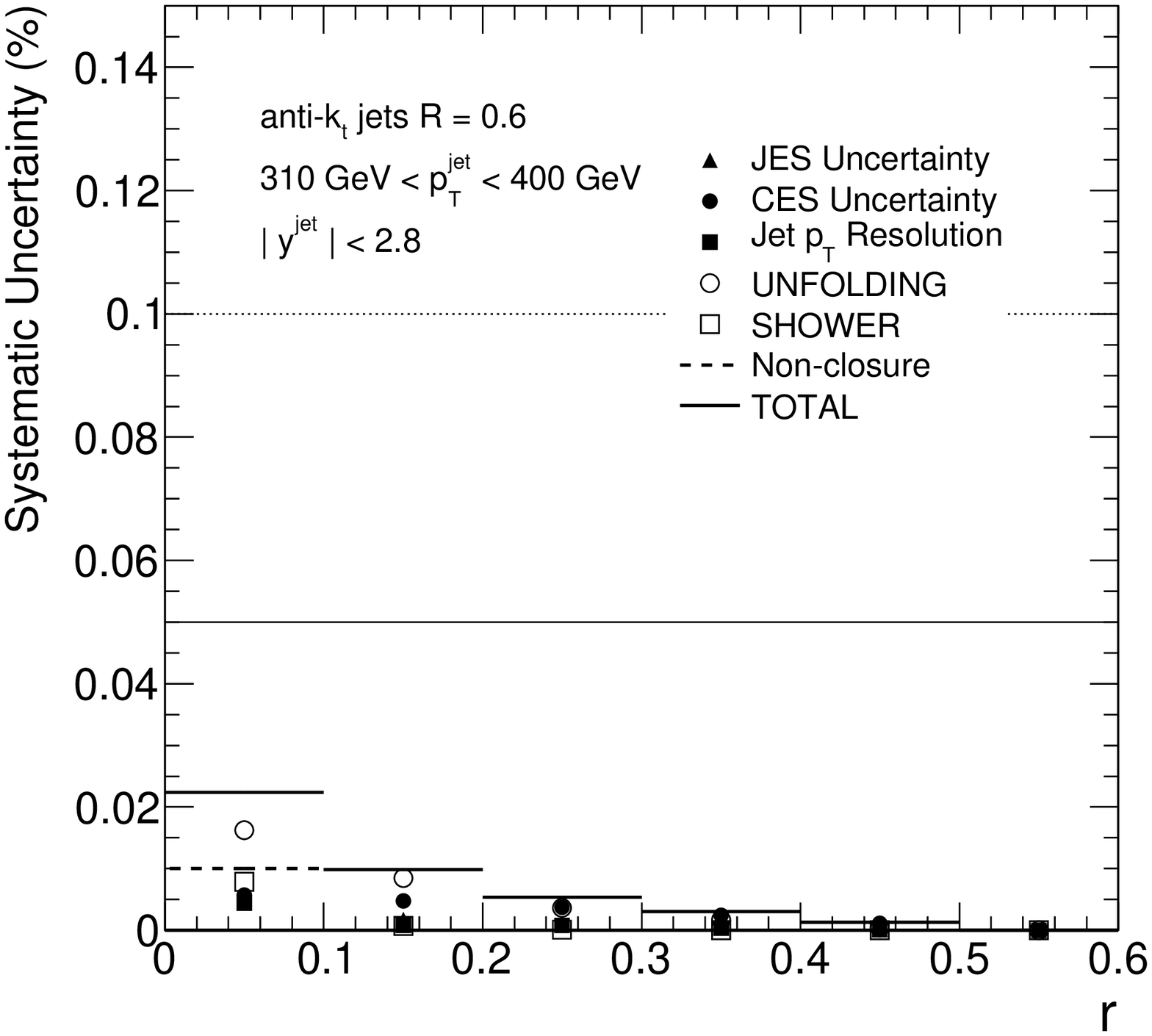}
\includegraphics[width=0.475\textwidth]{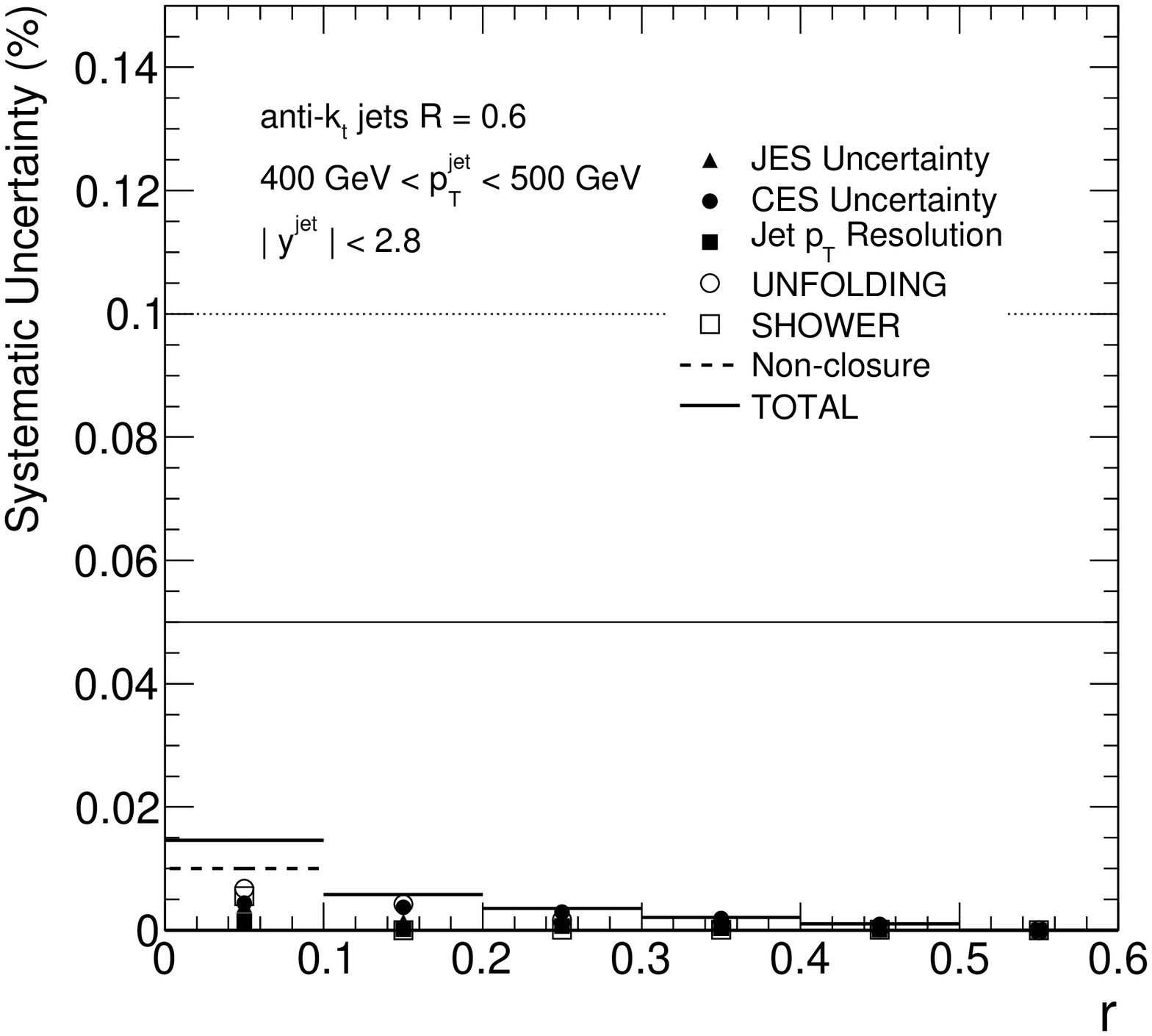}
}
\mbox{
\includegraphics[width=0.475\textwidth]{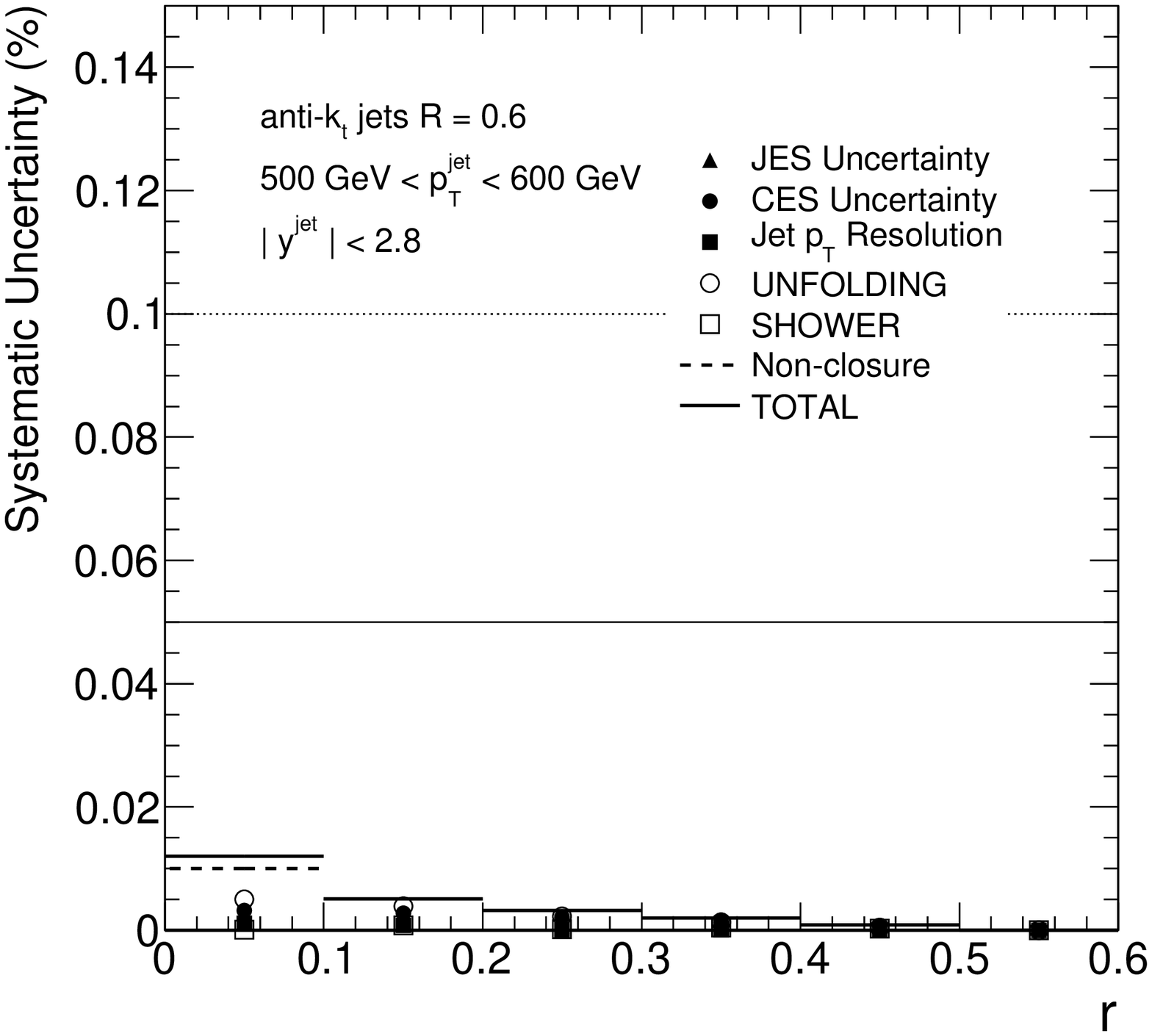}
}
\end{center}
\vspace{-0.7 cm}
\caption{\small
Summary of systematic uncertainties for the integrated jet shape measurements for jets
with $|\rapjet| < 2.8$ and $210 \ {\rm GeV} < \ptjet < 600  \ {\rm GeV}$.
}
\label{fig_int_sys2}
\end{figure}

\noindent
The different systematic uncertainties are added in
quadrature to the statistical uncertainty to obtain the final result.
The total systematic uncertainty for differential jet shapes decreases with
increasing $\ptjet$ and varies typically between 3$\%$ and 10$\%$ 
(10$\%$ and 20$\%$) at $r = 0.05$ ($r = 0.55$) (see Figures~\ref{fig_dif_sys1}~and~\ref{fig_dif_sys2}). 
The total uncertainty is dominated by
the systematic component, except at very large $\ptjet$ where the
measurements are still statistically limited.  In the case of the
integrated measurements, the total systematic uncertainty varies
between 10$\%$ and $2\%$ ($4\%$ and $1\%$) at $r=0.1$ ($r=0.3$) as
$\ptjet$ increases, and vanishes as $r$ approaches
the edge of the jet cone (see Figures~\ref{fig_int_sys1}~and~\ref{fig_int_sys2}). 

\clearpage

\clearpage

\clearpage

\clearpage

\clearpage

\clearpage

\clearpage

\clearpage

\clearpage

\clearpage

\clearpage

\clearpage

\subsection{Cross-checks using other detector objects}

The jet shape analysis is also performed using tracks inside the cone of jets reconstructed using 
calorimeter clusters. The inner detector tracks are required to pass selection criteria 
derived in the analysis of charged particle multiplicities in ATLAS~\cite{trk_mbts}:

\begin{itemize}
\item $\ptjet > 100$~MeV and $|\eta| < 2.5$
\item a hit in the first layer (layer-0) of the Pixel detector if one is expected 
(meaning that the layer-0 module is not out of the configuration and the track is not close to the module edge)
\item a minimum of one Pixel hit in any of the 3 layers of this detector
\item at least two (if $\ptjet > 100$~MeV), four(if $\ptjet > 200$~MeV) or six (if $\ptjet > 300$~MeV) SCT hits
\item transverse and longitudinal impact parameters calculated with respect to the event primary vertex 
$|d0_{PV}| < 1.5$~mm and $|z0_{PV}\times sin(\theta_{PV})| < 1.5$~mm respectively
\item $\chi^{2}$ probability $> 0.01$ for reconstructed tracks with $\ptjet > 10$~GeV, to remove high $\ptjet$ miss-measured tracks
\end{itemize}

\begin{figure}[tbh]
\begin{center}
\mbox{
\includegraphics[width=0.495\textwidth]{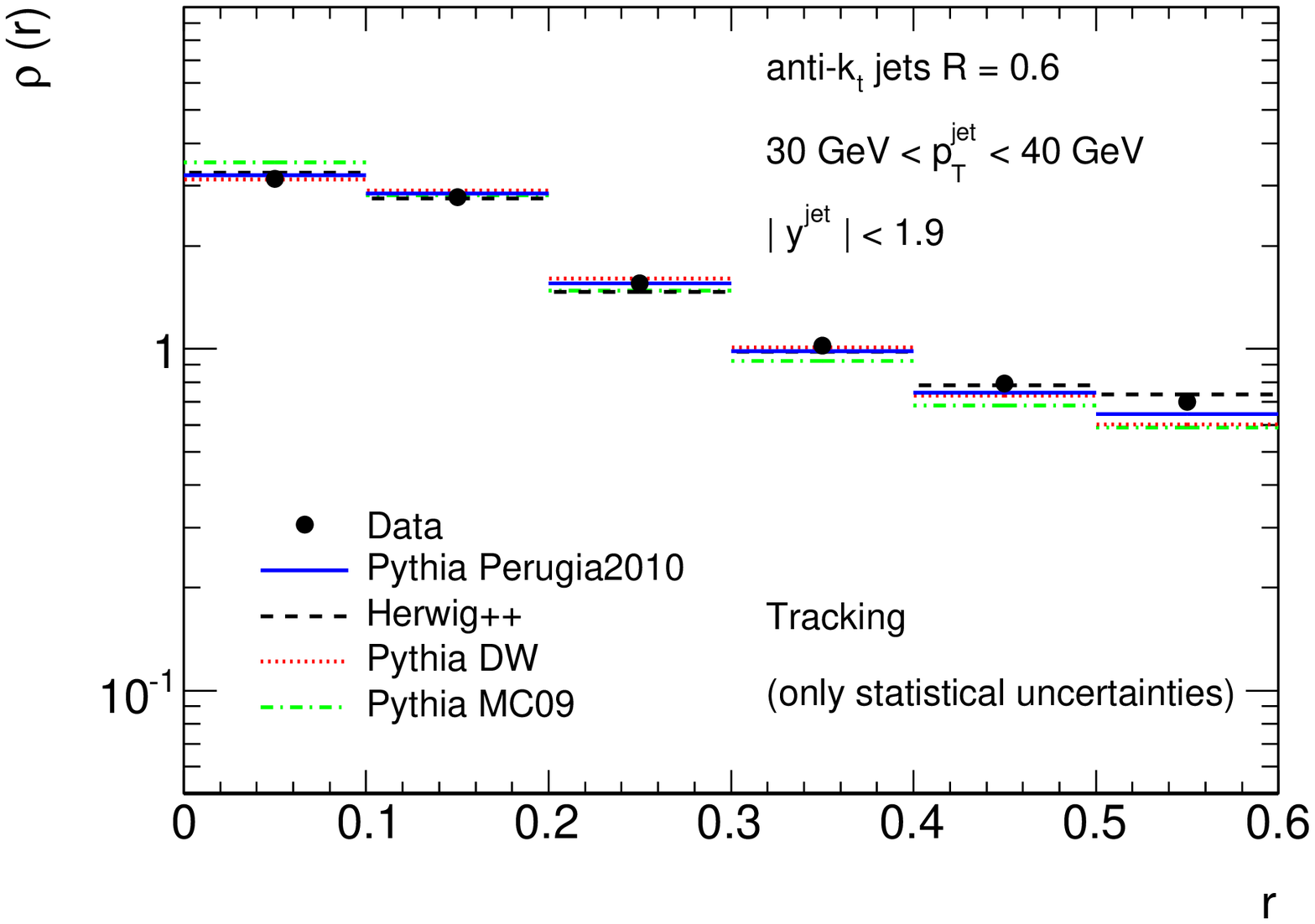}
\includegraphics[width=0.495\textwidth]{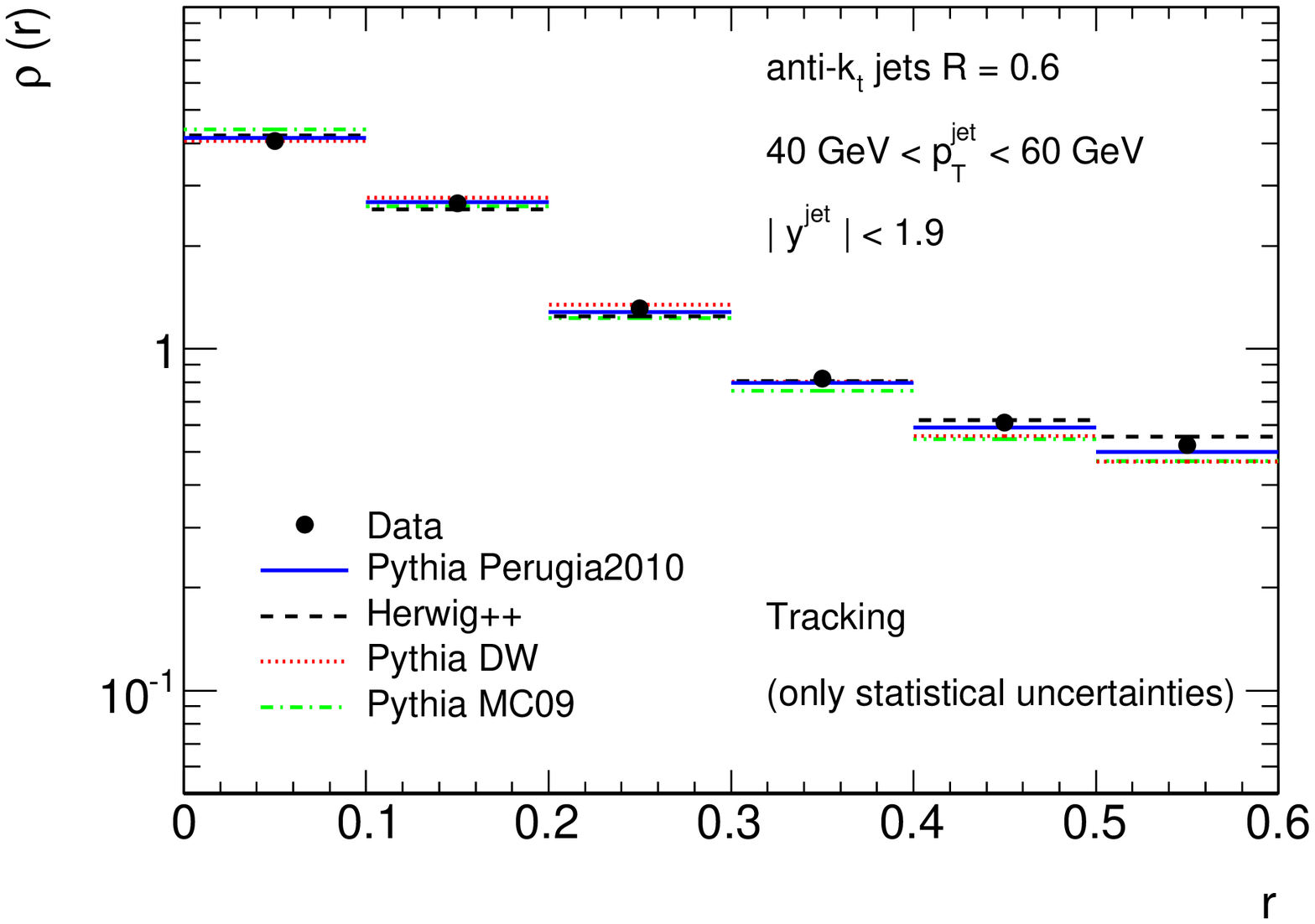}
}
\mbox{
\includegraphics[width=0.495\textwidth]{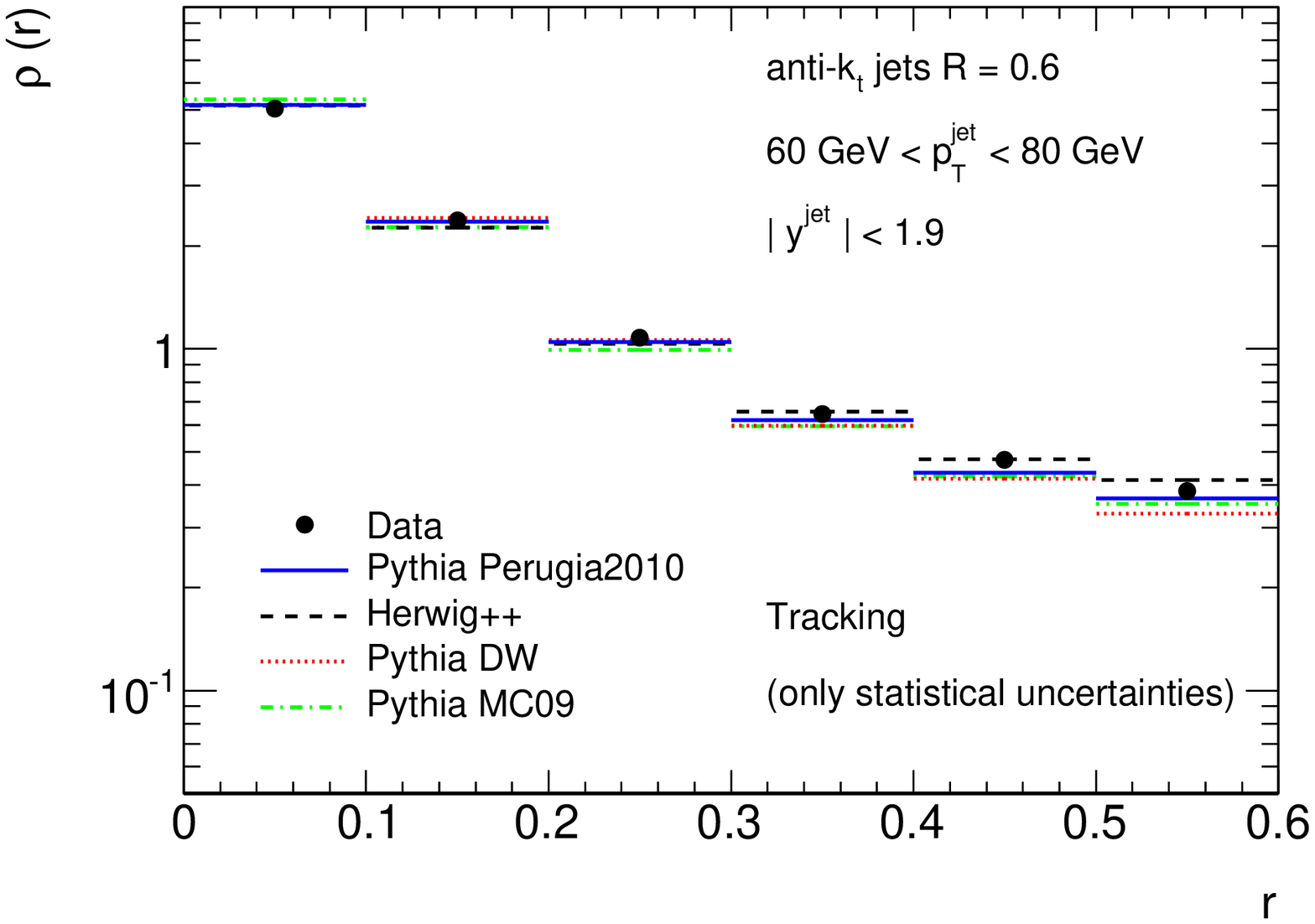}
\includegraphics[width=0.495\textwidth]{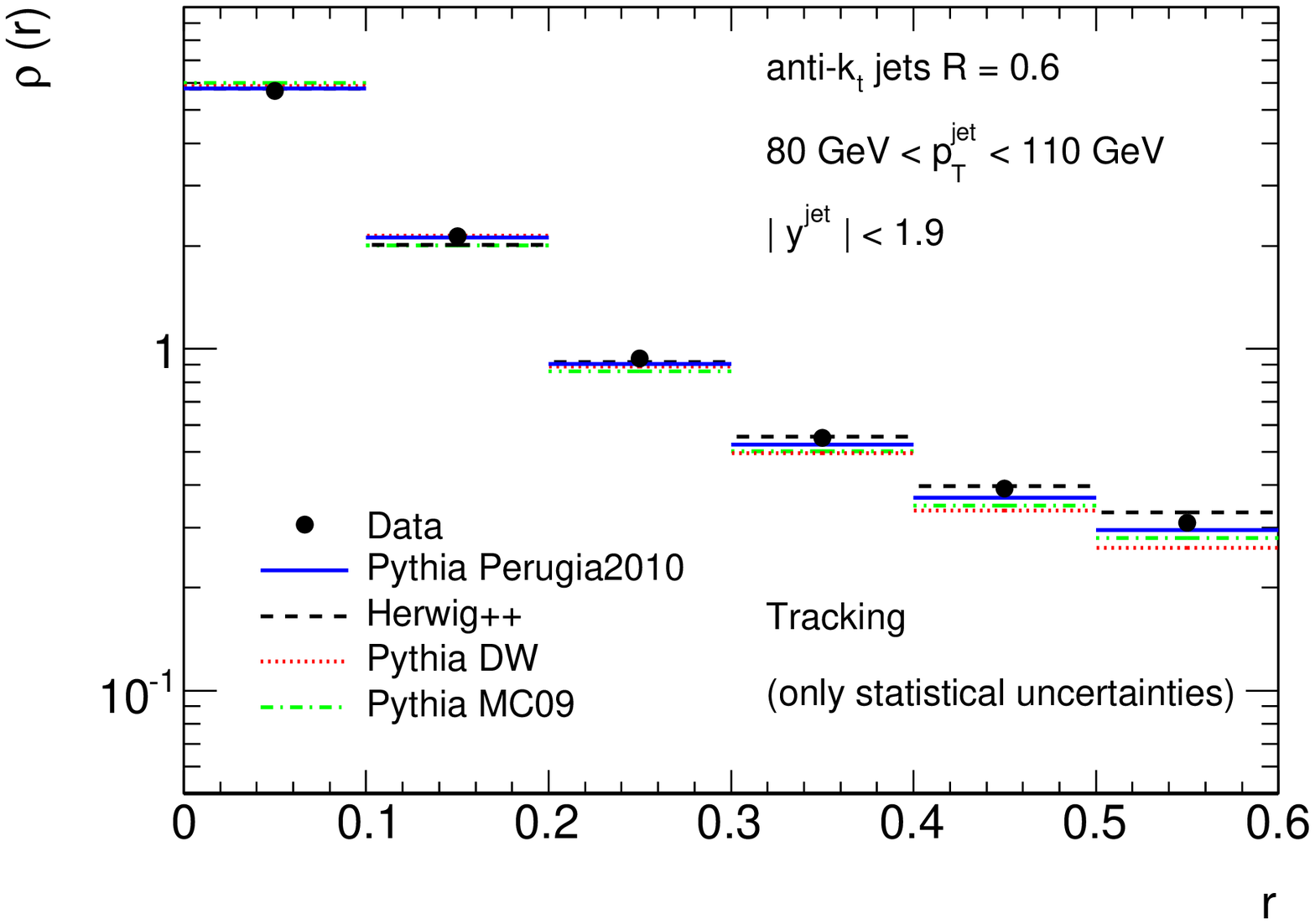}
}
\mbox{
\includegraphics[width=0.495\textwidth]{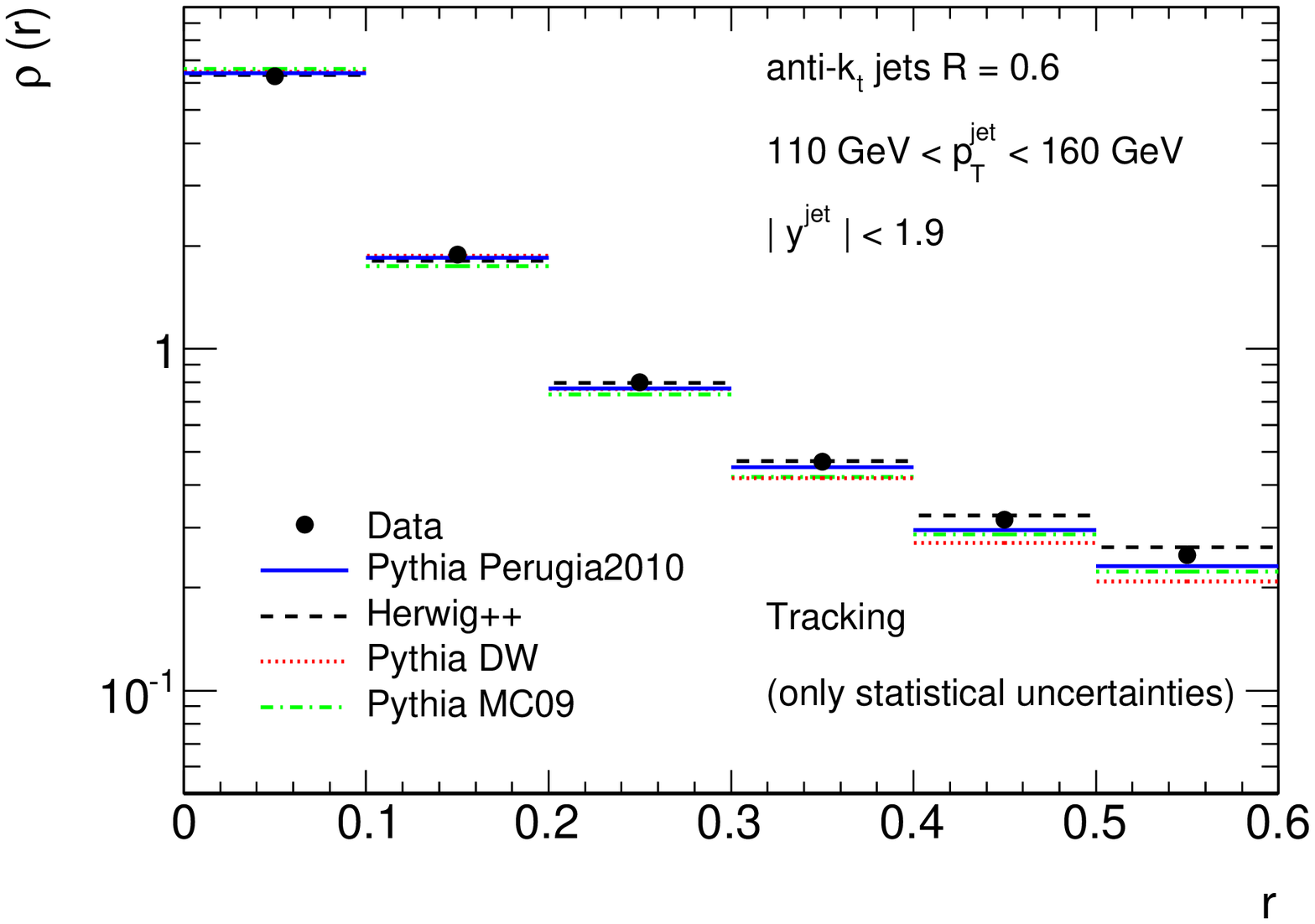}
\includegraphics[width=0.495\textwidth]{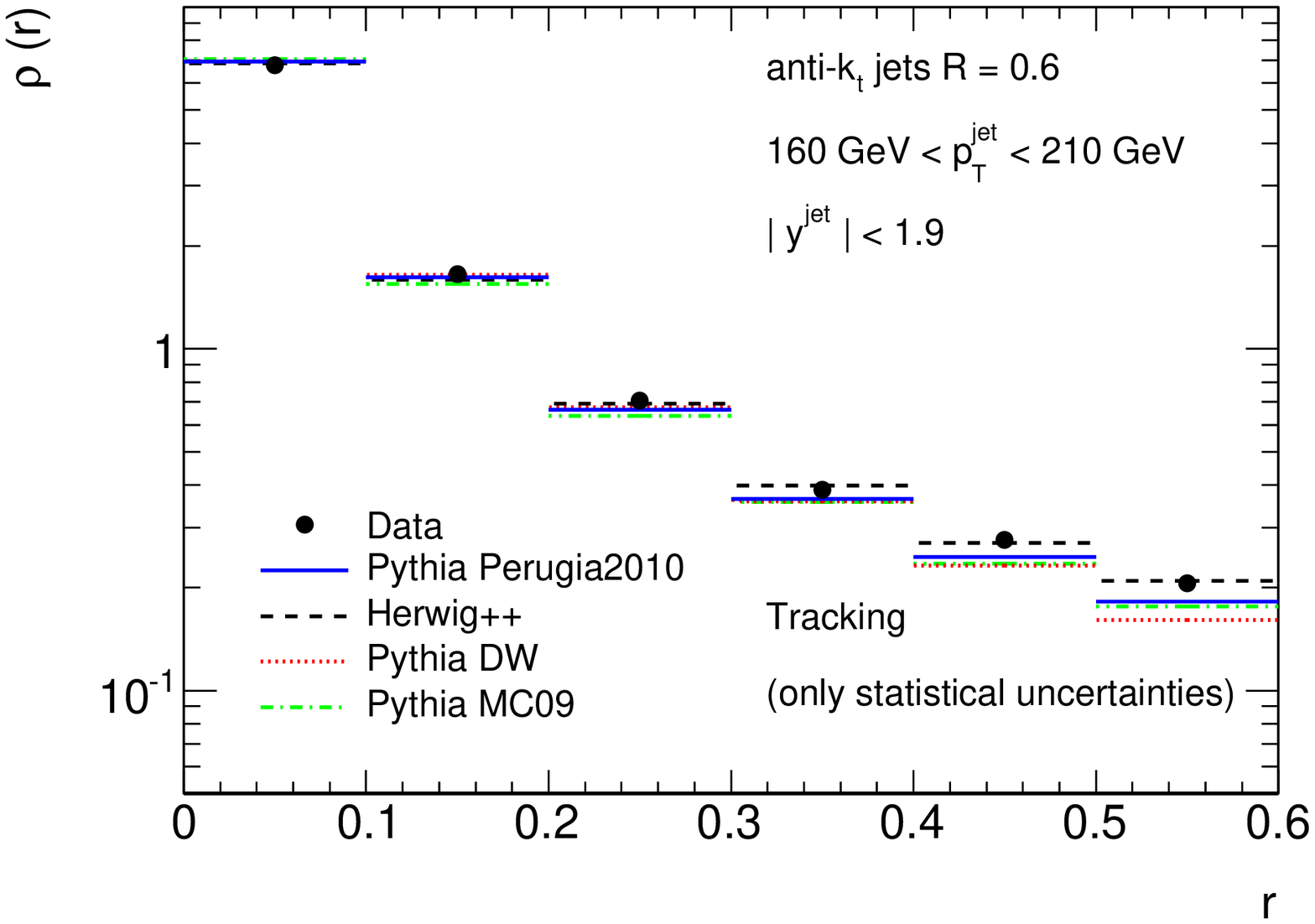}
}
\end{center}
\vspace{-0.7 cm}
\caption{\small
Measured differential jet shapes using tracks inside jets for jets 
with $|\rapjet| < 1.9$ and $30 \ {\rm GeV} < \ptjet < 210  \ {\rm GeV}$.
}
\label{fig_trk}
\end{figure}

\begin{figure}[tbh]
\begin{center}
\mbox{
\includegraphics[width=0.495\textwidth]{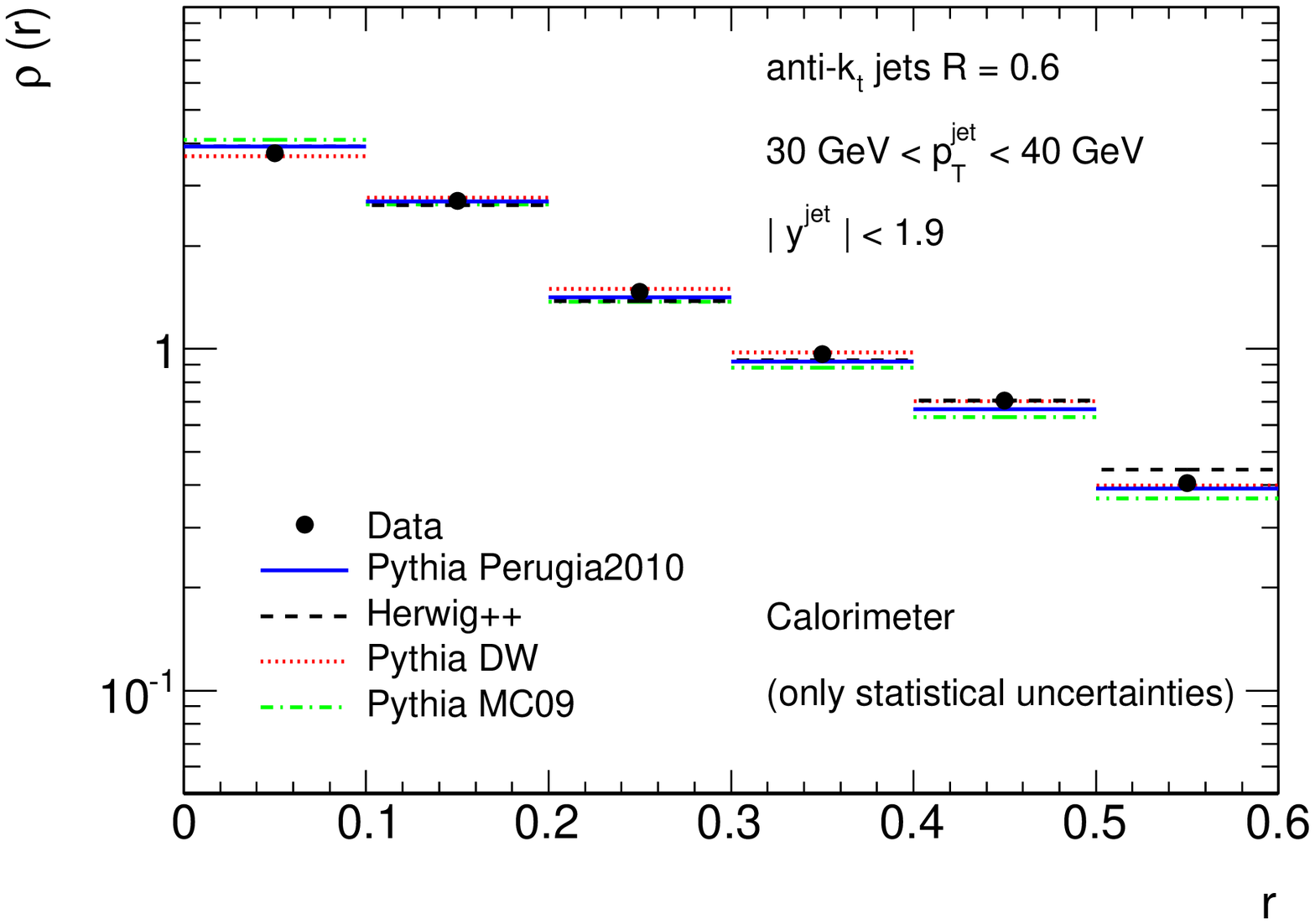}
\includegraphics[width=0.495\textwidth]{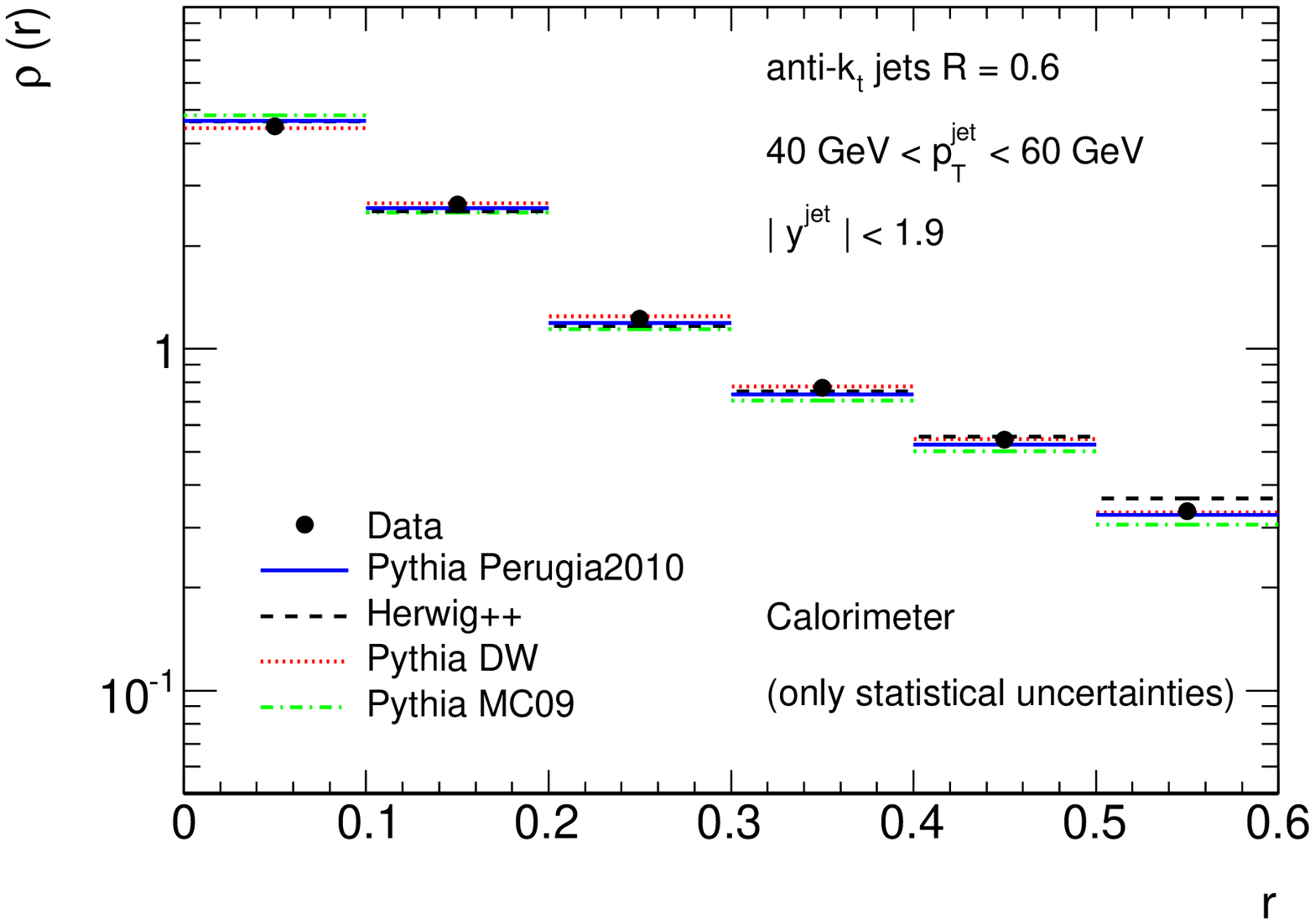}
}
\mbox{
\includegraphics[width=0.495\textwidth]{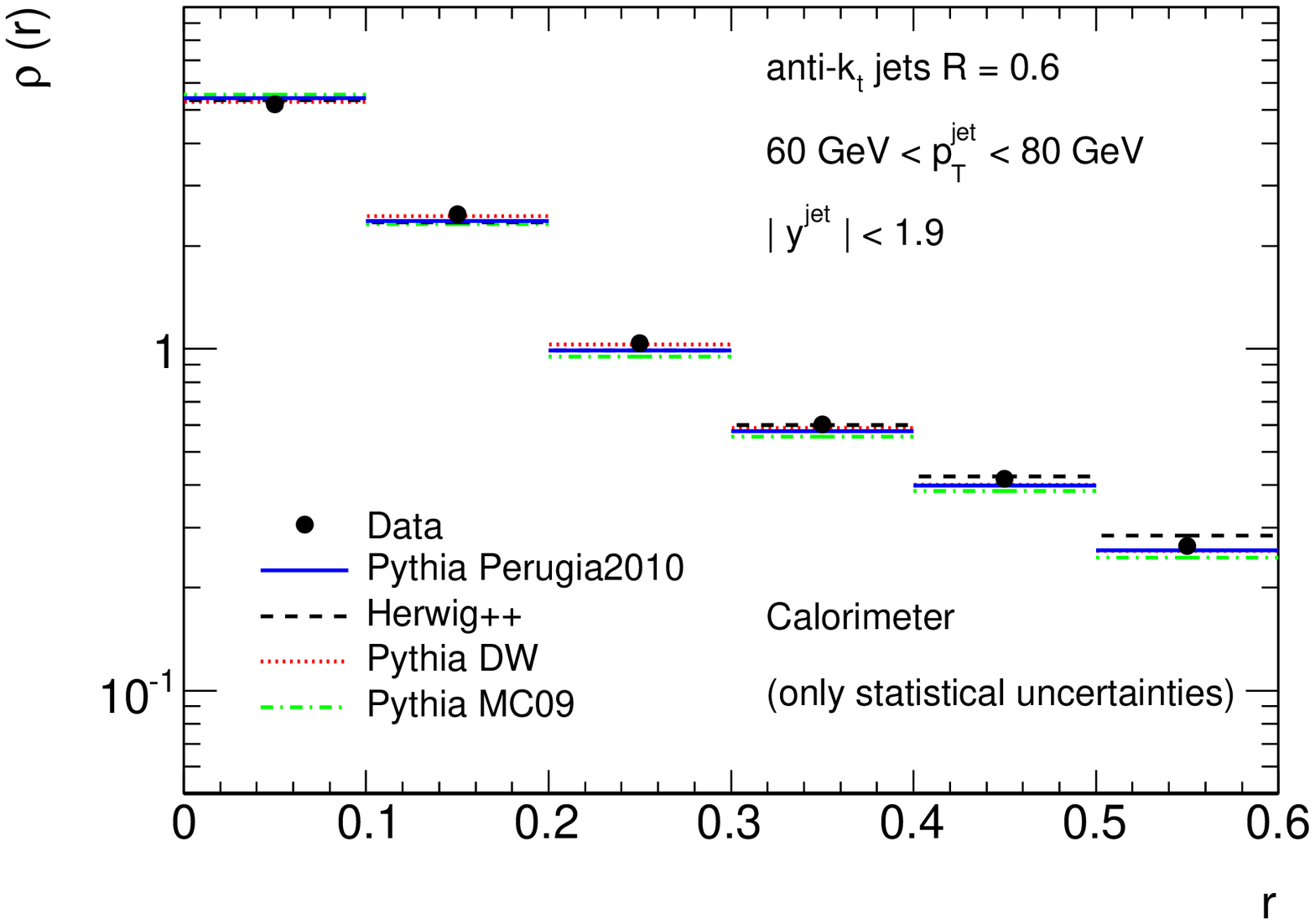}
\includegraphics[width=0.495\textwidth]{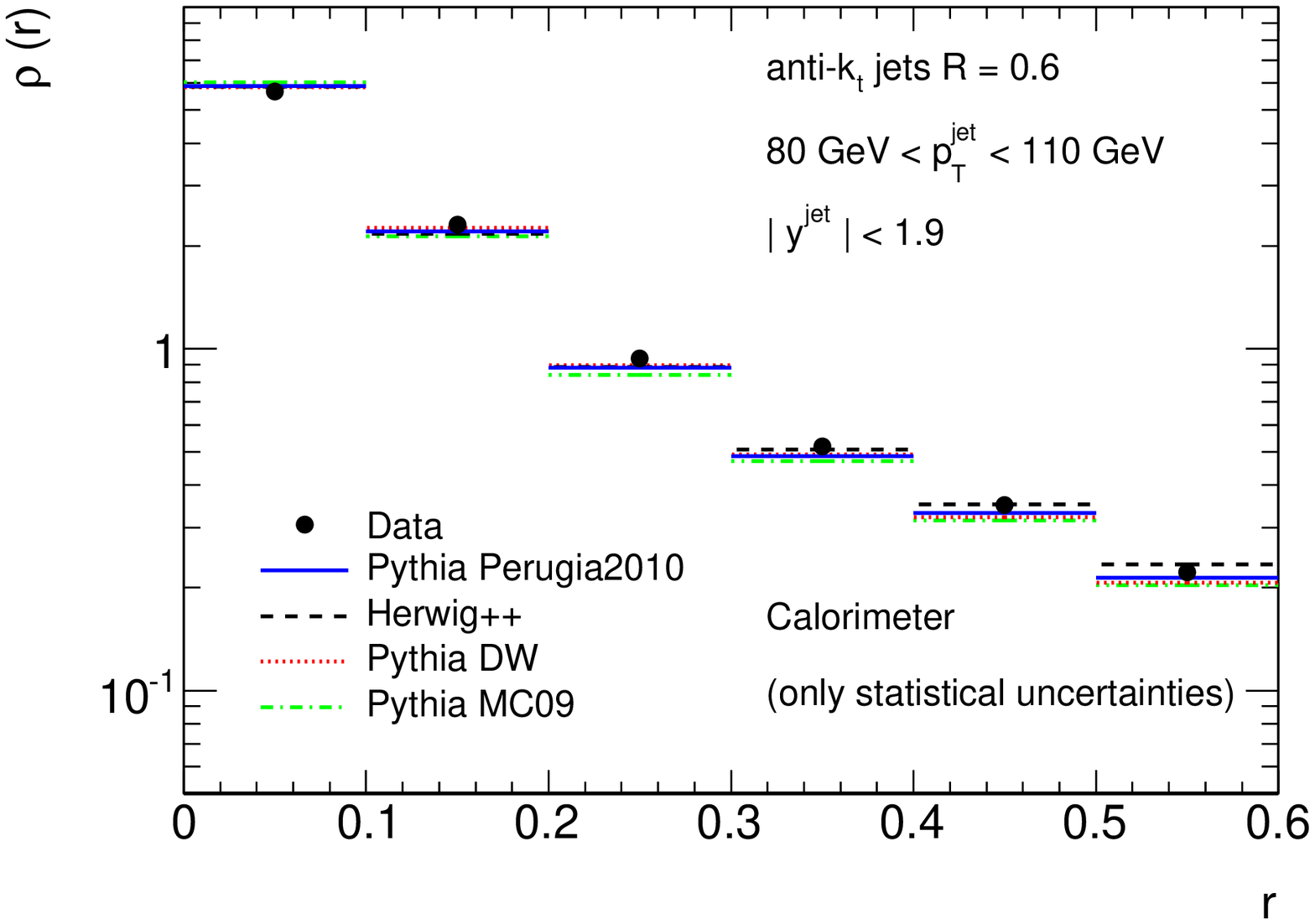}
}
\mbox{
\includegraphics[width=0.495\textwidth]{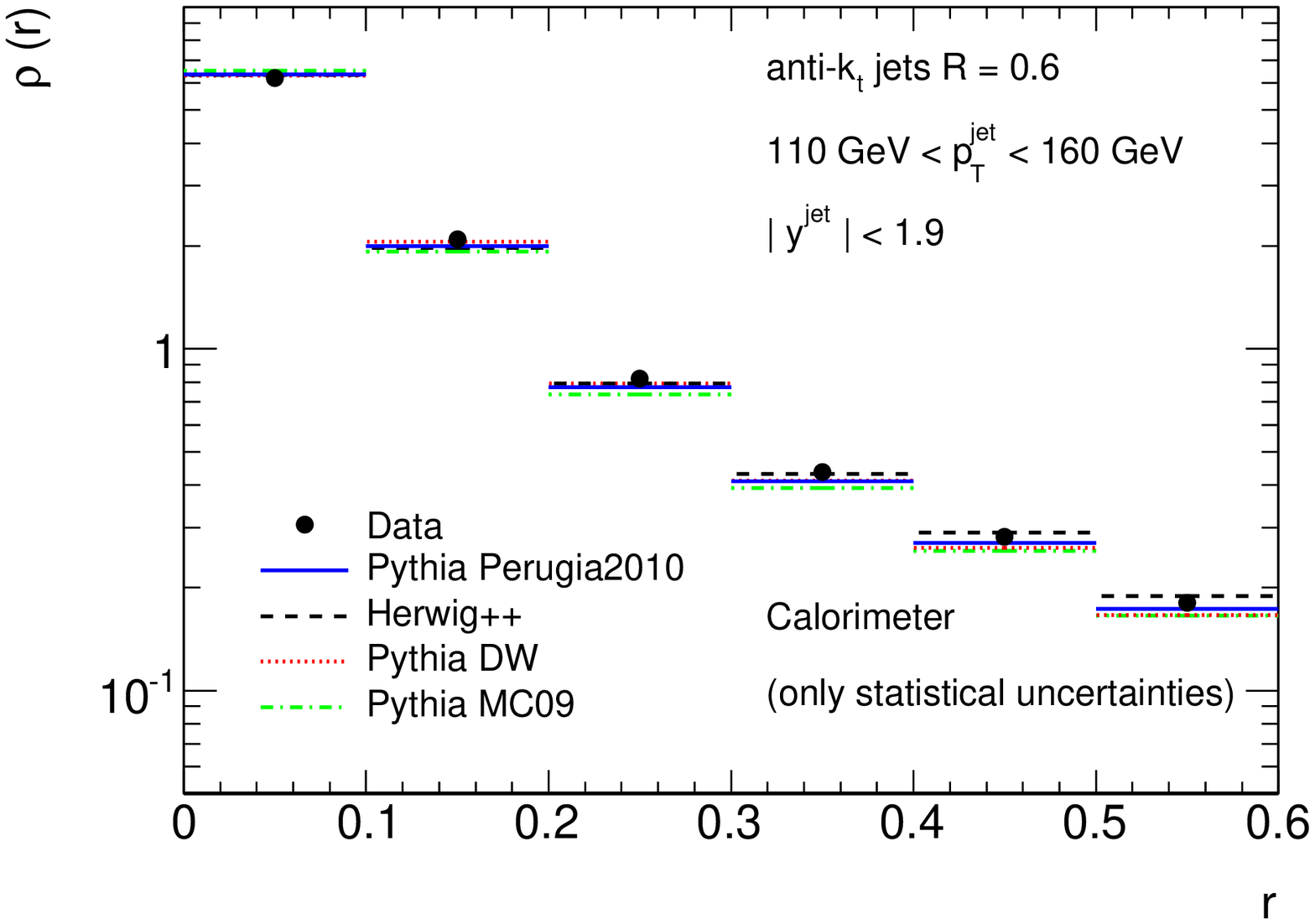}
\includegraphics[width=0.495\textwidth]{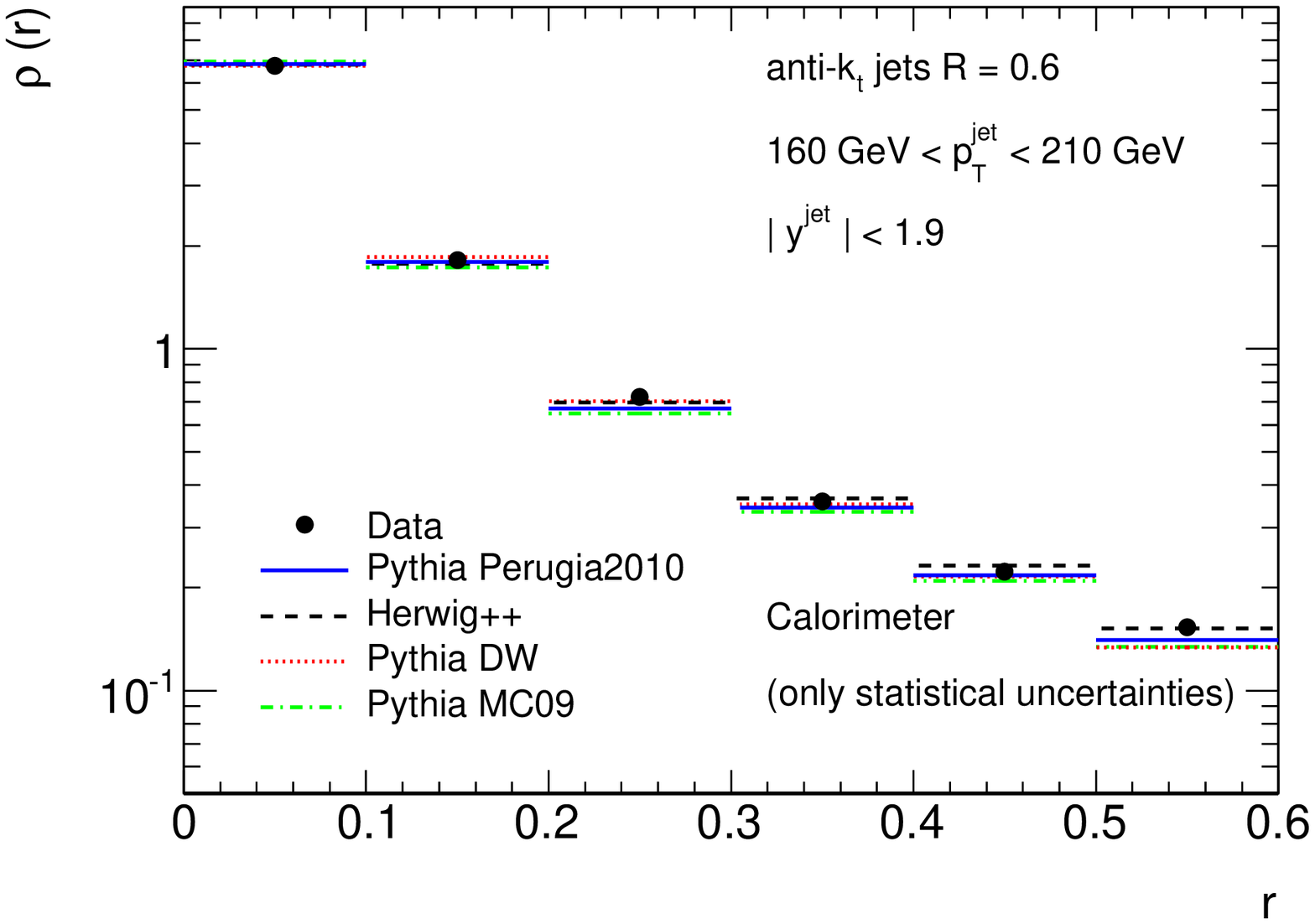}
}
\end{center}
\vspace{-0.7 cm}
\caption{\small
Measured differential jet shapes using calorimeter clusters for jets 
with $|\rapjet| < 1.9$ and $30 \ {\rm GeV} < \ptjet < 210  \ {\rm GeV}$.
}
\label{fig_cluster}
\end{figure}

\begin{figure}[tbh]
\begin{center}
\mbox{
\includegraphics[width=0.495\textwidth]{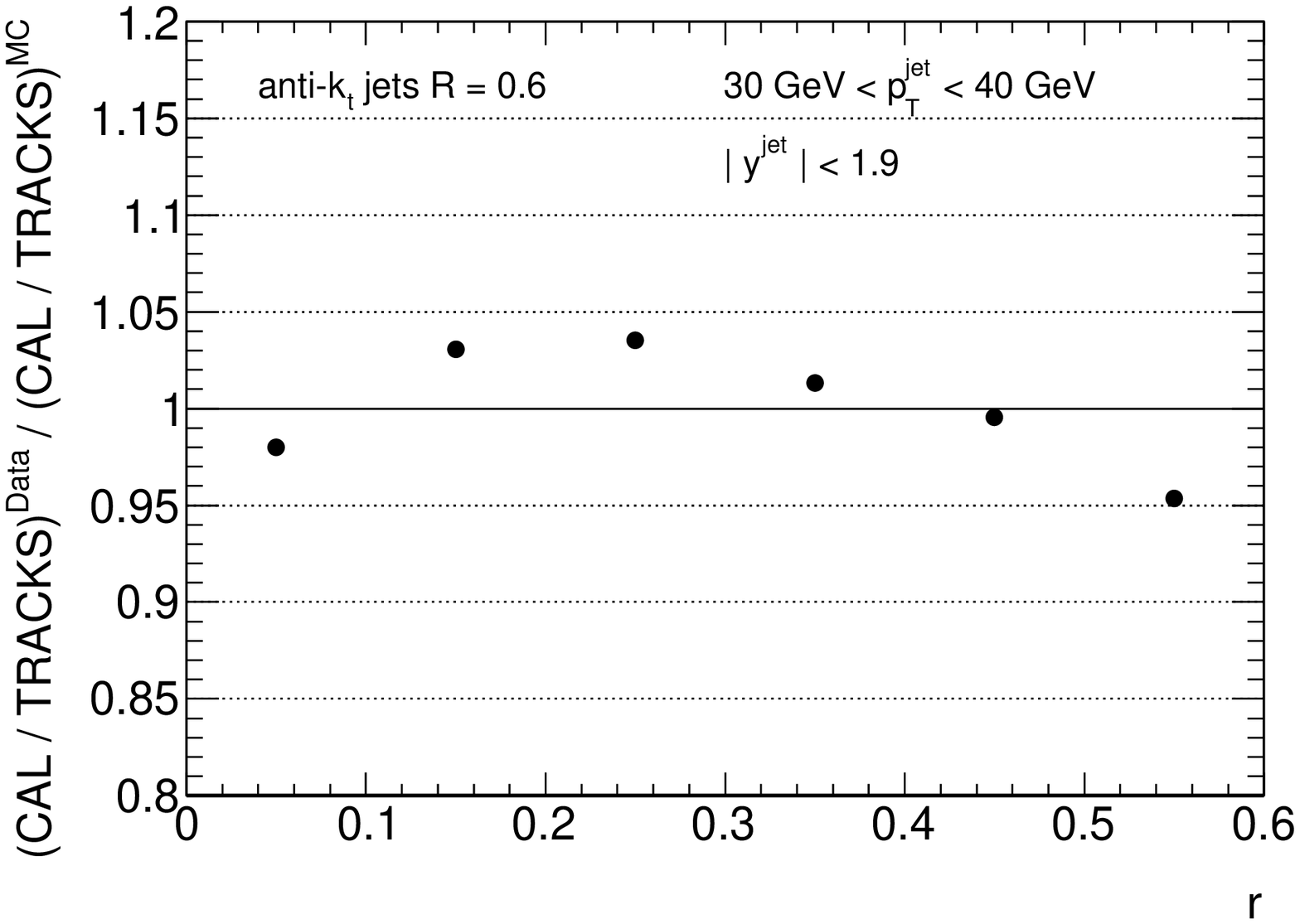}
\includegraphics[width=0.495\textwidth]{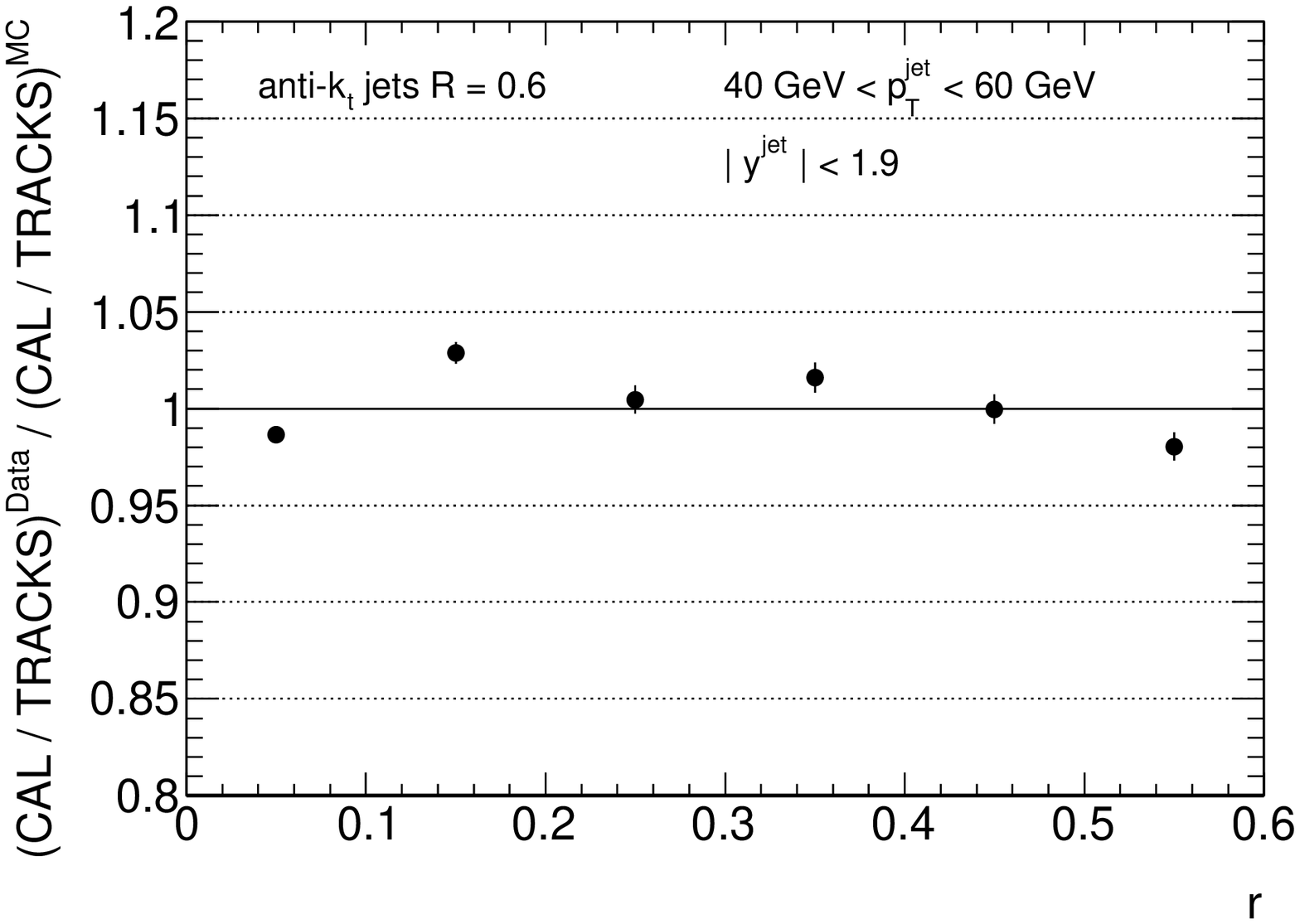}
}
\mbox{
\includegraphics[width=0.495\textwidth]{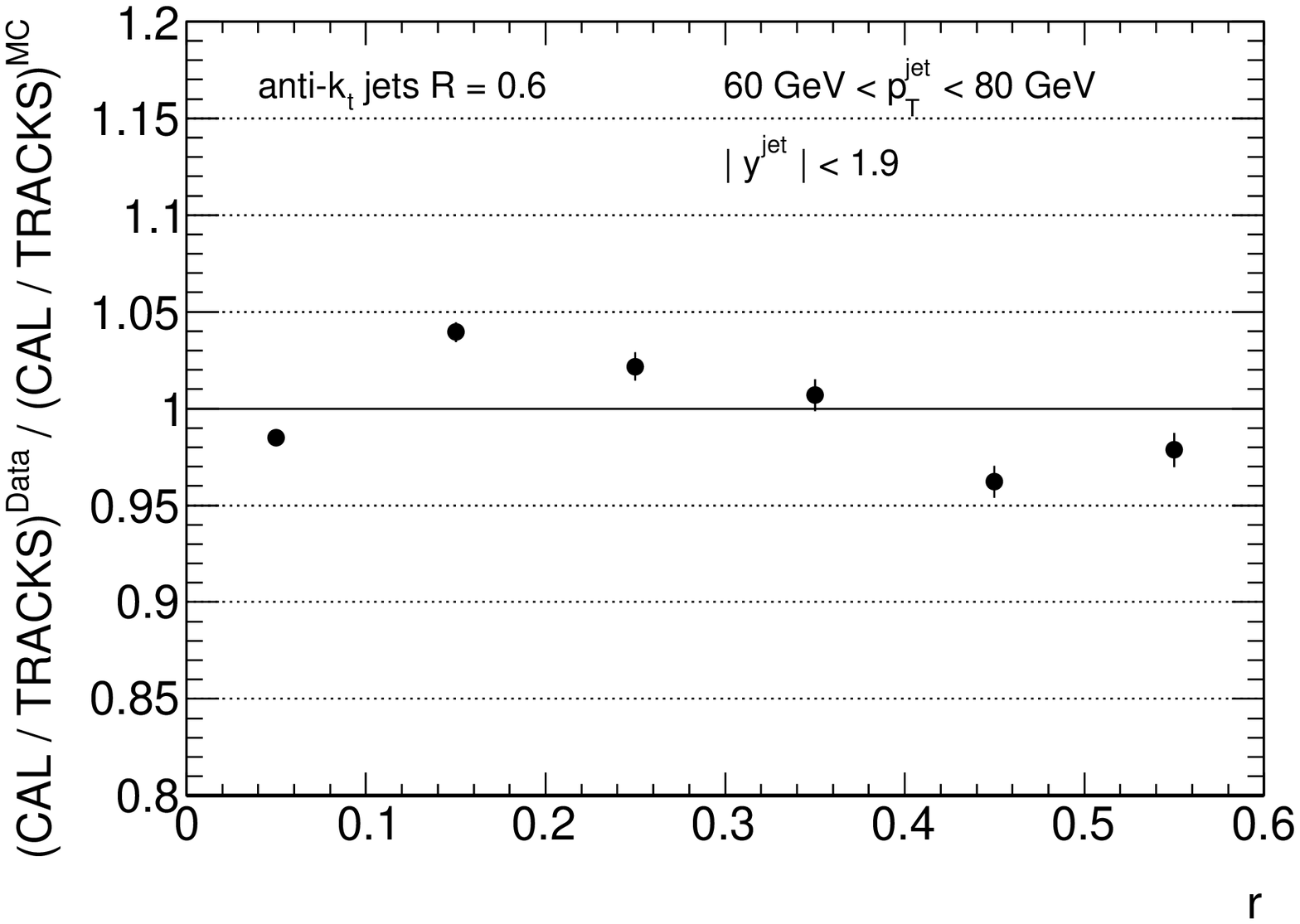}
\includegraphics[width=0.495\textwidth]{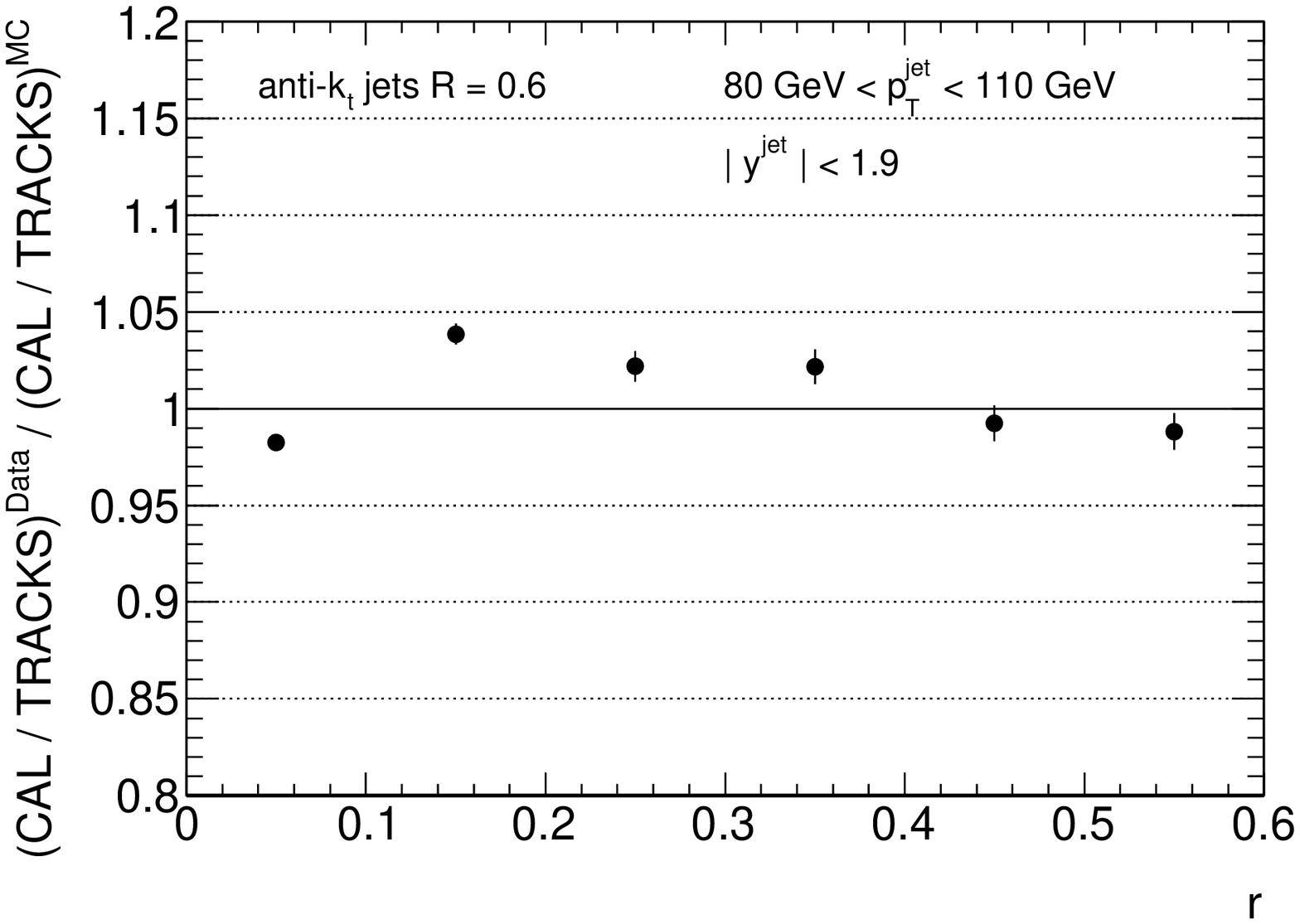}
}
\mbox{
\includegraphics[width=0.495\textwidth]{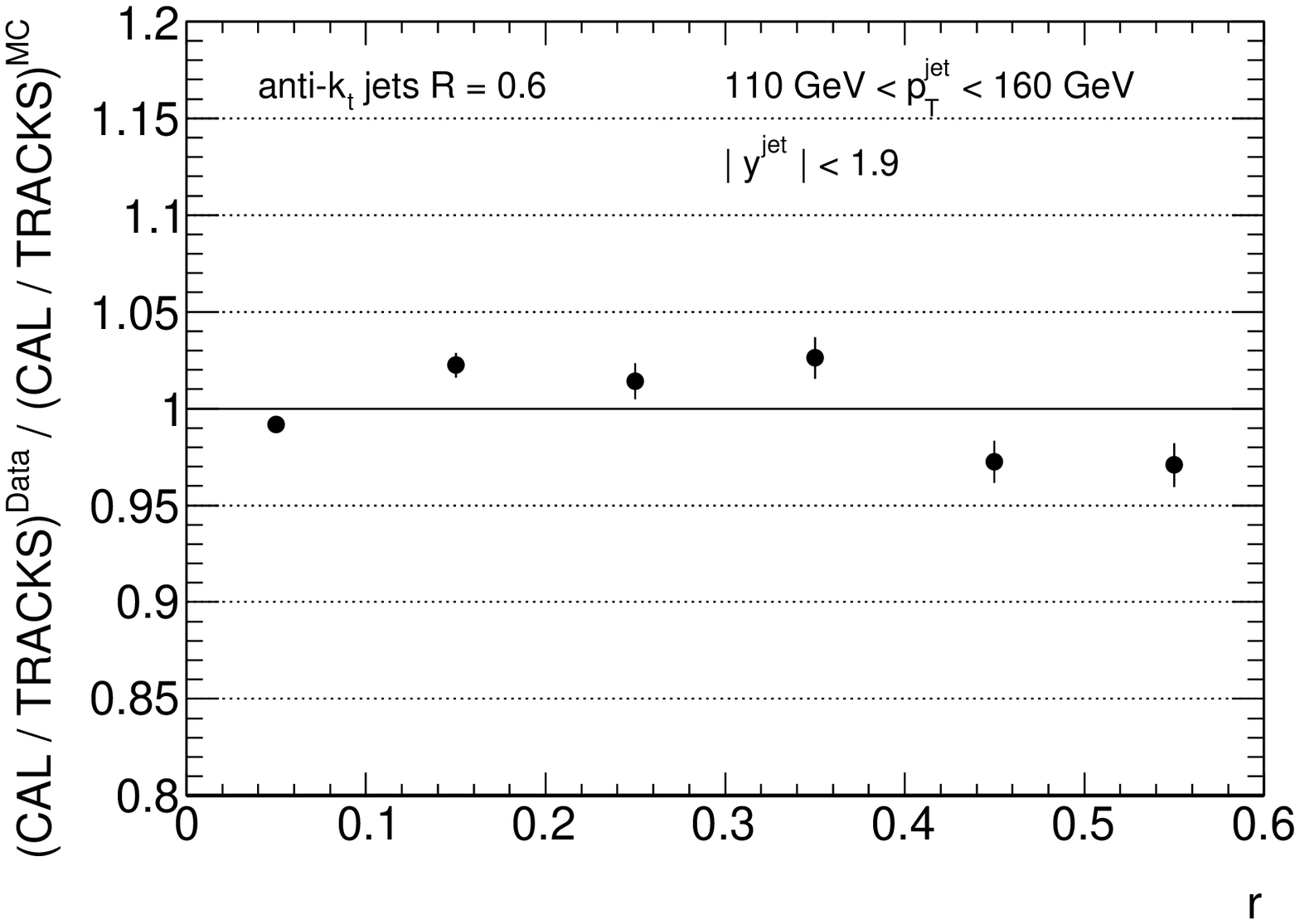}
\includegraphics[width=0.495\textwidth]{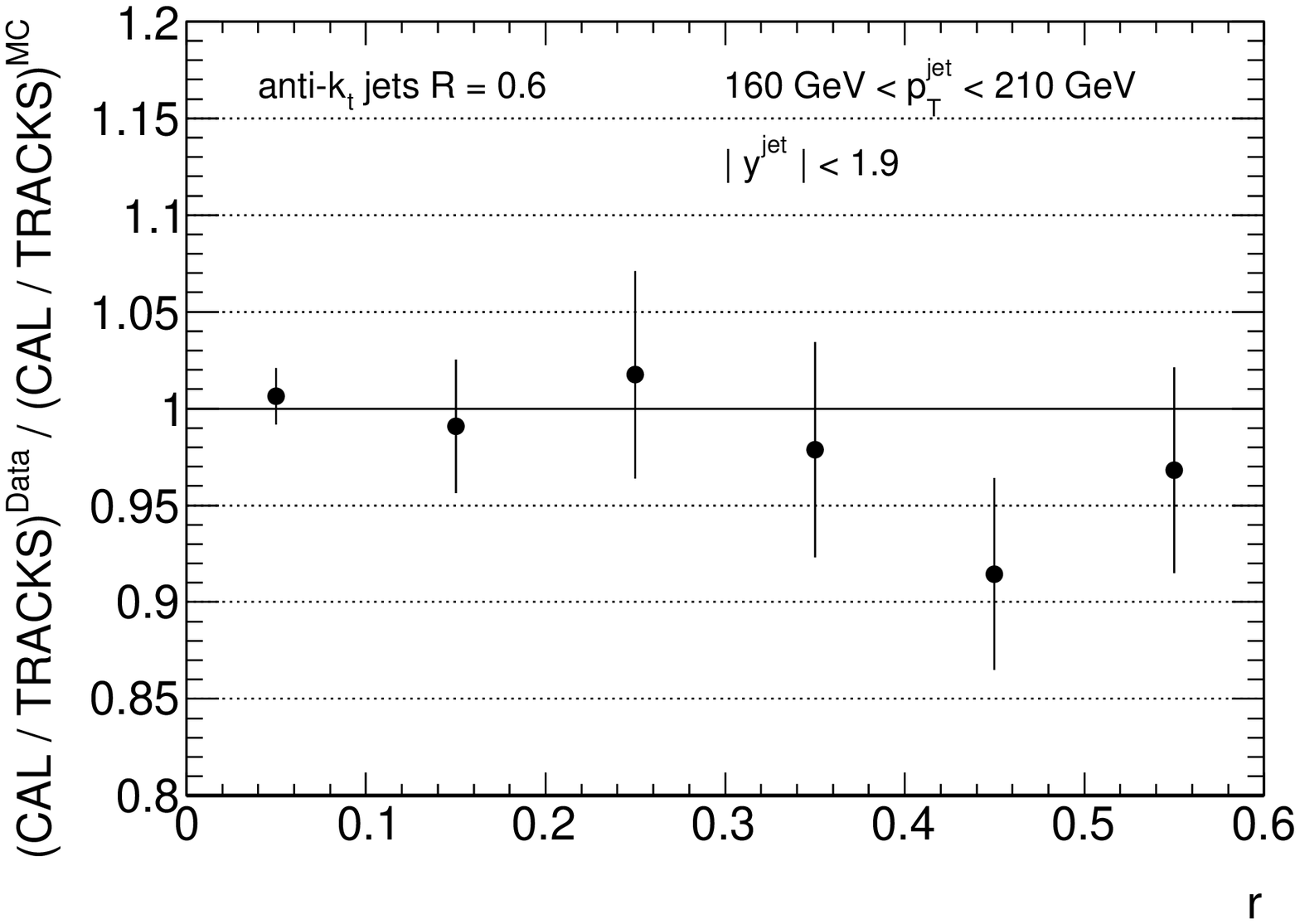}
}
\end{center}
\vspace{-0.7 cm}
\caption{\small
Double-ratio of the differential jet shapes derived by comparing calorimeter and tracking 
ratios of results in data and Monte Carlo simulations 
for jets with $|\rapjet| < 1.9$ and $30 \ {\rm GeV} < \ptjet < 210  \ {\rm GeV}$.
}
\label{fig_double_ratio}
\end{figure}

This provides an alternative method to measure the internal 
structure of jets based on charged particles. The measurements are limited to jets with $|\rapjet|<1.9$, 
as dictated by the tracking coverage and the chosen size of the jet. Figures~\ref{fig_trk} and~\ref{fig_cluster} 
show the differential jet shapes in this rapidity range using tracks and clusters respectively. The ratio of 
the jet shapes using clusters and tracks in data are compared to that in the Monte Carlos simulation (Figure~\ref{fig_double_ratio}). 
Maximum deviations from one of these ratios are about 5$\%$, well within the quoted systematic uncertainties.

\begin{figure}[tbh]
\begin{center}
\mbox{
\includegraphics[width=0.495\textwidth]{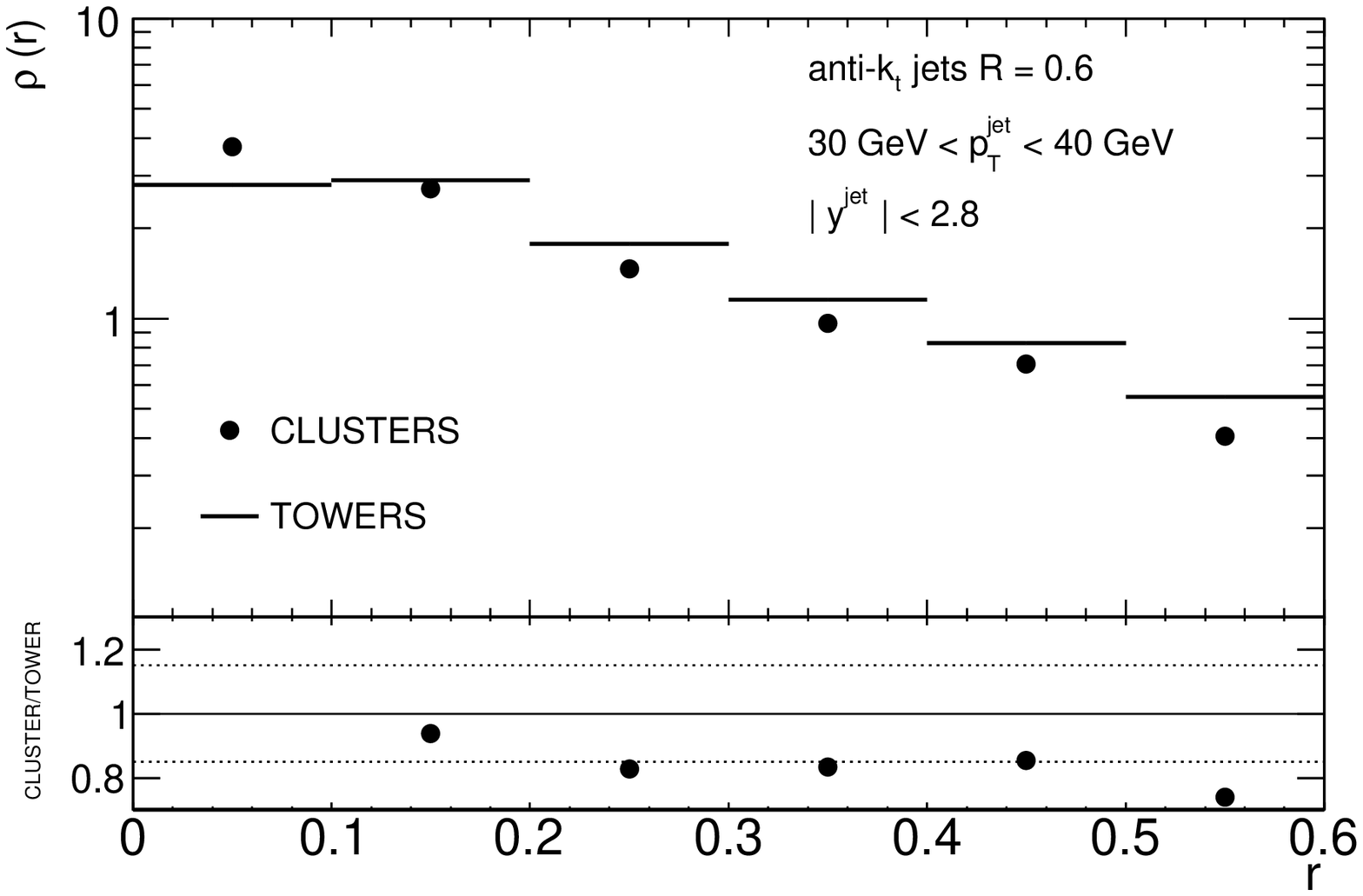}
\includegraphics[width=0.495\textwidth]{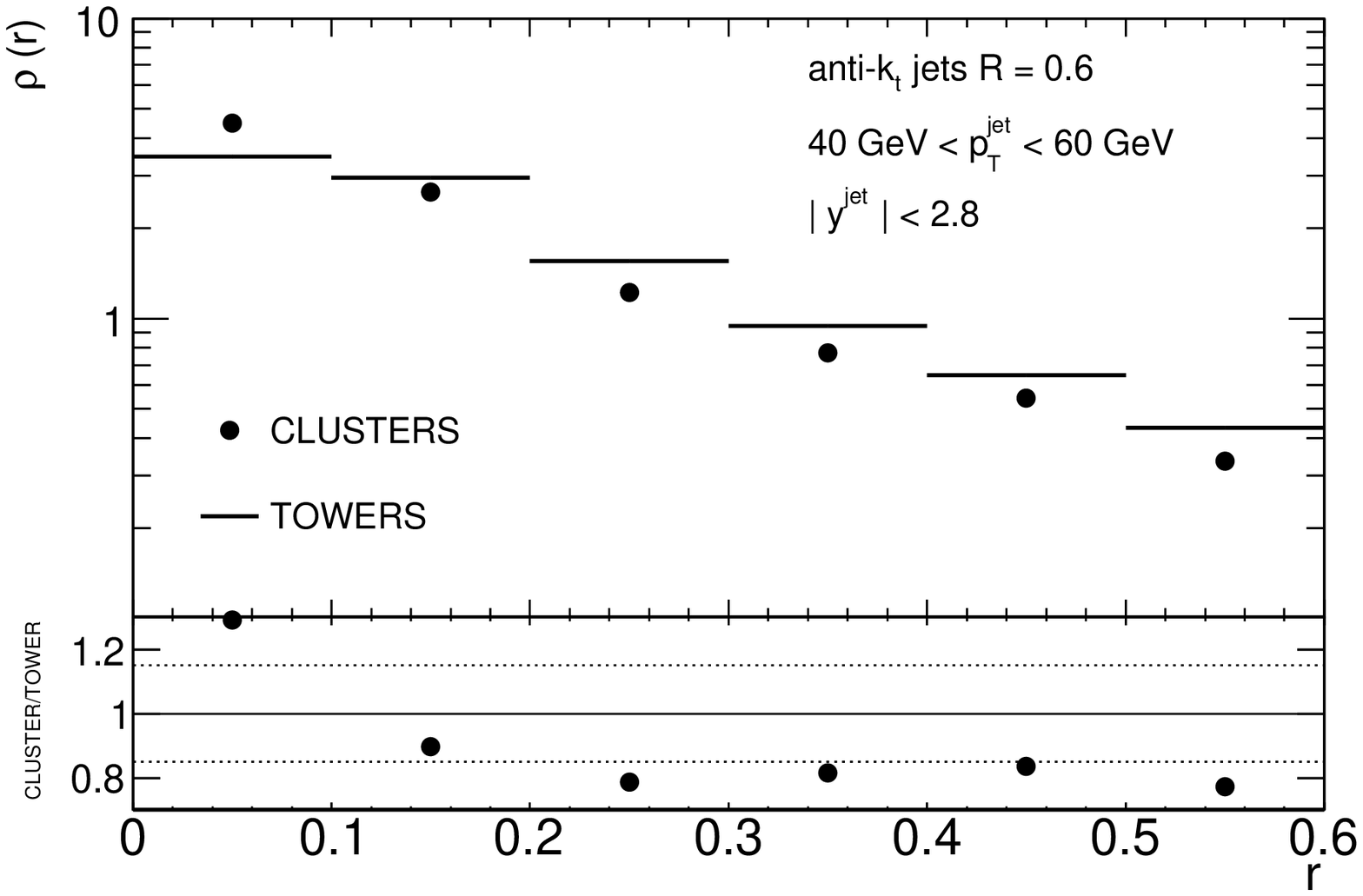}
}
\mbox{
\includegraphics[width=0.495\textwidth]{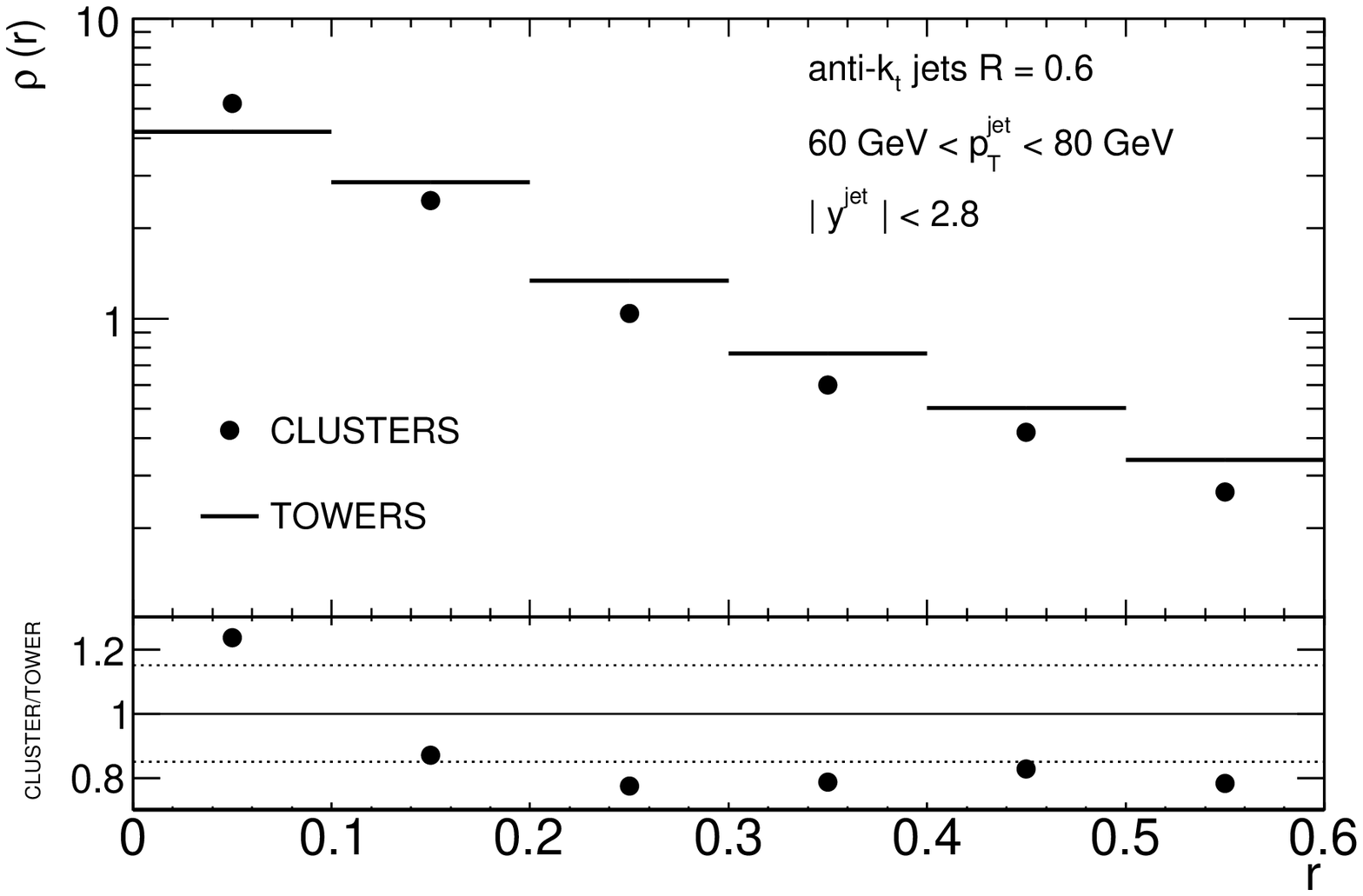}
\includegraphics[width=0.495\textwidth]{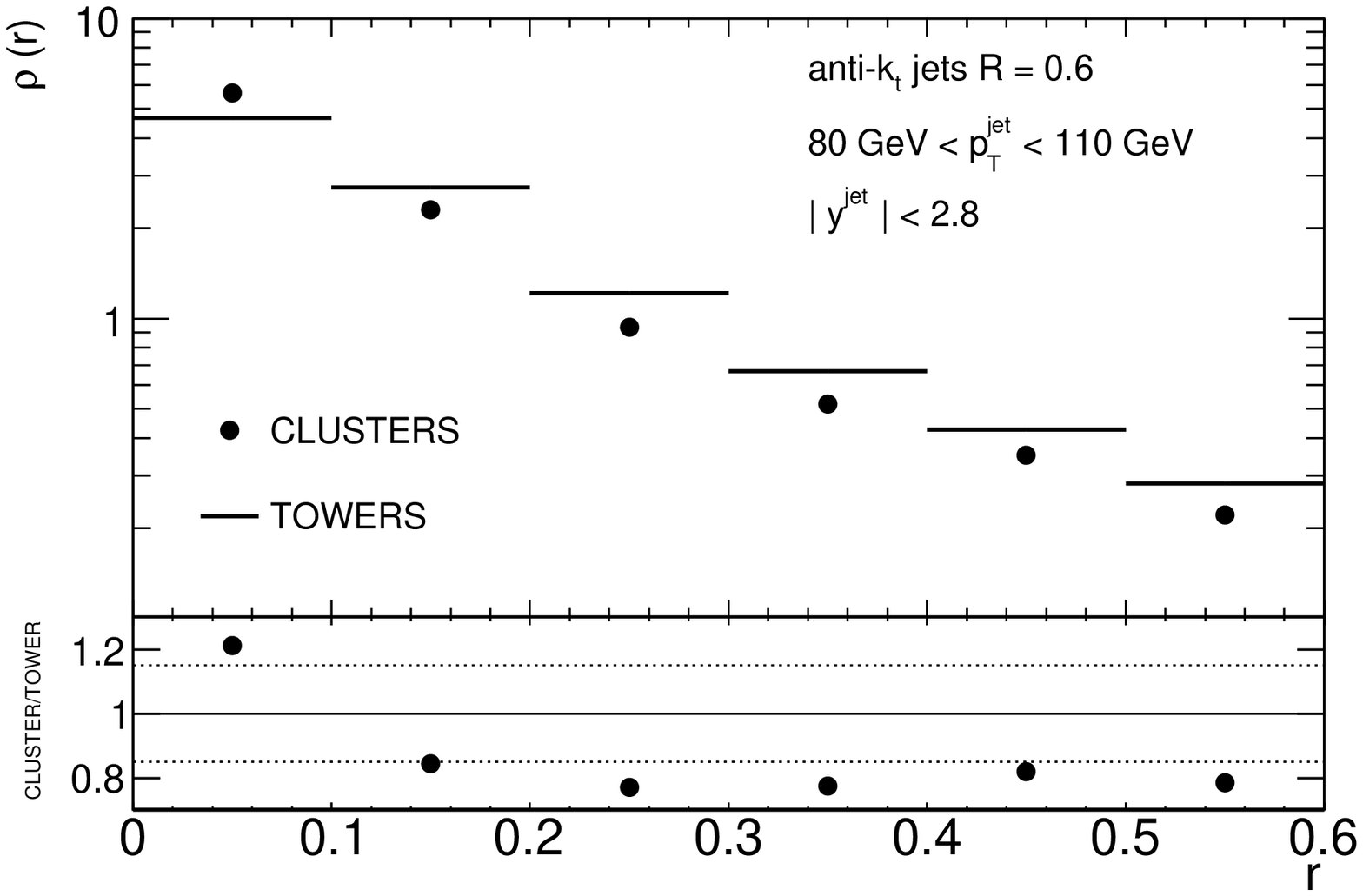}
}
\mbox{
\includegraphics[width=0.495\textwidth]{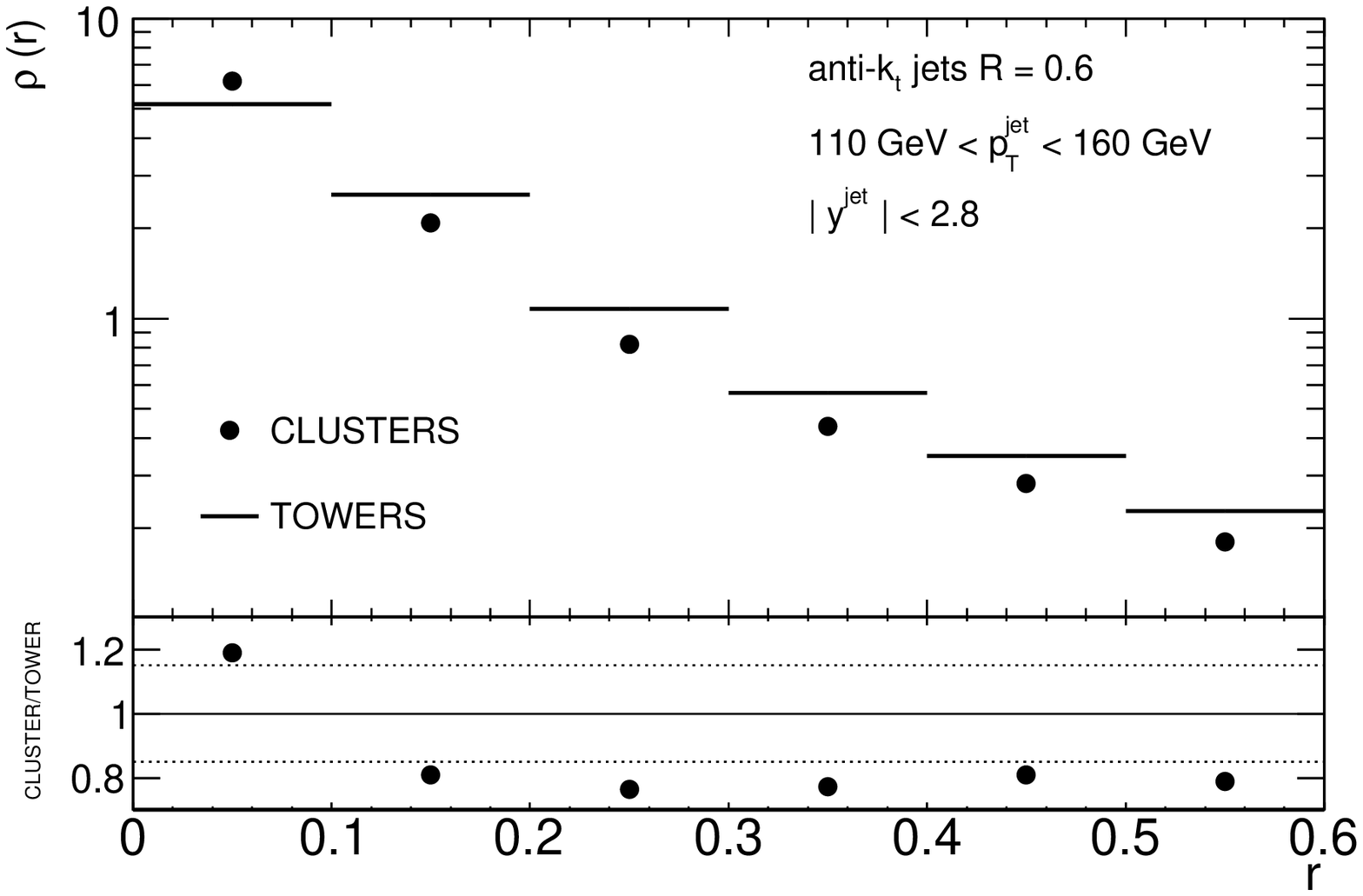}
\includegraphics[width=0.495\textwidth]{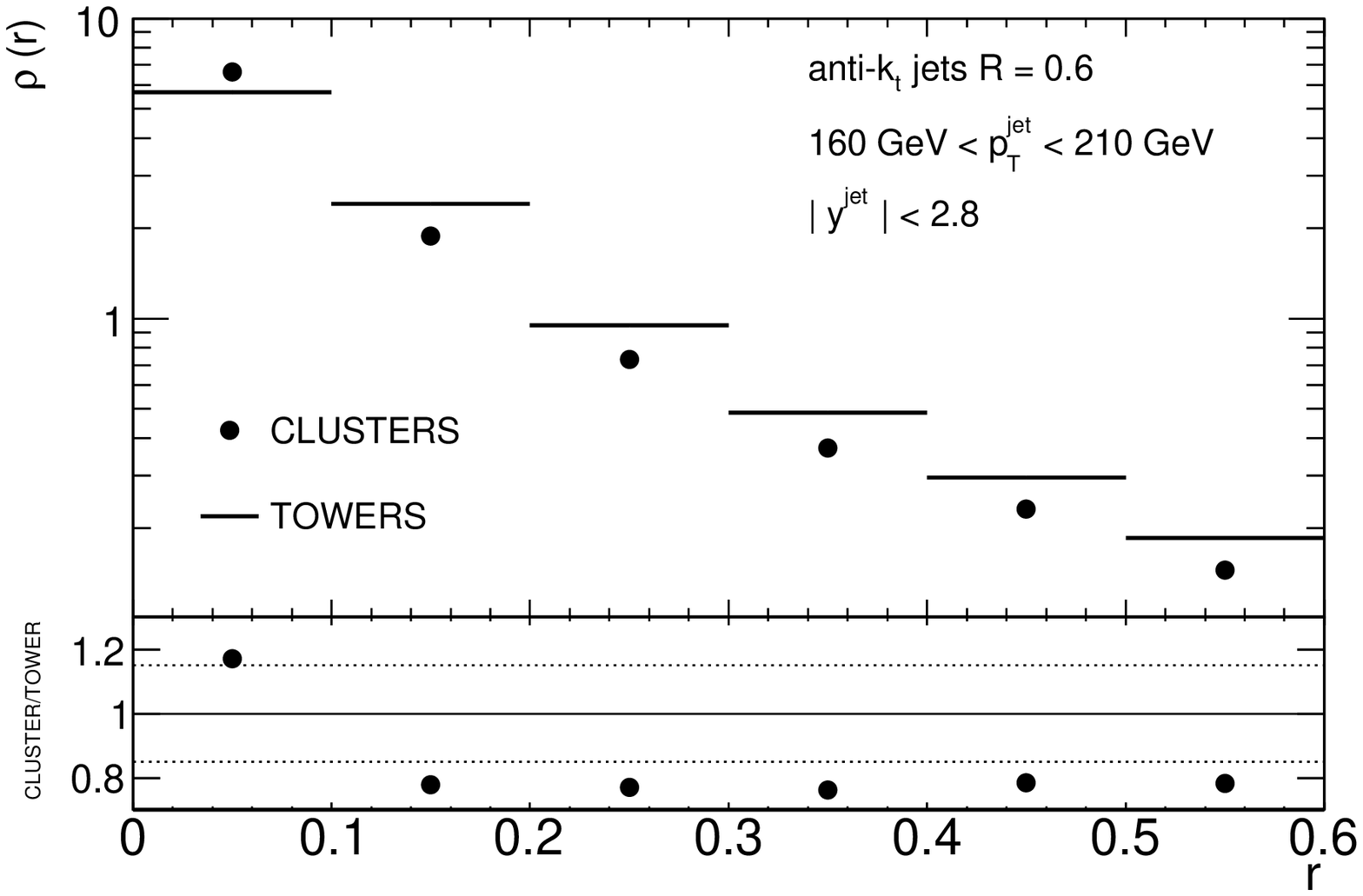}
}
\end{center}
\vspace{-0.7 cm}
\caption{\small
Comparison of differential jet shapes with calorimeter clusters and topo-towers before 
correcting for detector effects for jets with $|\rapjet| < 1.9$ and $30 \ {\rm GeV} < \ptjet < 210  \ {\rm GeV}$.
}
\label{fig_calo_tower}
\end{figure}

\begin{figure}[tbh]
\begin{center}
\mbox{
\includegraphics[width=0.495\textwidth]{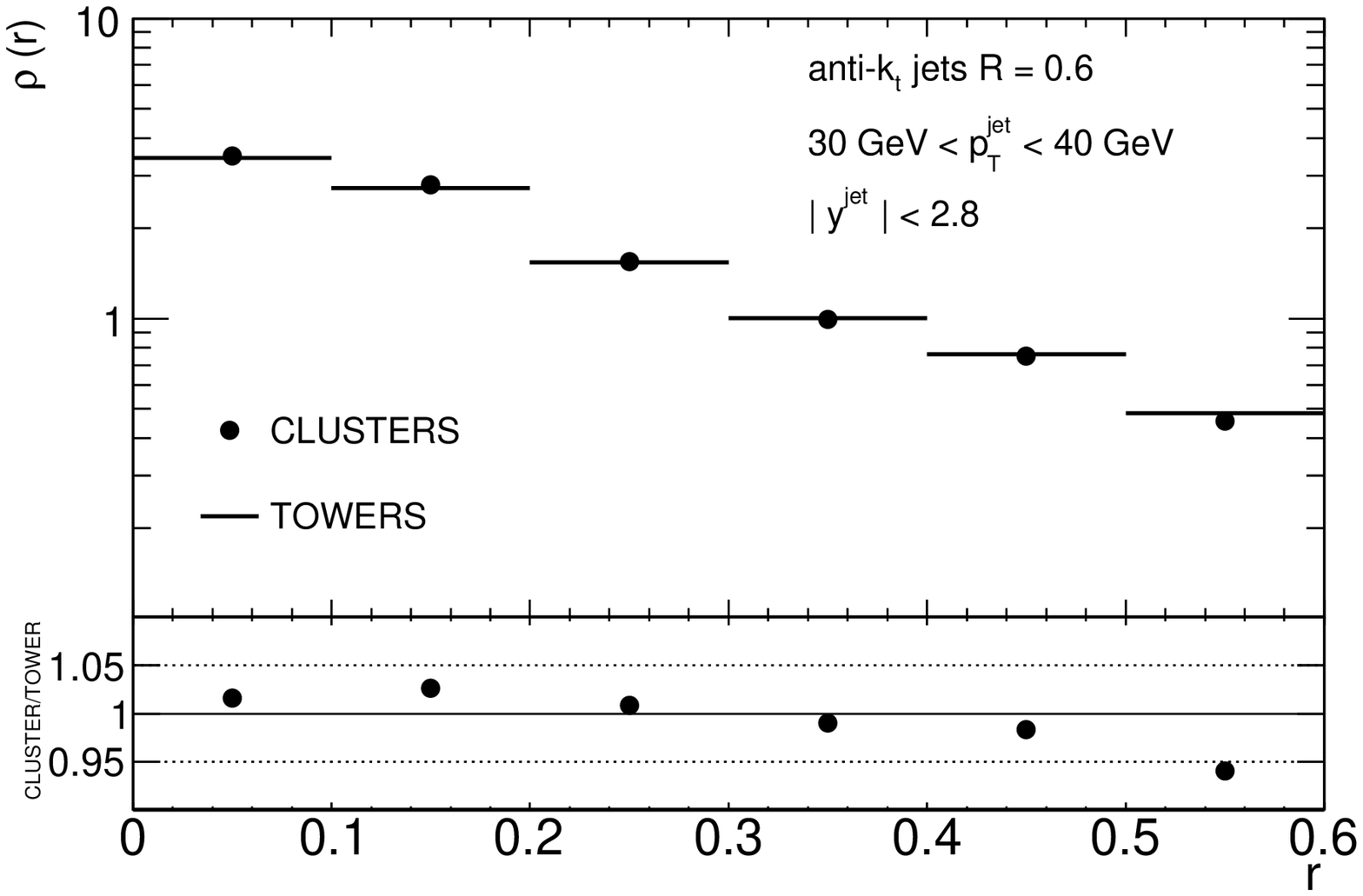}
\includegraphics[width=0.495\textwidth]{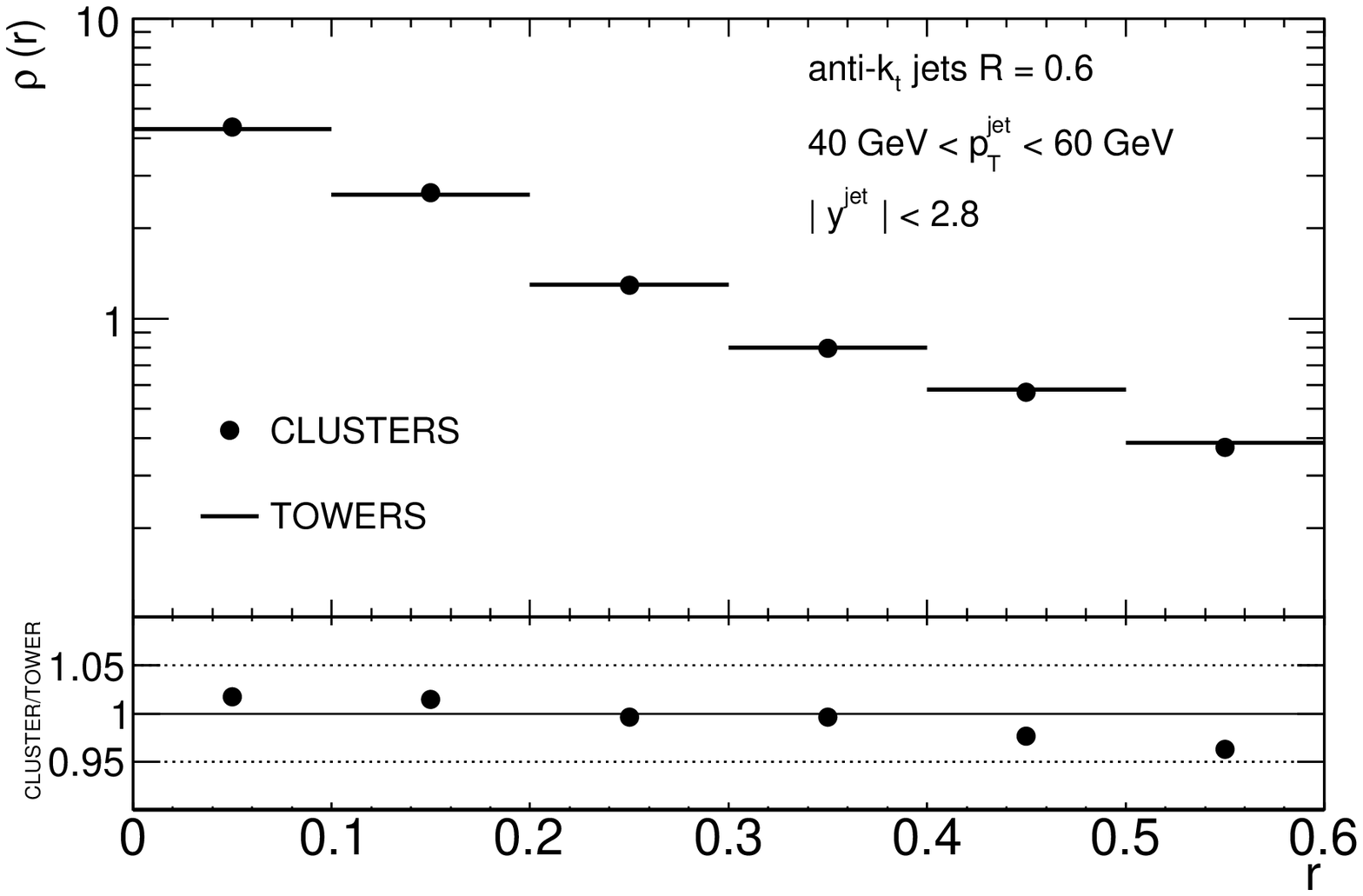}
}
\mbox{
\includegraphics[width=0.495\textwidth]{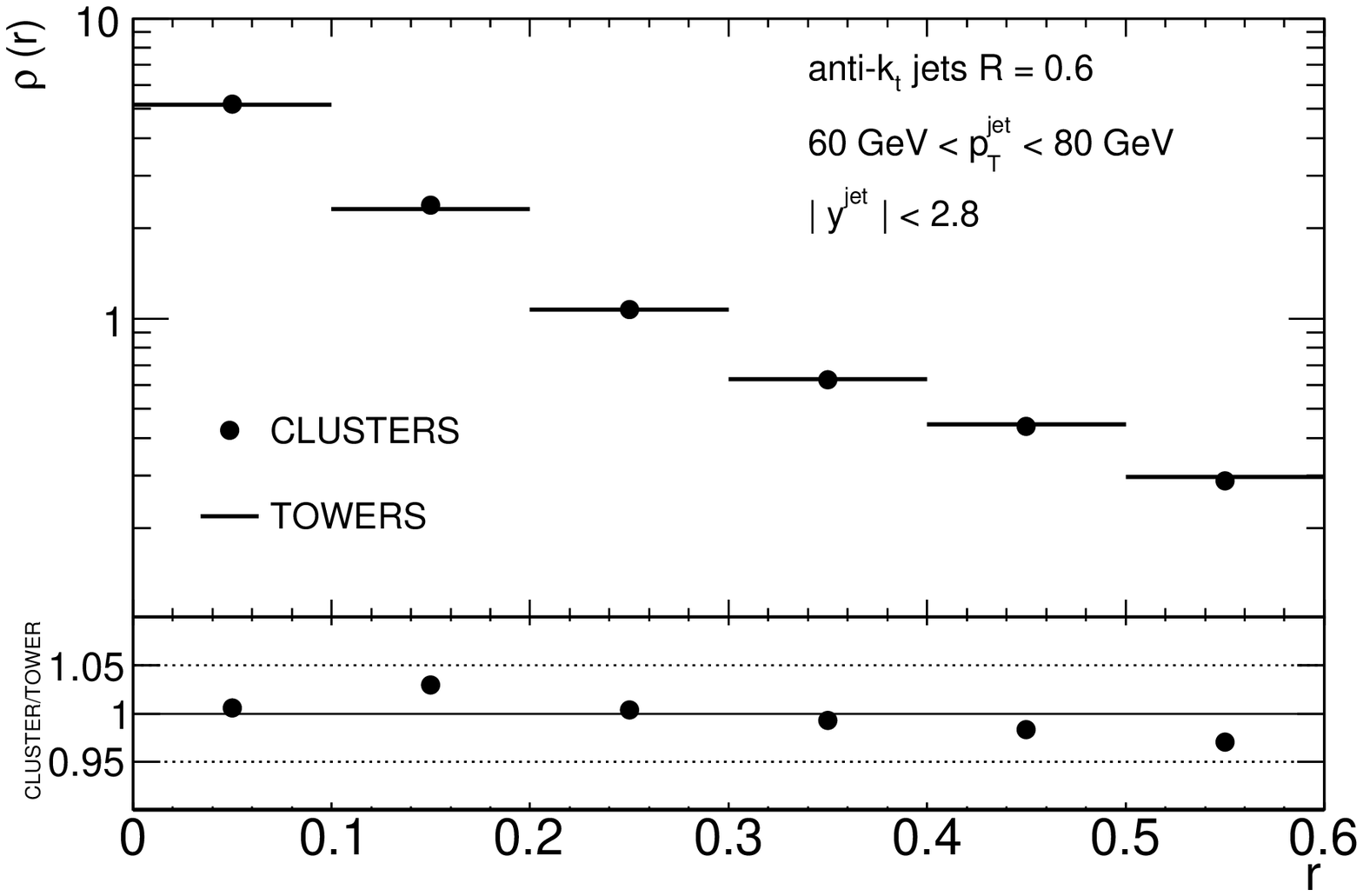}
\includegraphics[width=0.495\textwidth]{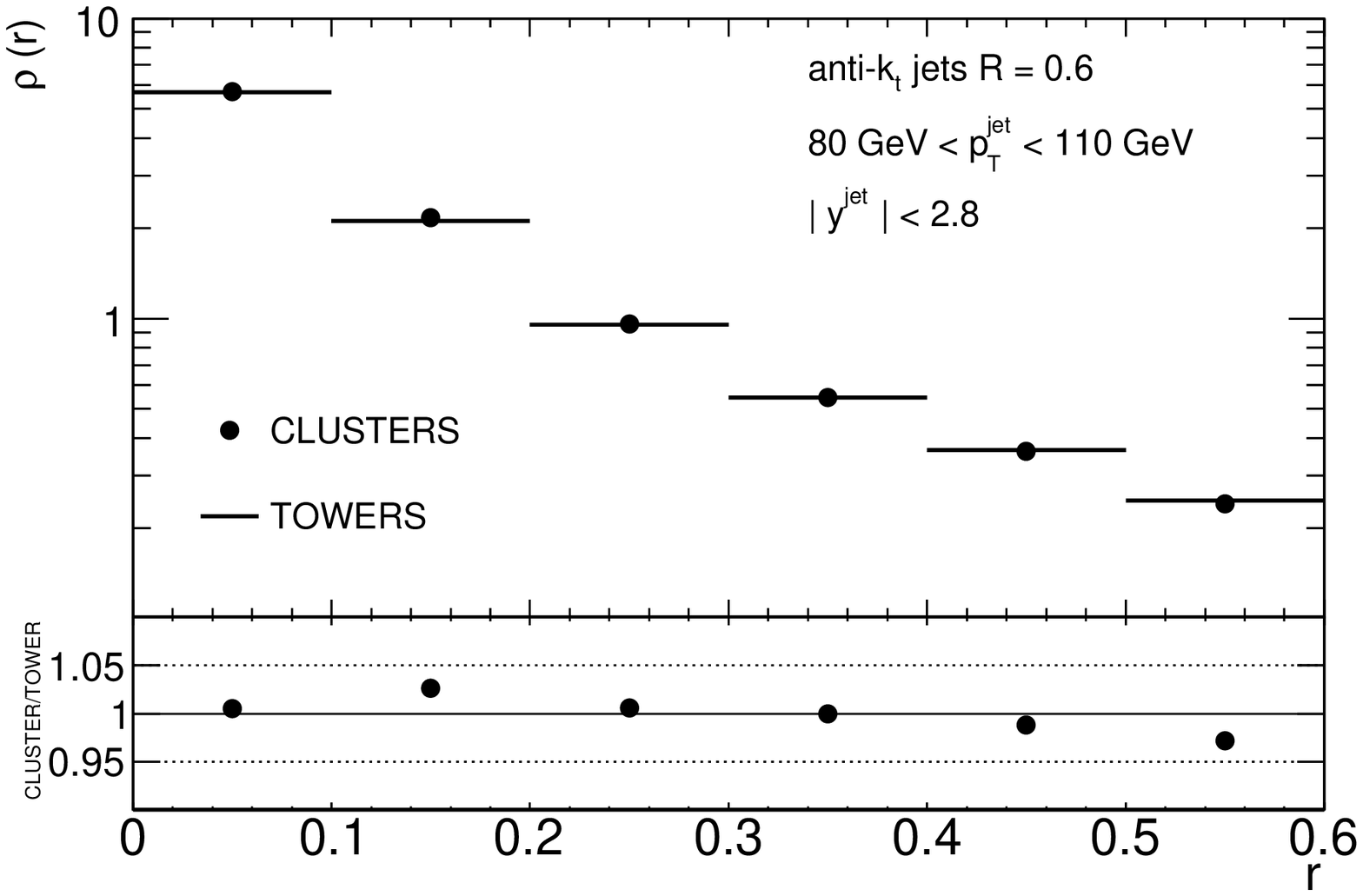}
}
\mbox{
\includegraphics[width=0.495\textwidth]{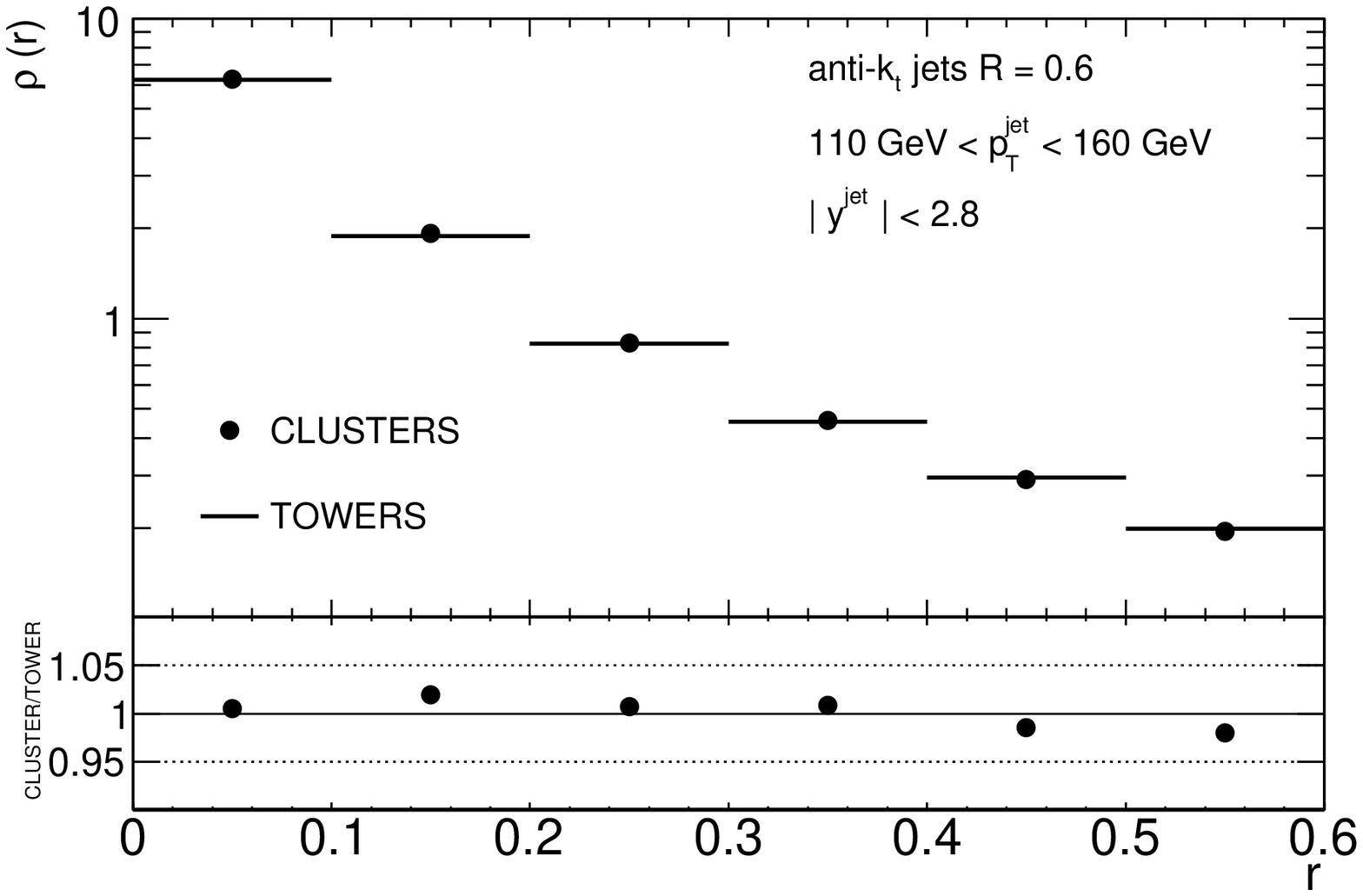}
\includegraphics[width=0.495\textwidth]{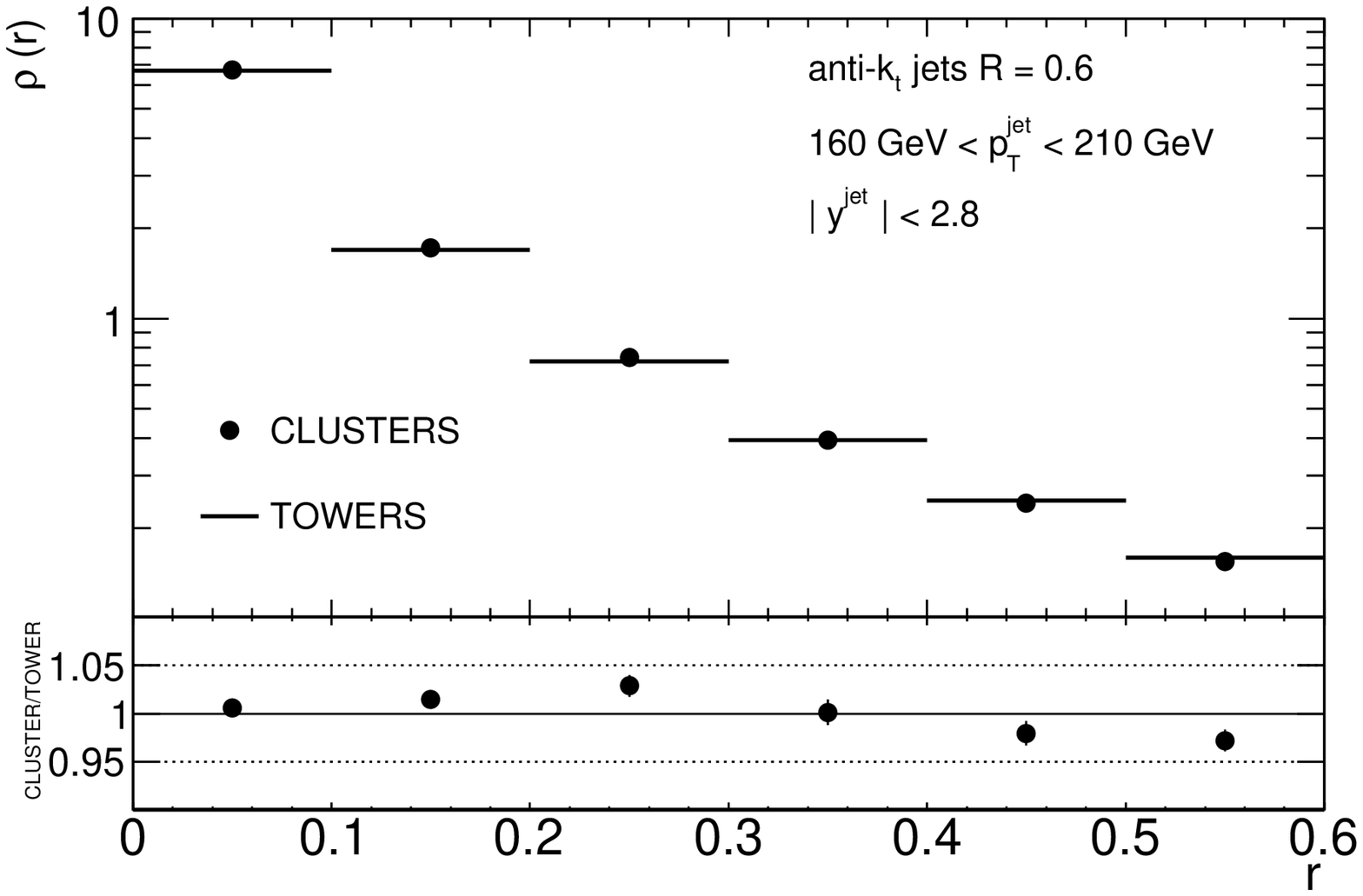}
}
\end{center}
\vspace{-0.7 cm}
\caption{\small
Comparison of differential jet shapes with calorimeter clusters and topo-towers after
correcting for detector effects for jets with $|\rapjet| < 1.9$ and $30 \ {\rm GeV} < \ptjet < 210  \ {\rm GeV}$.
}
\label{fig_hadron_tower}
\end{figure}

Finally, the jet shapes results are obtained using calorimeter towers of fixed size $0.1 \times 0.1$ ($y - \phi$ space) instead of 
topological clusters as input to the jet reconstruction algorithm. 
In order to suppress noise, the towers are built only with cells that belong to a cluster. 
Figure~\ref{fig_calo_tower} show that tower jets are significantly 
broader than cluster jets before correcting for detector effects. Instead, differences reduce to less than 5$\%$ after applying 
the corrections, as shown in Figure~\ref{fig_hadron_tower}. 
Again, the observed differences are well inside the quoted uncertainty on the nominal results based on calorimeter clusters.

\clearpage

\clearpage

\clearpage

\clearpage
\clearpage
\section{Results}

The measurements presented in this Chapter refer to differential 
and integrated jet shapes, $\rho (r)$ 
and $\Psi (r)$,  corrected at the particle level and obtained for \akt jets with distance parameter 
$R=0.6$  in the region  $|\rapjet| < 2.8$ and 
$30 \  {\rm{ GeV}} < \ptjet < 600 \ \rm GeV$. The measurements are presented in separate bins of 
 $\ptjet$ and $|\rapjet|$. Tabulated values
of the results are available in Ref.~\cite{tab}.
 
\begin{figure}[tbh]
\begin{center}
\mbox{
\includegraphics[width=0.5\textwidth,height=0.52\textwidth]{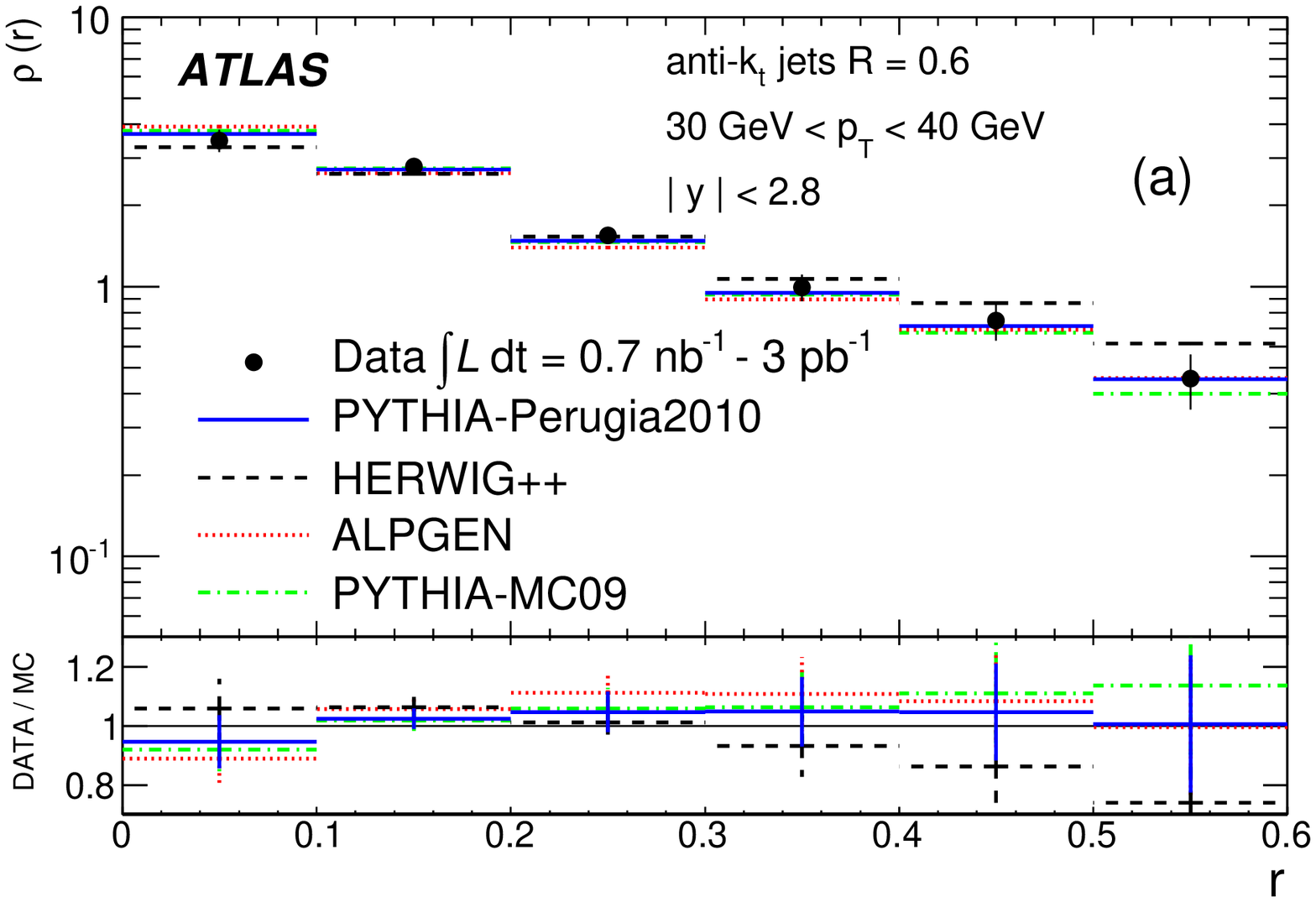} 
\includegraphics[width=0.5\textwidth,height=0.52\textwidth]{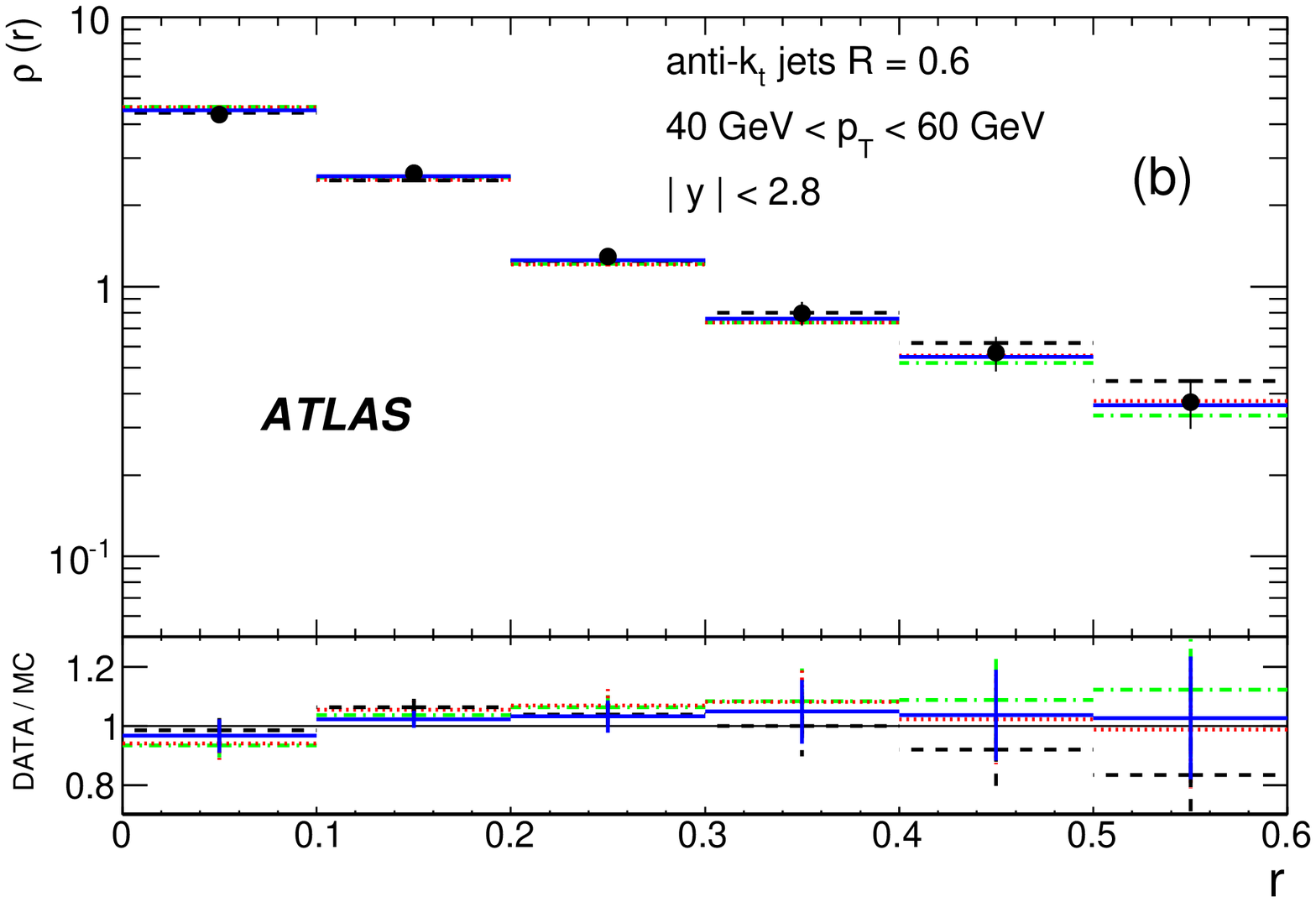}
}\vspace{-0.2cm}
\mbox{
\includegraphics[width=0.5\textwidth,height=0.52\textwidth]{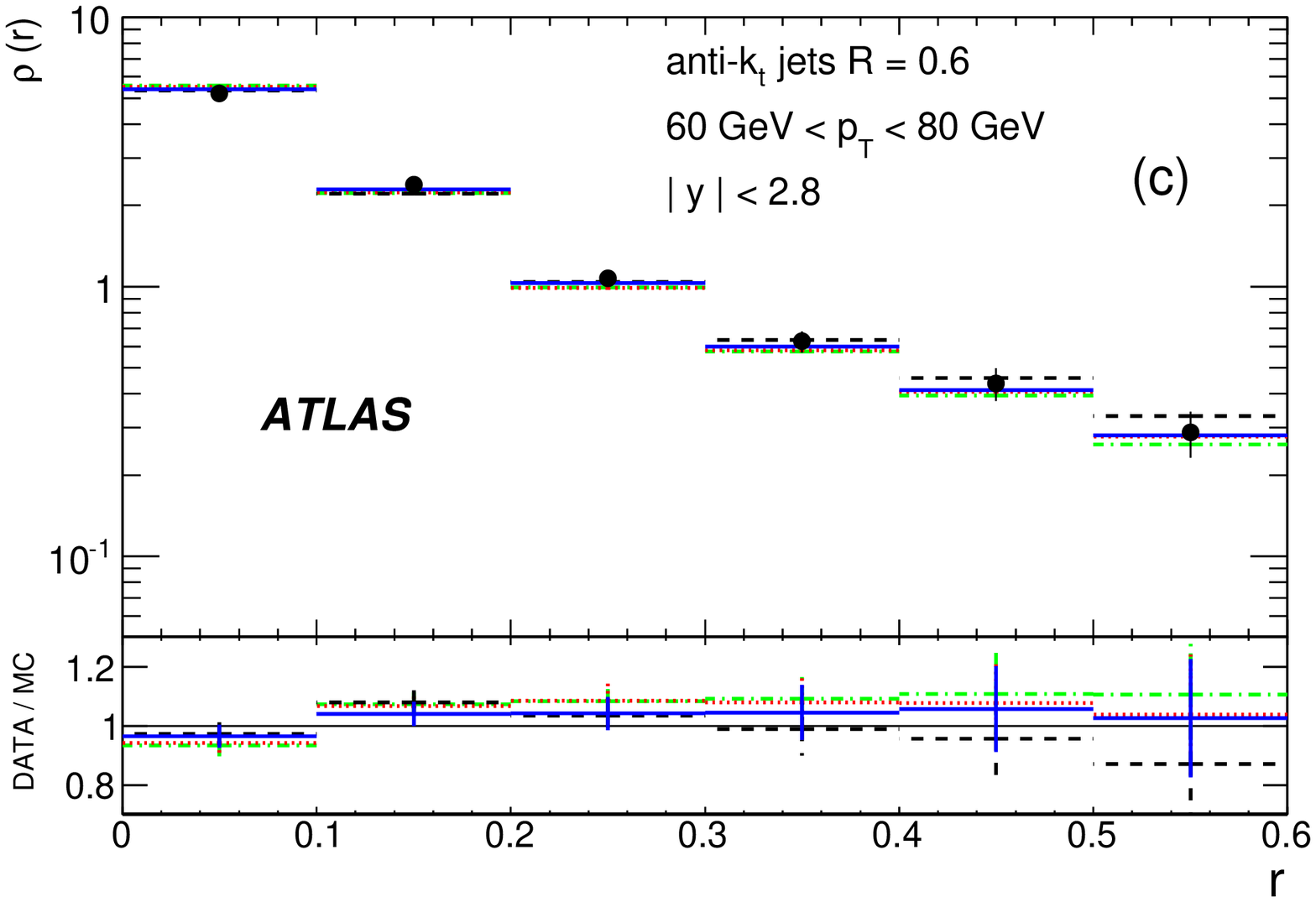}
\includegraphics[width=0.5\textwidth,height=0.52\textwidth]{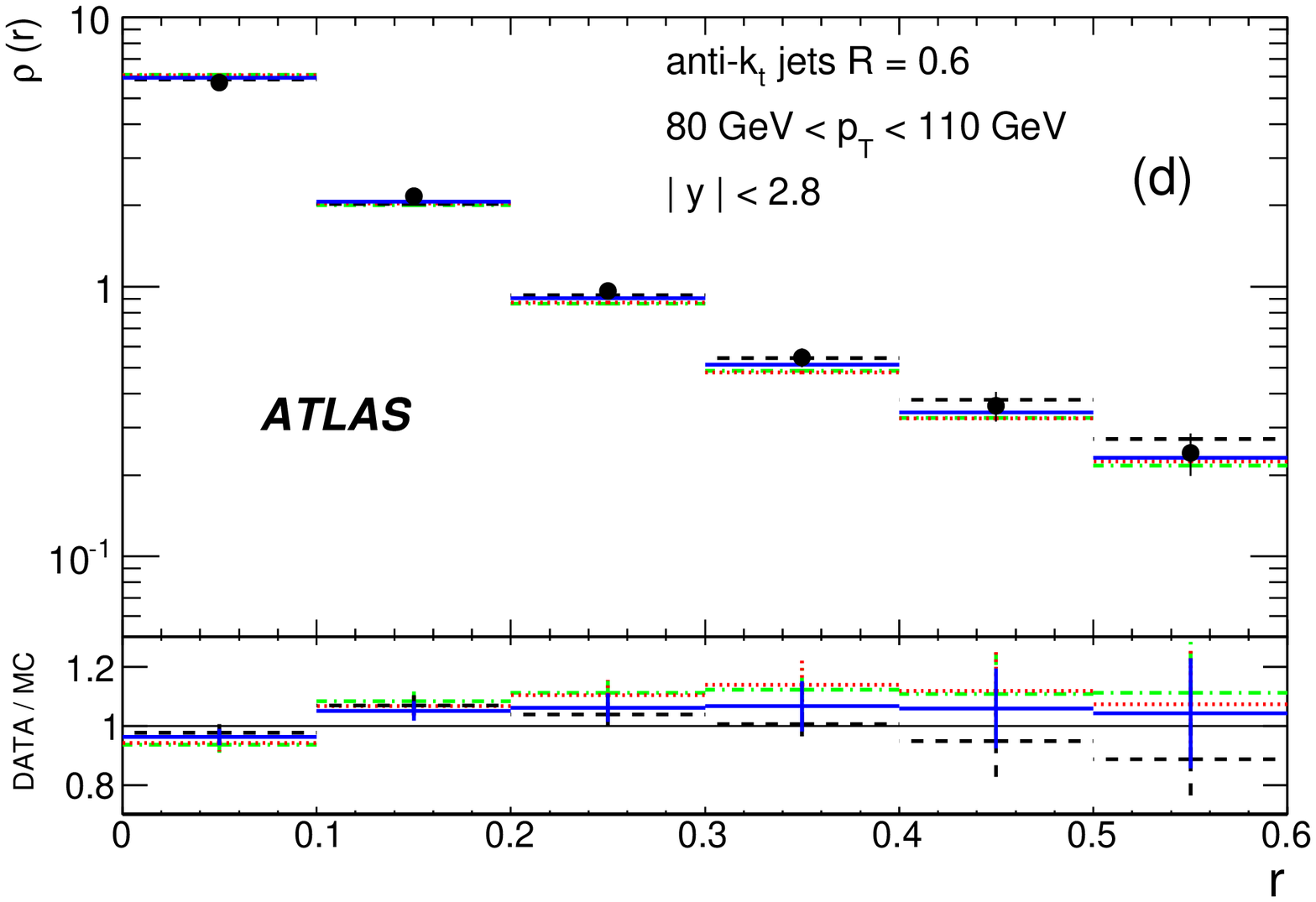}
}
\end{center}
\vspace{-0.7 cm}
\caption[The measured differential jet shape, $\rho(r)$, in inclusive jet production for jets
with $|\rapjet| < 2.8$ and $30 \ {\rm GeV} < \ptjet < 110  \ {\rm GeV}$]
{\small
The measured differential jet shape, $\rho(r)$, in inclusive jet production for jets 
with $|\rapjet| < 2.8$ and $30 \ {\rm GeV} < \ptjet < 110  \ {\rm GeV}$   
is shown in different $\ptjet$ regions. Error bars indicate the statistical and systematic uncertainties added in quadrature.
The predictions of   PYTHIA-Perugia2010 (solid lines),   HERWIG++ (dashed lines),   ALPGEN interfaced with HERWIG and JIMMY (dotted lines), and 
  PYTHIA-MC09 (dashed-dotted lines) are shown for comparison.} 
\label{fig1}
\end{figure}

\begin{figure}[tbh]
\begin{center}
\mbox{
\includegraphics[width=0.5\textwidth,height=0.52\textwidth]{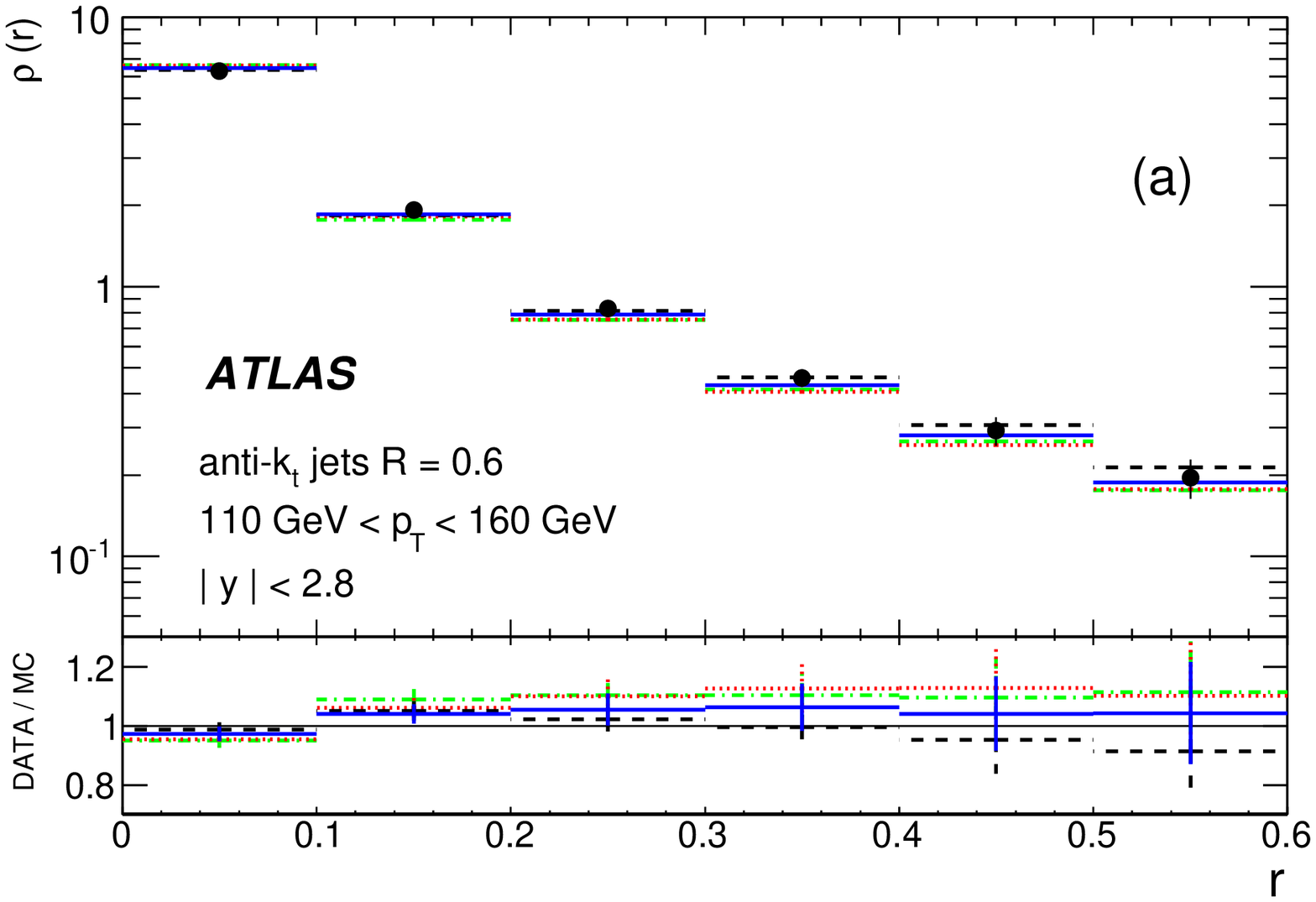} 
\includegraphics[width=0.5\textwidth,height=0.52\textwidth]{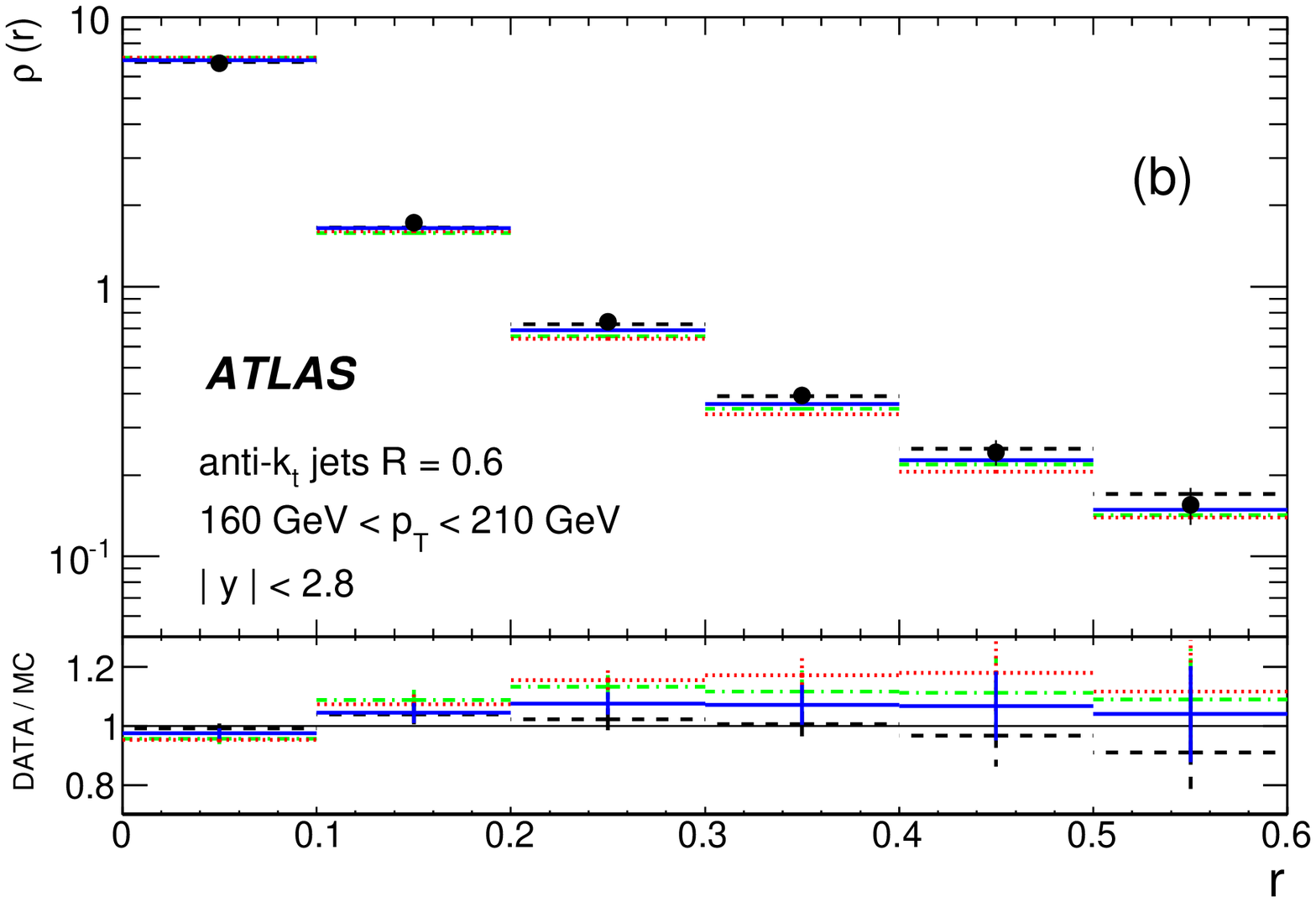}
}\vspace{-0.2cm}
\mbox{
\includegraphics[width=0.5\textwidth,height=0.52\textwidth]{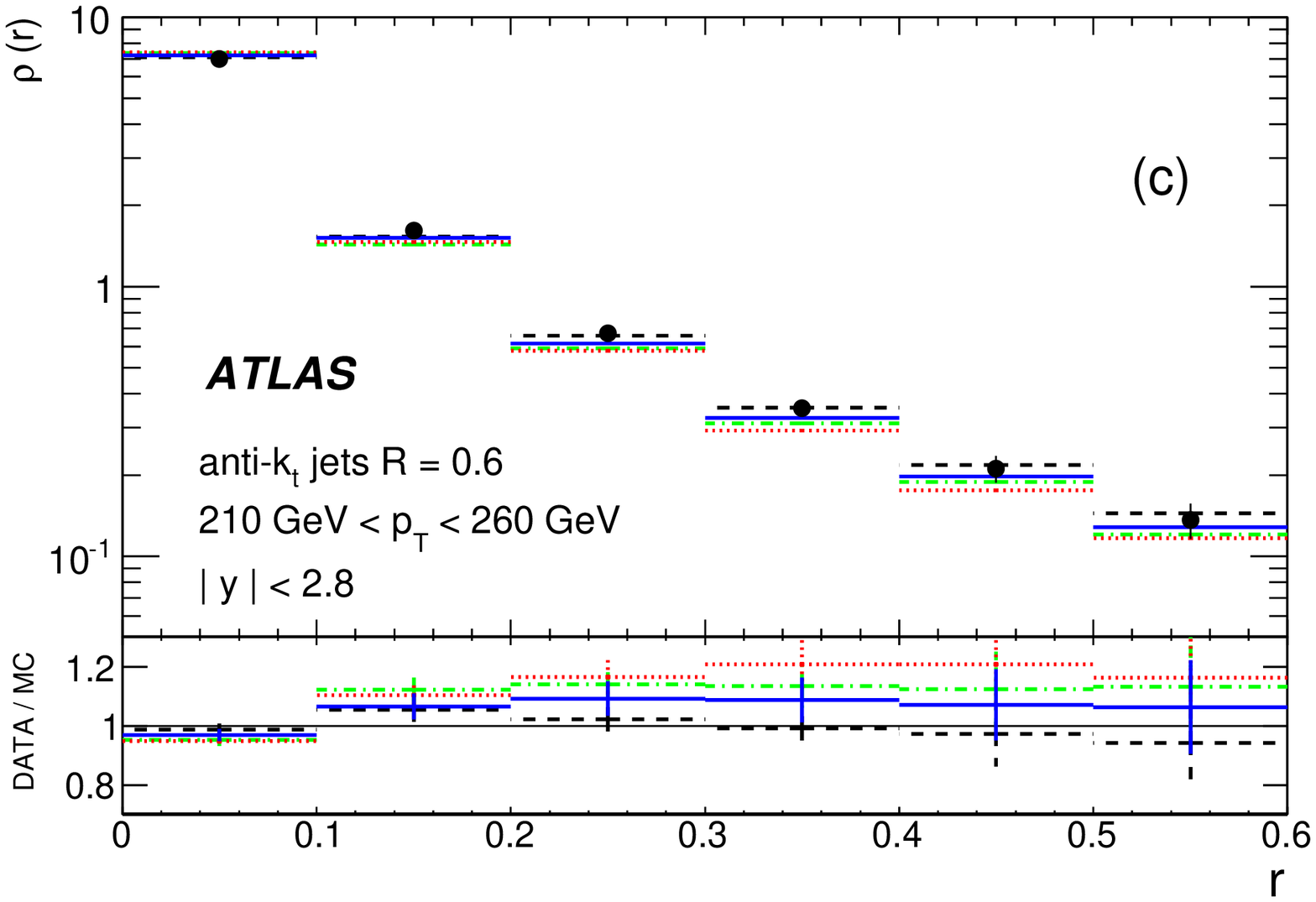} 
\includegraphics[width=0.5\textwidth,height=0.52\textwidth]{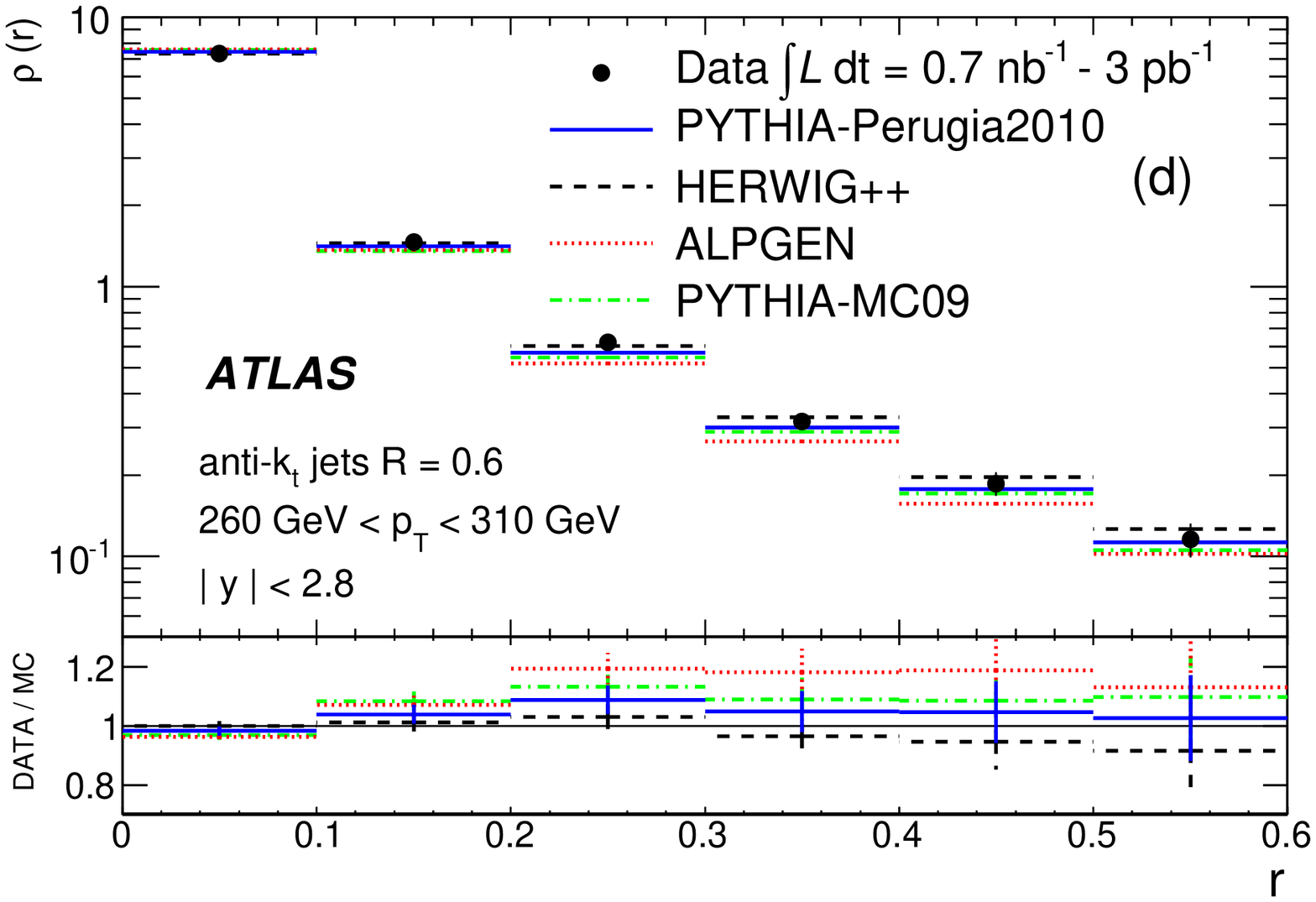} 
}
\end{center}
\vspace{-0.7 cm}
\caption[The measured differential jet shape, $\rho(r)$, in inclusive jet production for jets
with $|\rapjet| < 2.8$ and $110 \ {\rm GeV} < \ptjet < 310  \ {\rm GeV}$]
{\small
The measured differential jet shape, $\rho(r)$, in inclusive jet production for jets 
with $|\rapjet| < 2.8$ and $110 \ {\rm GeV} < \ptjet < 310  \ {\rm GeV}$   
is shown in different $\ptjet$ regions. Error bars indicate the statistical and systematic uncertainties added in quadrature.
The predictions of   PYTHIA-Perugia2010 (solid lines),   HERWIG++ (dashed lines),   ALPGEN interfaced with HERWIG and JIMMY (dotted lines), and 
  PYTHIA-MC09 (dashed-dotted lines) are shown for comparison.} 
\label{fig1b}
\end{figure}

\begin{figure}[tbh]
\begin{center}
\mbox{
\includegraphics[width=0.5\textwidth,height=0.52\textwidth]{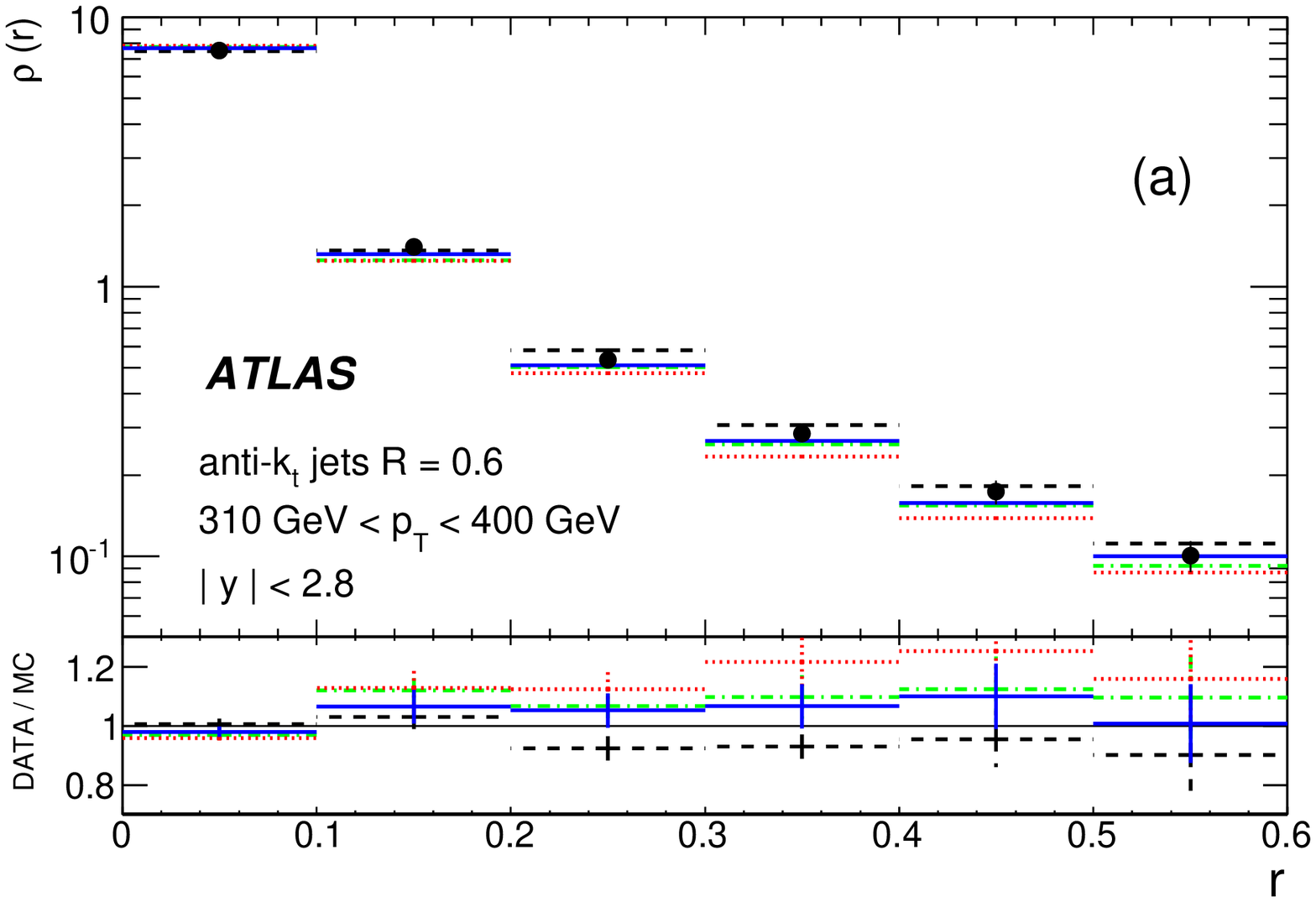} 
\includegraphics[width=0.5\textwidth,height=0.52\textwidth]{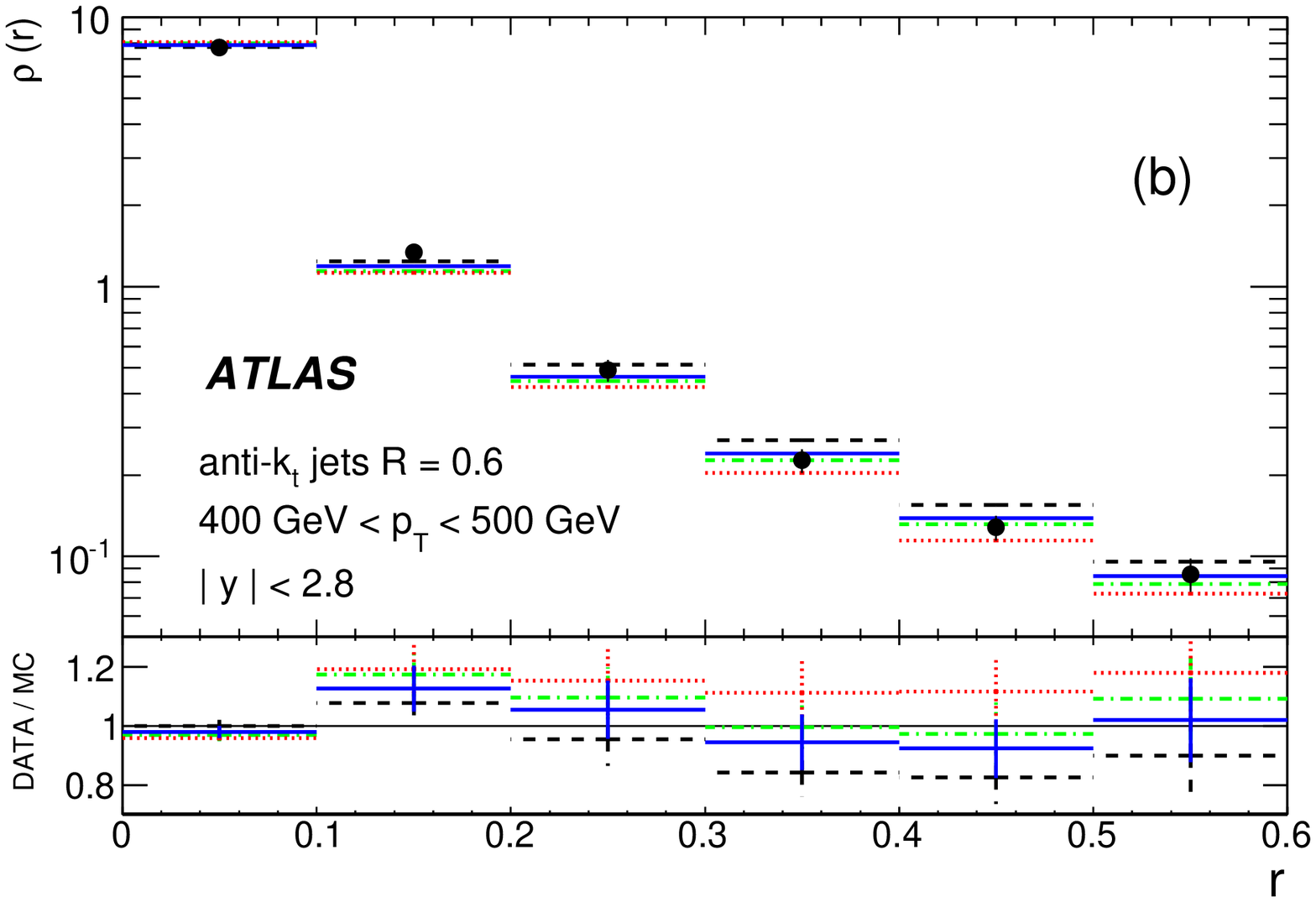}
}\vspace{-0.2 cm}
\mbox{
\includegraphics[width=0.5\textwidth,height=0.52\textwidth]{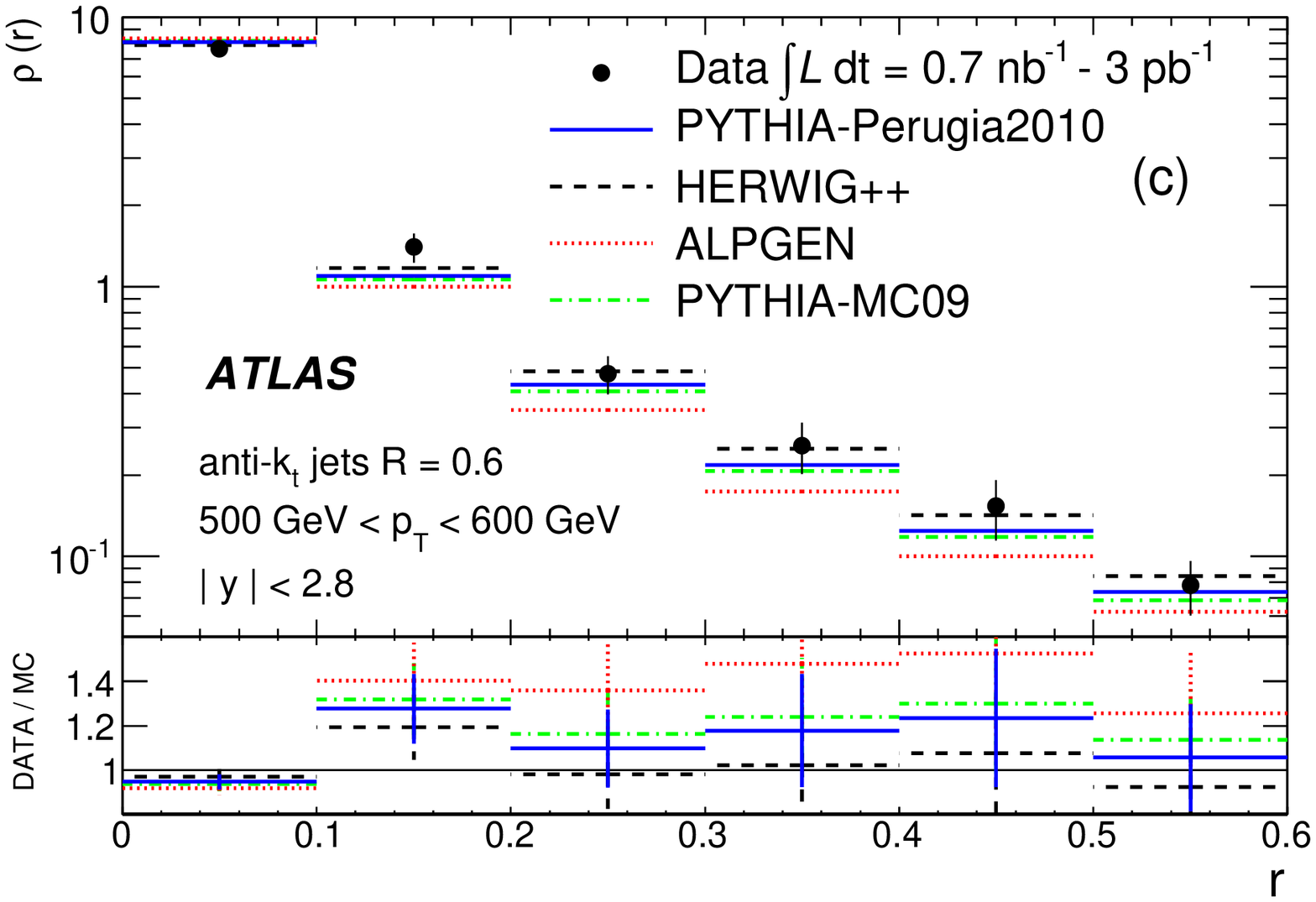}
}\vspace{-0.7 cm}
\caption[The measured differential jet shape, $\rho(r)$, in inclusive jet production for jets
with $|\rapjet| < 2.8$ and $310 \ {\rm GeV} < \ptjet < 600  \ {\rm GeV}$]
{\small
The measured differential jet shape, $\rho(r)$, in inclusive jet production for jets 
with $|\rapjet| < 2.8$ and $310 \ {\rm GeV} < \ptjet < 600  \ {\rm GeV}$   
is shown in different $\ptjet$ regions. Error bars indicate the statistical and systematic uncertainties added in quadrature.
The predictions of   PYTHIA-Perugia2010 (solid lines),   HERWIG++ (dashed lines),   ALPGEN interfaced with HERWIG and JIMMY  (dotted lines), and   PYTHIA-MC09 (dashed-dotted lines) are shown for comparison.
} 
\label{fig2}
\end{center}
\end{figure}

\begin{figure}[tbh]
\begin{center}
\mbox{
\includegraphics[width=0.5\textwidth,height=0.52\textwidth]{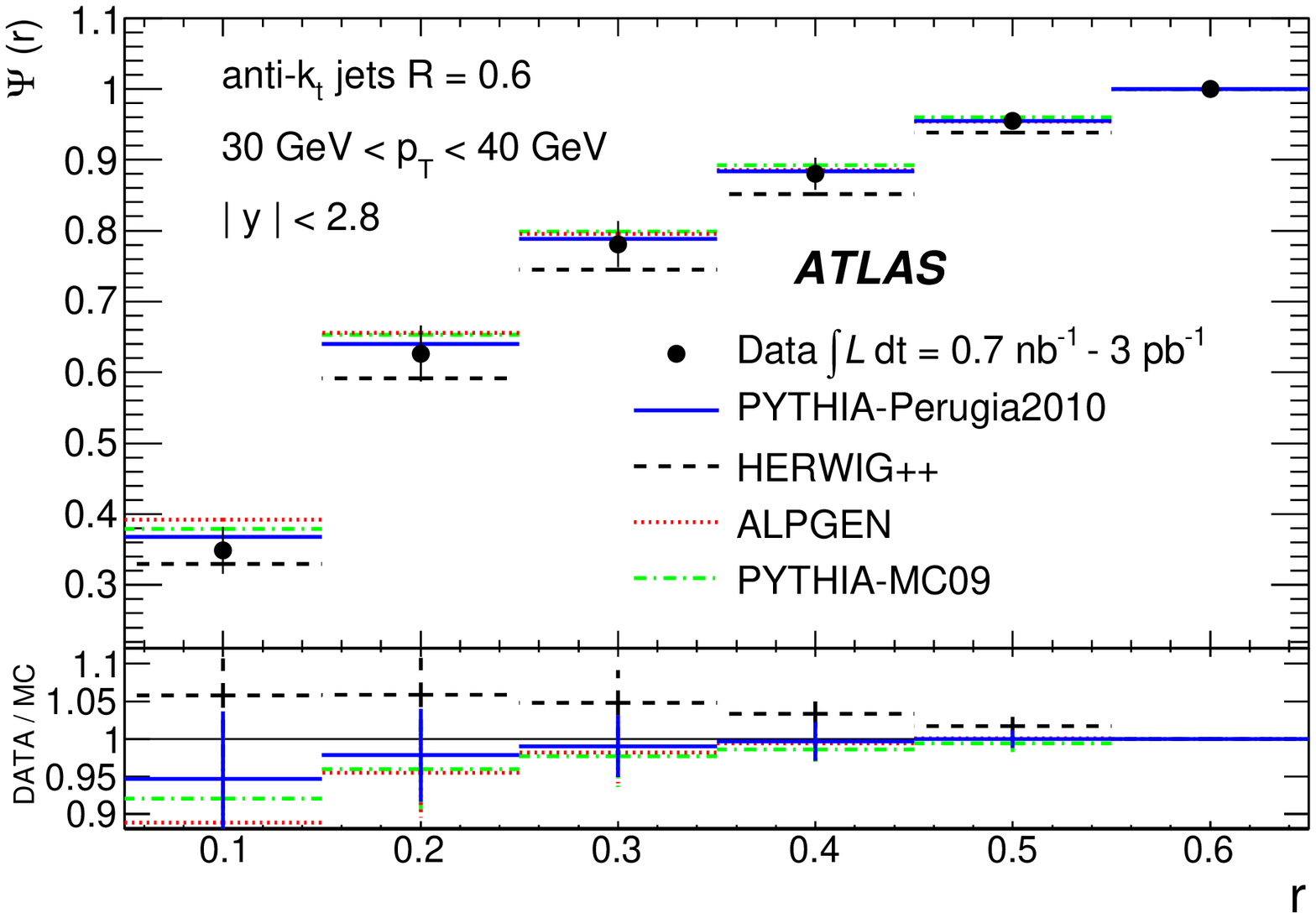}
\includegraphics[width=0.5\textwidth,height=0.52\textwidth]{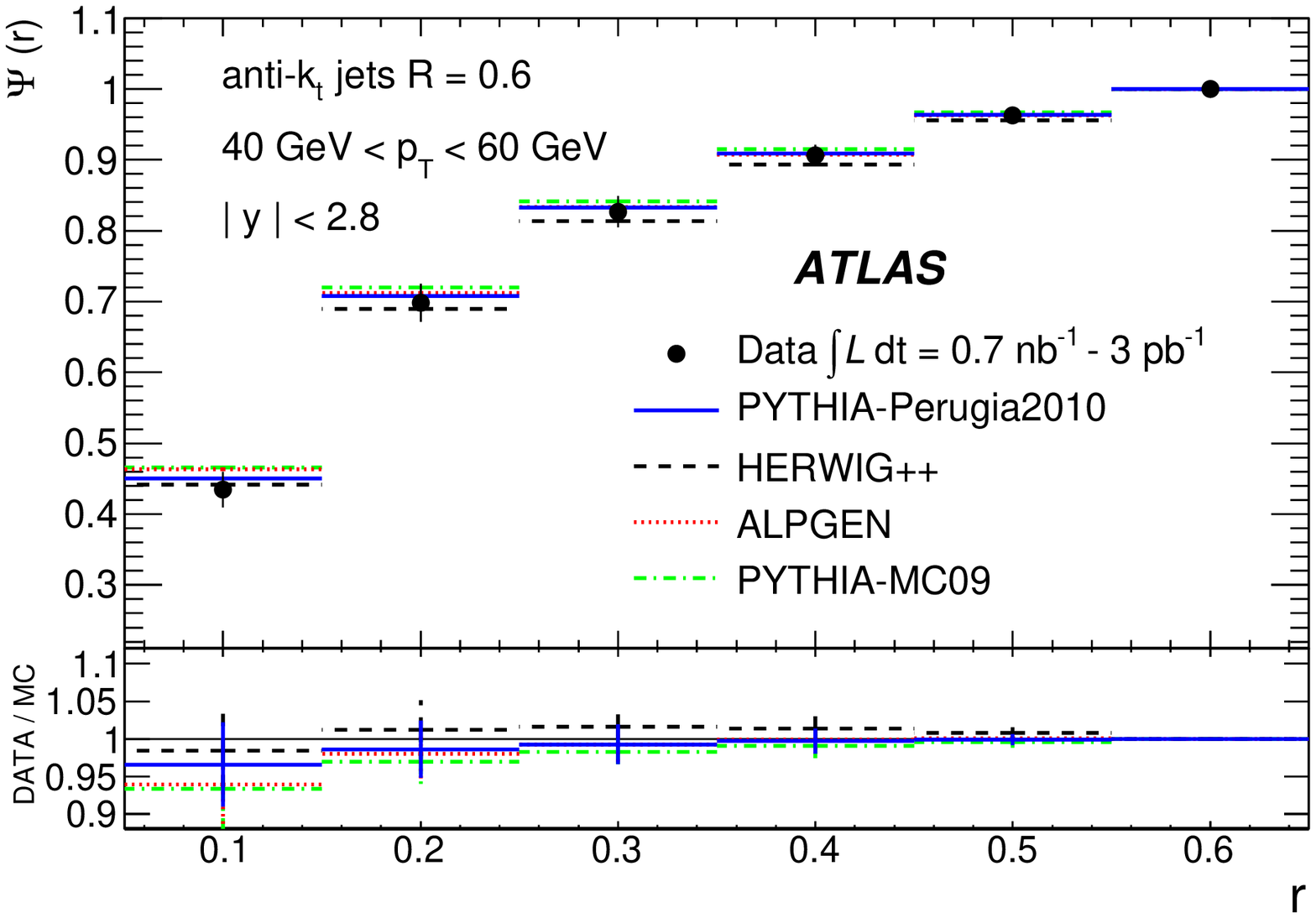}
}\vspace{-0.2cm}
\mbox{
\includegraphics[width=0.5\textwidth,height=0.52\textwidth]{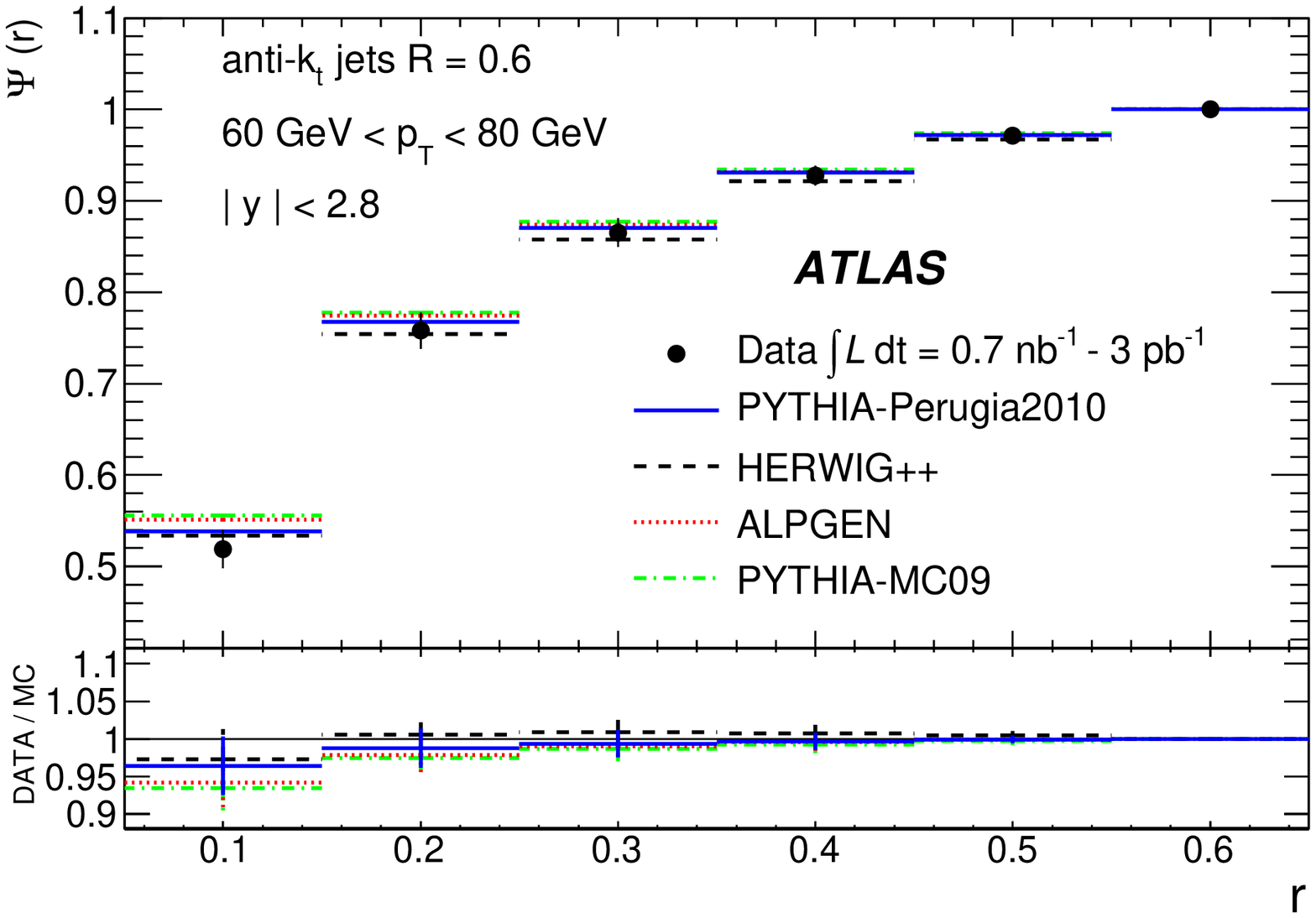}
\includegraphics[width=0.5\textwidth,height=0.52\textwidth]{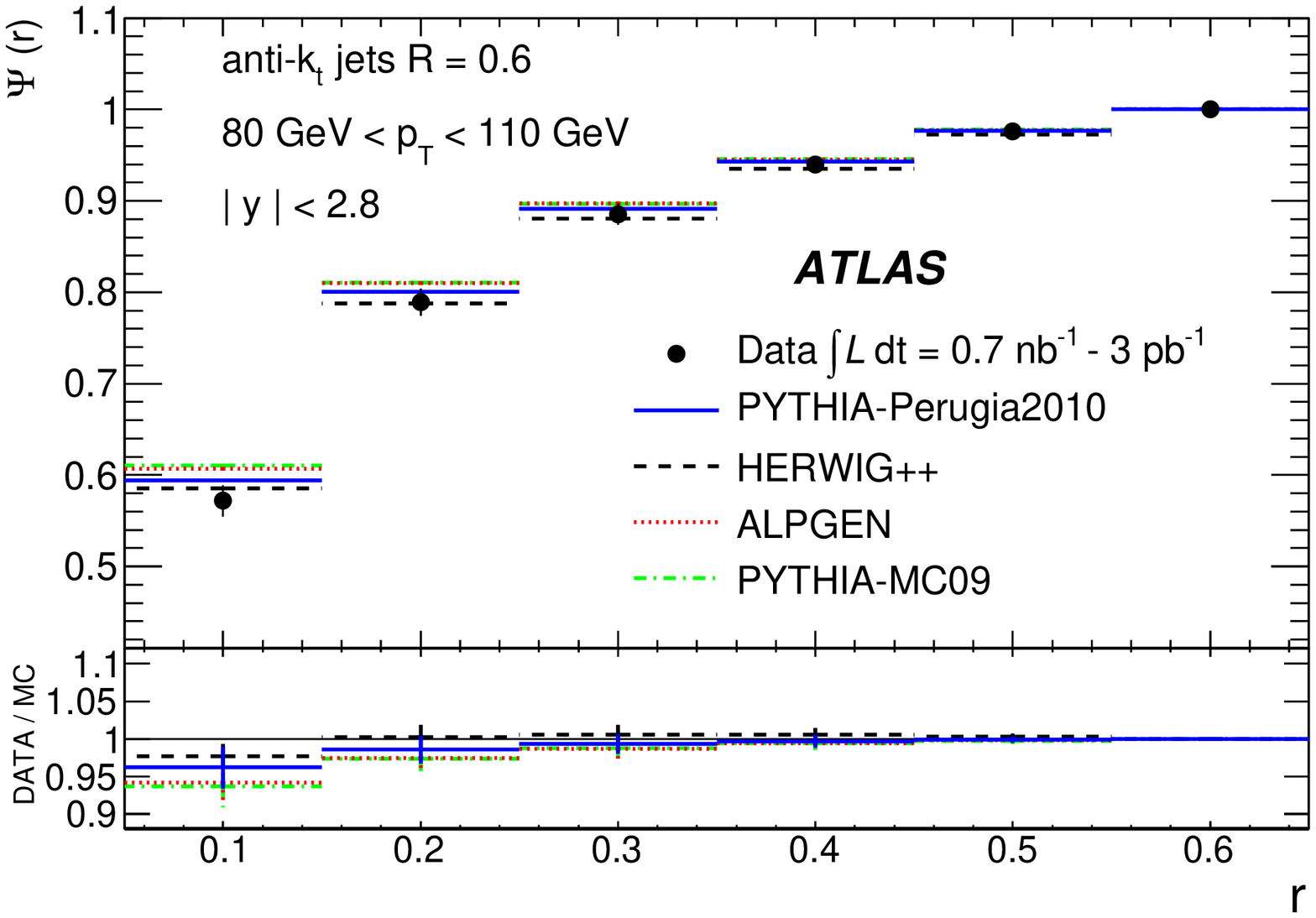}
}
\end{center}
\vspace{-0.7 cm}
\caption[The measured integrated jet shape, $\Psi(r)$, in inclusive jet production for jets
with $|\rapjet| < 2.8$ and $30 \ {\rm GeV} < \ptjet < 110  \ {\rm GeV}$]
{\small
The measured integrated jet shape, $\Psi(r)$, in inclusive jet production for jets
with $|\rapjet| < 2.8$ and $30 \ {\rm GeV} < \ptjet < 110  \ {\rm GeV}$
is shown in different $\ptjet$ regions. Error bars indicate the statistical and systematic uncertainties added in quadrature.
The predictions of   PYTHIA-Perugia2010 (solid lines),   HERWIG++ (dashed lines),   ALPGEN interfaced with HERWIG and JIMMY (dotted lines), and
  PYTHIA-MC09 (dashed-dotted lines) are shown for comparison.}
\label{figaux1}
\end{figure}

\begin{figure}[tbh]
\begin{center}
\mbox{
\includegraphics[width=0.5\textwidth,height=0.52\textwidth]{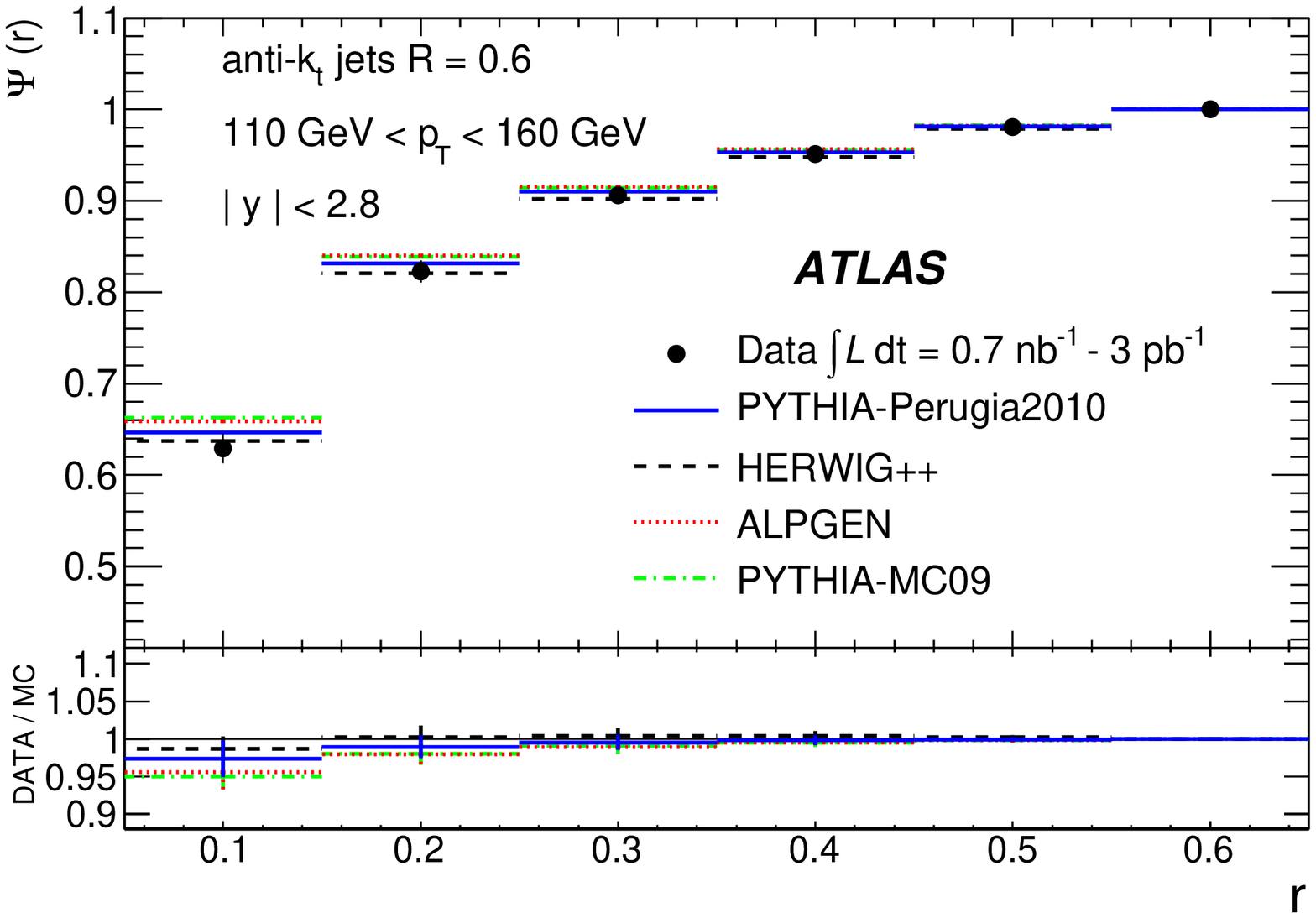}
\includegraphics[width=0.5\textwidth,height=0.52\textwidth]{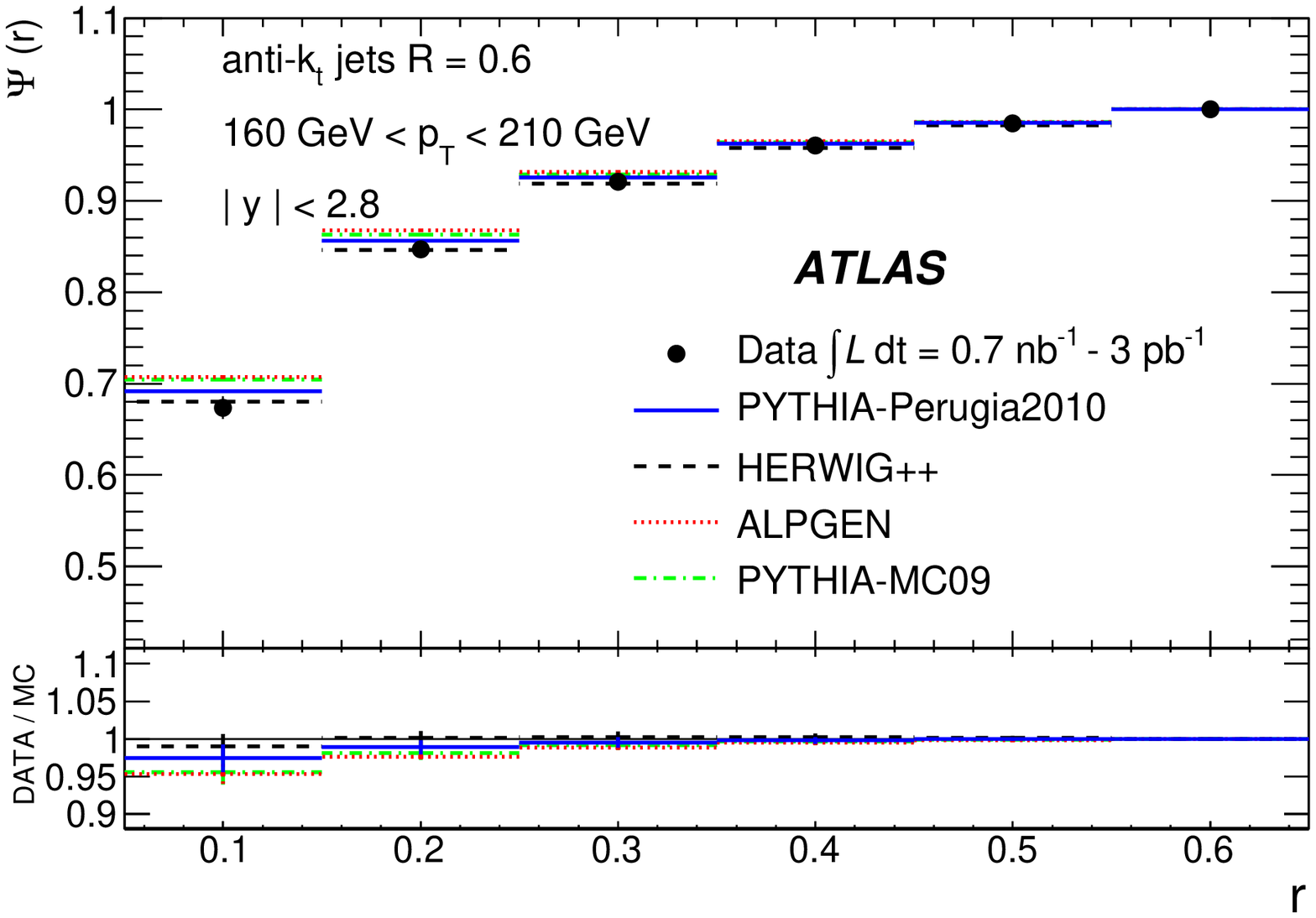}
}\vspace{-0.2cm}
\mbox{
\includegraphics[width=0.5\textwidth,height=0.52\textwidth]{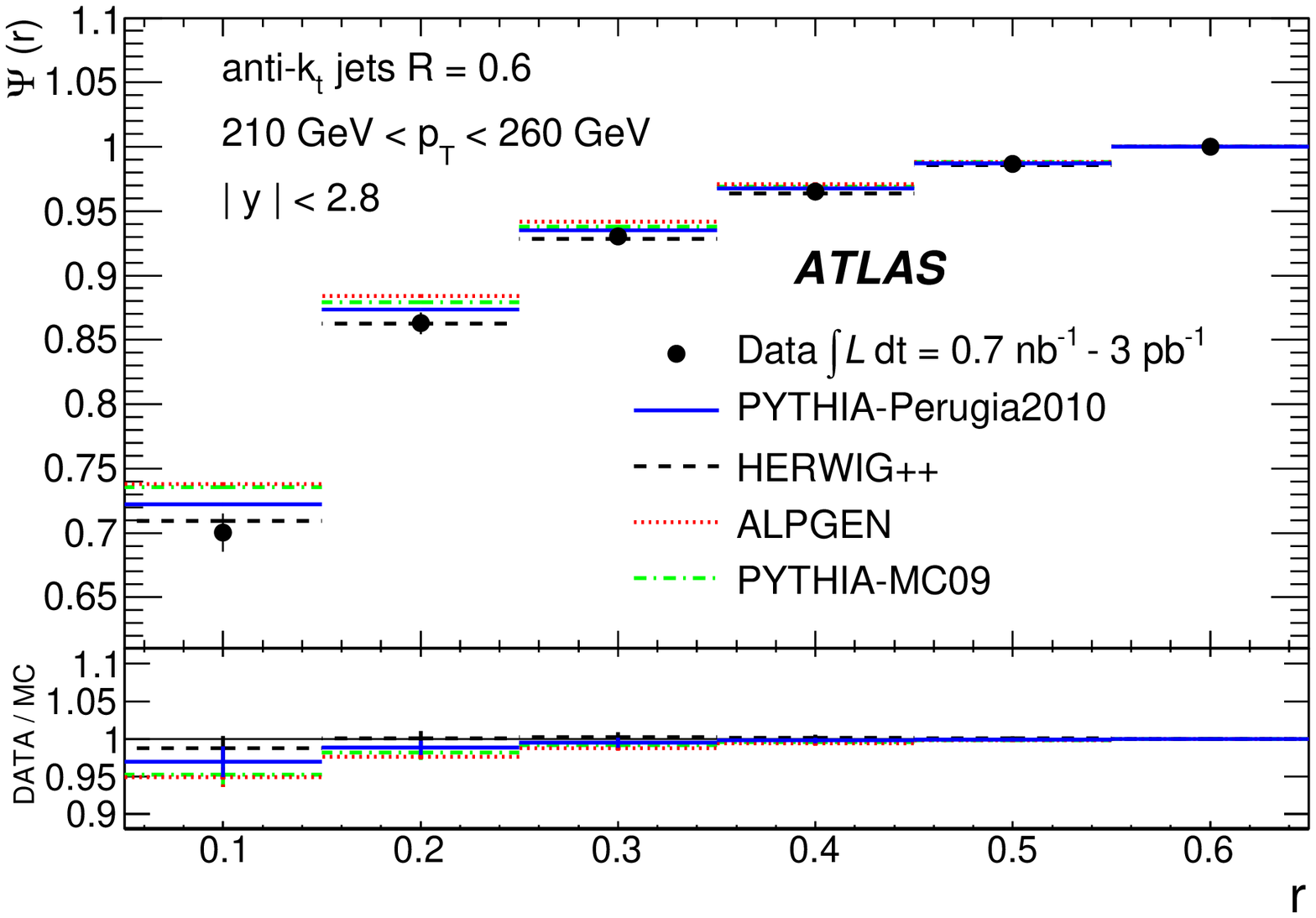}
\includegraphics[width=0.5\textwidth,height=0.52\textwidth]{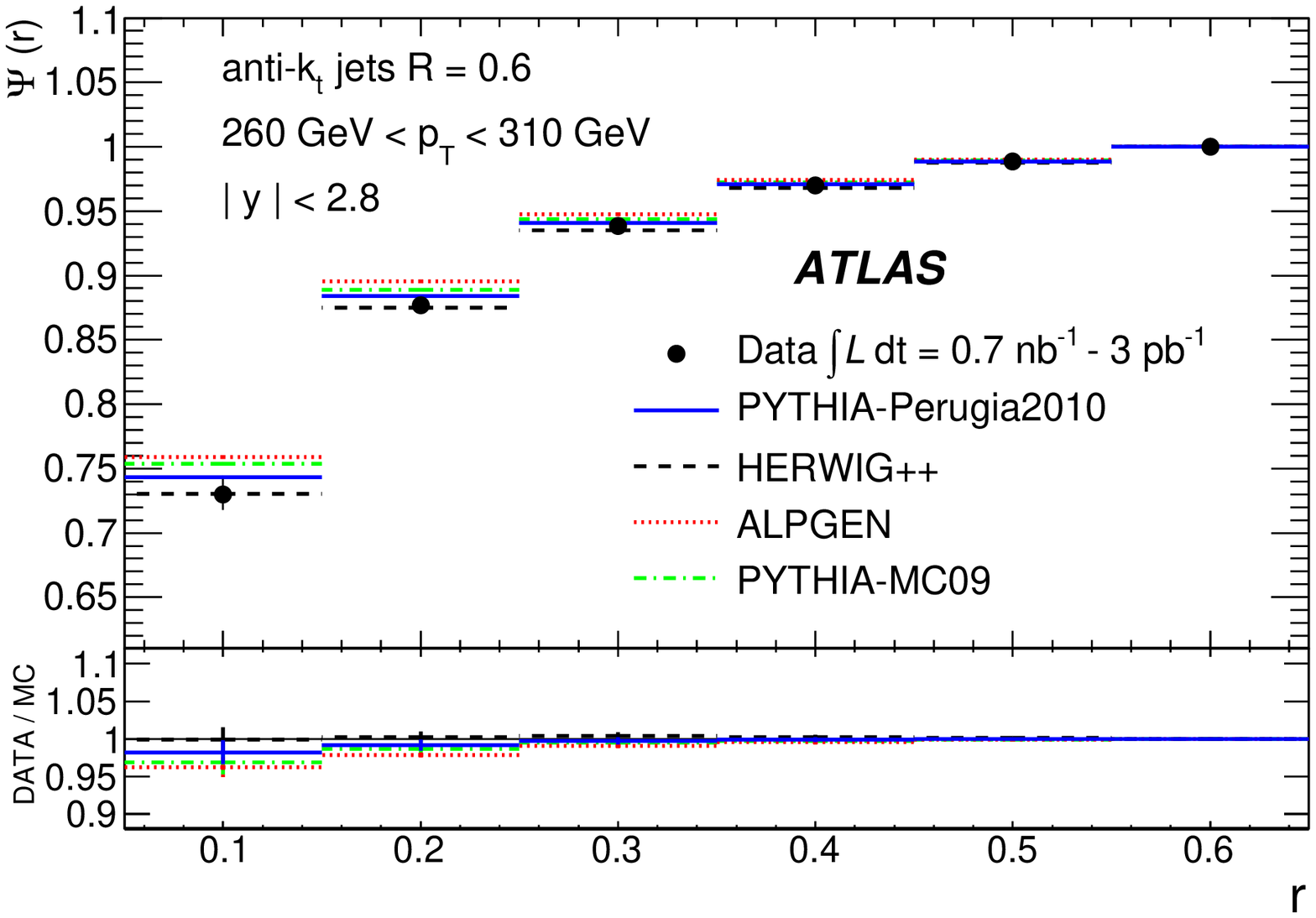}
}
\end{center}
\vspace{-0.7 cm}
\caption[The measured integrated jet shape, $\Psi(r)$, in inclusive jet production for jets
with $|\rapjet| < 2.8$ and $110 \ {\rm GeV} < \ptjet < 310  \ {\rm GeV}$]
{\small
The measured integrated jet shape, $\Psi(r)$, in inclusive jet production for jets
with $|\rapjet| < 2.8$ and $110 \ {\rm GeV} < \ptjet < 310  \ {\rm GeV}$
is shown in different $\ptjet$ regions. Error bars indicate the statistical and systematic uncertainties added in quadrature.
The predictions of   PYTHIA-Perugia2010 (solid lines),   HERWIG++ (dashed lines),   ALPGEN interfaced with HERWIG and JIMMY (dotted lines), and
  PYTHIA-MC09 (dashed-dotted lines) are shown for comparison.}
\label{figaux1b}
\end{figure}

\begin{figure}[tbh]
\begin{center}
\mbox{
\includegraphics[width=0.5\textwidth,height=0.52\textwidth]{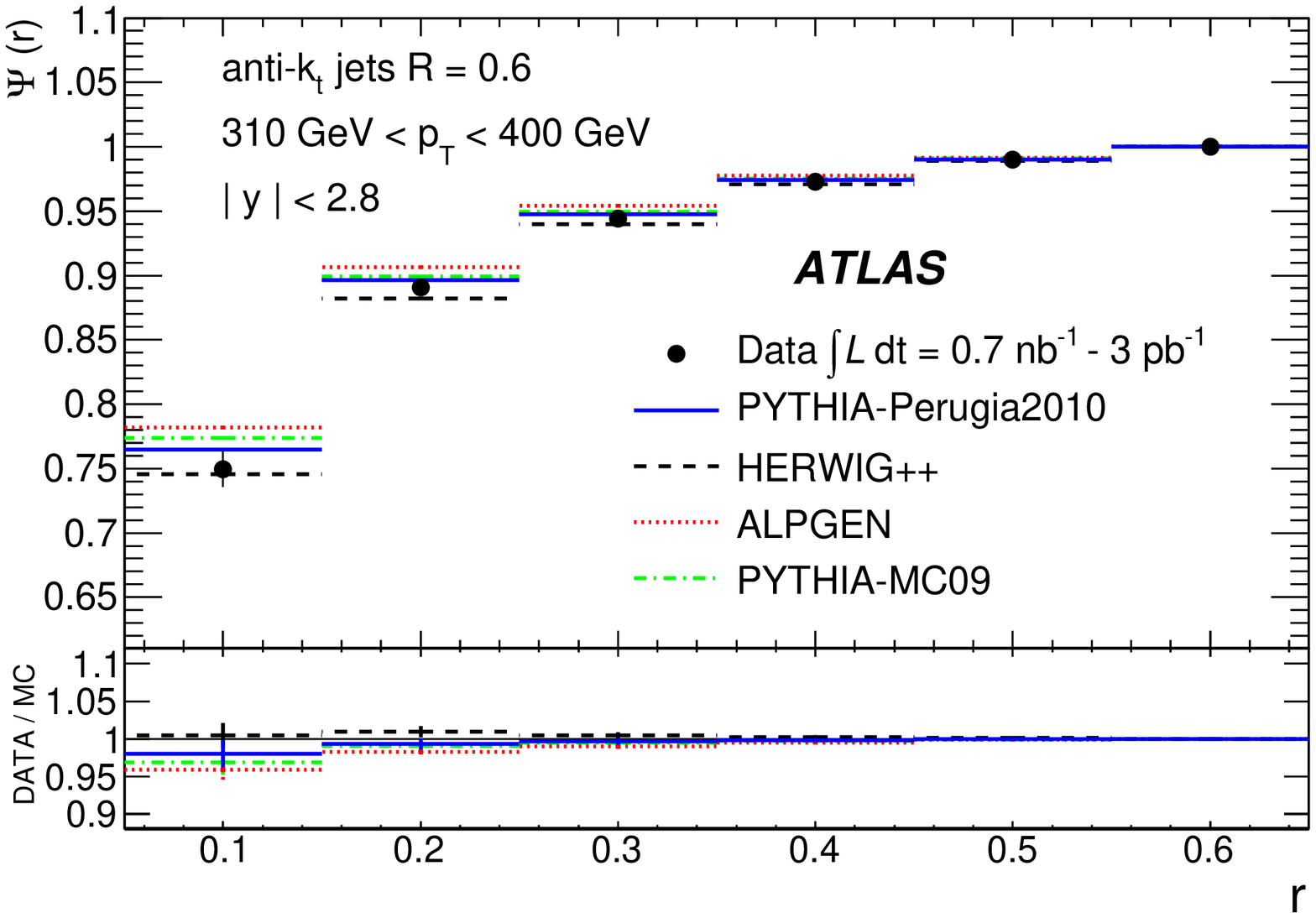}
\includegraphics[width=0.5\textwidth,height=0.52\textwidth]{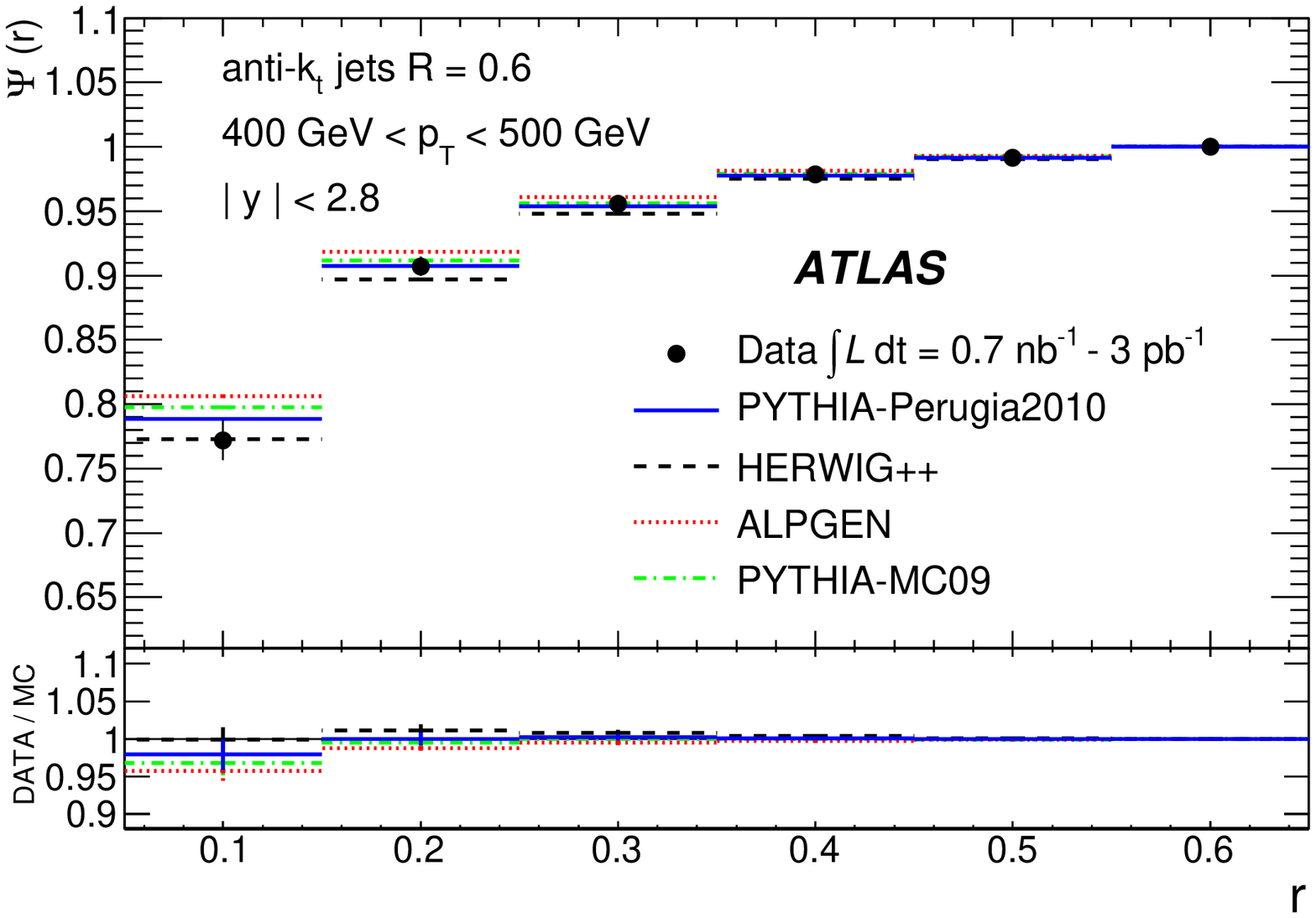}
}\vspace{-0.2 cm}
\mbox{
\includegraphics[width=0.5\textwidth,height=0.52\textwidth]{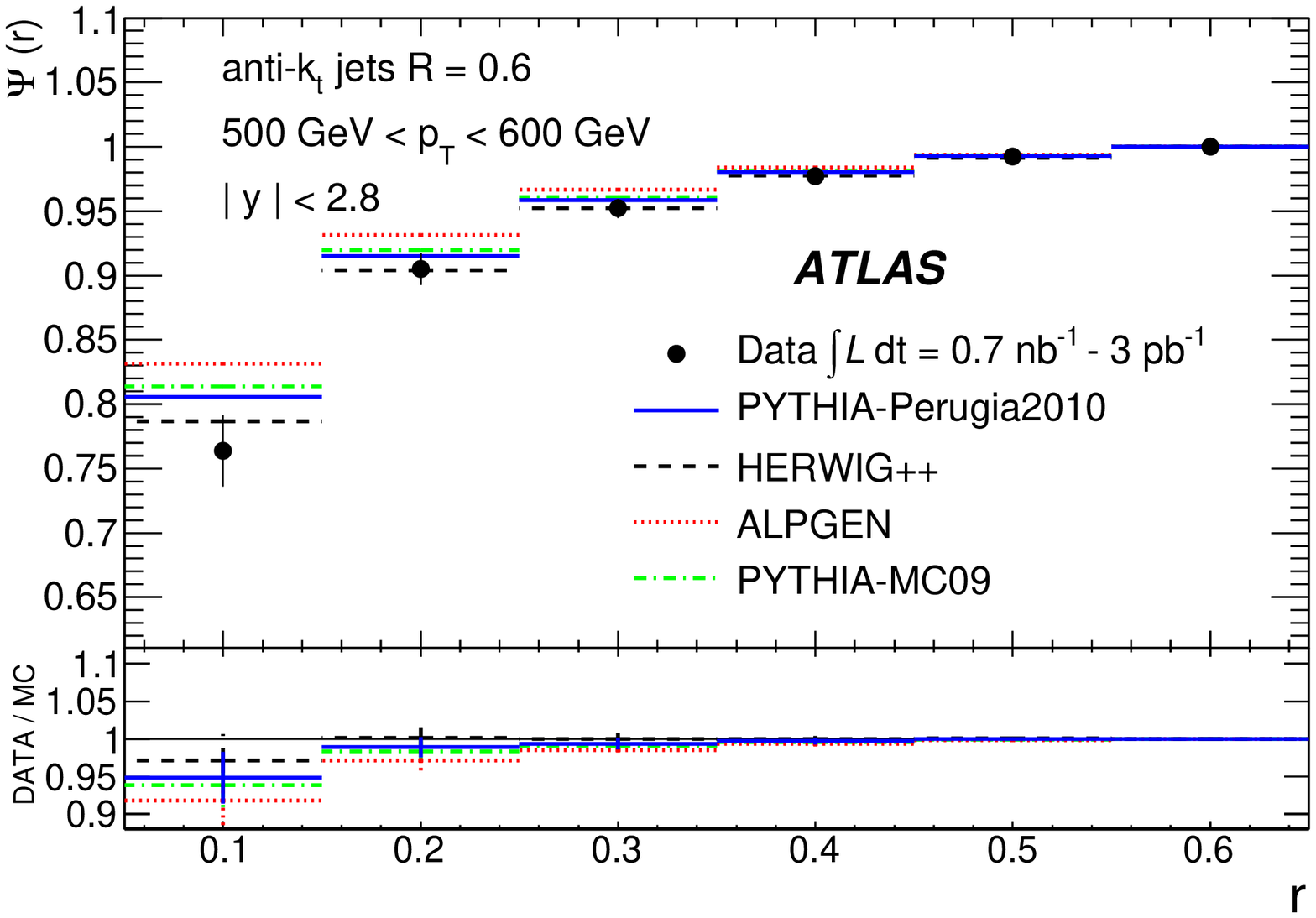}
}\vspace{-0.7 cm}
\caption[The measured integrated jet shape, $\Psi(r)$, in inclusive jet production for jets
with $|\rapjet| < 2.8$ and $310 \ {\rm GeV} < \ptjet < 600  \ {\rm GeV}$]
{\small
The measured integrated jet shape, $\Psi(r)$, in inclusive jet production for jets
with $|\rapjet| < 2.8$ and $310 \ {\rm GeV} < \ptjet < 600  \ {\rm GeV}$
is shown in different $\ptjet$ regions. Error bars indicate the statistical and systematic uncertainties added in quadrature.
The predictions of   PYTHIA-Perugia2010 (solid lines),   HERWIG++ (dashed lines),   ALPGEN interfaced with HERWIG and JIMMY  (dotted lines), and   PYTHIA-MC09 (dashed-dotted lines) are shown for comparison.
}
\label{figaux2}
\end{center}
\end{figure}

\begin{figure}[tbh]
\begin{center}
\mbox{
\includegraphics[width=0.8\textwidth]{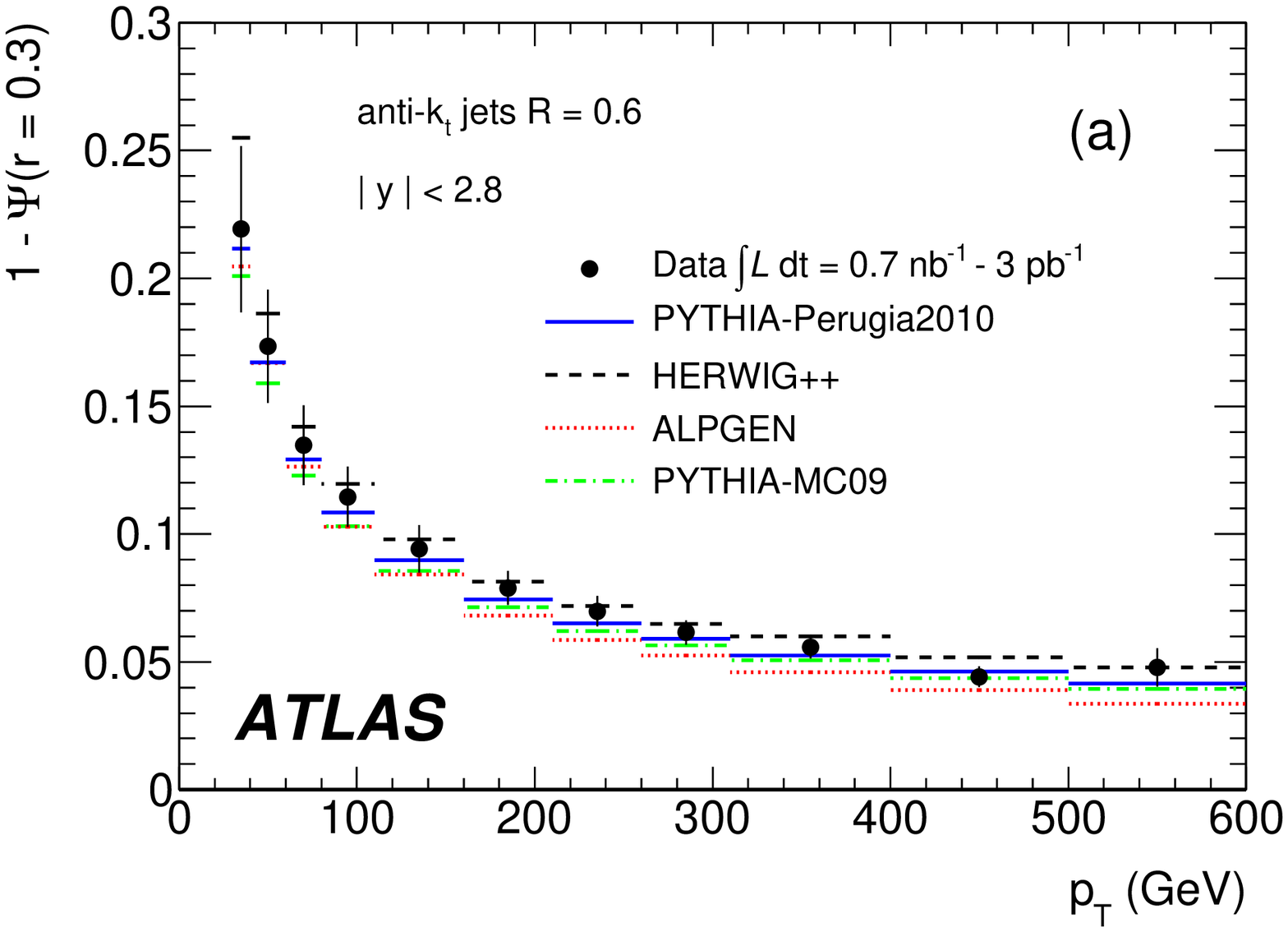} 
} \\
\mbox{
\includegraphics[width=0.8\textwidth]{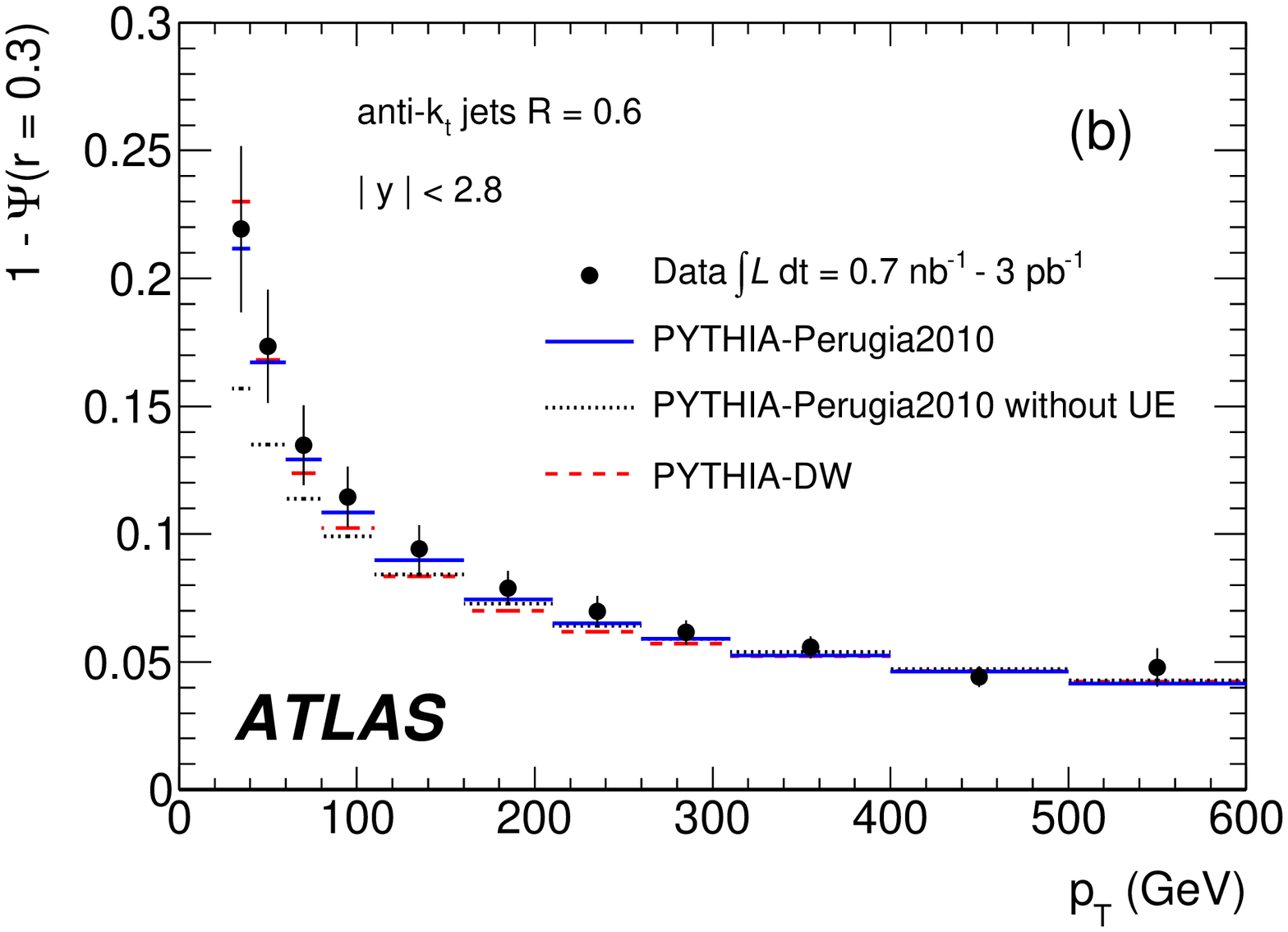} 
}
\end{center}
\vspace{-0.7 cm}
\caption[The measured integrated jet shape, $1 - \Psi(r=0.3)$, as a function of $\ptjet$
for jets with $|\rapjet| < 2.8$ and $30 \ {\rm GeV} < \ptjet < 600 \ {\rm GeV}$.]
{\small
The measured integrated jet shape, $1 - \Psi(r=0.3)$, as a function of $\ptjet$ 
for jets with $|\rapjet| < 2.8$ and $30 \ {\rm GeV} < \ptjet < 600 \ {\rm GeV}$.
Error bars indicate the statistical and systematic uncertainties added in quadrature.
The data are compared to the predictions of:   
(a) PYTHIA-Perugia2010 (solid lines),   HERWIG++ (dashed lines),   ALPGEN interfaced with HERWIG and JIMMY (dotted lines), and   PYTHIA-MC09 (dashed-dotted lines); (b) PYTHIA-Perugia2010 (solid lines),   PYTHIA-Perugia2010 without UE   (dotted lines), and   PYTHIA-DW (dashed lines).
} 
\label{fig3}
\end{figure}

Figures~\ref{fig1} to~\ref{fig2} (\ref{figaux1} to~\ref{figaux2}) show the measured differential (integrated)
jet shapes as a function of $r$ in different $\ptjet$ ranges. 
The dominant peak at small $r$ indicates that the 
majority of the jet momentum is concentrated close to the jet axis.  
At low  $\ptjet$,  more than 80$\%$ of the transverse
momentum is contained within a cone of radius $r=0.3$ around the jet direction.
This  fraction increases up to 95$\%$ at very high $\ptjet$, showing 
that jets become narrower as $\ptjet$ increases.  
This is also observed in Figure~\ref{fig3}, where the measured $1-\Psi(0.3)$, the 
fraction of the jet transverse momentum outside a fixed radius $r=0.3$, decreases as a function of $\ptjet$.  


The data are compared to predictions from   HERWIG++, ALPGEN,   PYTHIA-Perugia2010, and 
PYTHIA-MC09 in Figure~\ref{fig1} to Fig.~\ref{fig3}(a); and to predictions from 
PYTHIA-DW and PYTHIA-Perugia2010 with and without UE contributions in Figure~\ref{fig3}(b).
The jet shapes predicted by PYTHIA-Perugia2010 provide a reasonable description of the data,
while HERWIG++ predicts broader jets than the data at low and very high $\ptjet$.
The PYTHIA-DW predictions are in between PYTHIA-Perugia2010 and HERWIG++ at low $\ptjet$ and produce jets which are 
slightly narrower  at high $\ptjet$. 
ALPGEN is similar to PYTHIA-Perugia2010 at low $\ptjet$, but produces  
jets significantly narrower than the data  
at high $\ptjet$.  PYTHIA-MC09 tends to produce narrower jets  
than the data in the whole kinematic range under study. 
The latter may be  attributed to an
inadequate modeling of the soft gluon radiation and UE
contributions in   PYTHIA-MC09 samples, in agreement with previous
observations of the particle flow activity in the final state~\cite{mbts}. 
Finally, Figure~\ref{fig3}(b) shows that 
PYTHIA-Perugia2010 without UE contributions predicts jets much narrower than the 
data at low $\ptjet$. This  confirms the sensitivity of jet shape observables in the region 
$\ptjet < 160$~GeV to a proper description of the  UE activity in the final state~\footnote{These conclusions 
are supported by energy flow results at detector level, reported in Appendix~\ref{appendix_eflow}.}.
 
\begin{figure}[tbh]
\begin{center}
\mbox{
\includegraphics[width=0.495\textwidth]{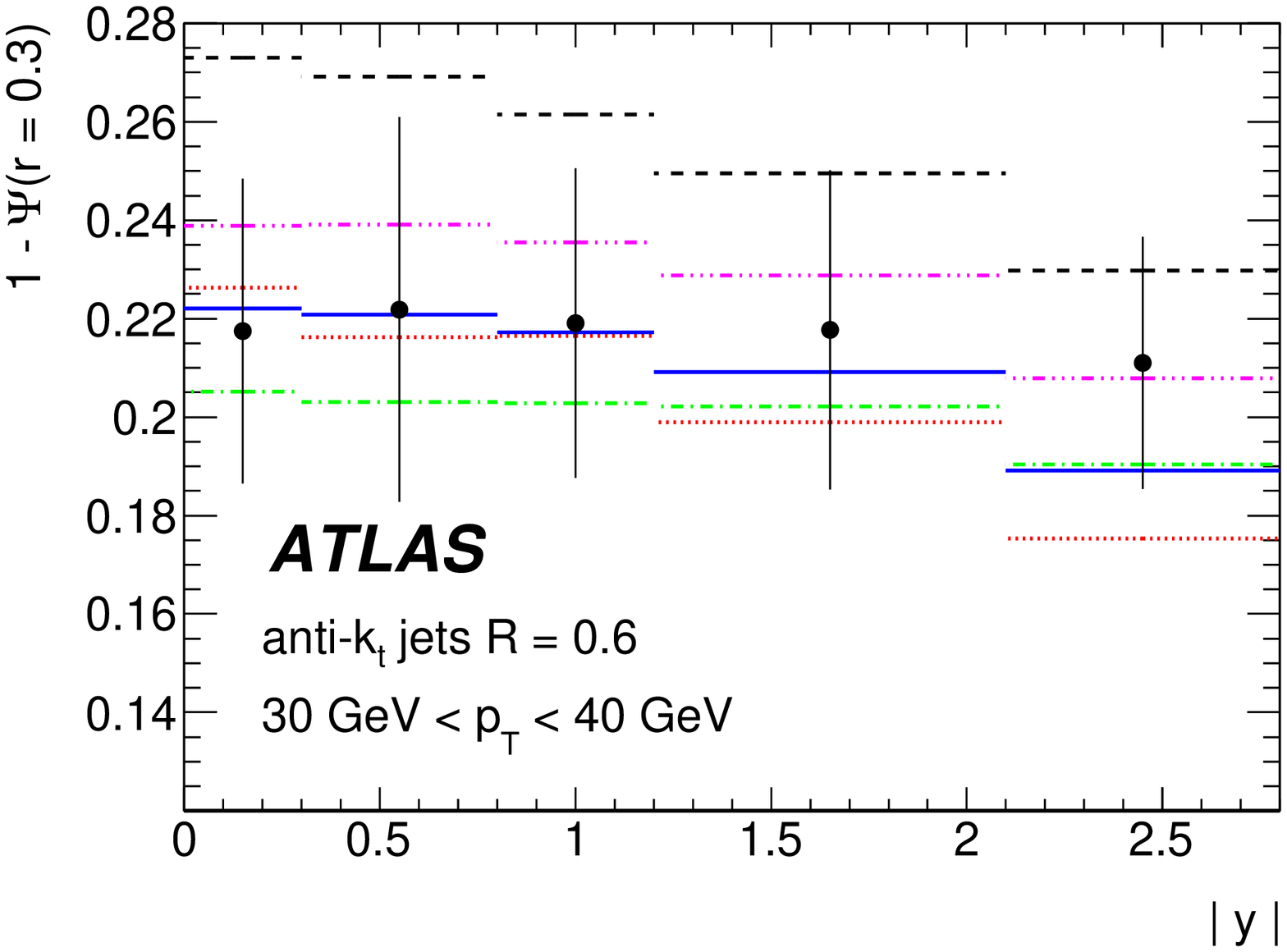} 
\includegraphics[width=0.495\textwidth]{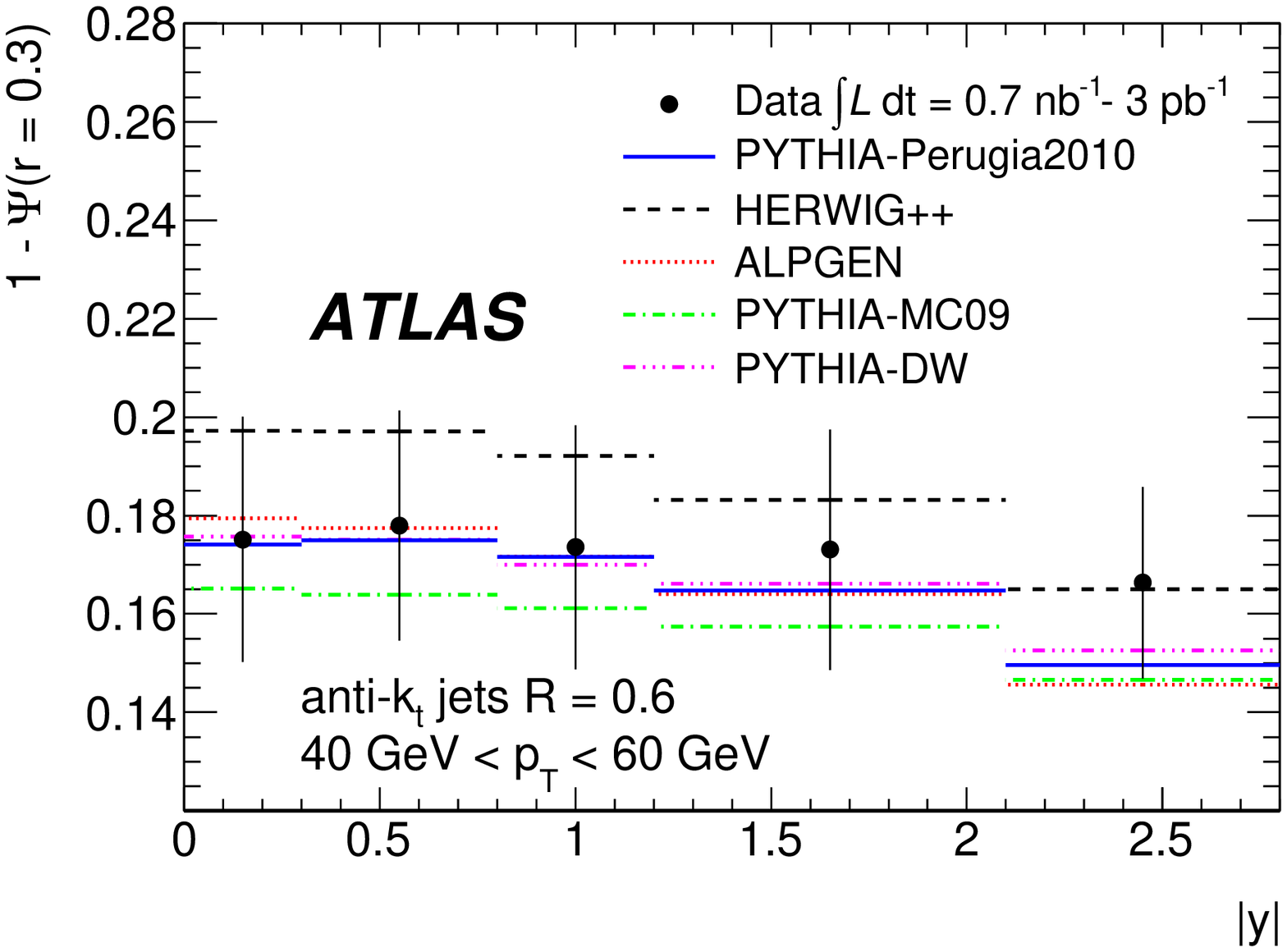} 
}
\mbox{
\includegraphics[width=0.495\textwidth]{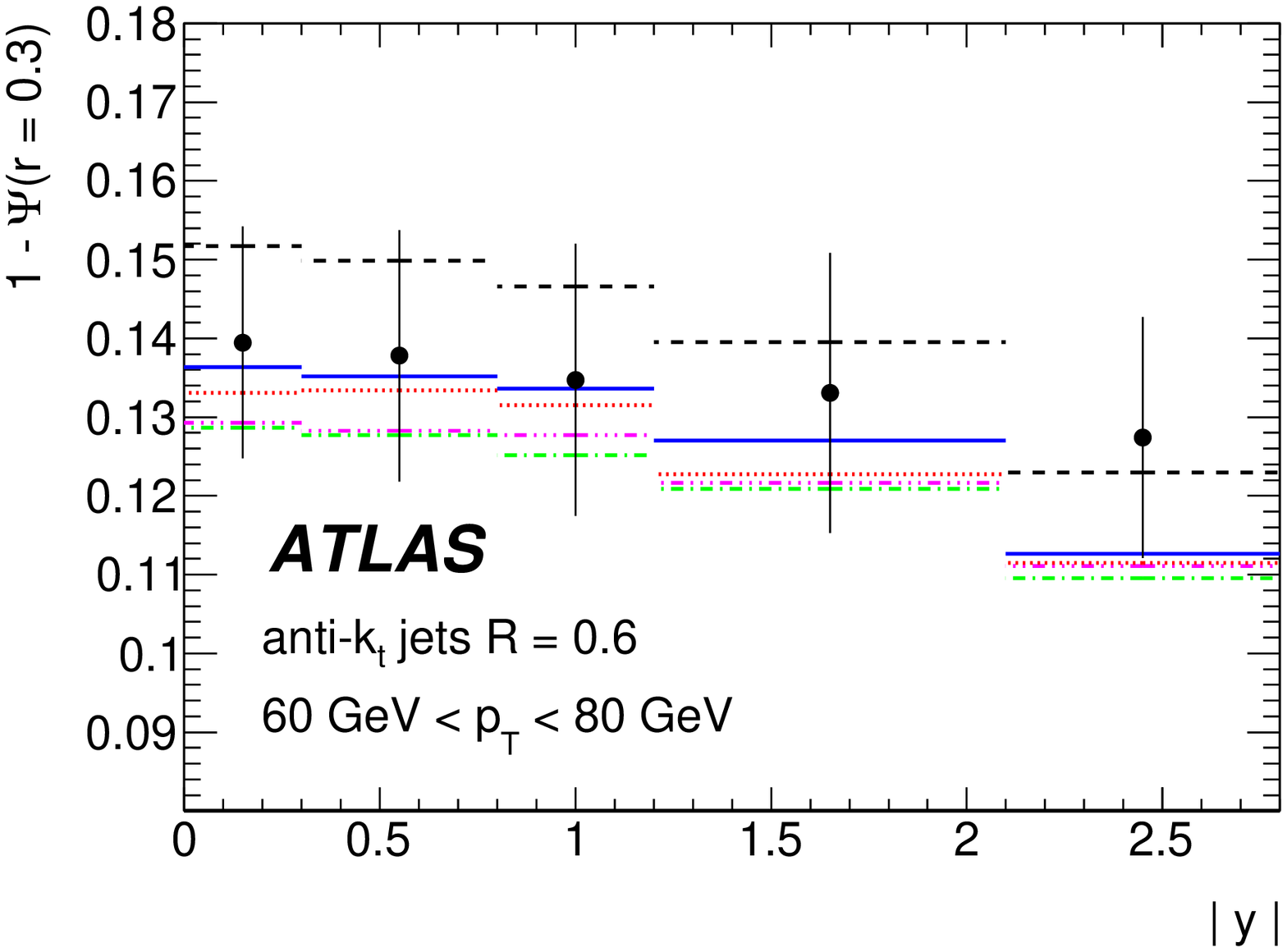}
\includegraphics[width=0.495\textwidth]{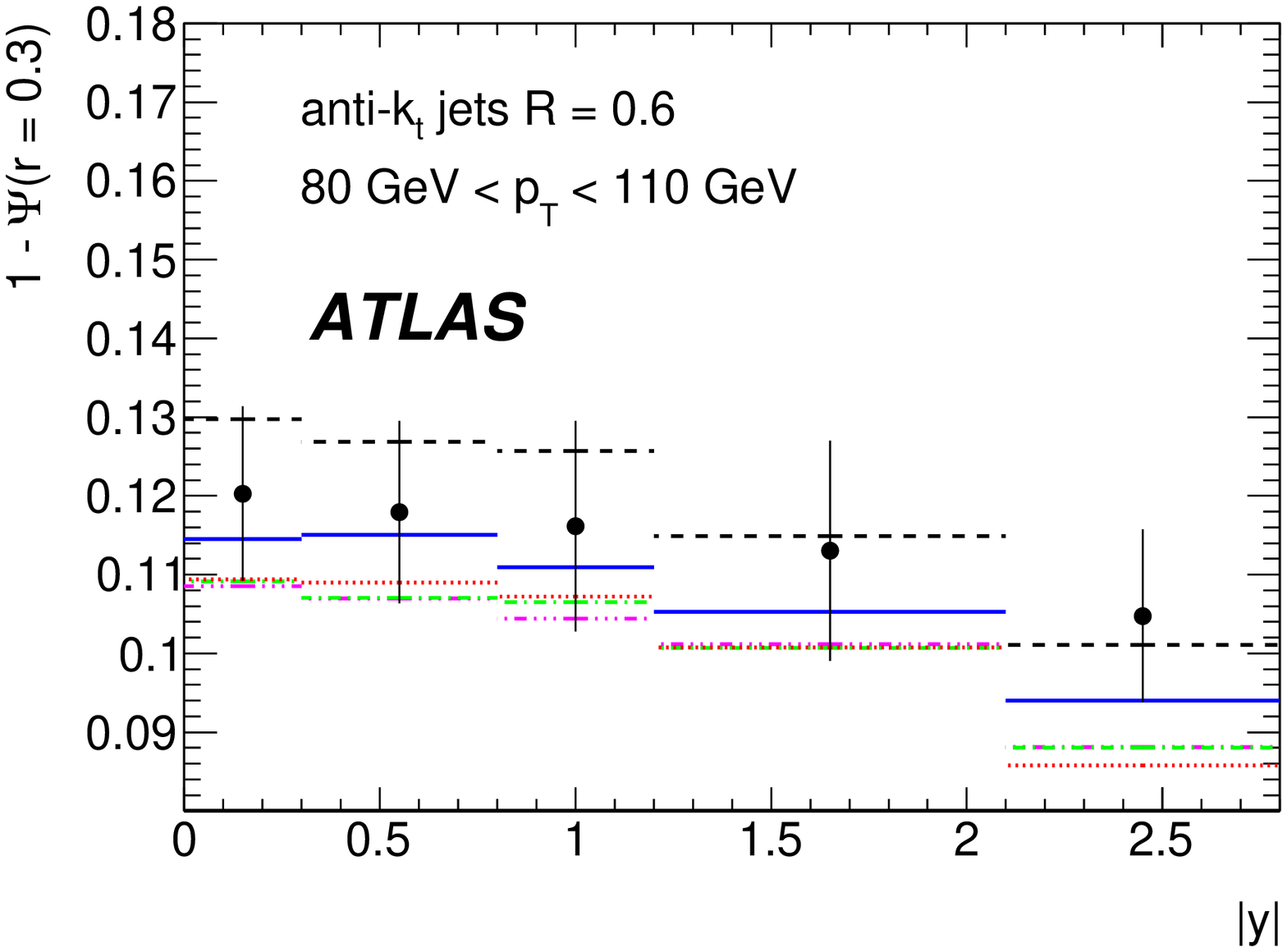} 
} 
\mbox{
\includegraphics[width=0.495\textwidth]{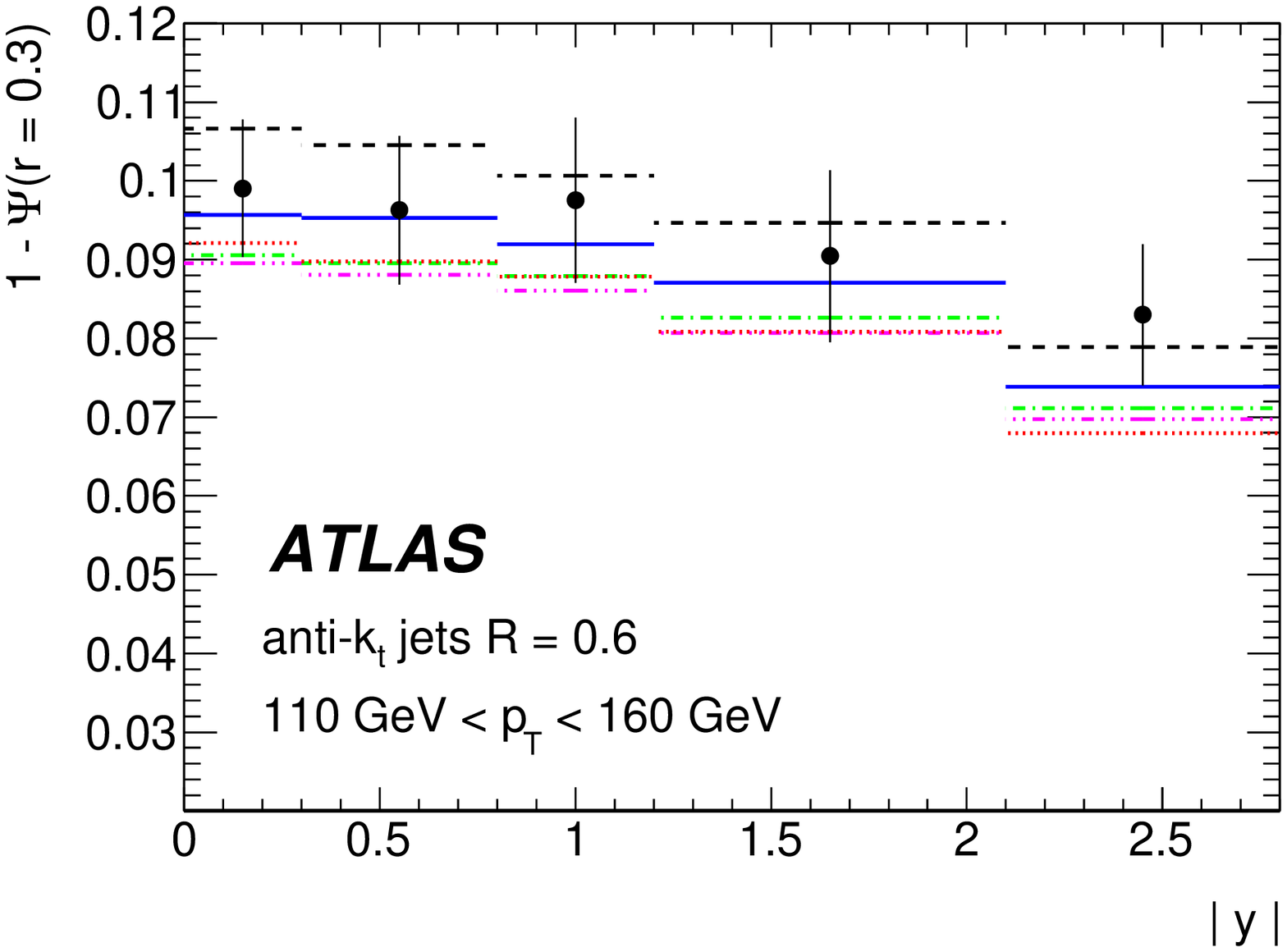} 
\includegraphics[width=0.495\textwidth]{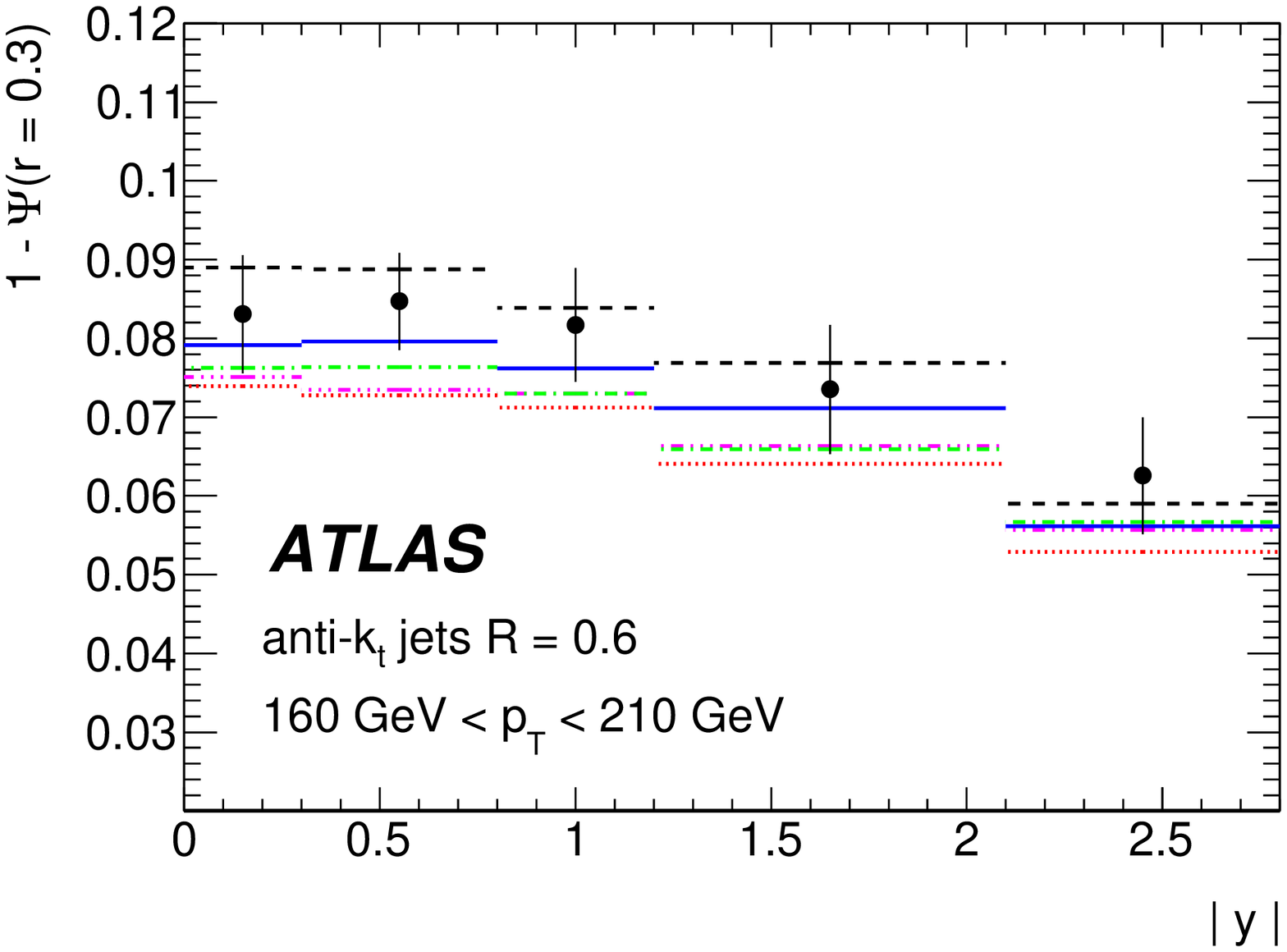}
}
\end{center}
\vspace{-0.7 cm}
\caption[The measured integrated jet shape, $1 - \Psi(r=0.3)$, as a function of $|\rapjet|$
for jets with $|\rapjet| < 2.8$ and $30 \ {\rm GeV} < \ptjet < 210 \ {\rm GeV}$ in different $y$ bins]
{\small
The measured integrated jet shape, $1 - \Psi(r=0.3)$, as a function of $|\rapjet|$ 
for jets with $|\rapjet| < 2.8$ and $30 \ {\rm GeV} < \ptjet < 210 \ {\rm GeV}$.
Error bars indicate the statistical and systematic uncertainties added in quadrature. 
The predictions of   PYTHIA-Perugia2010 (solid lines),   HERWIG++ (dashed lines),   ALPGEN interfaced with HERWIG and JIMMY (dotted lines), PYTHIA-MC09 (dashed-dotted lines), and PYTHIA-DW (dashed-dotted-dotted lines) are shown for comparison.
} 
\label{fig4}
\end{figure}

\begin{figure}[tbh]
\begin{center}
\mbox{
\includegraphics[width=0.495\textwidth]{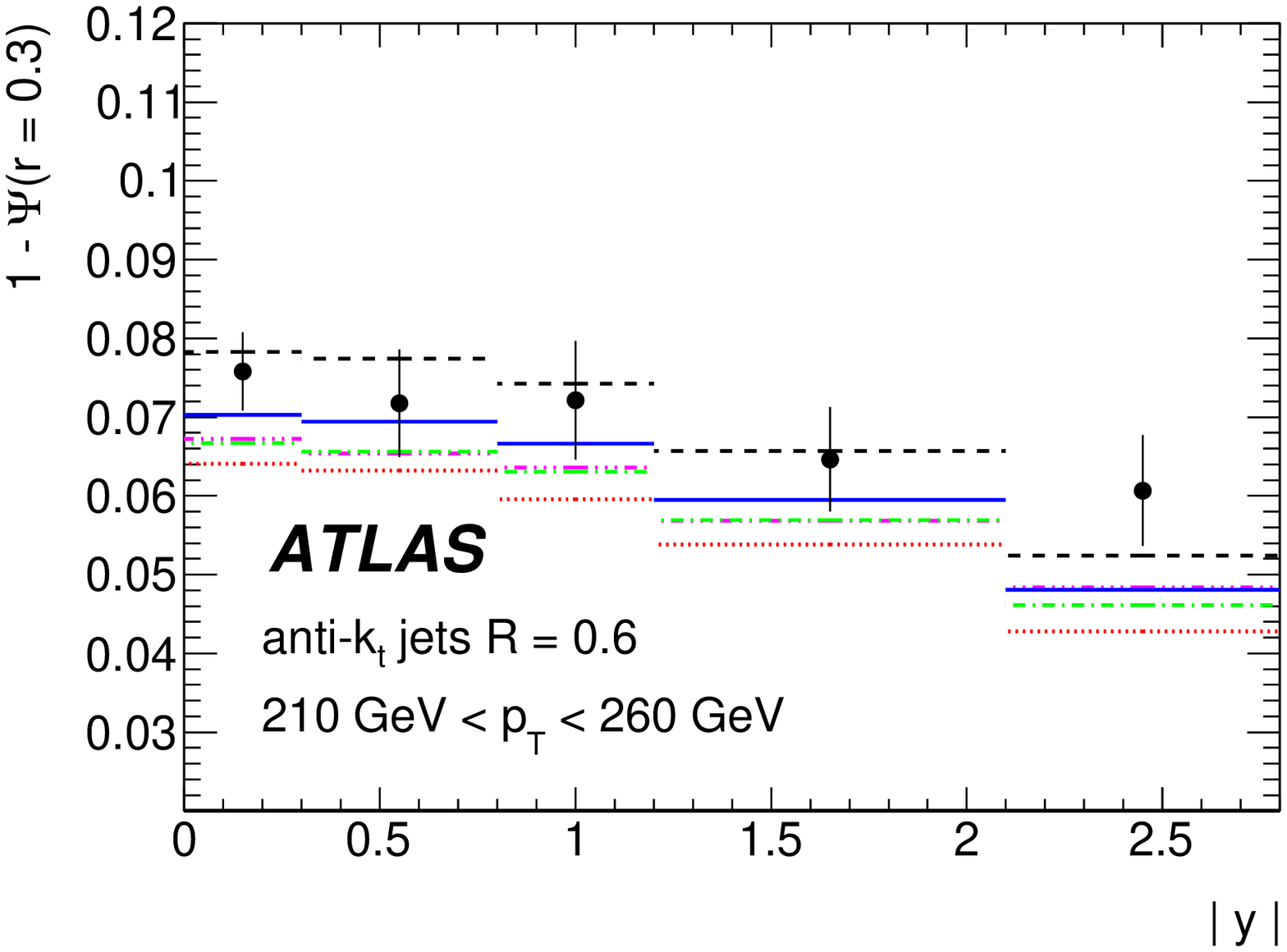}
\includegraphics[width=0.495\textwidth]{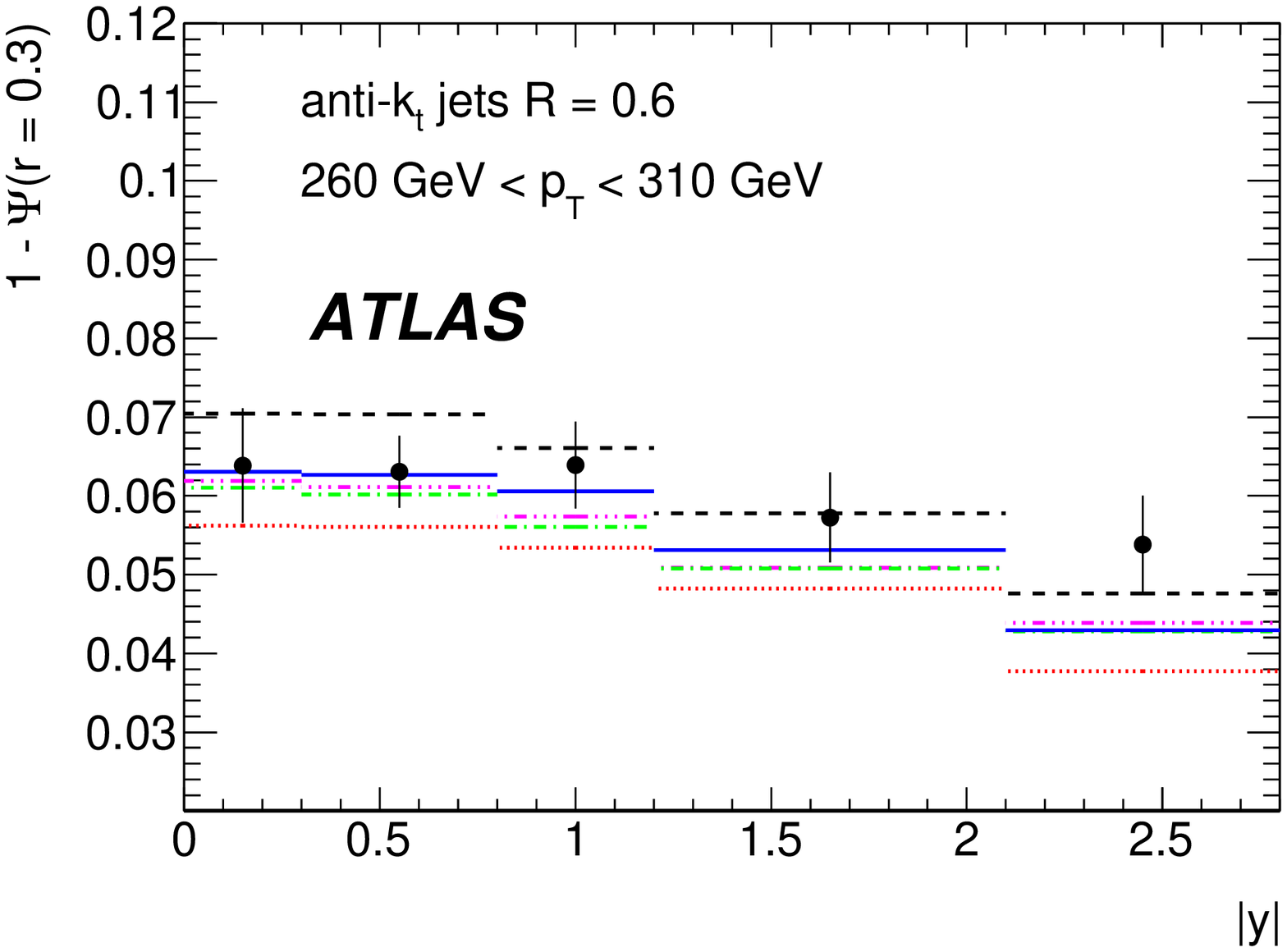}
}
\mbox{
\includegraphics[width=0.495\textwidth]{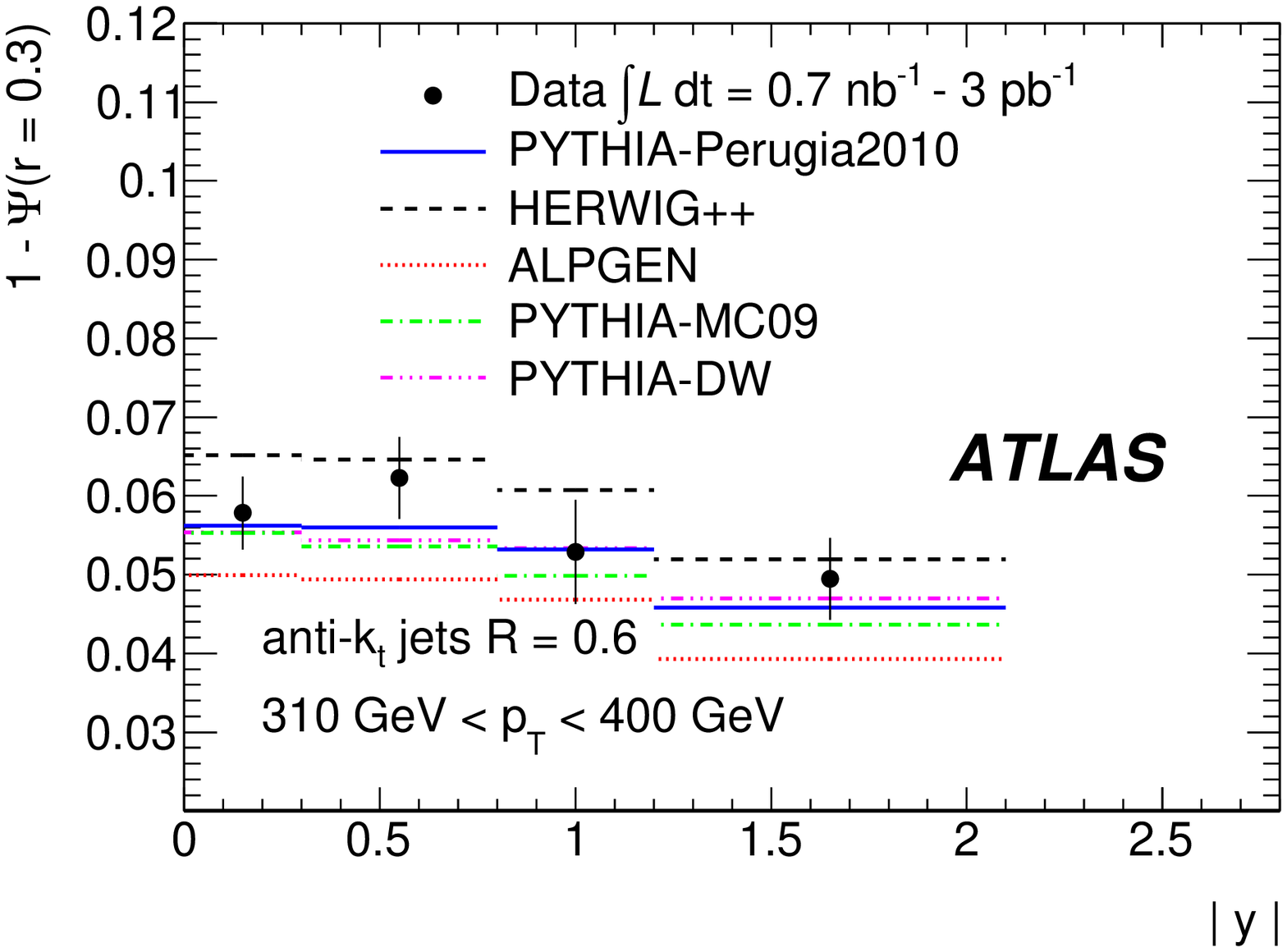}
\includegraphics[width=0.495\textwidth]{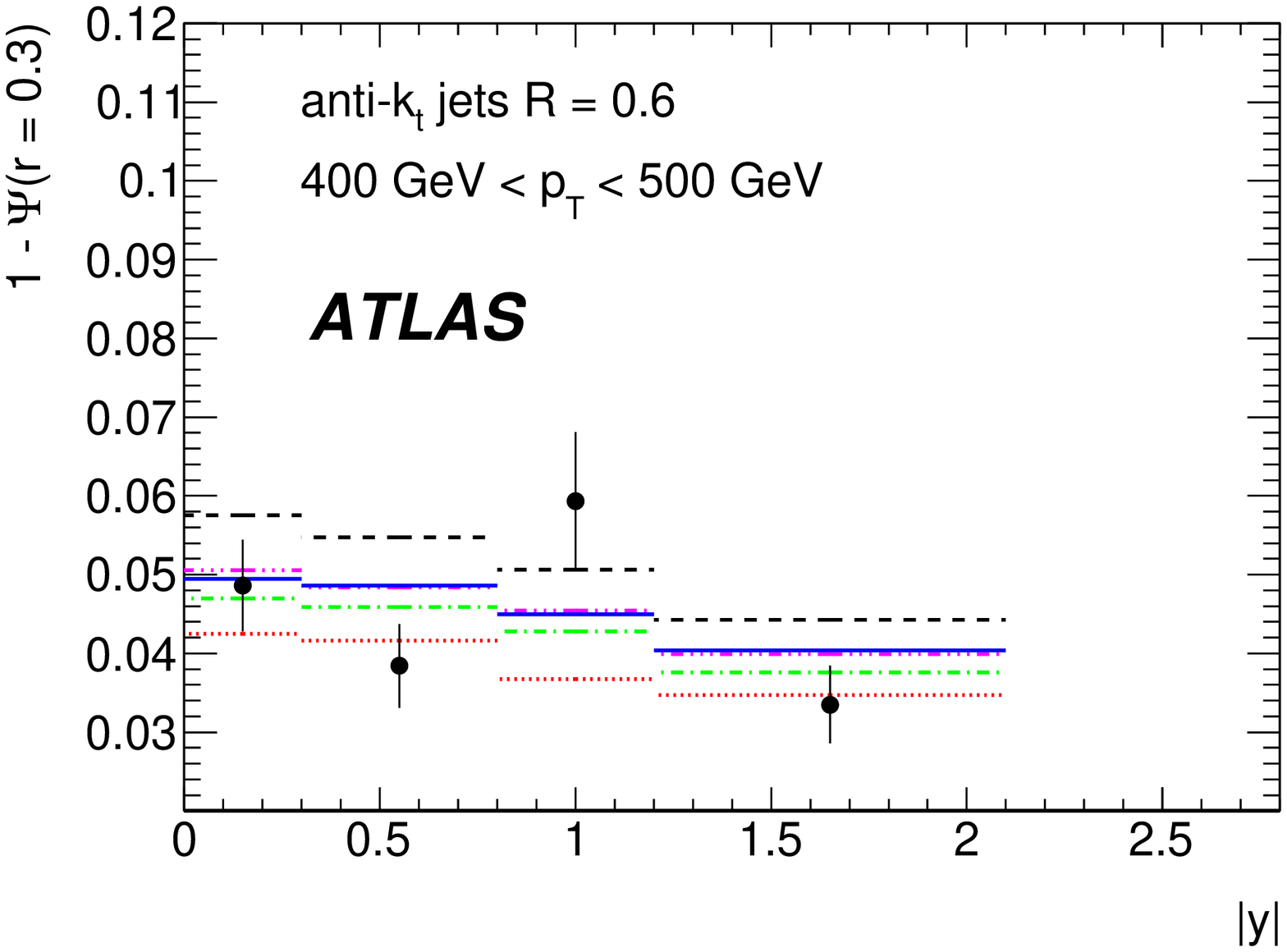}
}
\end{center}
\vspace{-0.7 cm}
\caption[The measured integrated jet shape, $1 - \Psi(r=0.3)$, as a function of $|\rapjet|$
for jets with $|\rapjet| < 2.8$ and $210 \ {\rm GeV} < \ptjet < 500 \ {\rm GeV}$ in different $y$ bins]
{\small
The measured integrated jet shape, $1 - \Psi(r=0.3)$, as a function of $|\rapjet|$
for jets with $|\rapjet| < 2.8$ and $210 \ {\rm GeV} < \ptjet < 500 \ {\rm GeV}$.
Error bars indicate the statistical and systematic uncertainties added in quadrature.
The predictions of   PYTHIA-Perugia2010 (solid lines),   HERWIG++ (dashed lines),   ALPGEN interfaced with HERWIG and JIMMY (dotted lines), PYTHIA-MC09 (dashed-dotted lines), and PYTHIA-DW (dashed-dotted-dotted lines) are shown for comparison.
}
\label{figaux4}
\end{figure}

\begin{figure}[tbh]
\begin{center}
\mbox{
\includegraphics[width=0.95\textwidth]{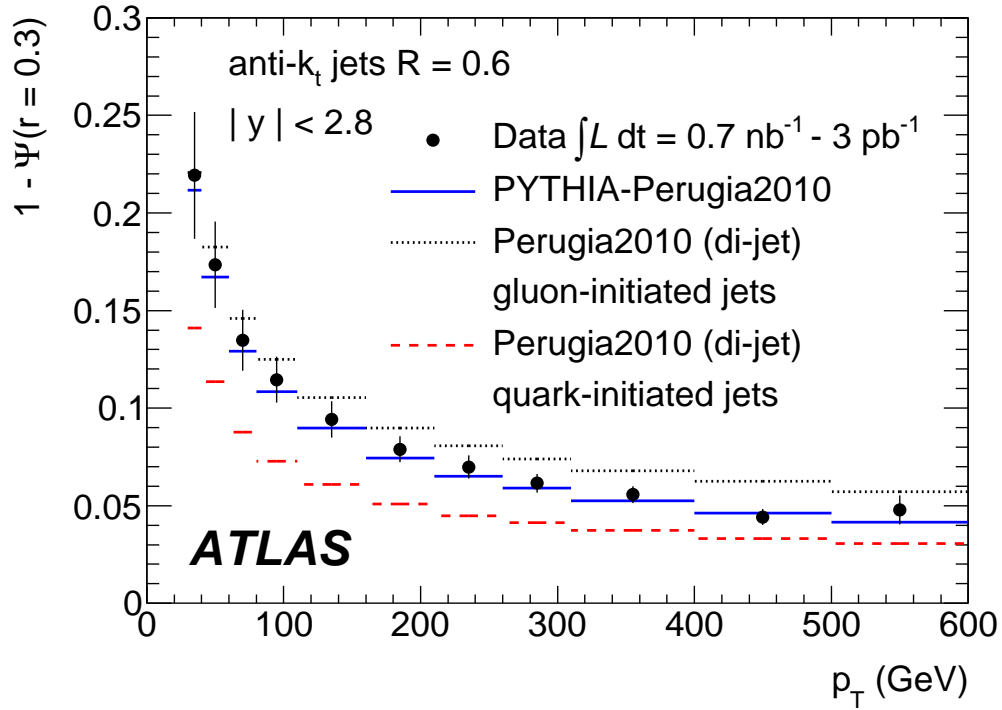}
}
\end{center}
\vspace{-0.7 cm}
\caption[The measured integrated jet shape, $1 - \Psi(r=0.3)$, as a function of $\ptjet$
for jets with $|\rapjet| < 2.8$ and $30 \ {\rm GeV} < \ptjet < 600 \ {\rm GeV}$ (compared to a different set of MCs)]
{\small
The measured integrated jet shape, $1 - \Psi(r=0.3)$, as a function of $\ptjet$ 
for jets with $|\rapjet| < 2.8$ and $30 \ {\rm GeV} < \ptjet < 600 \ {\rm GeV}$.
Error bars indicate the statistical and systematic uncertainties added in quadrature. 
The predictions of   PYTHIA-Perugia2010 (solid line) are shown for comparison, together with 
the prediction separately for  quark-initiated  (dashed lines) and
gluon-initiated jets (dotted lines) in dijet events.
} 
\label{fig5}
\end{figure}

\begin{figure}[tbh]
\begin{center}
\mbox{
\includegraphics[width=0.495\textwidth]{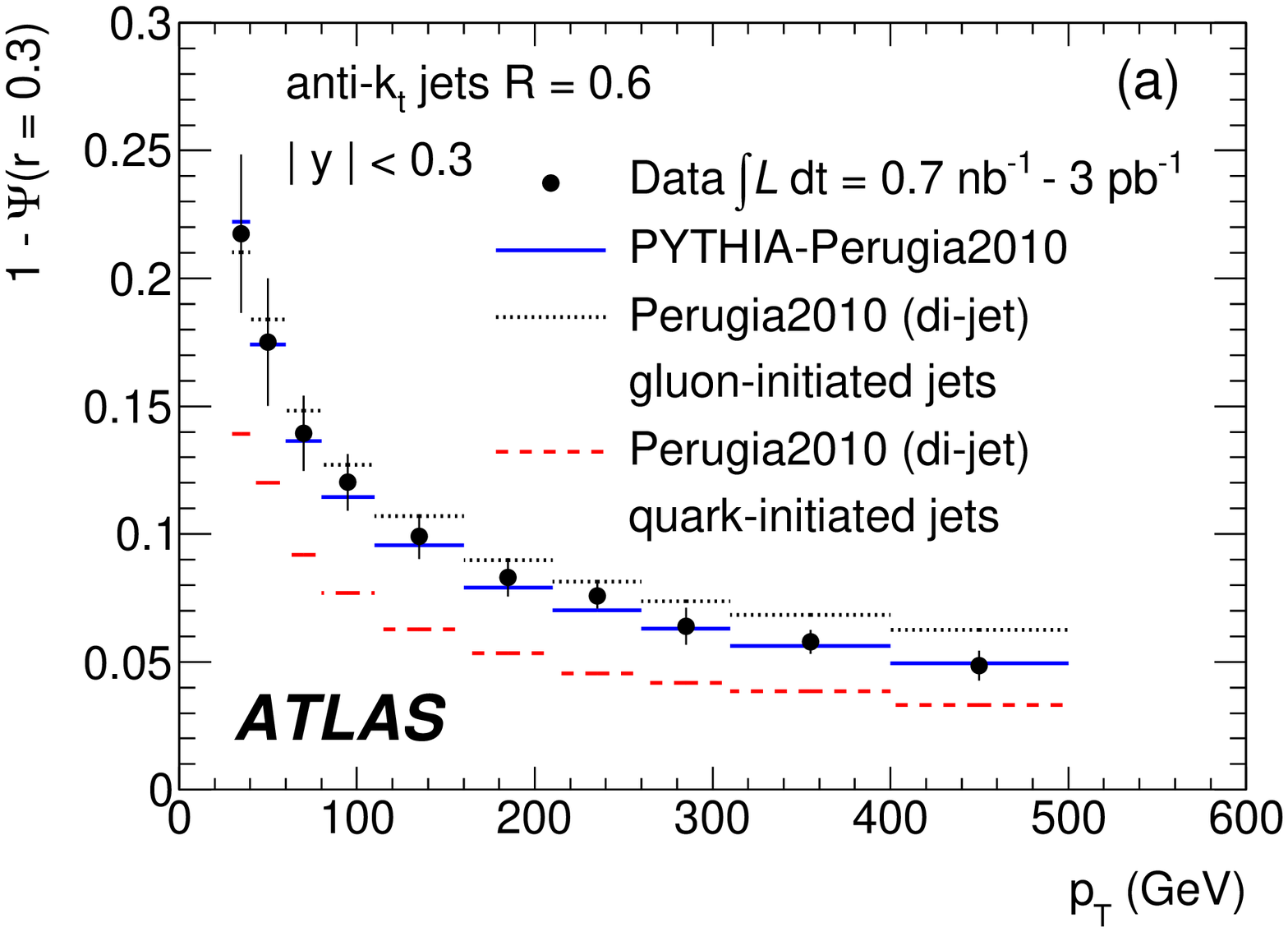}
\includegraphics[width=0.495\textwidth]{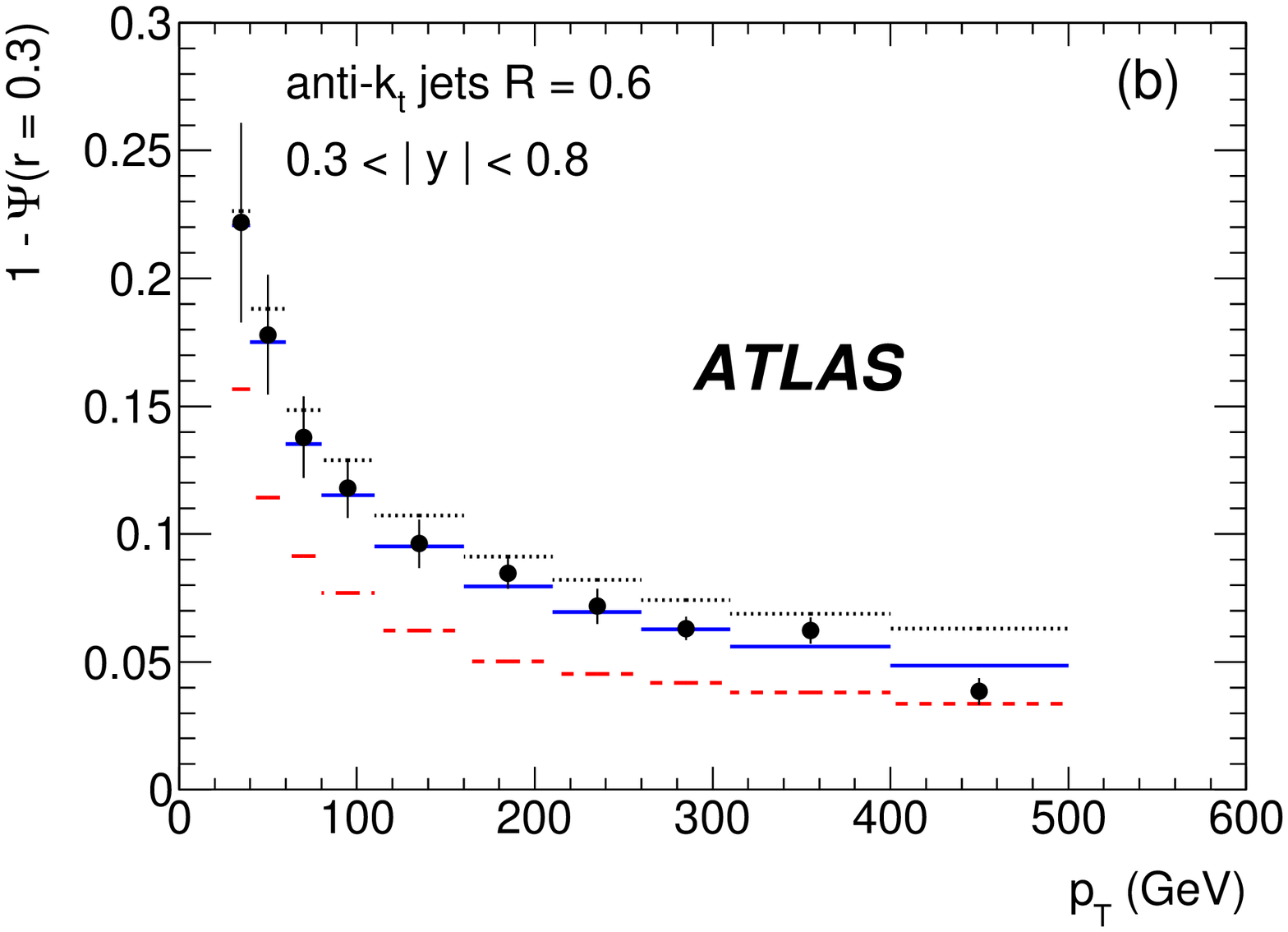}
}
\mbox{
\includegraphics[width=0.495\textwidth]{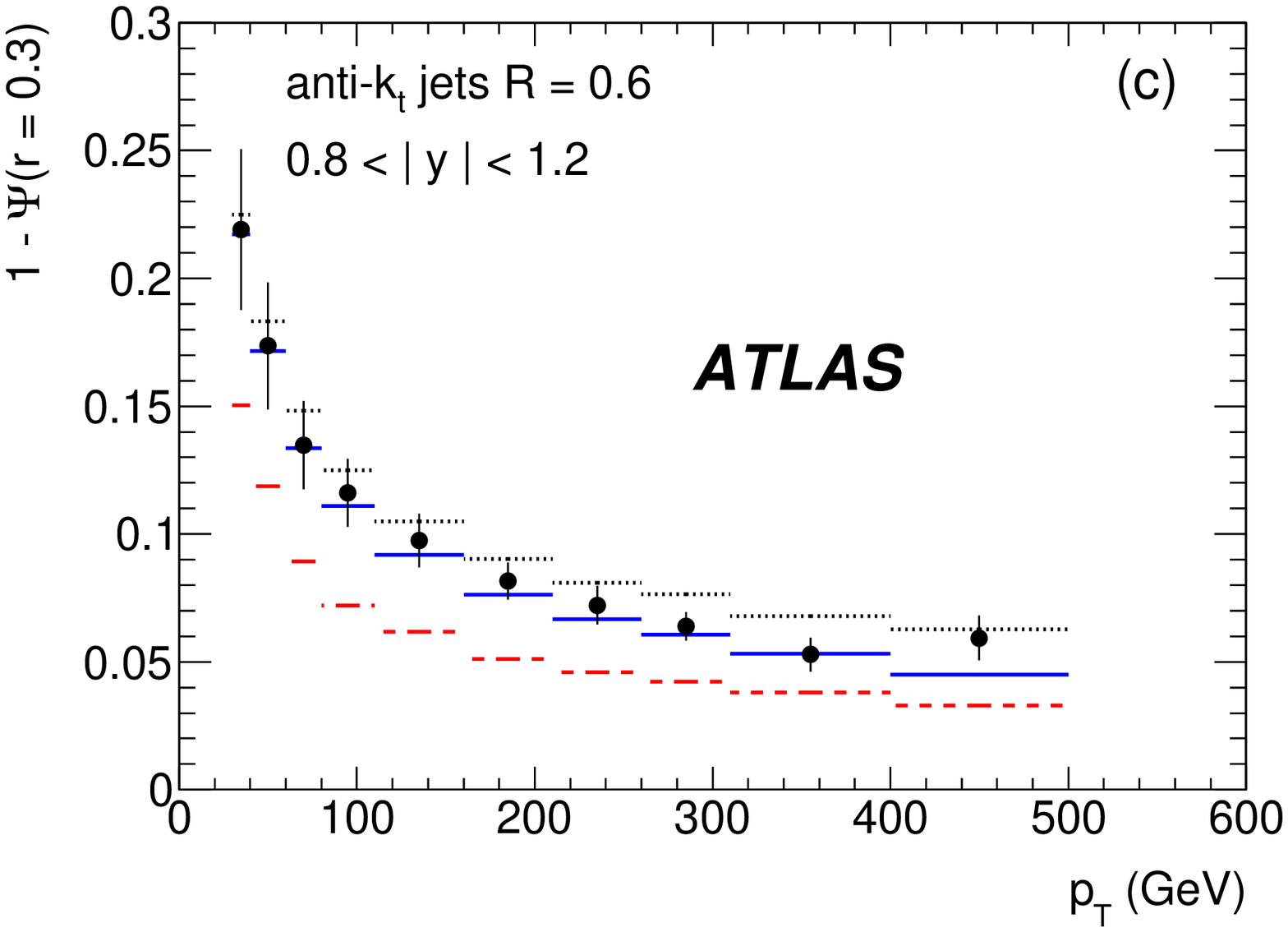} 
\includegraphics[width=0.495\textwidth]{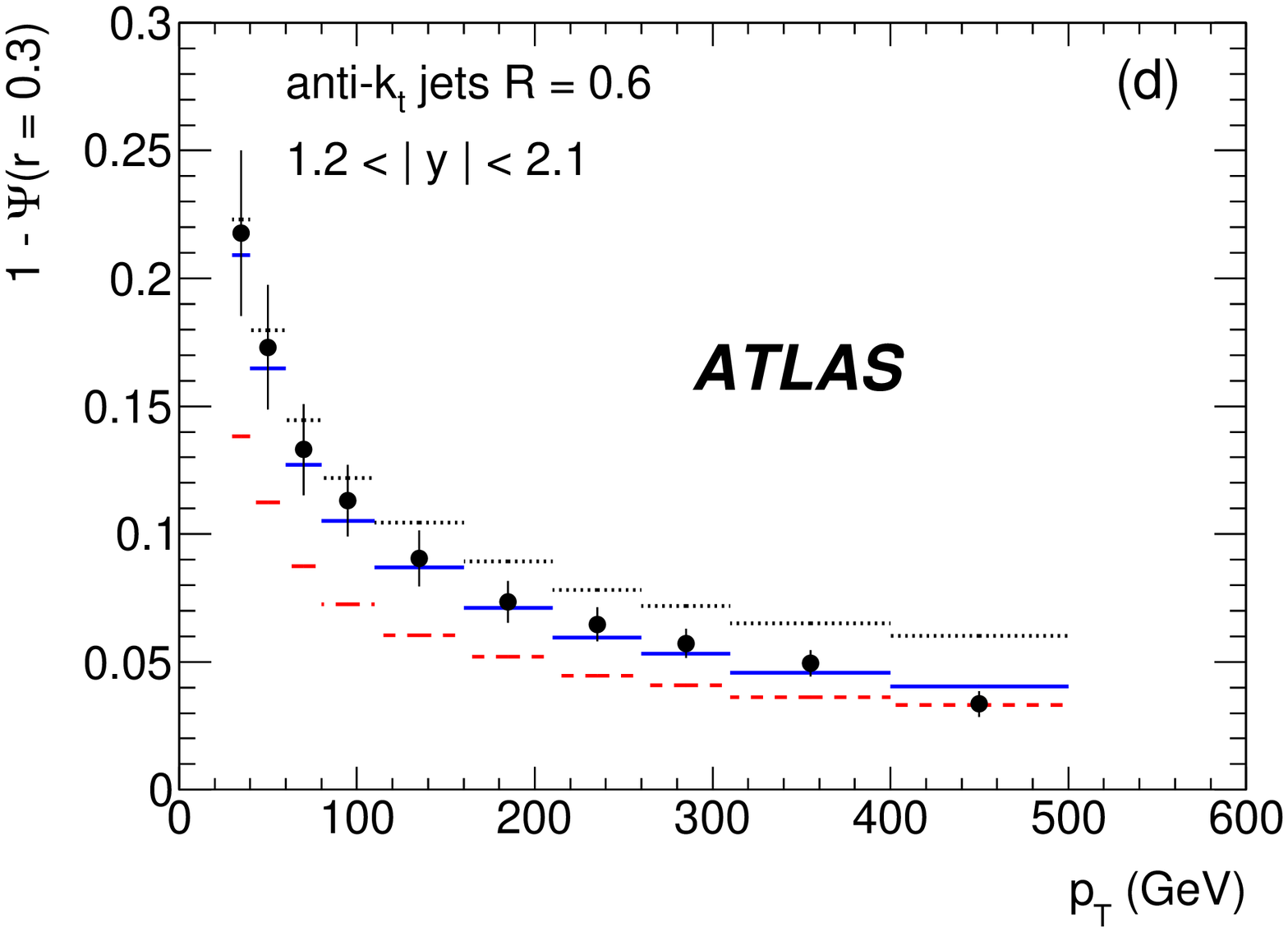}
}
\mbox{
\includegraphics[width=0.495\textwidth]{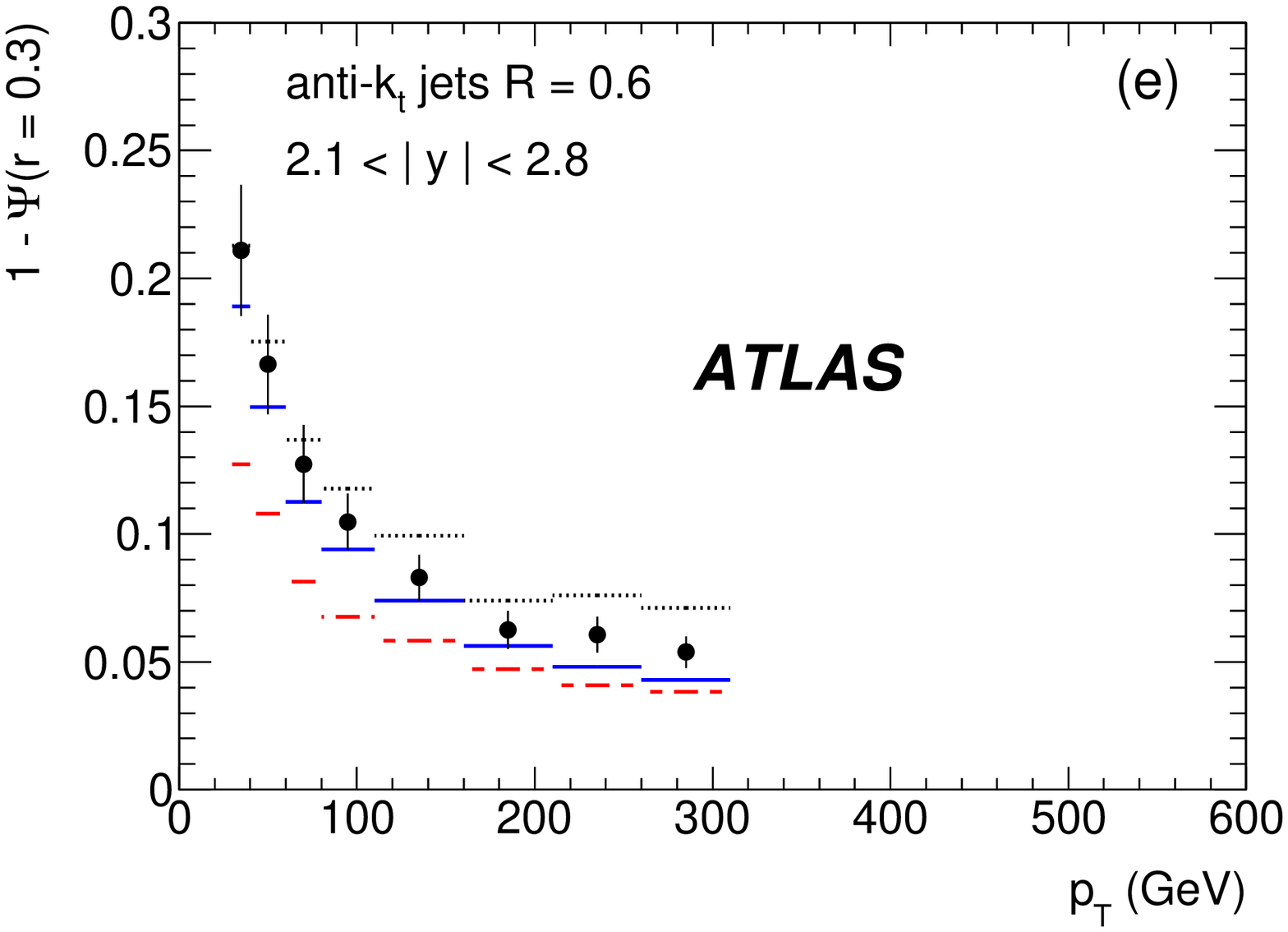}
}
\end{center}
\vspace{-0.7 cm}
\caption[The measured integrated jet shape, $1 - \Psi(r=0.3)$, as a function of $\ptjet$ in different jet rapidity regions
for jets with $|\rapjet| < 2.8$ and $30 \ {\rm GeV} < \ptjet < 500 \ {\rm GeV}$]
{\small
The measured integrated jet shape, $1 - \Psi(r=0.3)$, as a function of $\ptjet$ in different jet rapidity regions 
for jets with $|\rapjet| < 2.8$ and $30 \ {\rm GeV} < \ptjet < 500 \ {\rm GeV}$.
Error bars indicate the statistical and systematic uncertainties added in quadrature. 
The predictions of   PYTHIA-Perugia2010 (solid line) are shown for comparison, together with 
the prediction separately for  quark-initiated  (dashed lines) and
gluon-initiated jets (dotted lines) in dijet events.
} 
\label{fig6}
\end{figure}

The dependence on $|\rapjet|$ is shown in 
Figure~\ref{fig4} and \ref{figaux4}, where the measured jet shapes are presented 
separately in five different jet rapidity regions and 
different $\ptjet$ bins, for jets with $\ptjet < 500$~GeV. 
At high $\ptjet$, the measured  $1-\Psi(0.3)$ shape presents a mild $|\rapjet|$ dependence, indicating that the 
jets become slightly narrower in the forward regions. This tendency is observed also in the various 
MC samples. Similarly, Figures~\ref{fig5} and \ref{fig6} present the measured
$1-\Psi(0.3)$ as a function of $\ptjet$ in the different  $|\rapjet|$ regions compared 
to PYTHIA-Perugia2010 predictions.

Finally, and only for illustration, the typical shapes of   
quark- and gluon-initiated jets, as determined using events generated with PYTHIA-Perugia2010, are also shown in Figures~\ref{fig5} and ~\ref{fig6}.  
For this purpose, MC events are selected with at least two particle-level jets with $\ptjet > 30$~GeV and $|\rapjet|<2.8$ 
in the final state. 
The two leading jets in this dijet sample are classified as quark-initiated or gluon-initiated jets 
by matching (in $y - \phi$ space) their direction with one of the outgoing partons from the QCD 
$2 \rightarrow 2$ hard process. At low $\ptjet$ the measured jet shapes are 
similar to those from 
gluon-initiated jets, as expected from the dominance of hard processes with gluons in the final state.
At high $\ptjet$,  where the impact of the UE contributions becomes smaller (see Figure~\ref{fig3}(b)),  the 
observed trend  with $\ptjet$ in the data is mainly attributed to a changing quark- and gluon-jet mixture 
in the final state, convoluted with  perturbative QCD effects related to the running of the strong coupling.

The jet shapes results presented in this Chapter were obtained with the first $pp$ collisions at $\sqrt{s} = 7$~TeV, 
in which the pile-up can be neglected, and were published in~\cite{pp}. They indicate the
potential of jet shape measurements at the LHC
to constrain  the phenomenological models for soft gluon radiation, underlying event (UE)
activity, and non-perturbative fragmentation processes in the final state.
After the publication of these results, additional comparisons to predictions from 
different MC programs have been performed~\cite{bib_pub}. They are reported in the next Chapter, which  
includes $\chi^{2}$ tests to the data points for all the MC models considered.  
\clearpage


\clearpage


\clearpage


\clearpage


\clearpage

\clearpage


\clearpage

\clearpage


\clearpage

\newpage
$\,$
\clearpage
\newpage

\chapter{Jet Shapes in  ATLAS and Monte Carlo modeling}
\label{chap5}

In this Chapter the jet shapes results published in~\cite{pp}  
are compared to predictions of different MC programs, 
that include recent tuned sets of parameters to describe the ATLAS minimum bias (MB) and jet data.


\section{Monte Carlo samples}

The MC generators considered in this Chapter are listed below. 

\begin{itemize}
%
%
\item The PYTHIA 6.4.23 generator is used to  produce QCD-jet inclusive samples    
with both AMBT1 and AUET2~\cite{mc10} tunes, as determined by ATLAS, and with the Perugia2011 tune~\cite{Perugia2010}. 
PYTHIA-AMBT1 is based on MRST LO* parton density functions (PDFs) inside the proton, and uses as input results 
from the ATLAS charged particle MB analysis at $\sqrt{s} = 0.9$~TeV and $\sqrt{s} = 7$~TeV. 
PYTHIA-AUET2 is based on MRST LO** PDFs, uses also as input ATLAS UE data~\cite{ATLASue}, and includes a modified final-state parton shower 
from initial-state radiated partons to describe the CDF~\cite{ppbar} and ATLAS jet shape measurements.
The PYTHIA-Perugia2011 tune uses 
CTEQ5L PDFs and constitutes an updated version of the PYTHIA-Perugia2010 tune, as 
inspired by the comparison with early LHC data.  
Samples are also generated using the PYTHIA 8.145 
program~\cite{pythia8} which, among other updates, includes a fully interleaved $p_{\rm T}$-ordered evolution 
for initial- and final-state gluon radiation and multiple parton interactions (MPI).
The Pythia~8 samples are produced with the 
4C tune~\cite{pythia8} for the UE modeling, based on CTEQ6L1 PDFs and using Tevatron and ATLAS 
data as input.
%
%

 \item QCD-jet inclusive samples are generated with the Herwig++~2.4.2 and Herwig++~2.5.1 programs.
The Herwig++ 2.4.2 predictions used in~\cite{pp} suffered
from a wrong setting in the ATLAS generation of the MC samples, related to the parameters that control the contribution from
MPI, which {\it a priori} would affect the comparison with the data at low $\ptjet$.   In this Chapter,
the same Herwig++ 2.4.2 predictions already presented in the previous Chapter (here denoted as Herwig++ 2.4.2 bug), are
compared with  correctly generated HERWIG 2.4.2 predictions.
Herwig++~2.5.1 includes a modified algorithm for color reconnection between clusters before hadronization and
new color connections between soft scatters and the beam remnants that improve the description of the
MB data and UE-related observables at the LHC. In the case of  Herwig++ 2.4.2
the LO${}^*$ tune~\cite{hw++tune} is employed, while the UE7 tune~\cite{ue7}
is used for Herwig++2.5.1 predictions.
In addition, samples are produced with HERWIG 6.510 interfaced with JIMMY~4.31
to model the UE contributions with both AUET1-LO${}^*$~\cite{hwtune} and  AUET2-LO${}^{**}$~\cite{mc10}
tunes, as determined by ATLAS using data at 0.9~TeV and 7~TeV.
%
%
\item QCD-jet events are generated using the ALPGEN 2.13 program 
interfaced with PYTHIA 6.423 for the parton shower and fragmentation into hadrons. The samples are produced with CTEQ6L1 PDFs and the Perugia6 
tune to model the UE contributions. 
\item The Sherpa 1.2.3 and Sherpa 1.3.0 programs~\cite{sherpa} are used to generate QCD-jet samples.  
Default settings include CTEQ66 PDFs and MPI tuned to Tevatron and ATLAS data.
The Sherpa 1.3.0 version contains a modified parton shower evolution via pre-factors  within $\alpha_s(p_{\rm T})$, 
as motivated by next-to-leading-logarithm (NLL) perturbative QCD (pQCD) calculations, that improve the description of the 
measured Z boson $\ptjet$ in ATLAS.  Samples with different matrix element multiplicities  are 
produced, with the aim to illustrate the impact
of higher-order contributions to the description of the jet shapes. Samples with only $2 \to 2$ hard 
processes  and with up to $2 \to 6$ partons in the final state are considered, and  
interfaced with parton showers.     
%
%
\item QCD-jet samples are generated with the POWHEG-BOX r302 program, which includes next-to-leading order 
(NLO) pQCD predictions for inclusive jet production,  interfaced either with 
PYTHIA~6.423 or with HERWIG~6.510 plus JIMMY~4.31 for parton shower,  fragmentation into hadrons, and to model 
the underlying event. For the latter, the  AMBT1 or AUET1-LO${}^{*}$ tunes are used, respectively.  

\end{itemize}

\noindent
The simulated event samples were produced using the ATLAS interface to the  generators and the 
Rivet system was used to analyze the samples and produce the relevant jet shape observables.

\section{Results}

Different observables in data are  
compared to the MC predictions. This includes the measured 
differential jet shape $\rho(r)$ for jets with $|\rapjet| < 2.8$ in eight $\ptjet$ bins within the range
$30 < \ptjet < 310$~GeV, and the integrated jet shape $1 - \Psi(0.3)$ as a function of $|\rapjet|$ in different 
$\ptjet$ regions and  as a function of $\ptjet$ in different 
$|\rapjet|$ bins.


\subsection{Comparison with PYTHIA}

\begin{figure}[tbh]
\begin{center}
\mbox{
\includegraphics[width=0.495\textwidth,height=0.495\textwidth]{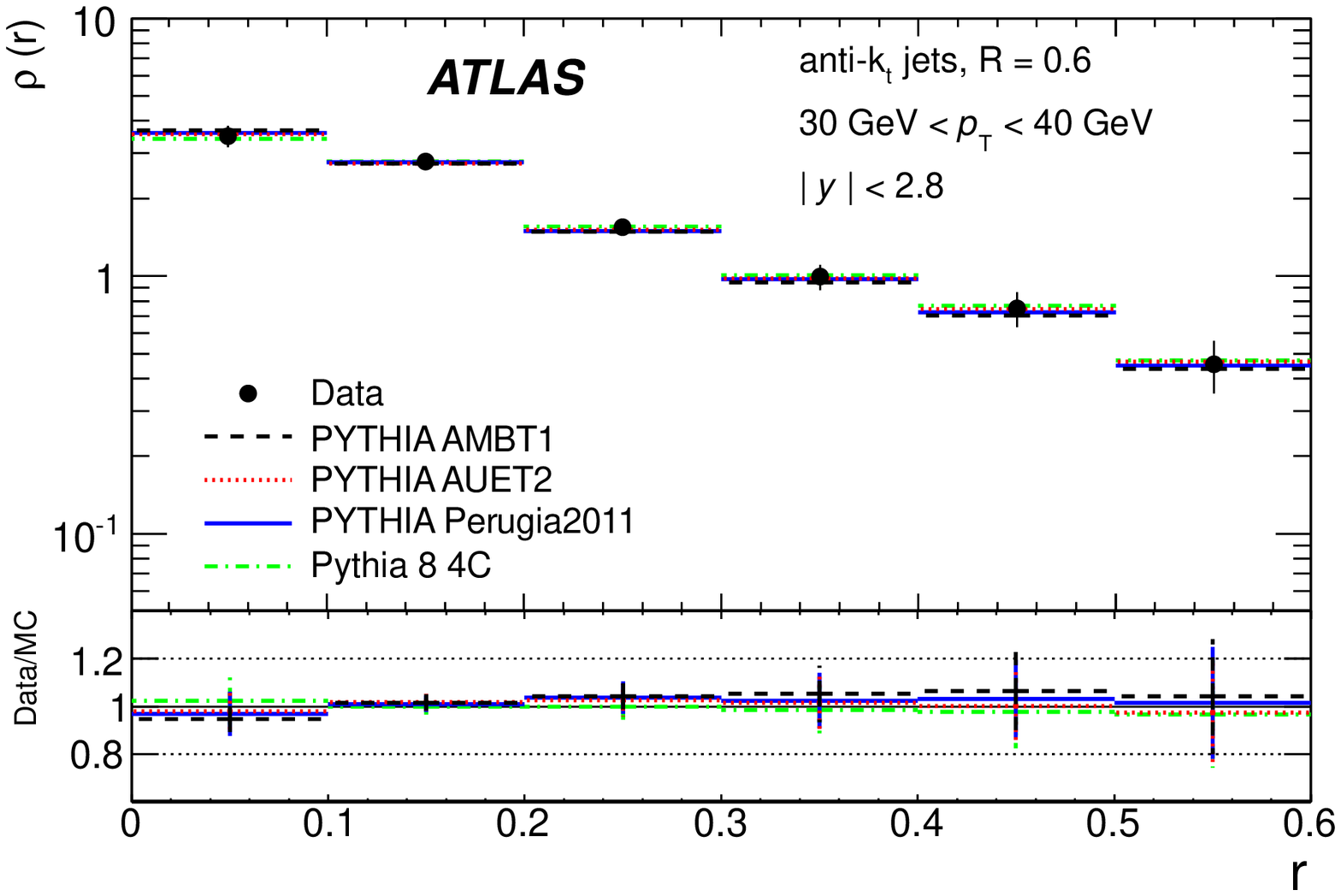} 
\includegraphics[width=0.495\textwidth,height=0.495\textwidth]{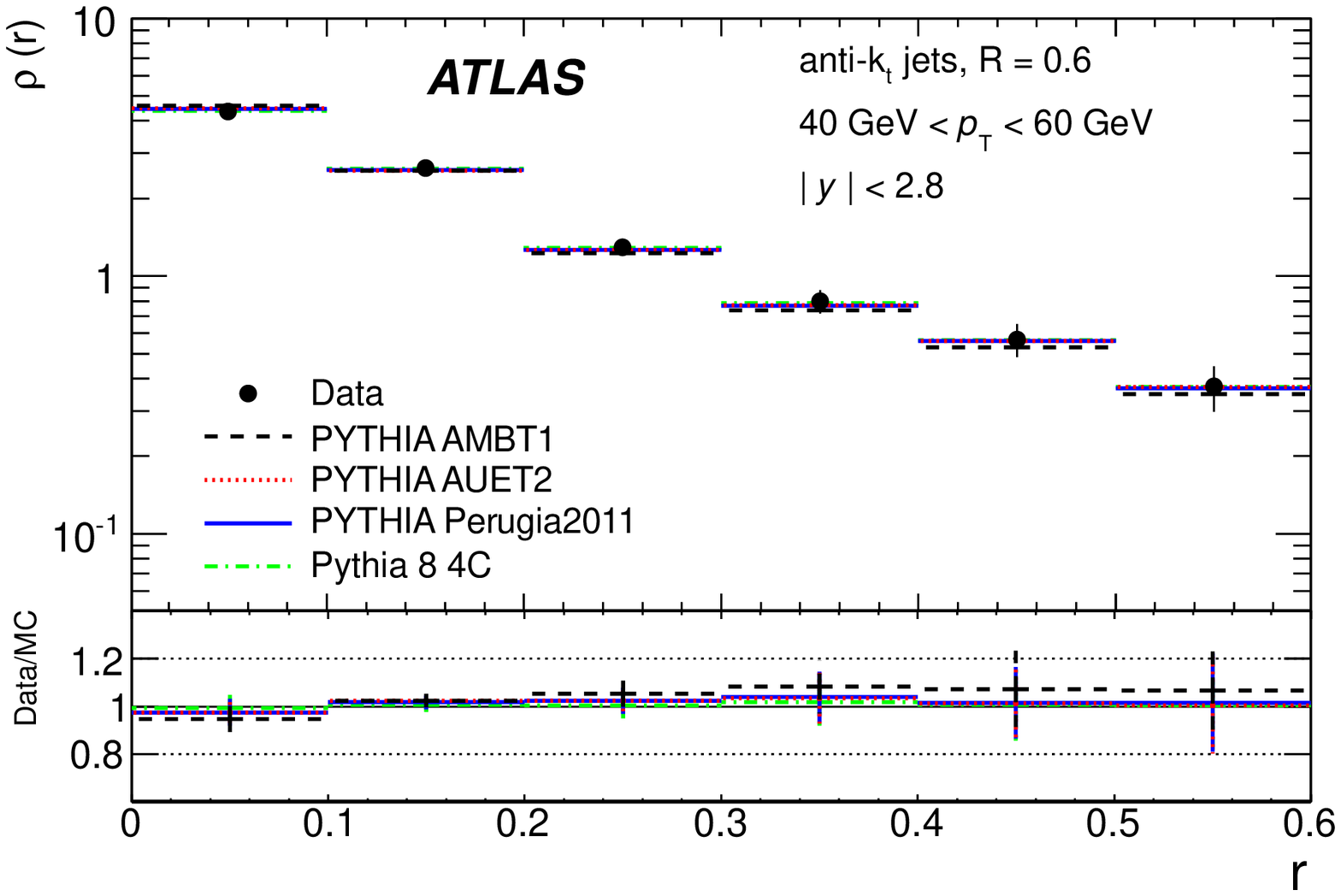}
}\vspace{-0.2cm}
\mbox{
\includegraphics[width=0.495\textwidth,height=0.495\textwidth]{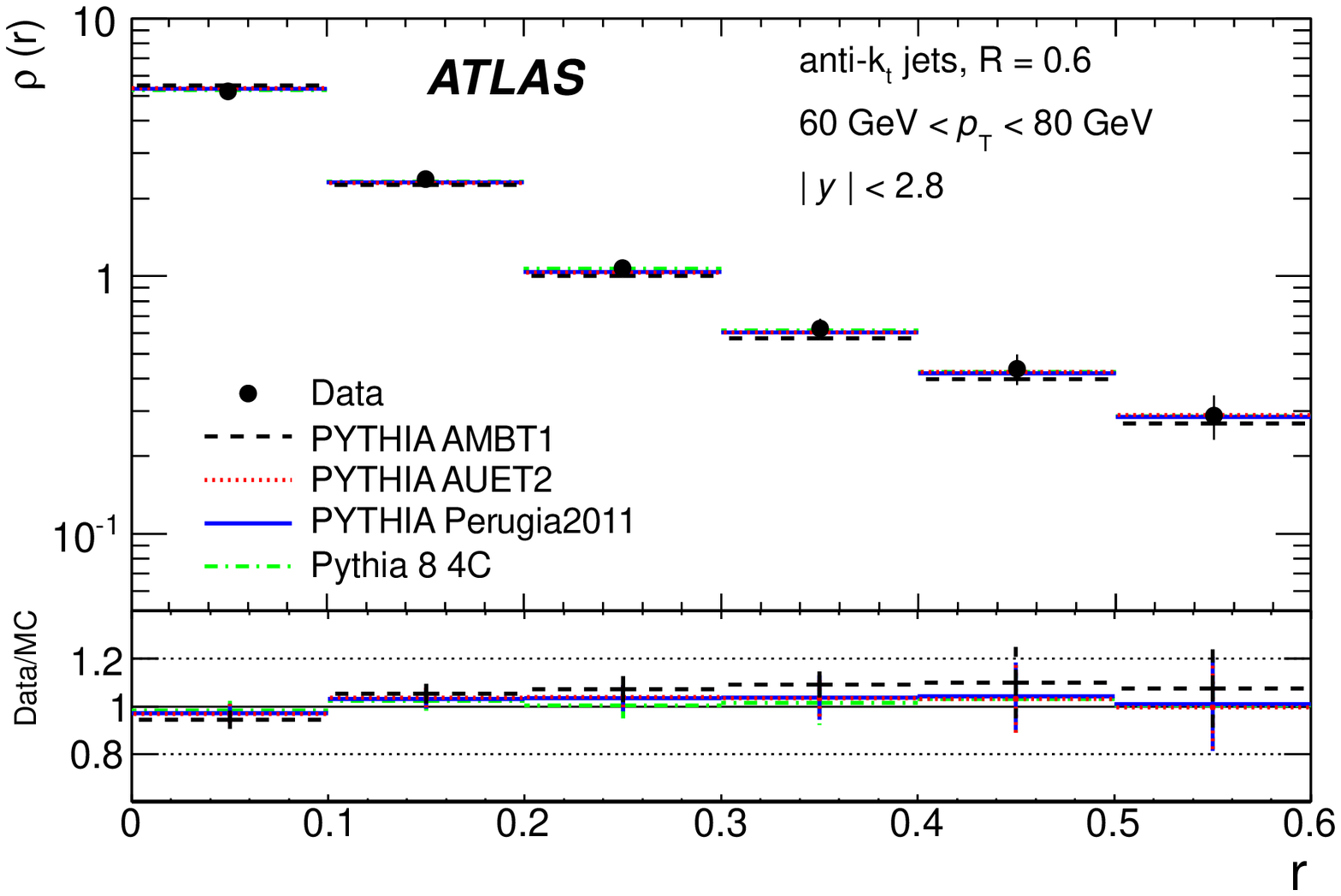}
\includegraphics[width=0.495\textwidth,height=0.495\textwidth]{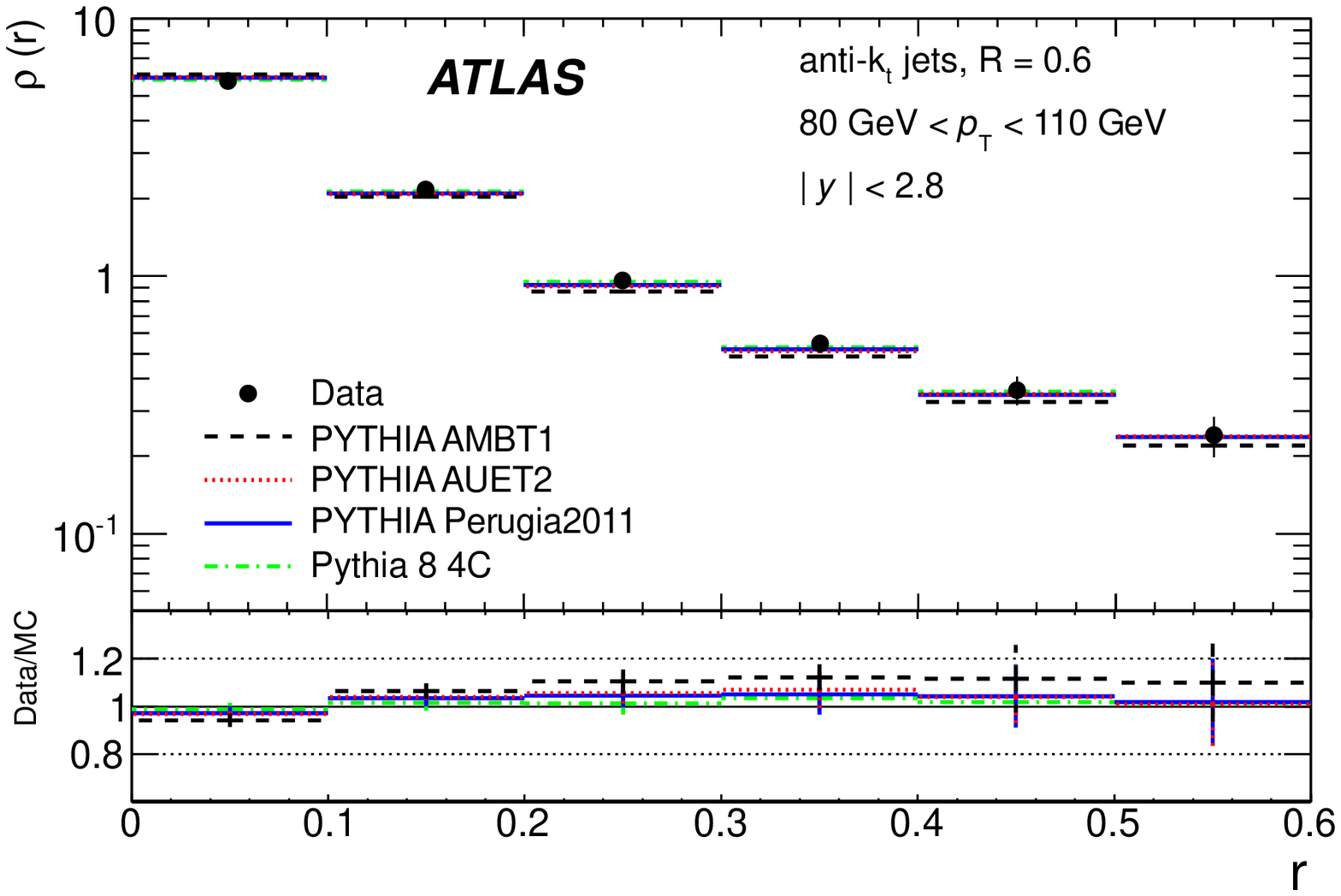}
}
\end{center}
\vspace{-0.7 cm}
\caption{\small
The measured differential jet shape, $\rho(r)$, in inclusive jet production for jets 
with $|\rapjet| < 2.8$ and $30 \ {\rm GeV} < \ptjet < 110  \ {\rm GeV}$   
is shown in different $\ptjet$ regions. Error bars indicate the statistical and systematic uncertainties added in quadrature.
The predictions of   PYTHIA-Perugia2011 (solid lines),   PYTHIA-AUET2 (dotted lines),   PYTHIA-AMBT1 (dashed lines), and Pythia~8-4C (dashed-dotted lines) are shown for comparison.}
\label{fig:pyt1}
\end{figure}

\begin{figure}[tbh]
\begin{center}
\mbox{
\includegraphics[width=0.495\textwidth,height=0.495\textwidth]{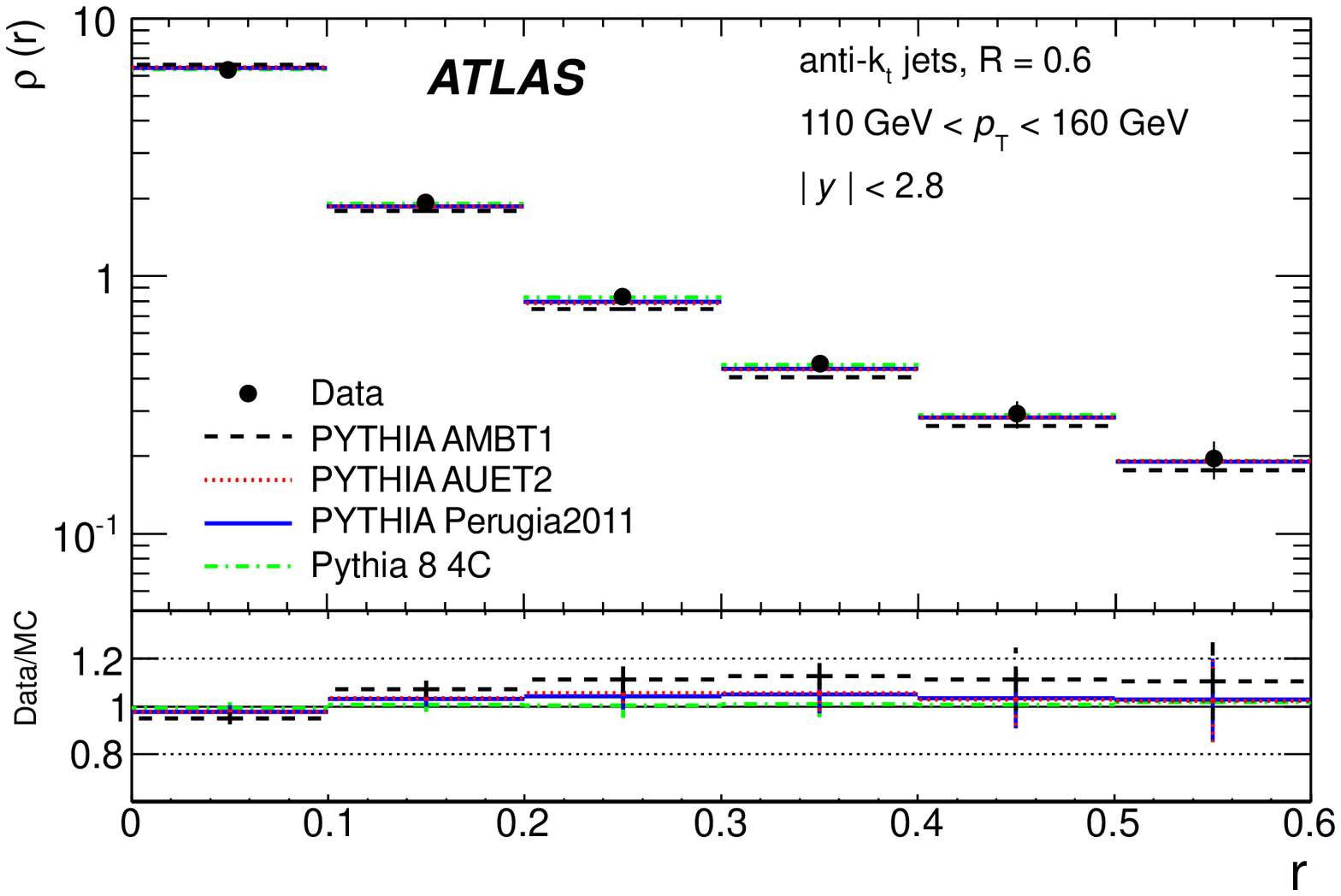} 
\includegraphics[width=0.495\textwidth,height=0.495\textwidth]{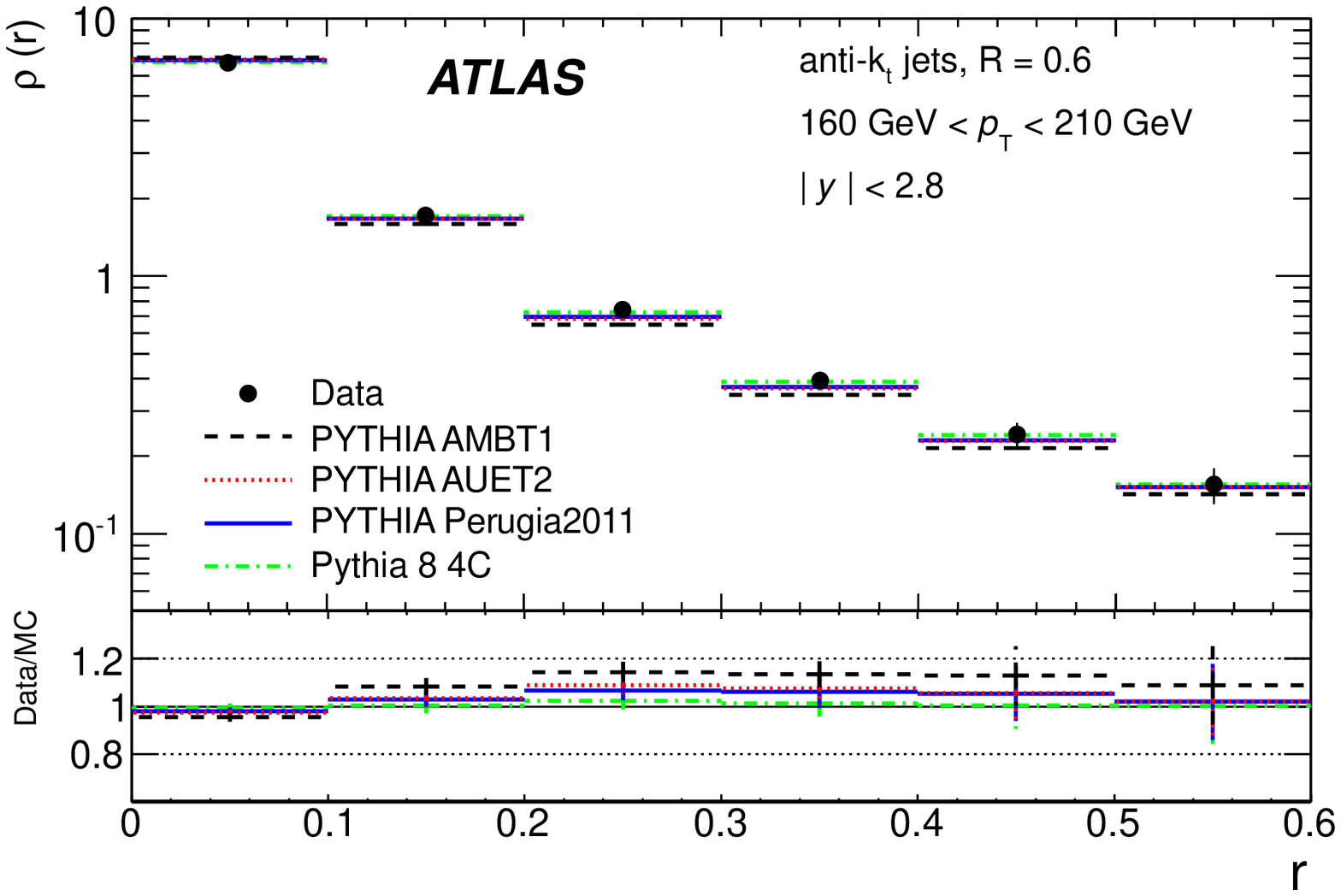}
}\vspace{-0.2cm}
\mbox{
\includegraphics[width=0.495\textwidth,height=0.495\textwidth]{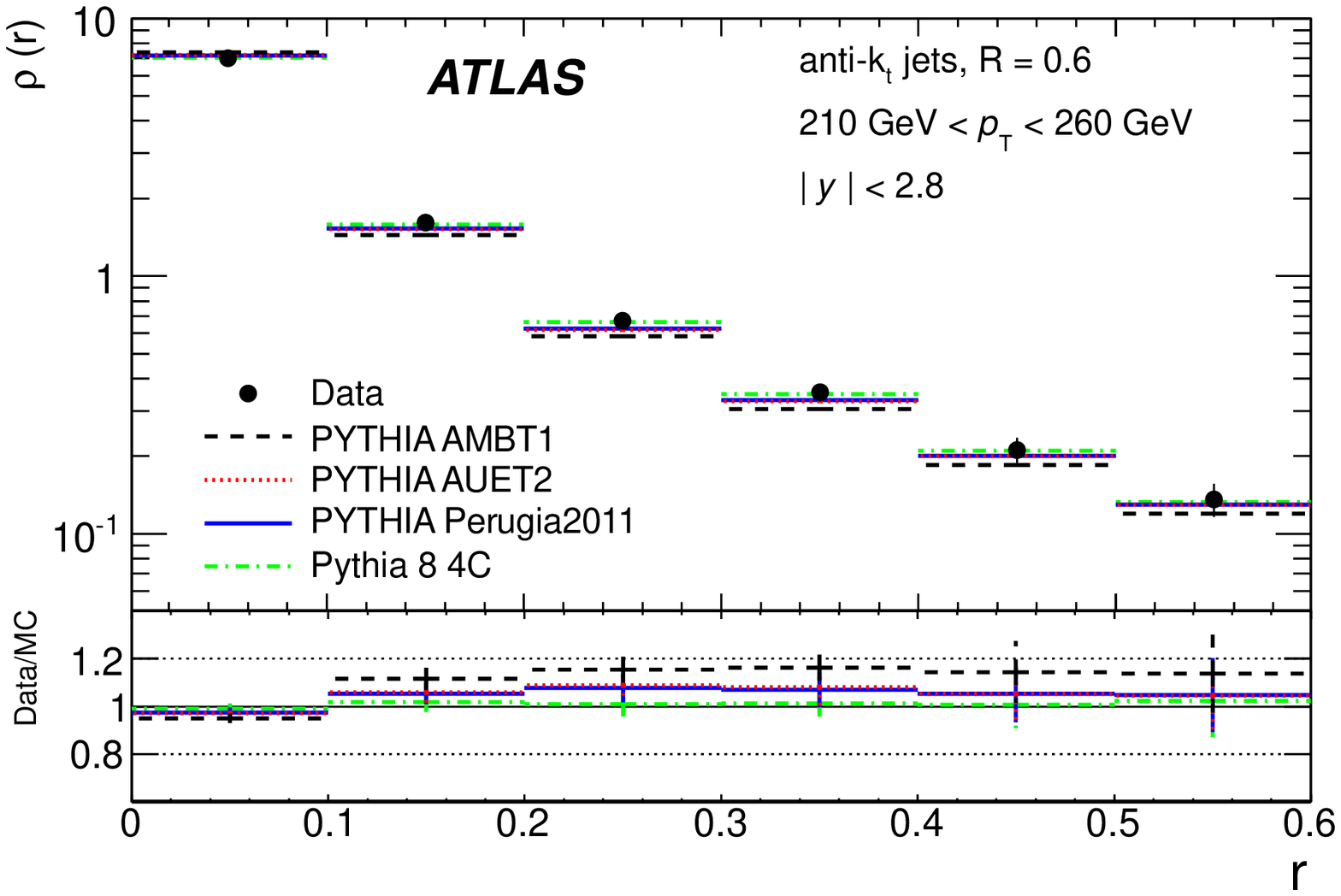}
\includegraphics[width=0.495\textwidth,height=0.495\textwidth]{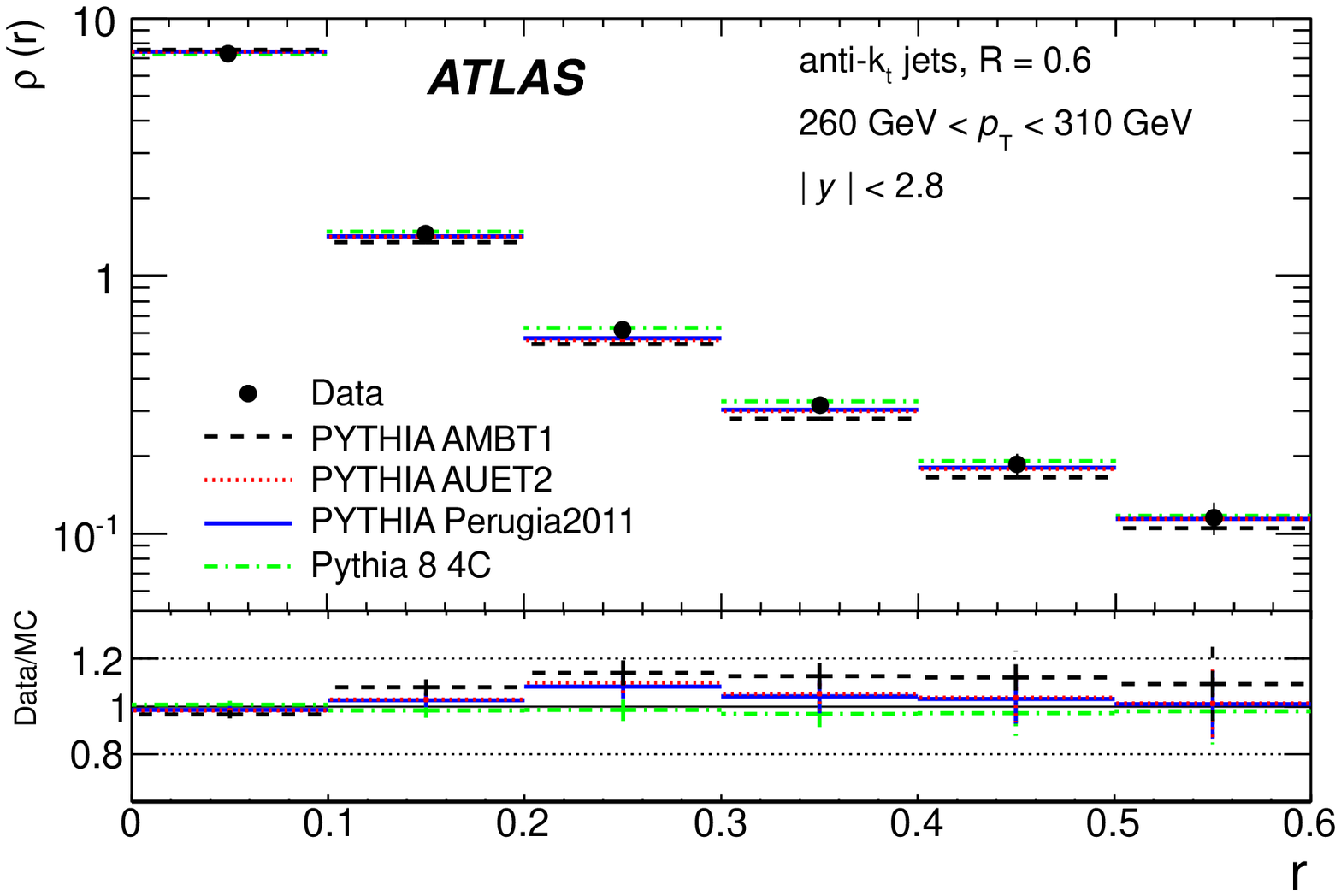}
}
\end{center}
\vspace{-0.7 cm}
\caption{\small
The measured differential jet shape, $\rho(r)$, in inclusive jet production for jets 
with $|\rapjet| < 2.8$ and $110 \ {\rm GeV} < \ptjet < 310  \ {\rm GeV}$   
is shown in different $\ptjet$ regions. Error bars indicate the statistical and systematic uncertainties added in quadrature.
The predictions of   PYTHIA-Perugia2011 (solid lines),   PYTHIA-AUET2 (dotted lines),   PYTHIA-AMBT1 (dashed lines), and Pythia~8-4C (dashed-dotted lines) are shown for comparison.
} 
\label{fig:pyt2}
\end{figure}

\begin{figure}[tbh]
\begin{center}
\mbox{
\includegraphics[width=0.495\textwidth]{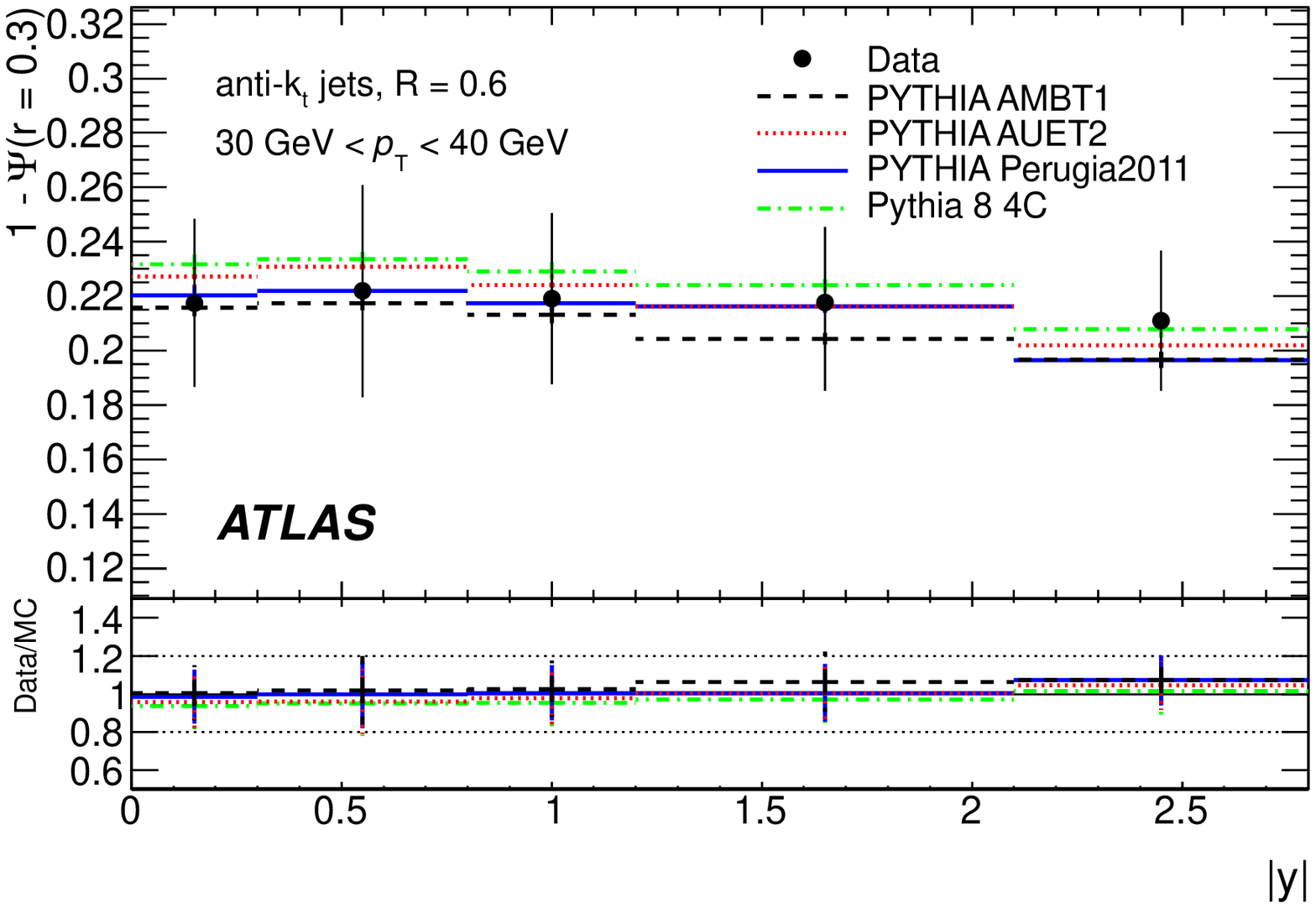}
\includegraphics[width=0.495\textwidth]{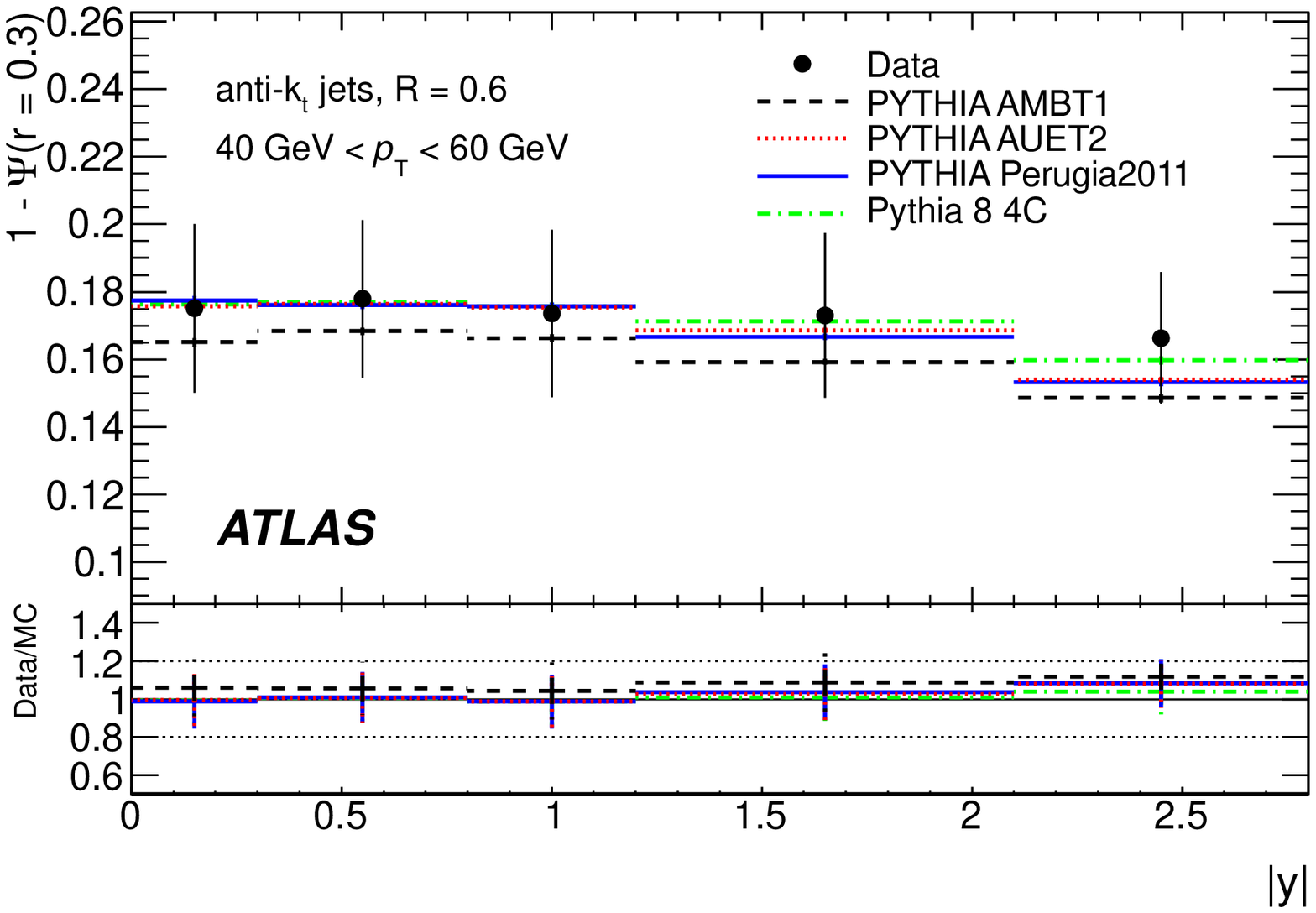}
}
\mbox{
\includegraphics[width=0.495\textwidth]{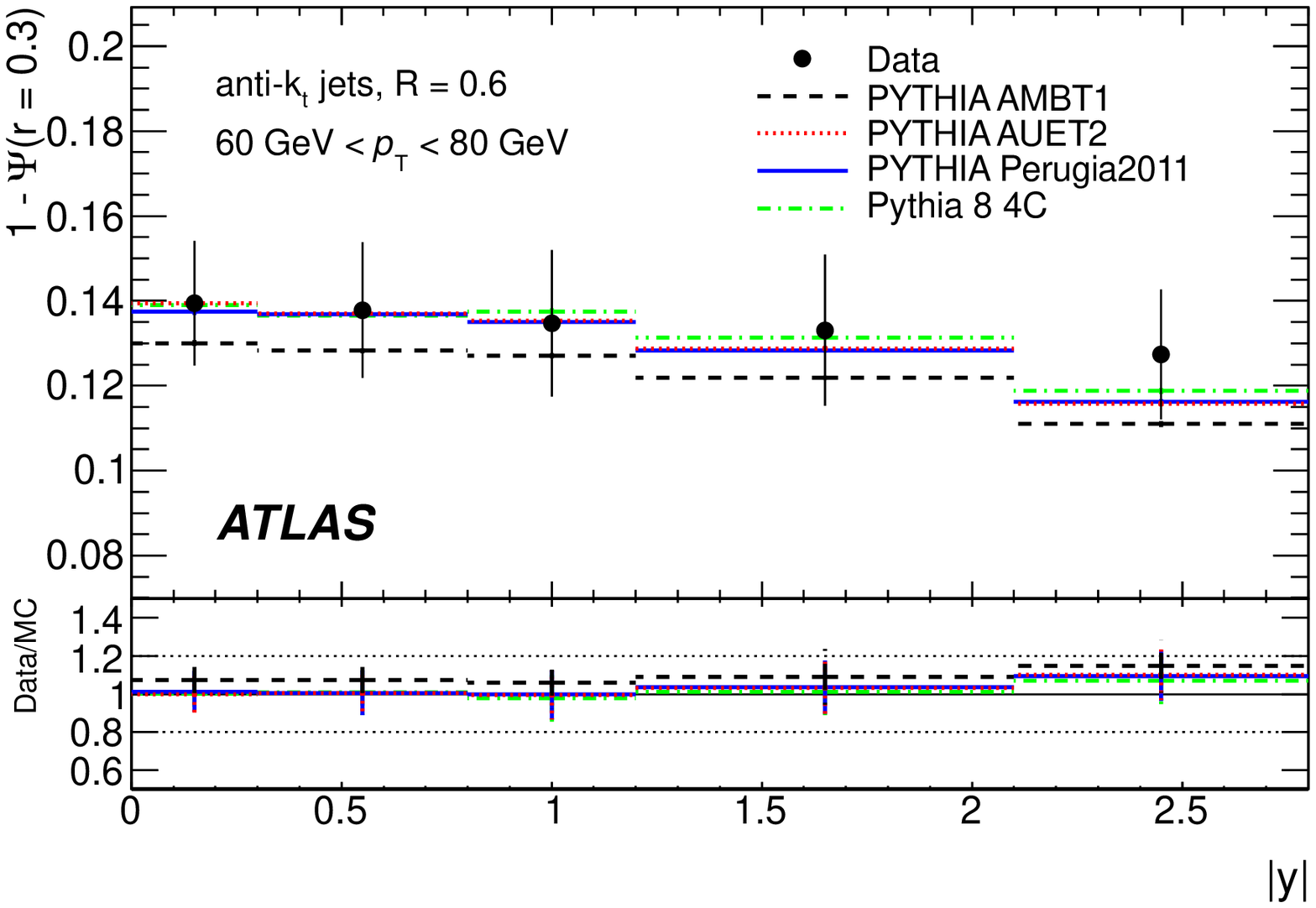} 
\includegraphics[width=0.495\textwidth]{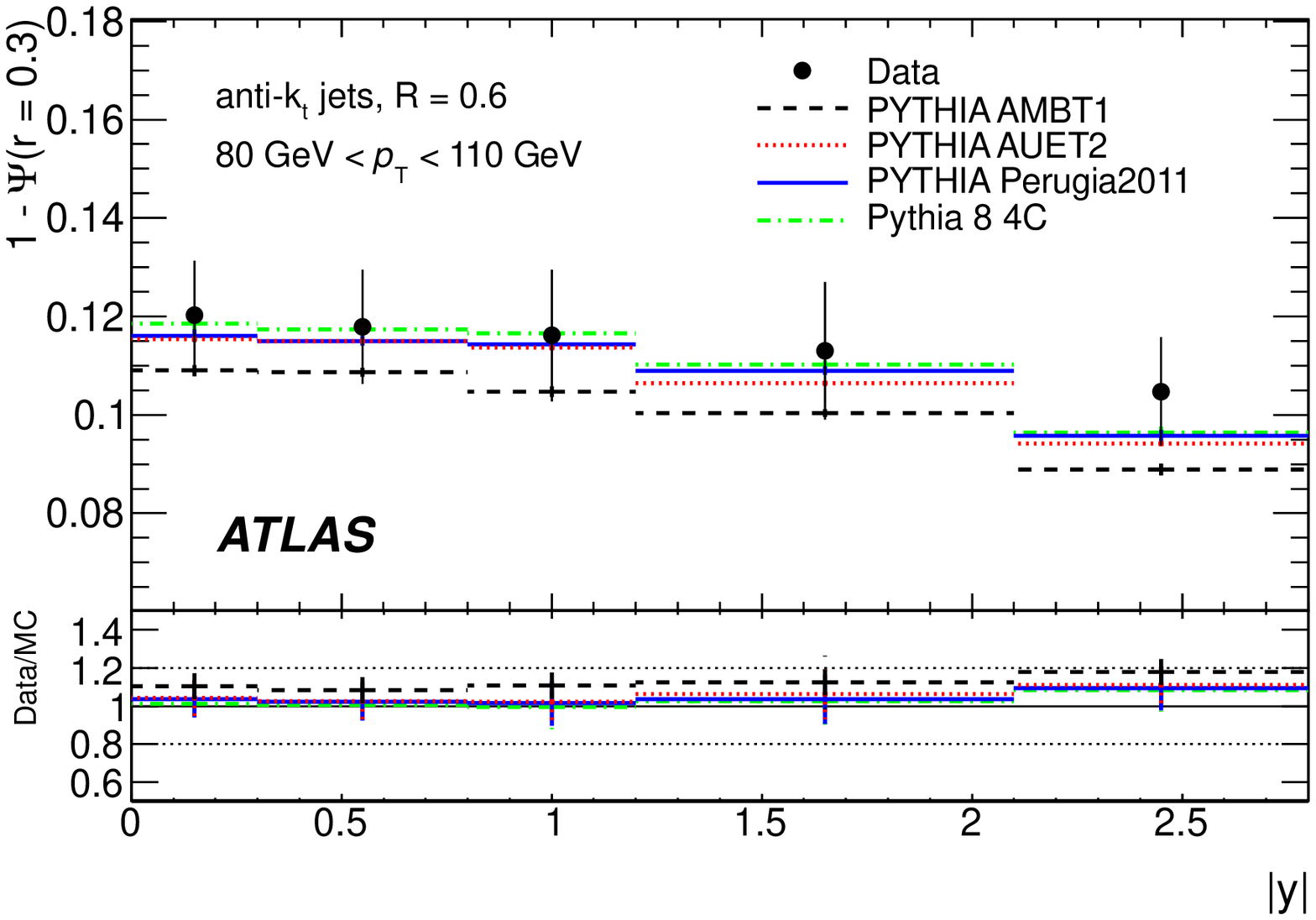}
}
\end{center}
\vspace{-0.7 cm}
\caption{\small
The measured integrated jet shape, $1 - \Psi(r=0.3)$, as a function of $|\rapjet|$ in different jet $\ptjet$ regions 
for jets with $|\rapjet| < 2.8$ and $30 \ {\rm GeV} < \ptjet < 110 \ {\rm GeV}$.
Error bars indicate the statistical and systematic uncertainties added in quadrature. 
The predictions of   PYTHIA-Perugia2011 (solid lines),   PYTHIA-AUET2 (dotted lines),   PYTHIA-AMBT1 (dashed lines), and Pythia~8-4C (dashed-dotted lines) are shown for comparison.
} 
\label{fig:pyt3}
\end{figure}

\begin{figure}[tbh]
\begin{center}
\mbox{
\includegraphics[width=0.495\textwidth]{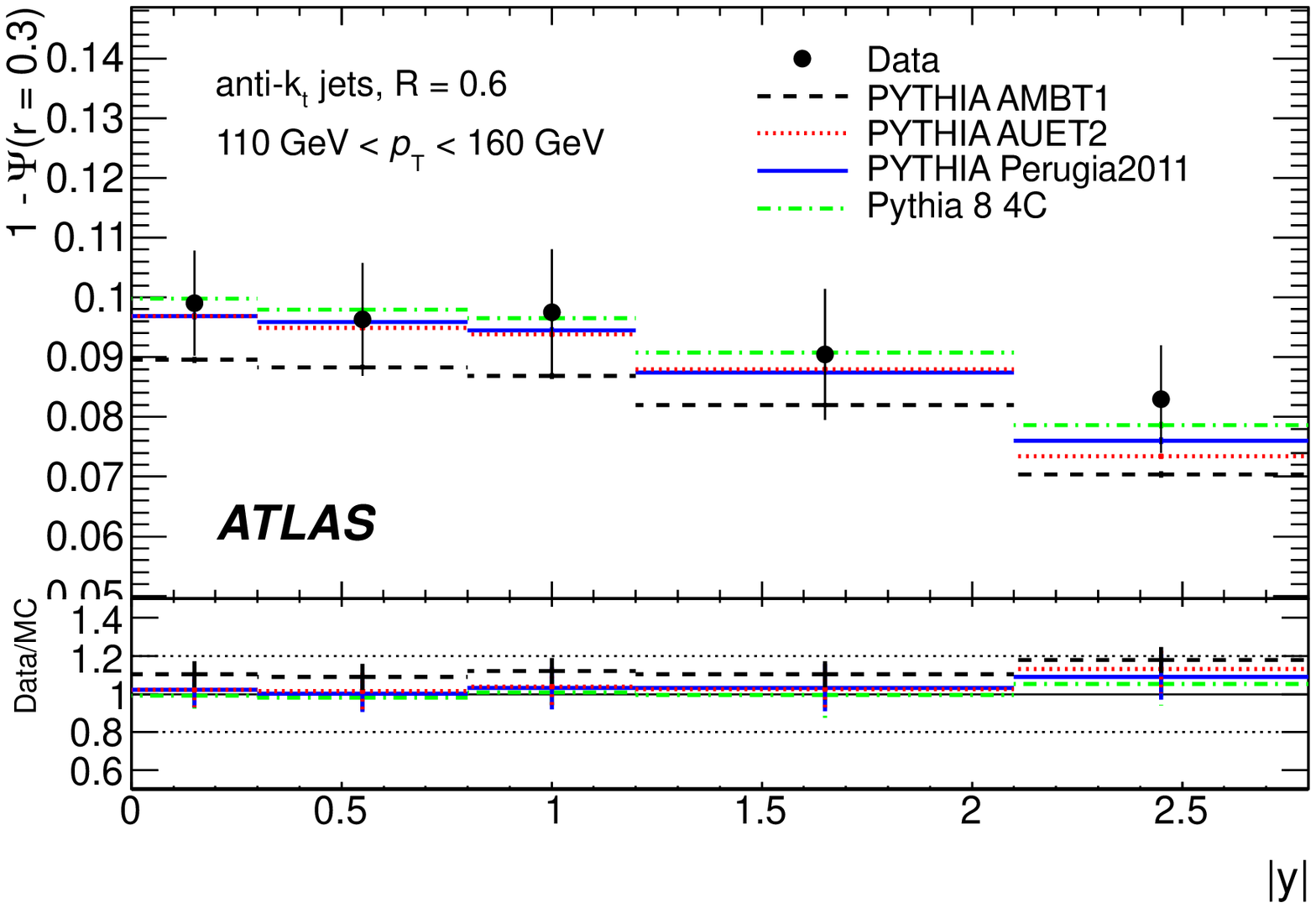}
\includegraphics[width=0.495\textwidth]{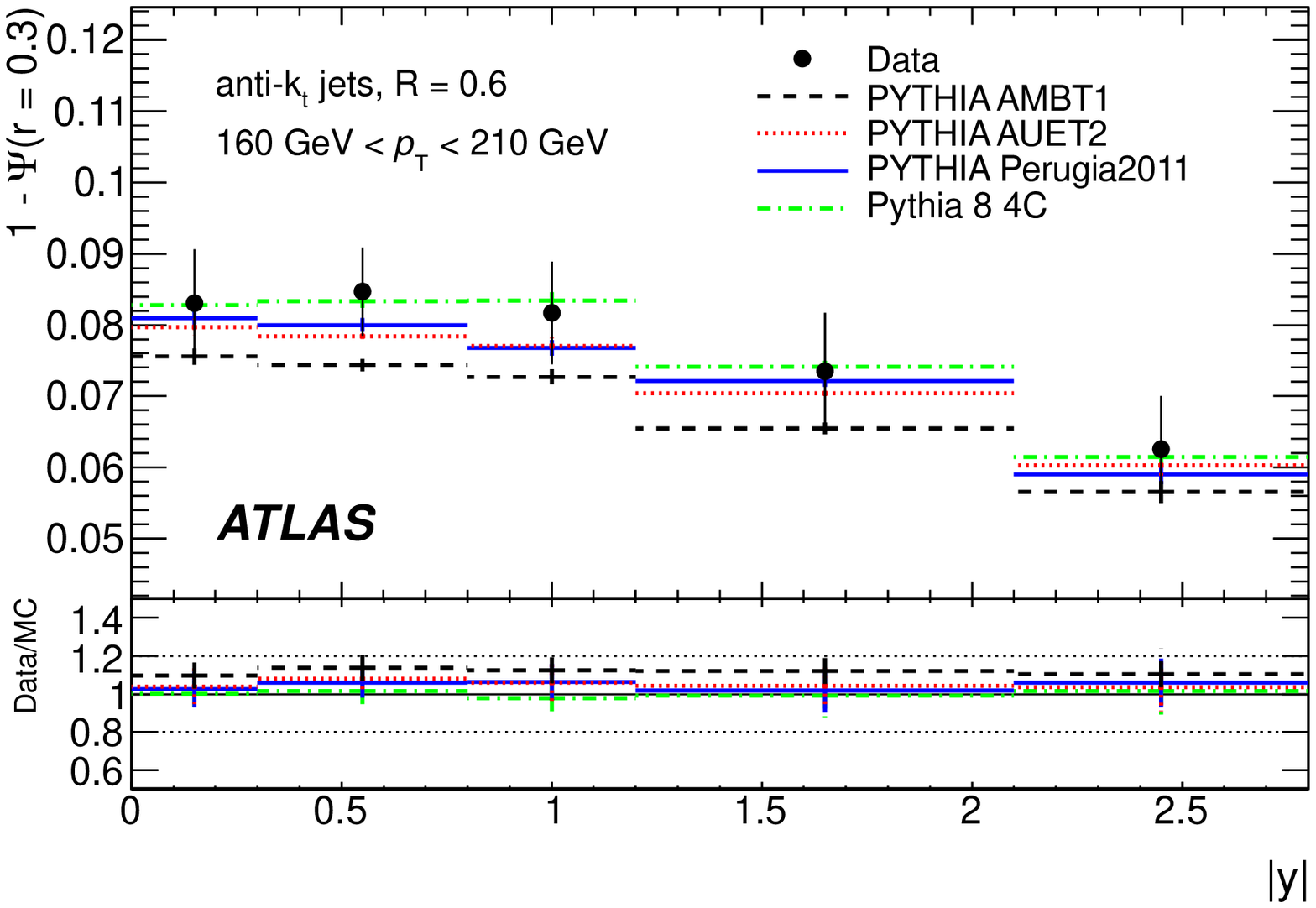}
}
\mbox{
\includegraphics[width=0.495\textwidth]{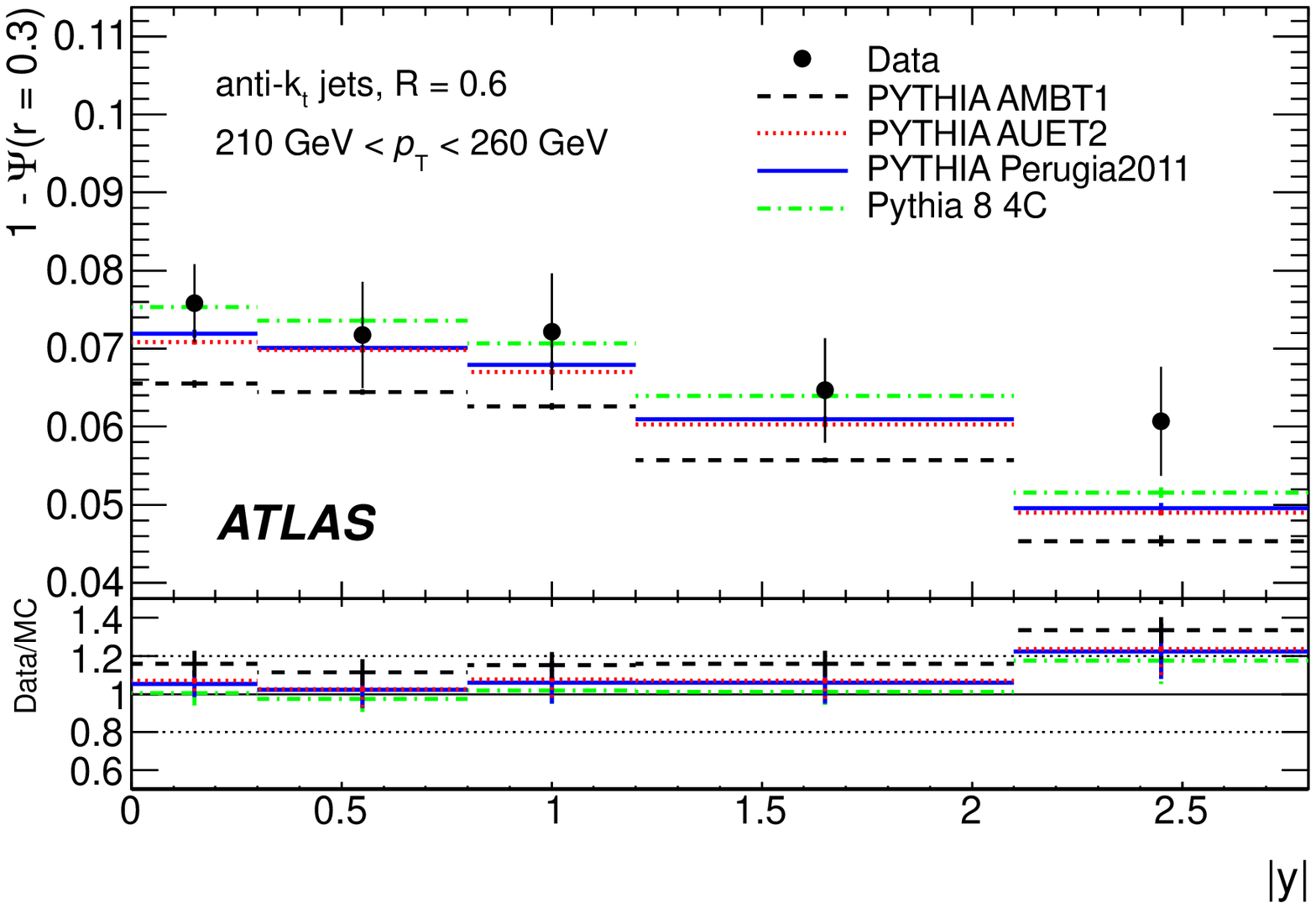} 
\includegraphics[width=0.495\textwidth]{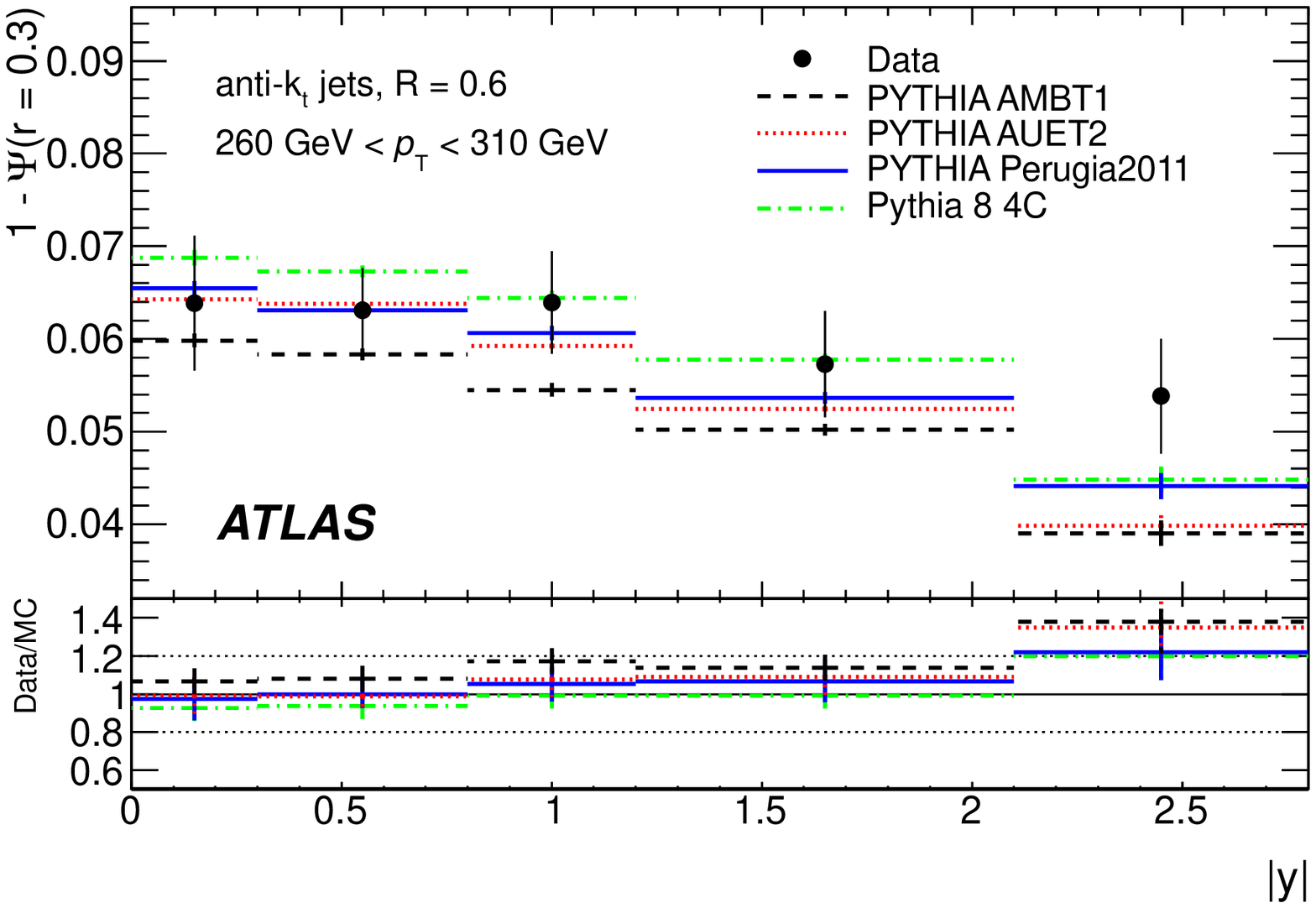}
}
\end{center}
\vspace{-0.7 cm}
\caption{\small
The measured integrated jet shape, $1 - \Psi(r=0.3)$, as a function of $|\rapjet|$ in different jet $\ptjet$ regions 
for jets with $|\rapjet| < 2.8$ and $110 \ {\rm GeV} < \ptjet < 310 \ {\rm GeV}$.
Error bars indicate the statistical and systematic uncertainties added in quadrature. 
The predictions of   PYTHIA-Perugia2011 (solid lines),   PYTHIA-AUET2 (dotted lines),   PYTHIA-AMBT1 (dashed lines), and Pythia~8-4C (dashed-dotted lines) are shown for comparison.
} 
\label{fig:pyt4}
\end{figure}

The measured jet shapes are compared to the different PYTHIA predictions in Figures~\ref{fig:pyt1}~to~\ref{fig:pyt5}.
As expected, PYTHIA6-AMBT1 tends to produce jets slightly narrower than the data as it underestimates the 
UE activity in dijet events~\cite{mc10} and lacks a tuned final-state parton shower from the initial-state 
radiation. 
PYTHIA6-AUET2, PYTHIA6-Perugia2011, and Pythia~8-4C provide a good 
description of the measured jet shapes in the whole jet kinematic range considered~\footnote{The reader should note that the  ATLAS jet shapes measurements are input to the PYTHIA6-AUET2 tune.}. 
       

\subsection{Comparison with Herwig++ and HERWIG/JIMMY}

The measurements are compared to the predictions from Herwig++ 2.4.2, Herwig++ 2.5.1, and HERWIG/JIMMY programs 
in Figure~\ref{fig:hrw1} to Figure~\ref{fig:hrw5}. As shown in the Figures,  
the effect of the wrong setting in Herwig++ 2.4.2 bug (see Section~2) 
is only visible for jets with $\ptjet < 40$~GeV. Herwig++ 2.4.2 provides a reasonable description of the data in the whole jet kinematic range considered.  HERWIG/JIMMY, which includes the AUET2 tune, describes the data best. Finally, Herwig++ 2.5.1 tends to produce jets narrower than the data.


\subsection{Comparison with ALPGEN and Sherpa}

The data are compared to the predictions from Sherpa 1.2.3 ($2 \to 2$ process), Sherpa 1.2.3 (up to $2 \to 6$ process), Sherpa 1.3.0 ($2 \to 2$ process), and ALPGEN interfaced with 
PYTHIA in Figure~\ref{fig:meps1} to Figure~\ref{fig:meps5}. Comparisons with ALPGEN interfaced with 
HERWIG and JIMMY were presented in ~\cite{pp}. The different Sherpa predictions are similar and provide a 
reasonable description of the data. This indicates that the presence of additional partons from 
higher-order matrix elements contributions do not affect the predicted jet shapes, mainly dictated by the 
soft radiation in the parton shower. The comparison between Sherpa 1.2.3 and Sherpa 1.3.0 shows that the 
NLL-inspired corrections included in the latter for the  parton shower  do not impact significantly 
the predicted jet shapes. ALPGEN interfaced with PYTHIA predicts too-narrow jets and does not describe the data. This was already the case for ALPGEN interfaced with HERWIG and JIMMY~\cite{pp} and requires further 
investigation to determine whether the disagreement observed with the data 
can be completely attributed to the  UE modeling in the MC samples or it is 
also related to the prescription followed by ALPGEN in merging the partons from the 
matrix elements with the parton showers in the final state.


\subsection{Comparison with POWHEG}

The impact of the presence of higher-order matrix elements contributions on the predicted jet shapes is 
further studied in Figure~\ref{fig:pow1} to Figure~\ref{fig:pow5}, where the data are compared to the predictions 
from POWHEG interfaced with PYTHIA or  HERWIG/JIMMY,  as well as to standalone PYTHIA
and HERWIG/JIMMY predictions.  For these comparisons, AMBT1 and AUET1-LO${}^{*}$ tuned sets of parameters are 
used for PYTHIA and HERWIG/JIMMY, respectively.  
In general, the  jet shapes predicted by 
POWHEG follow the trend of the corresponding
standalone predictions. POWHEG interfaced with HERWIG/JIMMY provides a reasonable description of the data while
the interface with PYTHIA predicts too-narrow jets.
This confirms that the shape of the jet is mainly dictated
by the parton shower implementation and the details of the UE modeling in the final state.

\clearpage


\begin{figure}[tbh]
\begin{center}
\mbox{
\includegraphics[width=0.495\textwidth]{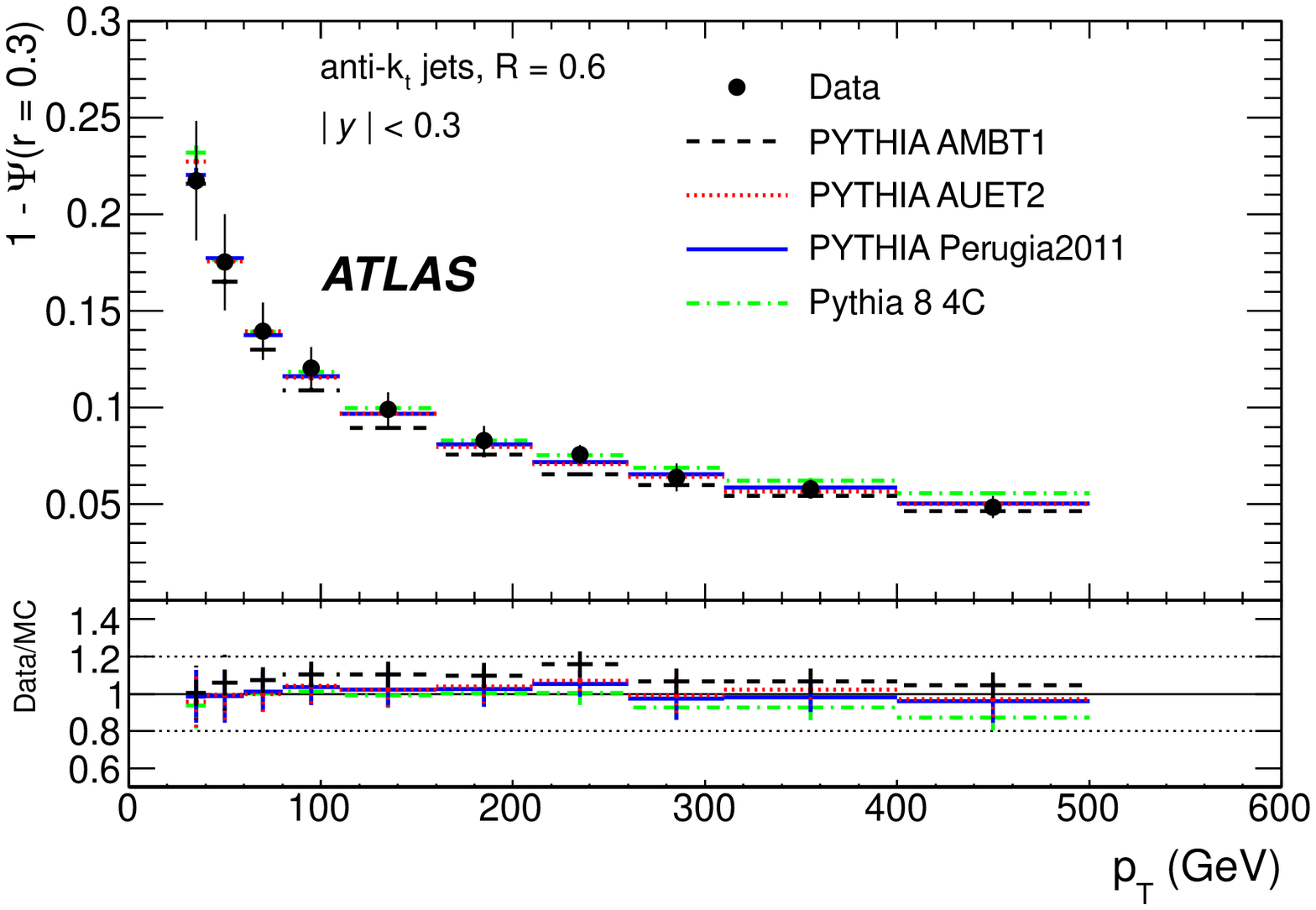}
\includegraphics[width=0.495\textwidth]{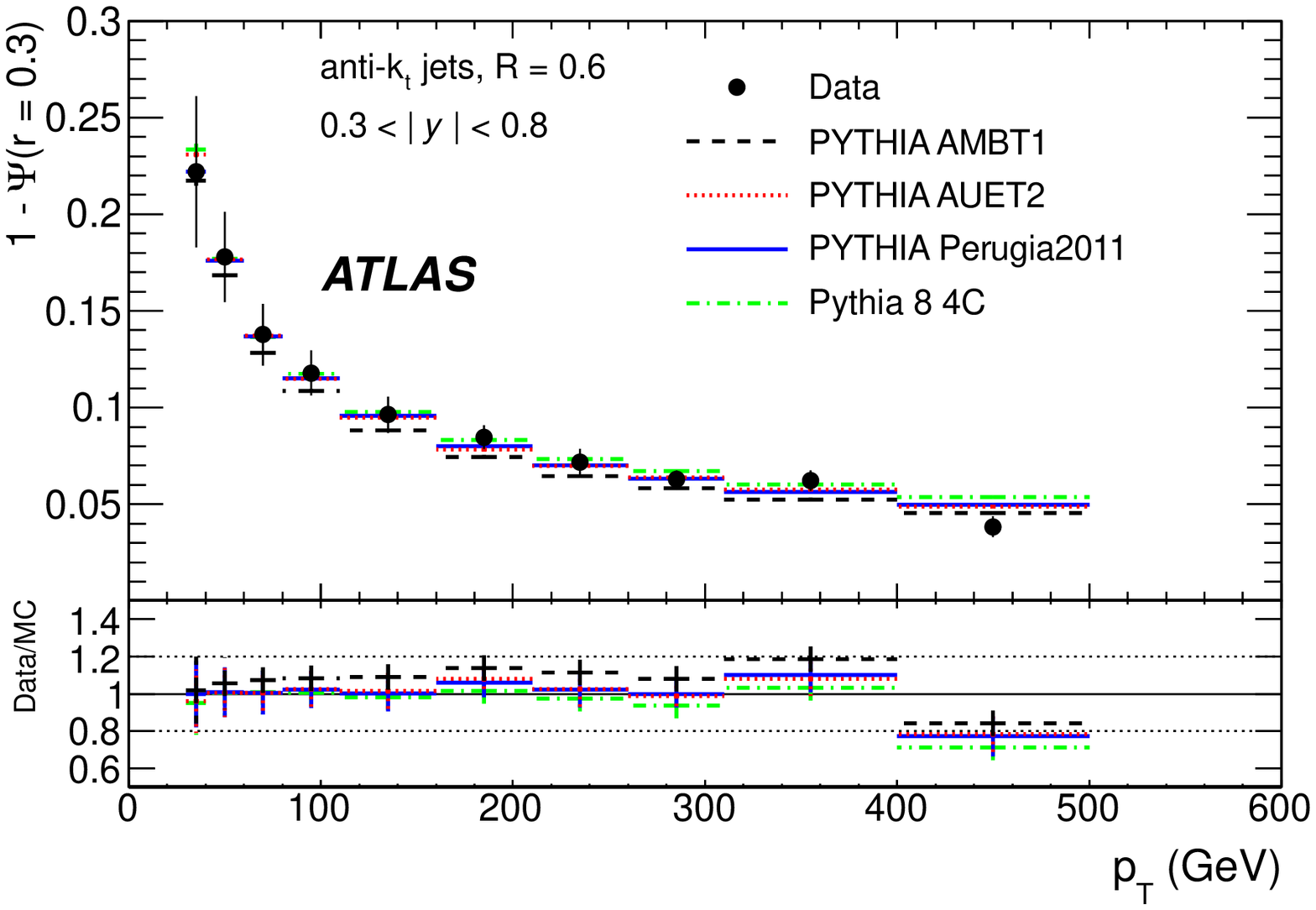}
}
\mbox{
\includegraphics[width=0.495\textwidth]{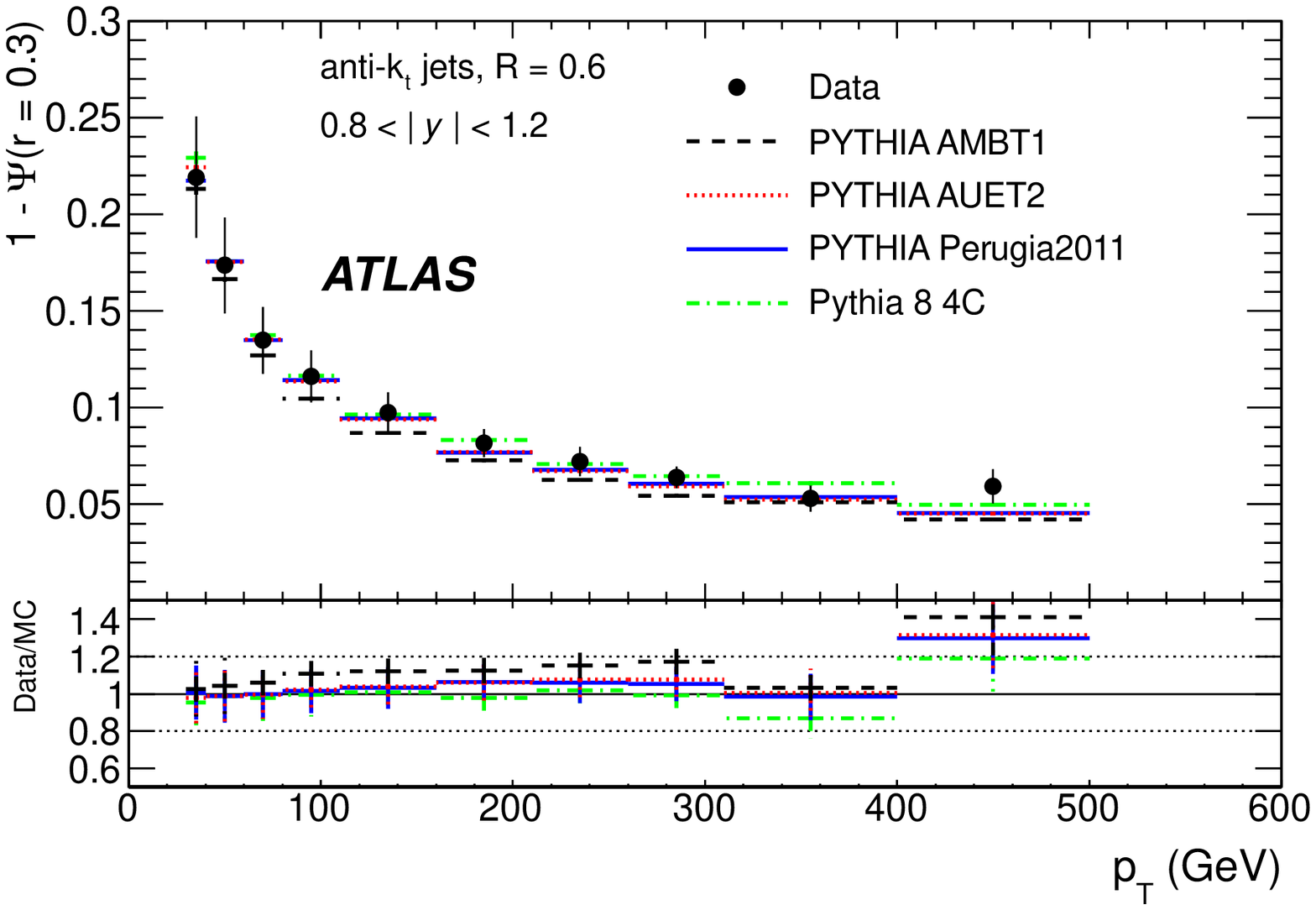} 
\includegraphics[width=0.495\textwidth]{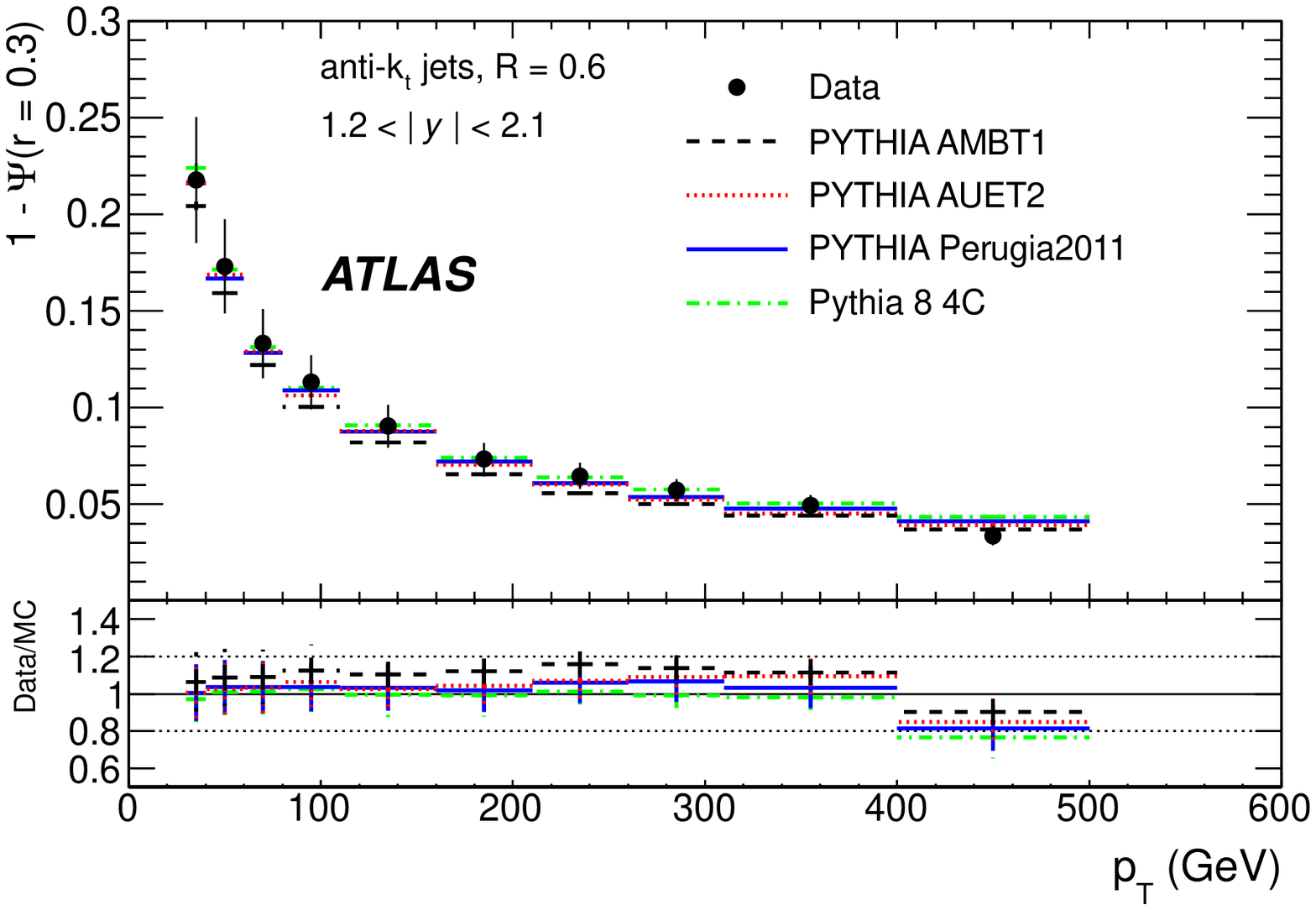}
}
\mbox{
\includegraphics[width=0.495\textwidth]{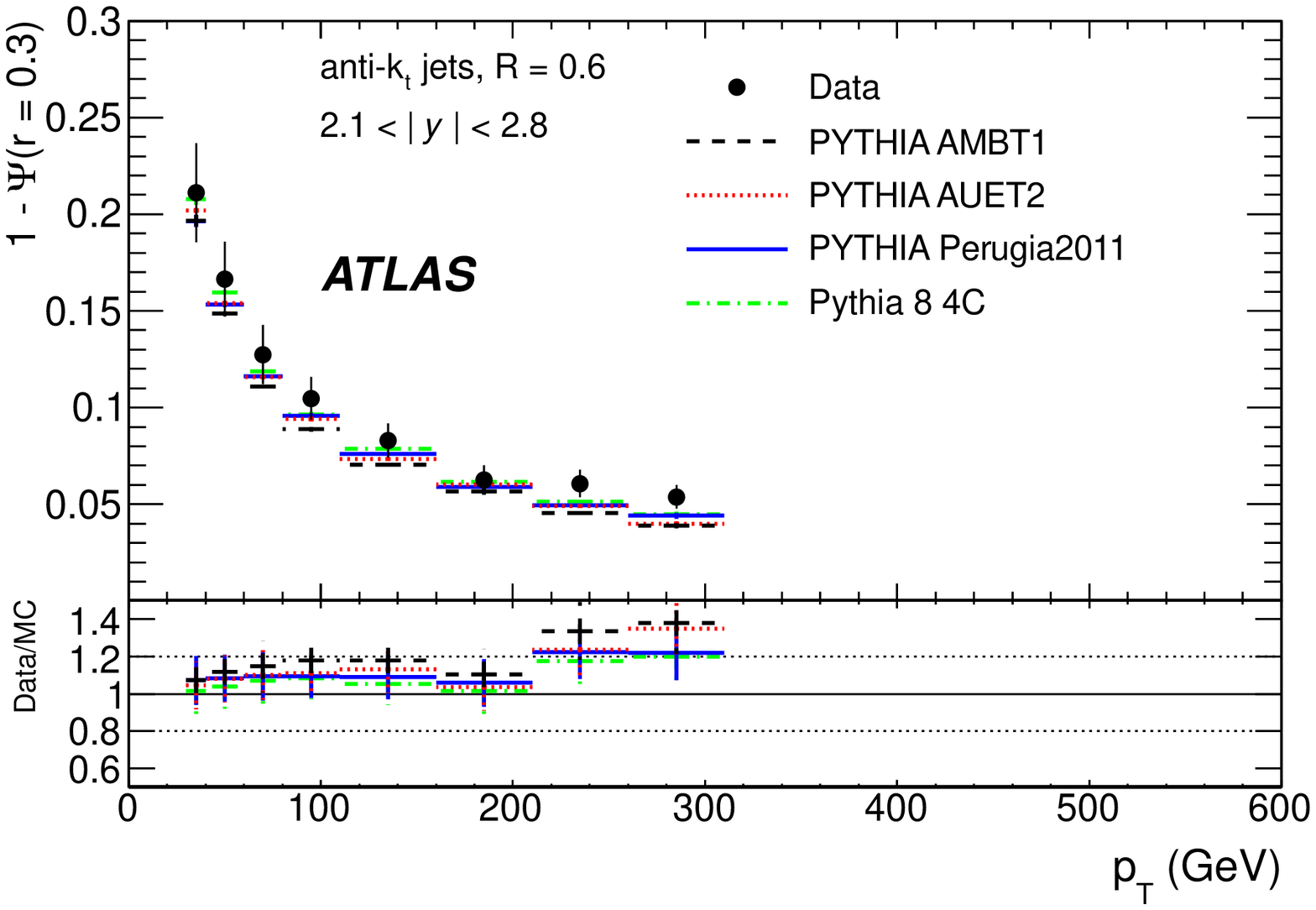}
}
\end{center}
\vspace{-0.7 cm}
\caption{\small
The measured integrated jet shape, $1 - \Psi(r=0.3)$, as a function of $\ptjet$ in different jet rapidity regions 
for jets with $|\rapjet| < 2.8$ and $30 \ {\rm GeV} < \ptjet < 500 \ {\rm GeV}$.
Error bars indicate the statistical and systematic uncertainties added in quadrature. 
The predictions of   PYTHIA-Perugia2011 (solid lines),   PYTHIA-AUET2 (dotted lines),   PYTHIA-AMBT1 (dashed lines), and Pythia~8-4C (dashed-dotted lines) are shown for comparison.
} 
\label{fig:pyt5}
\end{figure}



\begin{figure}[tbh]
\begin{center}
\mbox{
\includegraphics[width=0.495\textwidth,height=0.495\textwidth]{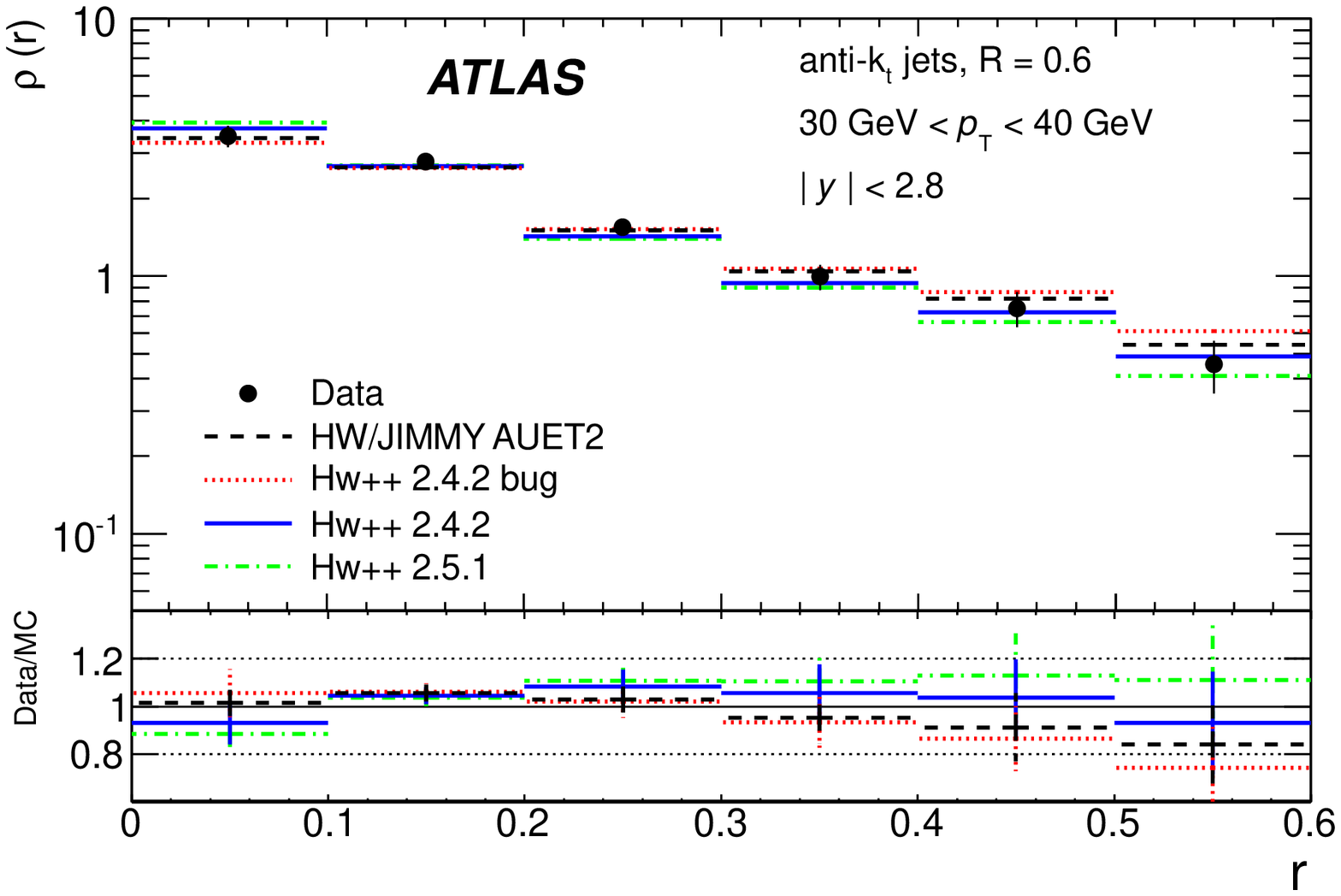} 
\includegraphics[width=0.495\textwidth,height=0.495\textwidth]{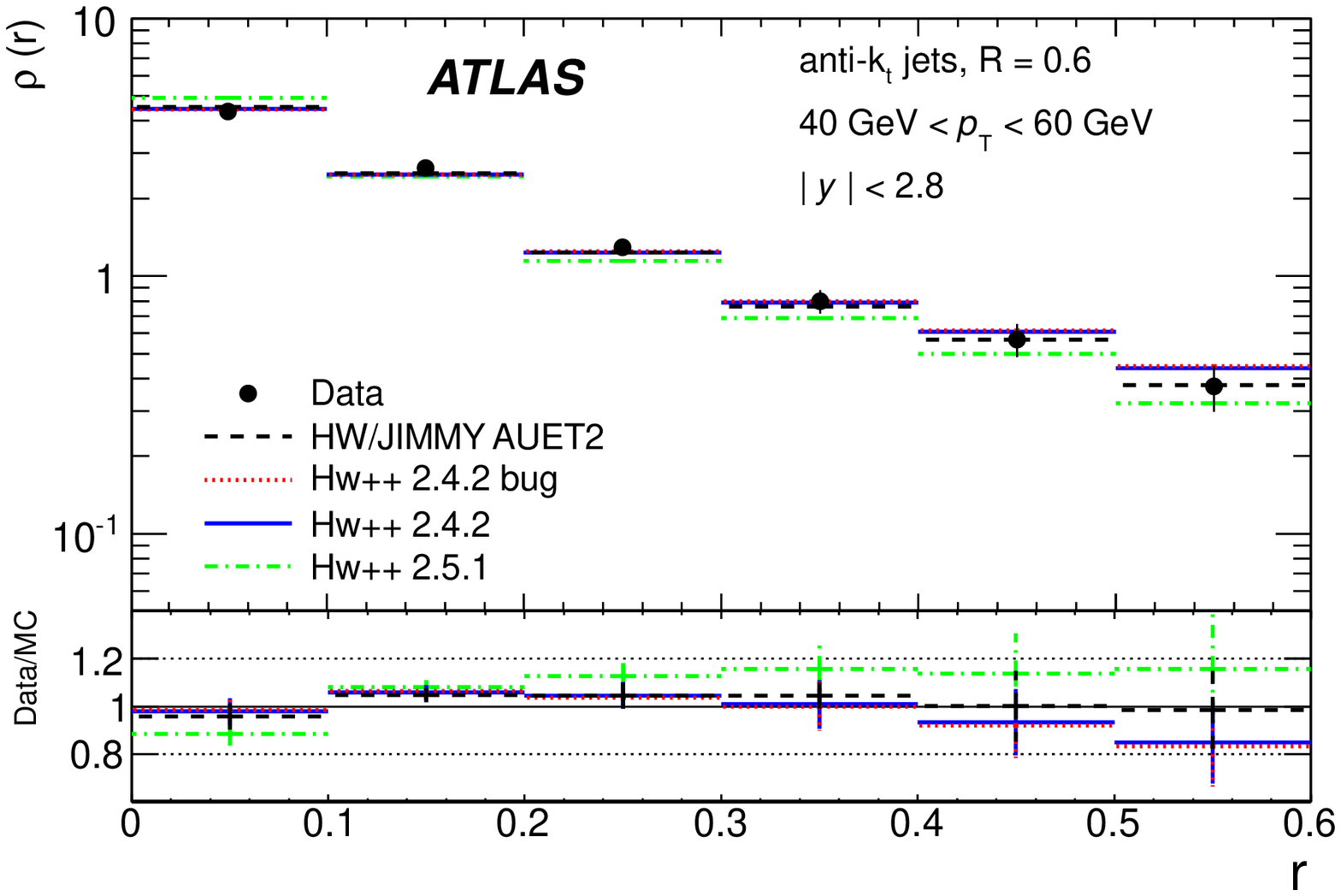}
}\vspace{-0.2cm}
\mbox{
\includegraphics[width=0.495\textwidth,height=0.495\textwidth]{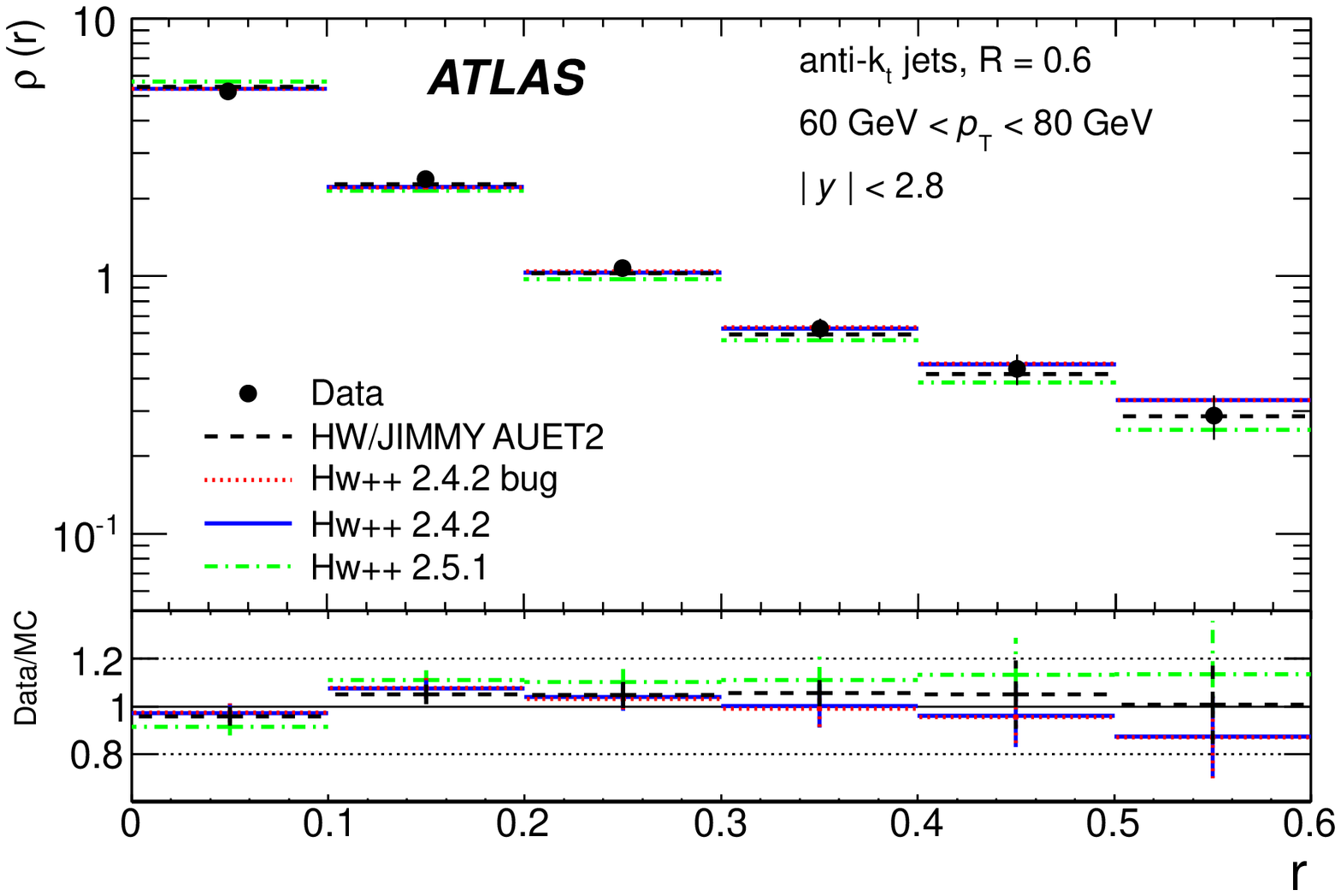}
\includegraphics[width=0.495\textwidth,height=0.495\textwidth]{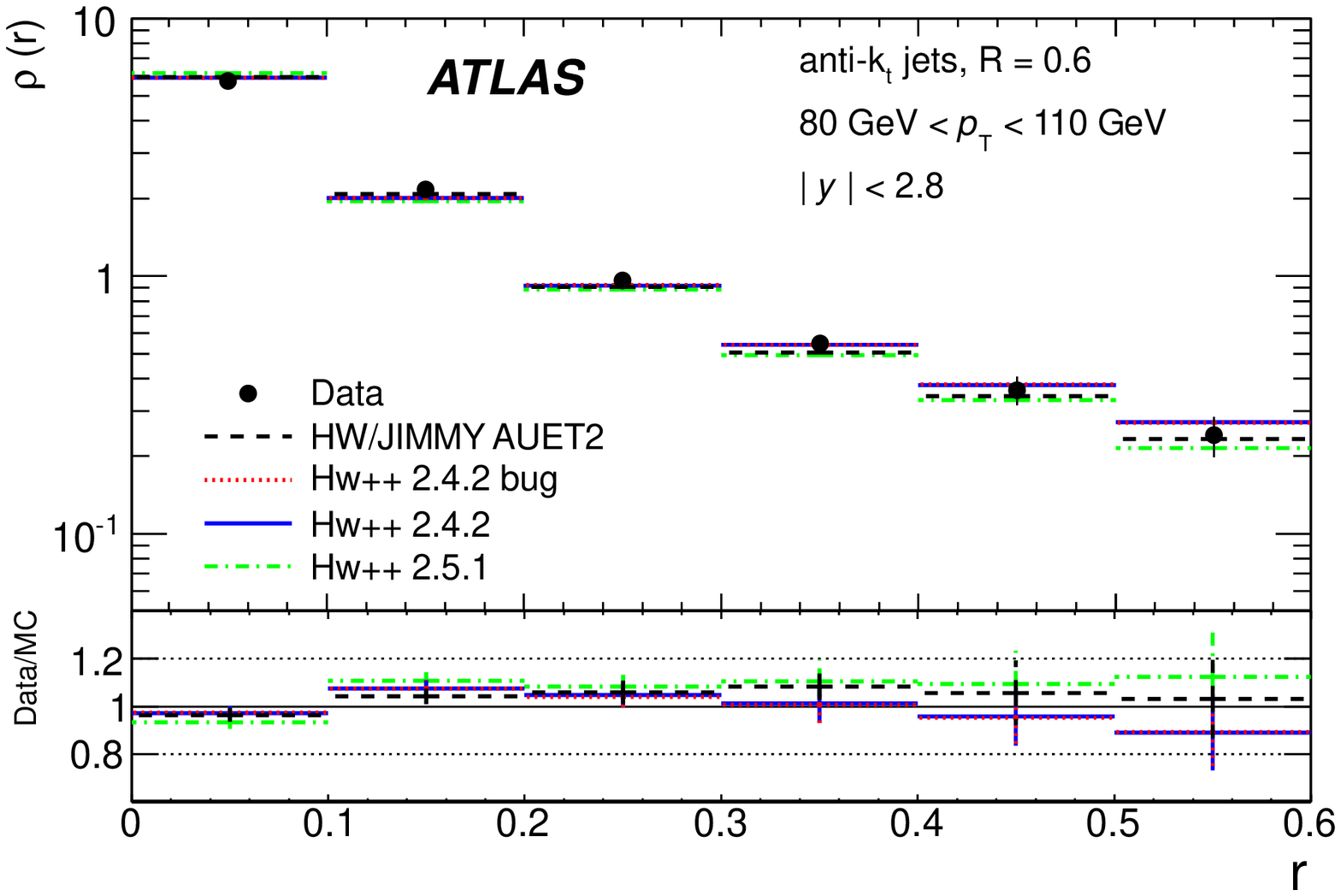}
}
\end{center}
\vspace{-0.7 cm}
\caption{\small
The measured differential jet shape, $\rho(r)$, in inclusive jet production for jets 
with $|\rapjet| < 2.8$ and $30 \ {\rm GeV} < \ptjet < 110  \ {\rm GeV}$   
is shown in different $\ptjet$ regions. Error bars indicate the statistical and systematic uncertainties added in quadrature.
The predictions of   Herwig++2.4.2 (solid lines),   Herwig++2.4.2 bug (dotted lines),   Herwig++ 2.5.1 (dashed-dotted lines), and HERWIG/JIMMY-AUET2 (dashed lines) are shown for comparison.} 
\label{fig:hrw1}
\end{figure}

\begin{figure}[tbh]
\begin{center}
\mbox{
\includegraphics[width=0.495\textwidth,height=0.495\textwidth]{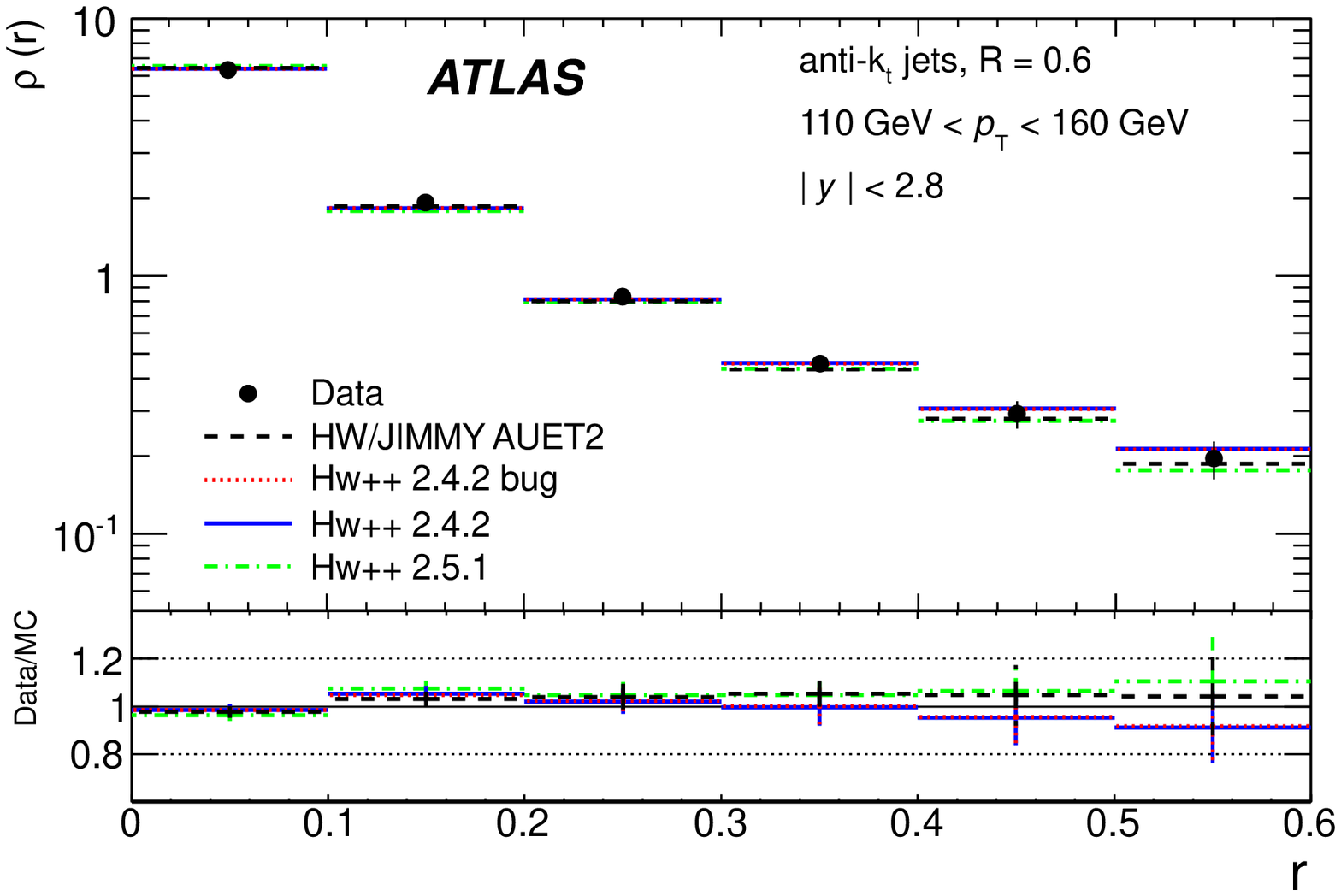} 
\includegraphics[width=0.495\textwidth,height=0.495\textwidth]{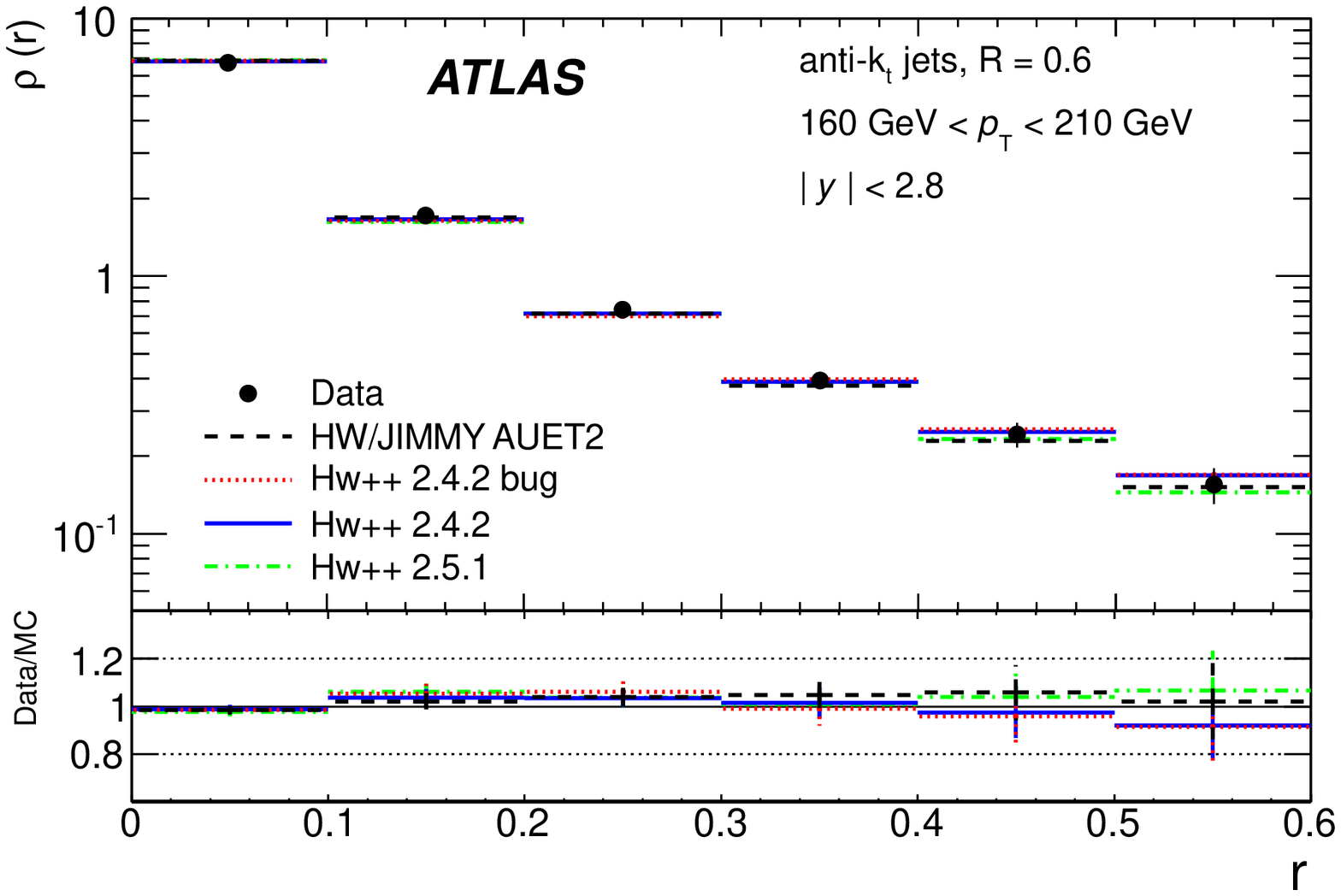}
}\vspace{-0.2cm}
\mbox{
\includegraphics[width=0.495\textwidth,height=0.495\textwidth]{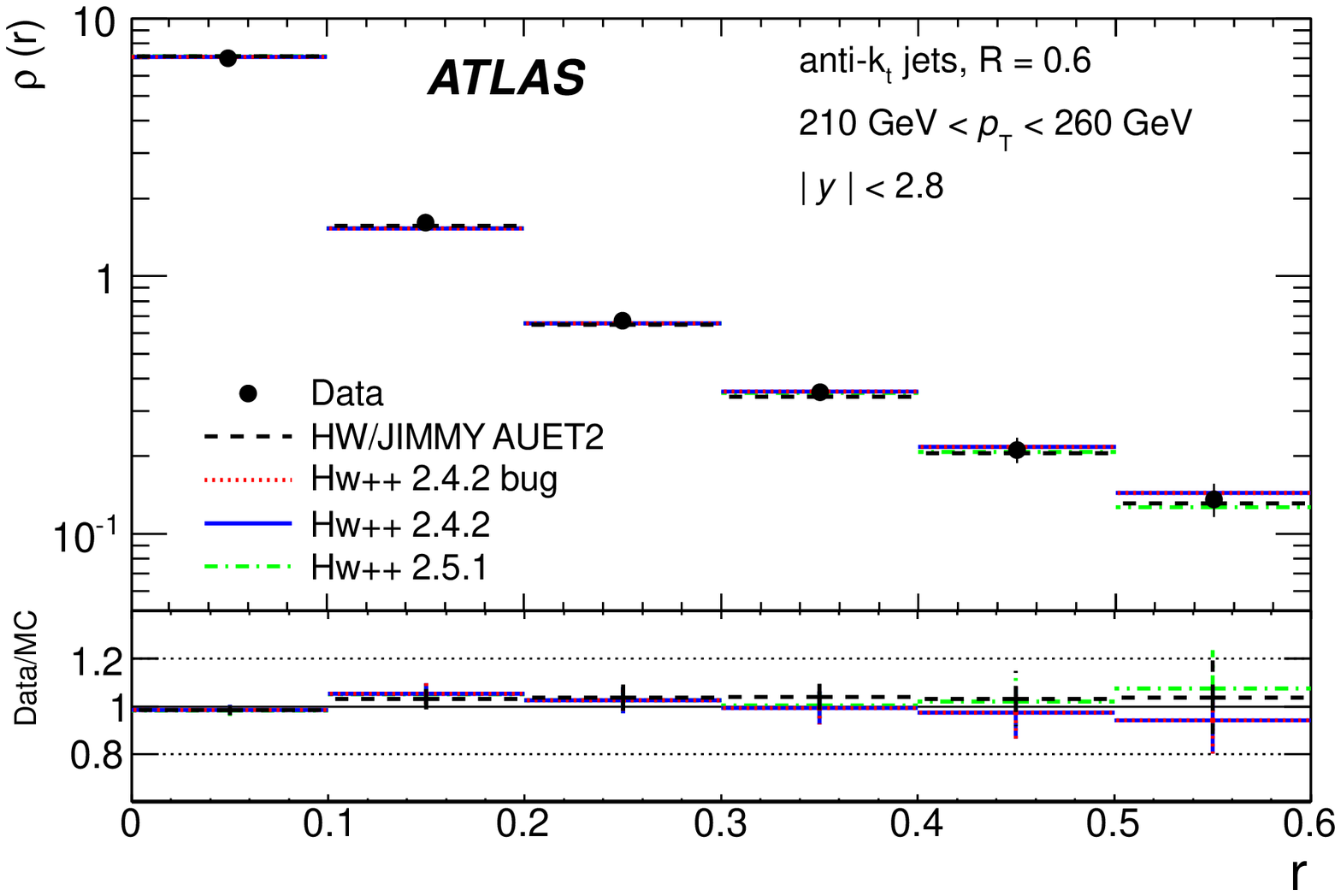}
\includegraphics[width=0.495\textwidth,height=0.495\textwidth]{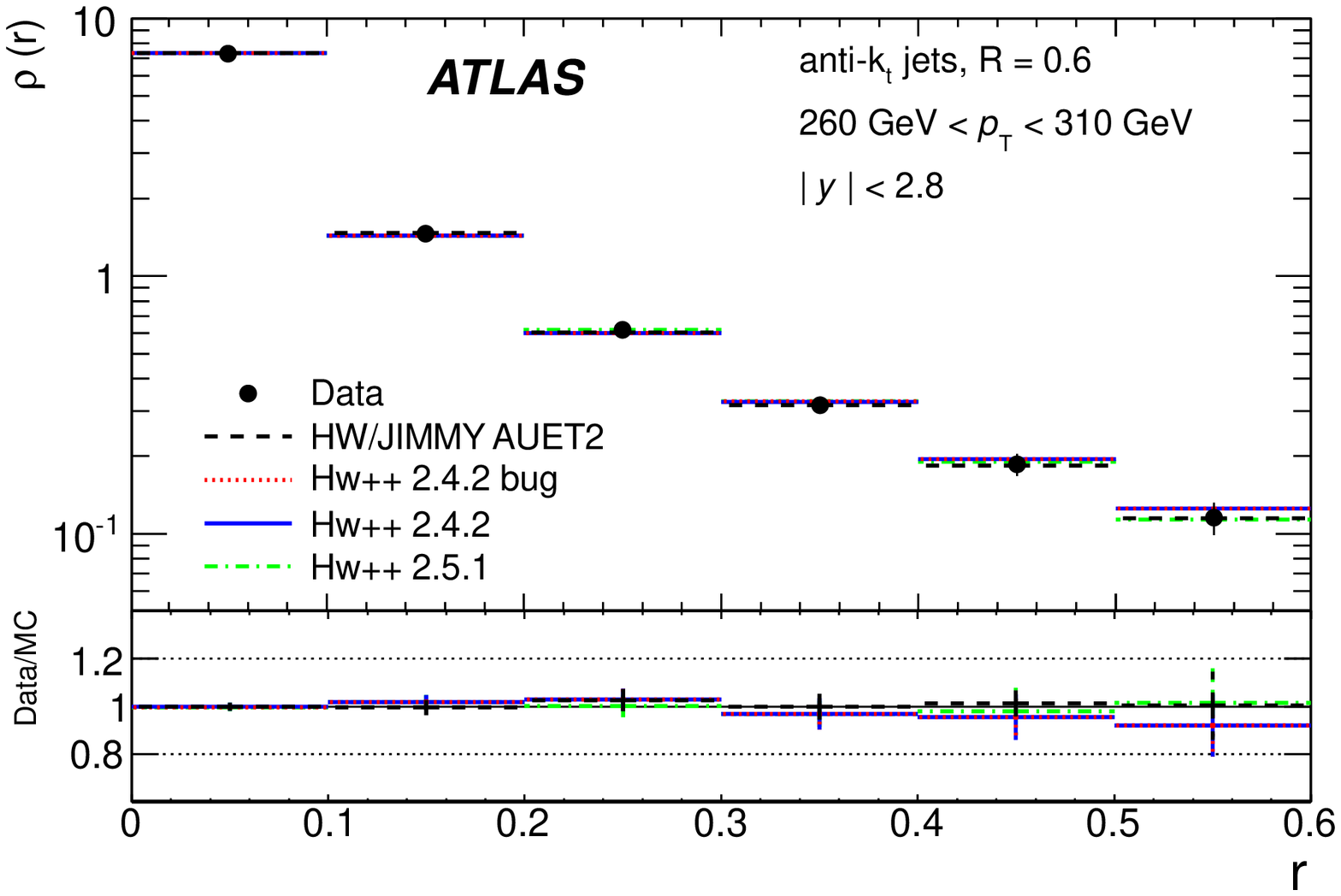}
}
\end{center}
\vspace{-0.7 cm}
\caption{\small
The measured differential jet shape, $\rho(r)$, in inclusive jet production for jets 
with $|\rapjet| < 2.8$ and $110 \ {\rm GeV} < \ptjet < 310  \ {\rm GeV}$   
is shown in different $\ptjet$ regions. Error bars indicate the statistical and systematic uncertainties added in quadrature.
The predictions of   Herwig++2.4.2 (solid lines),   Herwig++2.4.2 bug (dotted lines),   Herwig++ 2.5.1 (dashed-dotted lines), and HERWIG/JIMMY-AUET2 (dashed lines) are shown for comparison.} 
\label{fig:hrw2}
\end{figure}


\begin{figure}[tbh]
\begin{center}
\mbox{
\includegraphics[width=0.495\textwidth]{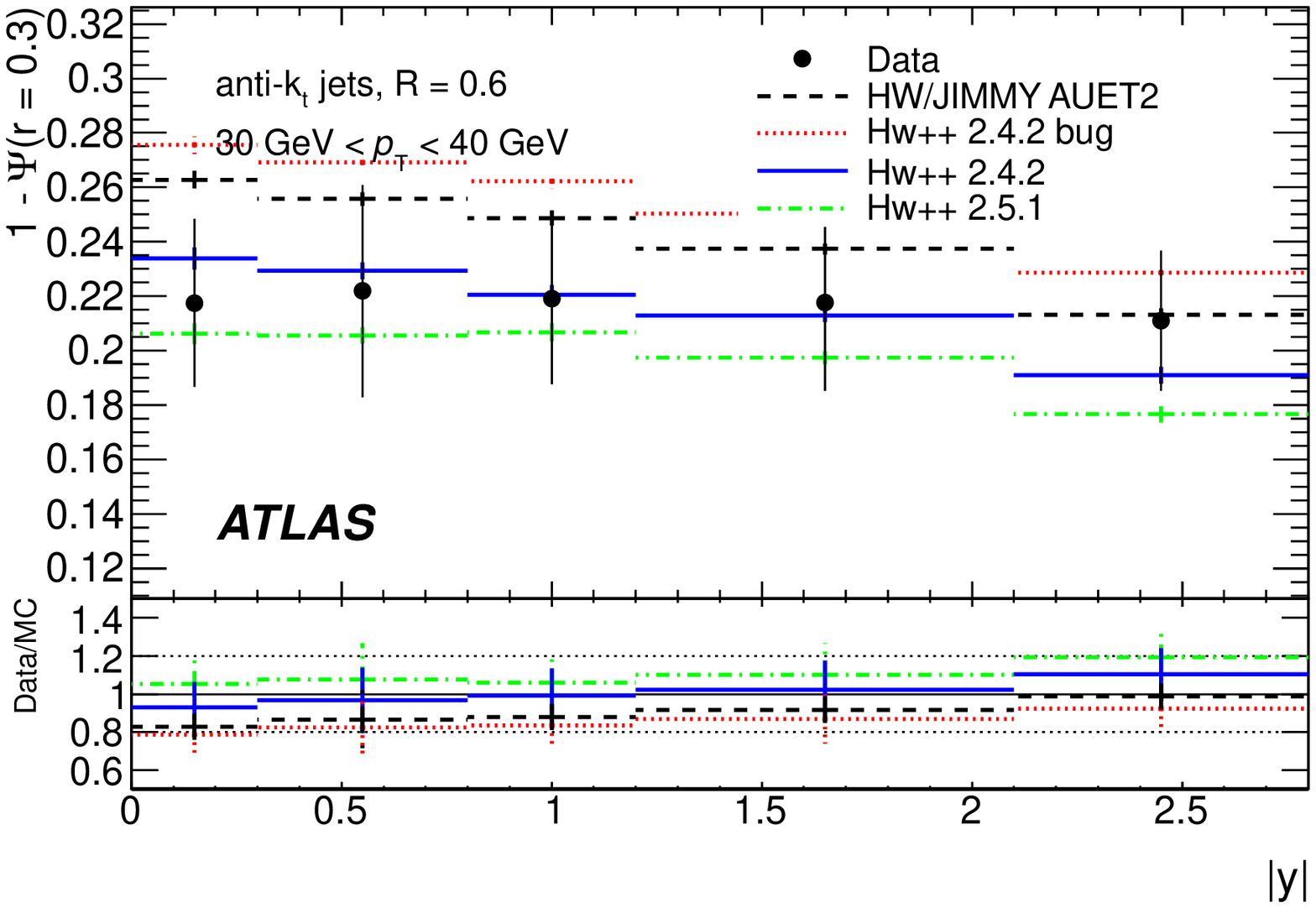}
\includegraphics[width=0.495\textwidth]{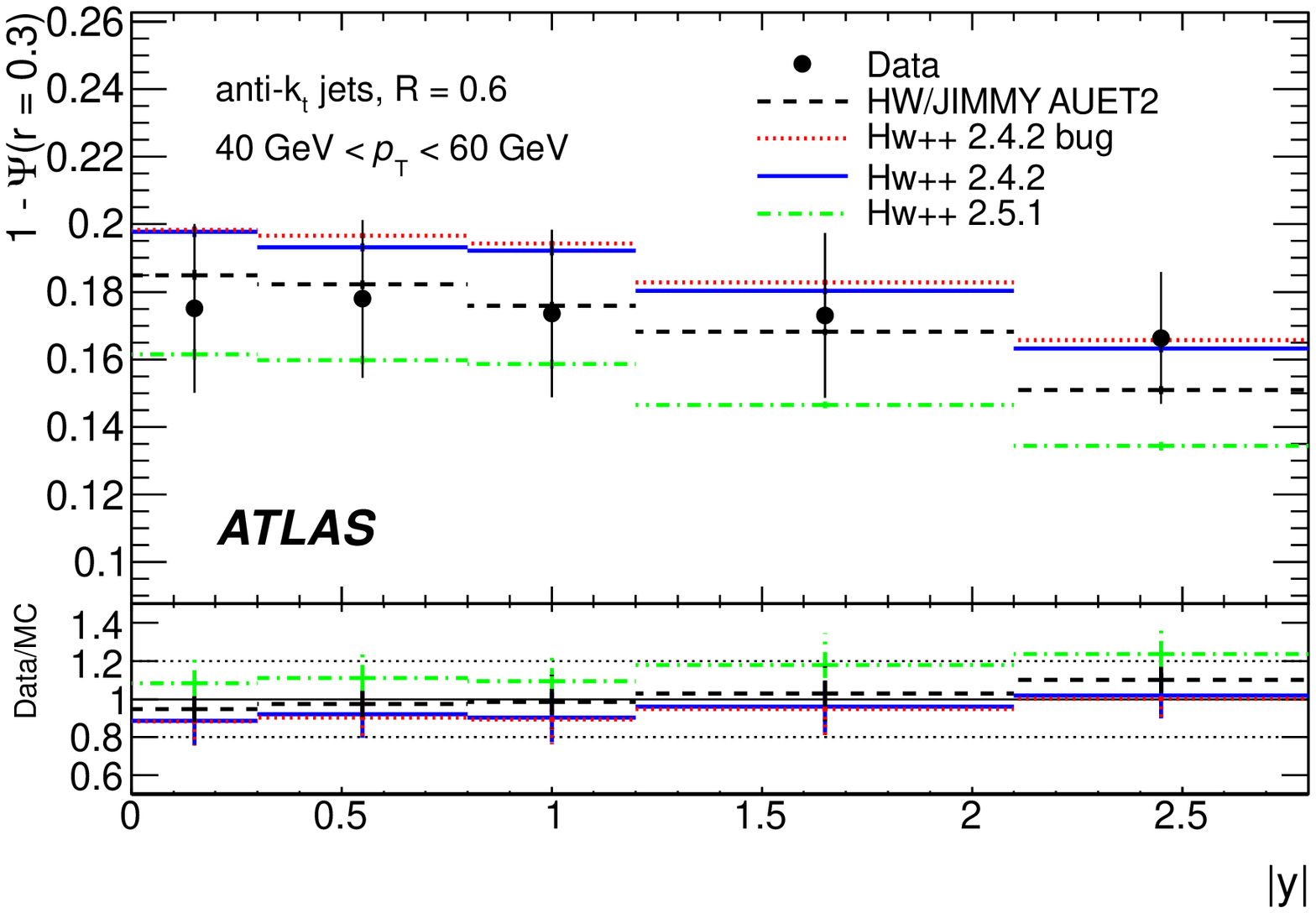}
}
\mbox{
\includegraphics[width=0.495\textwidth]{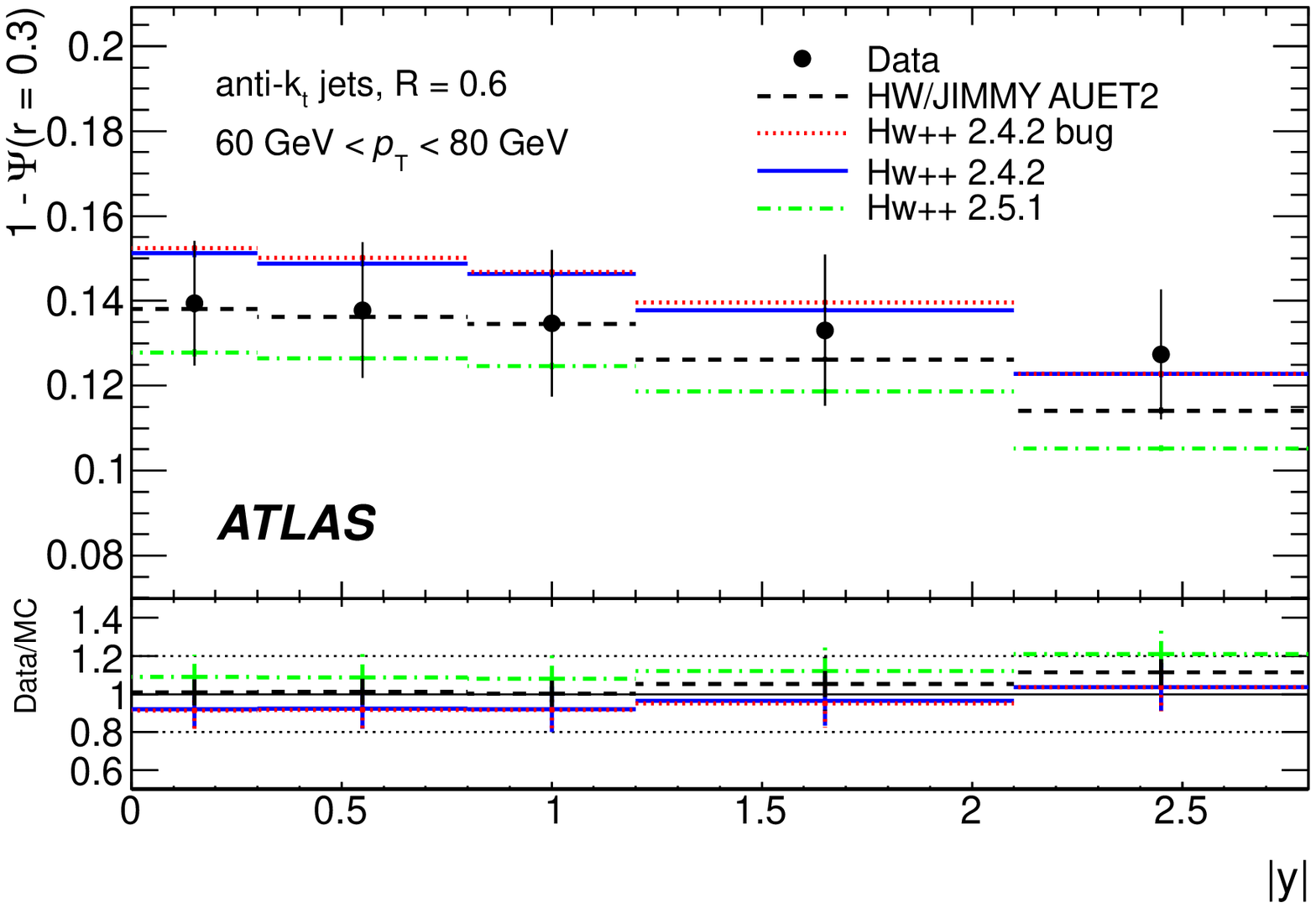} 
\includegraphics[width=0.495\textwidth]{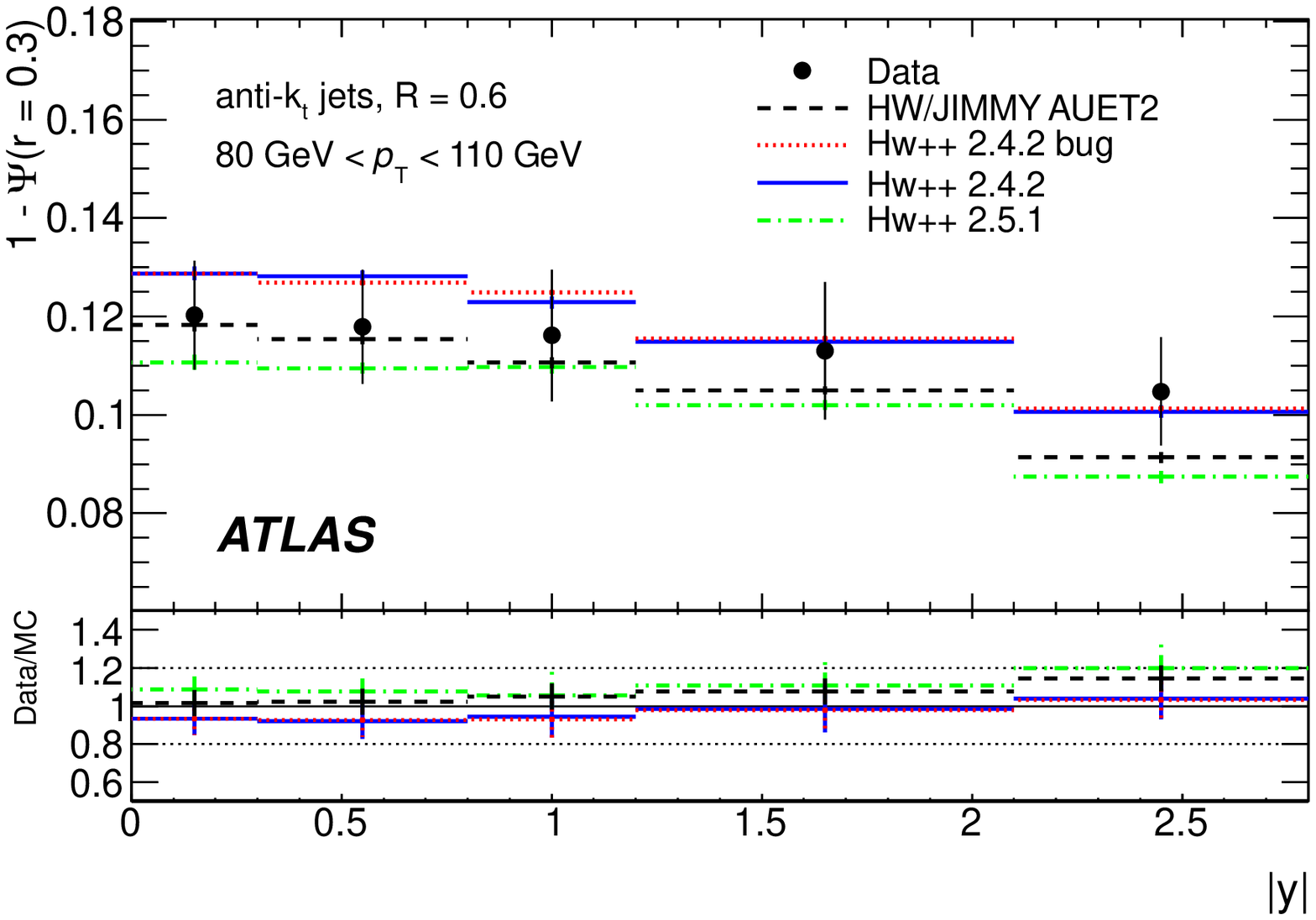}
}
\end{center}
\vspace{-0.7 cm}
\caption{\small
The measured integrated jet shape, $1 - \Psi(r=0.3)$, as a function of $|\rapjet|$ in different jet $\ptjet$ regions 
for jets with $|\rapjet| < 2.8$ and $30 \ {\rm GeV} < \ptjet < 110 \ {\rm GeV}$.
Error bars indicate the statistical and systematic uncertainties added in quadrature. 
The predictions of   Herwig++2.4.2 (solid lines),   Herwig++2.4.2 bug (dotted lines),   Herwig++ 2.5.1 (dashed-dotted lines), and HERWIG/JIMMY-AUET2 (dashed lines) are shown for comparison.
} 
\label{fig:hrw3}
\end{figure}

\begin{figure}[tbh]
\begin{center}
\mbox{
\includegraphics[width=0.495\textwidth]{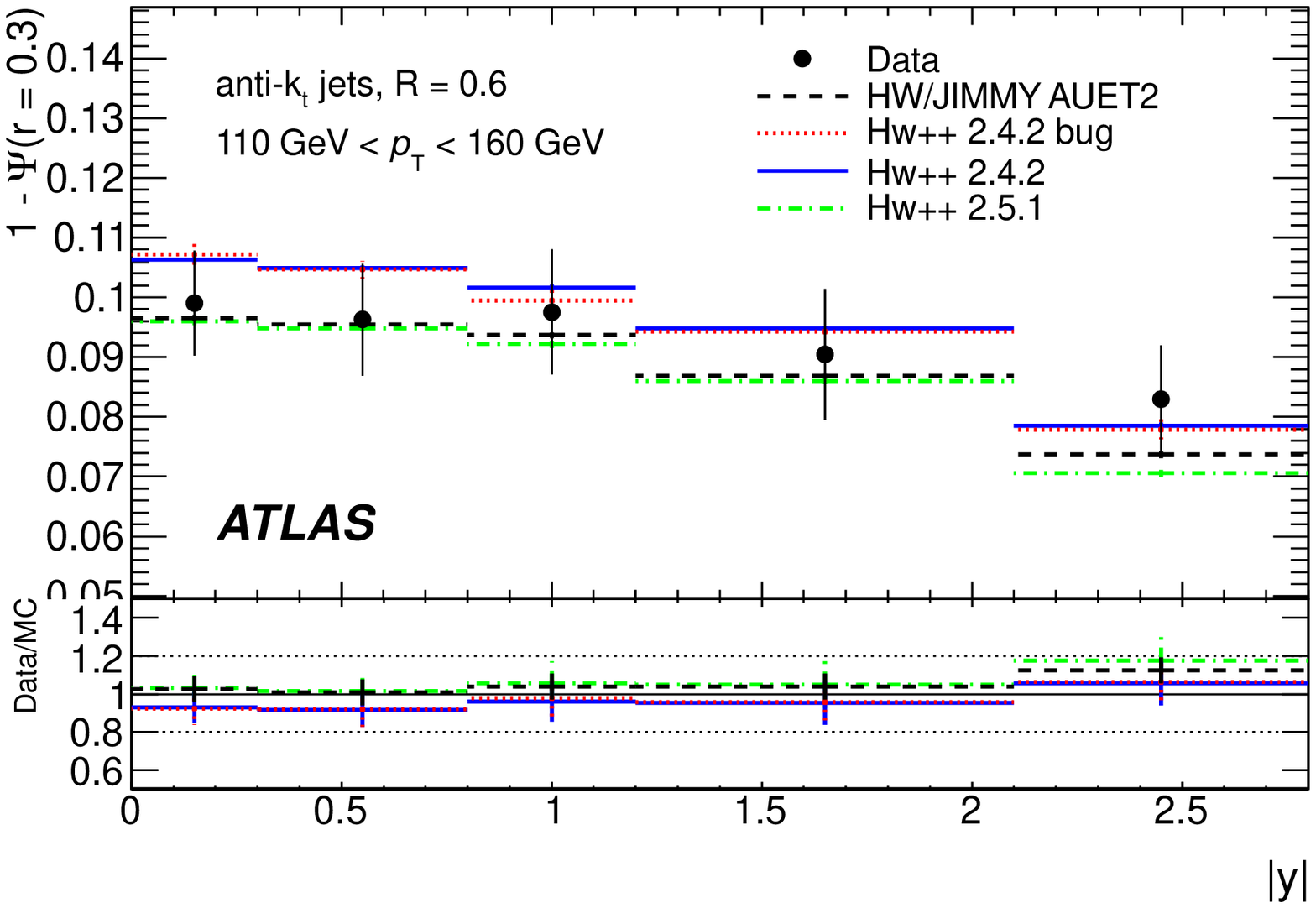}
\includegraphics[width=0.495\textwidth]{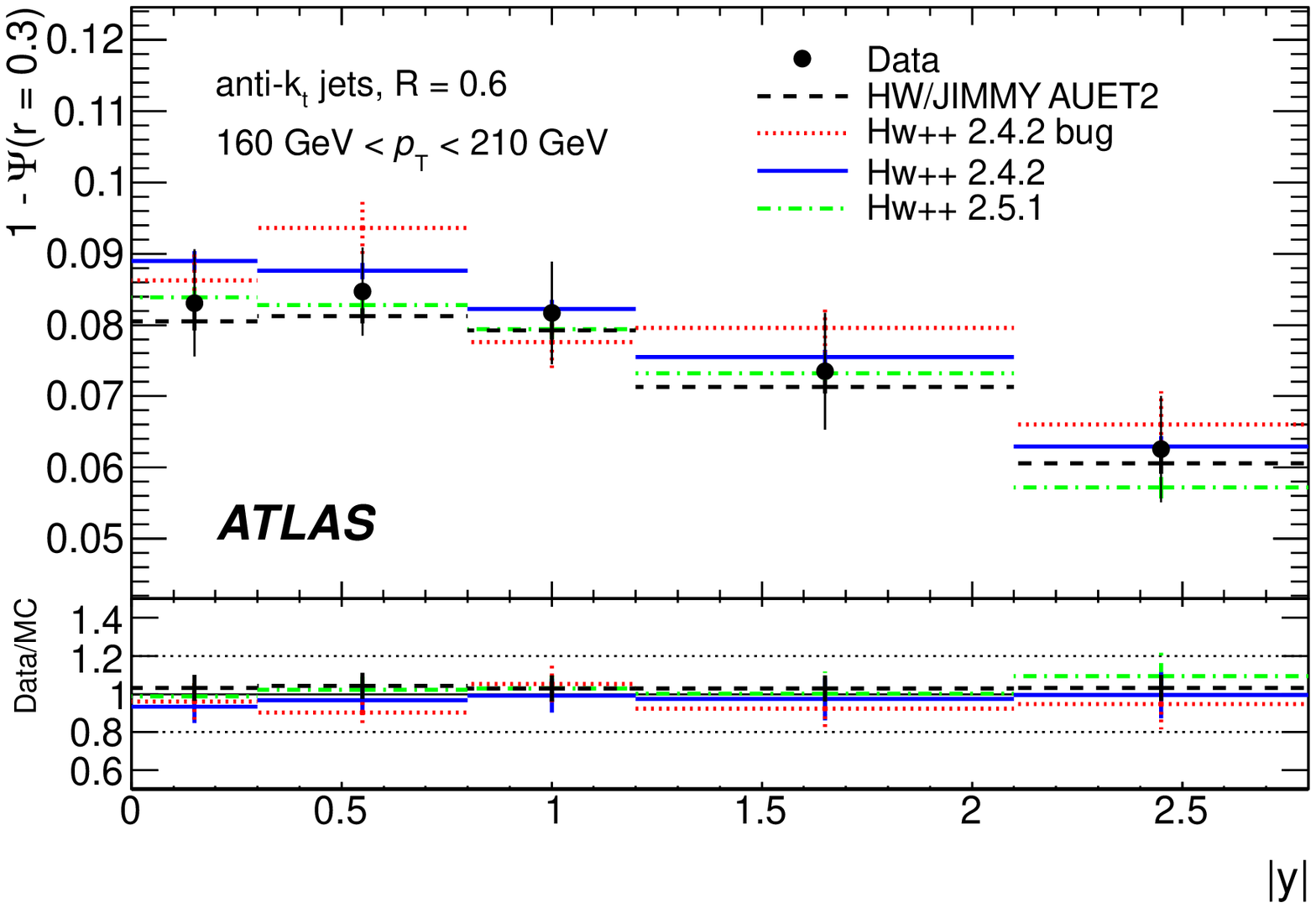}
}
\mbox{
\includegraphics[width=0.495\textwidth]{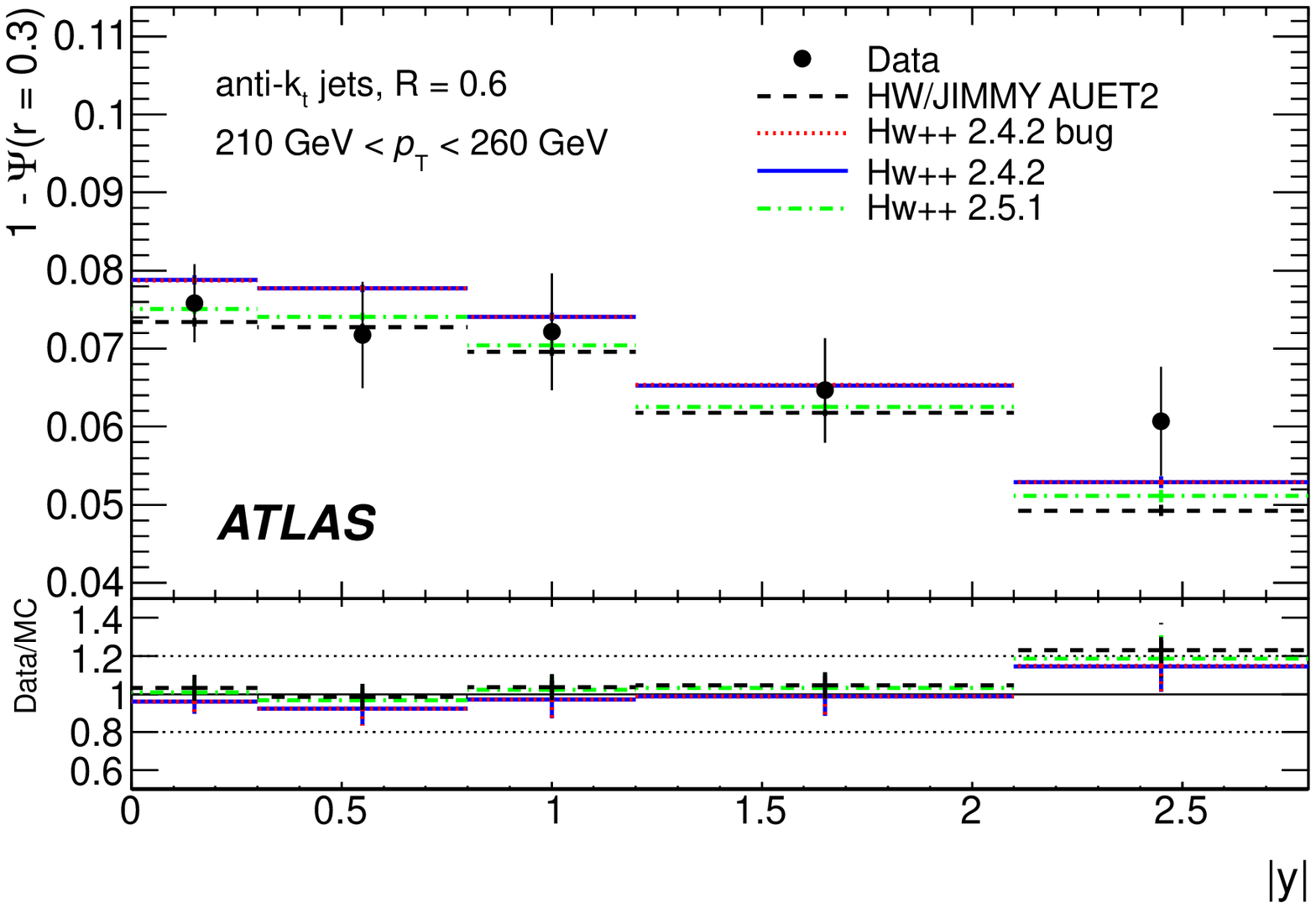} 
\includegraphics[width=0.495\textwidth]{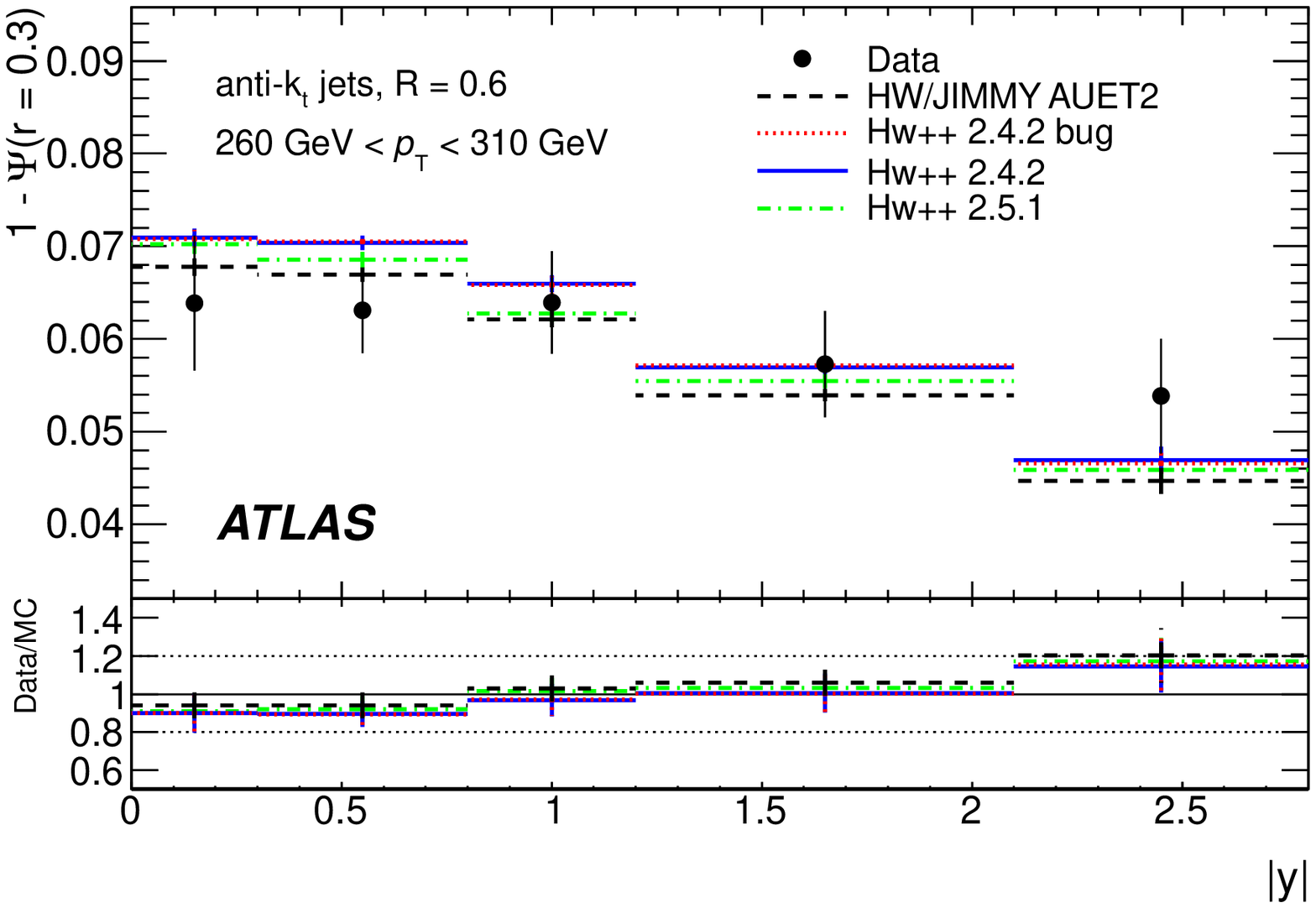}
}
\end{center}
\vspace{-0.7 cm}
\caption{\small
The measured integrated jet shape, $1 - \Psi(r=0.3)$, as a function of $|\rapjet|$ in different jet $\ptjet$ regions 
for jets with $|\rapjet| < 2.8$ and $110 \ {\rm GeV} < \ptjet < 310 \ {\rm GeV}$.
Error bars indicate the statistical and systematic uncertainties added in quadrature. 
The predictions of   Herwig++2.4.2 (solid lines),   Herwig++2.4.2 bug (dotted lines),   Herwig++ 2.5.1 (dashed-dotted lines), and HERWIG/JIMMY-AUET2 (dashed lines) are shown for comparison.
} 
\label{fig:hrw4}
\end{figure}


\begin{figure}[tbh]
\begin{center}
\mbox{
\includegraphics[width=0.495\textwidth]{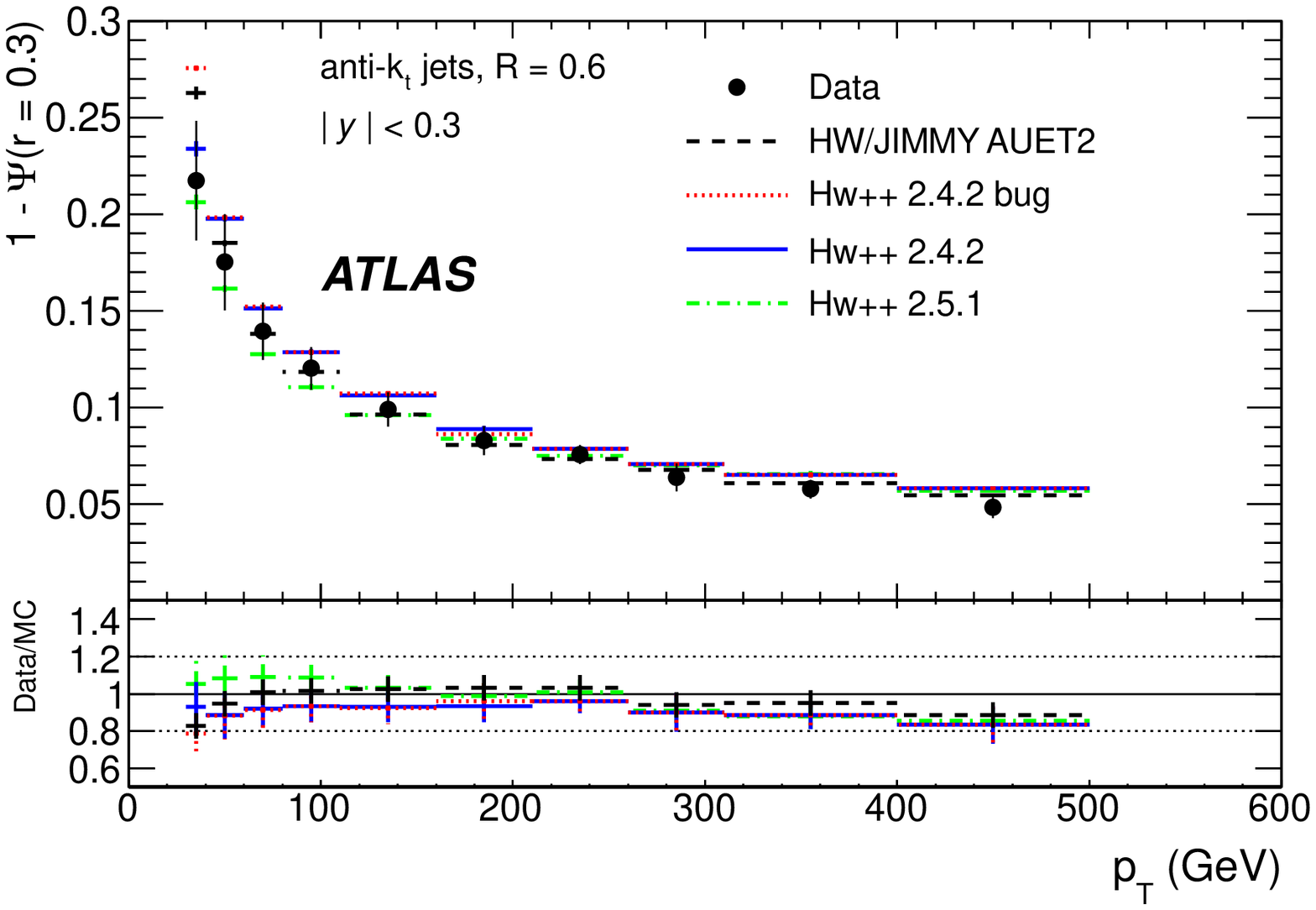}
\includegraphics[width=0.495\textwidth]{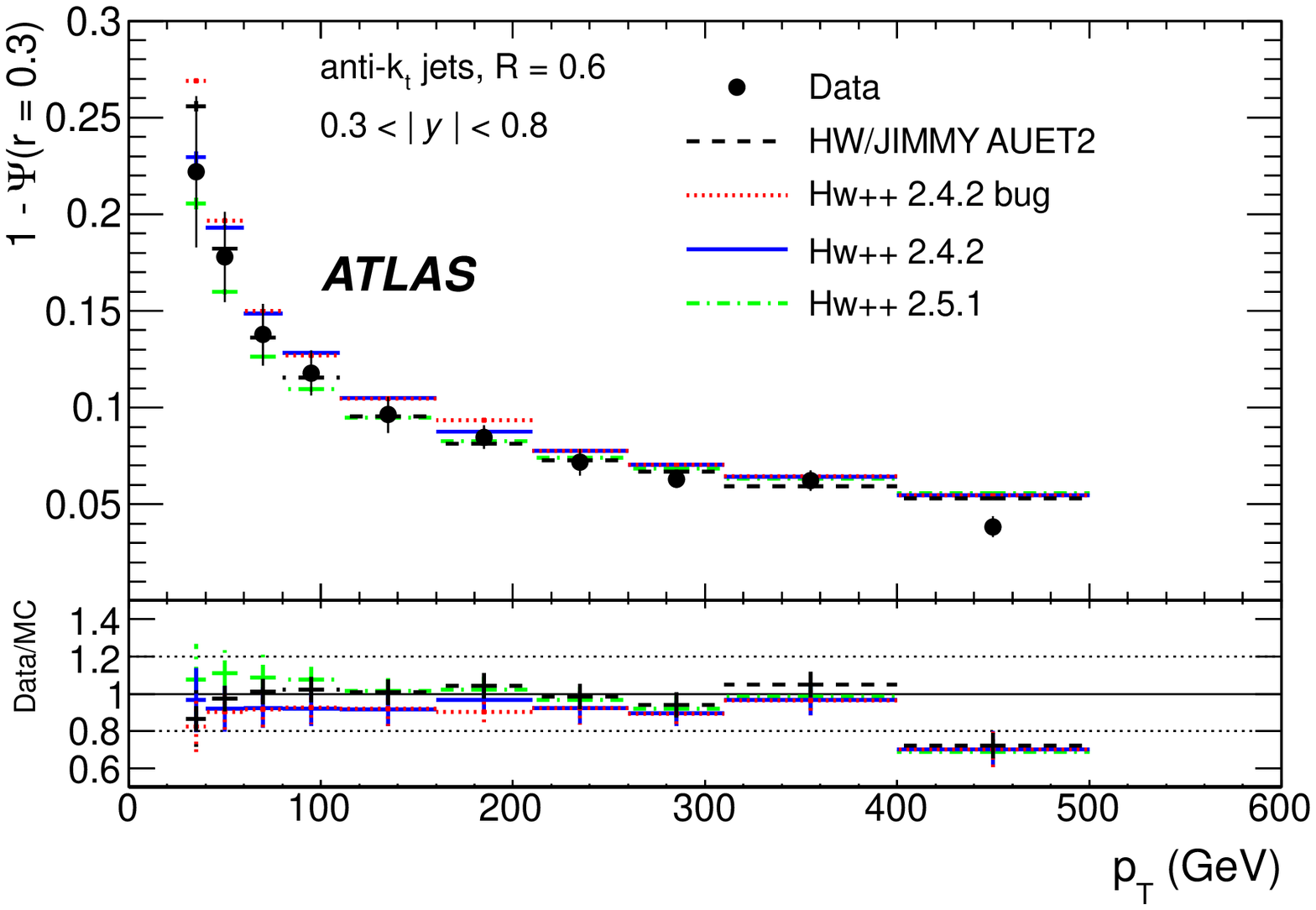}
}
\mbox{
\includegraphics[width=0.495\textwidth]{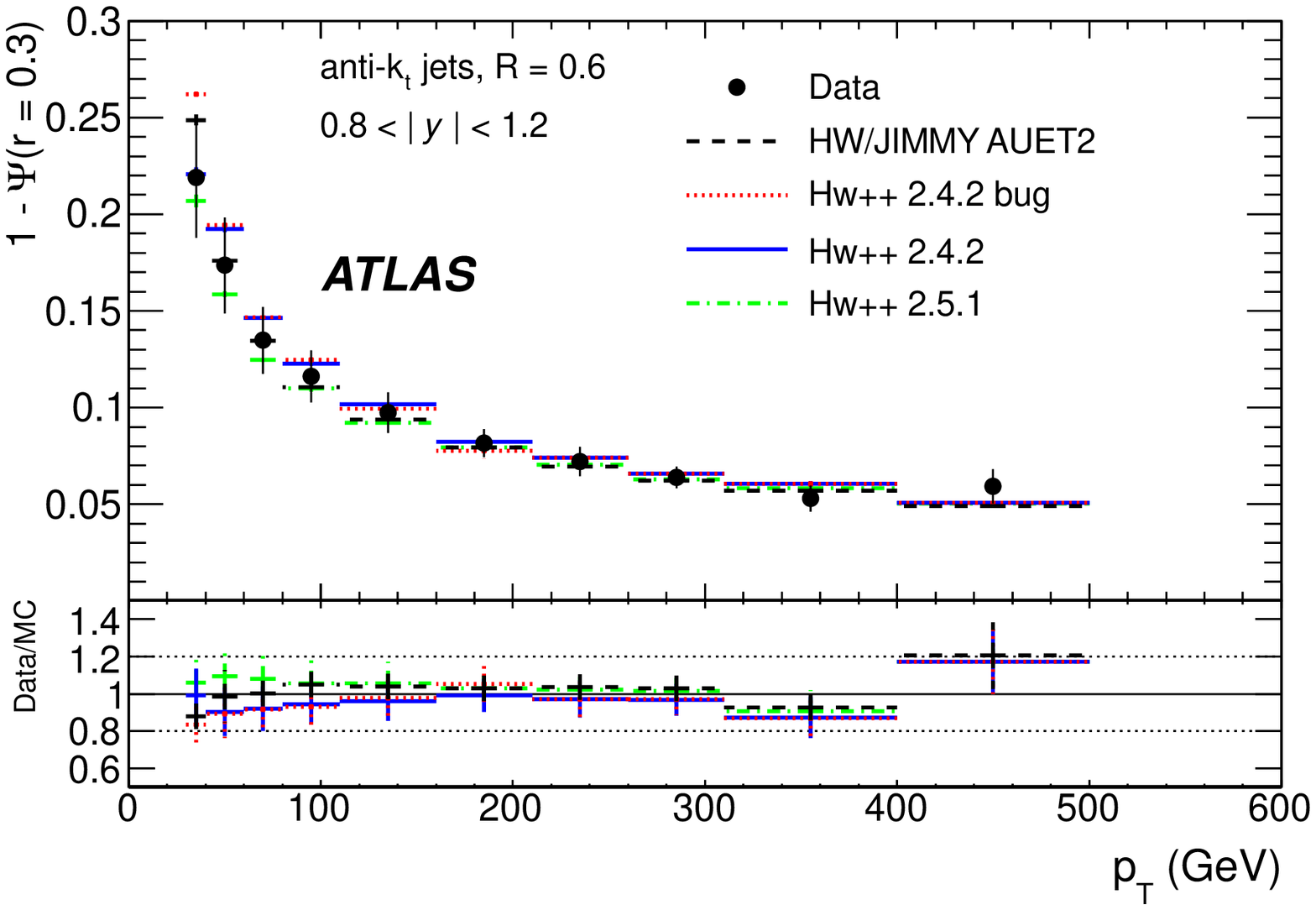} 
\includegraphics[width=0.495\textwidth]{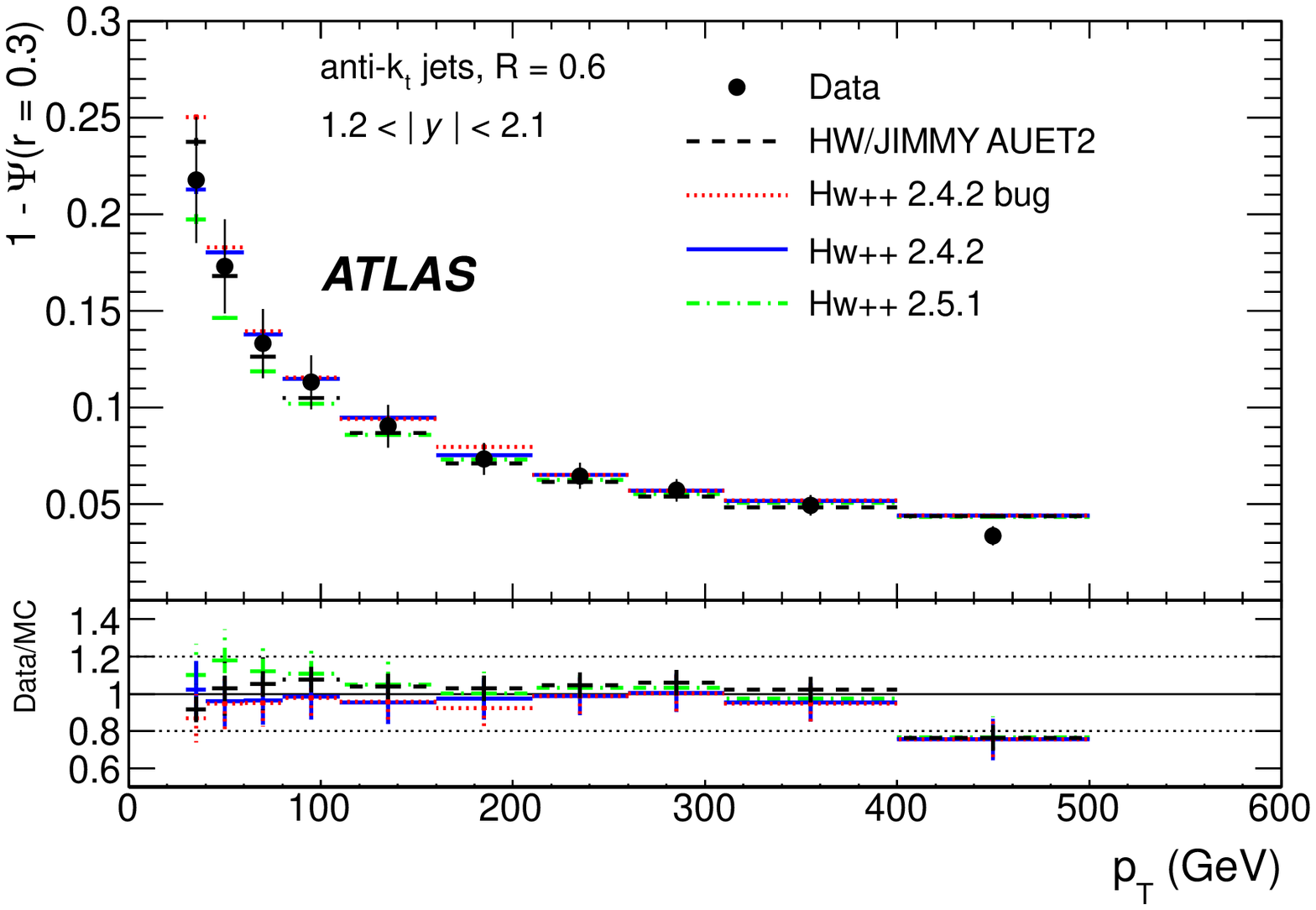}
}
\mbox{
\includegraphics[width=0.495\textwidth]{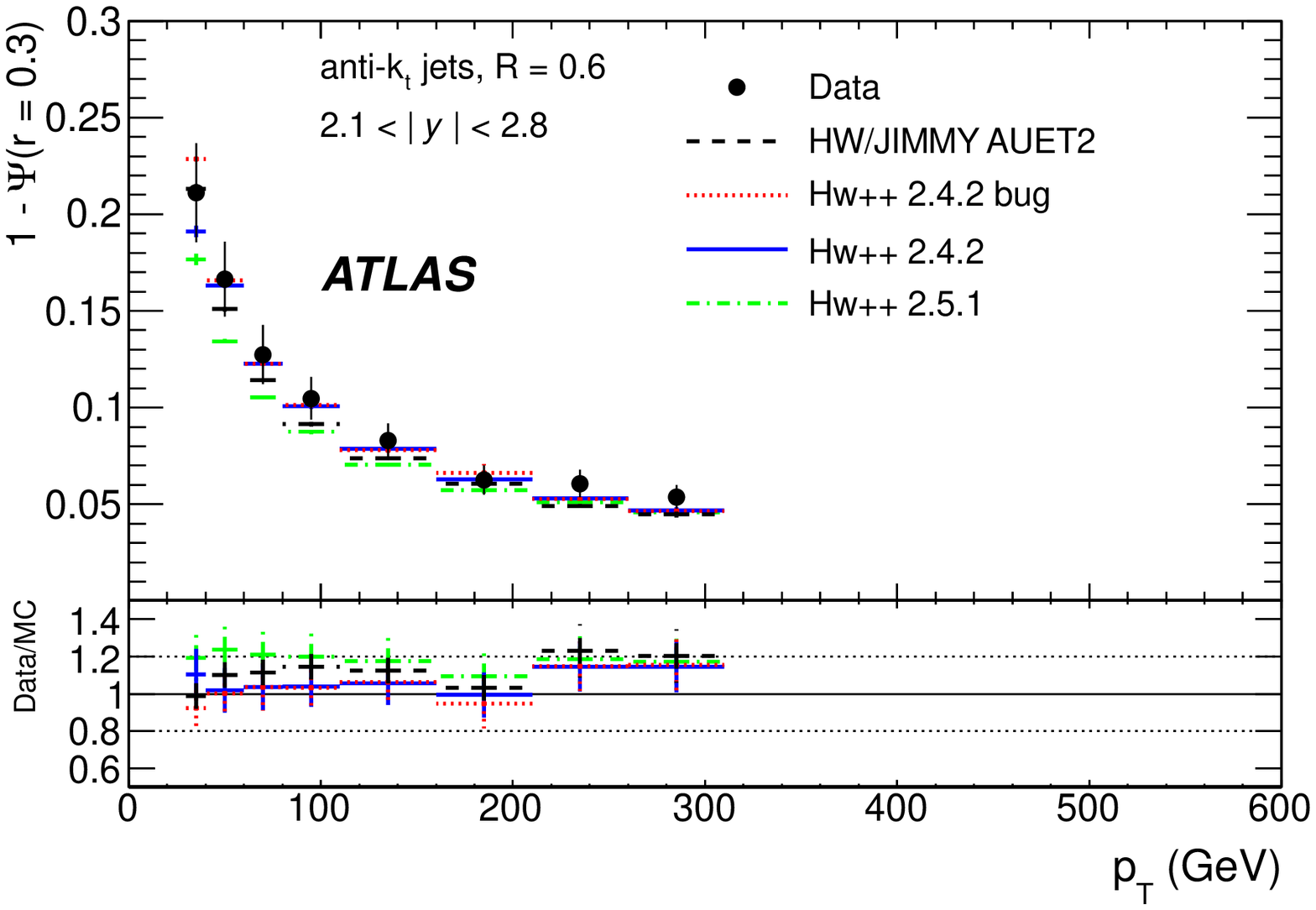}
}
\end{center}
\vspace{-0.7 cm}
\caption{\small
The measured integrated jet shape, $1 - \Psi(r=0.3)$, as a function of $\ptjet$ in different jet rapidity regions 
for jets with $|\rapjet| < 2.8$ and $30 \ {\rm GeV} < \ptjet < 500 \ {\rm GeV}$.
Error bars indicate the statistical and systematic uncertainties added in quadrature. 
The predictions of   Herwig++2.4.2 (solid lines),   Herwig++2.4.2 bug (dotted lines),   Herwig++ 2.5.1 (dashed-dotted lines), and HERWIG/JIMMY-AUET2 (dashed lines) are shown for comparison.
} 
\label{fig:hrw5}
\end{figure}



\begin{figure}[tbh]
\begin{center}
\mbox{
\includegraphics[width=0.495\textwidth,height=0.495\textwidth]{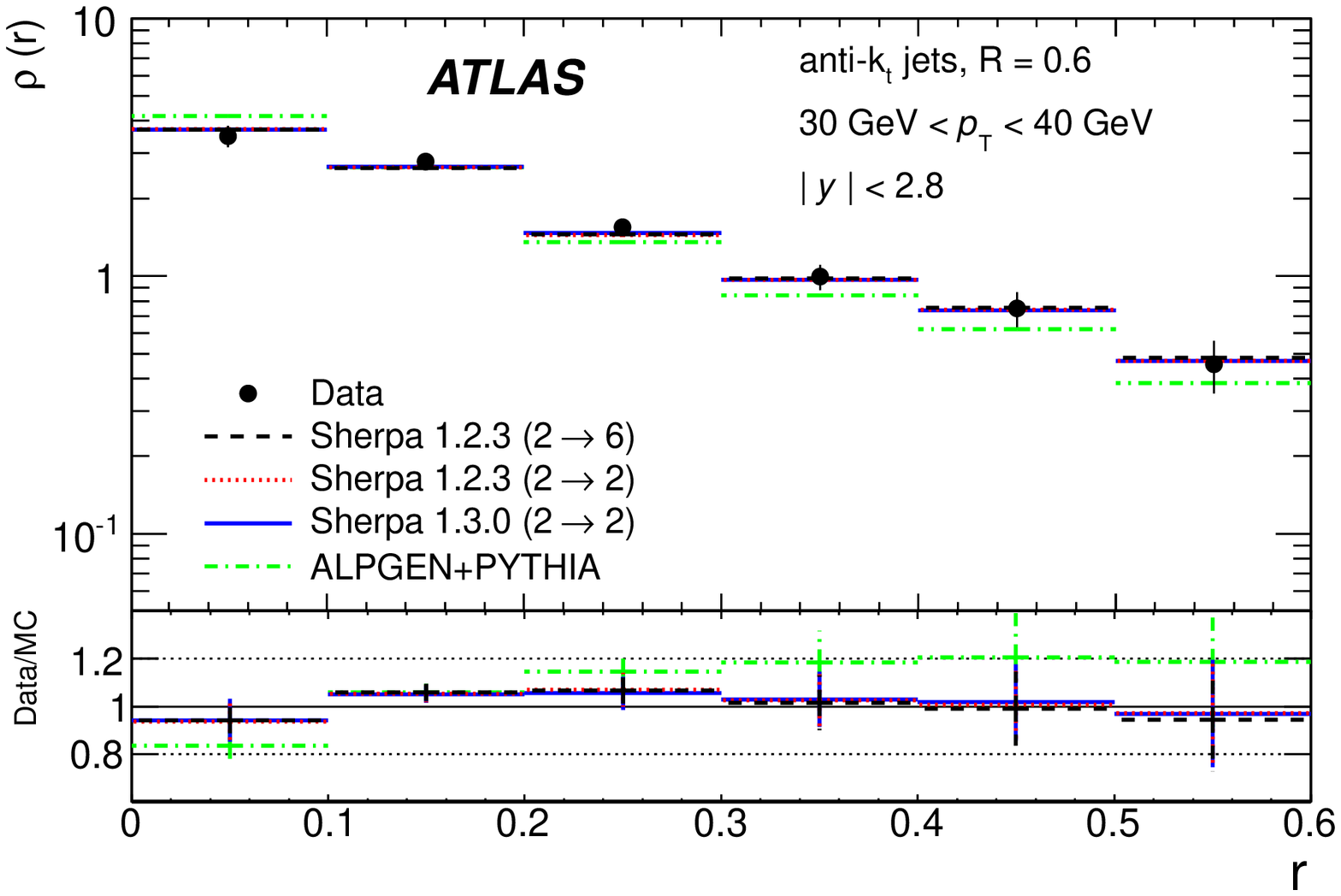} 
\includegraphics[width=0.495\textwidth,height=0.495\textwidth]{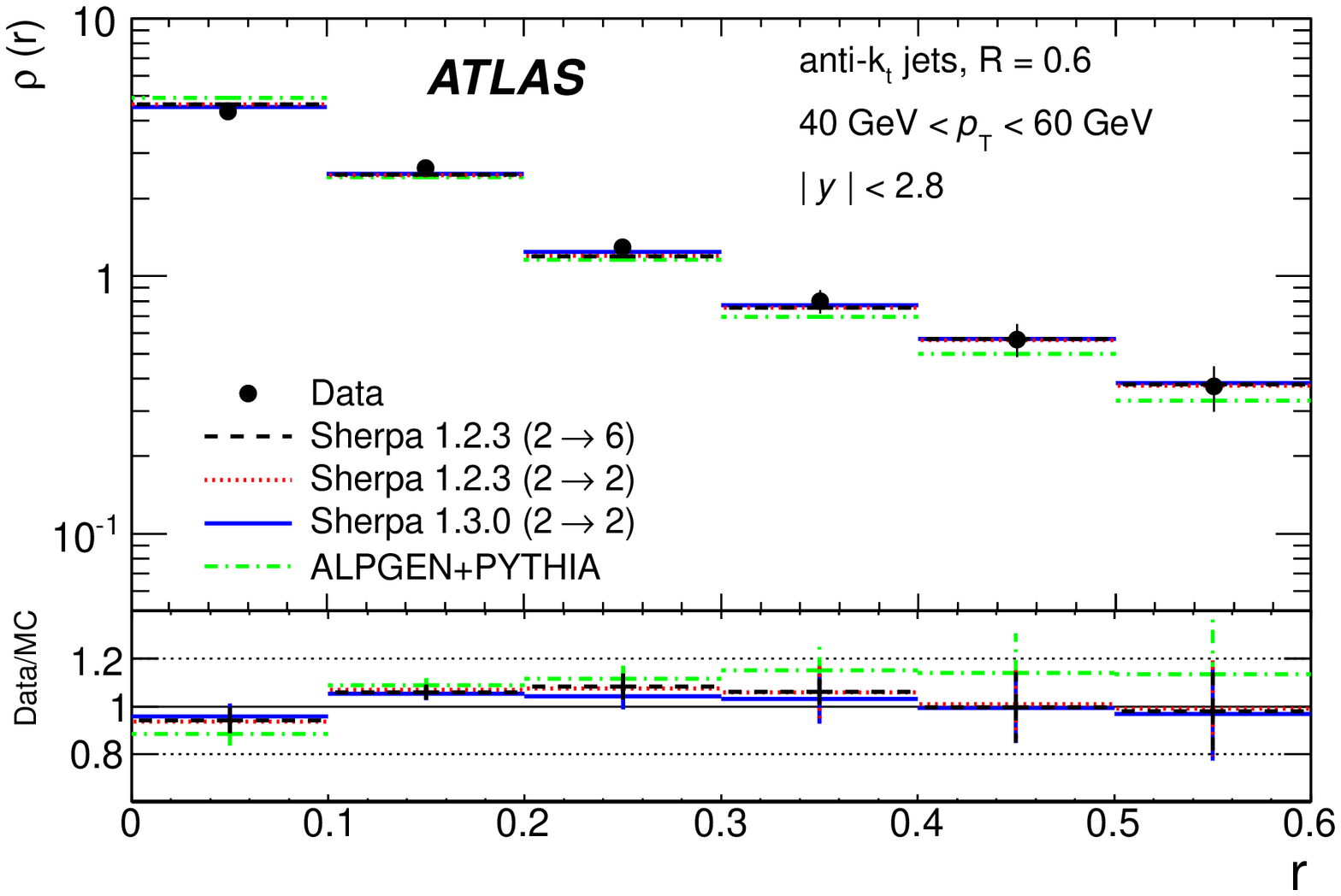}
}\vspace{-0.2cm}
\mbox{
\includegraphics[width=0.495\textwidth,height=0.495\textwidth]{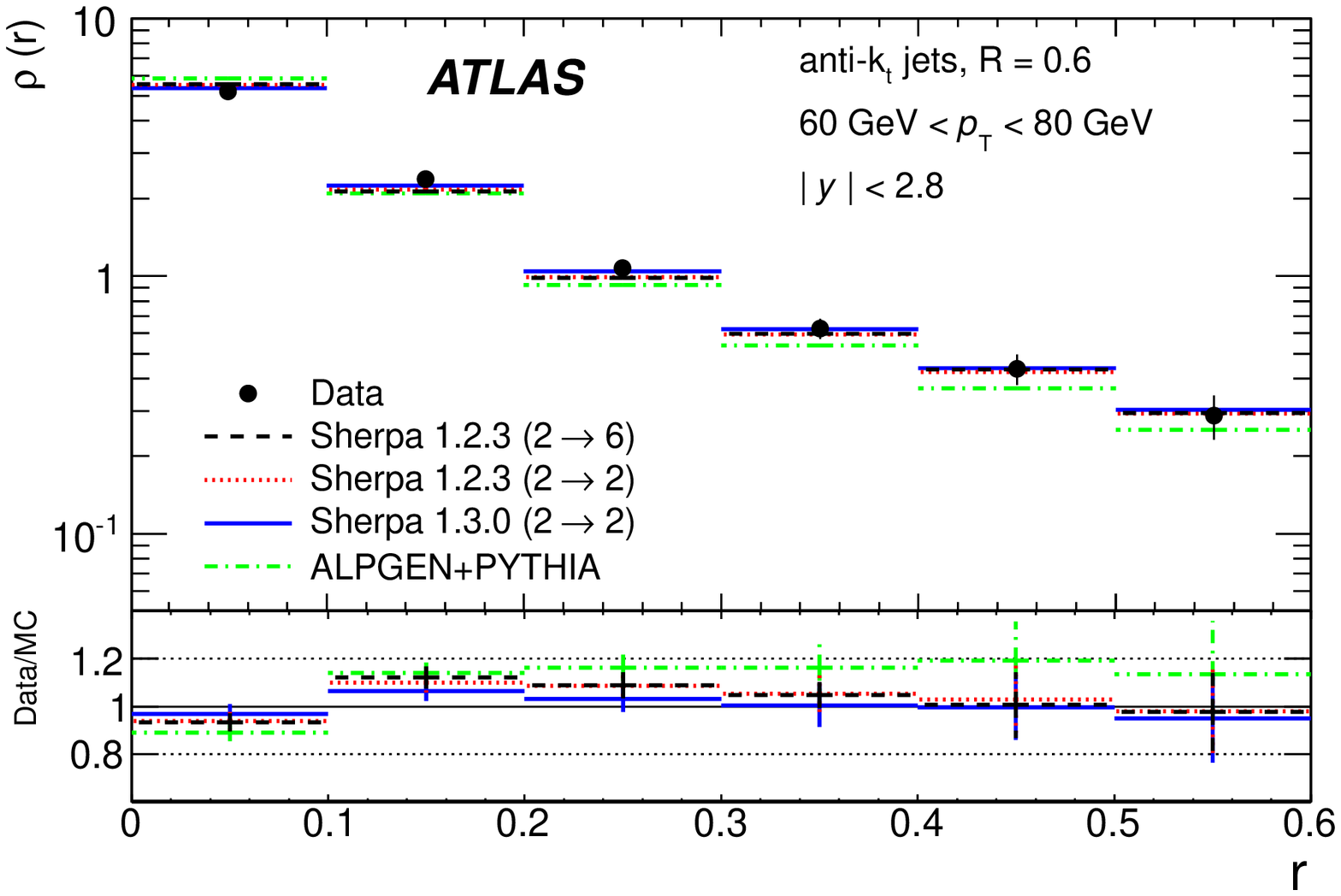}
\includegraphics[width=0.495\textwidth,height=0.495\textwidth]{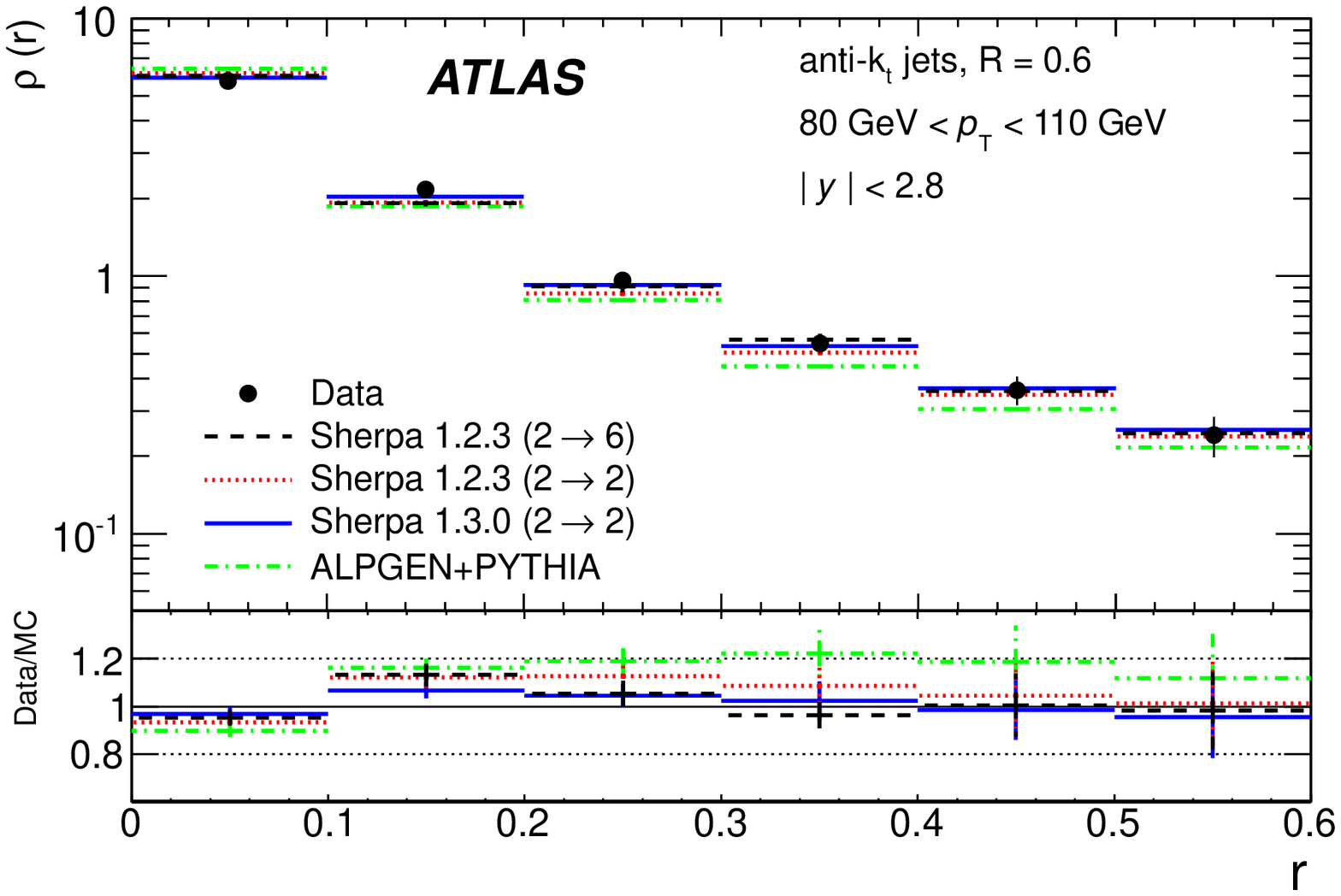}
}
\end{center}
\vspace{-0.7 cm}
\caption{\small
The measured differential jet shape, $\rho(r)$, in inclusive jet production for jets 
with $|\rapjet| < 2.8$ and $30 \ {\rm GeV} < \ptjet < 110  \ {\rm GeV}$   
is shown in different $\ptjet$ regions. Error bars indicate the statistical and systematic uncertainties added in quadrature.
The predictions of   Sherpa 1.3.0 $(2 \to 2)$(solid lines),   Sherpa 1.2.3 $(2 \to 2)$ (dotted lines),  Sherpa (up to $2 \to 6$) (dashed lines), and ALPGEN interfaced to PYTHIA (dashed-dotted lines) are shown for comparison.} 
\label{fig:meps1}
\end{figure}

\clearpage

\begin{figure}[tbh]
\begin{center}
\mbox{
\includegraphics[width=0.495\textwidth,height=0.495\textwidth]{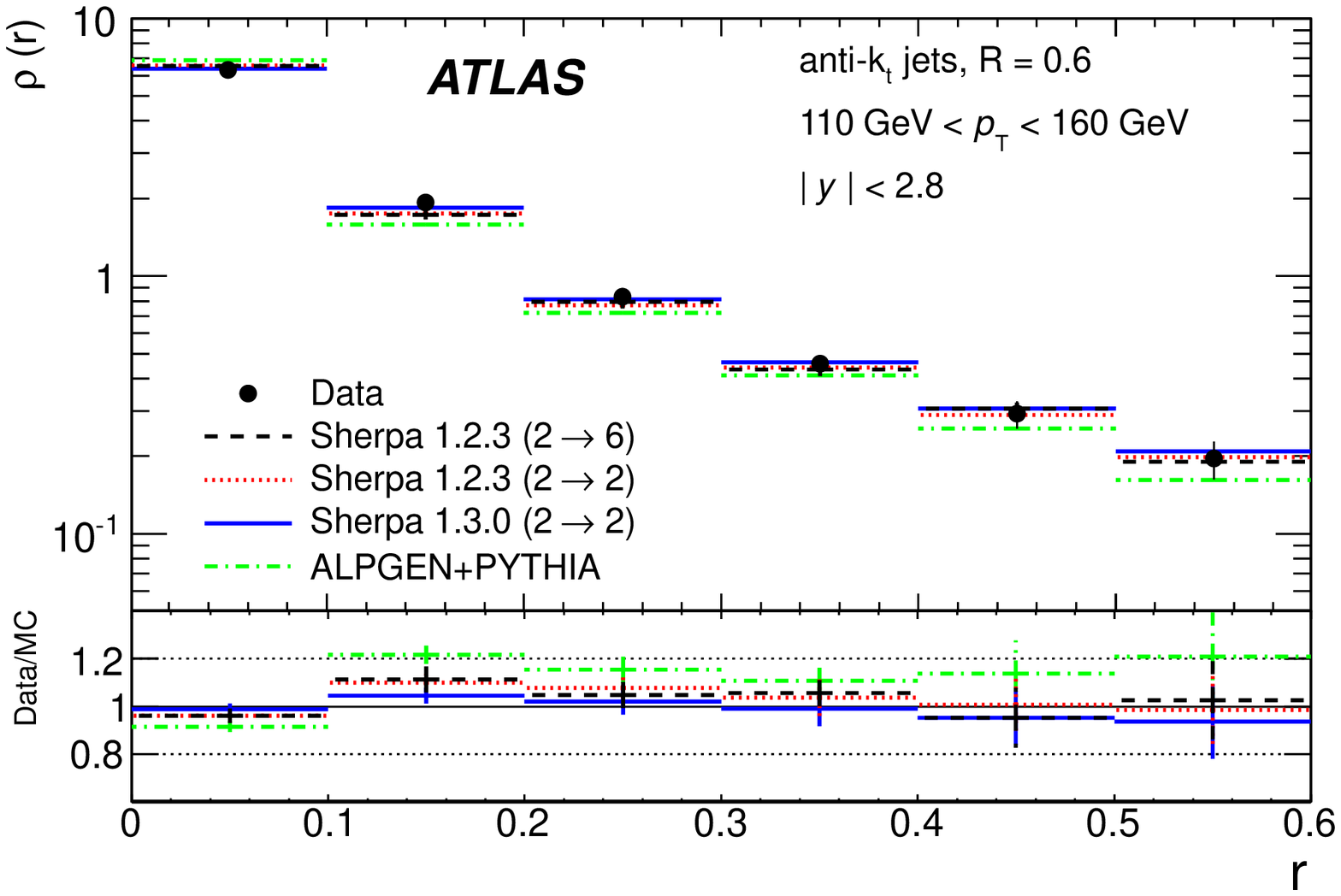} 
\includegraphics[width=0.495\textwidth,height=0.495\textwidth]{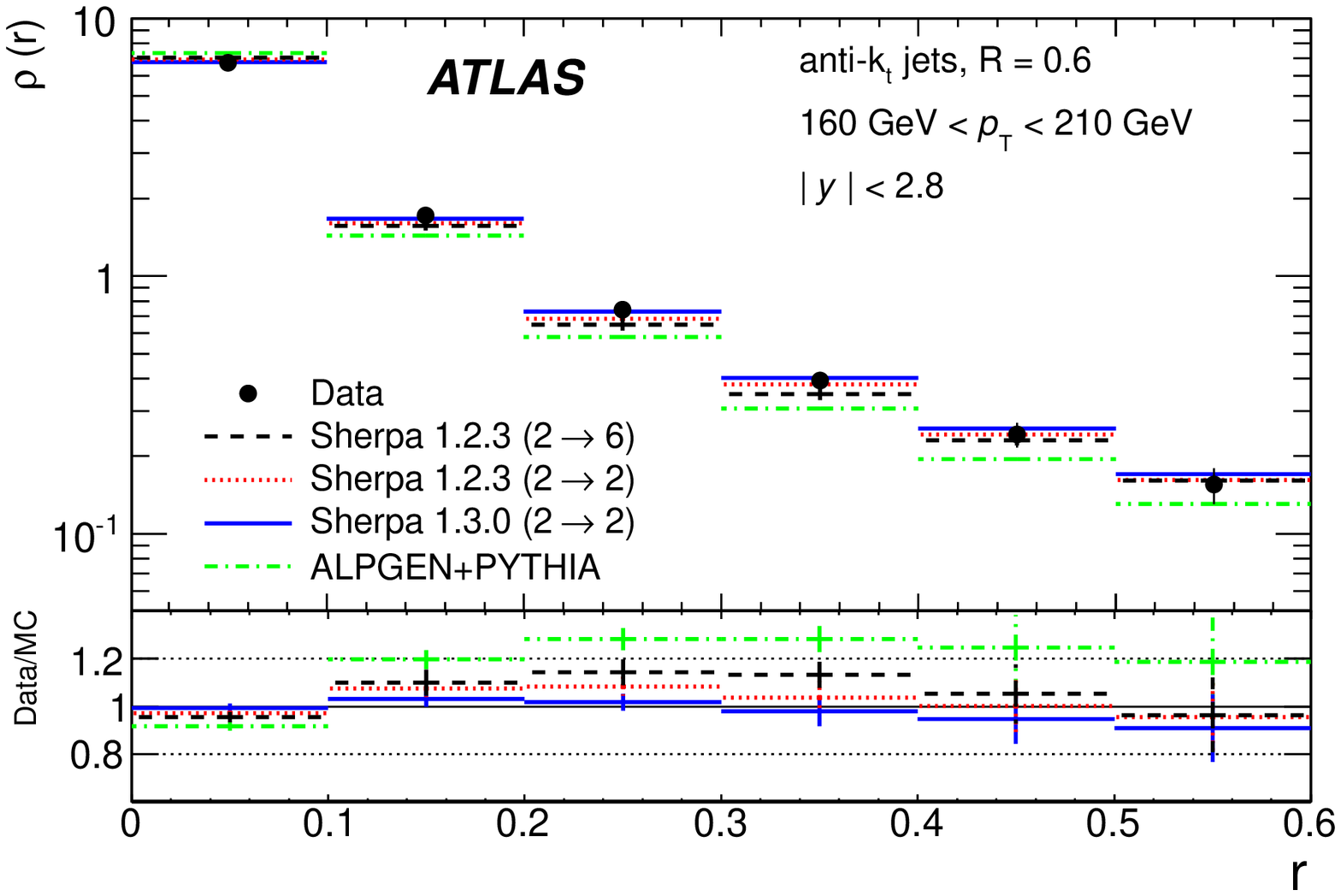}
}\vspace{-0.2cm}
\mbox{
\includegraphics[width=0.495\textwidth,height=0.495\textwidth]{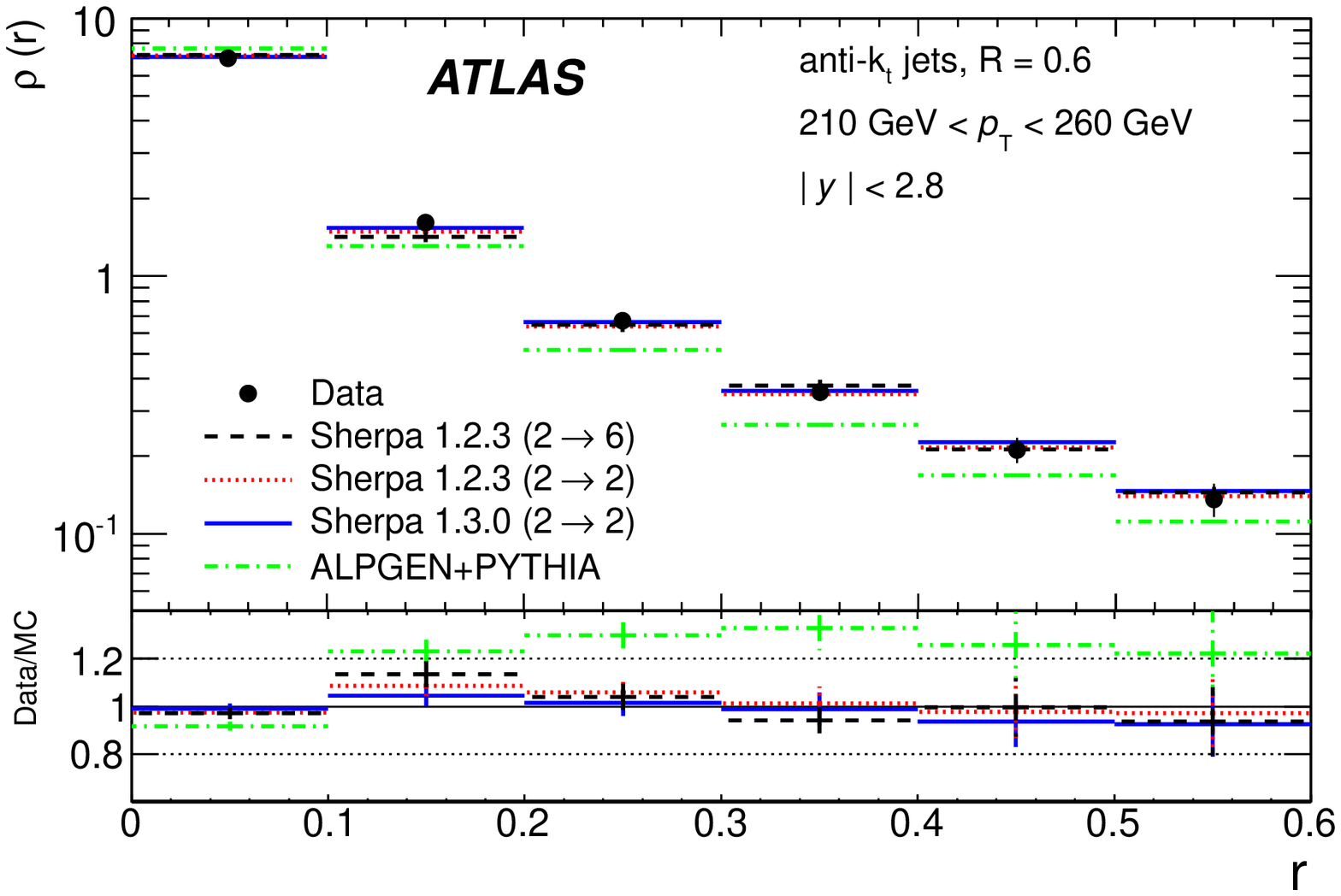}
\includegraphics[width=0.495\textwidth,height=0.495\textwidth]{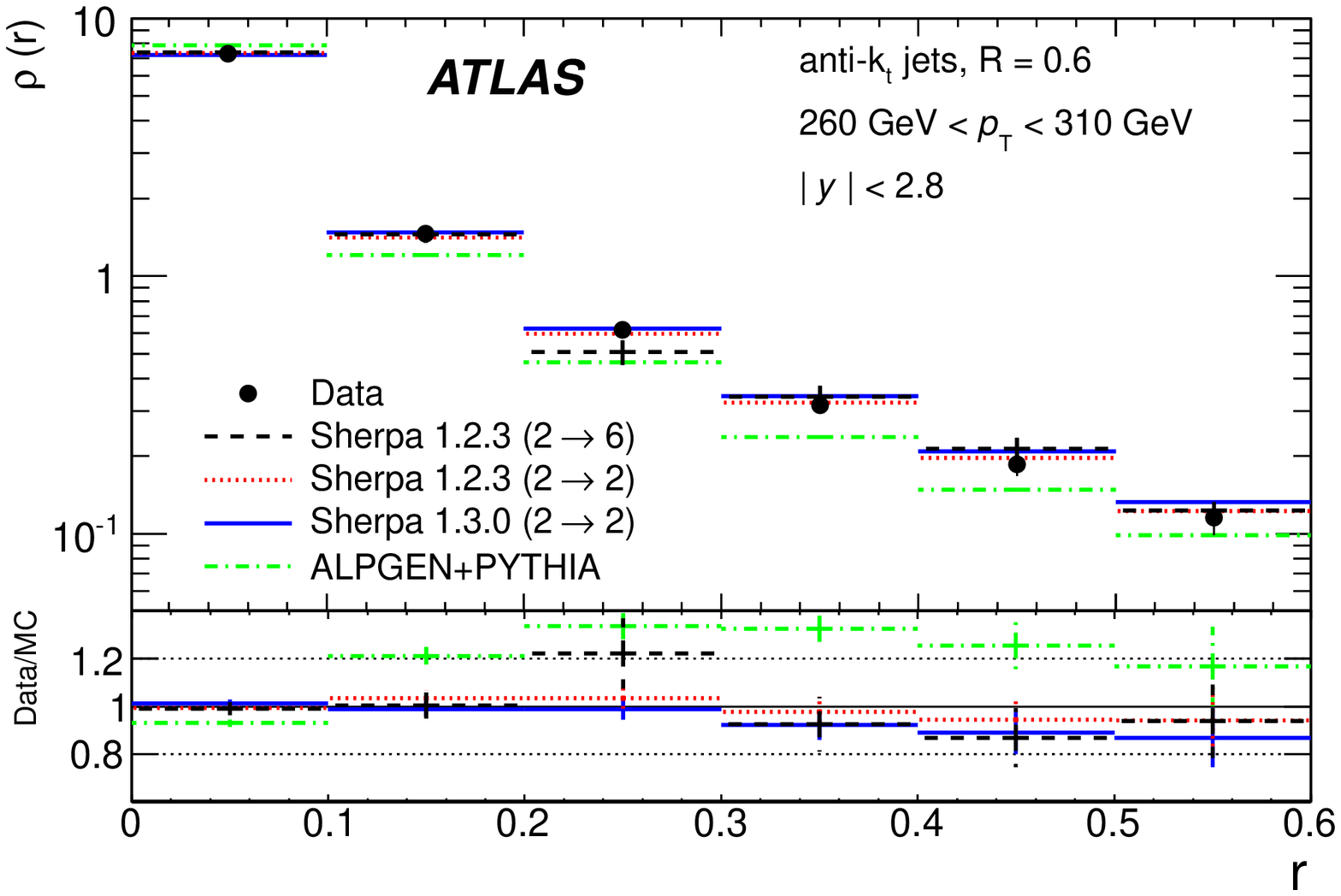}
}
\end{center}
\vspace{-0.7 cm}
\caption{\small
The measured differential jet shape, $\rho(r)$, in inclusive jet production for jets 
with $|\rapjet| < 2.8$ and $110 \ {\rm GeV} < \ptjet < 310  \ {\rm GeV}$   
is shown in different $\ptjet$ regions. Error bars indicate the statistical and systematic uncertainties added in quadrature.
The predictions of   Sherpa 1.3.0 $(2 \to 2)$(solid lines),   Sherpa 1.2.3 $(2 \to 2)$ (dotted lines),  Sherpa (up to $2 \to 6$) (dashed lines), and ALPGEN interfaced to PYTHIA (dashed-dotted lines) are shown for comparison.} 
\label{fig:meps2}
\end{figure}


\begin{figure}[tbh]
\begin{center}
\mbox{
\includegraphics[width=0.495\textwidth]{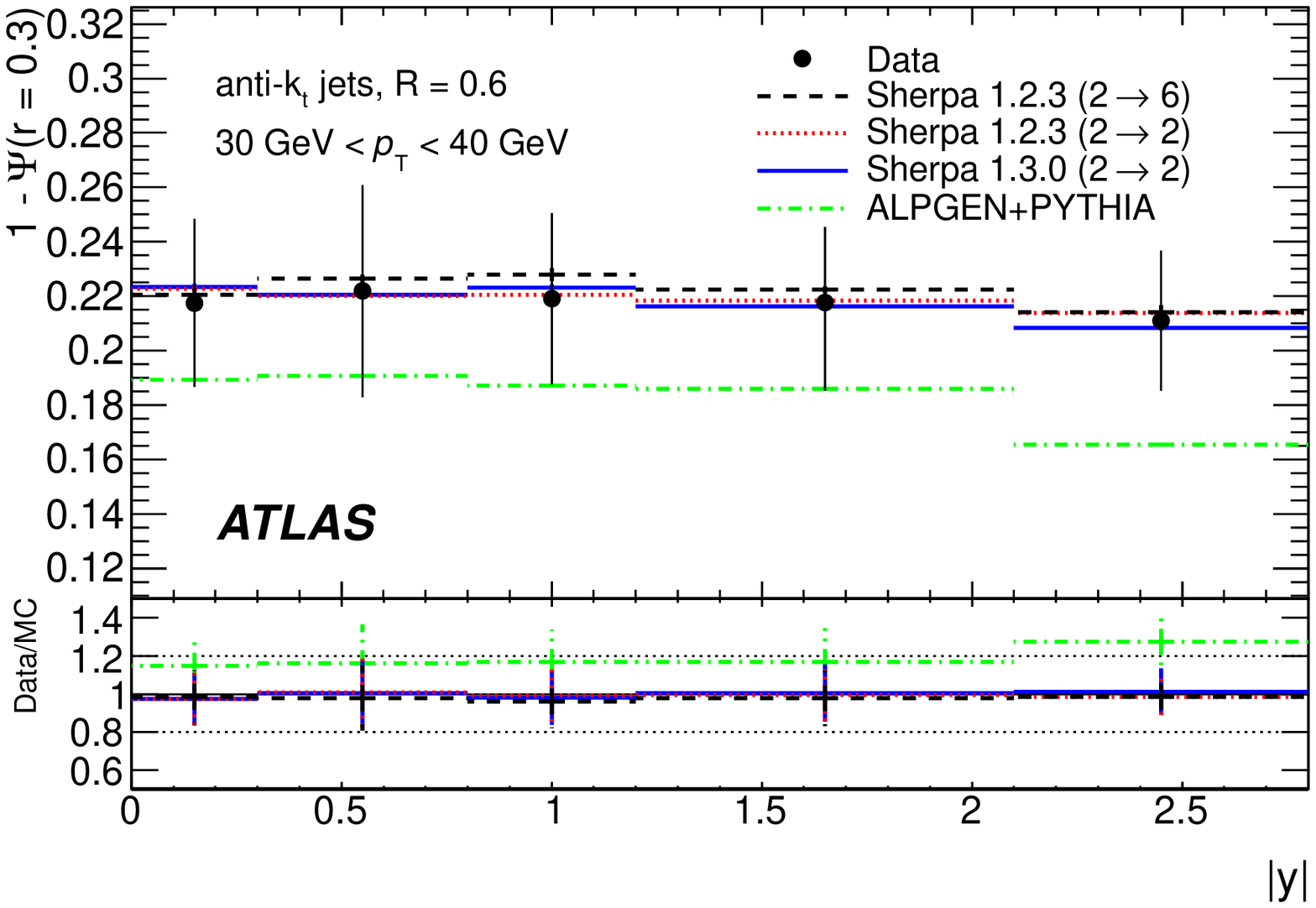}
\includegraphics[width=0.495\textwidth]{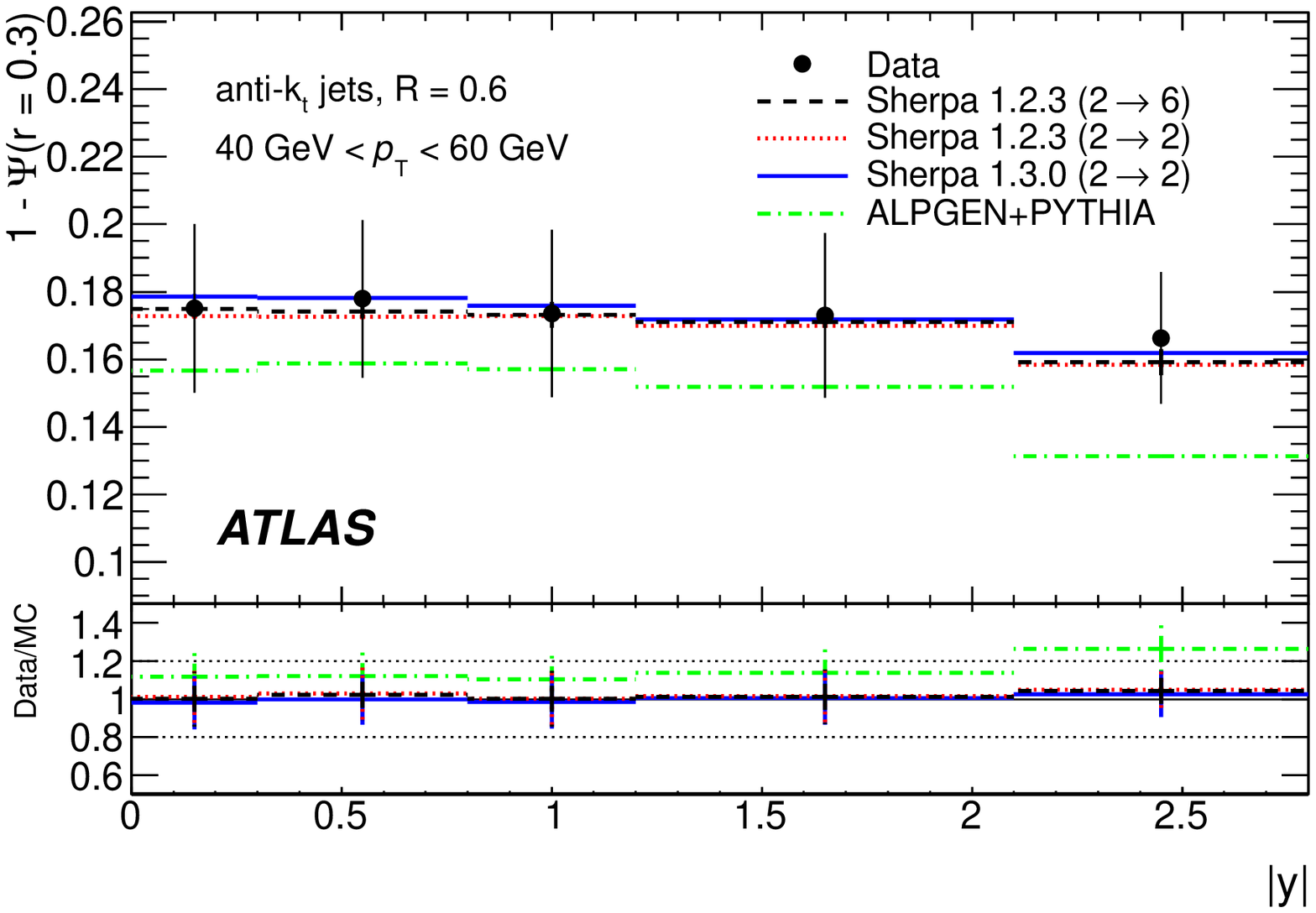}
}
\mbox{
\includegraphics[width=0.495\textwidth]{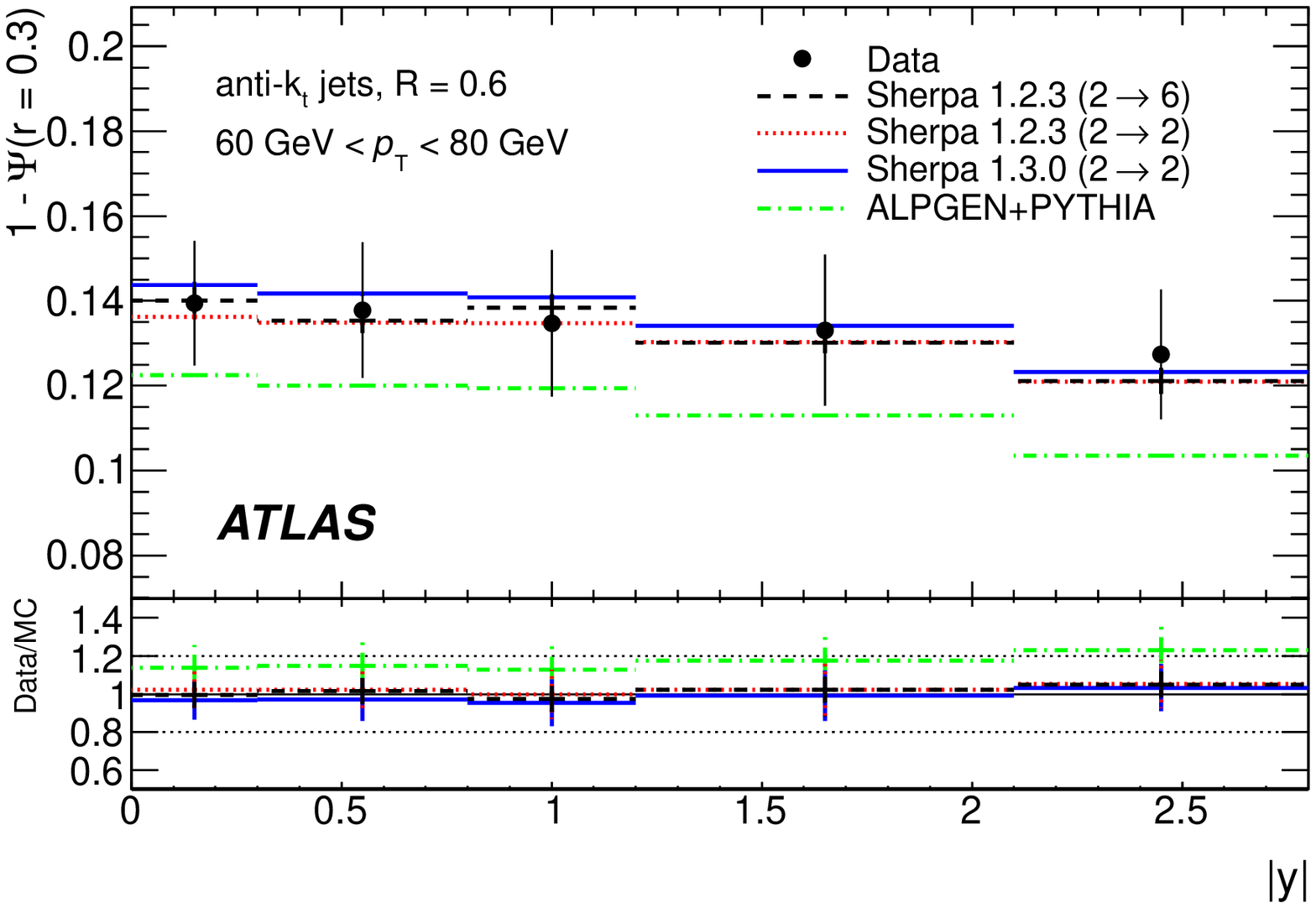} 
\includegraphics[width=0.495\textwidth]{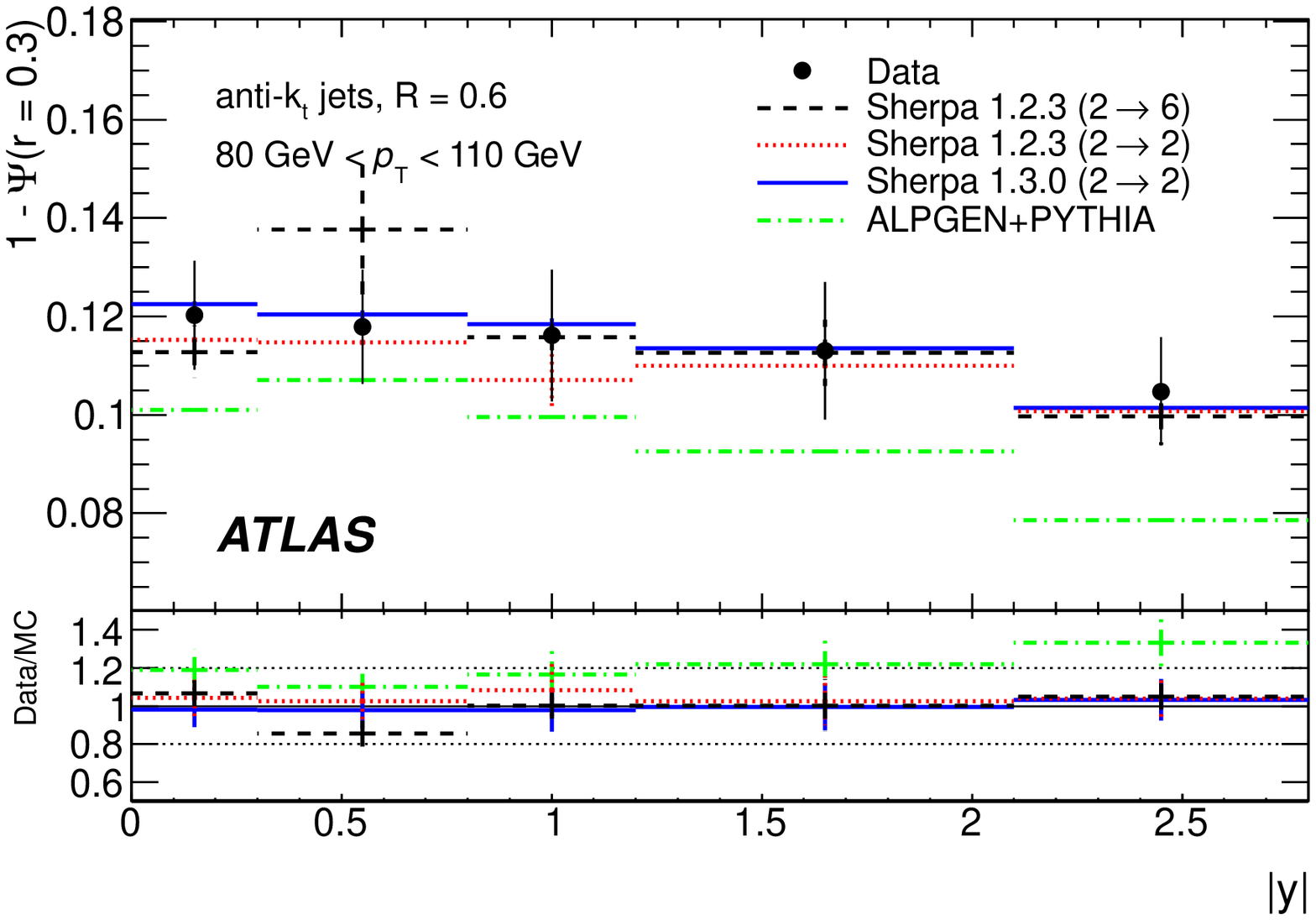}
}
\end{center}
\vspace{-0.7 cm}
\caption{\small
The measured integrated jet shape, $1 - \Psi(r=0.3)$, as a function of $|\rapjet|$ in different jet $\ptjet$ regions 
for jets with $|\rapjet| < 2.8$ and $30 \ {\rm GeV} < \ptjet < 110 \ {\rm GeV}$.
Error bars indicate the statistical and systematic uncertainties added in quadrature. 
The predictions of   Sherpa 1.3.0 $(2 \to 2)$(solid lines),   Sherpa 1.2.3 $(2 \to 2)$ (dotted lines),  Sherpa (up to $2 \to 6$) (dashed lines), and ALPGEN interfaced to PYTHIA (dashed-dotted lines) are shown for comparison.}
\label{fig:meps3}
\end{figure}

\begin{figure}[tbh]
\begin{center}
\mbox{
\includegraphics[width=0.495\textwidth]{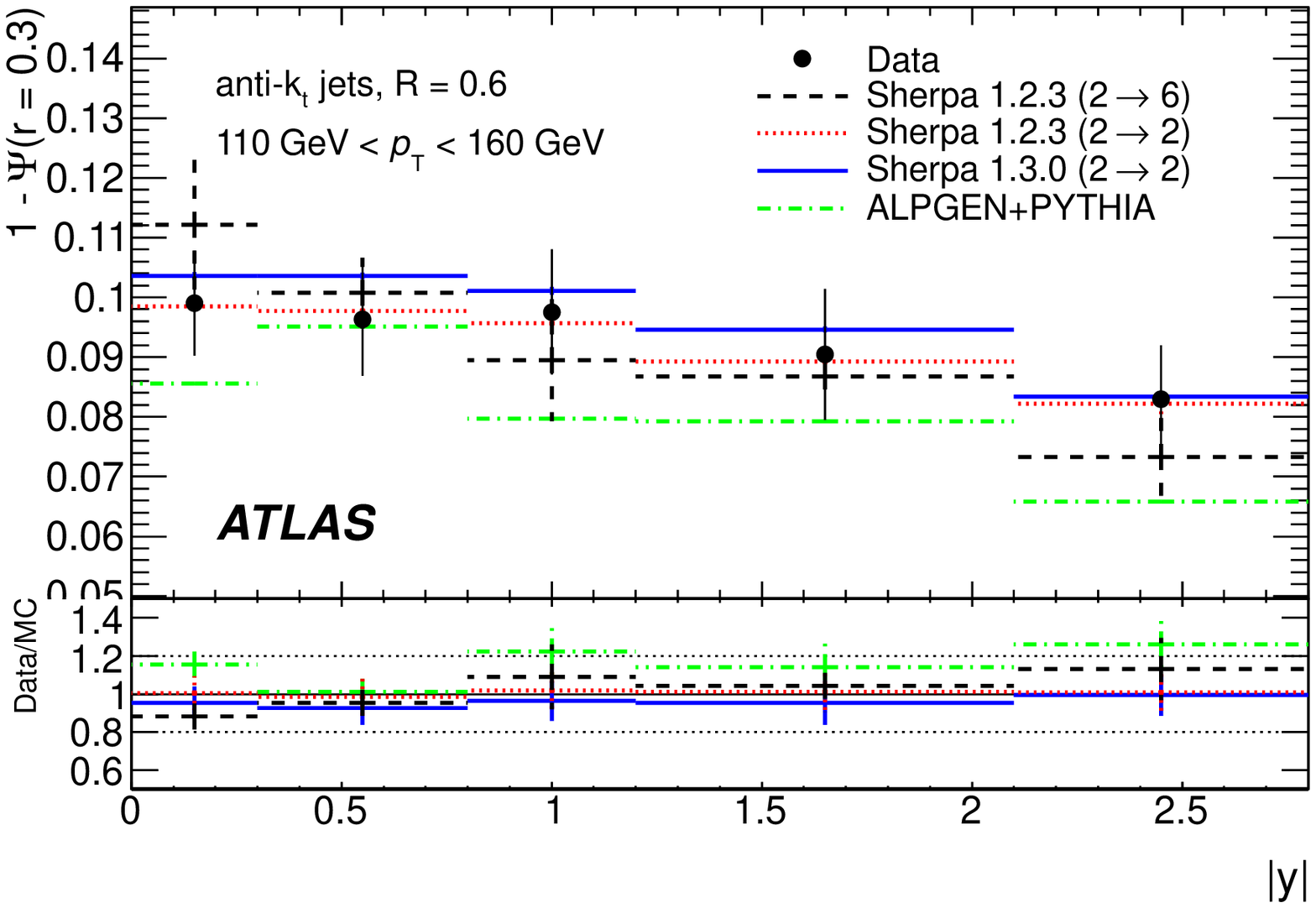}
\includegraphics[width=0.495\textwidth]{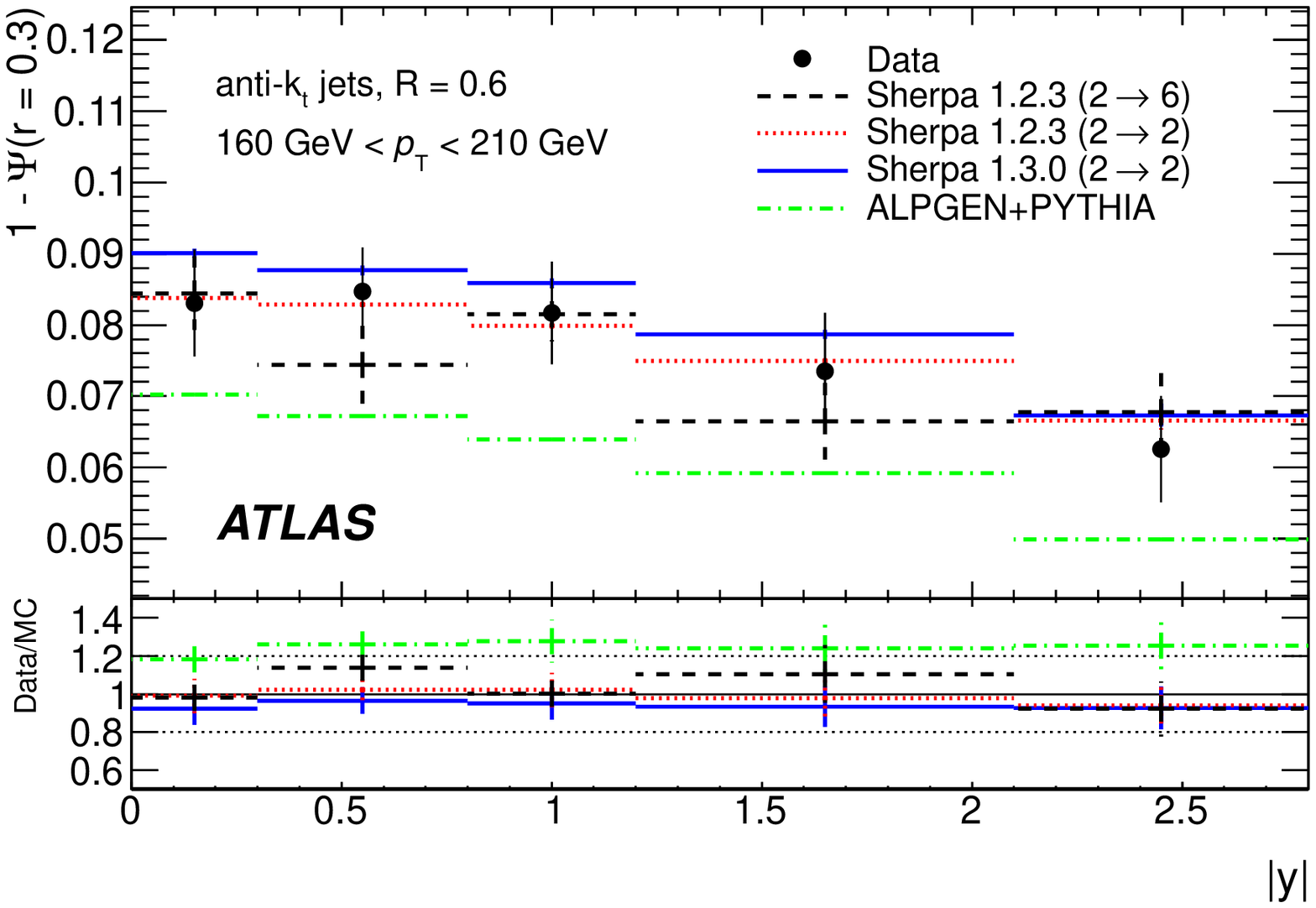}
}
\mbox{
\includegraphics[width=0.495\textwidth]{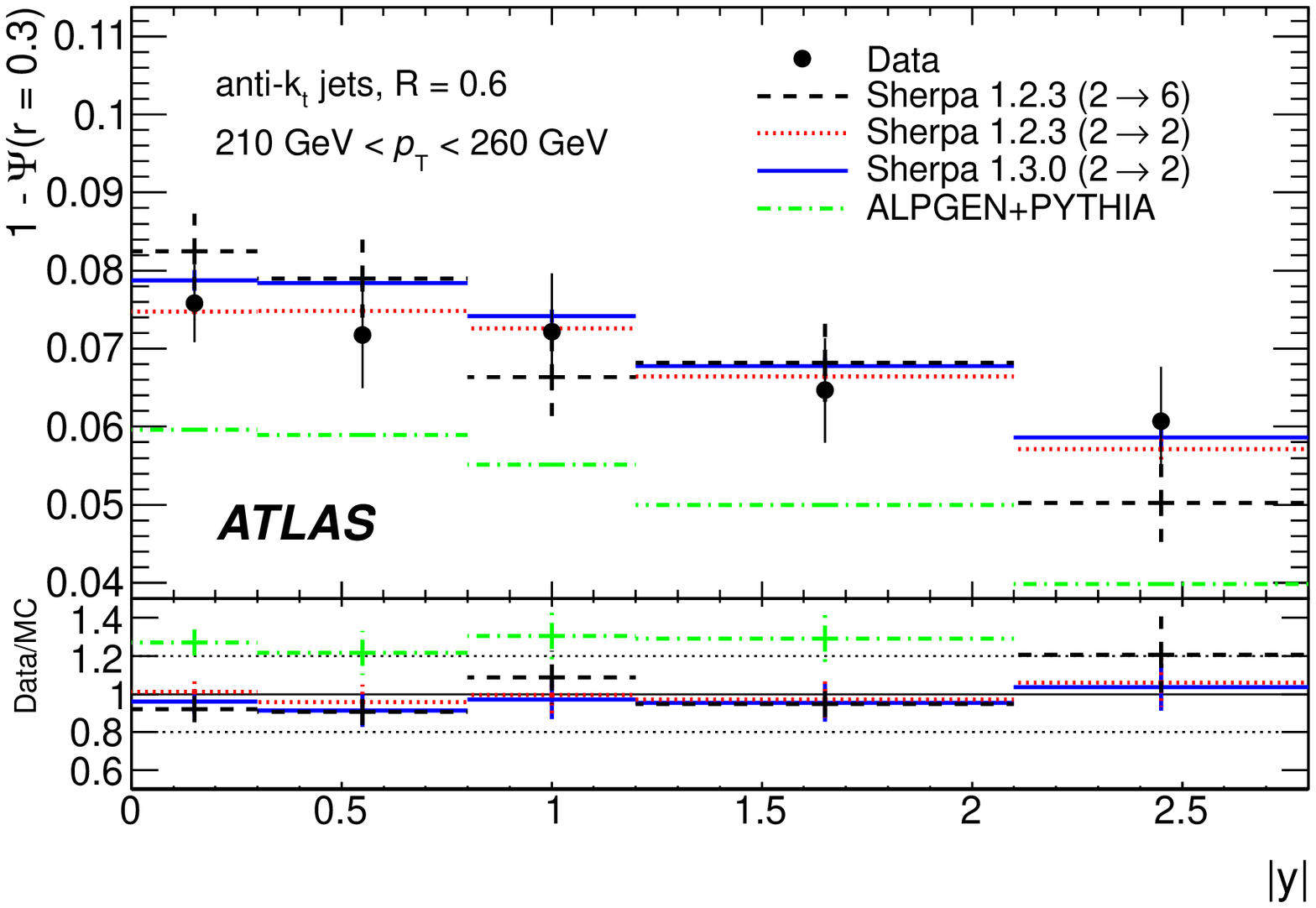} 
\includegraphics[width=0.495\textwidth]{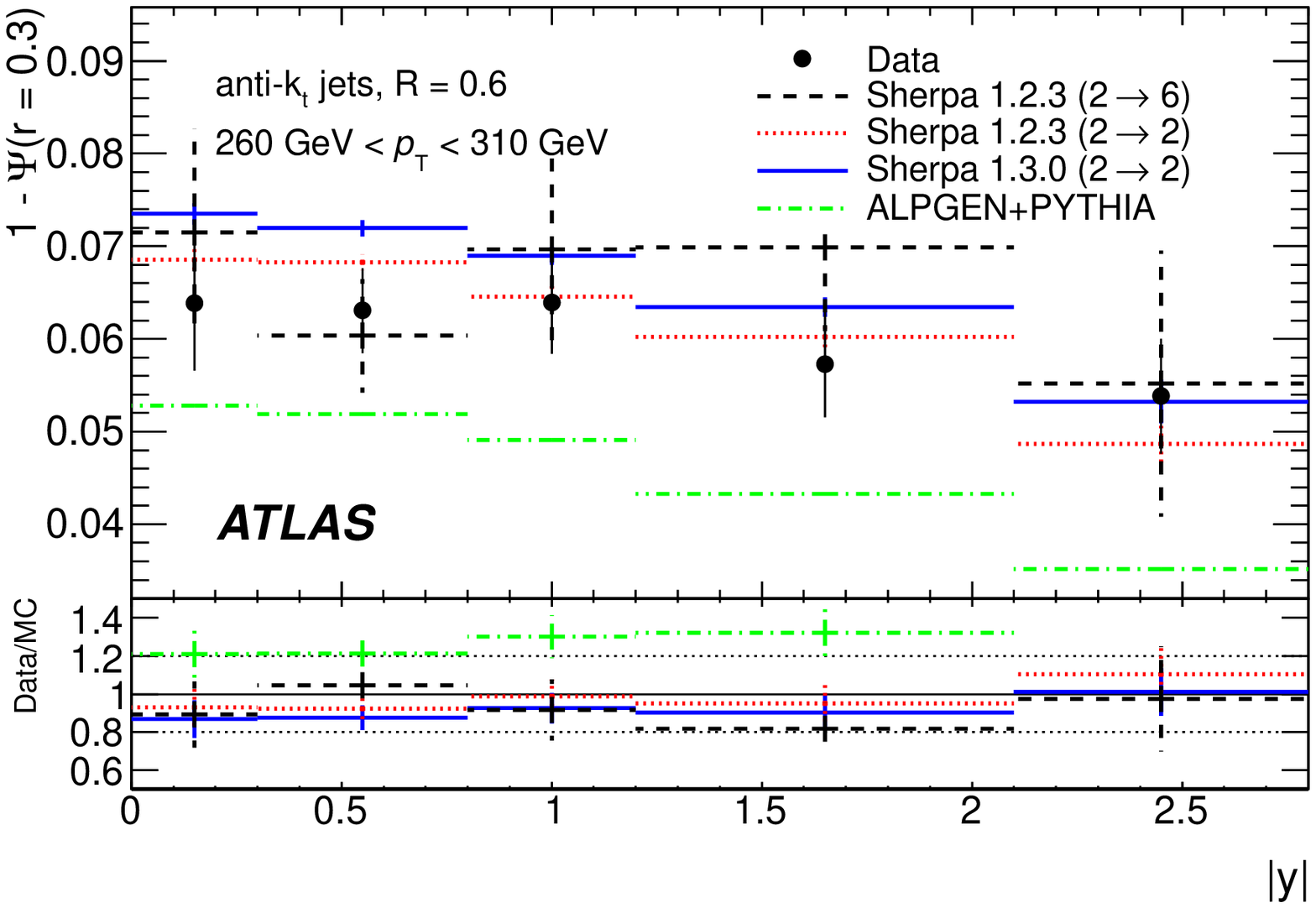}
}
\end{center}
\vspace{-0.7 cm}
\caption{\small
The measured integrated jet shape, $1 - \Psi(r=0.3)$, as a function of $|\rapjet|$ in different jet $\ptjet$ regions 
for jets with $|\rapjet| < 2.8$ and $110 \ {\rm GeV} < \ptjet < 310 \ {\rm GeV}$.
Error bars indicate the statistical and systematic uncertainties added in quadrature. 
The predictions of   Sherpa 1.3.0 $(2 \to 2)$(solid lines),   Sherpa 1.2.3 $(2 \to 2)$ (dotted lines),  Sherpa (up to $2 \to 6$) (dashed lines), and ALPGEN interfaced to PYTHIA (dashed-dotted lines) are shown for comparison.} 
\label{fig:meps4}
\end{figure}


\begin{figure}[tbh]
\begin{center}
\mbox{
\includegraphics[width=0.495\textwidth]{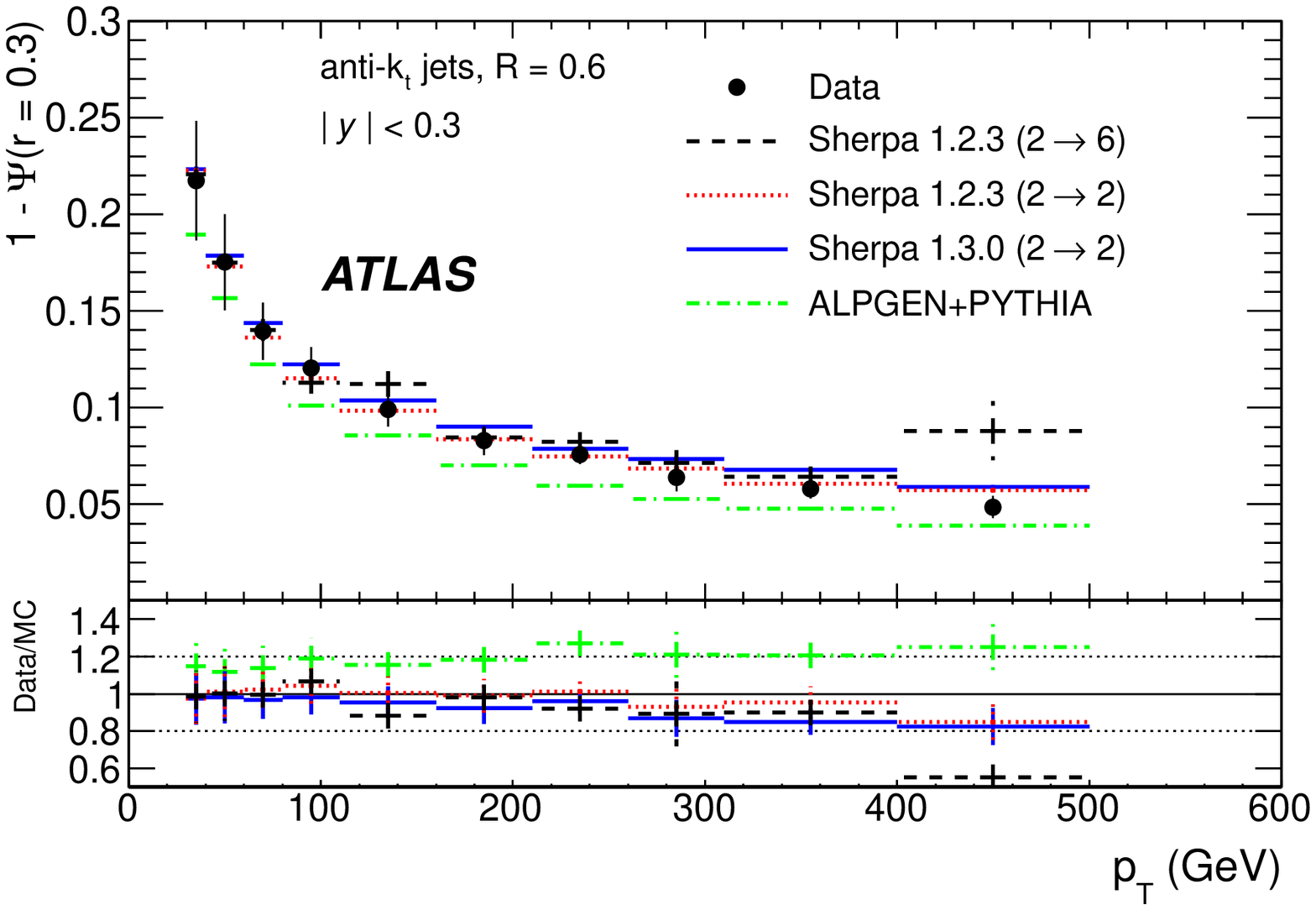}
\includegraphics[width=0.495\textwidth]{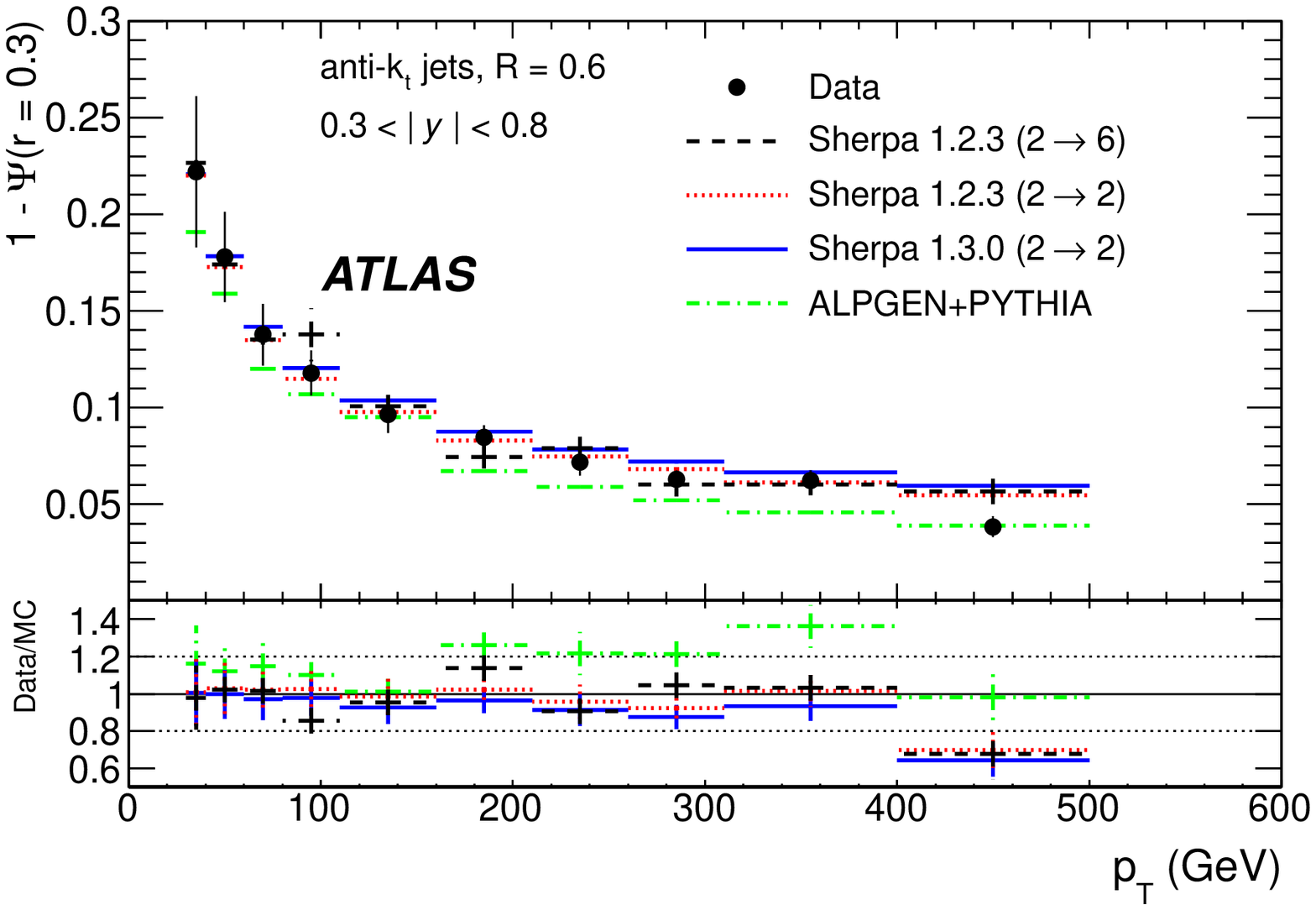}
}
\mbox{
\includegraphics[width=0.495\textwidth]{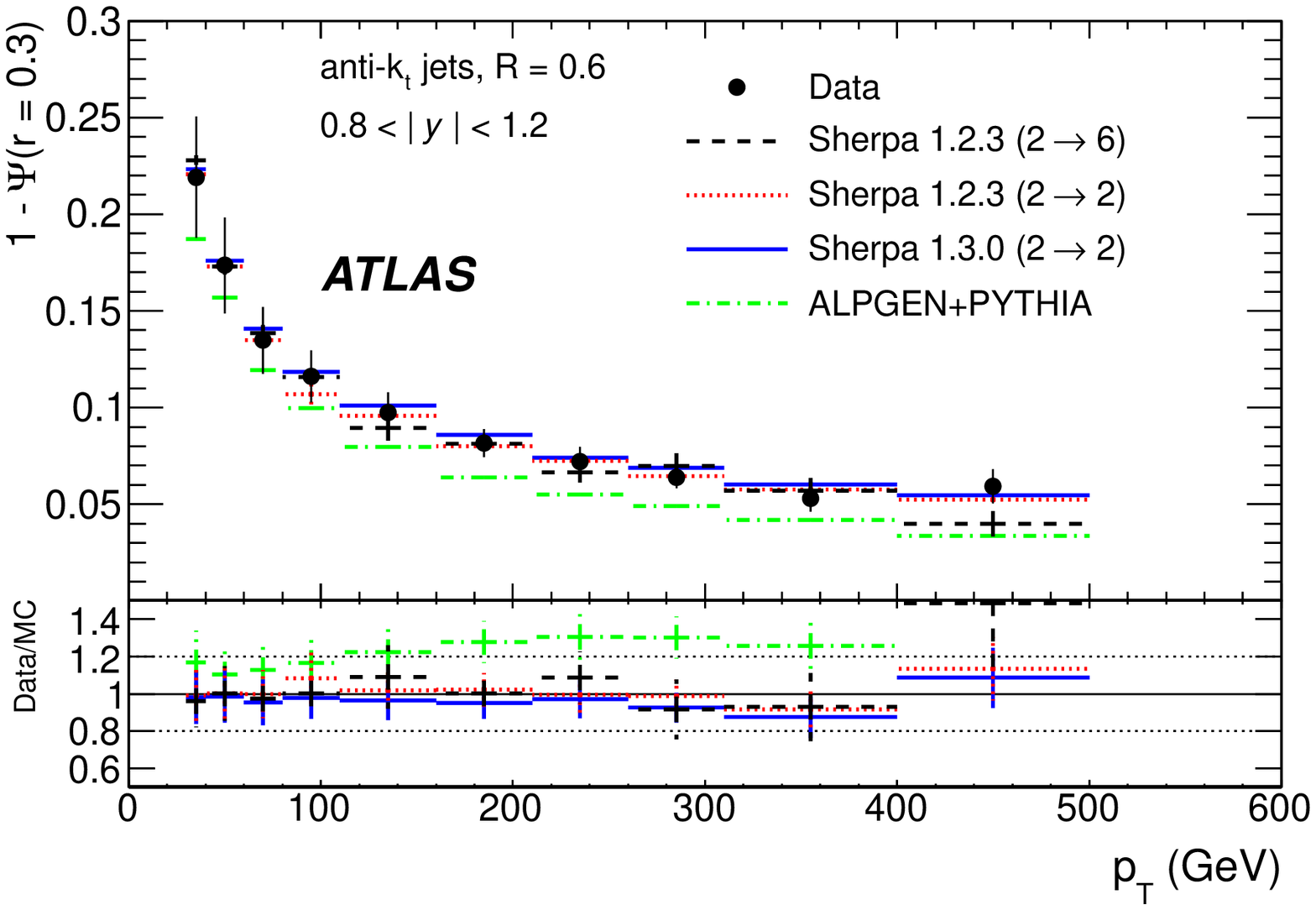} 
\includegraphics[width=0.495\textwidth]{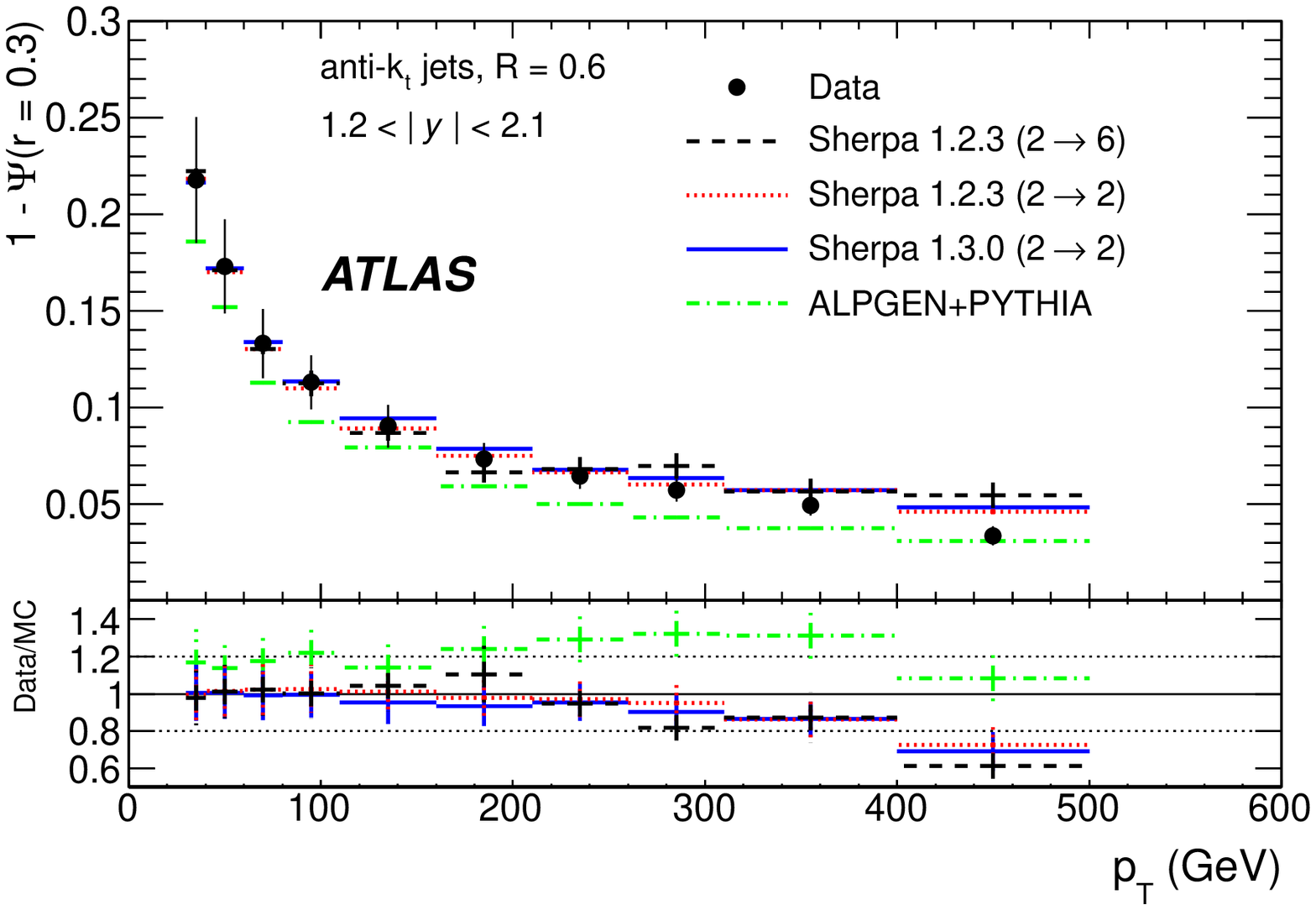}
}
\mbox{
\includegraphics[width=0.495\textwidth]{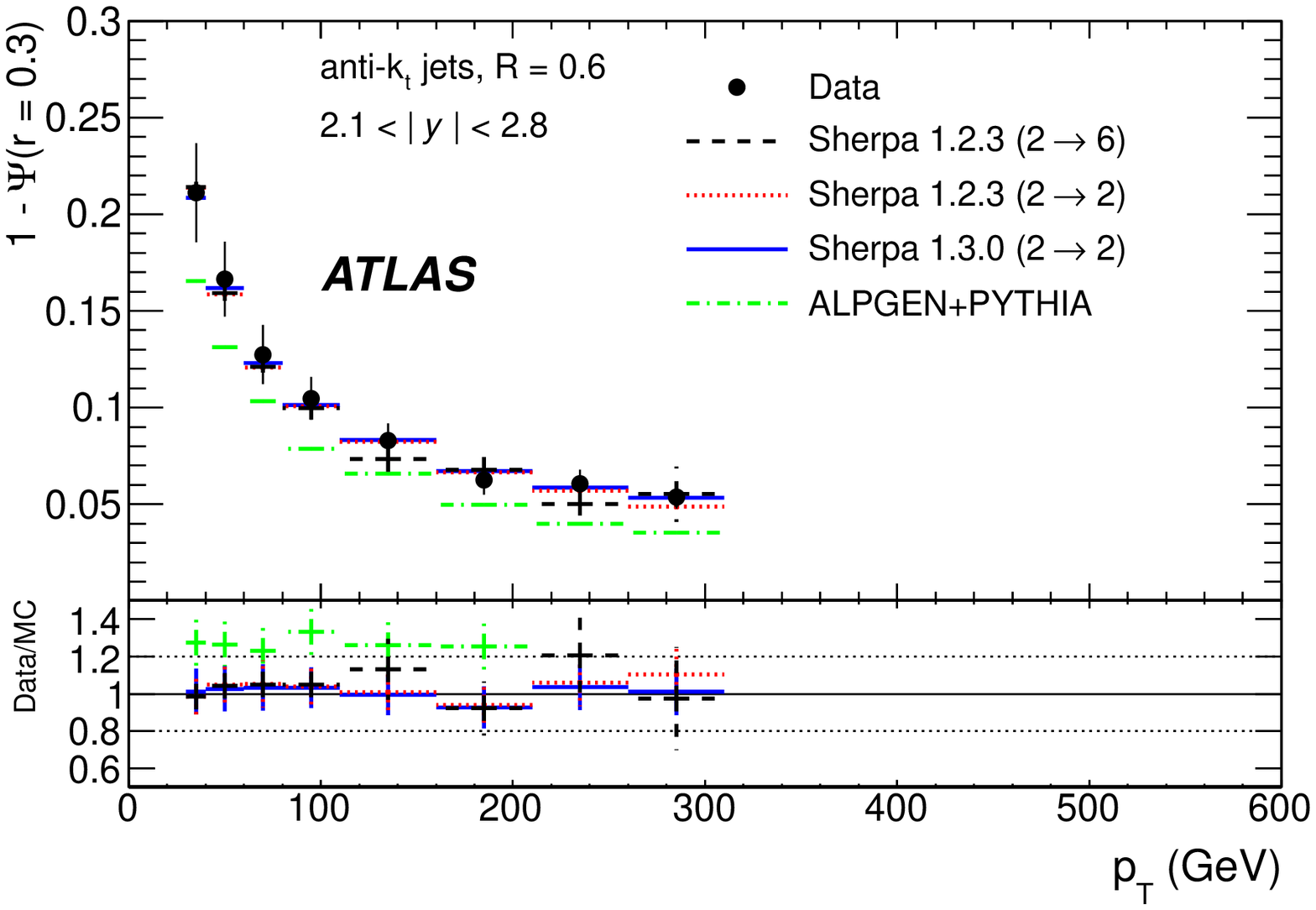}
}
\end{center}
\vspace{-0.7 cm}
\caption{\small
The measured integrated jet shape, $1 - \Psi(r=0.3)$, as a function of $\ptjet$ in different jet rapidity regions for jets with $|\rapjet| < 2.8$ and $30 \ {\rm GeV} < \ptjet < 500 \ {\rm GeV}$.
Error bars indicate the statistical and systematic uncertainties added in quadrature. 
The predictions of   Sherpa 1.3.0 $(2 \to 2)$(solid lines),   Sherpa 1.2.3 $(2 \to 2)$ (dotted lines),  Sherpa (up to $2 \to 6$) (dashed lines), and ALPGEN interfaced to PYTHIA (dashed-dotted lines) are shown for comparison.}  
\label{fig:meps5}
\end{figure}



\begin{figure}[tbh]
\begin{center}
\mbox{
\includegraphics[width=0.495\textwidth,height=0.495\textwidth]{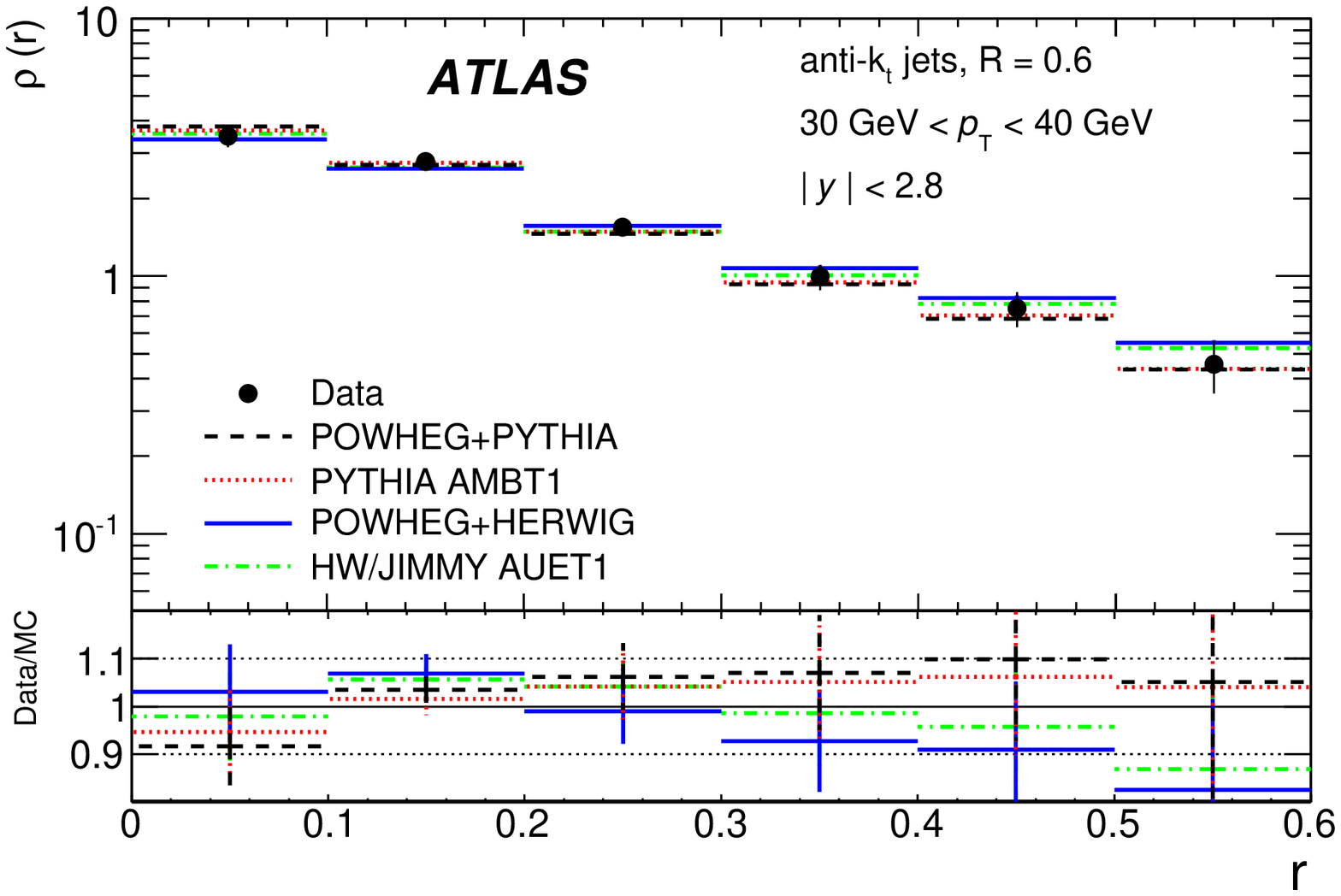} 
\includegraphics[width=0.495\textwidth,height=0.495\textwidth]{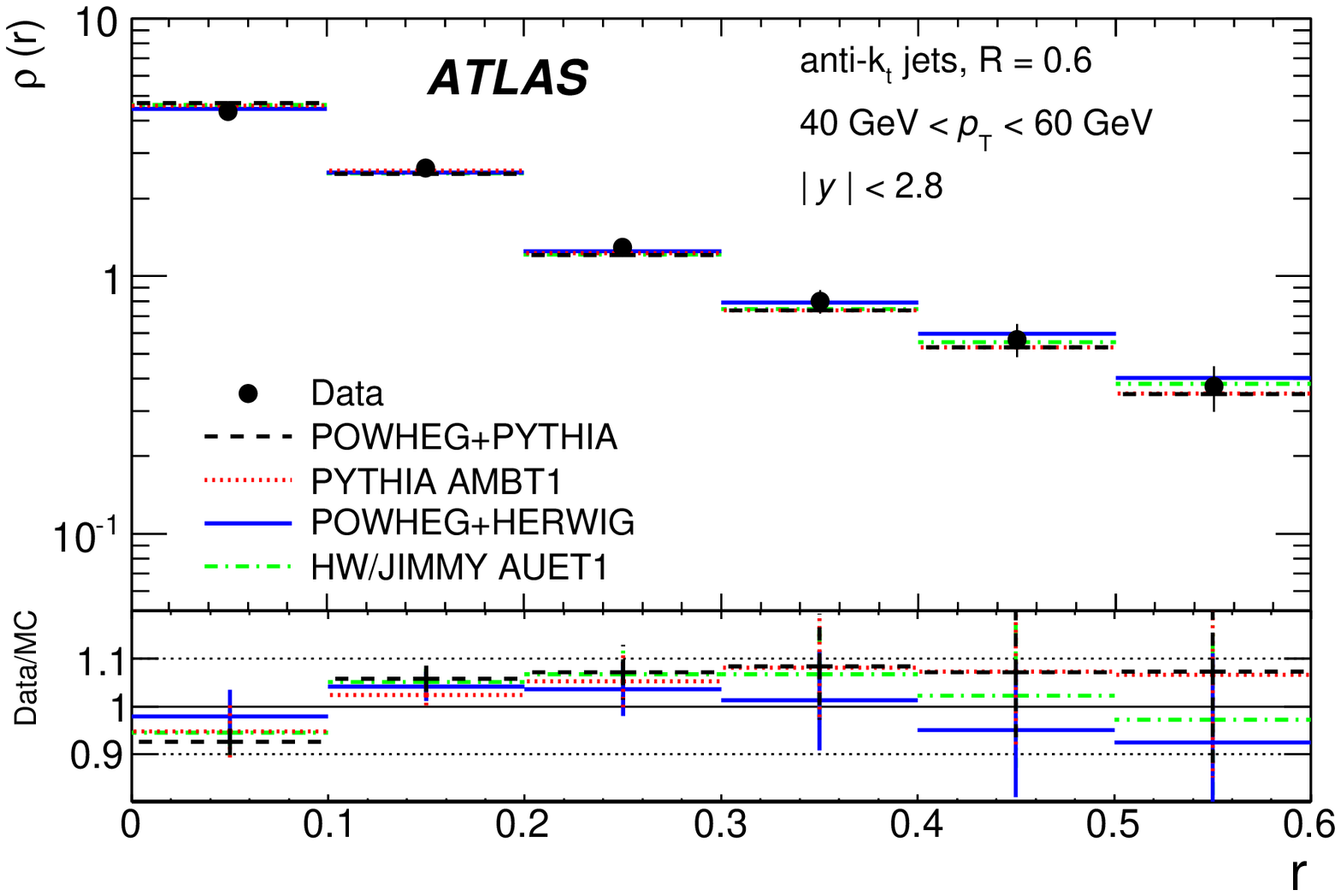}
}\vspace{-0.2cm}
\mbox{
\includegraphics[width=0.495\textwidth,height=0.495\textwidth]{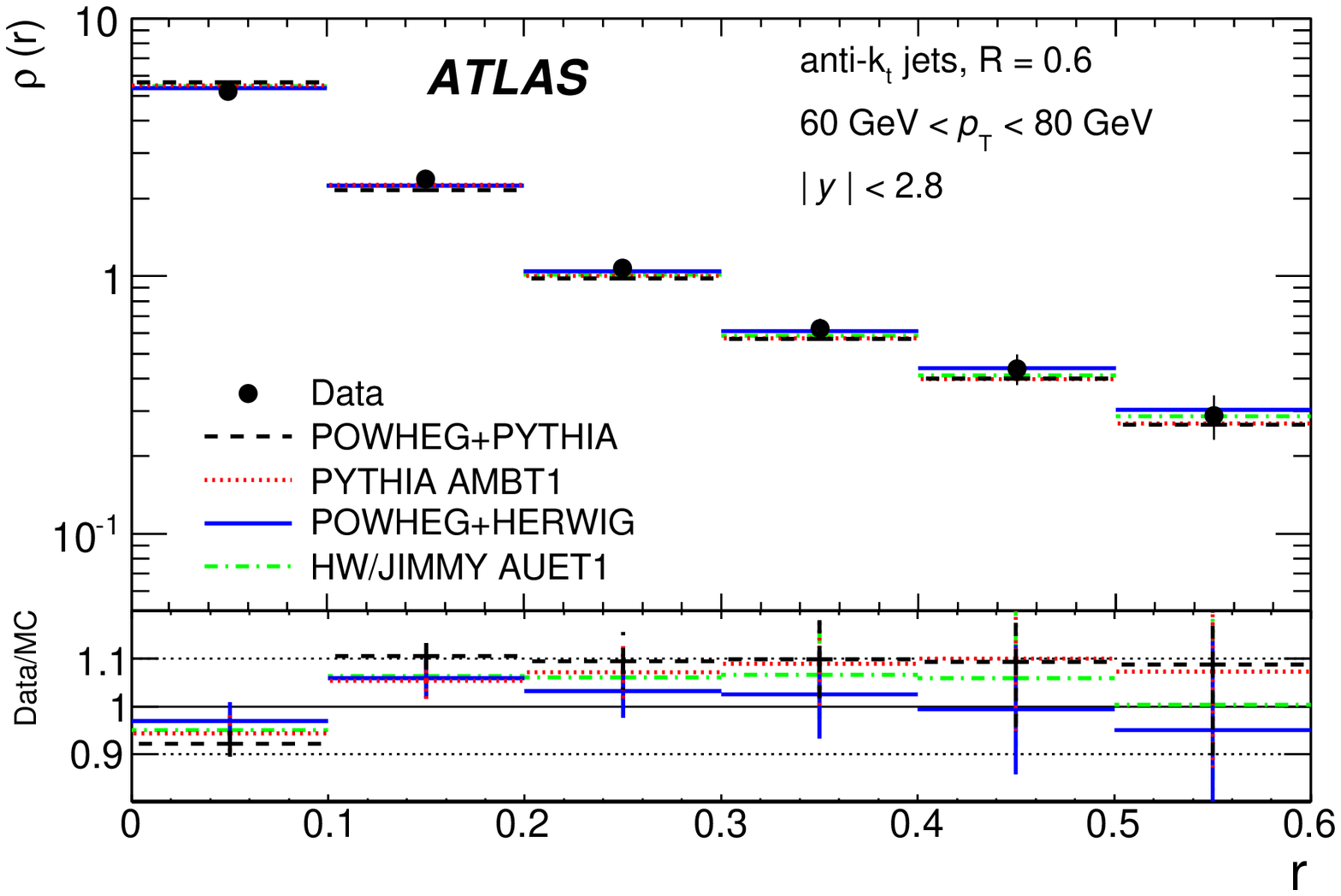}
\includegraphics[width=0.495\textwidth,height=0.495\textwidth]{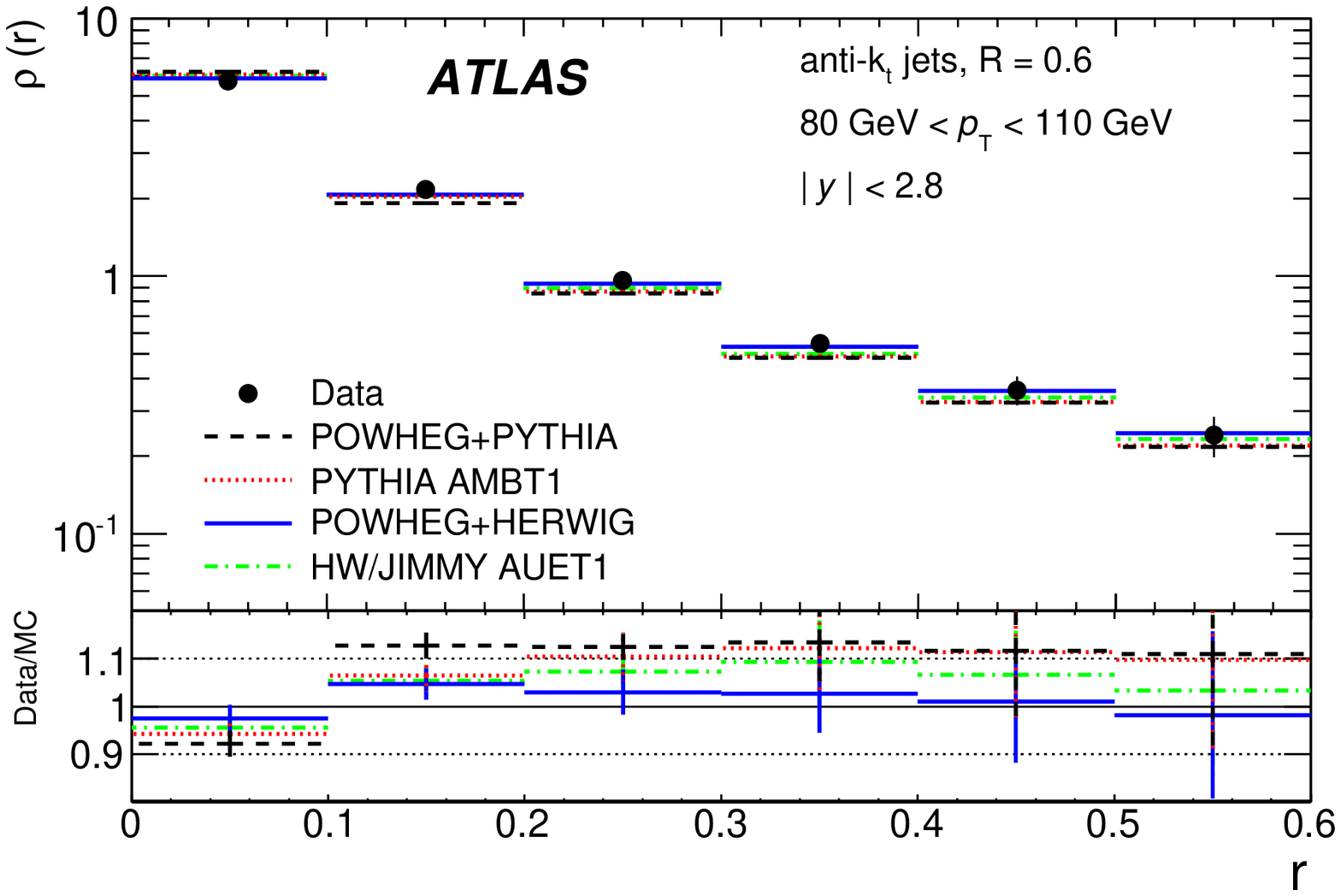}
}
\end{center}
\vspace{-0.7 cm}
\caption{\small
The measured differential jet shape, $\rho(r)$, in inclusive jet production for jets 
with $|\rapjet| < 2.8$ and $30 \ {\rm GeV} < \ptjet < 110  \ {\rm GeV}$   
is shown in different $\ptjet$ regions. Error bars indicate the statistical and systematic uncertainties added in quadrature.
The predictions of   POWHEG interfaced with PYTHIA-AMBT1 (dashed lines),   POWHEG interfaced with HERWIG/JIMMY-AUET1 (solid lines),  PYTHIA-AMBT1  (dotted lines), and HERWIG/JIMMY-AUET1 (dashed-dotted lines) are shown for comparison.} \label{fig:pow1}
\end{figure}

\begin{figure}[tbh]
\begin{center}
\mbox{
\includegraphics[width=0.495\textwidth,height=0.495\textwidth]{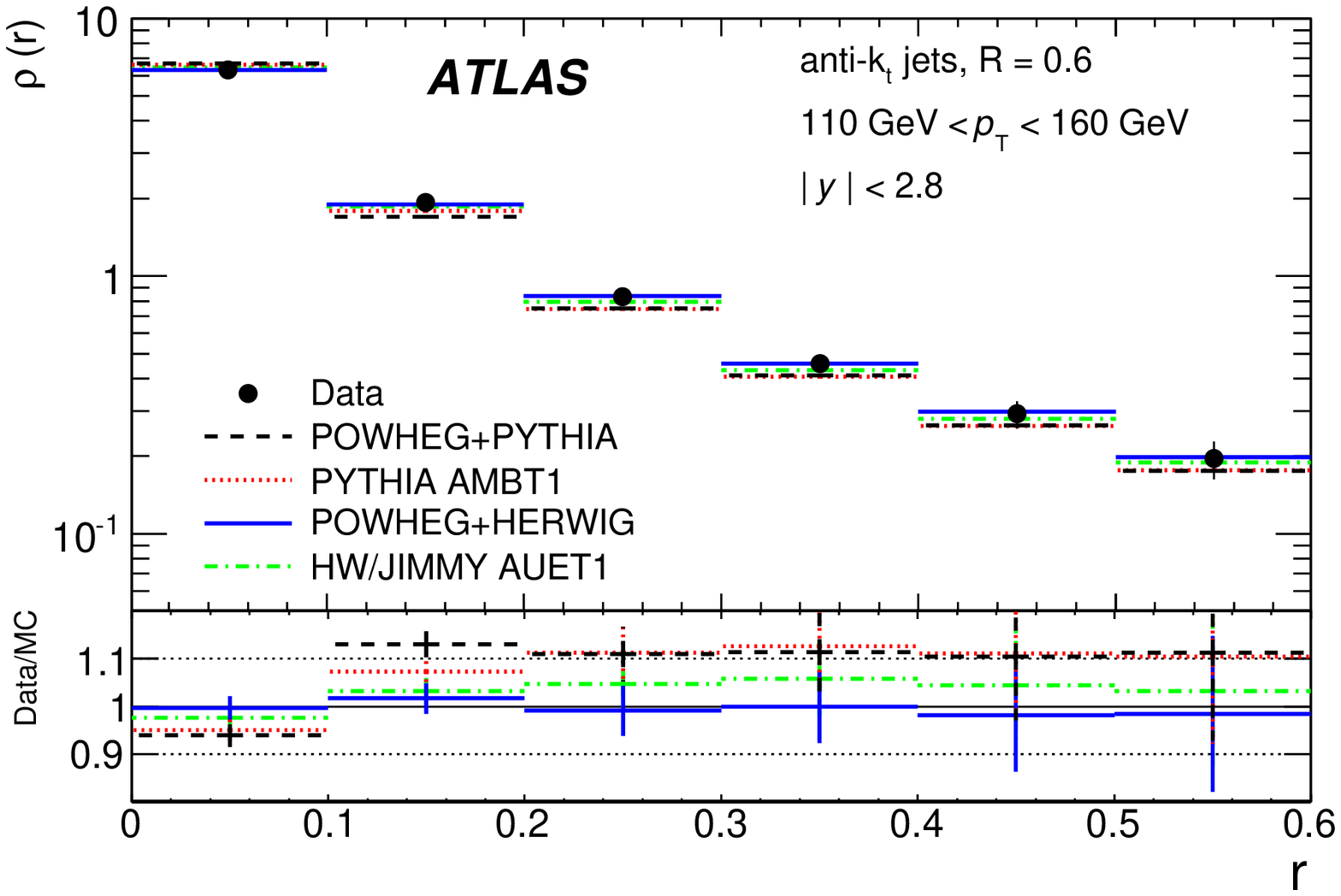} 
\includegraphics[width=0.495\textwidth,height=0.495\textwidth]{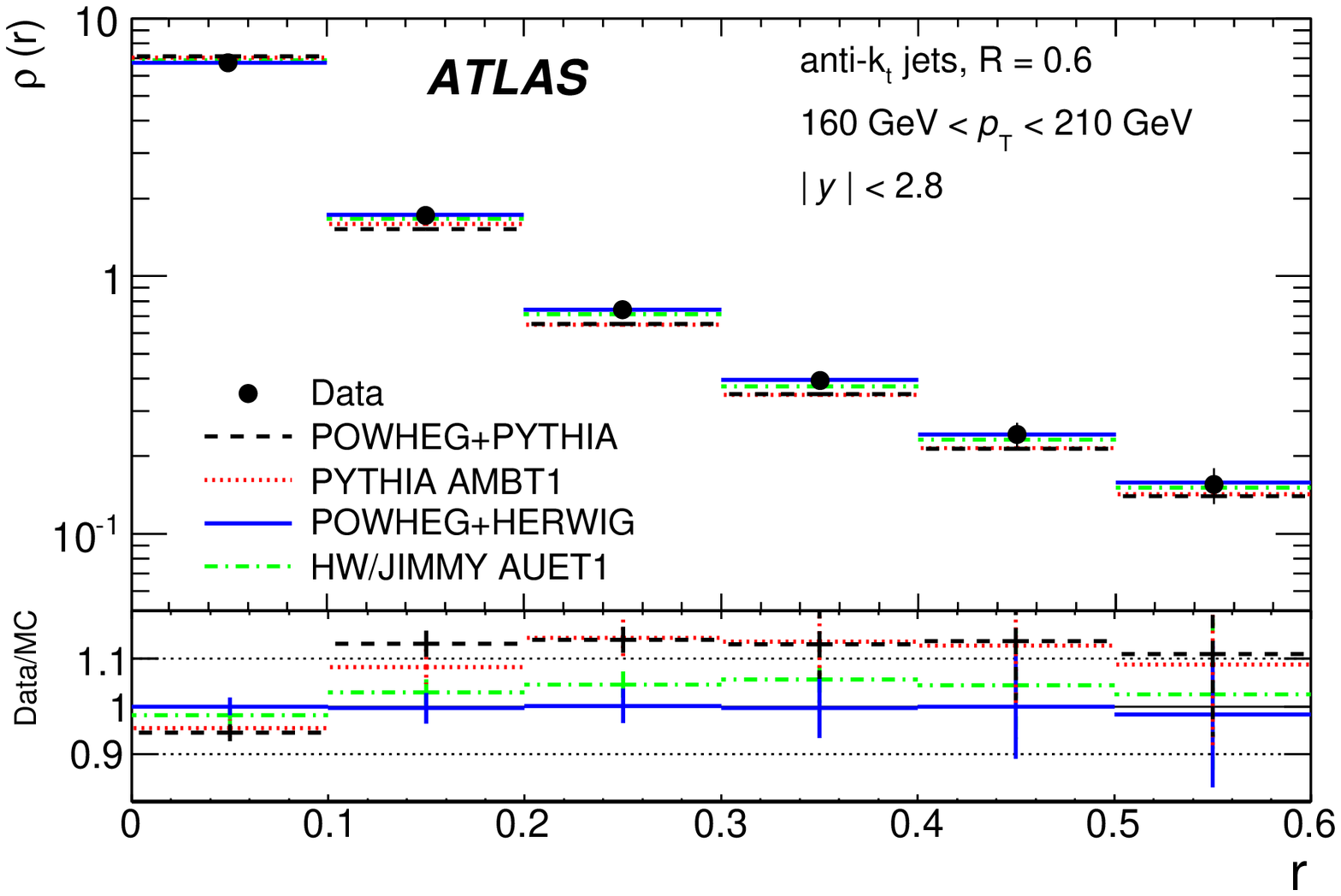}
}\vspace{-0.2cm}
\mbox{
\includegraphics[width=0.495\textwidth,height=0.495\textwidth]{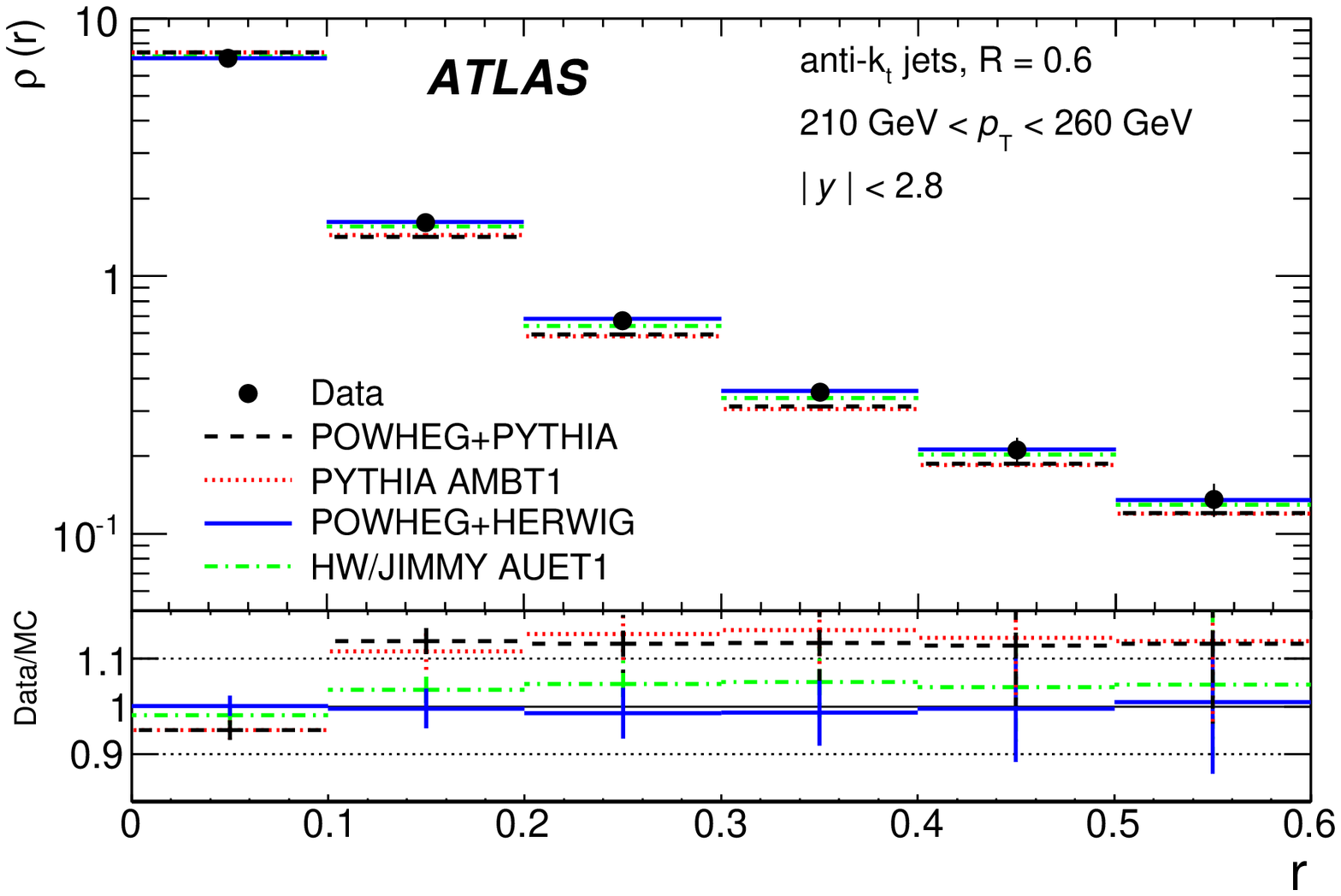}
\includegraphics[width=0.495\textwidth,height=0.495\textwidth]{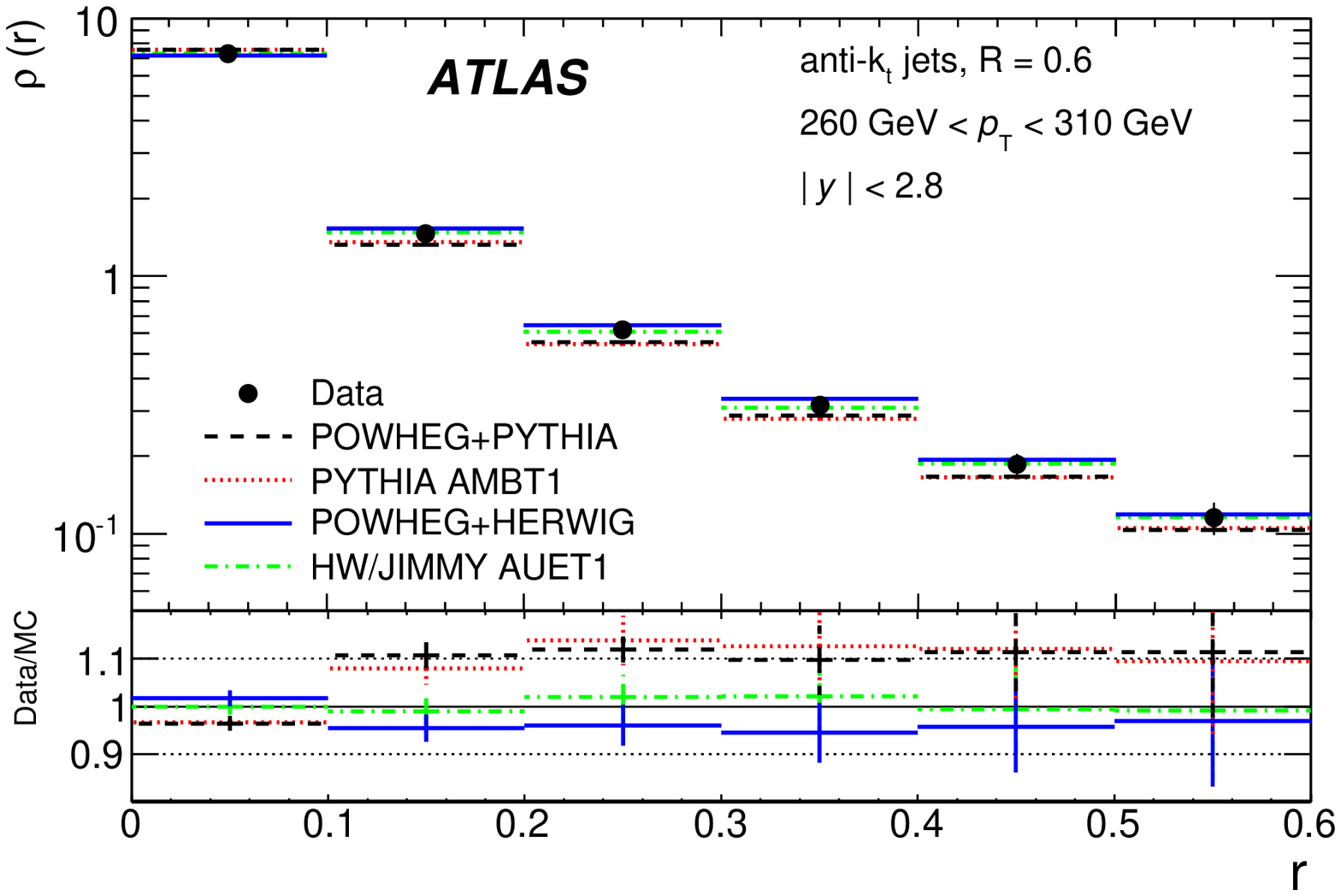}
}
\end{center}
\vspace{-0.7 cm}
\caption{\small
The measured differential jet shape, $\rho(r)$, in inclusive jet production for jets 
with $|\rapjet| < 2.8$ and $110 \ {\rm GeV} < \ptjet < 310  \ {\rm GeV}$   
is shown in different $\ptjet$ regions. Error bars indicate the statistical and systematic uncertainties added in quadrature.
The predictions of   POWHEG interfaced with PYTHIA-AMBT1 (dashed lines),   POWHEG interfaced with HERWIG/JIMMY-AUET1 (solid lines),  PYTHIA-AMBT1  (dotted lines), and HERWIG/JIMMY-AUET1 (dashed-dotted lines) are shown for comparison.}
\label{fig:pow2}
\end{figure}


\begin{figure}[tbh]
\begin{center}
\mbox{
\includegraphics[width=0.495\textwidth]{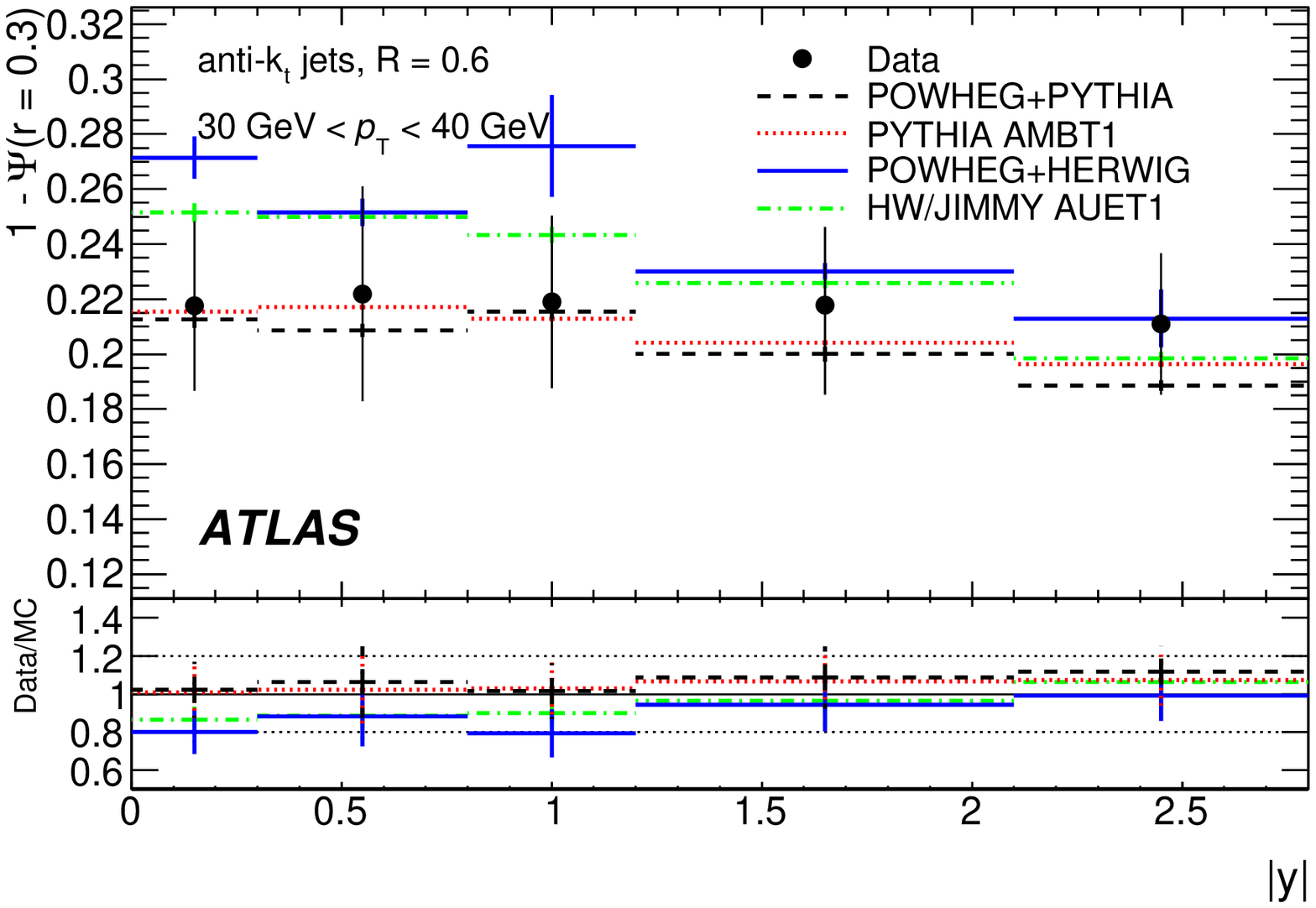}
\includegraphics[width=0.495\textwidth]{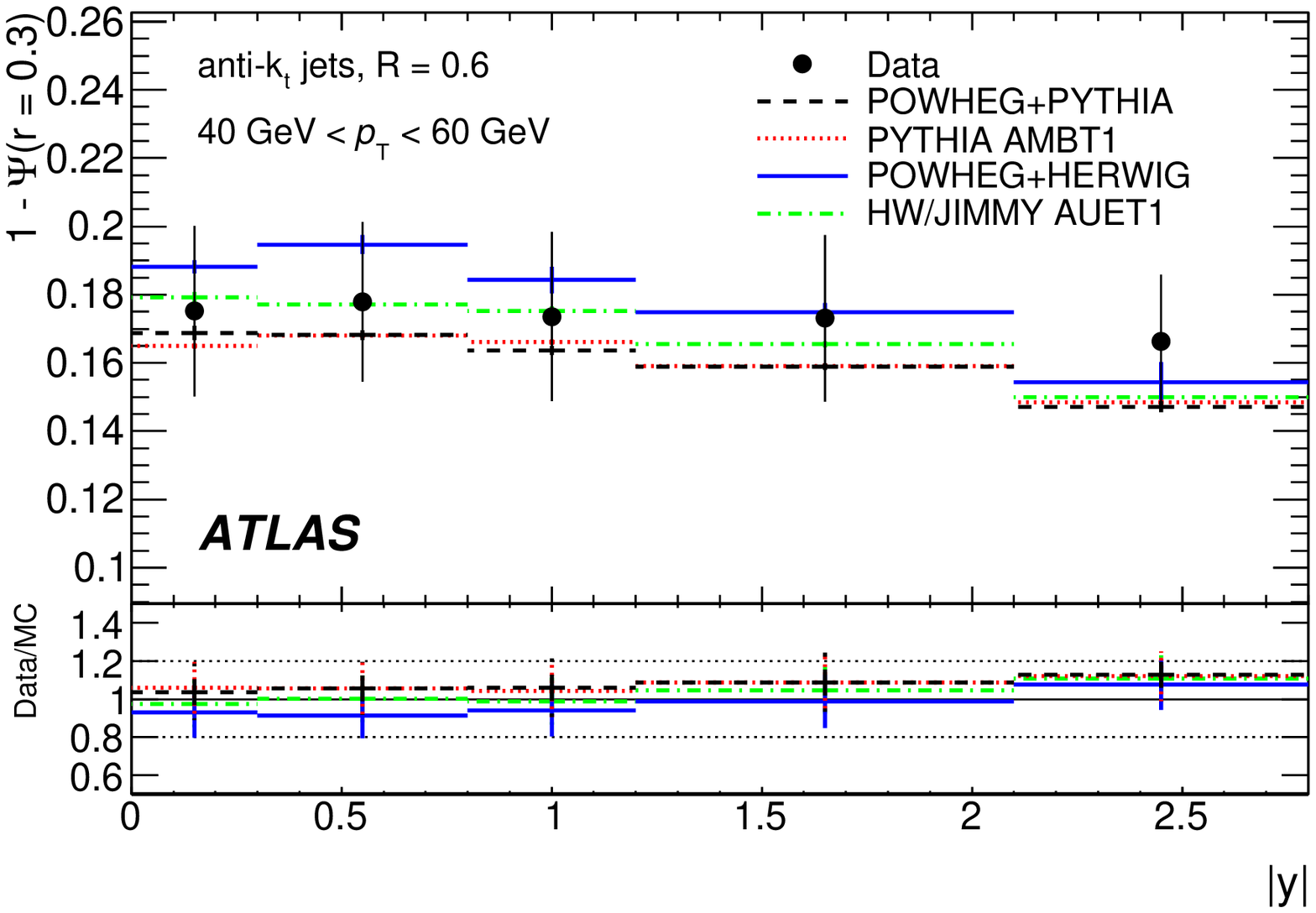}
}
\mbox{
\includegraphics[width=0.495\textwidth]{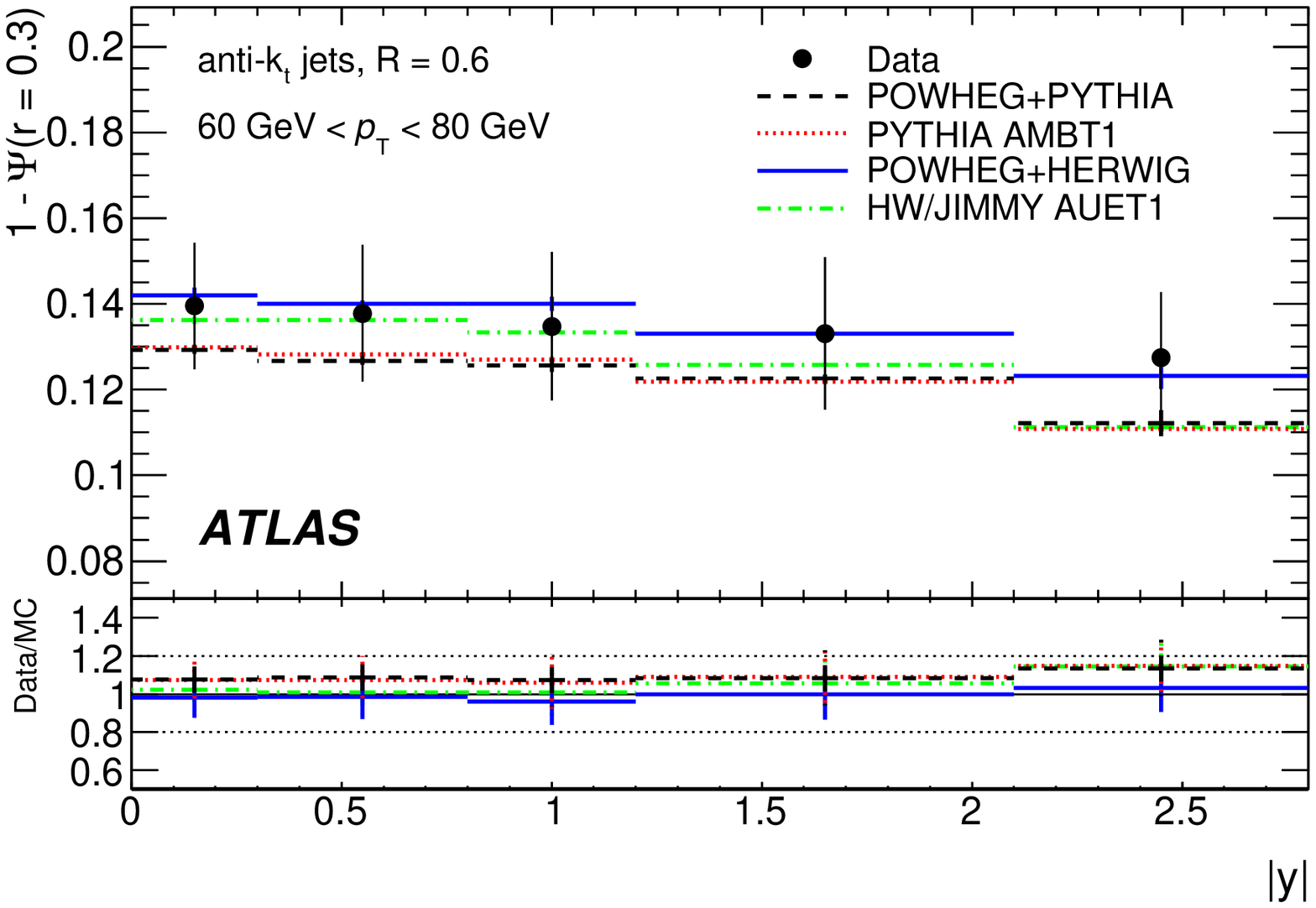} 
\includegraphics[width=0.495\textwidth]{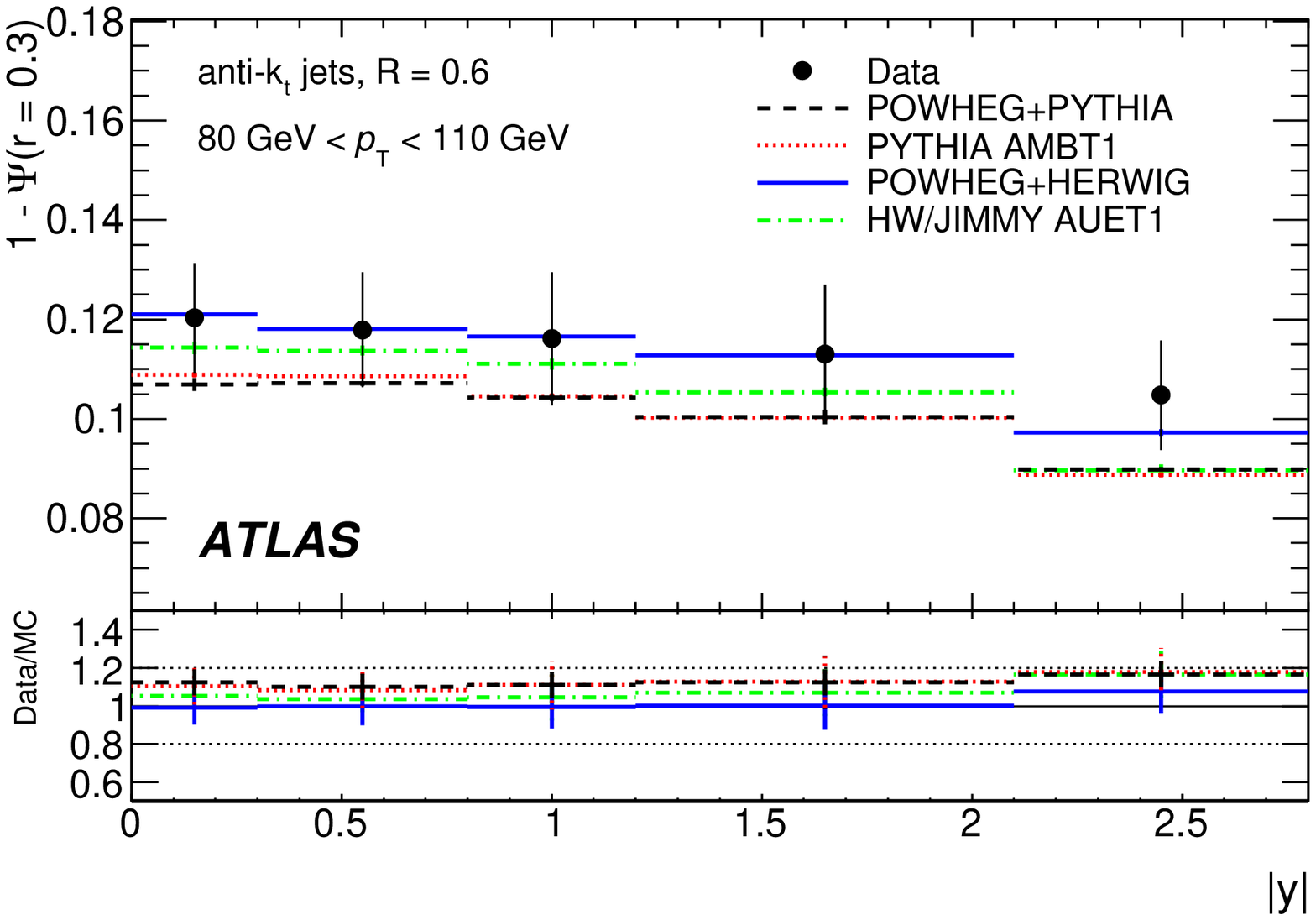}
}
\end{center}
\vspace{-0.7 cm}
\caption{\small
The measured integrated jet shape, $1 - \Psi(r=0.3)$, as a function of $|\rapjet|$ in different jet $\ptjet$ regions 
for jets with $|\rapjet| < 2.8$ and $30 \ {\rm GeV} < \ptjet < 110 \ {\rm GeV}$.
Error bars indicate the statistical and systematic uncertainties added in quadrature. 
The predictions of   POWHEG interfaced with PYTHIA-AMBT1 (dashed lines),   POWHEG interfaced with HERWIG/JIMMY-AUET1 (solid lines),  PYTHIA-AMBT1  (dotted lines), and HERWIG/JIMMY-AUET1 (dashed-dotted lines) are shown for comparison.}
\label{fig:pow3}
\end{figure}

\begin{figure}[tbh]
\begin{center}
\mbox{
\includegraphics[width=0.495\textwidth]{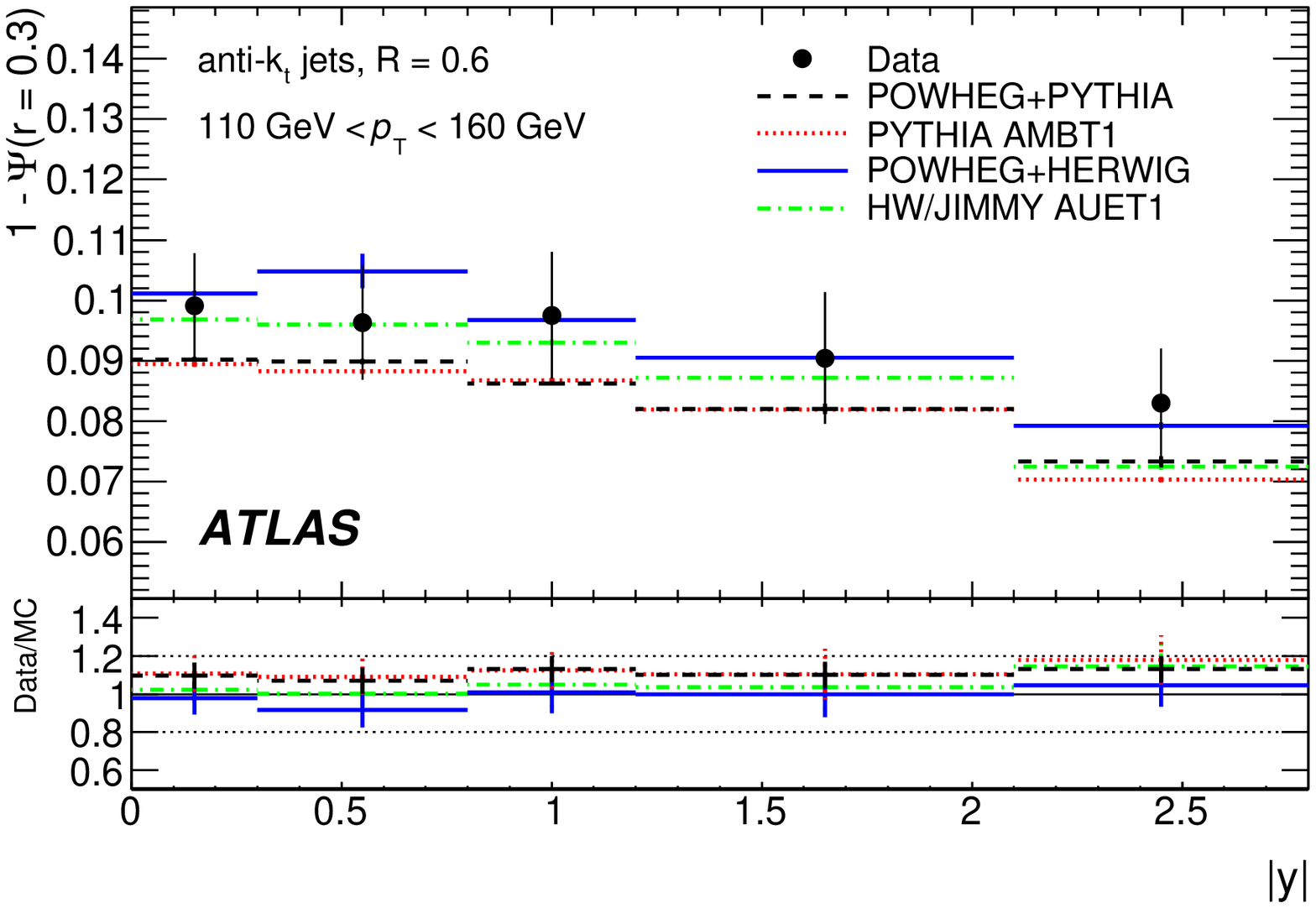}
\includegraphics[width=0.495\textwidth]{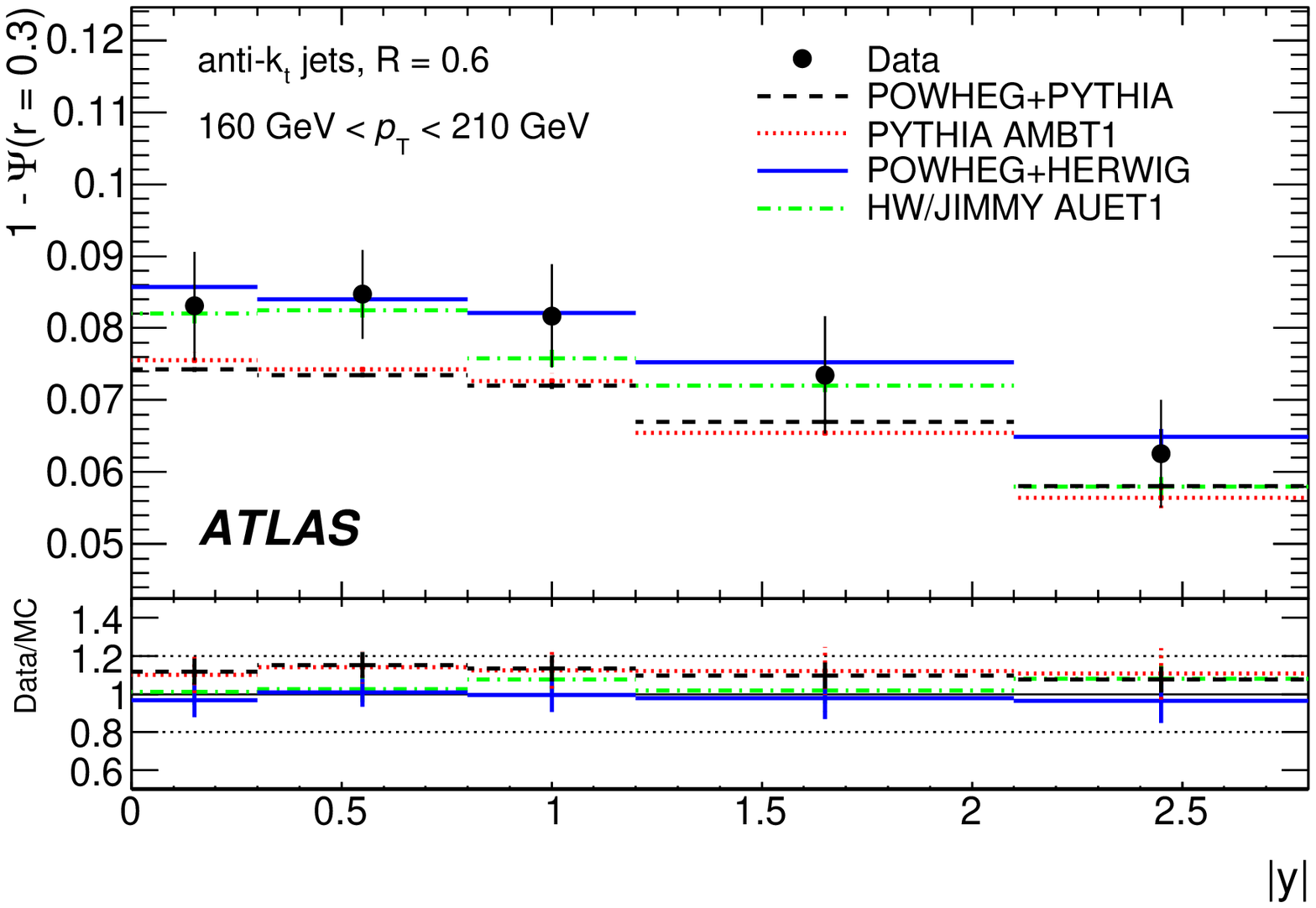}
}
\mbox{
\includegraphics[width=0.495\textwidth]{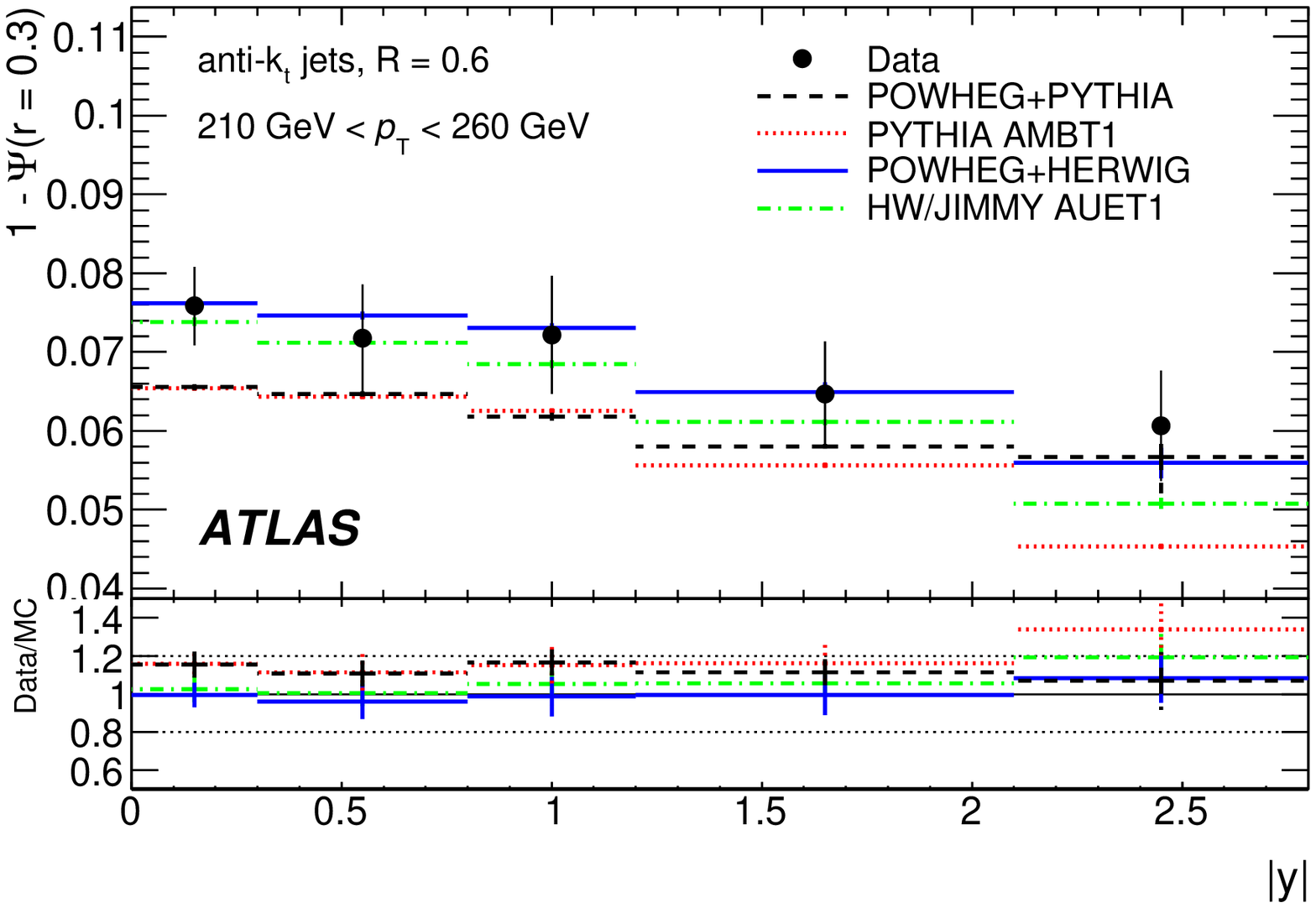} 
\includegraphics[width=0.495\textwidth]{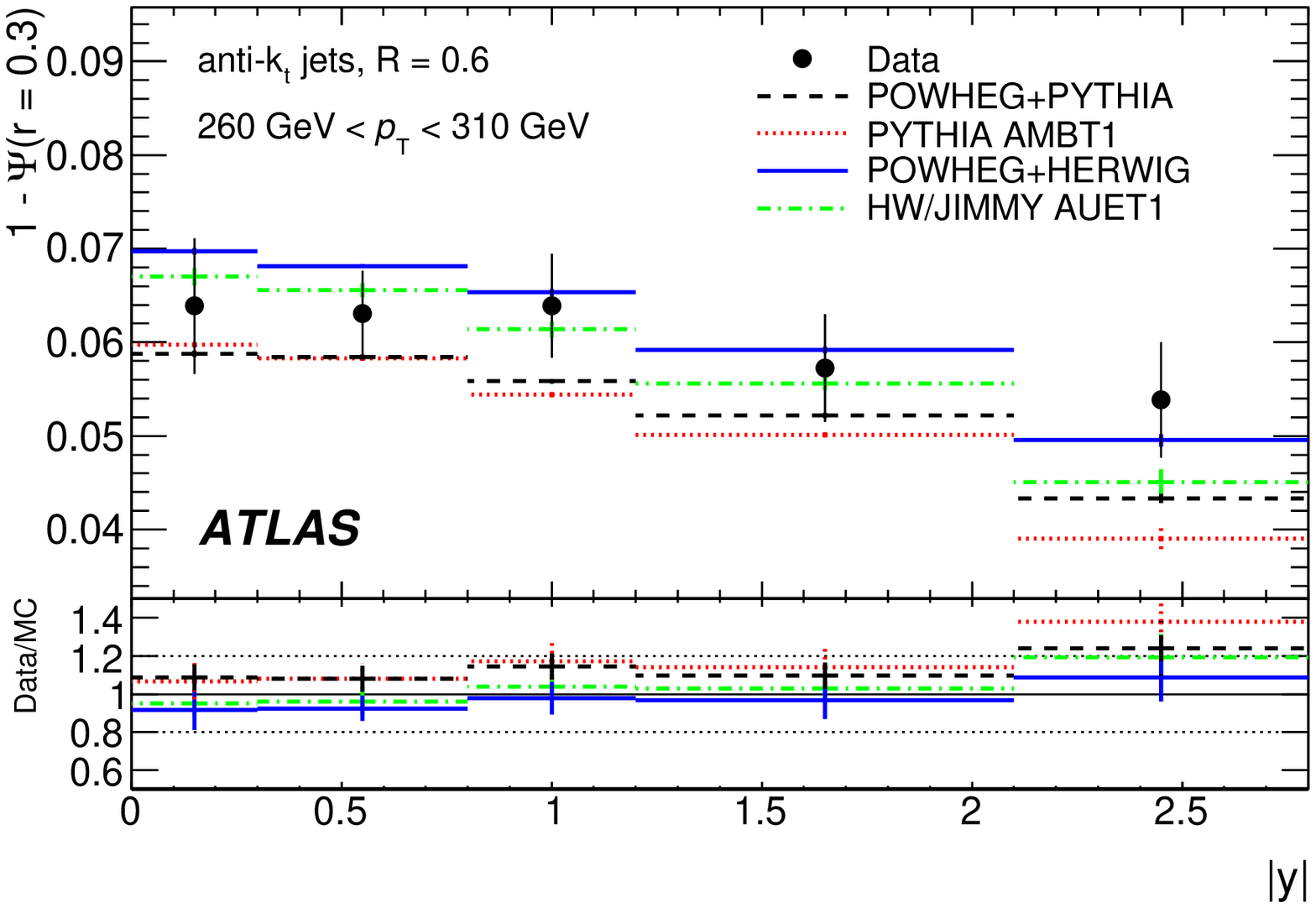}
}
\end{center}
\vspace{-0.7 cm}
\caption{\small
The measured integrated jet shape, $1 - \Psi(r=0.3)$, as a function of $|\rapjet|$ in different jet $\ptjet$ regions 
for jets with $|\rapjet| < 2.8$ and $110 \ {\rm GeV} < \ptjet < 310 \ {\rm GeV}$.
Error bars indicate the statistical and systematic uncertainties added in quadrature. 
The predictions of   POWHEG interfaced with PYTHIA-AMBT1 (dashed lines),   POWHEG interfaced with HERWIG/JIMMY-AUET1 (solid lines),  PYTHIA-AMBT1  (dotted lines), and HERWIG/JIMMY-AUET1 (dashed-dotted lines) are shown for comparison.}  
\label{fig:pow4}
\end{figure}


\begin{figure}[tbh]
\begin{center}
\mbox{
\includegraphics[width=0.495\textwidth]{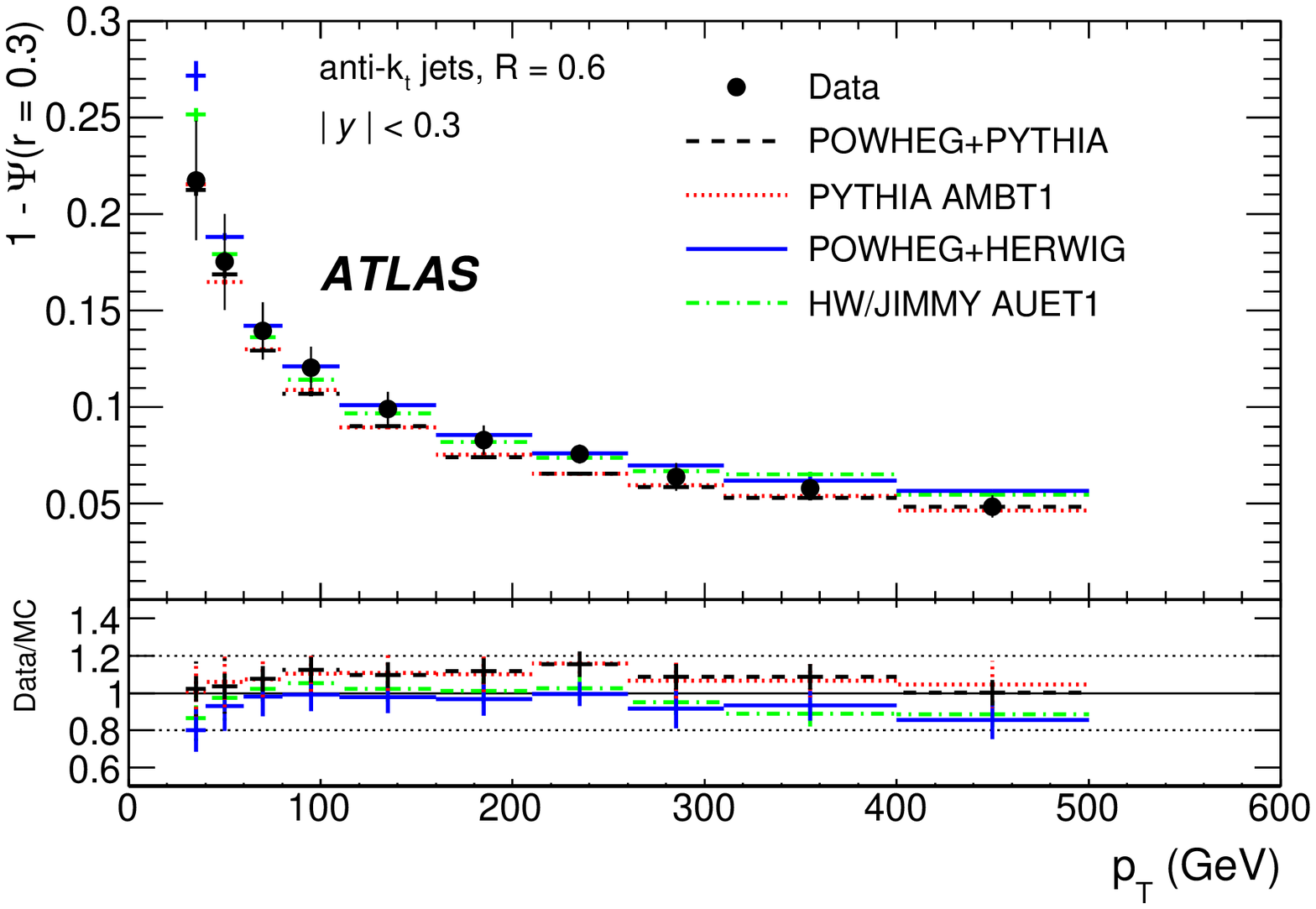}
\includegraphics[width=0.495\textwidth]{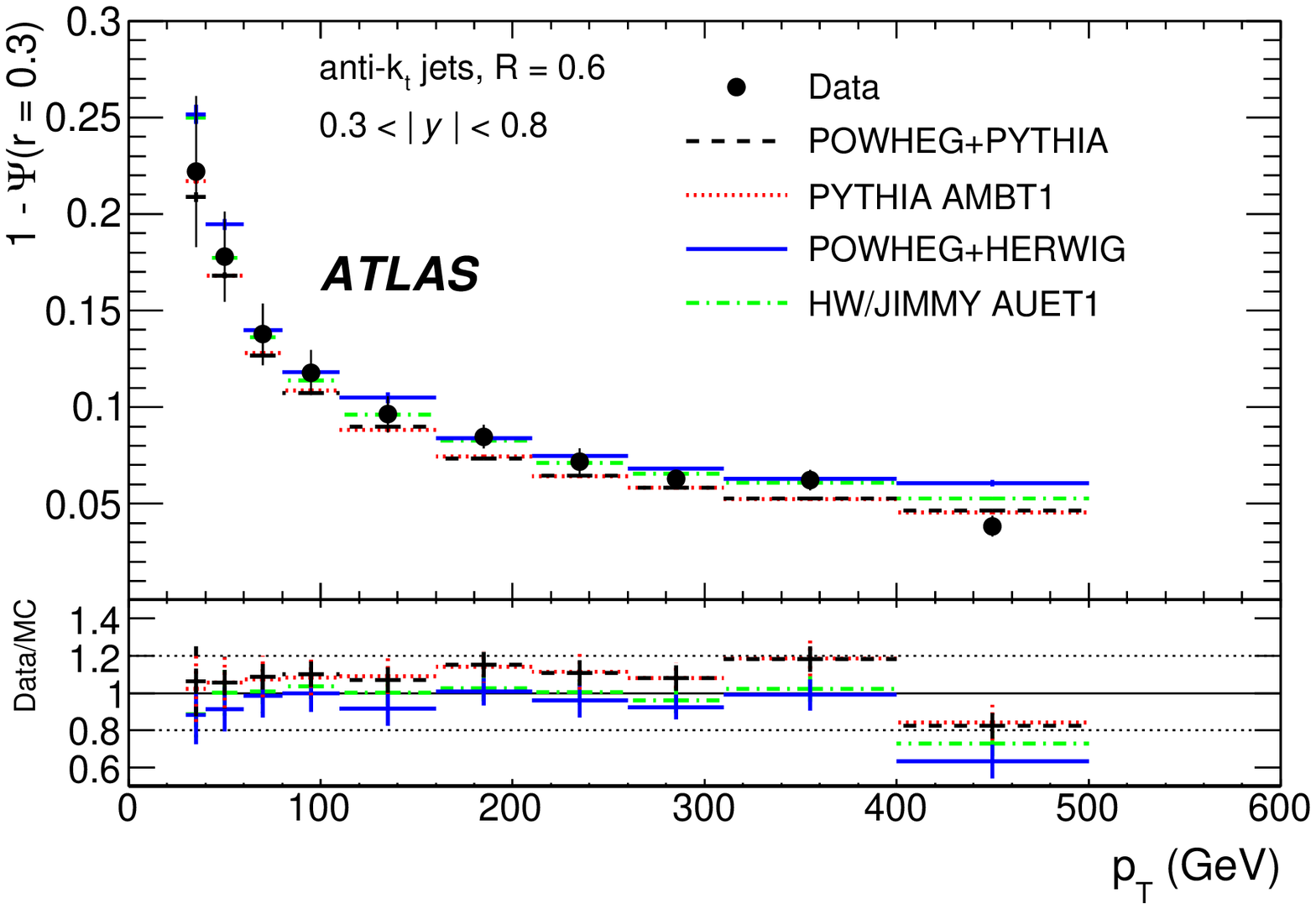}
}
\mbox{
\includegraphics[width=0.495\textwidth]{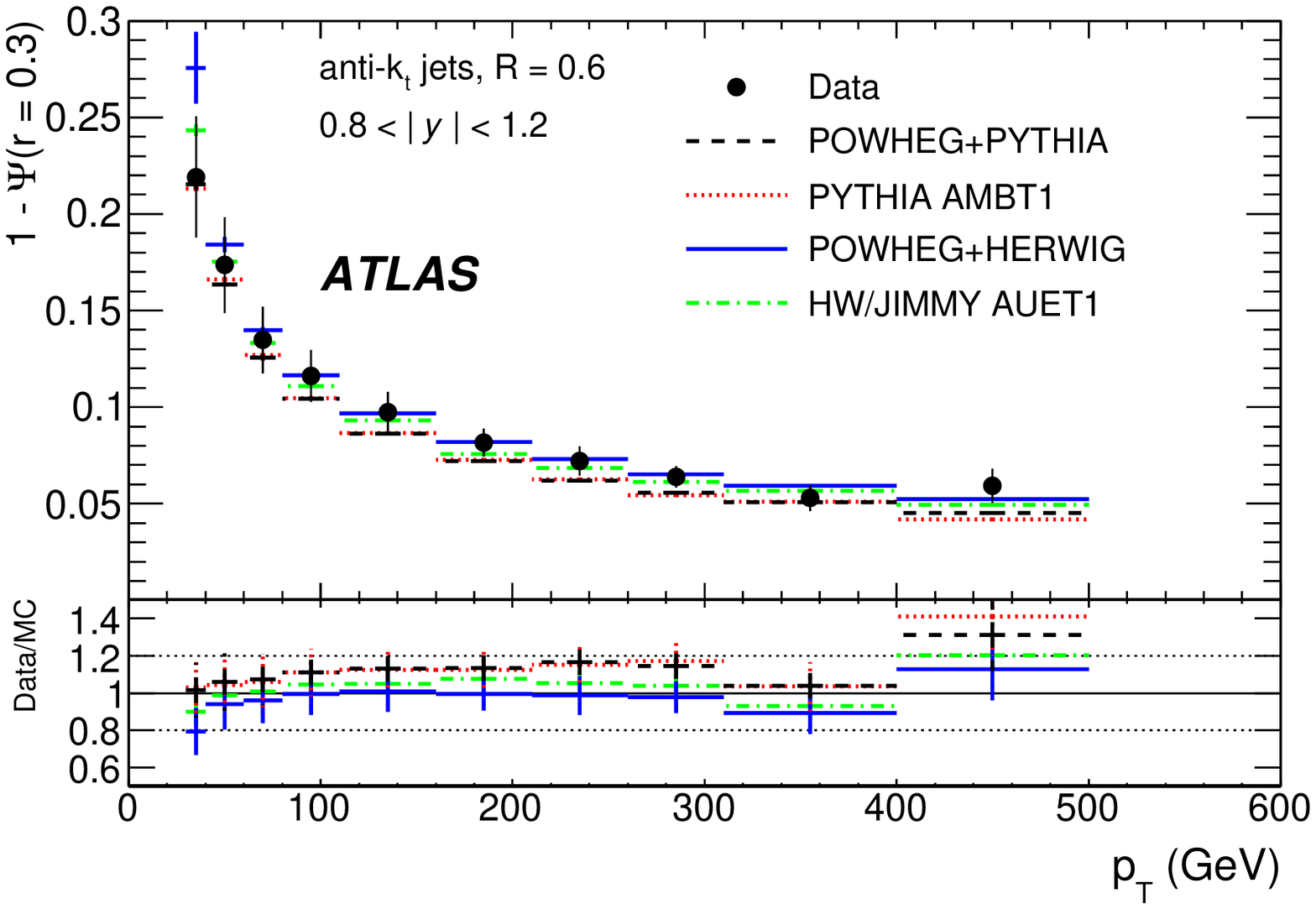} 
\includegraphics[width=0.495\textwidth]{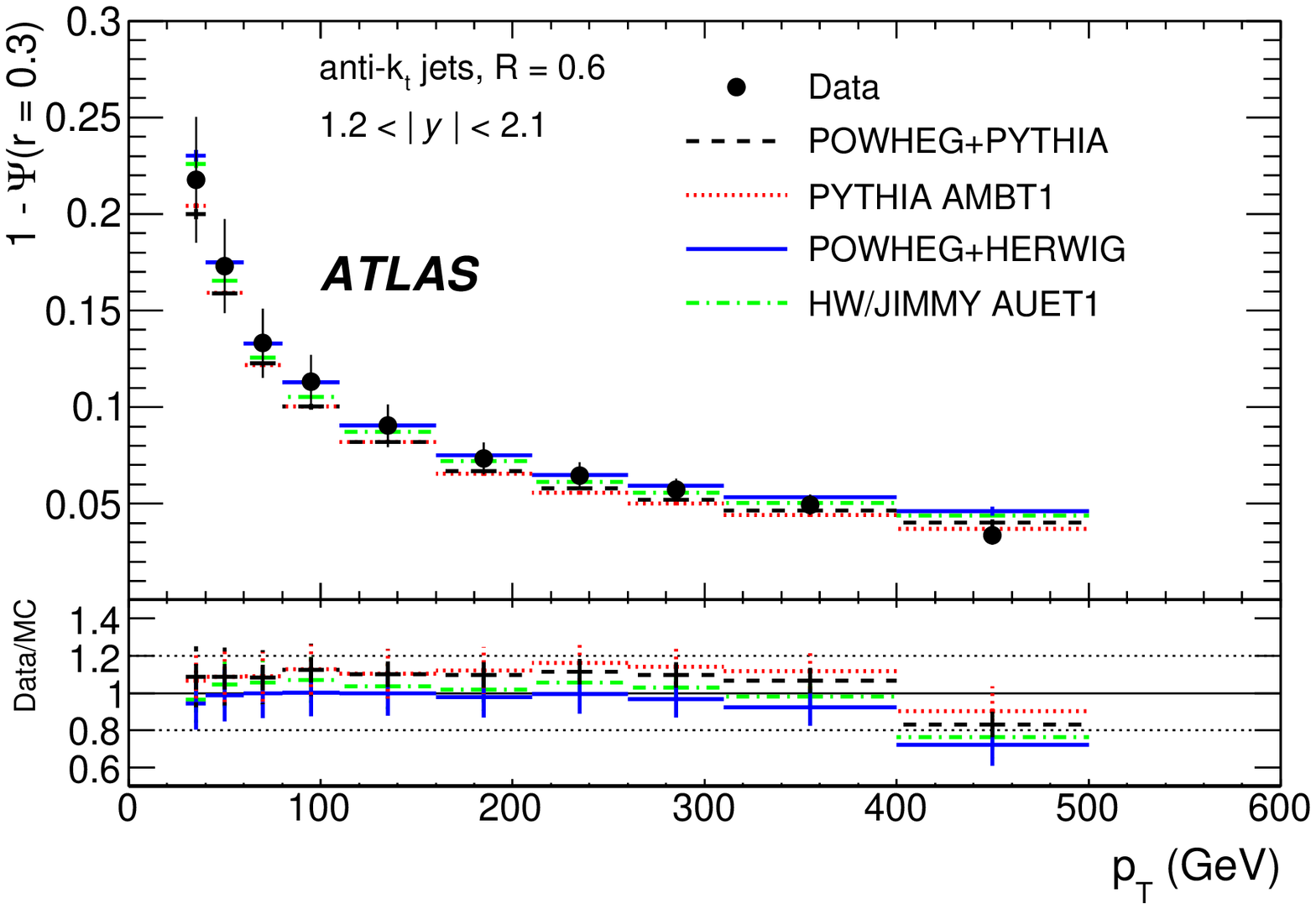}
}
\mbox{
\includegraphics[width=0.495\textwidth]{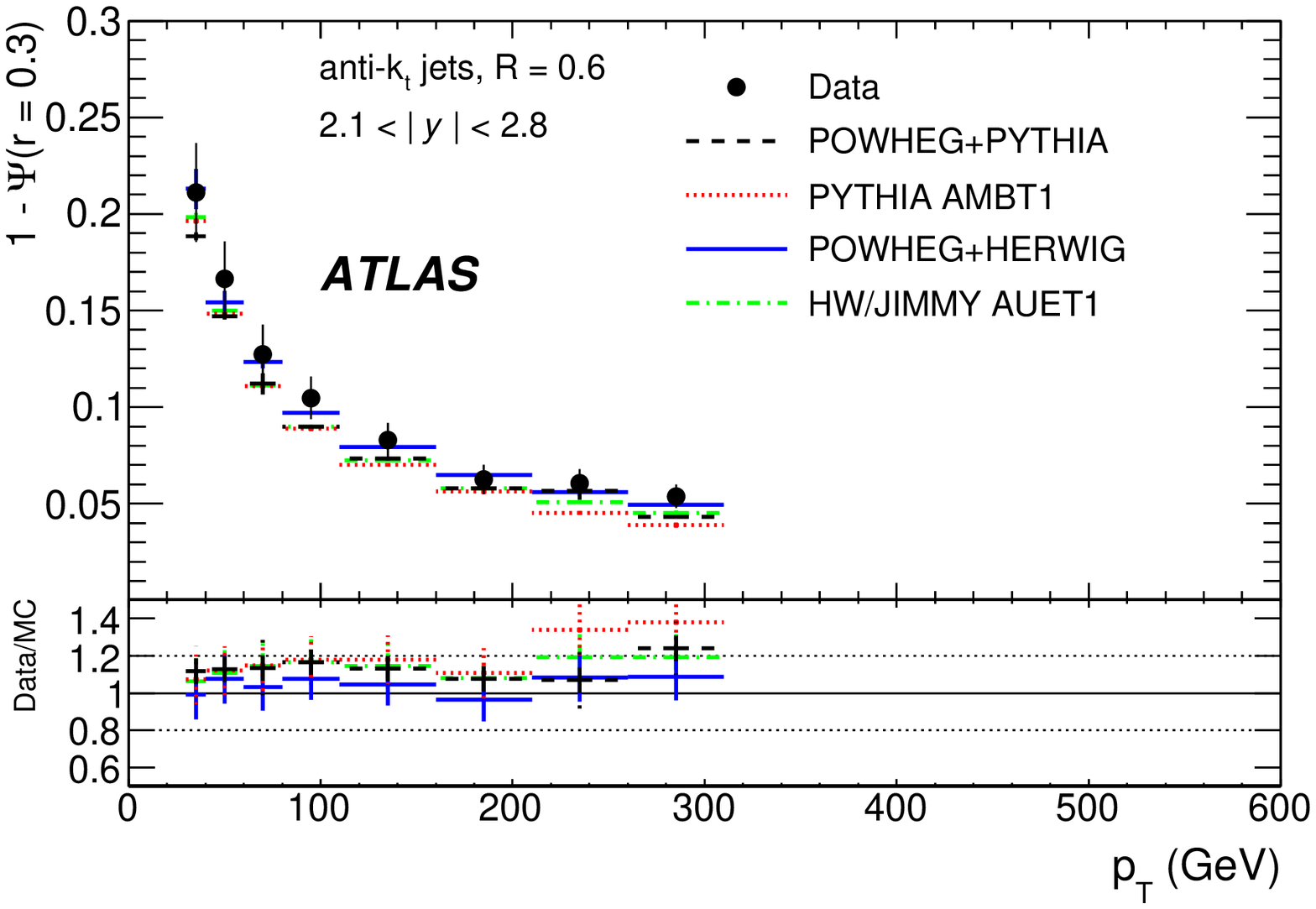}
}
\end{center}
\vspace{-0.7 cm}
\caption{\small
The measured integrated jet shape, $1 - \Psi(r=0.3)$, as a function of $\ptjet$ in different jet rapidity regions for jets with $|\rapjet| < 2.8$ and $30 \ {\rm GeV} < \ptjet < 500 \ {\rm GeV}$.
Error bars indicate the statistical and systematic uncertainties added in quadrature. 
The predictions of   POWHEG interfaced with PYTHIA-AMBT1 (dashed lines),   POWHEG interfaced with HERWIG/JIMMY-AUET1 (solid lines),  PYTHIA-AMBT1  (dotted lines), and HERWIG/JIMMY-AUET1 (dashed-dotted lines) are shown for comparison.}  \label{fig:pow5}
\end{figure}

%

\begin{figure}[tbh]
\begin{center}
\mbox{
\includegraphics[width=0.495\textwidth]{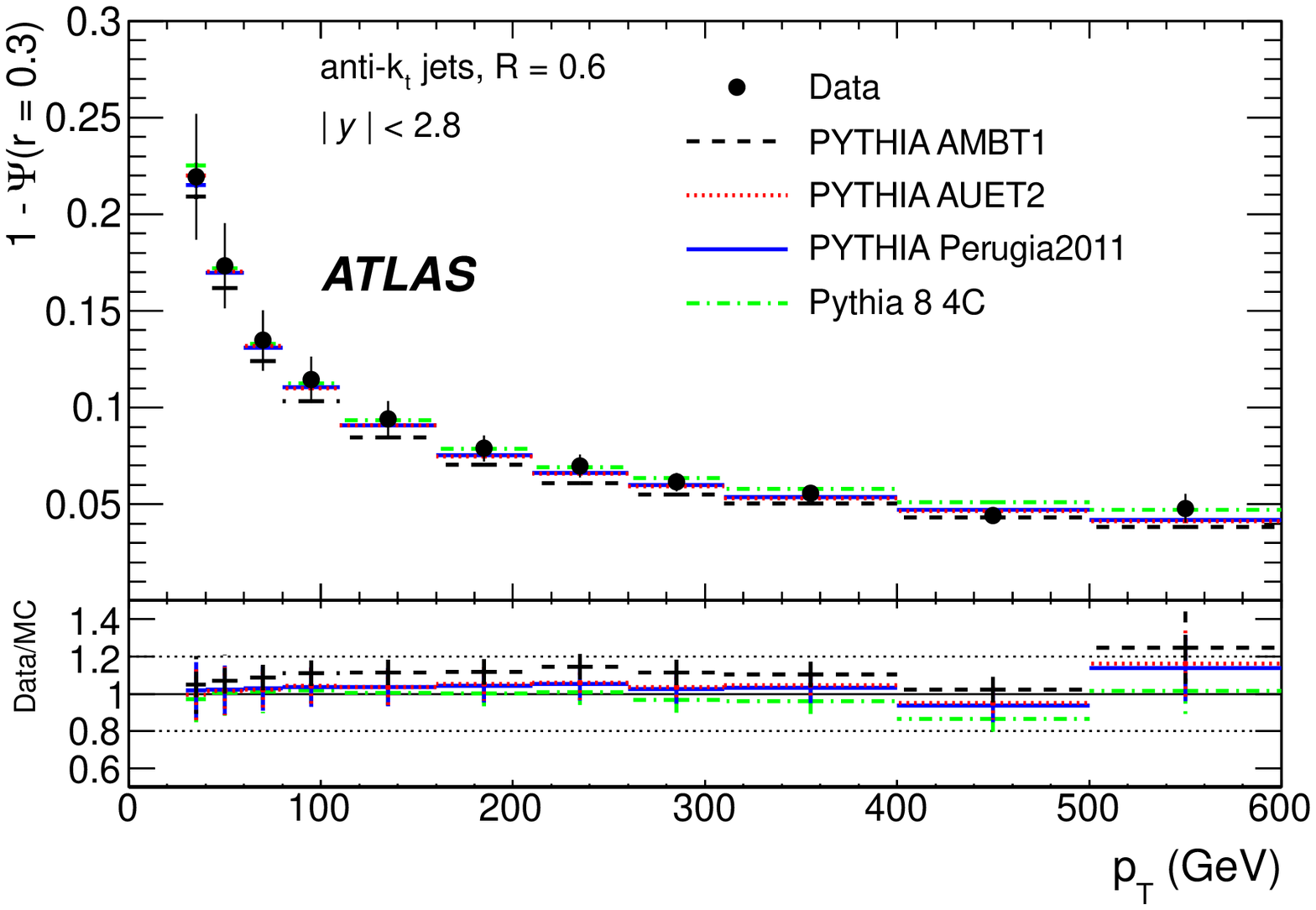}
\includegraphics[width=0.495\textwidth]{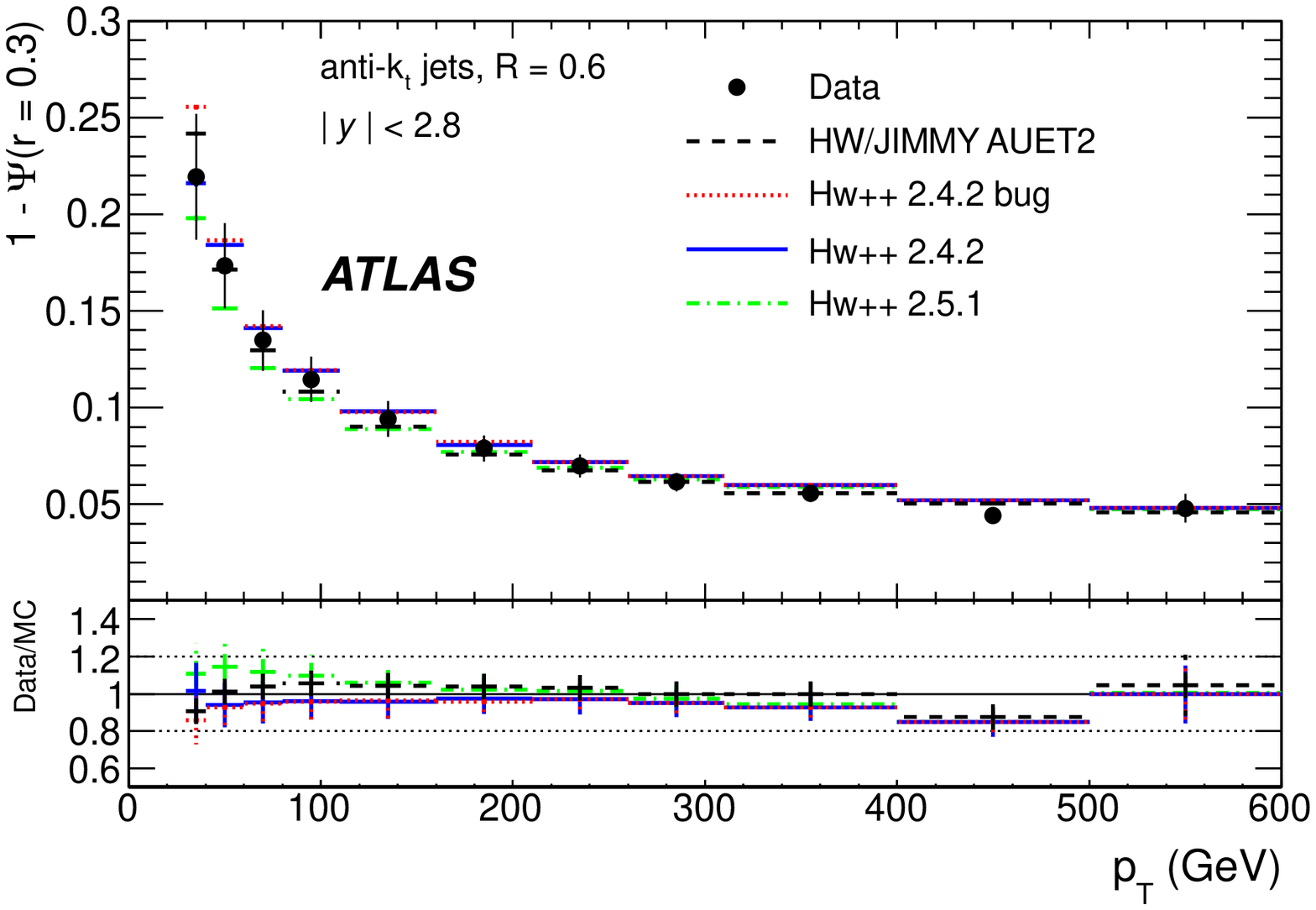}
}
\mbox{
\includegraphics[width=0.495\textwidth]{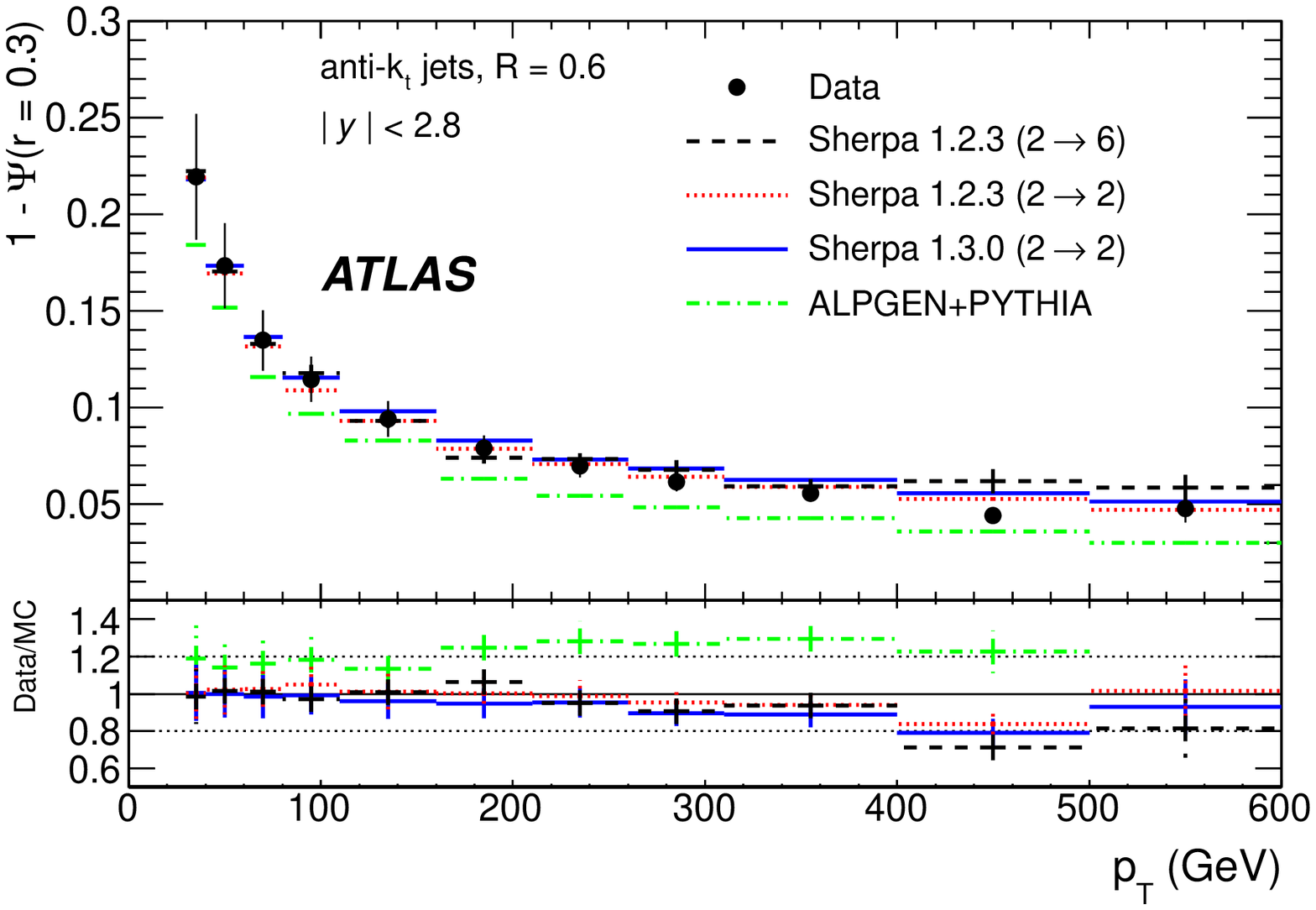}
\includegraphics[width=0.495\textwidth]{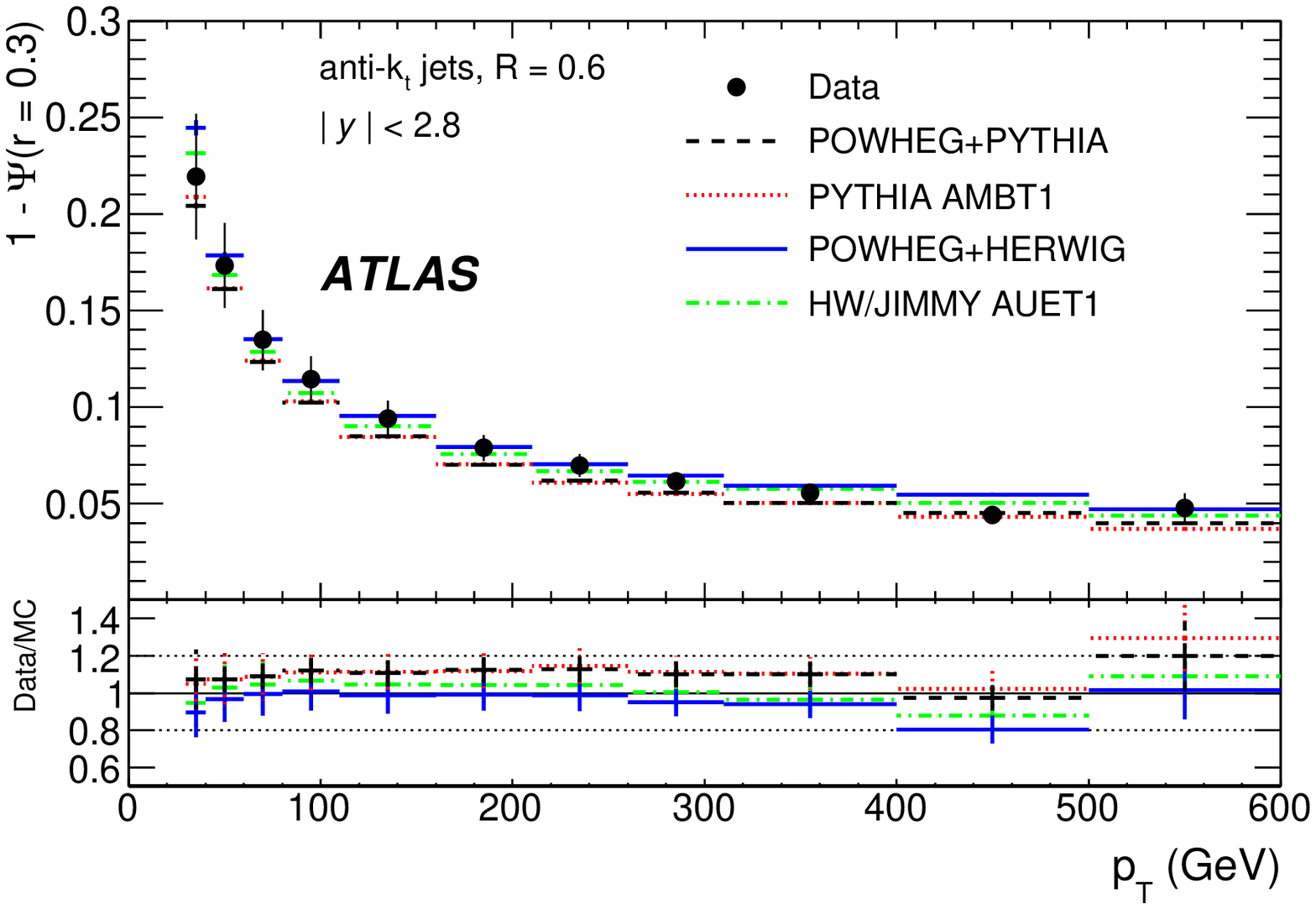}
}
\end{center}
\vspace{-0.7 cm}
\caption{\small
The measured integrated jet shape, $1 - \Psi(r=0.3)$, as a function of $\ptjet$ for jets with $|\rapjet| < 2.8$ and $30 \ {\rm GeV} < \ptjet < 600 \ {\rm GeV}$.
Error bars indicate the statistical and systematic uncertainties added in quadrature. 
The measurements are compared to the different MC predictions considered.}
\label{fig:sum}
\end{figure}

\clearpage


%

\subsection{$\chi^2$ statistical tests}

Finally, a $\chi^2$ test is performed to the data points 
in Figures~\ref{fig6}, \ref{fig:pyt5}, \ref{fig:hrw5}, \ref{fig:meps5}, and \ref{fig:pow5} 
with respect to a given MC prediction, separately in each rapidity region. 
The systematic uncertainties are considered independent and fully correlated across $\ptjet$ bins, 
and the test is carried out according to the formula

\begin{equation}
\chi^2 = \sum_{j=1}^{\ptjet \  \rm{bins}} \frac{[{  d}_j - {  mc}_j(\bar{s})]^2}{[\delta{  d}_j]^2 + [\delta{  mc}_j(\bar{s})]^2
 } 
+ \sum_{i=1}^{5} [s_i]^2 \ , 
\end{equation}
\noindent
where ${  d}_j$ is the measured data point in bin $j$,   ${  mc}_j(\bar{s})$ is 
the corresponding MC prediction, and  $\bar{s}$  
denotes the vector of  standard deviations, $s_i$, for the different independent sources of 
systematic uncertainty.
For each rapidity region considered,  the sums above  run over the total number of data points 
in $\ptjet$ and five independent sources of systematic uncertainty, and the $\chi^2$ is minimized with respect to $\bar{s}$.
Correlations among systematic uncertainties  
are taken into account in ${  mc}_j(\bar{s})$.  

The $\chi^2$ results for the different MC predictions shown in this and in the previous Chapter are collected in Table~5.1.


 \begin{table}
 \begin{footnotesize}
 \begin{center}
 \begin{tabular}{|c||c|c|c|c|c|} \hline
 \multicolumn{6}{|c|}{{\large{$\chi^2/\rm{d.o.f}$}}} \\ \hline\hline\hline
$|\rapjet|$ range  & $0.0 - 0.3$ &  $0.3 - 0.8$ & $0.8 - 1.2$ & $1.2 - 2.1$ & $2.1 - 2.8$ \\ \hline\hline
degrees of freedom (d.o.f)               &   10   &   10    &   10    &  10     & 8    \\ \hline 
PYTHIA-Perugia2010         & 0.6  &  1.8  &  2.4  &  1.4  &  1.4 \\
HERWIG++            & 2.2  &  2.3  &  3.1  &  1.8  &  4.0 \\
PYTHIA-MC09         & 1.0  &  2.5  &  2.4  &  1.5  &  3.2 \\
PYTHIA-DW           & 2.4  &  3.4  &  6.9 &  4.0  &   5.2 \\
ALPGEN              & 3.8  &  9.8  &   7.4 &   6.7 &   6.0  \\
PYTHIA-Perugia2010 (no UE) &  4.2 &   9.7 &   4.9 &   8.6 &   4.8 \\ \hline
PYTHIA6-AMBT1        &  1.6 &   2.9 &   4.4  &  2.4 &   4.4 \\
PYTHIA6-AUET2        &  0.8 &   2.1 &   2.3  &  1.3 &   3.3 \\ 
PYTHIA6-Perugia2011  &  0.5 &   2.2 &   1.7  &  1.3 &   1.6 \\
Pythia~8-4C           &  1.3 &   3.6 &   1.6  &  1.6 &   2.7 \\ \hline
Herwig++ 2.4.2       &  0.9  &  2.7 &   2.9 &   2.4 &   2.9 \\
Herwig++ 2.5.1       &  4.2  & 10.9 &   1.0 &   8.6 &   1.7 \\
HERWIG/JIMMY-AUET2   &  3.6  &  3.6 &   5.3 &   3.4 &   6.4 \\
HERWIG/JIMMY-AUET1   &  3.8  &  4.1 &   4.0 &   2.9 &   2.7 \\ \hline
Sherpa 1.2.3 (up to $2\to 6$) & 1.3 &   1.2  &  0.9 &   1.2  &  1.1 \\ 
Sherpa 1.2.3 ($2 \to 2$)      & 1.8 &   5.3  &  0.7 &   3.0  &  4.0 \\ 
Sherpa 1.3.0 ($2 \to 2$)      & 3.0 &   8.1  &  1.1 &   5.7  &  1.7 \\ 
ALPGEN+PYTHIA                 & 4.9 &   14.9 &  10.6 &  8.6  &  6.7 \\
POWHEG+PYTHIA                 & 1.9 &   3.6  &  4.4  &  1.7 &   1.1 \\
POWHEG+HERWIG                 & 4.7 &   9.8  &  9.4  &  2.2 &   2.4  \\ \hline\hline
\end{tabular}
\label{tab:chi2}
\caption{Results of $\chi^2$ tests to the data in Figures~\ref{fig6}, \ref{fig:pyt5}, \ref{fig:hrw5}, \ref{fig:meps5}, and \ref{fig:pow5} 
with respect to the different MC predictions.}
\end{center}
\end{footnotesize}
\end{table}

\newpage
$\,$
\clearpage
\newpage

\chapter*{Conclusions}
  \addcontentsline{toc}{chapter}{Conclusions}

This Thesis describes the measurements of the inclusive jet cross section and jet shapes in 
$pp$ collisions at $\sqrt{s} = 7$~TeV using the ATLAS detector. The $\akt$ algorithm with $R = 0.6$ 
is used to reconstruct jets. Both measurements are unfolded back to the hadron level and compared to 
theory predictions.
\\
\\
The inclusive jet cross section measurement has been performed using jets with $\ptjet > 20$~GeV 
and $|\rapjet| < 4.4$ in  a data sample corresponding to a total luminosity of $37$~pb${}^{-1}$ collected by ATLAS during 2010. 
The measurement constitute a stringent test of pQCD over ten orders of magnitude, 
covering a jet $\ptjet$ range up to 1.5~TeV.
The measurement expands by a factor of $\sim2$ the kinematic region in $\ptjet$ probed by the Tevatron, 
in a wider $y$ range.  
\\
\\
The data are compared with NLO predictions using several PDFs corrected for non-perturbative effects. 
A general agreement is found, but the predicted cross section tends to be larger than that in data 
at high $\ptjet$ in the forward region. This can be atributted to the current PDF of the gluon at high $x$, 
showing the potential of the measurement to constrain the parton distribution functions when used in global fits.
\\
\\
The measurement of jet shapes in inclusive jet production uses 3~pb${}^{-1}$ collected by ATLAS, 
for jets with $\ptjet > 30$~GeV and $|\rapjet| < 2.8$. It is important to notice that the measurement was 
performed with the very first $pp$ collisions at 7~TeV center-of-mass-energy, when the LHC luminosity 
was low enough to collect a data sample clean of pile-up.
The jet shape measurements can be used to constrain  the phenomenological models for soft gluon radiation, underlying event (UE)
activity, and non-perturbative fragmentation processes in the final state.
This is particularly important for searches of new physics that require a good control of the soft physics. 
The jet shape measurements constitute a first step towards the study of boosted topologies corresponding to the 
decay of heavy particles, that promise to play a central role in the serach for a light Higgs boson or in 
searches for new physics.
\\
\\
The integrated luminosity recorded by ATLAS has already increased above $1~fb^{-1}$ during 2011. The inclusive 
jet cross section is being updated with these new data, and further studies are being done in order to reduce the JES uncertainty. 
The large dataset being recorded will also allow to perform sensitive jet substructure analyses 
searching for new physics.  

\newpage

\appendix\chapter{Sensitivity of the jet shapes to the underlying event and to the pile-up using MC simulated events}
\label{appendix_MC}

Before the first $pp$ collisions were delivered by the LHC, studies on jet shapes were performed using 
PYTHIA QCD-dijet Monte Carlo simulated events in $pp$ collisions at $\sqrt{s} = 10$~TeV. 
The effect of the underlying event and the pileup on the jet shapes was investigated in 
jets reconstructed with the SISCone algorithm with R = 0.7 and splitting-merging fraction equal to 0.75.

\begin{figure}[tbh]
\begin{center}
\includegraphics[width=1.\textwidth]{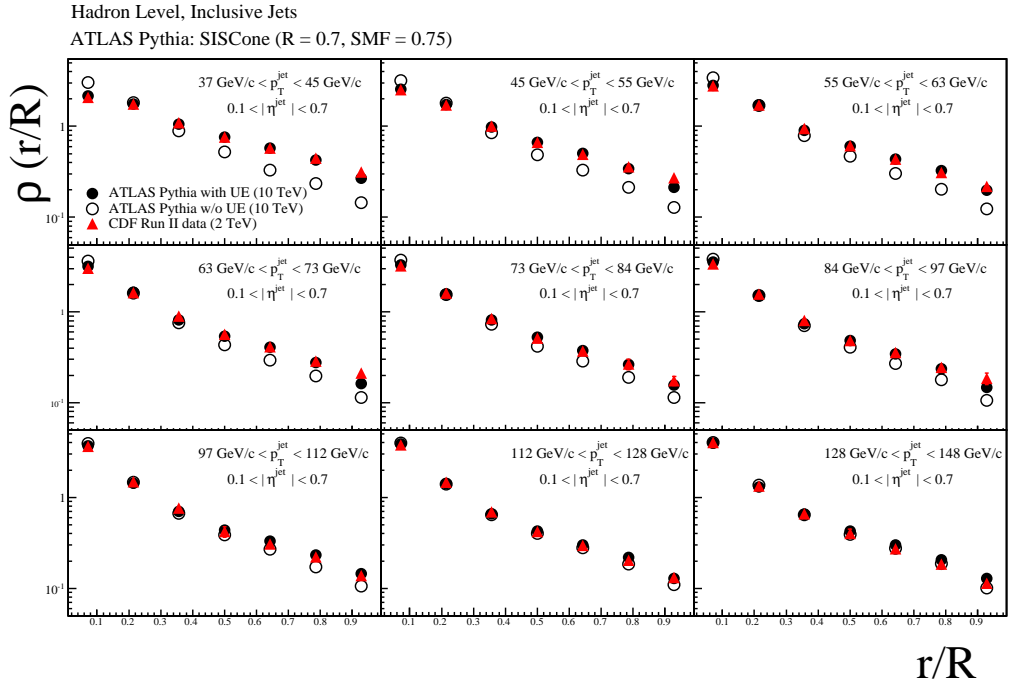}
\end{center}
\vspace{-0.7 cm}
\caption[Differential jet shapes for jets with $37~\mathrm{GeV} < \ptjet < 148~\mathrm{GeV}$ and $0.1 < |\eta| < 0.7$ 
for events with and without UE]
{\small
Differential jet shapes for jets with $37~\mathrm{GeV} < \ptjet < 148~\mathrm{GeV}$ and $0.1 < |\eta| < 0.7$ 
as reconstructed using the SISCone algorithm for events with (full dots) and without (open dots) UE 
contributions. In addition, results from CDF data (full triangles) are shown.
}
\label{fig_int_note1}
\end{figure}

\begin{figure}[tbh]
\begin{center}
\includegraphics[width=1.\textwidth]{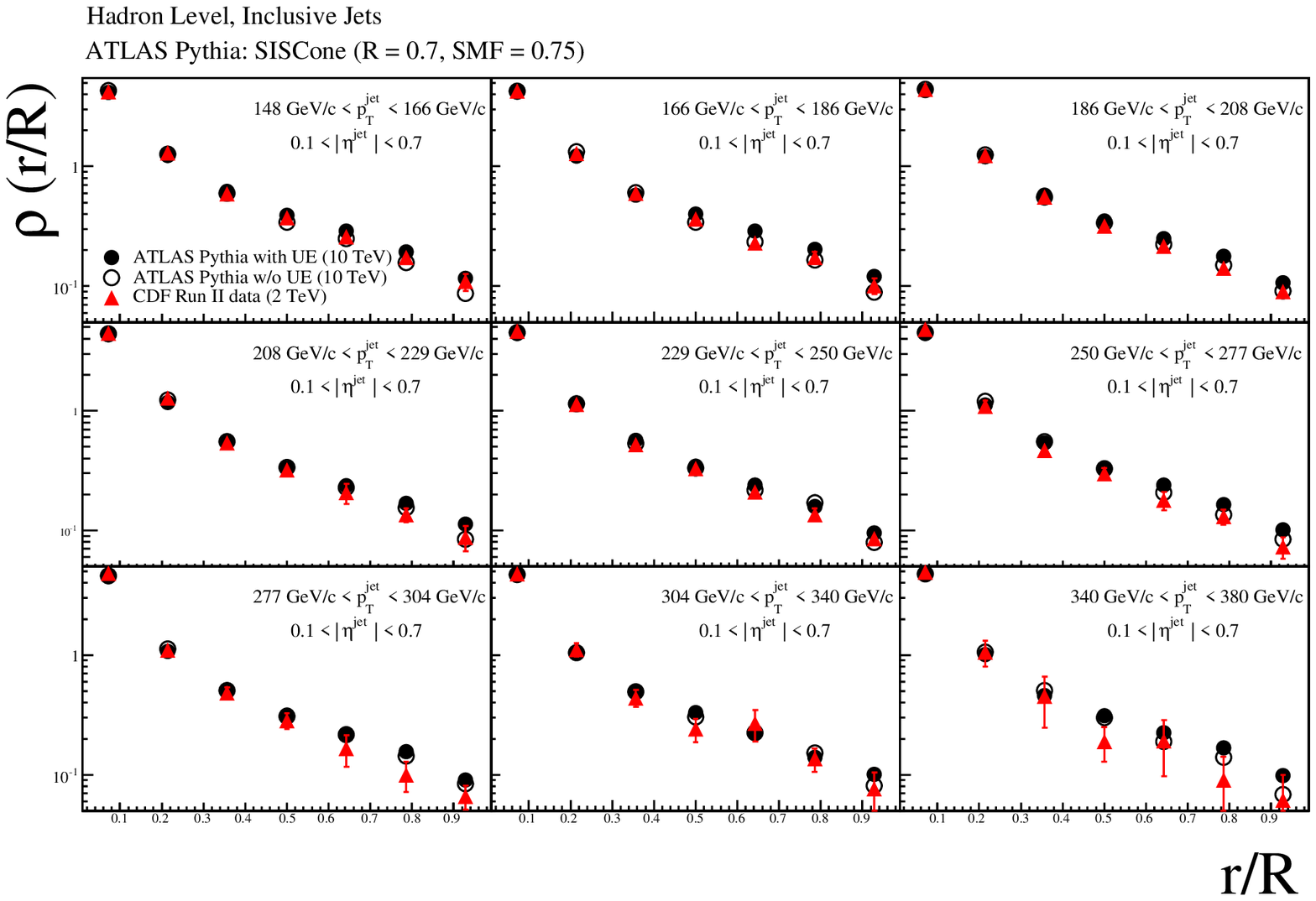}
\end{center}
\vspace{-0.7 cm}
\caption[Differential jet shapes for jets with $148~\mathrm{GeV} < \ptjet < 380~\mathrm{GeV}$ and $0.1 < |\eta| < 0.7$ 
for events with and without UE]
{\small
Differential jet shapes for jets with $148~\mathrm{GeV} < \ptjet < 380~\mathrm{GeV}$ and $0.1 < |\eta| < 0.7$ 
as reconstructed using the SISCone algorithm for events with (full dots) and without (open dots) UE 
contributions. In addition, results from CDF data (full triangles) are shown.
}
\label{fig_int_note2}
\end{figure}

\begin{figure}[tbh]
\begin{center}
\includegraphics[width=1.\textwidth]{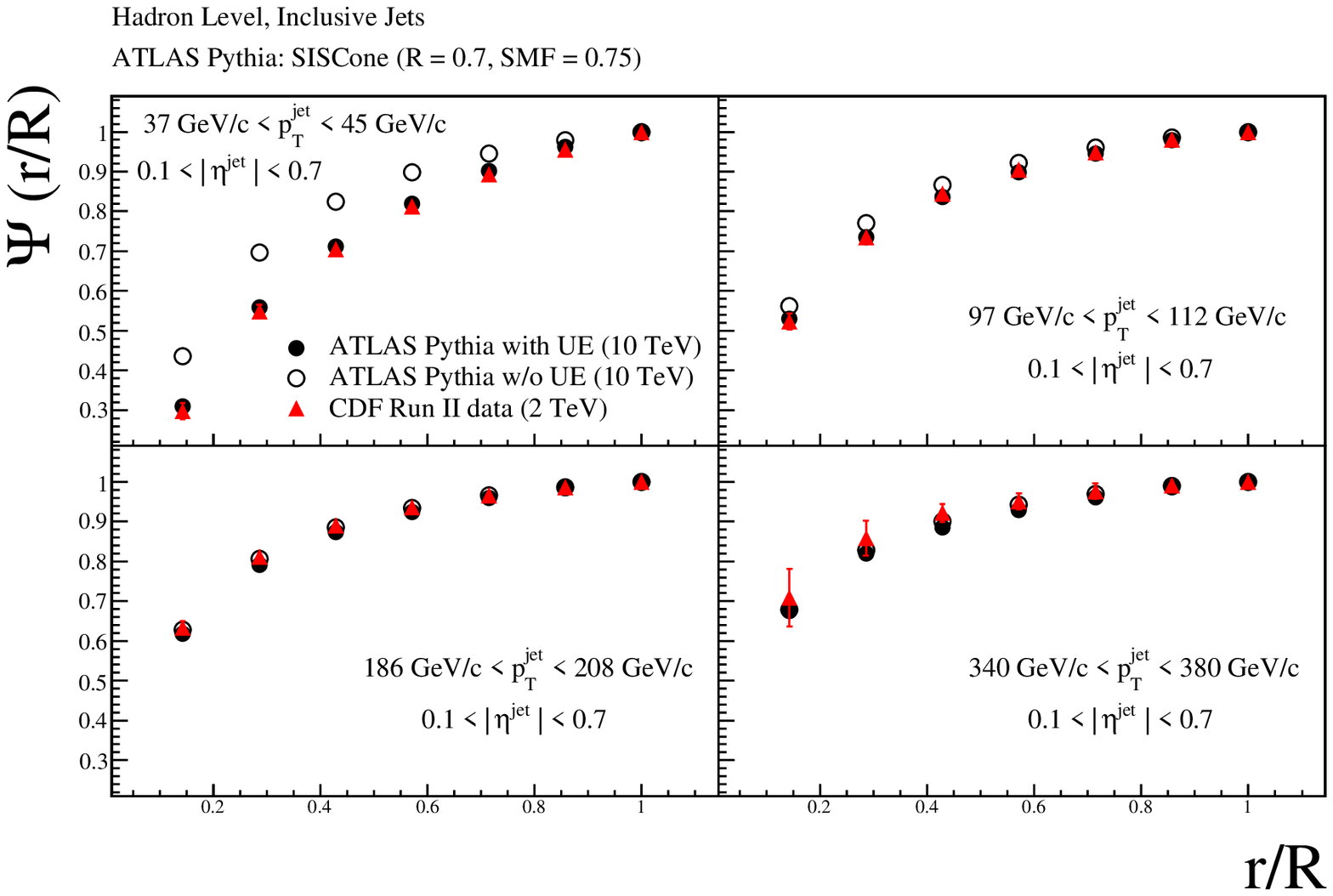}
\end{center}
\vspace{-0.7 cm}
\caption[Integrated jet shapes for jets with $37~\mathrm{GeV} < \ptjet < 380~\mathrm{GeV}$ and $0.1 < |\eta| < 0.7$ 
for events with and without UE]
{\small
Integrated jet shapes for jets with $37~\mathrm{GeV} < \ptjet < 380~\mathrm{GeV}$ and $0.1 < |\eta| < 0.7$
as reconstructed using the SISCone algorithm for events with (full dots) and without (open dots) UE
contributions. In addition, results from CDF data (full triangles) are shown.
}
\label{fig_int_note3}
\end{figure}

Figures~\ref{fig_int_note1}, \ref{fig_int_note2} and~\ref{fig_int_note3} show the comparison of jet shapes 
at hadron level in events with and without underlying event. 
As expected, the presence of the underlying event translates into broader jets at low $\ptjet$. 
The jet shapes results obtained in ATLAS Monte Carlo simulated events are compared with published results from the CDF experiment 
using a midpoint algorithm, corrected back to the hadron level. Given the difference in the 
quark/gluon-jet mixture, the different relative importance of the underlying event contributions in 
the Tevatron and the LHC energies and the differences in the jet reconstruction algorithms employed, 
one should not expect an a priori agreement between experiments 
and, therefore, no conclusions are made. However, it is remarkable to observe that SISCone simulated
jet shapes in ATLAS are so similar to those published by CDF. 

\begin{figure}[tbh]
\begin{center}
\includegraphics[width=1.\textwidth]{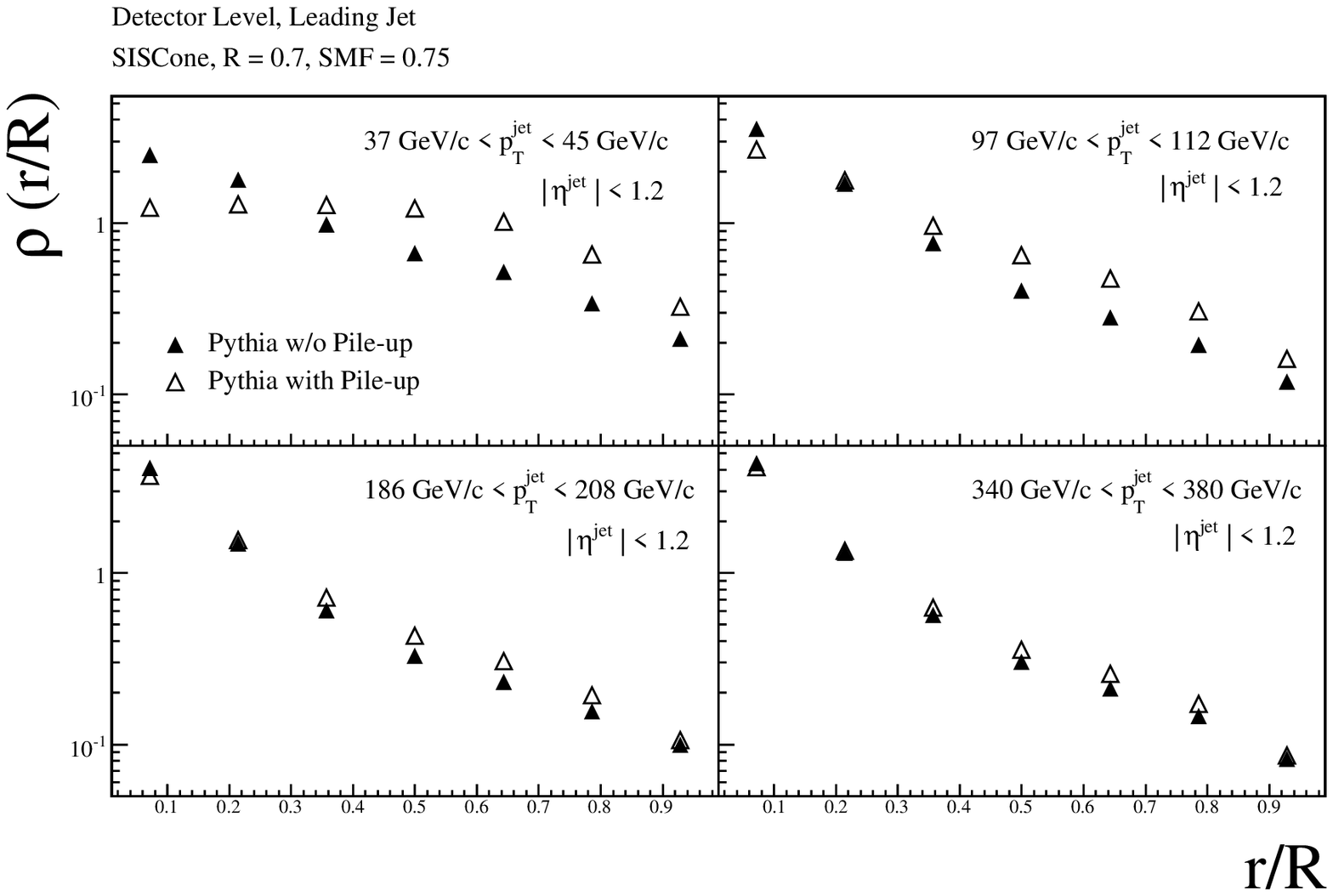}
\end{center}
\vspace{-0.7 cm}
\caption{\small
Differential jet shapes for the leading jet with $37~\mathrm{GeV} < \ptjet < 380~\mathrm{GeV}$ and $|\eta| < 1.2$ 
events with (open triangles) and without (full triangles) pile-up.
}
\label{fig_int_note4}
\end{figure}

\begin{figure}[tbh]
\begin{center}
\includegraphics[width=1.\textwidth]{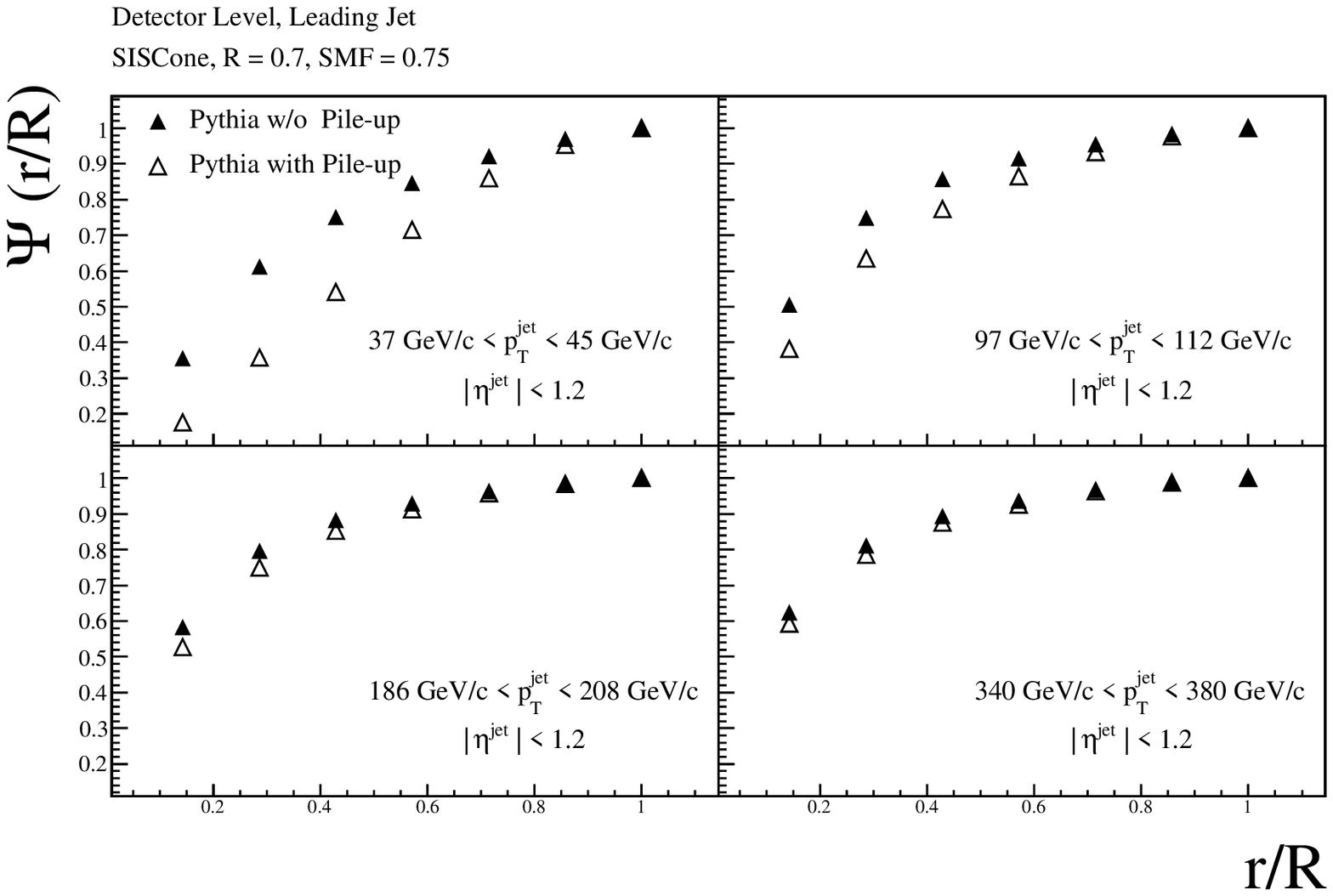}
\end{center}
\vspace{-0.7 cm}
\caption{\small
Integrated jet shapes for the leading jet with $37~\mathrm{GeV} < \ptjet < 380~\mathrm{GeV}$ and $|\eta| < 1.2$       
events with (open triangles) and without (full triangles) pile-up.
}
\label{fig_int_note5}
\end{figure}

\begin{figure}[tbh]
\begin{center}
\includegraphics[width=1.\textwidth]{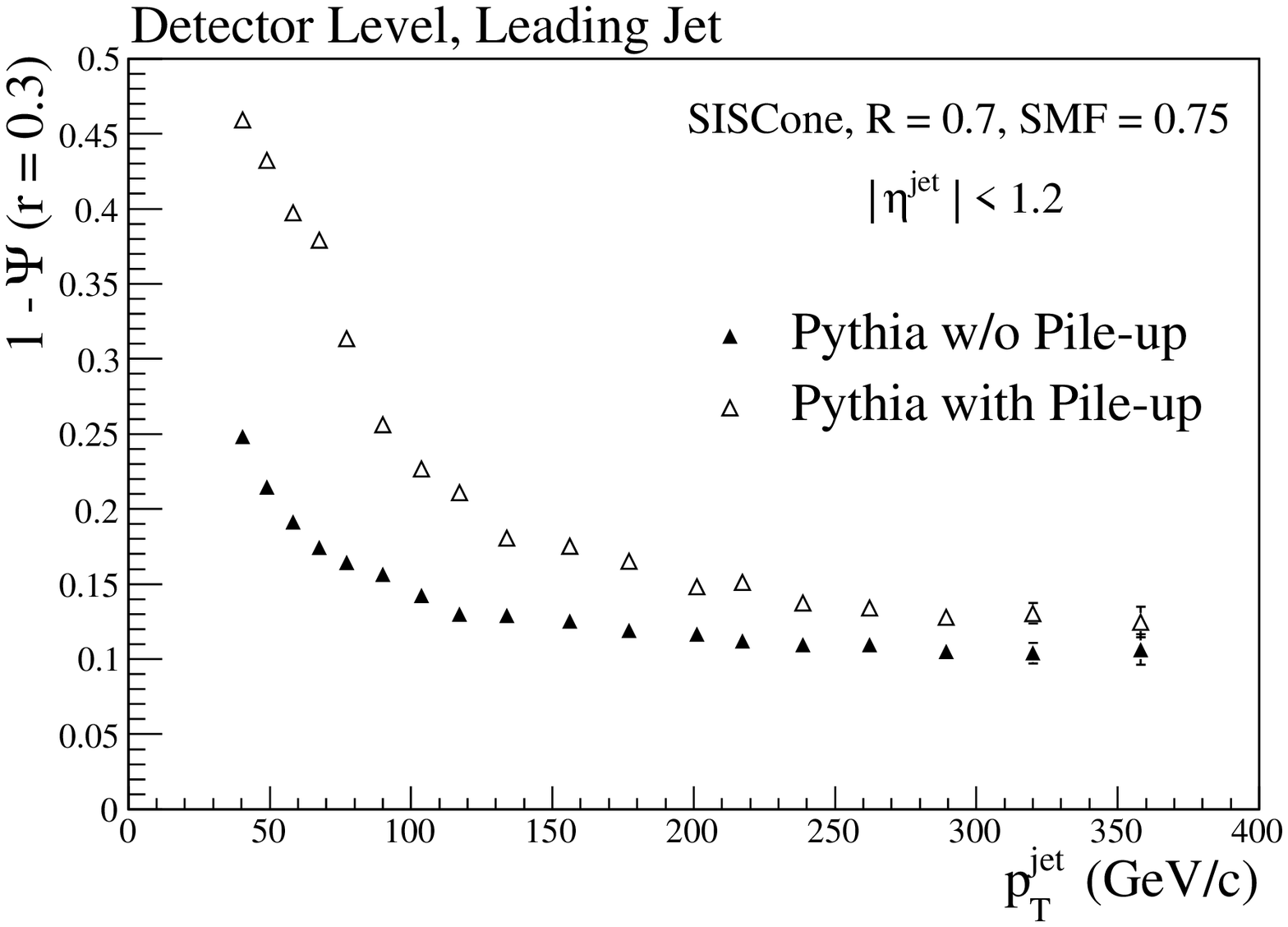}
\end{center}
\vspace{-0.7 cm}
\caption{\small
$1 - \Psi(0.3)$ for the leading jet with $37~\mathrm{GeV} < \ptjet < 380~\mathrm{GeV}$ and $|\eta| < 1.2$
events with (open triangles) and without (full triangles) pile-up.
}
\label{fig_int_note6}
\end{figure}

The sensitivity of the jet shapes to pile-up contributions is studied using a sample of QCD-dijet events at 
$\sqrt{s} = 10$~TeV, overlaid with minimum bias interactions corresponding to an instantaneous luminosity of 
$2 \times 10^{33}$~cm${}^{-2}$s${}^{-1}$, for which on average 4 additional interactions are expected. The contribution from 
cavern background events, arising from neutrons that interact before thermalisation, is also included. 
Figures~\ref{fig_int_note4}, \ref{fig_int_note5} and~\ref{fig_int_note6} show the comparison of jet shapes
at detector level in events with and without pile-up. 
Jets in events with pile-up are significantly broader, mainly at low $\ptjet$.

\newpage
$\,$
\clearpage
\newpage

\chapter{Jet shapes and energy flow in $pp$ collisions at $\sqrt{s} = 900$~GeV}
\label{appendix_900}

In this appendix, the measurement of jet shapes and energy flow at detector level 
in $pp$ collisions at $\sqrt{s} = 900$~GeV is presented. 
The data are compared to different Monte Carlo predictions including full ATLAS detector simulation.

\section{Event Selection and Monte Carlo Simulation}

The analysis is based on a minimum bias (MB) data sample
collected by the ATLAS experiment in 2009. Data was selected from luminosity blocks where
the  inner detector (ID) was in fully operational state, the data quality (DQ) flags for the calorimeter were good, and only
events in filled and paired bunch counter BCIDs were accepted.
The Minimum Bias trigger was employed. 
Events are then required to have a reconstructed primary vertex
with $z$-position within 10~cm of
the nominal interaction point and with at least three tracks pointing to it.
Jets are reconstructed from the energy deposits in the calorimeter
using the $\akt$ algorithm with $D=0.6$ and topotowers as input.

The events are required to have at least a jet with uncorrected 
transverse momentum $\ptjet$ above 7~GeV and rapidity in the range $|\rapjet|<2.6$. In addition, jets
affected by noise in the calorimeter are rejected.
The measurements are compared to Monte Carlo MB simulated events as generated using 
{\sc pythia} and {\sc phojet}~\cite{phojet} event generators,
interfaced with a full simulation of the ATLAS detector response to particles based on
{\sc geant4}. In the case of {\sc pythia}, different Monte Carlo samples with
slightly different tunes for the
parton shower and underlying event modeling in the final state have been considered.
Samples are generated with ATLAS MC09, DW and Perugia0 tunes.
The simulated events are passed through the same trigger selection and analysis chain as in the data.

\begin{figure}[h!]
\begin{center}
\includegraphics[width=0.75\linewidth]{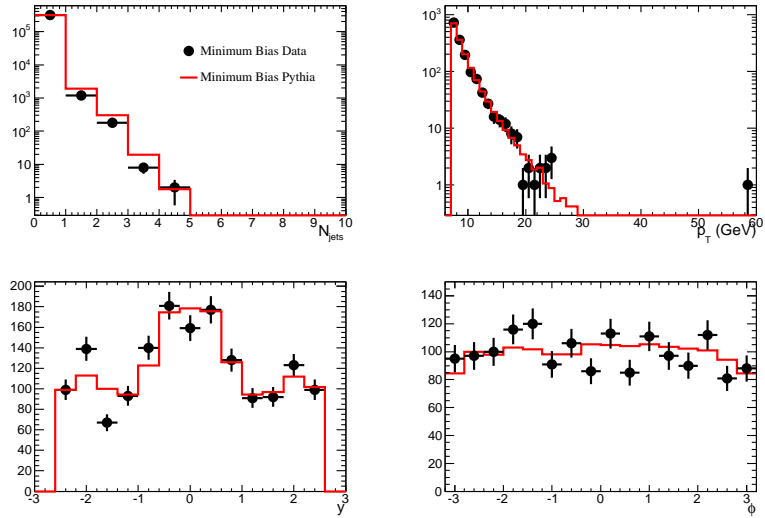}
\caption{
Measured basic jet kinematic distributions compared to Monte Carlo predictions ({\sc pythia}
tune ATLAS MC09),
normalized to the number of jets observed in data.
}
\label{fig:jetkin}
\end{center}
\end{figure}

Figure~\ref{fig:jetkin} shows the measured jet multiplicity, $\ptjet$, $\rapjet$ and $\phijet$
distributions compared to predictions from MB Monte Carlo events ({\sc pythia}
tune {\sc ATLAS MC09}). The measurements are reasonably well described by the simulation.

\subsection{Jet Shapes using Calorimeter Towers}

\begin{figure}[h!]
\begin{center}
\includegraphics[width=0.48\linewidth]{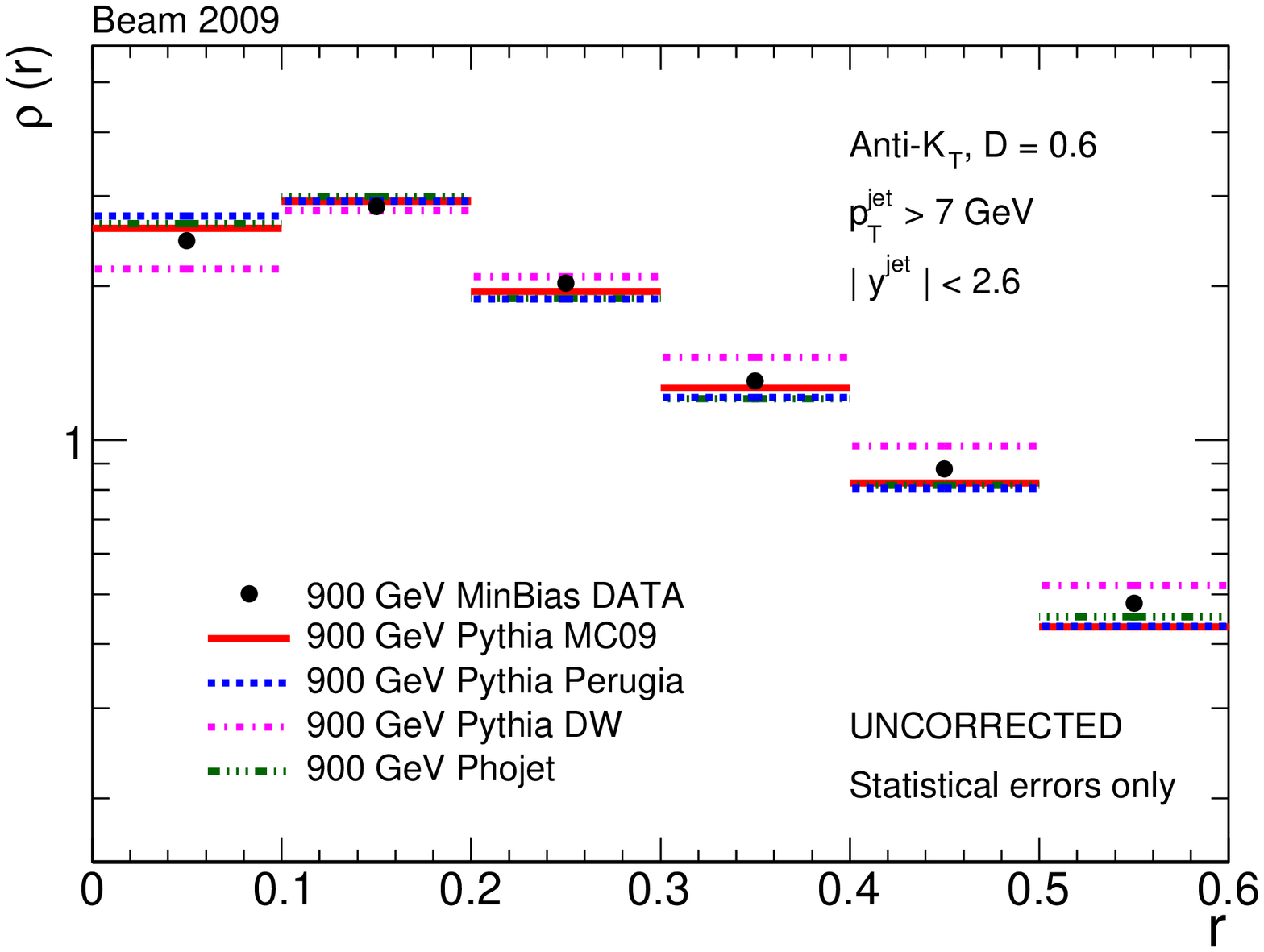}
\includegraphics[width=0.48\linewidth]{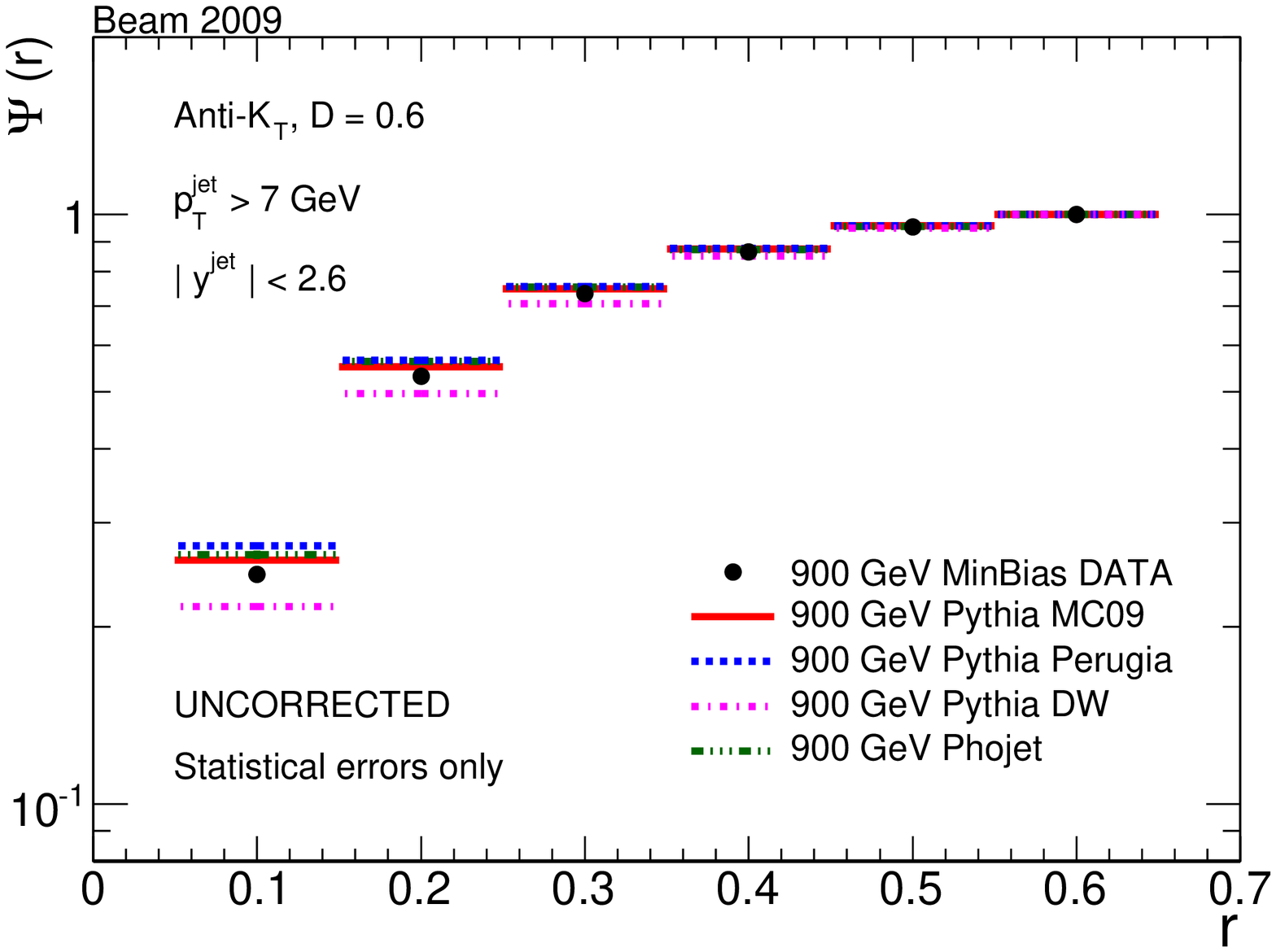}\\
\caption{Measured differential and integrated jet shapes using calorimeter towers
for jets with $\ptjet >7$~GeV and $|\rapjet|<2.6$. The
data are compared to various Monte Carlo simulations.}
\label{fig:all_incl}
\end{center}
\end{figure}

\begin{figure}[h!]
\begin{center}
\includegraphics[width=0.77\linewidth]{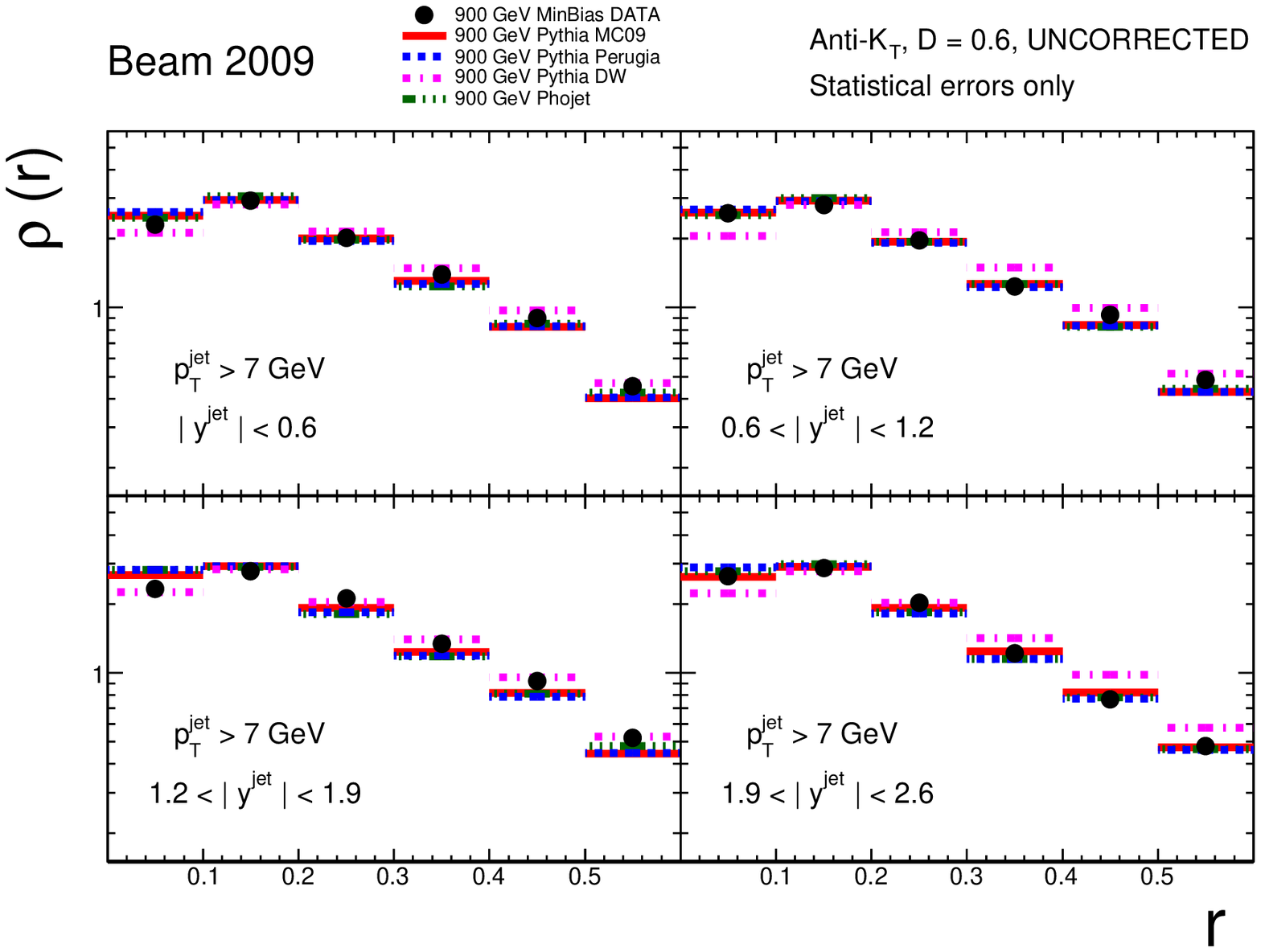}
\caption{Measured differential jet shapes using calorimeter towers
for jets with $\ptjet >7$~GeV as a function of $|\rapjet|$. The
data are compared to various Monte Carlo simulations.}
\label{fig:jet_dif_towers}
\end{center}
\end{figure}

\begin{figure}[h!]
\begin{center}
\includegraphics[width=0.77\linewidth]{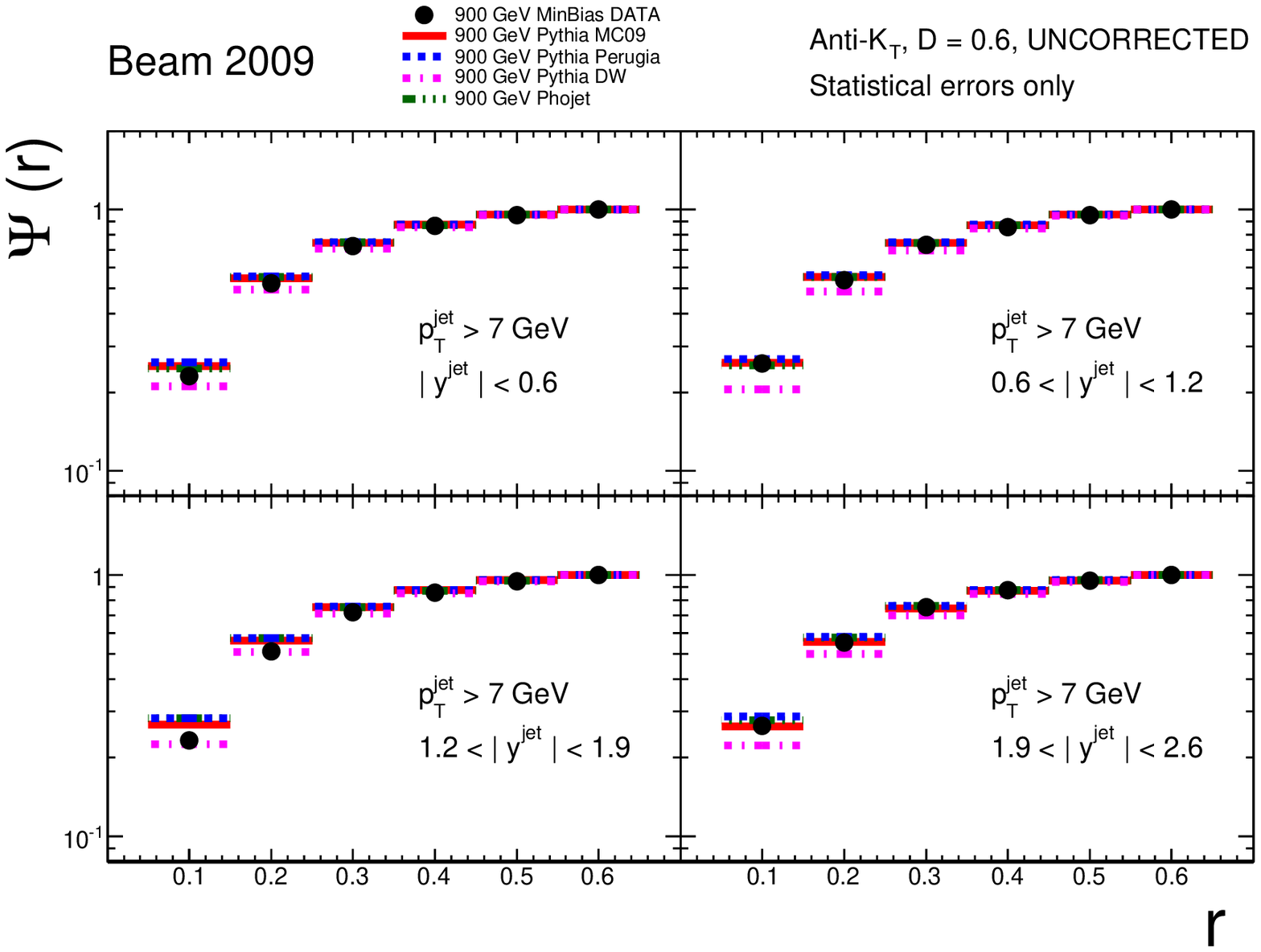}
\caption{Measured integrated jet shapes using calorimeter towers
for jets with $\ptjet >7$~GeV as a function of $|\rapjet|$. The
data are compared to various Monte Carlo simulations.}
\label{fig:jet_int_towers}
\end{center}
\end{figure}

\begin{figure}[h!]
\begin{center}
\includegraphics[width=0.77\linewidth]{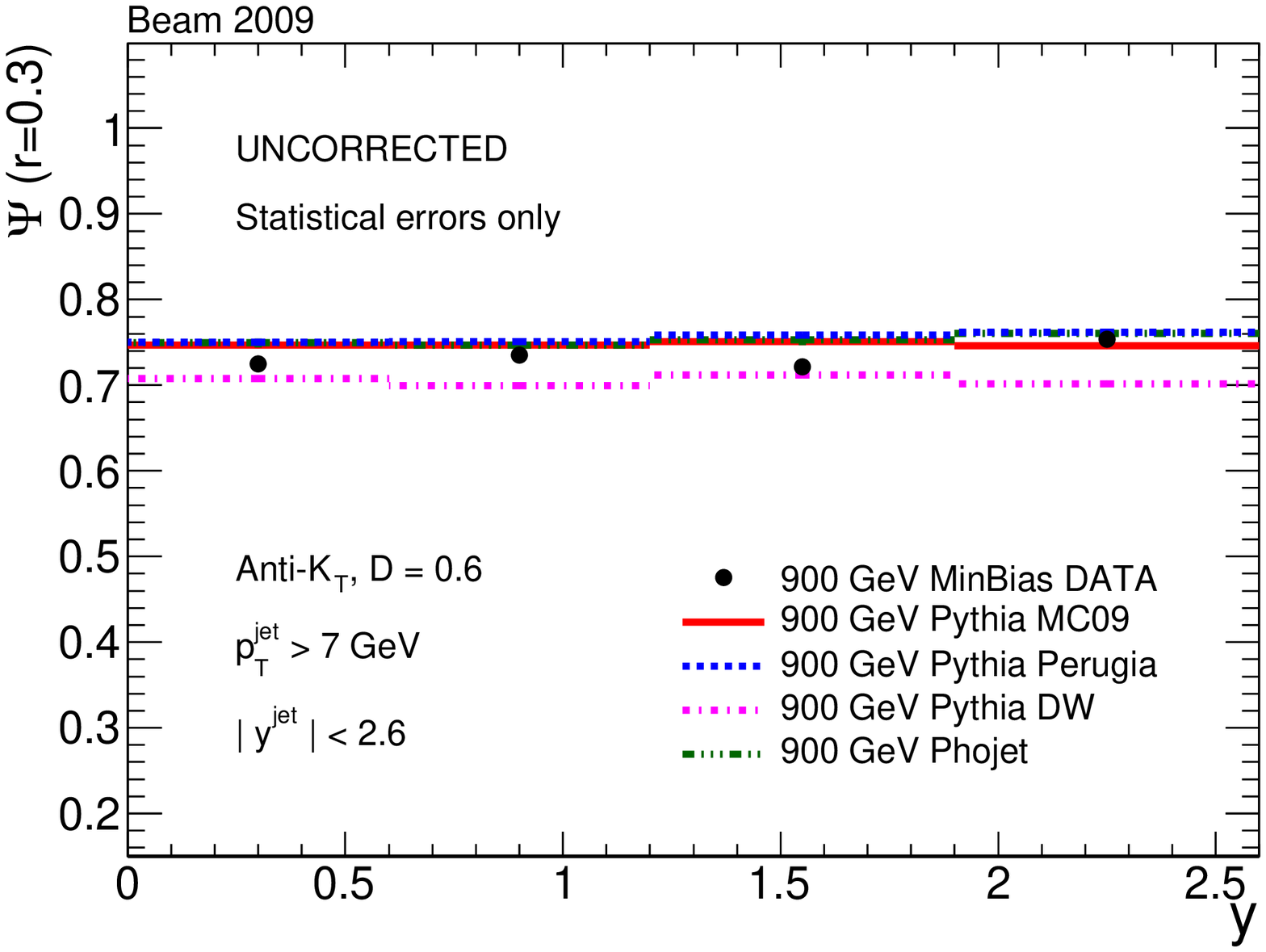}
\caption{Measured integrated jet shapes $\Psi(r=0.3)$
using calorimeter towers for jets with $\ptjet >7$~GeV as a function of $|\rapjet|$. The
data are compared to various Monte Carlo simulations.}
\label{fig:psi_03}
\end{center}
\end{figure}

Figure~\ref{fig:all_incl} shows the measured differential and integrated
jets shapes, as determined using calorimeter towers, for jets with
$\ptjet >7$~GeV and $|\rapjet|<2.6$. Similarly, Figs.~\ref{fig:jet_dif_towers} to~\ref{fig:jet_int_towers}
presents the measurements in separate bins of jet rapidity.
The profiles  present the expected shape
with a prominent peak at low $r$ which indicates that the
majority of the jet momentum is concentrated in the core of the jet. The measured jet shapes
do not present a significant rapidity dependence. This is clearly observed in Fig.~\ref{fig:psi_03}
where the measured integrated jet shape $\Psi(r=0.3)$ is presented as a function of $|\rapjet|$.
The data are compared with
Monte Carlo simulations from the different MB models. The predictions from {\sc pythia} MC09,
{\sc pythia} Perugia0 and {\sc phojet} are similar and
close to the measurements, although predict jets slightly narrower than the data. {\sc pythia} DW
produces jets slightly broader than the data.

\subsection{Jet Shapes using Tracks}

The tracking system provides an alternative approach to measure the internal structure of the jets based
on charged particles. Tracks are selected according to the following criteria:


\begin{itemize}
\item $p_T > 500$~MeV and $|\eta |  < 2.5$,
\item number of associated Pixel hits $\geq$ 1
\item number of associates SCT hits $\geq$ 6,
\item $|d0_{PV} | < 1.5$~mm,
\item $|z0_{PV} sin(\theta_{PV} )|  < 1.5$~mm,
\end{itemize}

\noindent
where $d0_{PV}$ and $z0_{PV}$
are the impact parameter and $z$
position at the perigee measured with respect to the primary vertex.

\begin{figure}[h!]
\begin{center}
\includegraphics[width=0.5\linewidth]{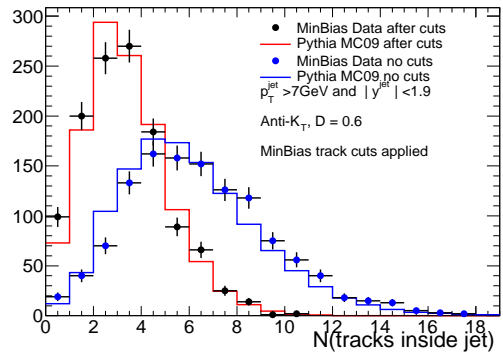}
\caption{Measured total number of tracks inside the jet before and after final track quality cuts,
for jets with $\ptjet >7$~GeV and $|\rapjet|<1.9$. The
data are compared to various Monte Carlo simulations.}
\label{fig:ntrack}
\end{center}
\end{figure}

\begin{figure}[h!]
\begin{center}
\includegraphics[width=0.48\linewidth]{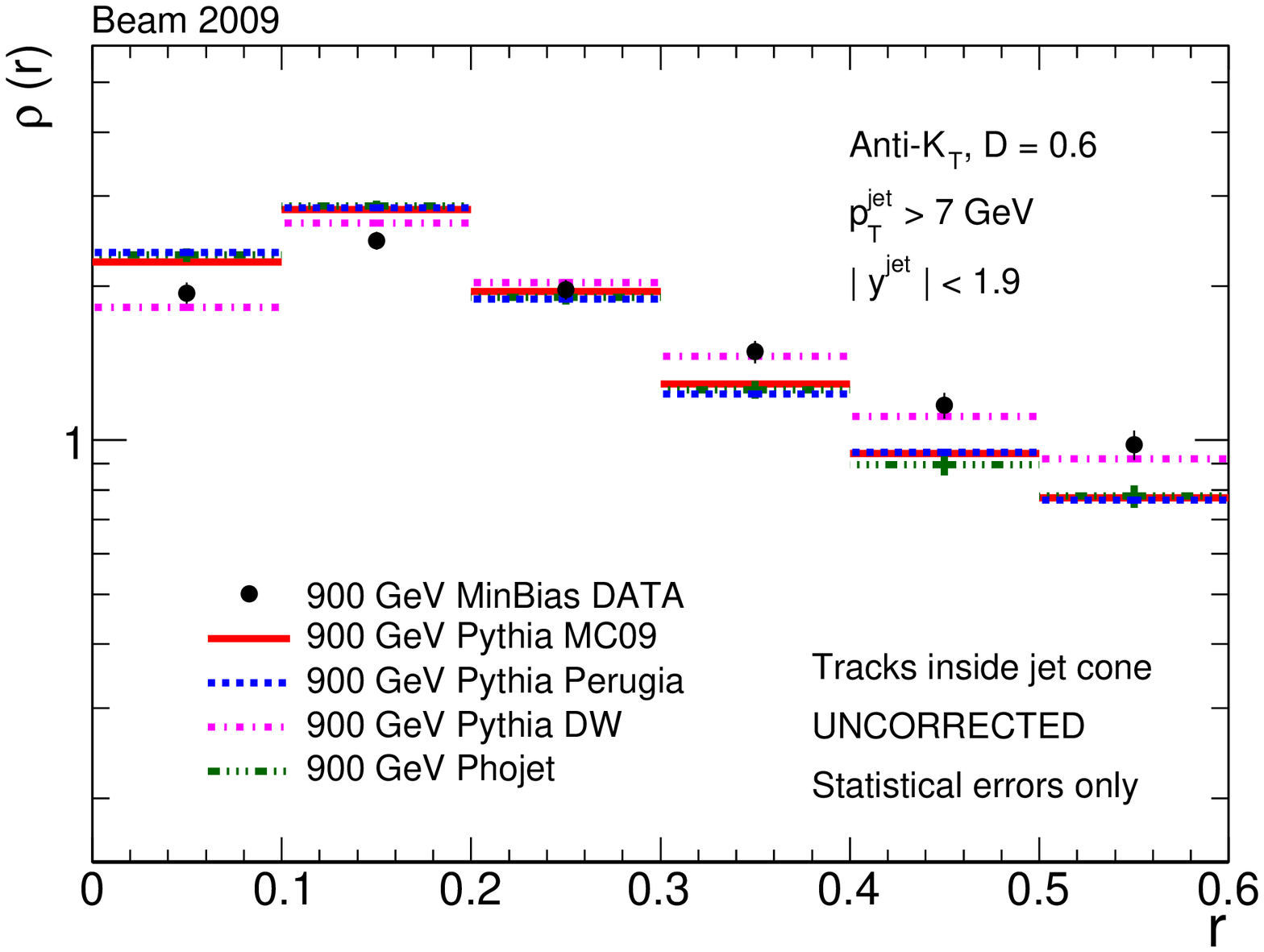}
\caption{Measured differential jet shapes using tracks
for jets with $\ptjet >7$~GeV and $|\rapjet|<1.9$. The
data are compared to various Monte Carlo simulations.}
\label{fig:all_incl_tracks}
\end{center}
\end{figure}

\begin{figure}[h!]
\begin{center}
\includegraphics[width=0.77\linewidth]{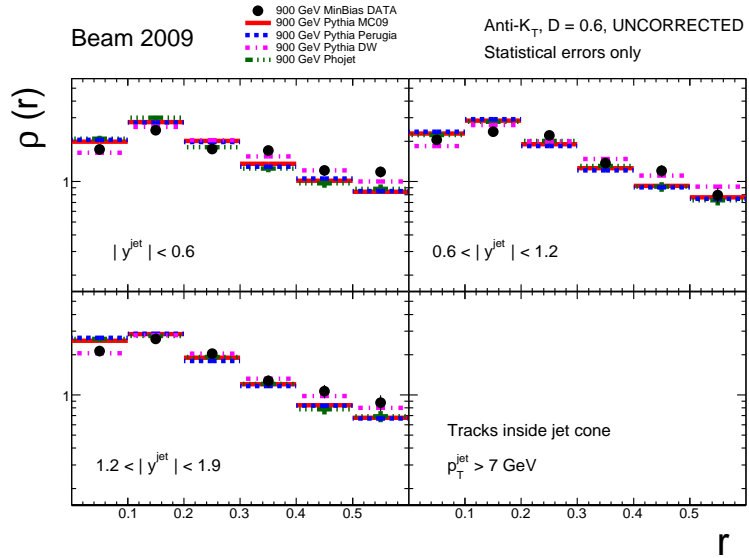}
\caption{Measured differential jet shapes using tracks
for jets with $\ptjet >7$~GeV as a function of $|\rapjet|$. The
data are compared to various Monte Carlo simulations.}
\label{fig:jet_dif_tracks}
\end{center}
\end{figure}

Figure~\ref{fig:ntrack} shows the
track multiplicity inside the jets compared to Monte Carlo predictions before and after applying final
track quality cuts. In both cases, the simulation provides a reasonable description of the data.
Figure~\ref{fig:all_incl_tracks}
shows the measured differential jet shapes and jet profiles, as determined
using tracks, for jets with $\ptjet >7$~GeV and $|\rapjet|<1.9$. The measurements are presented in
different rapidity bins in Fig.~\ref{fig:jet_dif_tracks}.
The conclusions are similar to those on the calorimeter quantities. The simulation provides a reasonable
description of the data although tends to produce jets slightly narrower than the data.

\clearpage

\section{Energy Flow}

The study of the particle flow away from the jet direction provides additional
information on the soft radiation and the underlying
event contribution in the final state, as well as testing the accuracy
of the Monte Carlo simulation of the detector response
to low energy particles.

\subsection{Energy flow in the azimuthal direction}

\begin{figure}[h!]
\begin{center}
\includegraphics[width=0.5\linewidth]{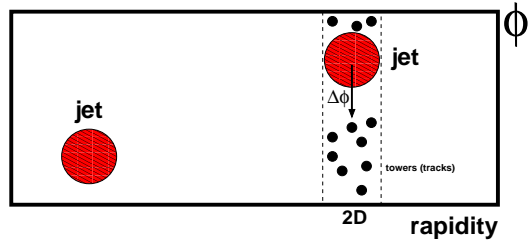}
\caption{Sketch of the particle flow as a function of the distance in azimuth to the jet axis.
}
\label{fig:eflows_phi}
\end{center}
\end{figure}

In this analysis, the hadronic activity outside the jet cone is studied  using either
calorimeter towers or tracks, and expressed in terms
of the average transverse momentum observed as a function of the distance to the jet axis in the azimuthal direction:

\begin{equation}
<\frac{  d p_{\rm T}}{|d \phi| d y }>_{\rm jets} = 
\frac{1}{ 2 D |\Delta \phi| }\frac{1}{\rm N_{jet}}\sum_{\rm jets} {  p_{\rm T}(|\phi-\Delta \phi/2|,|\phi+\Delta \phi/2|)},~0 \le \phi \le 
\pi  \ \ ,
\label{eq:eflows}
\end{equation}

\noindent
where  $p_{T}(|\phi-\Delta \phi/2|,|\phi+\Delta \phi/2|)$ is the scalar sum of the transverse momentum of the
constituents (calorimeter or tracks) at a given distance $\phi$ to the jet. Bins of $\Delta \phi = 0.2$  have
been used. 
As illustrated in Fig~\ref{fig:eflows_phi}, only constituents 
within the rapidity range occupied by the jet cone are considered, with the aim to limit the
effect of a different calorimeter response as a function of rapidity.

\begin{figure}[h!]
\begin{center}
\includegraphics[width=0.65\linewidth]{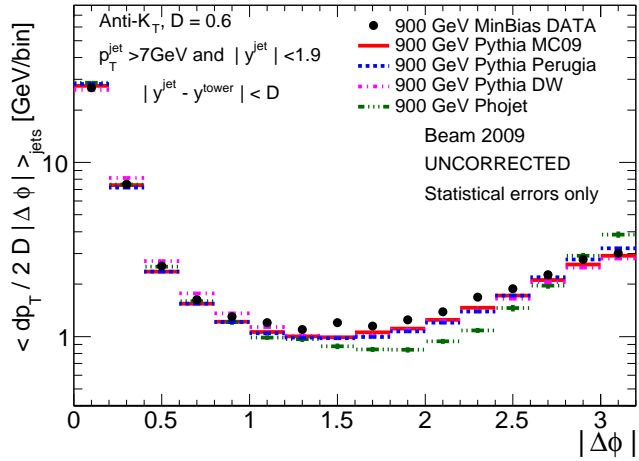}
\caption{Measured energy flow using calorimeter towers
as a function of $|\Delta \phi|$ with respect to the jet direction.
The measurements are compared to minimum bias Monte Carlo simulations.}
\label{fig:eflows_towers_tot}
\end{center}
\end{figure}

\begin{figure}[h!]
\begin{center}
\includegraphics[width=0.9\linewidth]{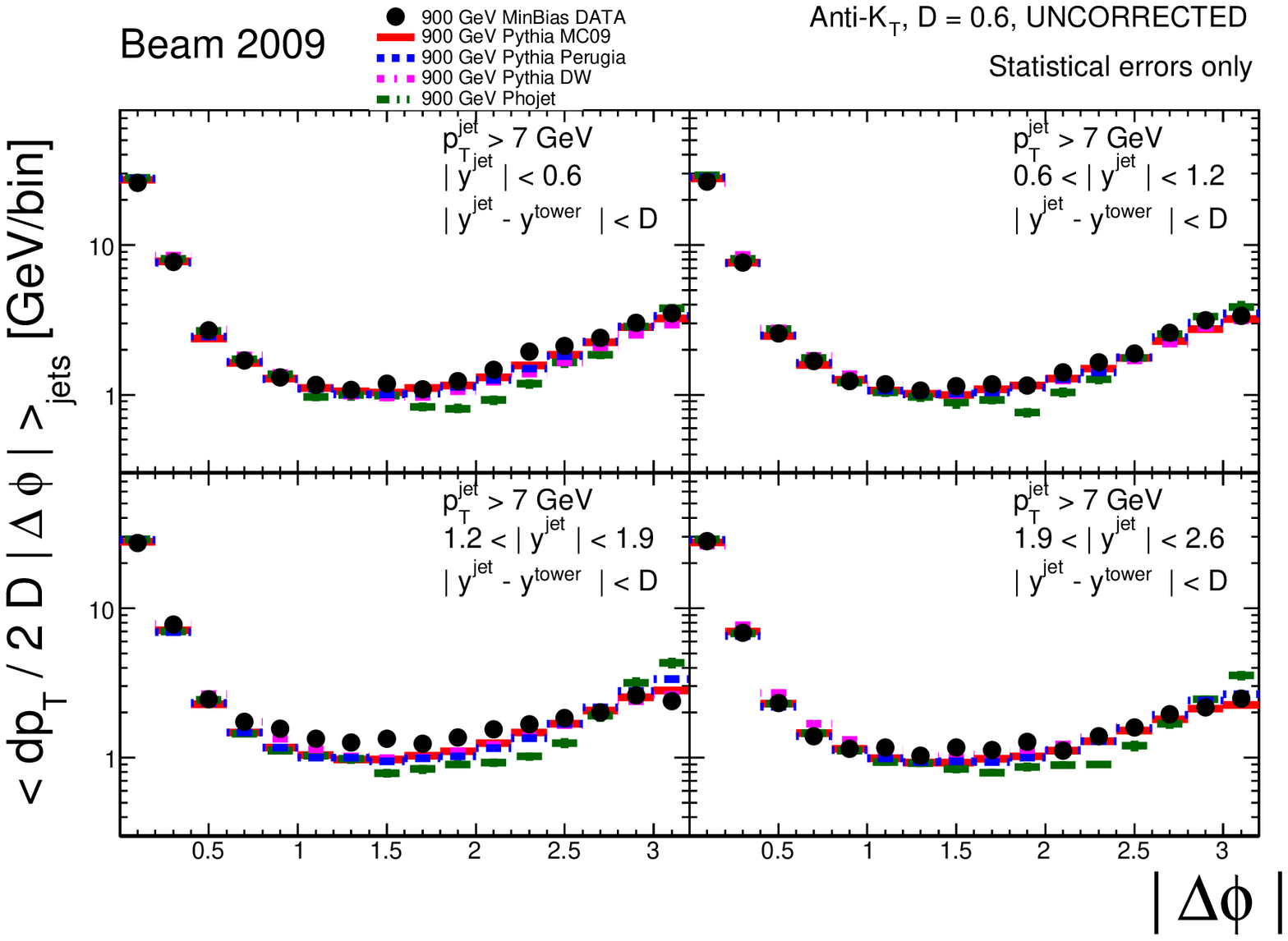}
\caption{Measured energy flow using calorimeter towers
as a function of $|\Delta \phi|$ with respect to the jet direction, in different jet rapidity regions.
The measurements are compared to minimum bias Monte Carlo simulations.}
\label{fig:eflows_towers}
\end{center}
\end{figure}

\begin{figure}[h!]
\begin{center}
\includegraphics[width=0.65\linewidth]{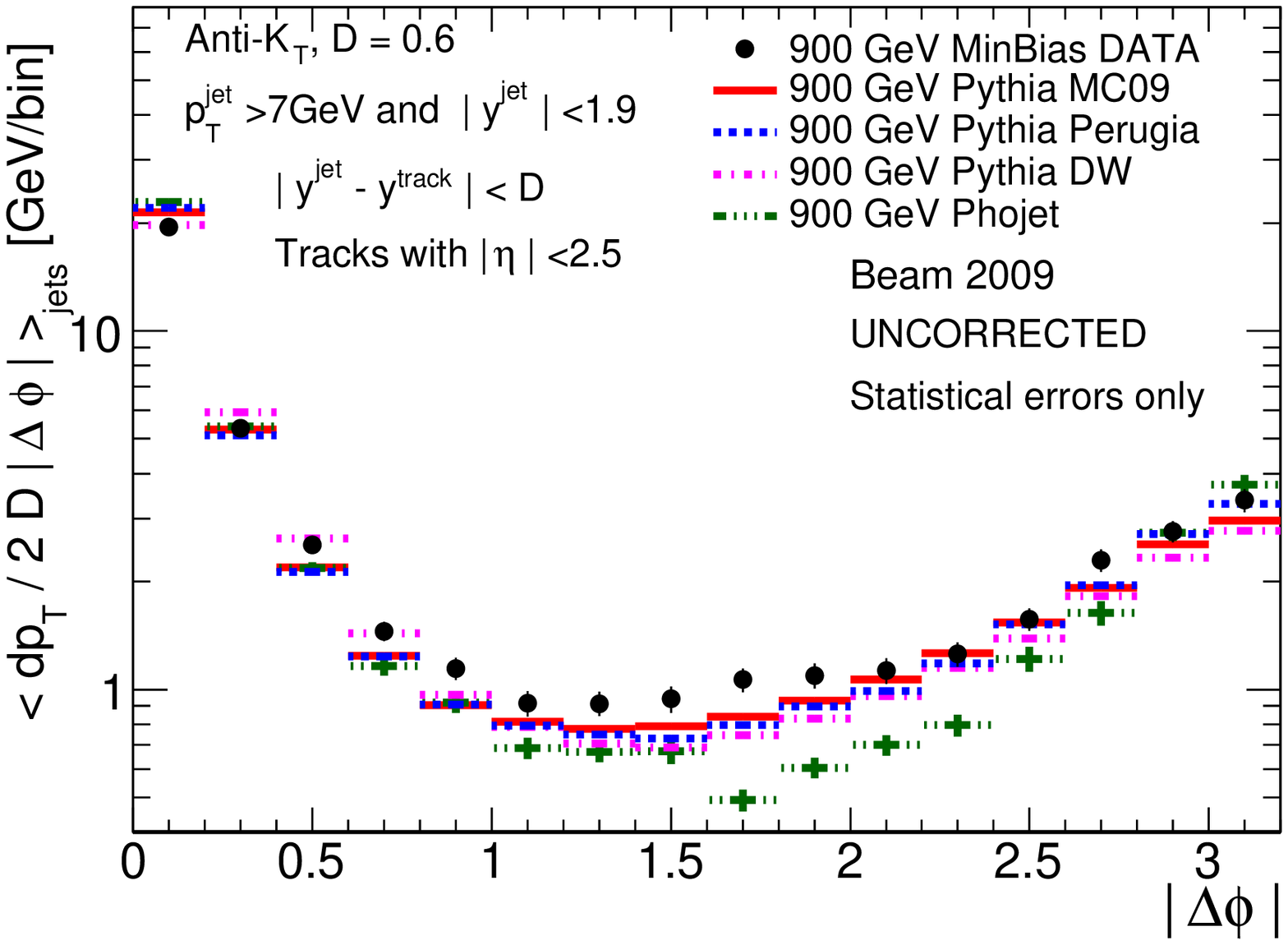}
\caption{Measured energy flow using tracks
as a function of $|\Delta \phi|$ with respect to the jet direction.
The measurements are compared to minimum bias Monte Carlo simulations.}
\label{fig:eflows_tracks_tot}
\end{center}
\end{figure}

\begin{figure}[h!]
\begin{center}
\includegraphics[width=0.9\linewidth]{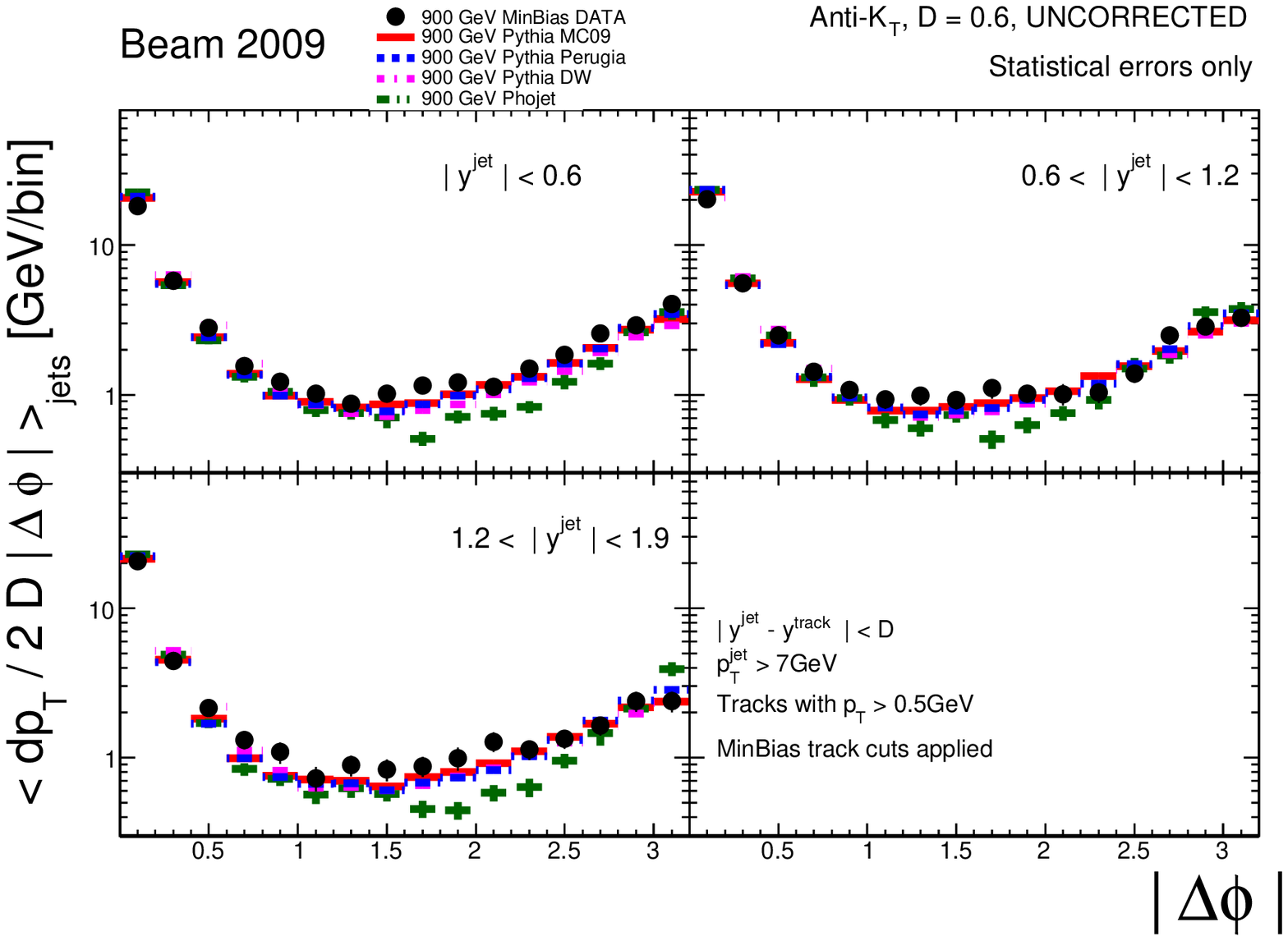}
\caption{Measured energy flow using tracks
as a function of $|\Delta \phi|$ with respect to the jet direction, in different jet rapidity regions.
The measurements are compared to minimum bias Monte Carlo simulations.}
\label{fig:eflows_tracks}
\end{center}
\end{figure}

\noindent
Figures~\ref{fig:eflows_towers_tot} and~\ref{fig:eflows_towers}  show the measured energy flow
using calorimeter towers for events
with at least one jet with $\ptjet>$7~GeV, in different jet rapidity regions up to $|\rapjet|=2.6$.
The measurements are compared to the predictions from different MB simulated samples. The Monte
Carlo simulation provides a reasonable description of  the core of the jet ($|\Delta \phi| < 0.6$)
but tends to slightly underestimate the region between jets with $|\Delta \phi | \sim \phi/2$,
dominated by soft hadronic activity. As $\Delta \phi$ increases, the
measured energy flow increases due to the presence of a second jet.
Similarly, Figs.~\ref{fig:eflows_tracks_tot} and~\ref{fig:eflows_tracks}  show the measured flow using tracks, in
events with jets with $\ptjet>$7~GeV and $|\rapjet|<1.9$, compared to various Monte Carlo predictions. The different
{\sc pythia} simulations
slightly overestimate the particle flow at the core of the jet while slightly underestimate
the activity in the region orthogonal to the jet direction. The latter effect is particularly significant
in the case of  {\sc phojet} samples in both track- and calorimeter-based measurements.


\subsubsection{Energy Flow in dijet events}

\begin{figure}[h!]
\begin{center}
\includegraphics[width=0.95\linewidth]{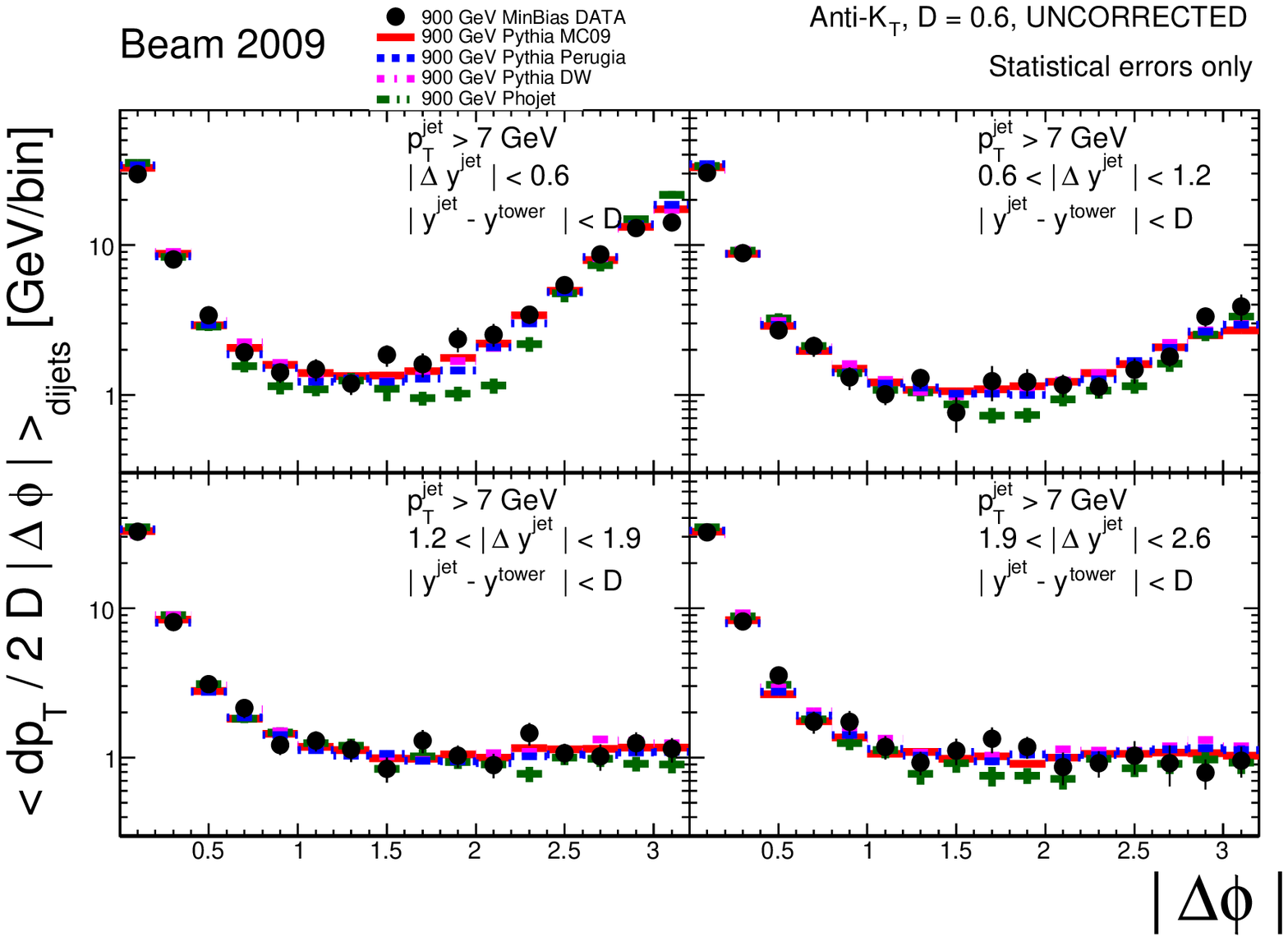}
\caption{Measured energy flow using calorimeter topotowers in dijet events and jets with
$\ptjet >7$~GeV and $|\rapjet |<2.6$,
as a function of $|\Delta \phi|$ with respect to the jet direction and the
rapidity separation between the two jets.
The measurements are compared to minimum bias Monte Carlo simulations.}
\label{fig:dijet_eflow_towers}
\end{center}
\end{figure}

\begin{figure}[h!]
\begin{center}
\includegraphics[width=0.95\linewidth]{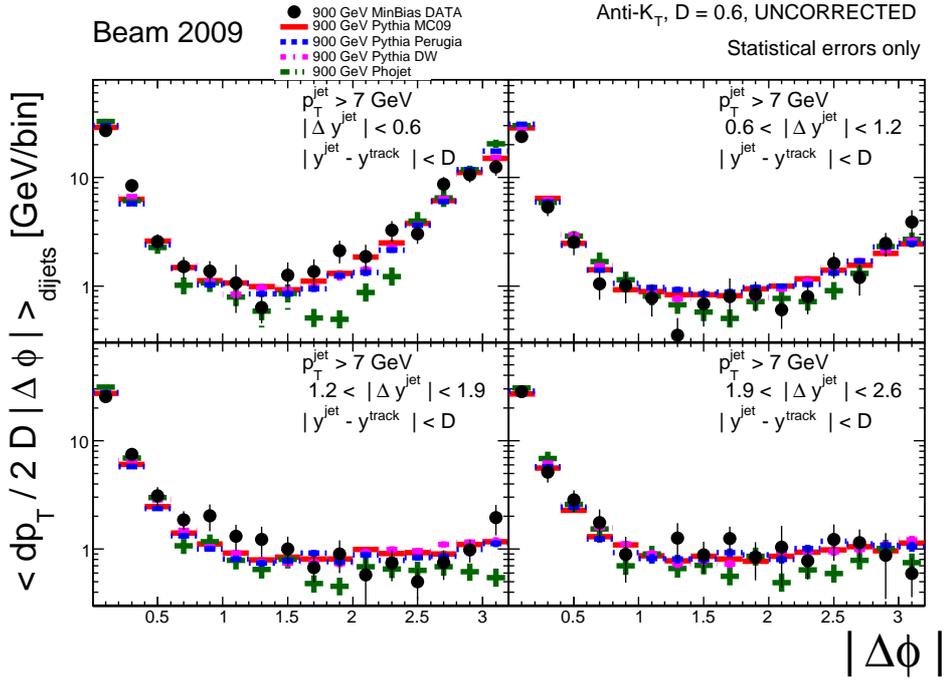}
\caption{Measured energy flow using calorimeter topotowers in dijet events and jets with
$\ptjet >7$~GeV and $|\rapjet |<1.9$,
as a function of $|\Delta \phi|$ with respect to the jet direction and the
rapidity separation between the two jets.
The measurements are compared to minimum bias Monte Carlo simulations.}
\label{fig:dijet_eflow_tracks}
\end{center}
\end{figure}

A subsample of events is selected with at least two jets with $\ptjet >7$~GeV and $|\rapjet|<2.6$, and the
energy flow $<\frac{  d p_{\rm T}}{|d \phi| d y }>_{\rm jets}$  are measured as a function of
the rapidity separation between the two jets $\Delta y^{jj}$. For large  $\Delta y^{jj}$,
the measured energy flow, as defined above in the azimuthal direction,
approximately decouples from the presence of a
second jet and gives a more direct
access to remaining underlying event contributions in the final state.  Figure~\ref{fig:dijet_eflow_towers}
shows the measured energy flow compared to Monte Carlo simulations.
As expected,  for $\Delta y^{jj} < 0.6$
the presence of the second jet at $|\Delta \phi| \sim \pi $ is very pronounced, while for  $\Delta y^{jj} > 1.2$
a plateau of hadronic activity at large $\Delta \phi$ is clearly observed.
Similarly, Fig.~\ref{fig:dijet_eflow_tracks} presents the measurements based
on tracks and only considering jets with  $|\rapjet |<1.9$. At large
$\Delta y^{jj}$ and large $|\Delta \phi |$ (in the plateau region) the different
{\sc pythia} simulated samples describe the data while the {\sc phojet} sample underestimates
the hadronic activity.

\clearpage
\subsection{Energy Flow in rapidity}

\begin{figure}[h!]
\begin{center}
\includegraphics[width=0.5\linewidth]{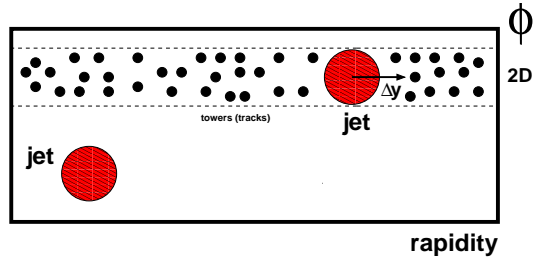}
\caption{Sketch of the particle flow as a function of the distance in rapidity to the jet axis.
}
\label{fig:eflows_rap}
\end{center}
\end{figure}

\begin{figure}[h!]
\begin{center}
\includegraphics[width=1.1\linewidth]{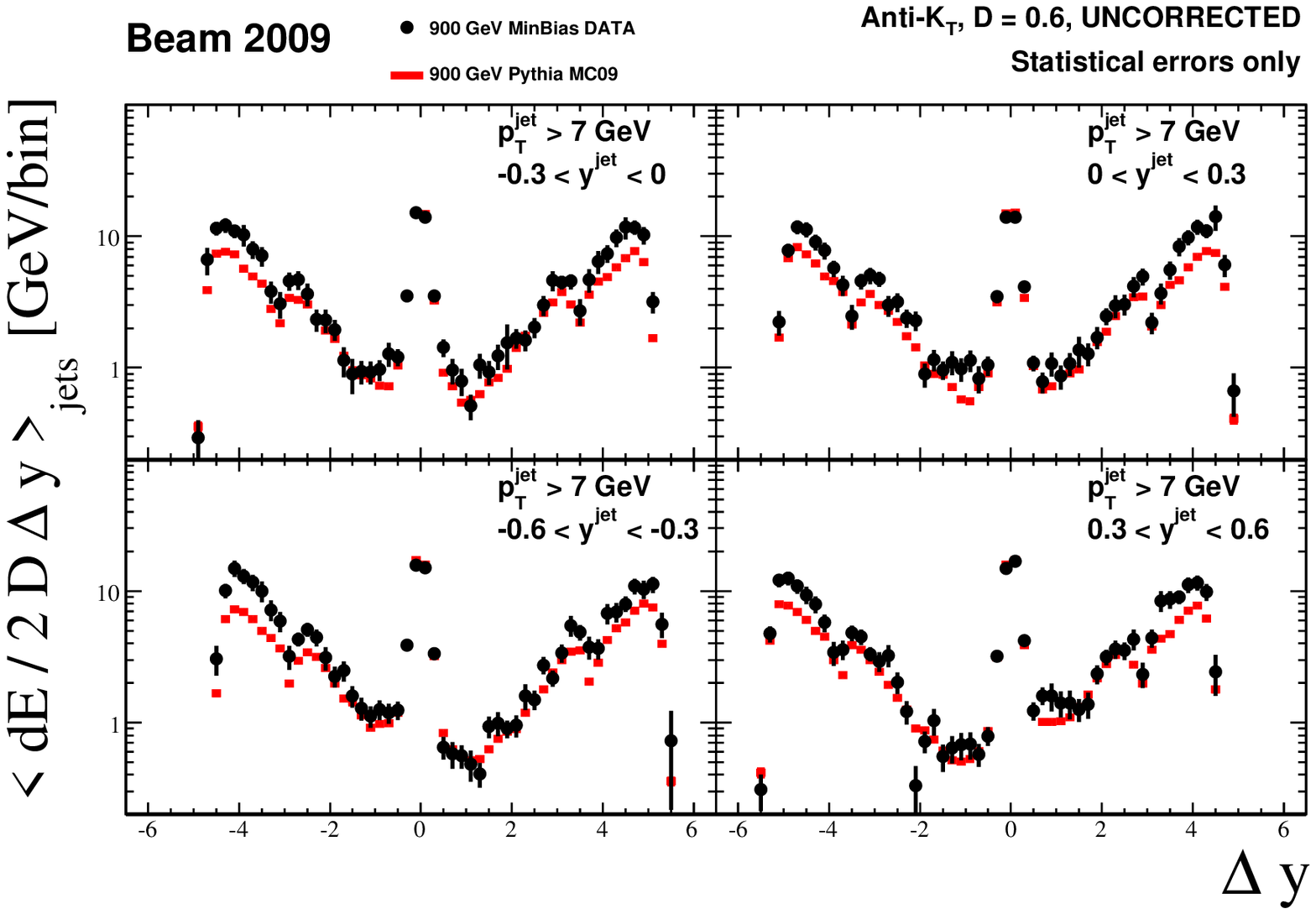}
\caption{Measured energy flow using calorimeter topotowers for jets with
$\ptjet >7$~GeV
as a function of $\Delta y$ in different jet rapidity regions.
The measurements are compared to minimum bias Monte Carlo simulations.}
\label{fig:eflow_rap1}
\end{center}
\end{figure}

\begin{figure}[h!]
\begin{center}
\includegraphics[width=1.1\linewidth]{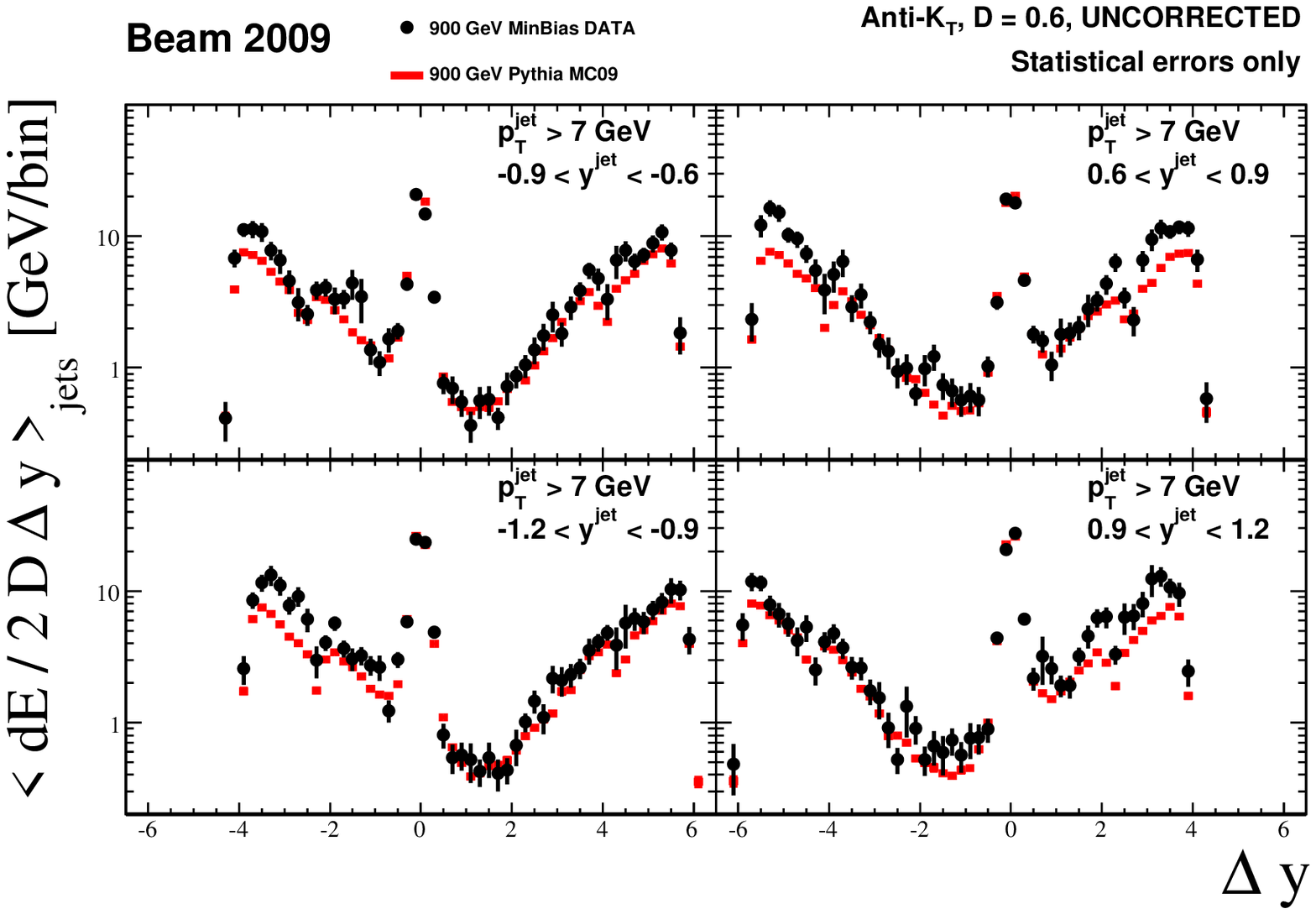}
\caption{Measured energy flow using calorimeter topotowers for jets with
$\ptjet >7$~GeV
as a function of $\Delta y$ in different jet rapidity regions.
The measurements are compared to minimum bias Monte Carlo simulations.}
\label{fig:eflow_rap2}
\end{center}
\end{figure}

\begin{figure}[h!]
\begin{center}
\includegraphics[width=1.1\linewidth]{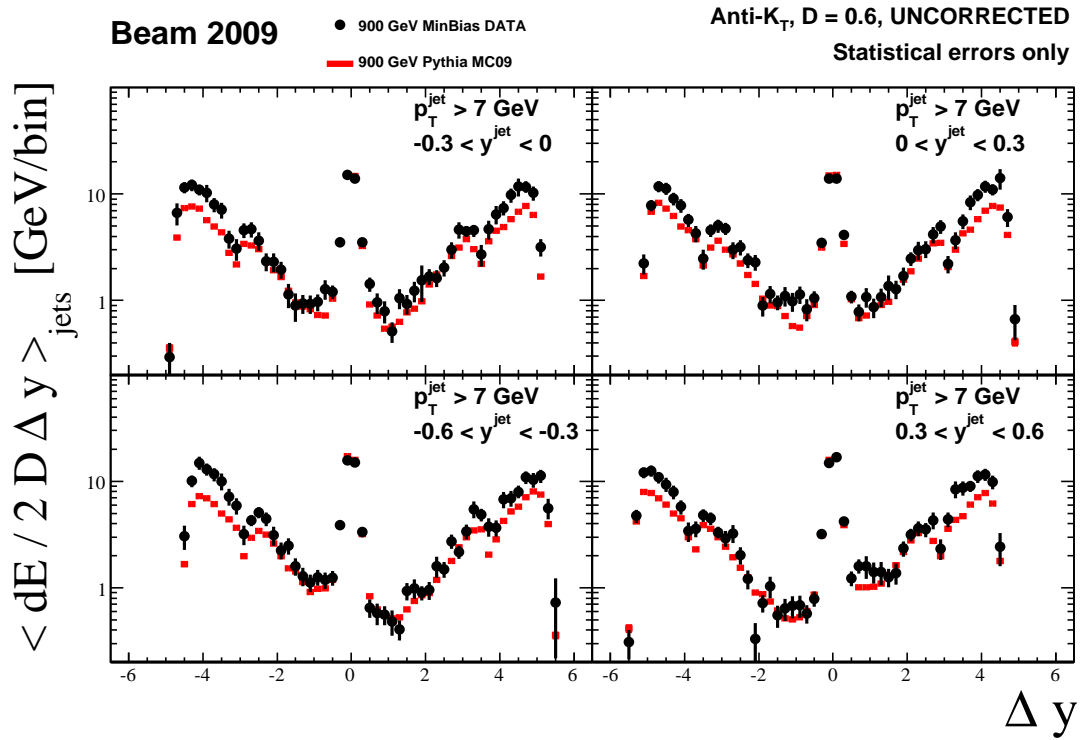}
\caption{Measured energy flow using calorimeter topotowers for jets with
$\ptjet >7$~GeV
as a function of $\Delta y$ in different jet rapidity regions.
The measurements are compared to minimum bias Monte Carlo simulations.}
\label{fig:eflow_rap3}
\end{center}
\end{figure}

The flow of energy around the jet direction is studied as a function of the distance in
rapidity $\Delta y$ to the jet axis.
In this case only towers within a azimuthal distance $\Delta \phi < D$
are considered (see Fig.~\ref{fig:eflows_rap}), and energy instead of transverse momentum
is computed to appreciate better the color flow between the jet and the direction of the
proton remnants:
\begin{equation}
<\frac{  d E}{d \phi d y }>_{\rm jets} = 
\frac{1}{ 2 D |\Delta y| }\frac{1}{\rm N_{jet}}\sum_{\rm jets} {  E(y-\Delta y/2,\phi+\Delta y/2)},
\label{eq:eflows_rap}
\end{equation}
\noindent
where  $ E(y-\Delta y/2,\phi+\Delta y/2)$ is the sum of the energy of the
constituents (calorimeter towers) at a given distance $y$ to the jet. Bins of $\Delta y = 0.2$  have
been considered. Measurements are performed for jets in different rapidity regions in the range
$-1.2 < \rapjet < 1.2$ (see Figs.~\ref{fig:eflow_rap1} and \ref{fig:eflow_rap2})
and compared to Monte Carlo simulations. Assuming a symmetric response of the calorimeter for
positive and negative rapidities, one can fold
both figures in a single one and define the observable  in terms of absolute
jet rapidity bins (see Fig~\ref{fig:eflow_rap3}) to reduce the statistical fluctuations.

\noindent
The measured energy flow present the expected peak around the jet axis, and outside the
jet cone the energy flow increases
as the distance to the jet increases.  The measurements show a structure in $\Delta y$
that can be correlated with the transition across different calorimeter
 subsystems with different response. The Monte Carlo simulation ({\sc pythia} MC09) provides a reasonable description
of the core of the jet while systematically underestimates the activity away from the jet. The
differences are particularly pronounced in the very forward region of the calorimeter and in
the crack regions between calorimeters, and can be attributed to deficiencies in the
simulation of the detector material and response to low energy particles.

\clearpage

\subsection{Energy Profiles beyond the cone of the jet}

Finally, the energy profile $<dp_{\rm T}/d r^2 >_{\rm jets}$ is defined in a similar way as the differential jet shape 
but normalized to the area $\Delta A$ in each annulus, and without dividing, jet-by-jet,  by the sum of the transverse momenta 
in $0 < r < D$:

\begin{equation}
<\frac{  d p_{\rm T}}{  d r^2}>_{jets} = 
\frac{1}{  \Delta A}\frac{1}{\rm N_{jets}}\sum_{\rm jets} {  p_{\rm T}(r-\Delta \ r/2,r+\Delta \ r/2)},~0 \le r \le D
\label{eq:jet_prof}
\end{equation}

\noindent Only central jets with rapidity in the region $|\rapjet|<0.3$ and
$|\rapjet|<0.6$ are considered and the measurements are carried out using tracks. The use of tracks instead of
calorimeter towers allows studying the particle flow  around the jet direction without any bias due to the jet search algorithm
employed, or the variations of the calorimeter response and the presence of calorimeter
cracks as a function of rapidity. The measurements are compared to Monte Carlo simulations in Fig.~\ref{fig:monk}.
As already mentioned, the Monte Carlo models overestimates the amount of transverse momentum close to the
core of the jet. The different {\sc pythia} samples provide a reasonable description of the activity in the
region away from the jet axis while {\sc phojet} underestimates it.

\begin{figure}[h!]
\begin{center}
\includegraphics[width=0.5\linewidth]{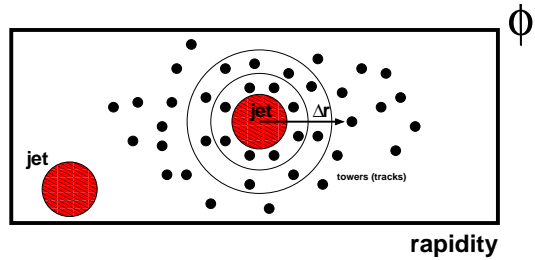}
\caption{Sketch of the particle flow as a function of the distance in radius to the jet axis.
}
\label{fig:eflows_rad}
\end{center}
\end{figure}

\begin{figure}[h!]
\begin{center}
\mbox{
\includegraphics[width=0.48\linewidth]{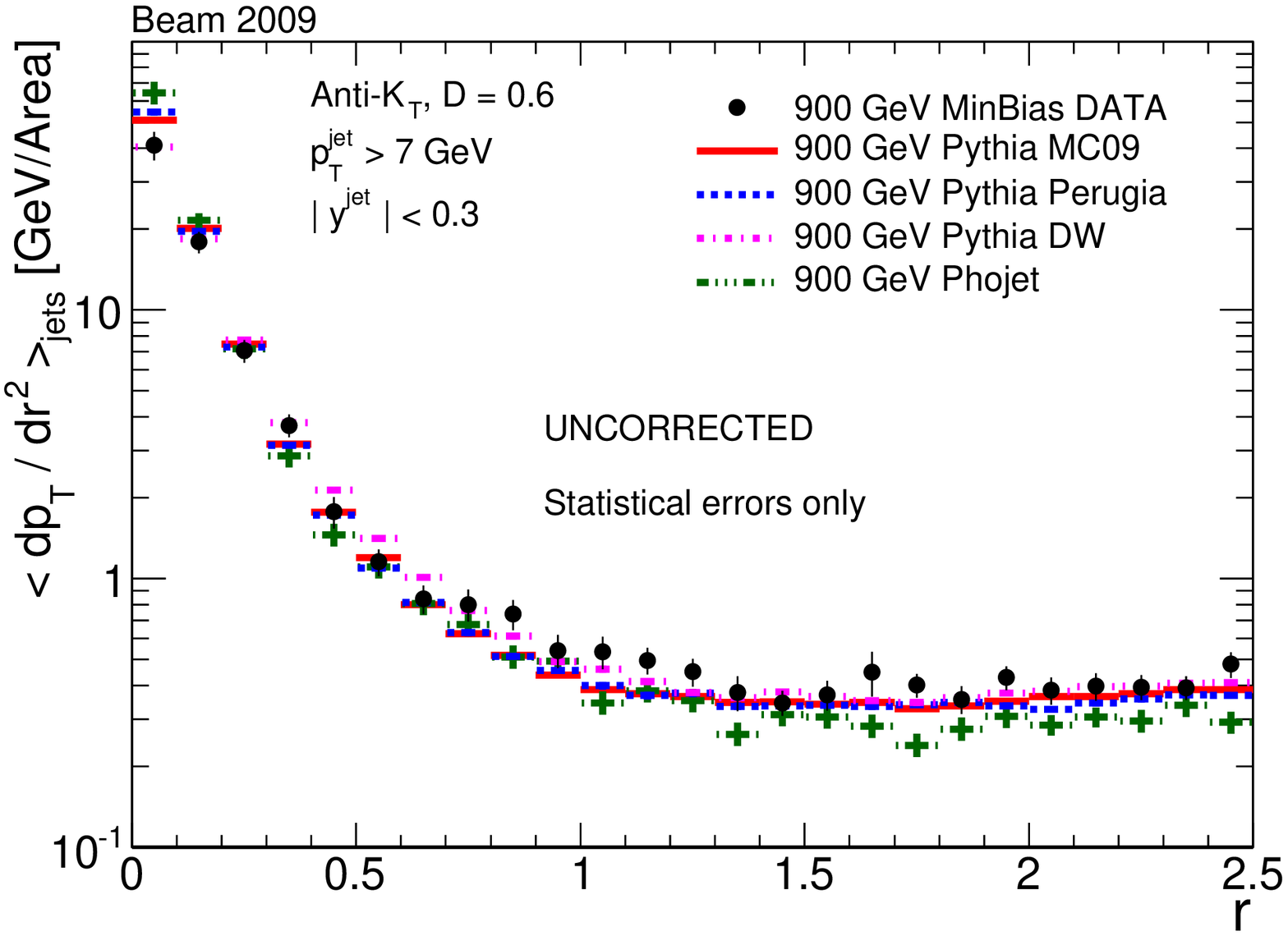}
\includegraphics[width=0.48\linewidth]{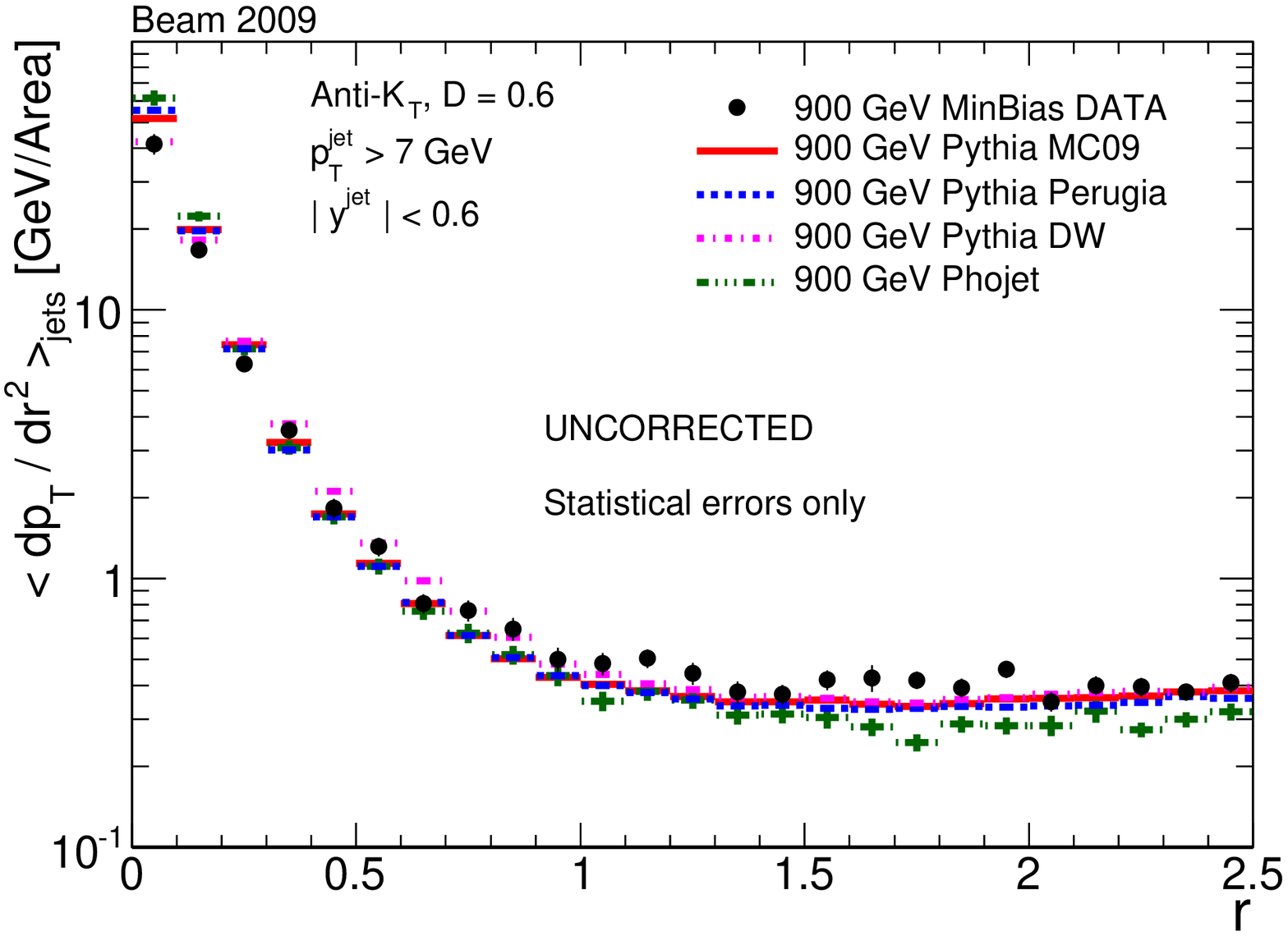}
}
\caption{Measured energy flow using tracks for jets with
$\ptjet >7$~GeV
as a function of $\Delta r$ for jets with $|\rapjet|<0.3$ and $|\rapjet|<0.6$.
The measurements are compared to minimum bias Monte Carlo simulations.}
\label{fig:monk}
\end{center}
\end{figure}

\newpage
\chapter{Energy Flow}
\label{appendix_eflow}

The study of the energy flow away from the jet direction as described in Appendix~\ref{appendix_900} 
was also performed in $pp$ collisions at $\sqrt{s} = 7$~TeV at detector level.

\begin{figure}[tbh]
\begin{center}
\mbox{
\includegraphics[width=0.495\textwidth]{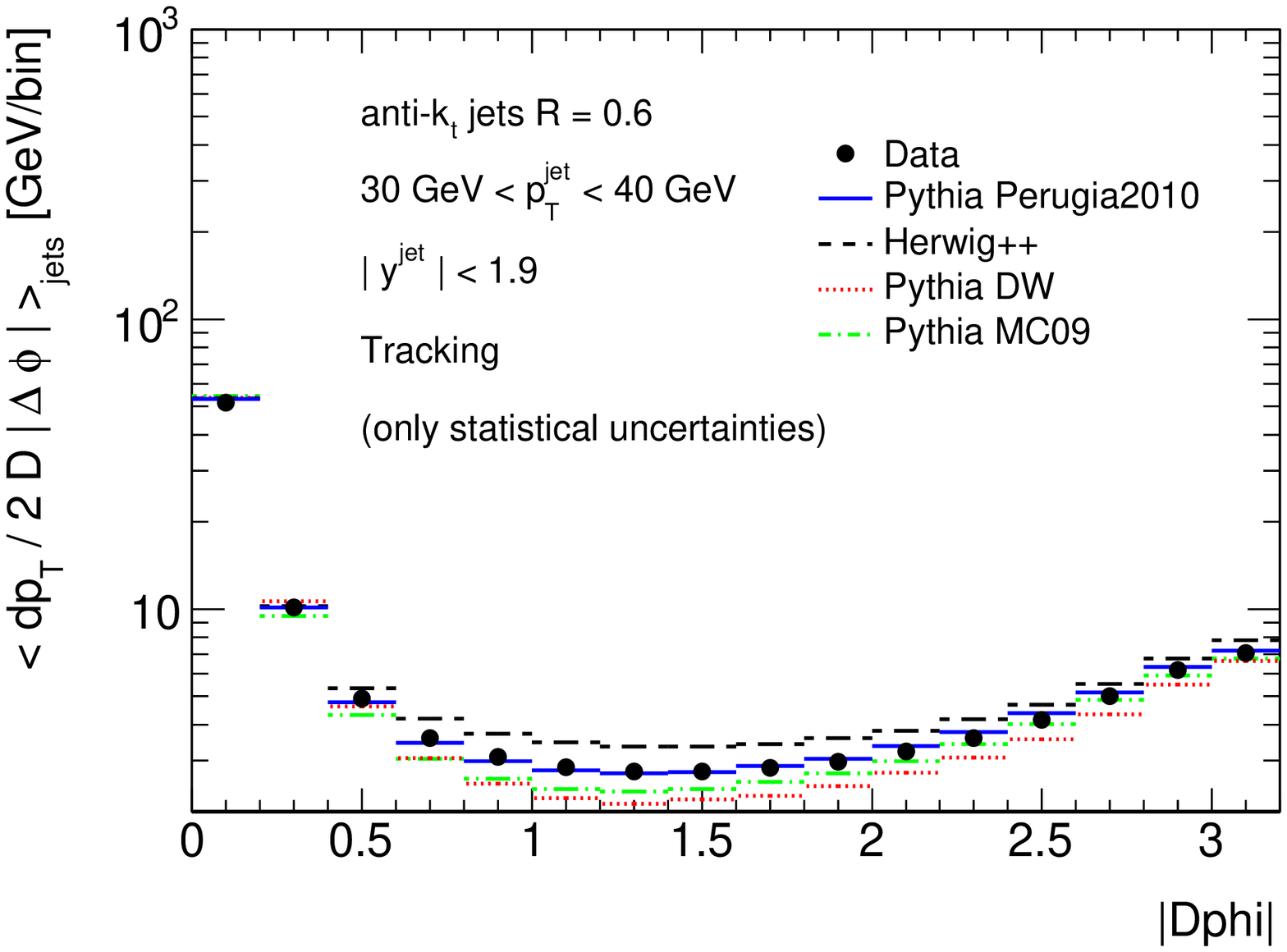}
\includegraphics[width=0.495\textwidth]{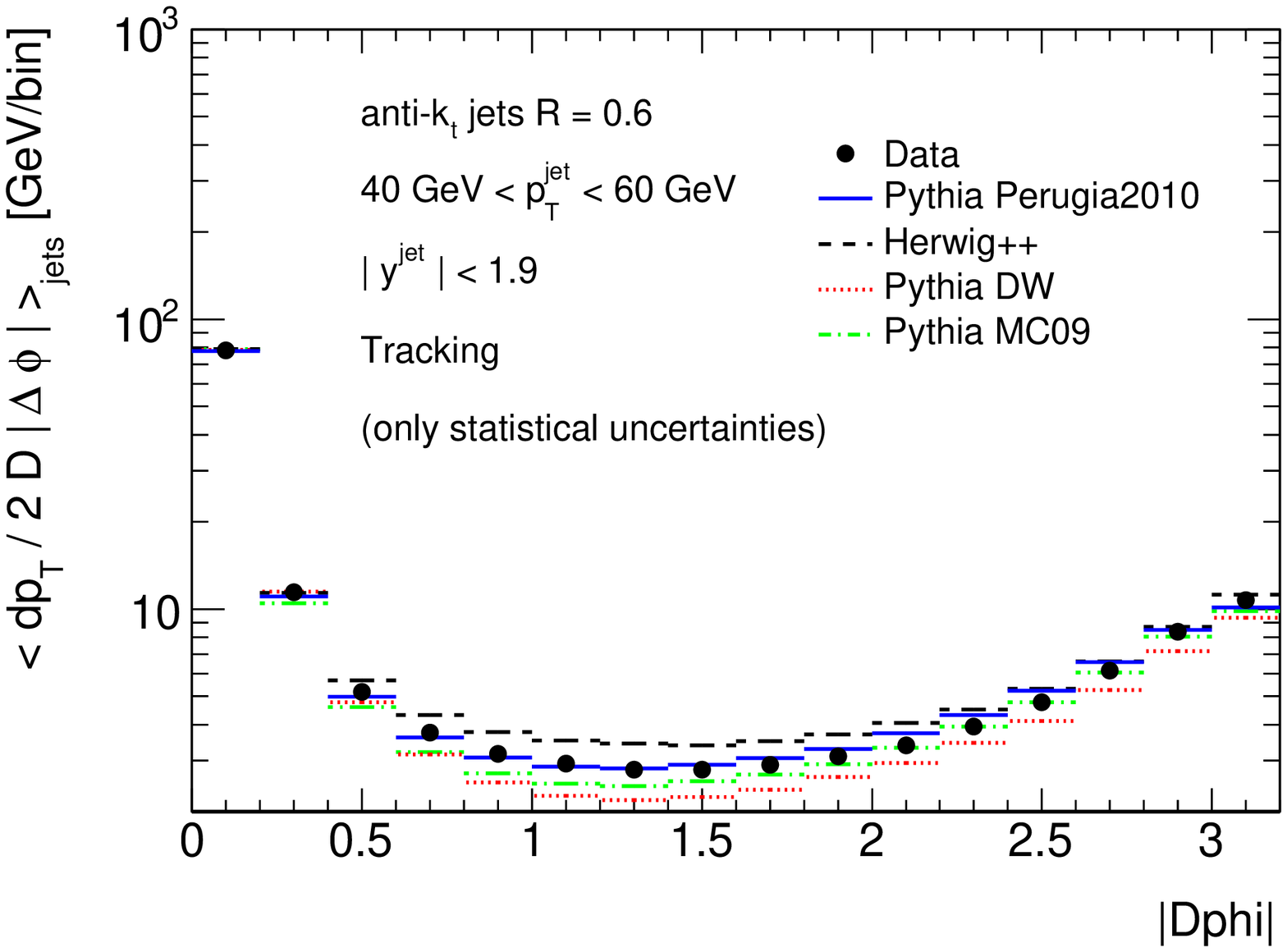}
}
\mbox{
\includegraphics[width=0.495\textwidth]{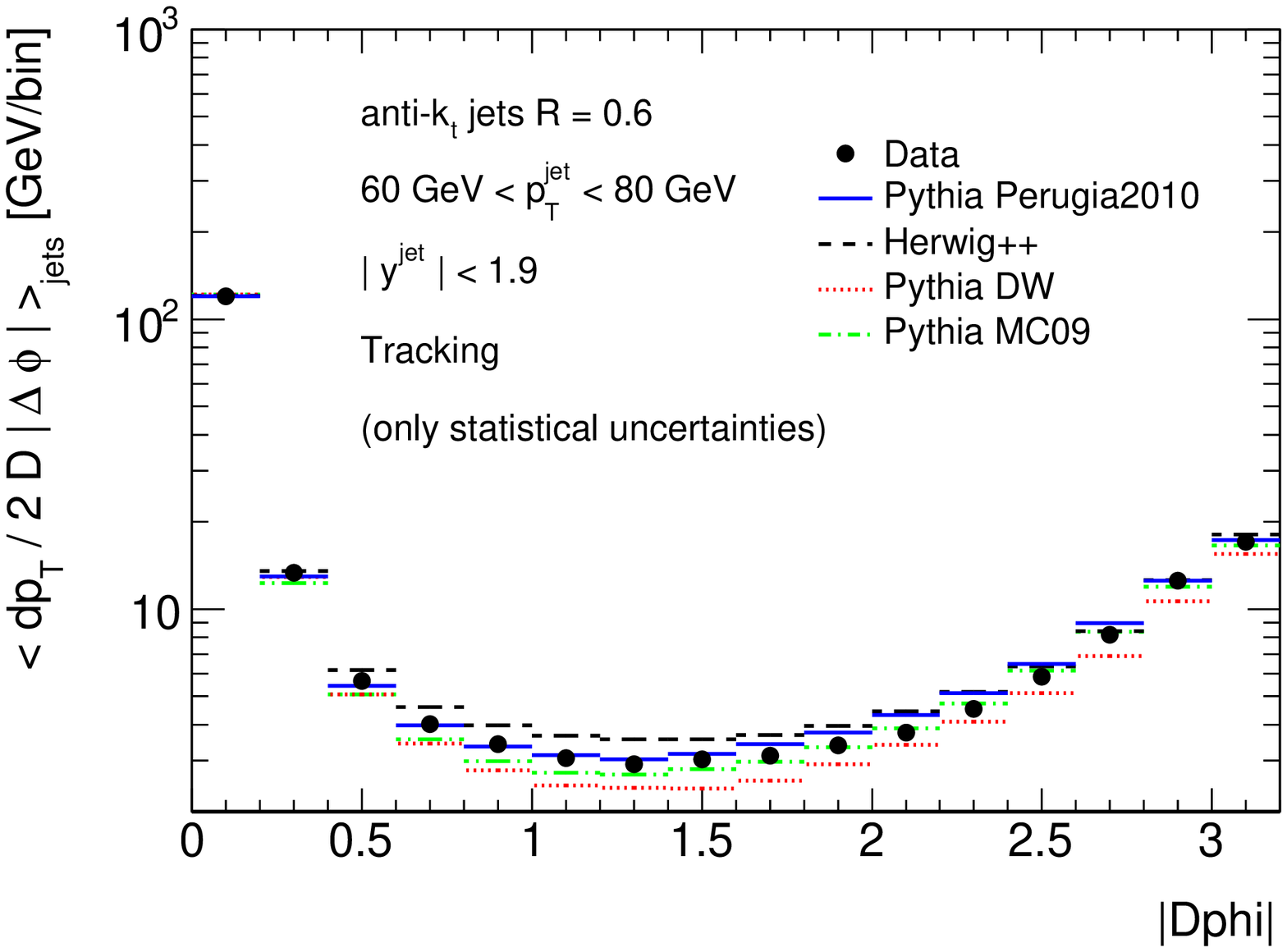}
\includegraphics[width=0.495\textwidth]{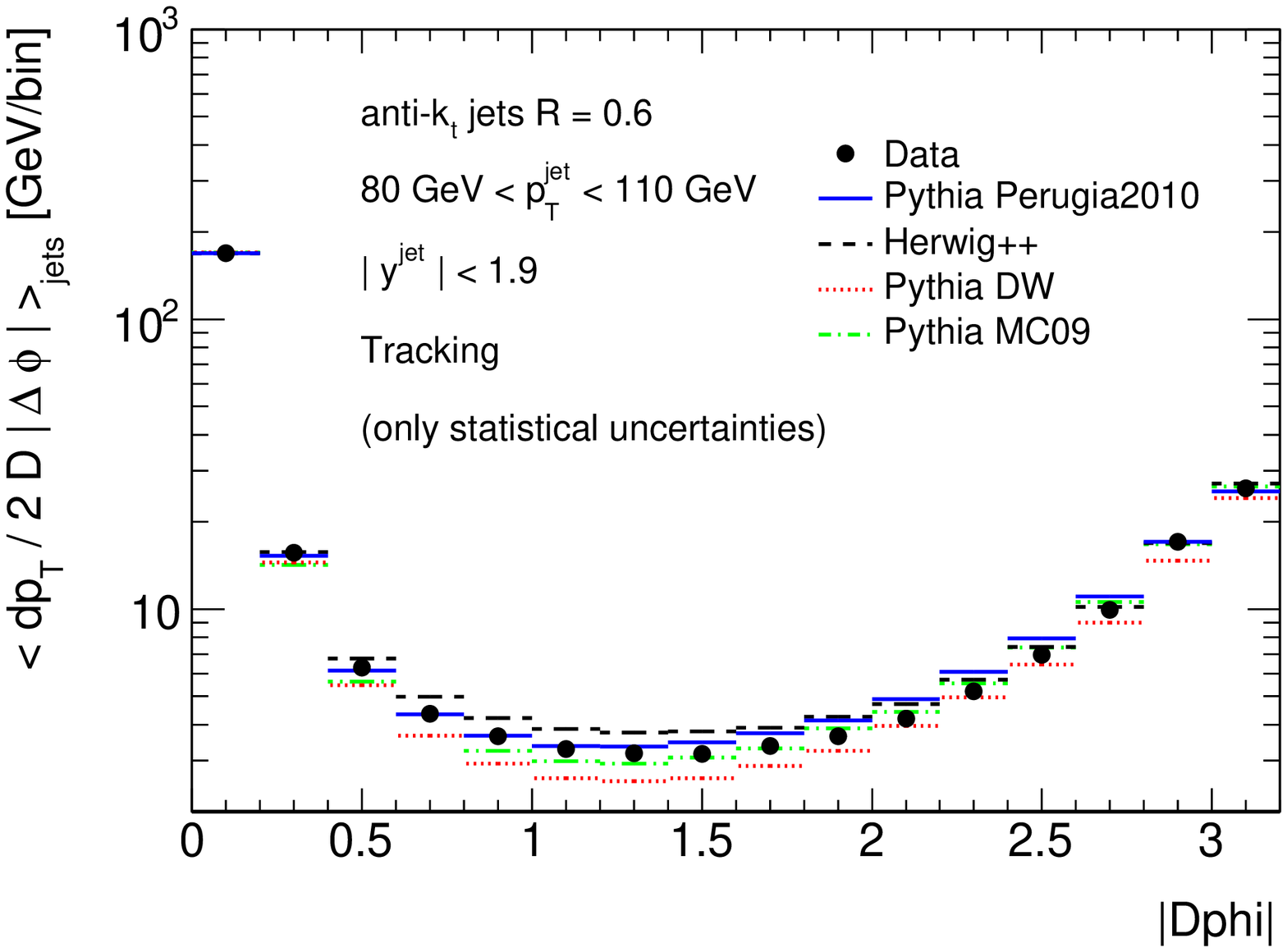}
}
\mbox{
\includegraphics[width=0.495\textwidth]{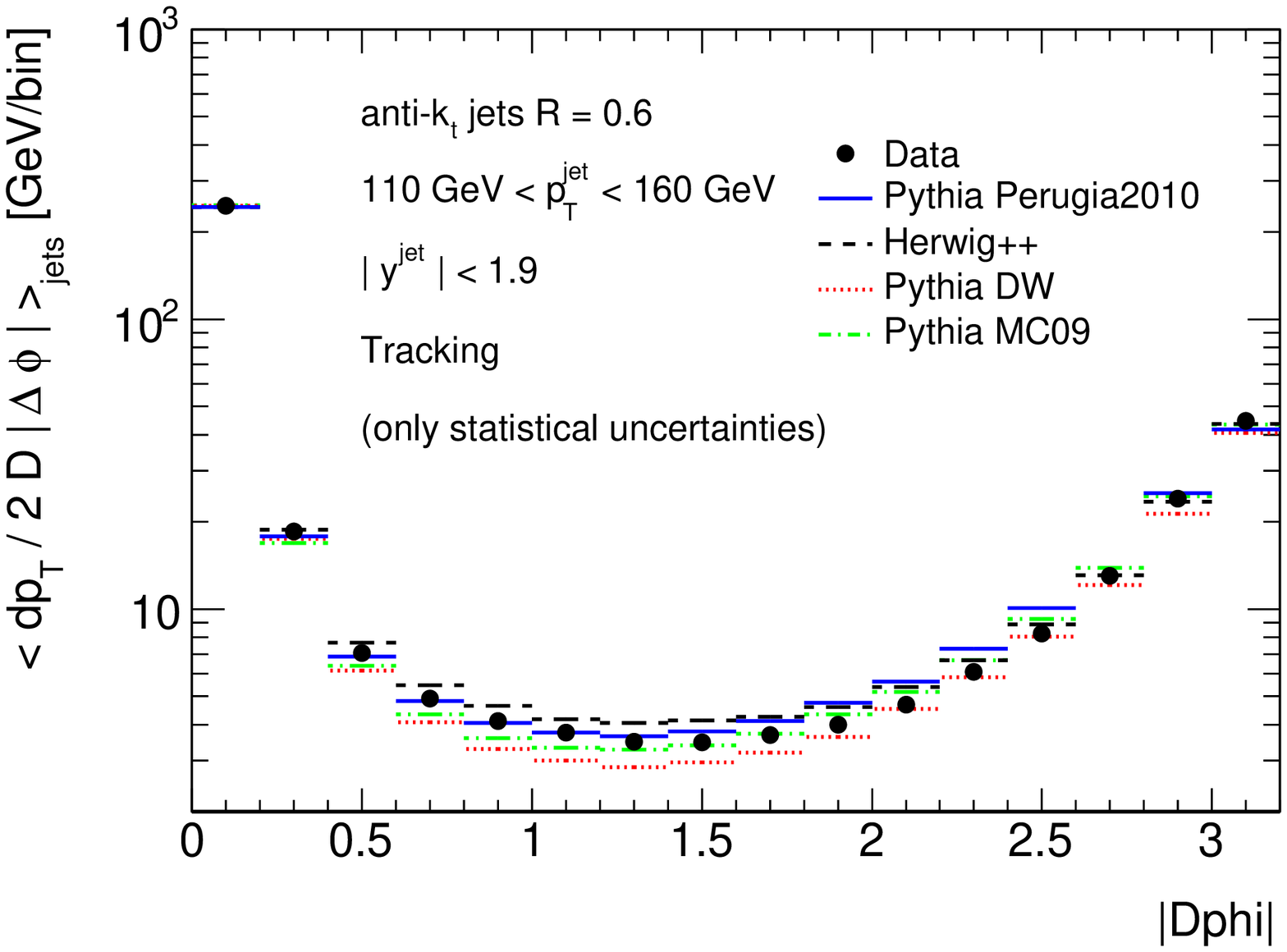}
\includegraphics[width=0.495\textwidth]{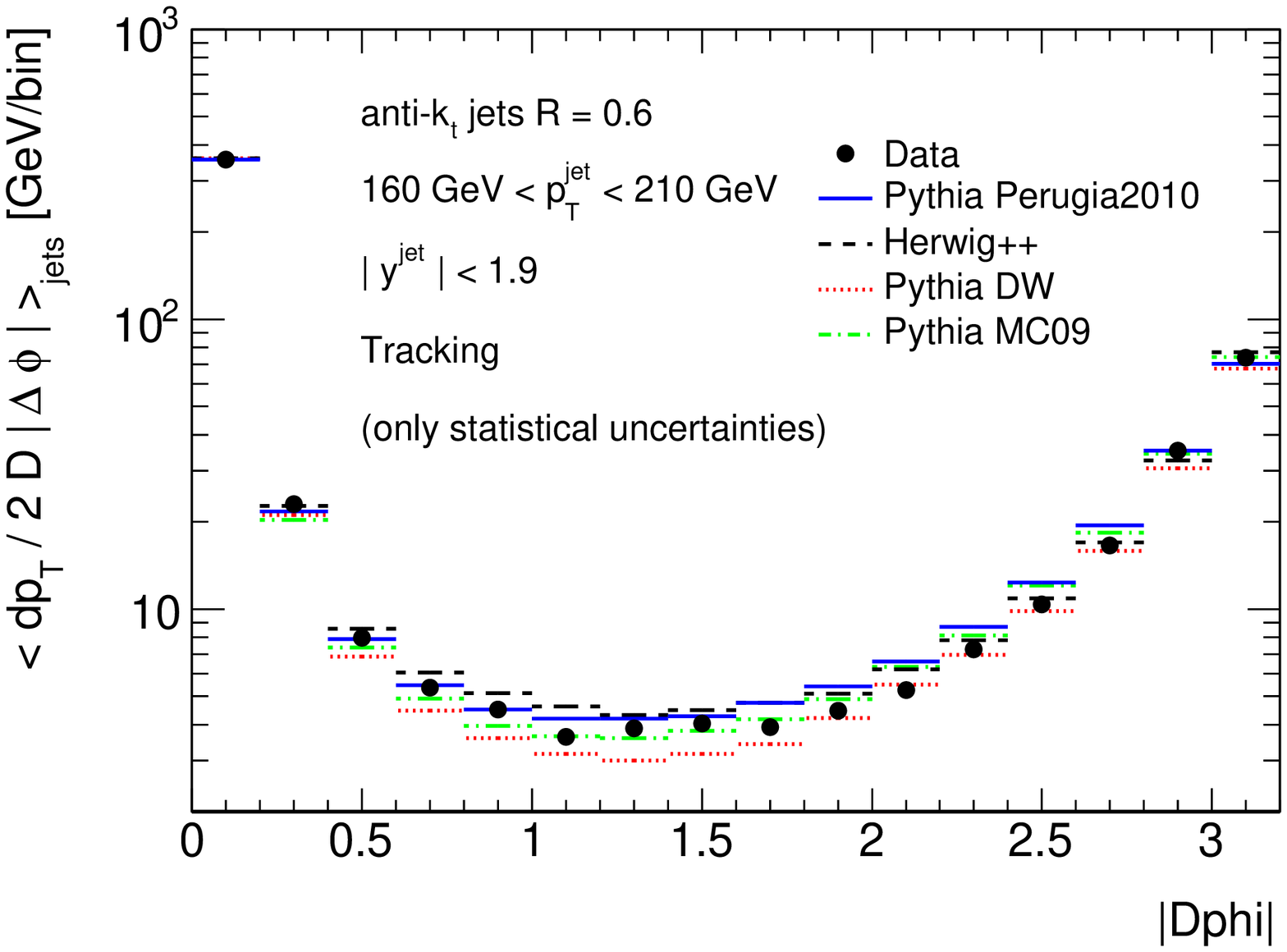}
}
\end{center}
\vspace{-0.7 cm}
\caption{\small
Energy flow using tracks as a function $|\Delta\phi|$ with respect to the jet direction
for jets with $|\rapjet| < 1.9$ and $30 \ {\rm GeV} < \ptjet < 210  \ {\rm GeV}$.
}
\label{fig_eflow_trk}
\end{figure}

\begin{figure}[tbh]
\begin{center}
\mbox{
\includegraphics[width=0.495\textwidth]{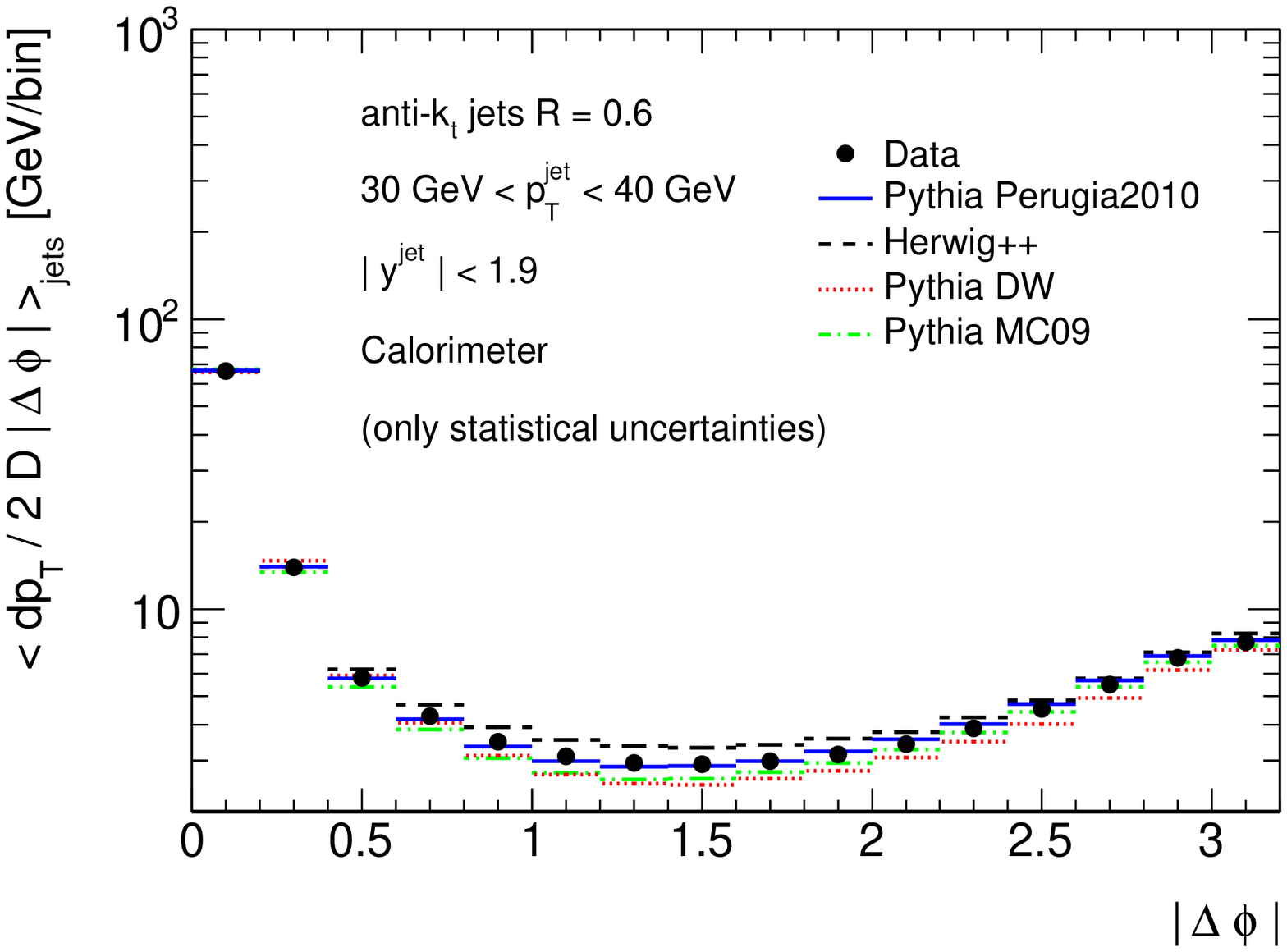}
\includegraphics[width=0.495\textwidth]{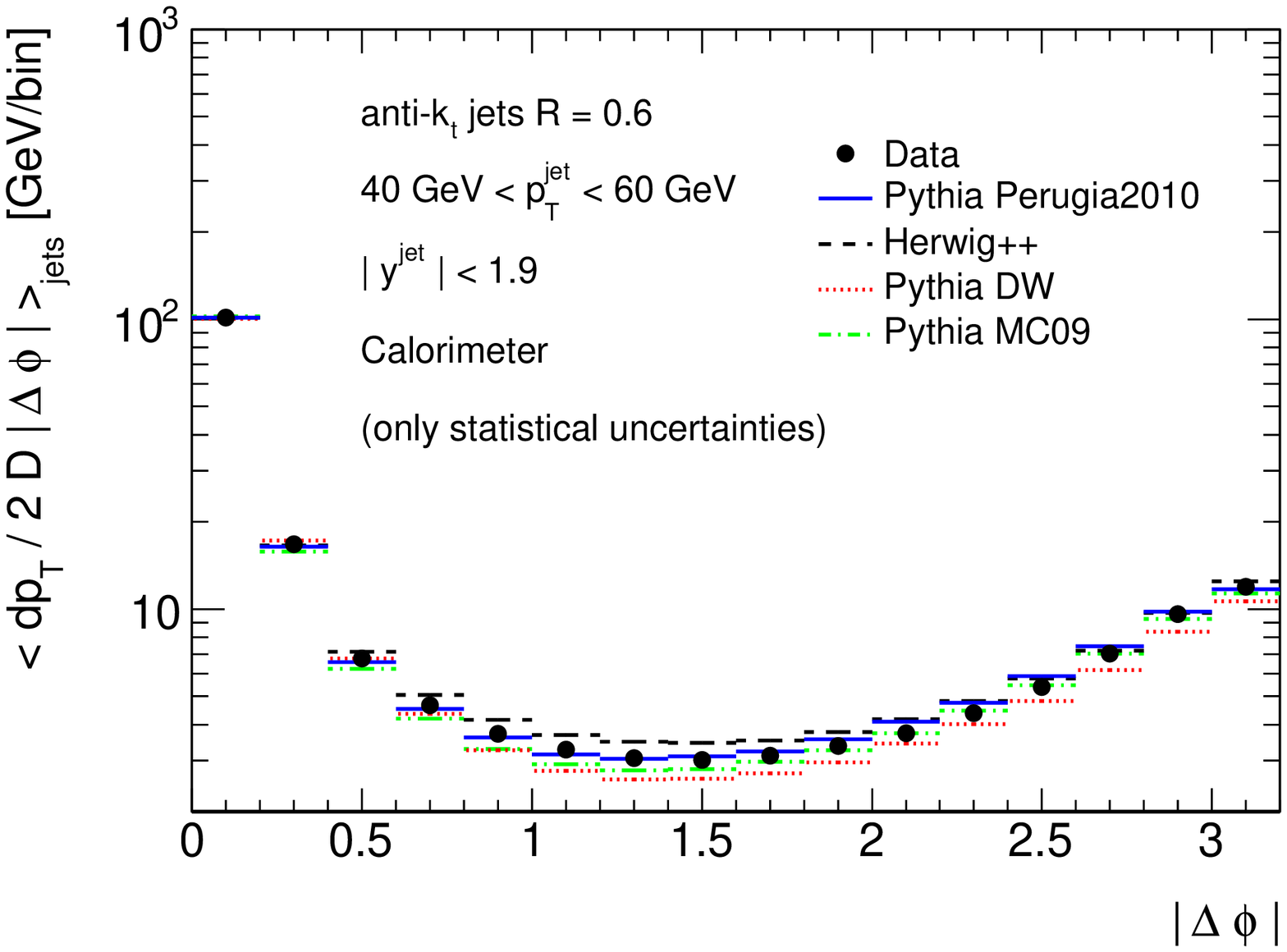}
}
\mbox{
\includegraphics[width=0.495\textwidth]{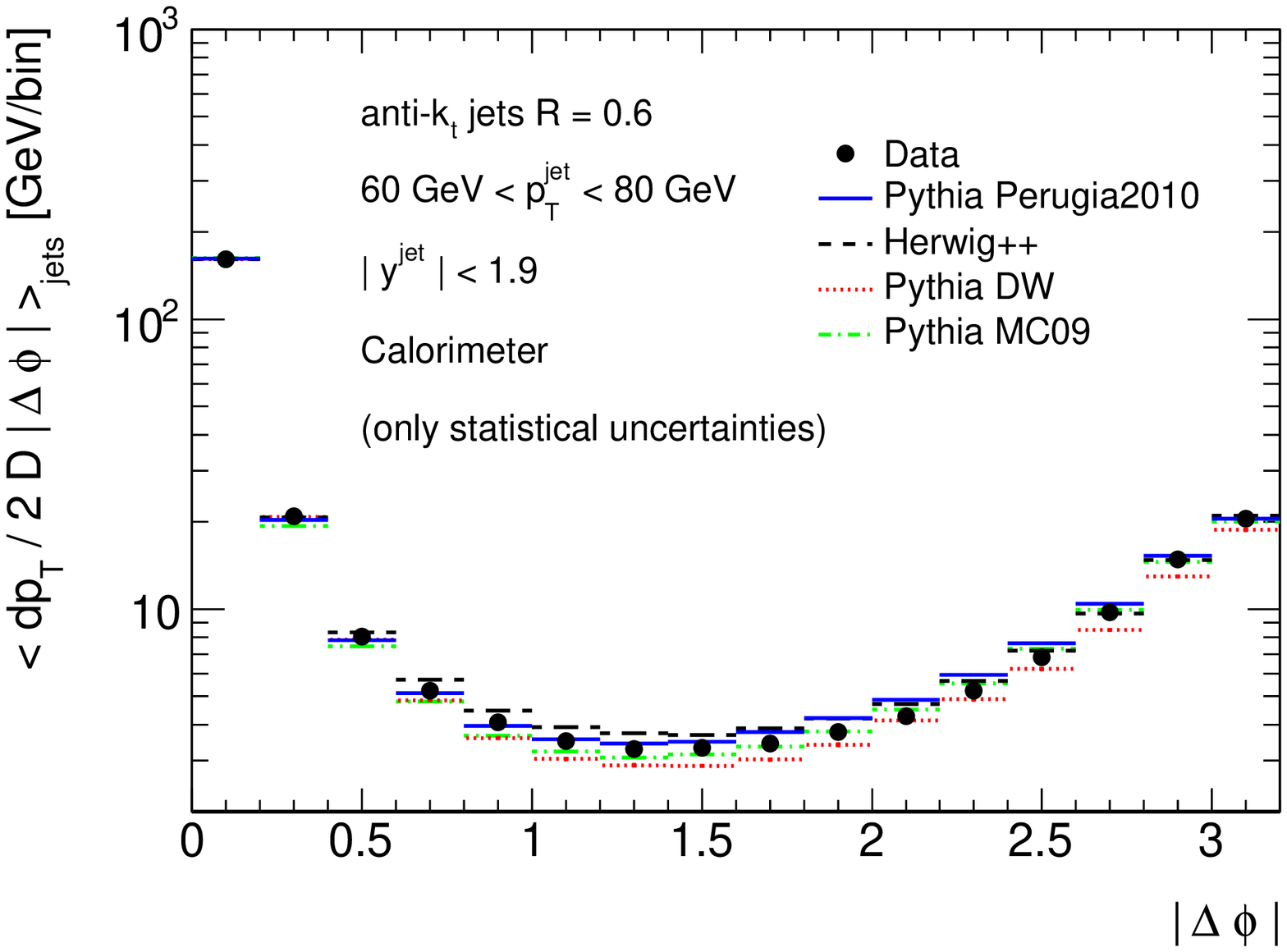}
\includegraphics[width=0.495\textwidth]{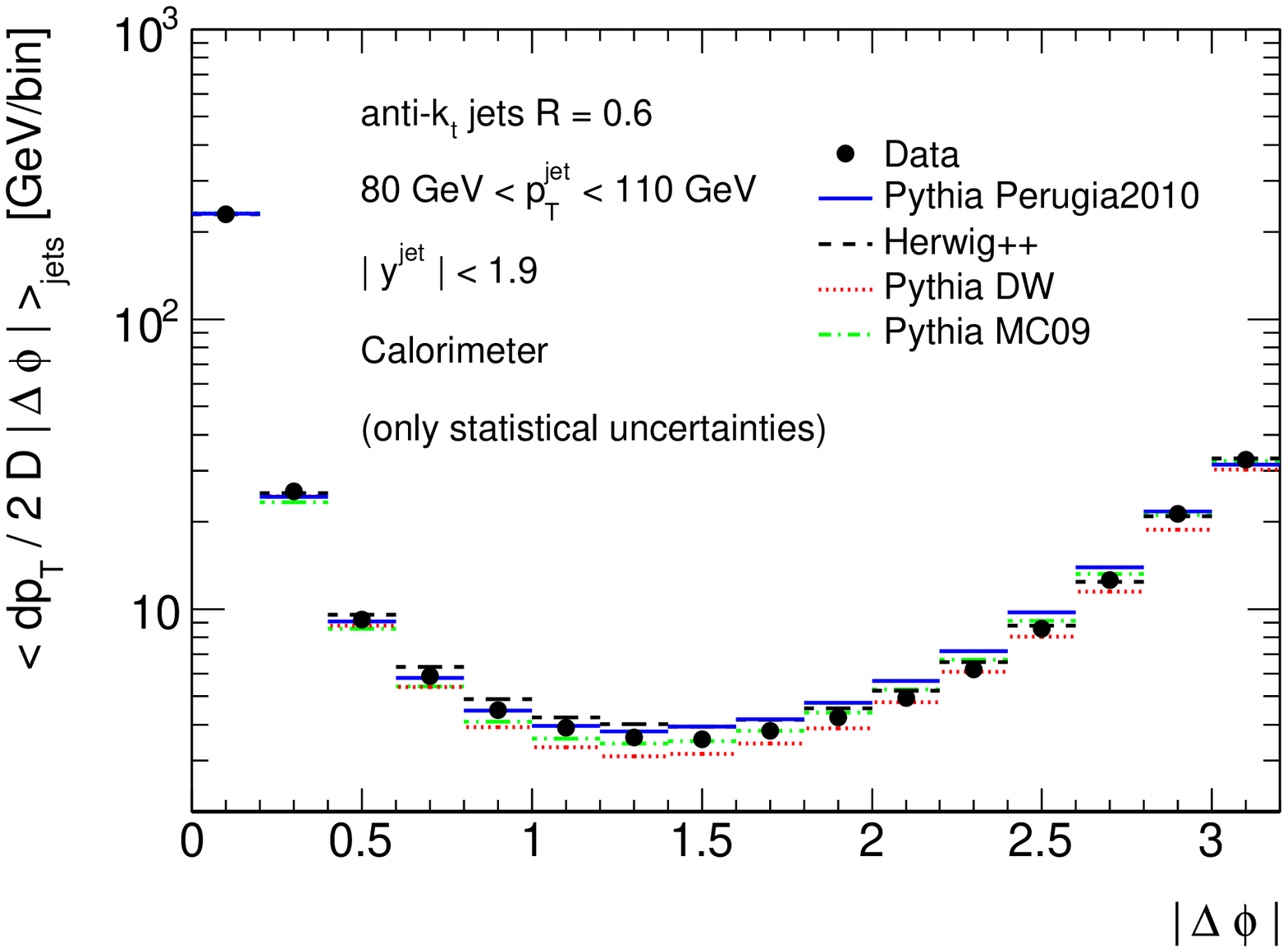}
}
\mbox{
\includegraphics[width=0.495\textwidth]{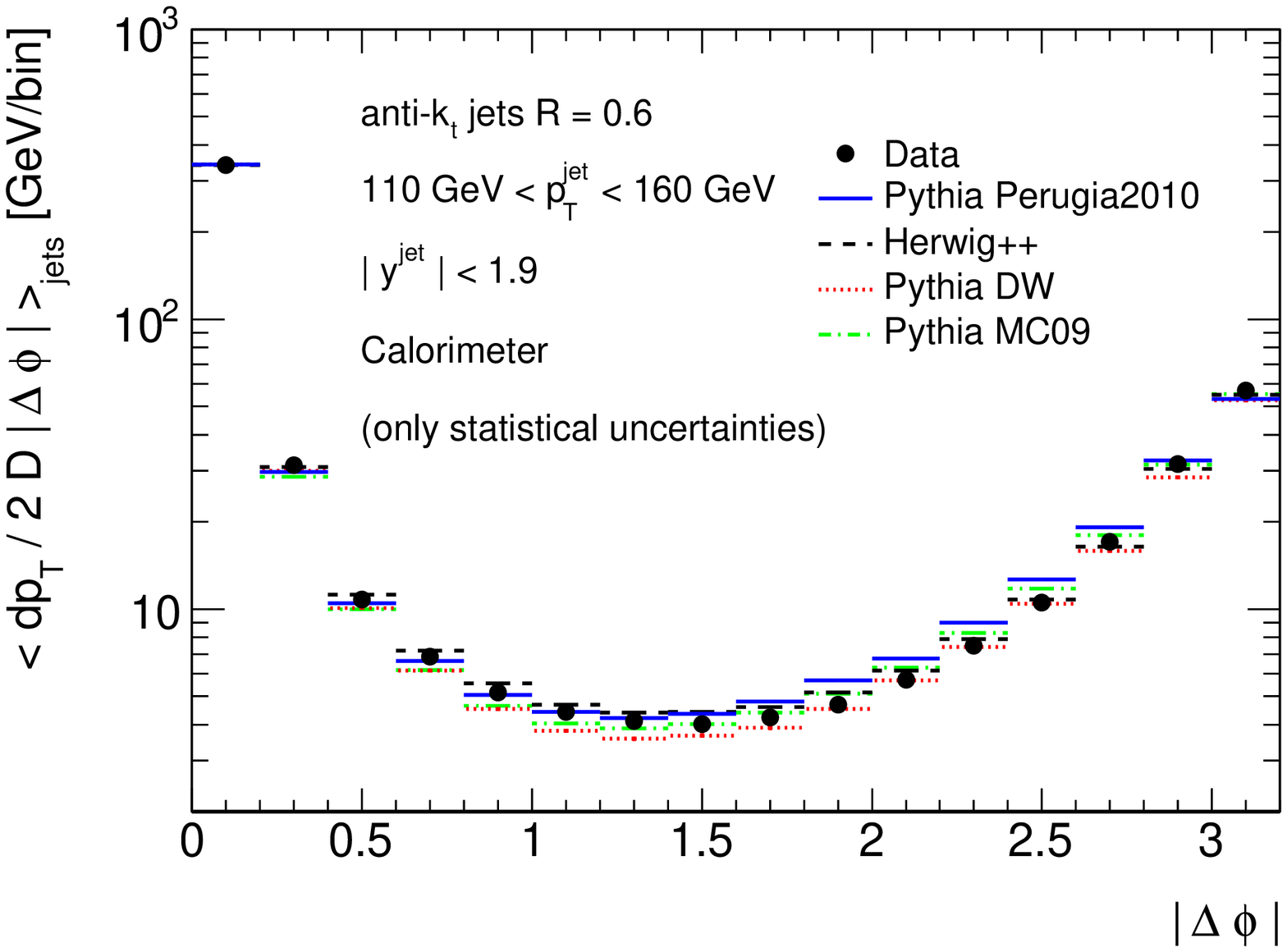}
\includegraphics[width=0.495\textwidth]{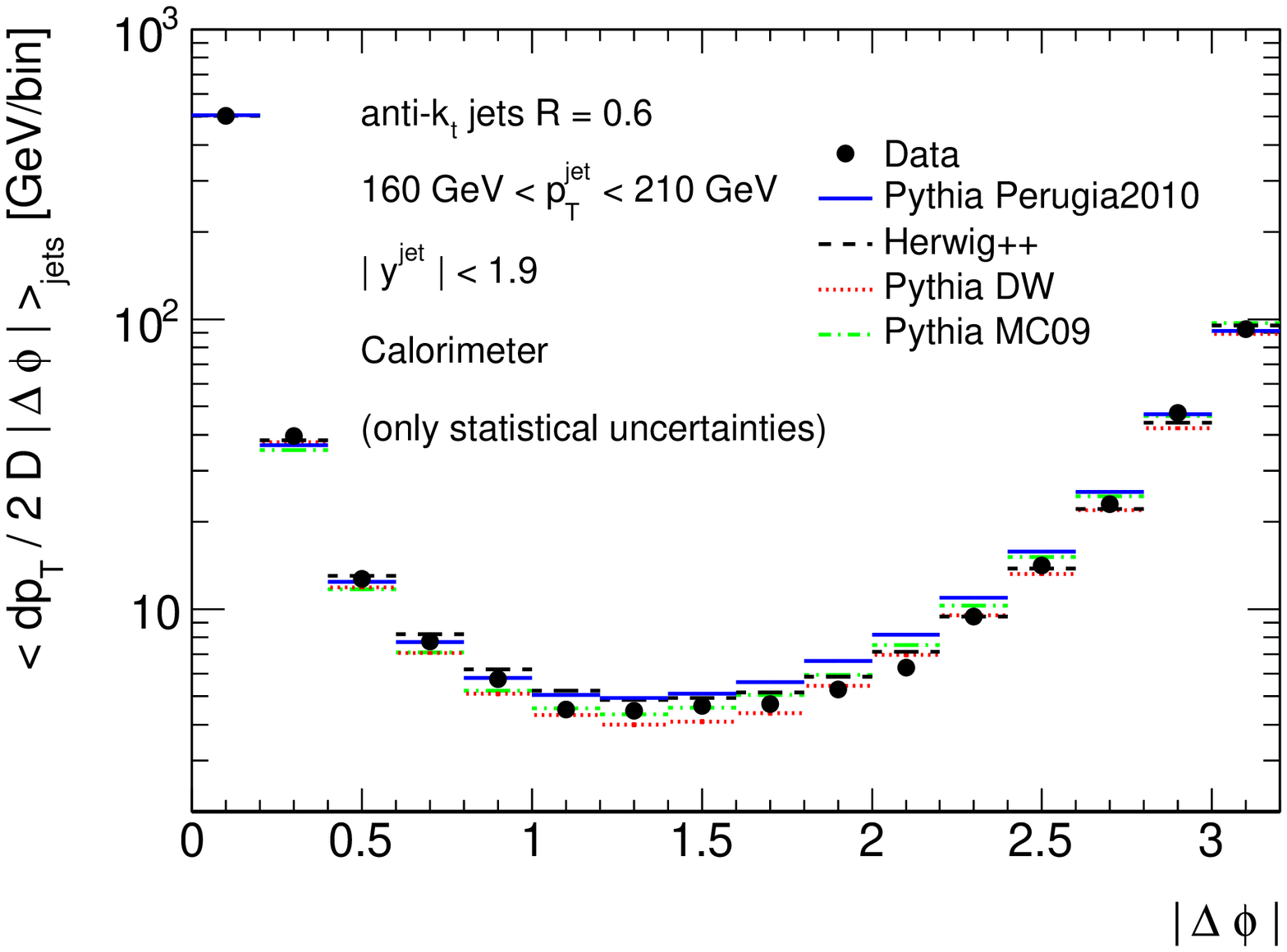}
}
\end{center}
\vspace{-0.7 cm}
\caption{\small
Energy flow using calorimeter clusters as a function $|\Delta\phi|$ with respect to the jet direction
for jets with $|\rapjet| < 1.9$ and $30 \ {\rm GeV} < \ptjet < 210  \ {\rm GeV}$.
}
\label{fig_eflow_calo1}
\end{figure}

\begin{figure}[tbh]
\begin{center}
\mbox{
\includegraphics[width=0.495\textwidth]{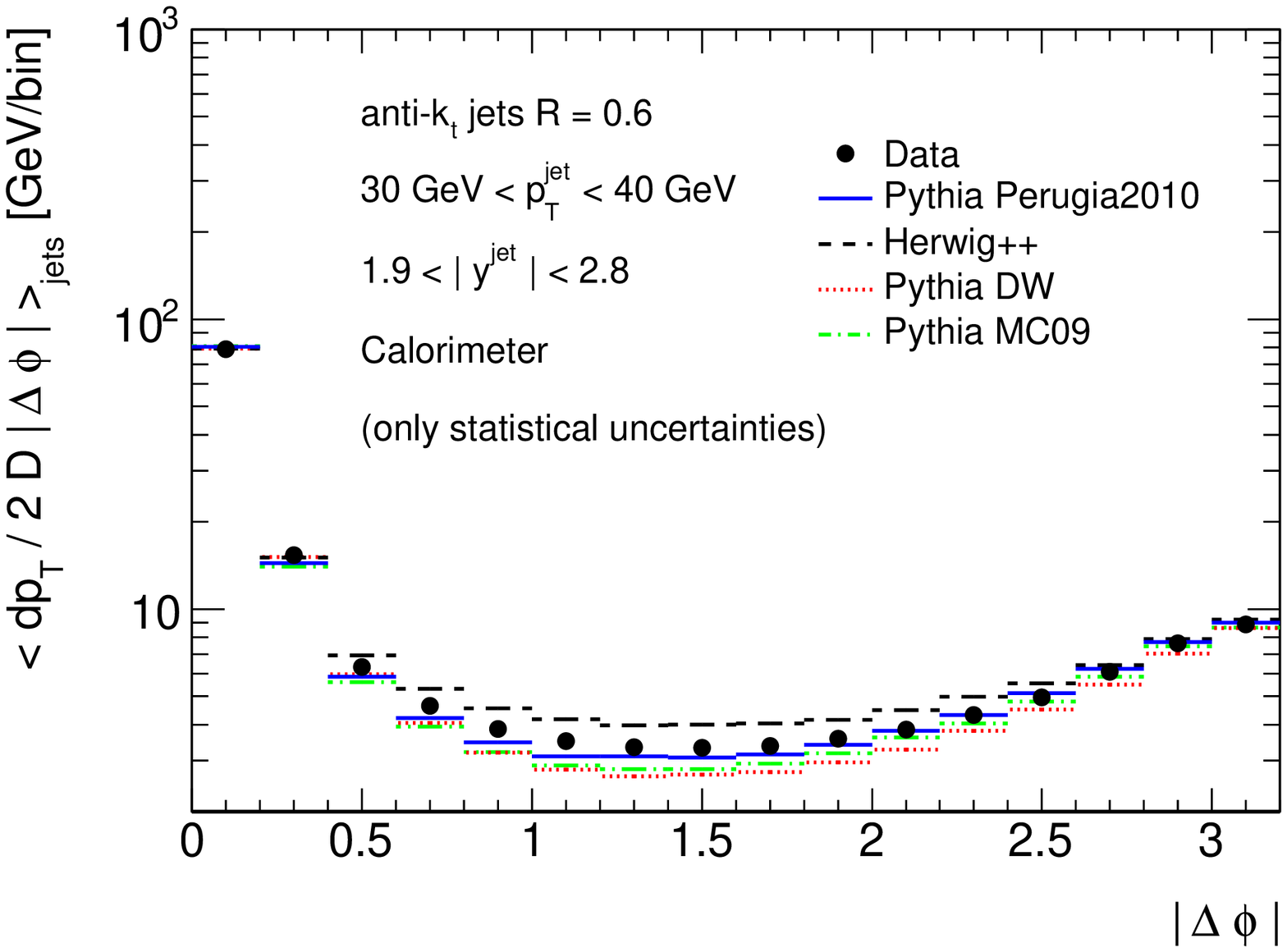}
\includegraphics[width=0.495\textwidth]{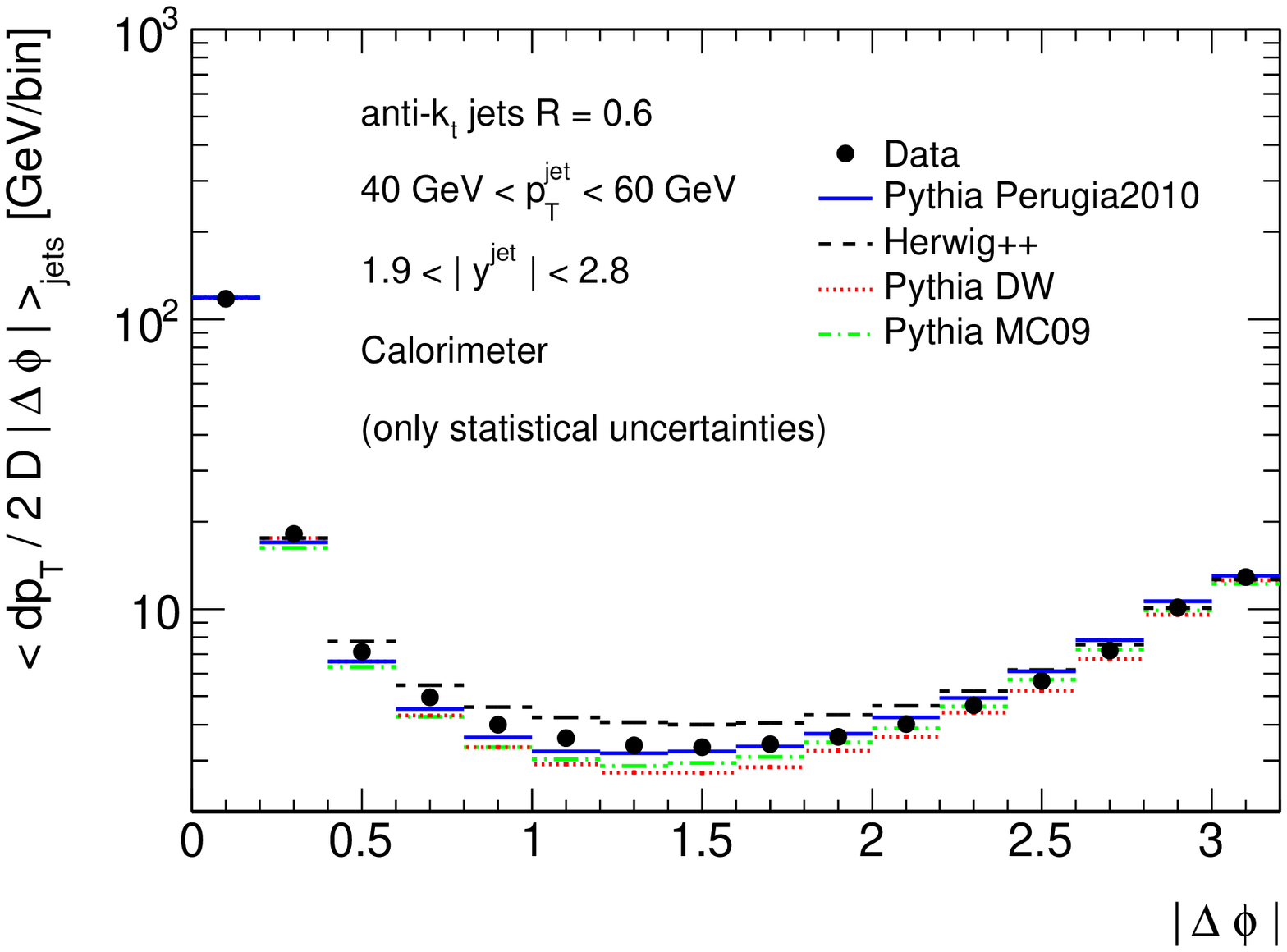}
}
\mbox{
\includegraphics[width=0.495\textwidth]{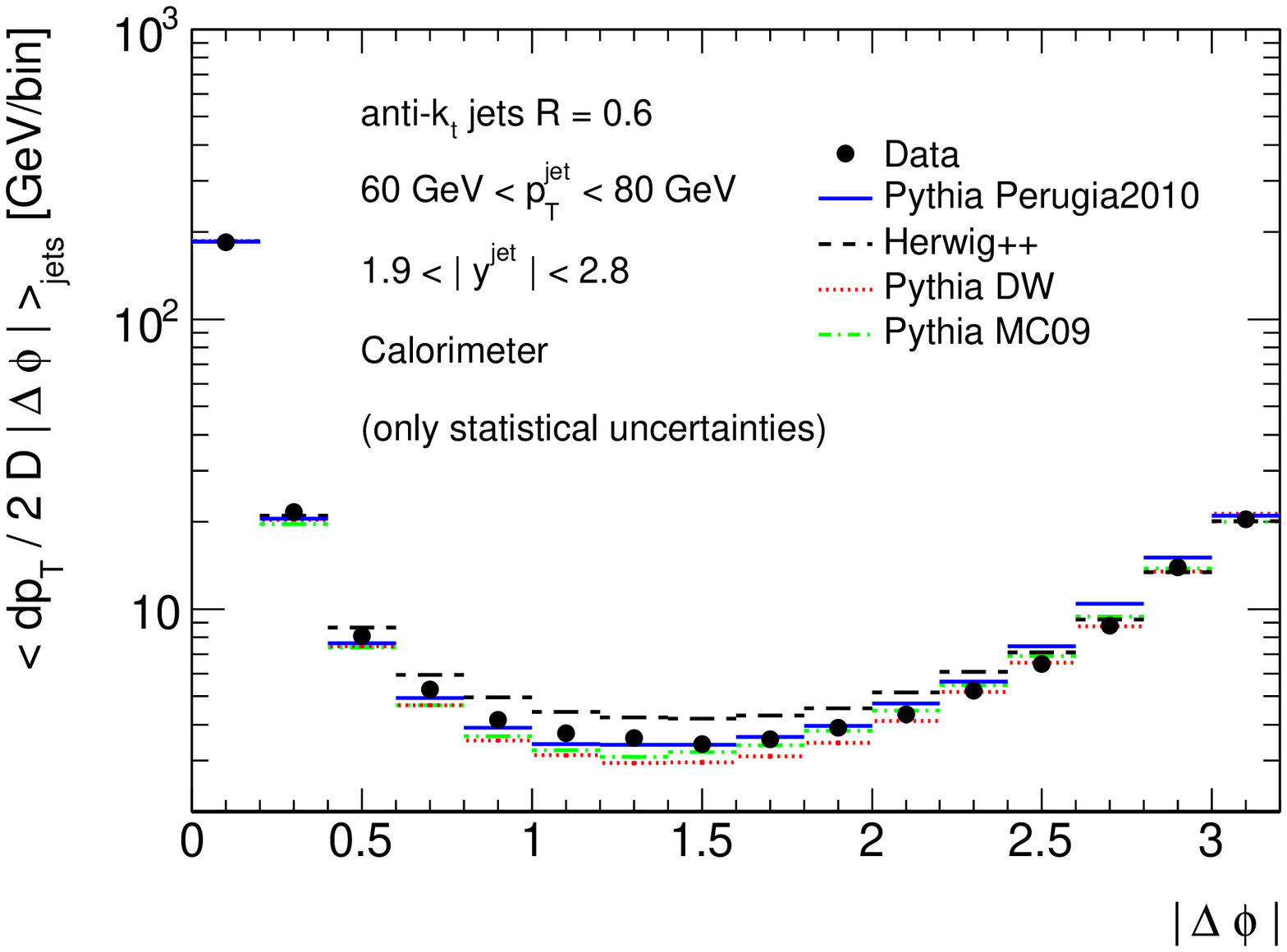}
\includegraphics[width=0.495\textwidth]{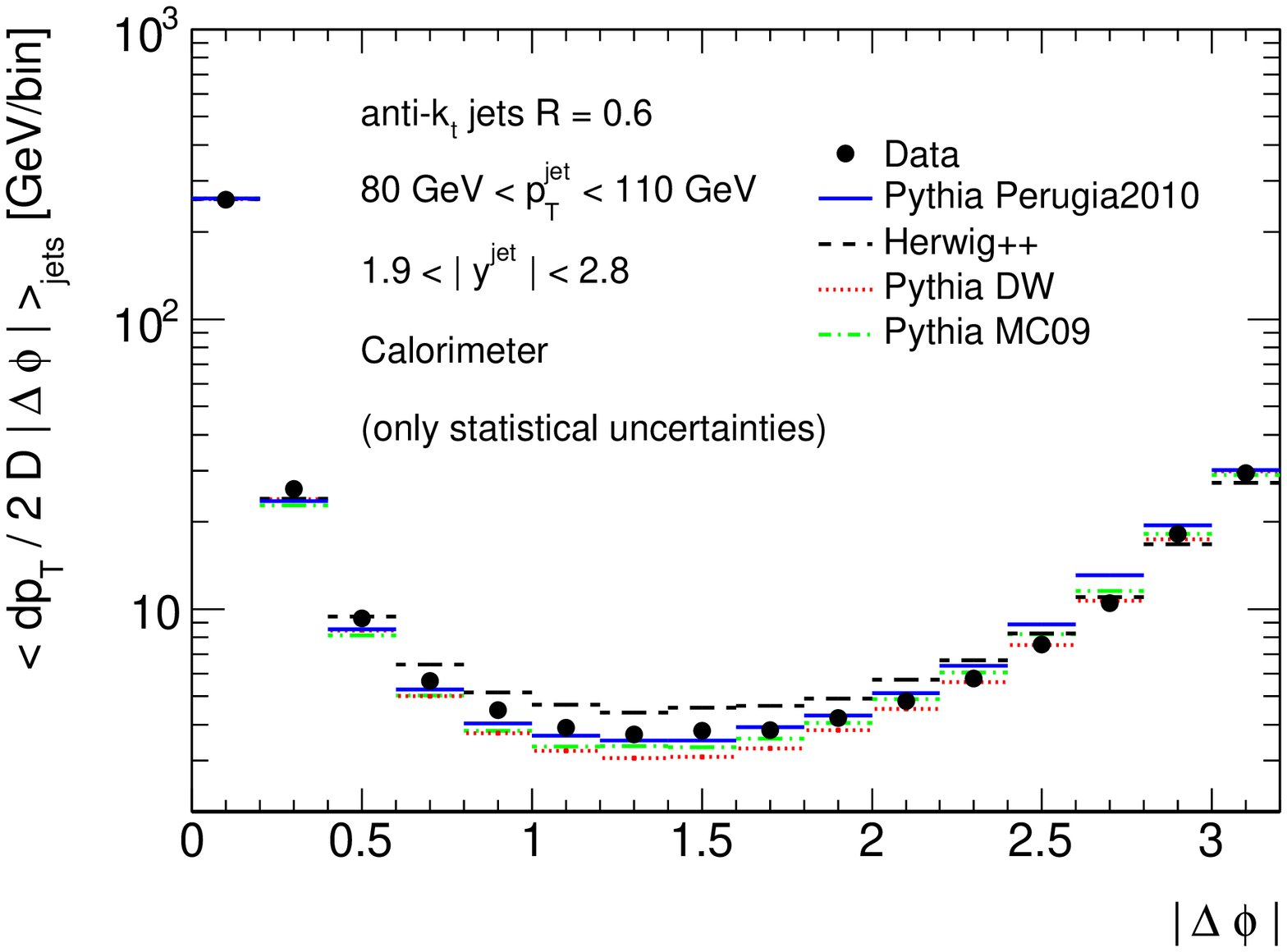}
}
\mbox{
\includegraphics[width=0.495\textwidth]{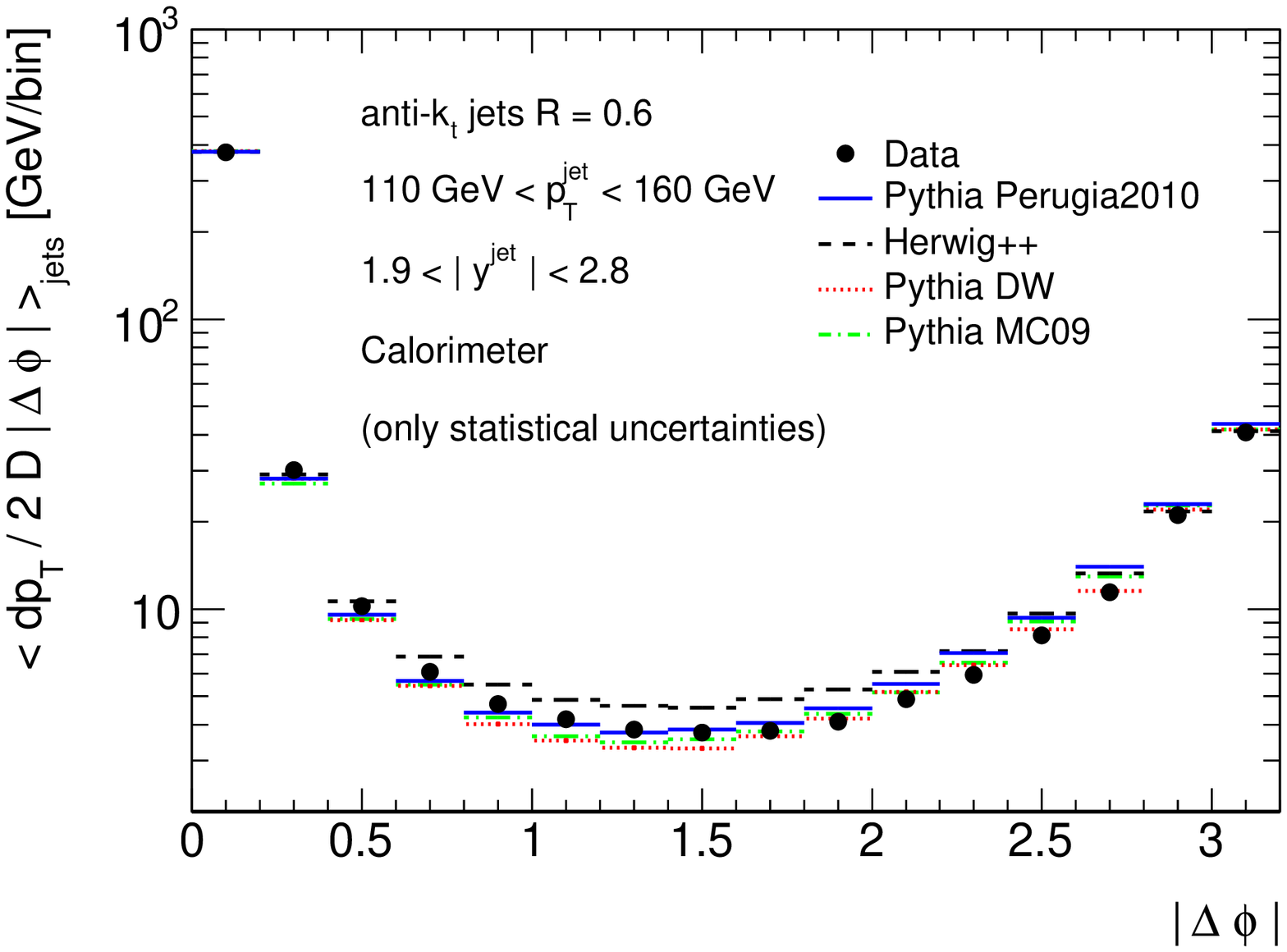}
\includegraphics[width=0.495\textwidth]{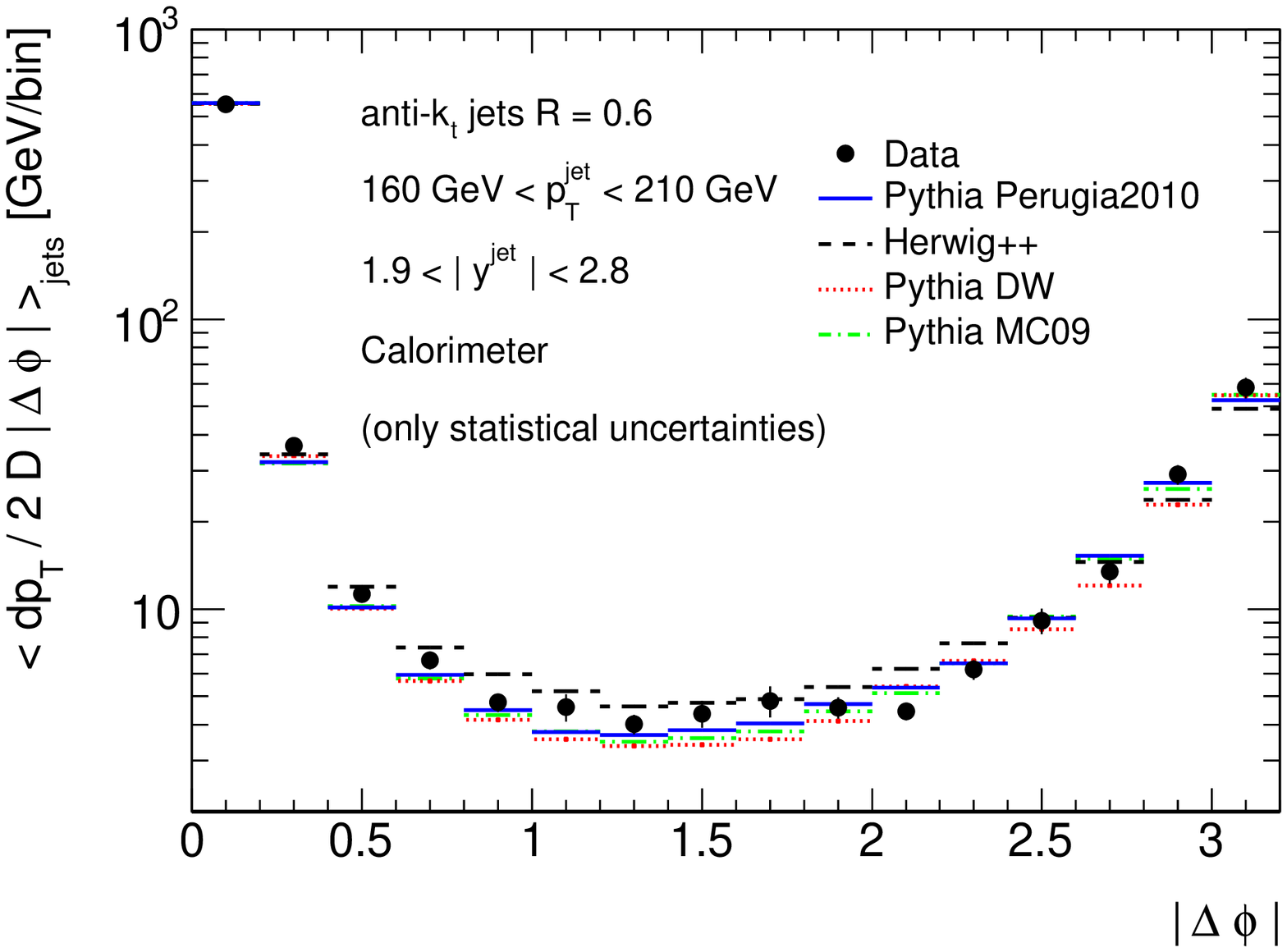}
}
\end{center}
\vspace{-0.7 cm}
\caption{\small
Energy flow using calorimeter clusters as a function $|\Delta\phi|$ with respect to the jet direction
for jets with $1.9 < |\rapjet| < 2.8$ and $30 \ {\rm GeV} < \ptjet < 210  \ {\rm GeV}$.
}
\label{fig_eflow_calo2}
\end{figure}

Figures~\ref{fig_eflow_trk} and~\ref{fig_eflow_calo1} show the measured energy flow using tracks and
calorimeter clusters respectively, for jets with $|\rapjet| < 1.9$ and $30 \ {\rm GeV} < \ptjet < 210  \ {\rm GeV}$.
Figure~\ref{fig_eflow_calo2} extend in rapidity the measurement using clusters up to $|\rapjet| = 2.8$.
As expected, the measured energy flow increases due to the presence of a second jet as $\Delta\phi$ tends to $\pi$.
The measurements are compared to different MC predictions. PYTHIA-PERUGIA20101 gives a good description of the data.
Both PYTHIA-MC09 and PYTHIA-DW tends to underestimate the activity in the region away from the jet axis, dominated by soft
hadronic activity, whereas HERWIG++ tend to overestimate it, specially in the forward region.

\clearpage

\clearpage

\begin{thebibliography}{99}
\bibitem{tab} A complete set of tables for differential and integrated measurements
as a function of $\ptjet$ and $|\rapjet|$
are available at the Durham HepData repository
\end{thebibliography}
\end{document}